%% file: thesis_final.tex
\documentclass[12pt,oneside,letterpaper]{book}

\usepackage[T1]{fontenc}
\usepackage[utf8]{inputenc}
\usepackage[english]{babel}
\usepackage{amsmath}
\usepackage{amssymb}
\usepackage{braket}
\usepackage{mathtools}
\usepackage{dsfont}
\usepackage{subcaption}
\usepackage{booktabs}
\usepackage{yfonts}
\usepackage{setspace}
\usepackage{cite}
\usepackage{verbatim}
\usepackage{fancyhdr}

\usepackage{amstext} 
\usepackage{array}   
\newcolumntype{C}{>{$}c<{$}} 

\usepackage{geometry}
\usepackage[usenames,dvipsnames]{xcolor}
\usepackage{hyperref}
\hypersetup{
  colorlinks,
  citecolor=Blue,
  linkcolor=Blue,
  urlcolor=Blue}

\clearpage

\def\be{\begin{equation}}
\def\ee{\end{equation}}
\def\rme{{\rm e}}
\newcommand{\non}{\nonumber}
\newcommand{\diff}{\mathrm{d}}
\newcommand{\ii}{\mathrm{i}} 

\def\Oa{\mathcal{O}(\alpha)}

\def\aa{\mathtt{a}}
\def\bb{\mathsf{b}}
\def\cc{\mathtt{c}}
\newcommand{\ext}[1]{\mathsf{#1}}

\def\poly{{\Psi}} 
\def\crho{{c}}
\def\mm{{\mathsf{m}}}
\def\kk{{\mathsf{k}}}
\def\mn{{\cal M}_N}
\def\ms{{\cal M}_S}

\def\pp{{p}}
\def\nn{{n}}
\def\ww{{w}}
\def\BR{I_1}
\newcommand{\RRe}{\mathrm{Re}}
\newcommand{\IIm}{\mathrm{Im}}
\newcommand{\dd}{\delta}

\begin{document}


\frontmatter
\begin{titlepage}
\vspace{5mm}
\begin{figure}[hbtp]
\centering
\includegraphics[scale=.5]{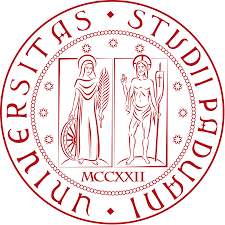}
\end{figure}
\vspace{3mm}
\begin{center}
{\LARGE{\textsc{\bf UNIVERSIT\`A DEGLI STUDI DI PADOVA}}\\}
\vspace{5mm}
{\Large{\bf Dipartimento di Fisica e Astronomia ``Galileo Galilei''}} \\
\vspace{5mm}
{\Large{\textsc{\bf PhD Course: PHYSICS}}}\\
\vspace{20mm}
{\Large{\textsc{\bf Doctoral Thesis}}}\\
\vspace{10mm}
\begin{spacing}{3}
{\Large\textbf{Black holes from the Gravitational Path Integral}}
\\
\vspace{-0.3cm}
{\large \textbf{Supersymmetric indices and precision holography}}
\\
\end{spacing}
\vspace{8mm}
\end{center}

\begin{spacing}{1.5}
\begin{tabular}{ l l}
{\bf Coordinator} & {\bf Prof. Giulio Monaco}\\
{\bf Supervisor} & {\large{\bf Dr. Davide Cassani }} \\
{\bf Co-supervisor} & {\bf Prof. Gianguido Dall'Agata}\\
\end{tabular}
\end{spacing}
\begin{tabular}{ l l}
\hspace{9cm}
{\bf Candidate} & {\large{\bf Enrico Turetta}} \\
\end{tabular}

\vspace{1cm}

\begin{center}
{\large{\bf XXXVIII Cycle}}
\end{center}
\end{titlepage}
\clearpage{\pagestyle{empty}\cleardoublepage}

\begin{titlepage}

\subsection*{Abstract}
%
The counting of microstates of certain supersymmetric black holes with anti-de Sitter or flat asymptotics is obtained by computing a supersymmetric index in a weakly coupled string theory or a dual superconformal field theory. These indices are protected observables, whose exact value can be reliably extrapolated from weak to strong coupling, where the gravitational description applies. 
In this Thesis, we review recent progress in formulating such protected observables directly within the gravitational theory, via the Euclidean path integral. It has been recently understood that, under suitable boundary conditions, the latter computes a gravitational index, providing a reliable counterpart to the microscopic count. We study the gravitational index in the semiclassical limit, where it reduces to a sum over complex Euclidean saddles weighted by their on-shell action. These saddles are supersymmetric but ``non-extremal'', and arise in both anti-de Sitter and flat spaces. 
In the holographic setting, we investigate four-derivative corrections to the thermodynamics of AdS$_5$ black holes. Using off-shell superconformal methods, we construct the corrected action of five-dimensional gauged supergravity, both with and without vector multiplet couplings, thereby providing an effective model that reproduces the ’t Hooft anomalies of generic holographic superconformal field theories, including the first subleading terms in the large-$N$ expansion. We then evaluate the corrected on-shell action of supersymmetric AdS$_5$ black holes and find exact agreement with a Cardy-like limit of the superconformal index of the dual conformal field theory. Crucially, the two-derivative solution is enough to perform this computation. From a Legendre transform of the action we obtain the corrected microcanonical entropy of supersymmetric and extremal black holes, and, in the minimal theory, we independently confirm this result by applying Wald’s formula to the corrected near-horizon geometry.
We then turn to the gravitational index with asymptotically flat boundary conditions. We uncover a broad family of saddles with U$(1)^3$ symmetry and present a general classification based on their rod structure, which characterizes their topology. These solutions may feature multiple horizons or three-dimensional bubbles with $S^3$, $S^2 \times S^1$, or lens space topology, and allowing for conical singularities yields further geometries, involving spindles and branched spheres. We focus on configurations with either a single horizon, or a horizon accompanied by an exterior bubble. We determine their on-shell actions and study their thermodynamics. For the simpler geometries, the on-shell action is computed using an odd-dimensional version of equivariant localization. These saddles interpolate, by tuning the inverse temperature $\beta$ and performing suitable analytic continuations, between supersymmetric extremal black holes (spherical black holes, black rings and black lenses) and horizonless bubbling geometries. In both limits their on-shell action, which is independent of $\beta$, remains finite and well-defined.

\end{titlepage}

\begin{titlepage}
%
%
%



\subsection*{Based on the contributions of the author during the PhD}

\vspace{1cm}

\subsubsection*{List of Publications}
\begin{enumerate}
\item[\cite{Cassani:2023vsa}] D. Cassani, A. Ruip\'erez and E. Turetta, \textit{Boundary terms and conserved charges in higher-derivative gauged supergravity}, JHEP \textbf{06} (2023) 203, 
\\
e-Print:  \href{https://arxiv.org/abs/2304.06101}{\textbf{[hep-th/2304.06101]}},
DOI:  \href{https://doi.org/10.1007/JHEP06(2023)203}{10.1007/JHEP06(2023)203}
\item[\cite{Cassani:2024tvk}] D. Cassani, A. Ruip\'erez and E. Turetta, \textit{Higher-derivative corrections to flavoured BPS black hole thermodynamics and holography}, JHEP \textbf{05} (2024) 276, 
\\
e-Print: \href{https://arxiv.org/abs/2403.02410}{\textbf{[hep-th/2403.02410]}}, 
DOI: \href{https://doi.org/10.1007/JHEP05(2024)276}{10.1007/JHEP05(2024)276}
\item[\cite{Cassani:2024kjn}] D. Cassani, A. Ruip\'erez and E. Turetta, \textit{Localization of the 5D supergravity action and Euclidean saddles for the black hole index}, JHEP \textbf{12} (2024) 086, 
\\
e-Print:  \href{https://arxiv.org/abs/2409.01332}{\textbf{[hep-th/2409.01332]}}, 
DOI: \href{https://doi.org/10.1007/JHEP12(2024)086}{10.1007/JHEP12(2024)086}
\item[\cite{Cassani:2025iix}] D. Cassani, A. Ruip\'erez and E. Turetta, \textit{Bubbling saddles of the gravitational index}, SciPost Phys. \textbf{19} (2025) 134,
\\
e-Print:  \href{https://arxiv.org/abs/2507.12650}{\textbf{[hep-th/2507.12650]}}, 
DOI: \href{https://scipost.org/10.21468/SciPostPhys.19.5.134}{10.21468/SciPostPhys.19.5.134}
\end{enumerate}

\vspace{1cm}

\subsubsection*{Contributions submitted to the arXiv prior to the start of the PhD}

\begin{enumerate}
\item[\cite{Cassani:2022lrk}] D. Cassani, A. Ruip\'erez and E. Turetta, \textit{Corrections to AdS$_5$ black hole thermodynamics from higher-derivative supergravity}, JHEP \textbf{11} (2022) 059, 
\\
e-Print:  \href{https://arxiv.org/abs/2208.01007}{\textbf{[hep-th/2208.01007]}}, 
DOI: \href{https://doi.org/10.1007/JHEP11(2022)059}{10.1007/JHEP11(2022)059}

\end{enumerate}

\vspace{1cm}

\subsubsection*{Preprints submitted to the arXiv after the completion of the first version of this Thesis}

\begin{enumerate}

\item[\cite{Massai:2025nci}] S. Massai and E. Turetta, \textit{Three-charge black holes from the worldsheet},
\\
e-Print:  \href{https://arxiv.org/abs/2511.02499}{\textbf{[hep-th/2511.02499]}}
\end{enumerate}

\end{titlepage}

\tableofcontents

\begin{titlepage}

\vspace*{250px}

\vspace{250px}

\hspace{10cm} {\it dedicated to my grandmother.}
\end{titlepage}

\mainmatter


\pagenumbering{arabic} 
\input{introduction}

\input{background}

\input{research_one}
\input{research_two}

\input{research_three}

\input{research_four}

\input{research_five}

\input{conclusion}

\begin{appendix}
\input{appendix_A}

\end{appendix}

\bibliographystyle{JHEP}
\bibliography{phd_references}

\end{document}

%% file: introduction.tex
\part{Introduction}

\chapter{A brief history of black holes and supersymmetric indices}

\noindent
The theory of General Relativity, since its inception in 1915, has withstood a century of experimental observations, establishing itself as our most successful description of classical gravitation: matter tells the spacetime how to curve, and spacetime tells matter how to move. Yet, our understanding of gravitational interactions is incomplete until it fully accounts for its most enigmatic prediction: the existence of {\it black holes}. These are classical solutions of Einstein's field equations characterized by a curvature singularity shielded behind an event horizon. After crossing the horizon 
nothing, not even light, can escape the gravitational pull and avoid falling into the singularity. Sufficiently close to the singularity, the spacetime becomes so strongly curved that theoretical predictions from general relativity break down, leaving the fate of the infalling matter unresolved. 

The first black hole solution, found by \emph{Schwarzschild} in 1916, revealed the simplest, spherically symmetric black hole. Soon after, Reissner and Nordström extended this to include electric charge, while the rotating generalization, discovered by Kerr, had to wait until 1963. For an introduction to general relativity and black hole solutions see the textbook~\cite{Wald:1984rg} and the lecture notes~\cite{Townsend:1997ku}. Although these objects were long regarded as mathematical curiosities, a series of landmark observations has confirmed their physical reality. For instance, the first black hole binary system to be established via dynamical observations was Cygnus X-1, discovered at the dawn of X-ray astronomy in the mid-1960s; while more recently the direct detection of gravitational waves by the LIGO/Virgo Collaboration (2015) provided evidence of black hole mergers~\cite{LIGOScientific:2016aoc}. Finally, the Event Horizon Telescope achieved the first image of a black hole shadow in M87 (2019)~\cite{EventHorizonTelescope:2019dse} and Sgr$\text A$ (2022)~\cite{EventHorizonTelescope:2022wkp}. Together, these observations not only validate the (astro)physical reality of black holes but also motivate the demand for a framework (within or beyond general relativity) that can consistently account for their interior structure. 

\medskip

The breakthrough in our theoretical understanding of black holes occurred in the 1970s, when it was realized that, upon coupling general relativity semiclassically to a quantum field theory, black holes behave as genuine thermodynamical systems, endowed with a temperature, an entropy, and emitting thermal radiation~\cite{Bekenstein:1972tm,Bardeen:1973gs,Hawking:1975vcx} (see~\cite{Witten:2024upt} for a recent review). The most celebrated illustration of this thermodynamic behavior is a simple \emph{gedankenexperiment} proposed by Bekenstein: imagine tossing a cup of coffee into a black hole along a perfectly radial trajectory (so as not to produce any angular momentum). Once the coffee crosses the horizon, the outside universe loses precisely the energy and entropy carried by the cup. The black hole subsequently evolves to a new equilibrium configuration with higher mass, preserving the total energy, but it appears that the remaining entropy of the accessible universe has decreased, in violation of the second law of thermodynamics. 

Bekenstein’s resolution was to assign to the black hole itself an intrinsic entropy. Recalling that, in all physical processes, the area of the event horizon never decreases (that is the area theorem of black hole mechanics), he conjectured that this entropy must be proportional to the horizon area: with this assumption the entropy increase of the black hole after absorbing a cup of (sufficiently hot) coffee more than compensates for the loss of the entropy outside. Hawking’s later discovery that the quantum process of particle production at the event horizon causes black holes to emit thermal radiation at a characteristic temperature provided concrete proof of Bekenstein’s conjecture. These developments culminated in the celebrated Bekenstein-Hawking area formula:
\begin{equation}
{\cal S}_{\rm BH} = \frac{k_B c^3}{4\hbar G_N}{\cal A}_H\,.
\end{equation}
This formula combines elements with different origins: the Boltzmann constant $k_B$ originates from statistical physics, the speed of light $c$ from special relativity, Newton's constant $G_N$ from the theory of gravitation, the area ${\cal A}_H$ refers to some geometrical properties, and finally Planck's constant $\hbar$ brings quantum mechanics into the game. It became then clear by the end of the 1970s that to make sense of black holes we would need to combine elements from different branches of physics, including \emph{quantum mechanics}.

As for any statistical system, thermodynamics emerges as a macroscopic and coarse-grained description of an ensemble of microscopic states (\emph{microstates}), just like the entropy of a gas arises from the fact that it exists a large number of translational, vibrational and rotational degrees of freedom of its molecules. This is quantified by Boltzmann equation: 
\begin{equation}
{\cal S}= k_B \log d,
\end{equation}
relating a macroscopic entropy ${\cal S}$ to the logarithm of the degeneracy $d$ of underlying microstates. If black holes truly behave as an ordinary thermodynamic system, they must correspond to a vast ensemble of quantum microstates, invisible to general relativity, whose degeneracy reproduces the Bekenstein-Hawking entropy via Boltzmann’s formula. Moreover, the process of Hawking’s radiation suggests that black holes are intrinsically quantum objects, indicating that a quantum‑mechanical account of their microstates, bringing together quantum mechanics and general relativity into a theory of \emph{quantum gravity}, is required. 
Such a theory of quantum gravity must be capable of identifying, enumerating and describing  the quantum states that underlie the macroscopic black hole solutions of general relativity.

As in any ordinary quantum system, we expect that the thermodynamic and statistical properties of black holes are encoded in an appropriate Euclidean functional integral that computes a partition function. This idea has been introduced by Gibbons and Hawking in 1976~\cite{Gibbons:1976ue}, proposing that black holes appear as saddle points of a \emph{gravitational path integral} of the form
\begin{equation}\label{eq:grav_path_int}
Z_{\rm grav} = \int {\cal D}g_{\mu\nu}\,.\,.\,. \,{\rm e}^{-I[g,...]}\,,
\end{equation}
where $g_{\mu\nu}$ denotes the spacetime metric, $I[g,...]$ is the Euclidean action, and we have suppressed any additional quantum fields appearing in the theory, that would likewise be integrated over. This construction plays a crucial role in the developments reviewed in this Thesis; we will return to its details later in this Introduction. For the moment it suffices to know that the path integral in general relativity, when treated as an ordinary quantum field theory, suffers from ultraviolet (UV) divergences that cannot be absorbed into a finite number of independent couplings. In other words, general relativity is non‑renormalizable and thus unsuitable as a fine-grained, high‑energy description. 
This failure suggests that a consistent theory of quantum gravity would involve extra degrees of freedom going beyond general relativity, that must play a role in the full path integral. Nevertheless, as we shall see below, in certain controlled regimes one can still make sense of it and extract non‑trivial predictions for the quantum description of black holes.

\medskip

In this section, we argued for the necessity of a theory of quantum gravity to account for black hole microstates, motivated by Bekenstein’s thought experiment. Before moving to the next sections, we introduce a final guiding principle for a quantum description of gravity: the \emph{holographic principle}, formulated by ’t Hooft~\cite{tHooft:1993dmi} and Susskind~\cite{Susskind:1994vu} in the early 1990s. 

In an ordinary quantum field theory at finite temperature, the degeneracy of available states in a finite region of spacetime grows exponentially with the volume (an example being the entropy of a gas confined to a room). However, in a gravitational context, this cannot hold. Consider a region of space with fixed volume, filled with matter carrying energy just below the threshold for black hole formation (${\rm Energy}< {\rm Vol}^{1/3}$, up to factors of order one in natural units $k_B = c = \hbar =1$). Standard thermodynamic reasoning would assign to this configuration an entropy scaling with the volume, which would exceed the Bekenstein-Hawking entropy of a black hole occupying the same region. If we were to increase the energy slightly by throwing in matter and form a black hole, the resulting entropy would decrease, thereby violating the second law of thermodynamics. 
The holographic principle offers a resolution to this paradox: in a consistent theory of quantum gravity, all physical phenomena within a given spatial region should be described by a quantum theory residing at the boundary of that region. The interior volume, and the spacetime metric, should not be relevant to avoid overcounting of physical states. 
This idea naively resonates with the Gibbons-Hawking prescription for computing the gravitational path integral, which involves fixing boundary conditions at the asymptotic spacetime region and integrating over all bulk metrics (and topologies) that smoothly fill the interior. At the same time, it offers a compelling perspective on how quantum gravity might account for black hole microstates: via a holographically dual description.

Over the last decades, \emph{string theory} has emerged as the most promising candidate for a theory of quantum gravity. String theory not only accounts for black hole entropy and microstates (as demonstrated in the foundational works in mid-1990s by Sen~\cite{Sen:1995in} and Strominger-Vafa~\cite{Strominger:1996sh}) but also provides an explicit realization of the holographic principle, at least in certain regimes, through the AdS/CFT correspondence, conjectured by Maldacena in 1997~\cite{Maldacena:1997re}. 
In what follows, we will review in more detail how string theory and its holographic formulation offer a framework to address the long-standing problem of reproducing and counting black hole microstates. We will also outline the necessary background and highlight the open questions that motivate the recent developments to which this Thesis contributes.


\section{Black holes in string theory and microstate counting}

\medskip

String theory originated in the late 1960s as a model of the strong nuclear force, with the central idea that one dimensional objects, strings, propagate through spacetime spanning a two-dimensional surface known as the \emph{worldsheet}, and interact by splitting and joining. 
In the mid-1970s, Scherk and Schwarz recognized that certain vibrational modes of these strings yield a massless spin-two field, naturally interpreted as the spacetime metric; because strings have finite extent, their interactions remain well-behaved at arbitrarily high energies, suggesting that string theory might provide a UV-complete quantum theory of gravity.
During the 1980s, Green and Schwarz demonstrated anomaly cancellation in the ten‑dimensional supersymmetric version of string theory (superstring theory), establishing its mathematical consistency, and the following developments of the physics of compactifications linked the ten‑dimensional framework to four‑dimensional models. Supersymmetry, a spacetime symmetry exchanging bosons and fermions, played a central role in these developments. We will return to the concept of supersymmetry later in this Introduction, as it is also crucial for achieving a consistent counting of black hole microstates. 
Finally, in the mid‑1990s Polchinski’s identification of D‑branes as dynamical, non-perturbative objects, and Witten’s unification of the five available formulations of string theory into an eleven‑dimensional \emph{M‑theory} via dualities, set the stage for the first successful black hole microstate counts, as well as for Maldacena's AdS/CFT correspondence. Indeed, one of the major achievements of string theory has been the microscopic explanation for the Bekenstein-Hawking entropy for a class of \emph{supersymmetric black holes}, by identifying it with the statistical entropy of bound states of strings and branes, that are the fundamental dynamical objects in the theory~\cite{Sen:1995in,Strominger:1996sh}. These and the related developments established string theory as a leading candidate for a unified theory of quantum gravity.
The goal of this section is not to give a comprehensive review of string theory, but rather to highlight the key concepts underlying these results and their relevance to this Thesis. References for this part include~\cite{Polchinski:1998rq,Polchinski:1998rr,Becker:2006dvp} (see also~\cite{Murthy:2023mbc} for a discussion focused on black holes in string theory).

\medskip

In string theory there are two types of strings: \emph{closed} strings, which form continuous loops, and \emph{open strings}, that must terminate on higher-dimensional extended membranes, known as \emph{Dp-branes} (here, $p$ indicates the number of spatial dimensions the brane spans).

\begin{figure}[!htb]
	\centering
	\includegraphics[width=0.3\textwidth]{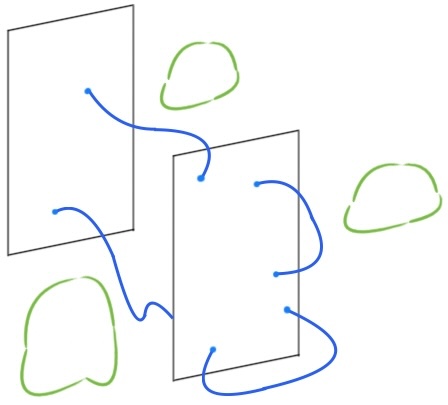}
	\caption{\it  Closed strings (in green) form closed loops, while open strings (blue) are forced to end on D-branes.}
	\label{fig:string_brane}
\end{figure} 
The specific spectrum of allowed Dp-branes depends on the formulation of string theory under consideration. For concreteness, we will consider Type IIB string theory, where only odd-dimensional branes (D$1$, D$3$, D$5$, etc.) appear as stable objects. 

String theory naturally incorporates two fundamental length scales: the Planck scale $\ell_{\rm P}$, setting the strength of gravitational interactions\footnote{The $D$-dimensional gravitational constant $G_N^{(D)}$ is related to the Planck scale as $G_N^{(D)} = \left( 2\pi \right) ^{D-3} \ell_{\rm P}^{D-2}$.}, and the \emph{string length} $\ell_{\rm s}$. These are combined into the dimensionless string coupling $g_{\rm s}$, 
\begin{equation}
g_{\rm s} = \left(\frac{\ell_{\rm P}}{\ell_{\rm s}}\right)^4\,,
\end{equation}
that controls the perturbative expansion of string theory. In the weak coupling regime, $g_{\rm s} \ll 1$ (or, equivalently, $\ell_{\rm s} \gg \ell_{\rm P}$) string loop corrections are suppressed, and a classical (tree-level) description for closed strings suffices. On the other hand, the perturbative expansion for the open string sector is really an expansion in $g_{\rm s} N$, where $N$ is the number of D-branes. 

The spectrum of string excitations exhibit a tower of quantized vibrational modes. Massive modes have mass quantized in units of the string length $M_{\rm s} \sim \ell_{\rm s}^{-1}$, while massless modes dominate the low-energy physics. In particular, if we just consider processes with an energy scale set by ${\cal E} \ll \ell_{\rm s}^{-1}$, such that for these processes strings can be effectively treated like point-particles and the worldsheet collapses to a worldline, massive modes behave as extremely heavy states and decouple. For open strings, the massless bosonic spectrum comprises a set of gauge fields and scalars living on the worldbrane, and their low-energy dynamics give rise to non‑gravitational, (supersymmetric) gauge theories, providing a natural mechanism for embedding gauge interactions within the string framework. On the other hand, the massless spectrum of closed strings includes a graviton $g_{\mu\nu}$, a two-form $B_{\mu\nu}$ and a dilaton $\Phi$. These fields constitute the bosonic sector of ten‑dimensional supergravity, which thus emerges as the \emph{low‑energy}, \emph{weakly coupled} limit of string theory.

A supergravity theory is a supersymmetric version of general relativity. The first such example has been first formulated in 1976 (thanks to the work of Freedman, Ferrara, van Nieuwenhuizen and Deser, Zumino) as a \emph{gauge theory of supersymmetry}, meaning it is invariant under local supersymmetry transformations. As in any gauge theory, gauging a symmetry requires introducing a gauge field whose infinitesimal transformation includes the spacetime derivative of a local symmetry parameter:
\begin{equation}
\delta \psi_\mu = \partial_\mu \epsilon + .\,.\,.\,\,,
\end{equation}
where the ellipsis denotes additional terms not relevant to this discussion. Since the supersymmetry parameter $\epsilon$ is a \emph{spinor}, the associated gauge field $\psi_\mu$ is a spin-3/2 (vector-spinor) field, known as the \emph{gravitino}. Its name reflects the fact that local supersymmetry necessarily requires the presence of a graviton $g_{\mu\nu}$, so any locally supersymmetric theory is a theory of (super)gravity. Beyond the universal sector given by the graviton and the gravitino, there can be additional fields, depending on the spacetime dimensions and the number of preserved supersymmetry. For an introduction to supergravity see~\cite{DallAgata:2021uvl,Freedman:2012zz}.
Supergravity theories naturally arise as infrared limits of string theory (for ${\cal E} \ll \ell_{\rm s}^{-1}$ and $g_{\rm s} \sim 0$) and form a crucial interface between string theory and low-energy physics. For example, they retain information about certain non-perturbative features of the full theory, such as D$p$-branes, which also appear as classical solitonic solutions to the supergravity field equations. We will use this fact later, as the supergravity perspective has often provided valuable insights into their dynamics. Importantly, when a supergravity theory arises from string theory, its solutions, such as black holes, can, at least in principle, be uplifted to full string backgrounds, providing an explicit embedding into a UV-complete framework.
There are two main classes of supergravity theories, both of which will play a role in the discussion that follows: \emph{ungauged} and \emph{gauged} supergravity. 
In the class of gauged supergravity theories of interest to us, a certain global R-symmetry is promoted to a local (gauged) one, rendering the gravitini charged under this symmetry (but more general gaugings are possible). This gauging induces a non-zero scalar potential, typically sourcing a negative cosmological constant. As a result, gauged supergravity theories admit anti-de Sitter (AdS) vacua and are central to the gravitational side of the AdS/CFT correspondence. By contrast, ungauged supergravity theories admit asymptotically flat solutions, which are more directly connected to the geometry of our nearly-flat Universe.
For these reasons, along with others that will emerge later, related to their powerful underlying supersymmetric algebra, supergravity theories provide an ideal laboratory for exploring quantum properties of black holes~\cite{Cassani:2025sim}.

\paragraph{Three charge black hole in string theory.}

Black holes in string theory arise as bound states of intersecting strings and D-branes wrapped on compact internal spaces. A D$p$-brane couples electrically to a RR $(p+1)$-form gauge field, then a stack of $N_p$ D$p$-branes carries $N_p$ units of RR charge. Stable D-branes in Type II string theory are half-BPS (Bogomol'nyi-Prasad-Sommerfield) objects, meaning that they preserve half of the total supersymmetries ($16$ out of $32$). The worldvolume theory on the brane is weakly coupled when $g_{\rm s} N_p \ll 1$. 

A stack of half-BPS D-branes can be described as an extremal black brane solution in classical supergravity. These solutions can be found in Type II ten-dimensional supergravity, where the appropriate $(p+1)$-form gauge potential supports the brane's electric charge. This is quantized in suitable units and reproduces the RR charge carried by the D-brane. These black brane solutions possess an event horizon that shields a singularity and saturate a certain BPS bound ensuring that half of the supersymmetries are preserved. This condition translates into a linear relation between brane's mass $E$ and electric charge
\begin{equation}
E = \frac{N_p}{\left( 2\pi\right)^p g_{\rm s} \ell_{\rm s}^{p+1}}\,.
\end{equation}
The classical supergravity description can be trusted as long as the curvature remains small compared to the string scale. 

The canonical example of system of branes describing a black hole is the \emph{D1-D5-P system} in Type IIB string theory compactified on $S^1 \times T^4$. This construction involves three types of objects: $N_5$ D$5$-branes wrapped on the full compact space $S^1 \times T^4$, $N_1$ D$1$-branes around $S^1$ and uniformly \emph{smeared} over $T^4$, together with $N_P$ units of momentum along $S^1$. This setup was considered in~\cite{Strominger:1996sh} and led to the first successful microscopic derivation of the Bekenstein-Hawking entropy formula from string theory.
\begin{table}[h!]
    \centering
    \begin{tabular}{|c|cccccccccc|}
    \hline
         \;& \;  $0$\;& \; $1$\;& \; $2$\;& \; $3$ \;& \;$4$ \;& \; $5$ \; & \; $6$ \; & \; $7$ \; & \; $8$ \; & \; $9$
         \\
         \hline 
         D5  \;& \;  $-$\;& \; $\cdot$\;& \; $\cdot$\;& \; $\cdot$ \;& \;$\cdot$ \;& \; $-$ \; & \; $-$\; & \; $-$ \; & \; $-$ \; & \; $-$
         \\
         D1  \;& \;  $-$\;& \; $\cdot$\;& \; $\cdot$\;& \; $\cdot$ \;& \;$\cdot$ \;& \; $-$ \; & \; $\sim$\; & \; $\sim$ \; & \; $\sim$ \; & \; $\sim$
         \\
         P  \;& \;  $-$\;& \; $\cdot$\;& \; $\cdot$\;& \; $\cdot$ \;& \;$\cdot$ \;& \; $-$ \; & \; $\sim$\; & \; $\sim$ \; & \; $\sim$ \; & \; $\sim$
         \\
         \hline
\end{tabular}
    \caption{\it The D1-D5-P system. The string background is decomposed as follows: the extermal spacetime dimensions in the compactification are denoted by $01234$, an $S^1$ lies along direction $5$, and $T^4$ extends along directions $6789$. The symbol "$-$" denotes a direction along which a brane extends, while "$\sim$" denotes smearing. All constituents appear as pointlike sources from the perspective of the external spatial directions.}
    \label{tab:D1-D5-P}
\end{table}

Each constituents breaks half of the supersymmetry, then the combined configuration preserves 4 out of the 32 supersymmetries, i.e. it is a 1/8-BPS system. When the size of the compactified directions is sufficiently small in string units, while the black brane's event horizon is large, the configuration can be described by a five-dimensional (ungauged) supergravity, where it corresponds to a supersymmetric black hole carrying three independent electric charges~\cite{Callan:1996dv}. Its entropy can be expressed as:
\begin{equation}
\label{eq:D1D5P_entropy}
{\cal S}_{\rm BH} =  2\pi \sqrt{N_5\,N_1\,N_P}\,.
\end{equation}
The supergravity approximation is valid when the curvature is small, which in turn requires all charges to be large. For instance, one must have
\begin{equation}
N_5\, >\, g_{\rm s} N_5\, \gg \,1\,,
\end{equation}
and similar relations for $N_{1,P}$. However, since the effective coupling for open strings are actually $g_{\rm s} N_1$ and $g_{\rm s} N_5$, in this regime open strings turn out to be strongly coupled.

This behavior is a general feature of D-brane systems. Typically, such systems are characterized by a quantum number, $N$ (which may correspond to brane number, a quantized flux, or a combination thereof), such that:
\begin{itemize}
\item \emph{Macroscopic picture}: 

as long as $N > g_{\rm s} N \gg 1$ we are in a low-curvature regime, in which the classical supergravity description remains valid. The effective description admits a black hole carrying a charge proportional to $N$;
\item  \emph{Microscopic picture}: 

when $g_{\rm s} N \ll 1$, the supergravity description is not valid anymore, but the system can be represented as a bound state of fluctuating strings and branes labelled by $N$ in weakly-coupled string theory.
\end{itemize}

\begin{figure}[!htb]
	\centering
	\includegraphics[width=0.7\textwidth]{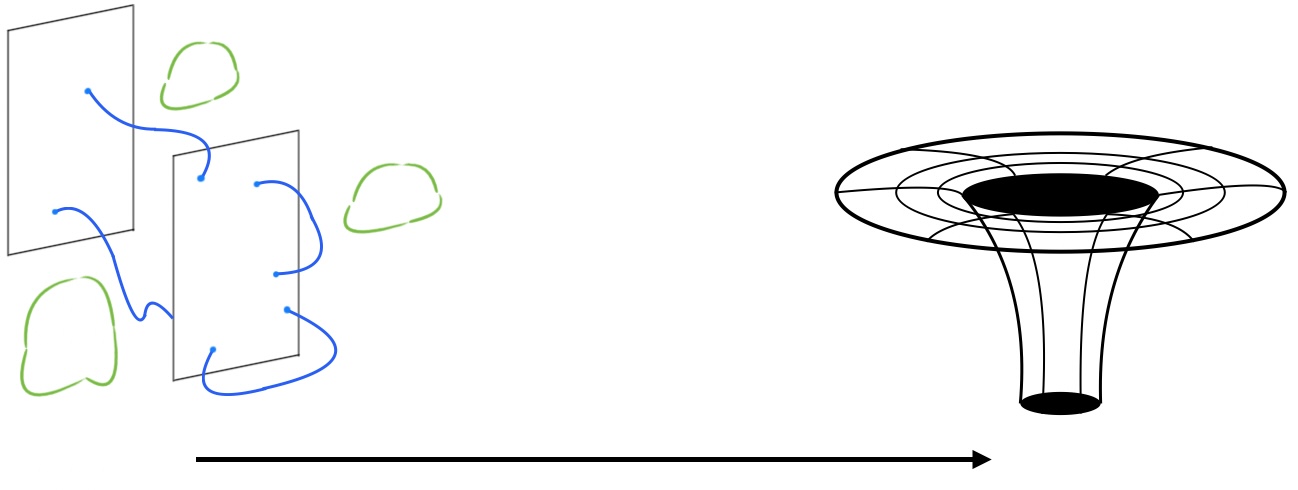}
	\caption{\it  The two pictures of a black hole in string theory: the horizontal axis represents the effective coupling $g_{\rm s} N$. When $g_{\rm s} N \ll 1$ (left) the effective description is that of fluctuating strings and branes, carrying the quantum number $N$, in a weakly coupled string theory; while for $g_{\rm s} N \gg 1$ (right) the interactions become strong and a black hole with charge $N$ is formed. The string theory is strongly coupled, but the supergravity description is reliable.}
	\label{fig:bh_string_theory}
\end{figure} 

The microscopic picture offers the right context for calculating the statistical entropy of the macroscopic black hole. However, for a given set of charges, the microscopic and macroscopic regimes are reliable for different values of the string coupling, so one might worry that the state count could differ as we tune $g_{\rm s}$. The breakthrough of~\cite{Sen:1995in,Strominger:1996sh} was to address the problem of microstate counting within the framework of \emph{supersymmetric compactifications} of string theory, which is the setup we reviewed so far. As we will review in detail in the next sections, this setting allows for the definition of certain \emph{protected} observables, that remain invariant as continuous couplings change their value. Among these we find the \emph{supersymmetric index}, which counts states preserving a fraction of the total supersymmetry, weighted so as to distinguish between bosonic and fermionic contributions. When performing this counting, it is important to identify the full ensemble of quantum states carrying the same charges and preserving the same supersymmetries as the macroscopic black hole.

Going back to the D1-D5-P system, the low-energy physics of the microscopic picture is captured by a two-dimensional conformal field theory (CFT) living on the common $S^1$ direction. A Cardy formula~\cite{Cardy:1986ie} can be applied to extract the asymptotic growth of supersymmetric ground states, i.e. a protected observable~\cite{Strominger:1996sh}:
\begin{equation}\label{eq:Strominger_Vafa_index}
\log d(N_1, N_5, N_P) = 2\pi \sqrt{N_1\,N_5\,N_P} + \,.\,.\,.\,,
\end{equation} 
with subleading corrections suppressed when the charges are taken to be large. Thanks to supersymmetry this results can be reliably extrapolated from weak to strong coupling, and compared to the supersymmetric Bekenstein-Hawking entropy \eqref{eq:D1D5P_entropy}, achieving a successful match via Boltzmann equation.  

Before concluding this section, let us note that \eqref{eq:Strominger_Vafa_index} is an asymptotic expression, valid in a large-charge limit. It then receives both subleading ${\cal O}(1/N)$ corrections and corrections that are logarithmic in the charges. Whenever such terms can be computed microscopically, they yield precise predictions for subleading corrections to the black hole entropy. 
It is important to emphasize that the Bekenstein-Hawking entropy itself is only an approximate formula, valid at leading order in a low-energy theory of gravity. Since supergravity arises as an effective low-energy approximation of a more fundamental theory, such as string theory, quantum gravity effects are expected to appear in the form of higher-derivative corrections to the two-derivative supergravity action.\footnote{In the presence of higher-derivative terms, black hole thermodynamics remains well-defined. The thermodynamic entropy is modified and must be computed using Wald’s formula~\cite{Wald:1993nt,Iyer:1994ys}, which generalizes the Bekenstein-Hawking area law. For extremal black holes, this framework is conveniently summarized by Sen’s entropy function formalism~\cite{Sen:2005wa,Sen:2007qy}.}
In string theory, these higher-derivative terms typically correspond to finite string length effects, as strings are extended rather than pointlike objects. Such corrections are controlled by powers of $\alpha' = \ell_{\rm s}^2$, and are known as $\alpha'$-corrections. Additionally, there are genuine quantum corrections which corresponds to loop effects in the gravitational theory (see~\cite{Mandal:2010cj} for a review).
When the \emph{quantum entropy} can be computed on the microscopic side, it becomes crucial to develop a corresponding macroscopic description. This allows for a \emph{precision} microstate counting, taking into account the full structure of quantum gravity effects. Higher-derivative corrections to the entropy, particularly in the context of holography, will play a central role in the developments presented in this Thesis.


\subsection{Supersymmetry and the Witten index}

Supersymmetry plays a crucial role in the context of microstate counting, as it allows for the definition of protected observables, quantities that remain invariant as continuous couplings are varied while preserving supersymmetry. For this reason, we will primarily focus in this Thesis on \emph{supersymmetric black holes}, which provide an ideal setting for elucidating the mechanisms of microstate counting.\footnote{Although supersymmetry does not appear to be realized at the energy scales currently accessible in our Universe, the insights gained from studying supersymmetric models are expected to be broadly applicable. The same fundamental physics should govern astrophysical black holes, but in the absence of supersymmetry, establishing precise analytic connections between macroscopic and microscopic descriptions becomes significantly more challenging.} 
In this section, we review the key features of the supersymmetric theories that are relevant for our purposes and discuss the basic properties of supersymmetric indices. The indices of interest typically take the form of refined versions of the \emph{Witten index}, originally introduced in~\cite{Witten:1982df} in the context of supersymmetric quantum mechanics. Remarkably, even in such a simple setting as one-dimensional quantum mechanics, the essential aspects of supersymmetry that guarantee the existence of protected observables are already manifest.

\medskip

The main ingredients we will need are a pair of complex conjugate fermionic (nilpotent) operators, ${\cal Q}$ and $\bar {\cal Q}$, known as \emph{supercharges}, along with a \emph{fermion number} operator $\rm F$, i.e. a $\mathbb Z_2$-valued operator that gives $0$($1$) when acting on bosons(fermions). The relevant (anti)commutation relations in the algebra take the form:
\begin{equation}
\label{eq:general_superalgebra}
\begin{aligned}
\left\{ {\cal Q},\,\bar {\cal Q}\right\} = H\,,\qquad \left[ {\cal Q},\,H\right] = \left[ \bar {\cal Q},\,H \right] =  0\,,
\\[1mm]
\left\{ {\cal Q},\,\left( - \right)^{\rm F} \right\} = 0\,.
\end{aligned}
\end{equation}
In the context of supersymmetric quantum mechanics, $H$ represents the Hamiltonian of the system. In more general theories, however, it may correspond to a suitable combination of bosonic symmetry generators. For concreteness and simplicity, we will take $H$ to be the Hamiltonian, using supersymmetric quantum mechanics as a toy model. In this setting, we consider the eigenstates of $H$, denoted as $\ket \psi$, with associated eigenvalues $E$, i.e., $H\ket \psi = E \ket \psi$.\footnote{To keep the presentation simple, we assume a discrete and gapped spectrum.} 
Below, we outline some consequences of the superalgebra \eqref{eq:general_superalgebra} (cf.~\cite{Razamat:2022gpm}): 
\begin{enumerate}
\item Each subset ${\cal H}_E$ of the total Hilbert space ${\cal H}$, for $E>0$, decomposes as ${\cal H}_E = {\cal H}_E^{\cal Q} \oplus {\cal H}_E^{\bar {\cal Q}}$, where ${\cal H}_E^{\cal Q}$ (${\cal H}_E^{\bar {\cal Q}}$) denotes the space of ${\cal Q}$ ($\bar {\cal Q}$)-exact states with energy $E$. Precisely, any state $\ket \psi \in {\cal H}_E$ can be written as a linear combination of a ${\cal Q}$-exact and a $\bar {\cal Q}$-exact state:
\begin{equation}
\ket \psi = \frac{\left\{ {\cal Q},\,\bar {\cal Q}\right\}}{E}\ket \psi = E^{-1}\Bigl[{\cal Q} \bar {\cal Q} \ket \psi + \bar {\cal Q}{\cal Q} \ket \psi \Bigr]\,.
\end{equation}
\item The supercharge ${\cal Q}$ defines a map between the subspaces 
\begin{equation}
{\cal Q} \,:\, {\cal H}_E^{\bar {\cal Q}} \rightarrow {\cal H}_E^{\cal Q}\,,
\end{equation}
with $\bar {\cal Q}$ acting as its inverse. This implies that states within each positive-energy sector ${\cal H}_E$ are \emph{paired} under the action of ${\cal Q}$. 
Furthermore, due to the second line of \eqref{eq:general_superalgebra}, paired states must have opposite statistics: if one is a \emph{bosonic} state $\ket b$, satisfying $\left( -\right)^{\rm F} \ket b = \ket b$, the other, $\ket f = {\cal Q} \ket b$, is a \emph{fermionic} state, such that $\left( -\right)^{\rm F} \ket f = -\ket f$. These paired states of opposite statistics are referred to as \emph{superpartners}, and together they form \emph{long representations} of the supersymmetry algebra.
\item Zero-energy states are special: they are annihilated by both supercharges
\begin{equation}
H\ket \psi = 0 \quad \implies \quad  {\cal Q} \ket \psi \,=\,\bar {\cal Q} \ket \psi\, =\, 0\,,
\end{equation}
and therefore are not required to be paired. Such states form \emph{short representations} of the superalgebra and are commonly referred to as \emph{BPS states}.
\end{enumerate}
In short, any theory realizing the superalgebra \eqref{eq:general_superalgebra} necessarily includes fermionic degrees of freedom, and its spectrum organizes into superpartner pairs. The only exception is the ground states, which are allowed to remain unpaired. This organization of the spectrum arises purely from the supersymmetry algebra and does not depend on the specific Lagrangian formulation of the theory. 

Suppose then the theory depends on a set of continuous couplings. Varying these couplings may deform the spectrum and shift energy levels, but cannot alter the pairing of states into superpartners, as this is dictated solely by the superalgebra. In particular, it may happen that a pair of superpartners transitions from a non-zero energy level to the ground state as the coupling is varied. However, the pairing remains intact, unless supersymmetry is spontaneously broken. Crucially, while the total number of BPS states (i.e., ground states) may change, the \emph{difference} between the number of bosonic and fermionic zero-energy states remains \emph{invariant}. 

This observation motivates the definition of an observable that captures only the unpaired BPS states, weighting bosonic and fermionic contributions with opposite signs. Such an object is the \emph{Witten index}~\cite{Witten:1982df}, defined as the trace over the full Hilbert space:
\begin{equation}\label{eq:Witten_index}
Z_{\rm W} = {\rm Tr} \left[ \left( - \right)^{\rm F} \,{\rm e}^{-\beta H} \right]\,.
\end{equation}
Here, the parameter $\beta$, with ${\rm Re} \beta >0$, serves as a regulator for the infinite sum, suppressing contributions from high-energy states to ensure convergence of the trace. Since all states with energy $E>0$ appear in boson-fermion pairs, their contributions to the trace cancel. As a result, only the unpaired states forming short representations contribute to the index:
\begin{equation}
Z_{\rm W} = {\rm Tr}_{{\cal H}_0}\left[ \left( - \right)^{\rm F} \right] = d_0^b - d_0^f\,,
\end{equation}
where $d_0^b$ and $d_0^f$ denote the number of bosonic and fermionic zero-energy states, respectively. Since the index receives contributions only from ground states, it is independent of $\beta$, as well as of any other continuous parameter in the theory. 
This invariance allows the index to be computed in a weakly coupled regime, where identifying BPS states is often more tractable, while ensuring that the result remains valid at strong coupling. As mentioned above, this is a main reason why indices of this type are valuable tools for counting black hole microstates. 
In more elaborate theories, especially in higher dimensions, the supersymmetry algebra typically involves multiple supercharges and has a more intricate structure. In such cases, BPS states are annihilated by at least one of the supercharges. More generally, they are classified according to the fraction of the total supercharges by which they are annihilated. Depending on the structure of the supersymmetry algebra, these states may carry non-zero energy. It is then possible to define refined supersymmetric indices, which capture contributions from specific classes of BPS states. These indices are still topological invariants, robust against continuous deformations of the theory, but their precise definition depends on the structure of the underlying supersymmetry algebra. A relevant example of such an index is given below. 

\paragraph{An index for asymptotically flat black holes.}

For asymptotically flat black holes, such as the D1-D5-P black hole, the relevant indices are \emph{helicity supertraces}. Since in five external spacetime dimensions the spatial rotation group is $SO(4) \cong SU(2)_+ \times SU(2)_-$, states can be labelled by two angular momenta, $J_{\pm}$, corresponding to the Cartan generator of each $SU(2)_\pm$ factor. The relevant index reads~\cite{Dabholkar:2010rm}: 
\begin{equation}\label{eq:BMPV_ent}
d\left(N_1,\,N_2,\,N_3,\,J_-\right) = {\rm Tr}\left[\left( - \right)^{\rm F} \left(2J_+\right)^n\right] = {\rm exp}\left[2\pi \sqrt{N_1\,N_5\,N_P - J_-^2} +\,.\,.\,.\right] \,,
\end{equation}
where $\left( - \right)^{\rm F} = {\rm e}^{2\pi \ii J_1}$ thanks to the spin-statistics theorem, and $n\in \mathbb Z$ is related to the number of fermionic zero modes that are present due to the broken supersymmetries. This index can be used to provide a microscopic account of the entropy of the supersymmetric rotating BMPV black hole~\cite{Breckenridge:1996is}, that generalizes the static ($J_-=0$) black hole considered in~\cite{Strominger:1996sh}. 
The insertion of powers of $2J_+$ in the trace (with $n=6$ for compactifications on $S^1 \times T^4$) is necessary to absorb the fermionic zero modes mentioned above. Without this insertion, the index would vanish. Once these modes are appropriately dealt with, the helicity supertrace effectively reduces to a standard Witten index. These zero modes can contribute when computing one-loop logarithmic corrections to the Bekenstein-Hawking entropy (see~\cite{Anupam:2023yns} for a recent discussion), and it is therefore important to treat them correctly in the full quantum description. However, in this Thesis we will not be concerned with such corrections, thus, for our purposes, there will be no distinction between helicity supertraces and Witten indices, and we will neglect the additional angular momentum insertions.\footnote{For a review of computations of such indices in compactifications of string theory with ${\cal N} = 4,8$ supersymmetry see~\cite{Mandal:2010cj}.} 

A central question we will explore later in this Thesis is how the index can be computed in the macroscopic picture, i.e. at strong string coupling. In particular, we aim to understand what is the gravitational definition of the index (for asymptotically flat black holes, but not only). Indeed, one might be concerned that the microscopic index, which counts states with a sign depending on their fermion number, should not be compared directly with the macroscopic entropy, which counts an absolute degeneracy of states. This raises the possibility that the agreement between the two could be accidental and may break down at subleading order.\footnote{For certain supersymmetric black holes, it has been shown that the microstates contributing to the entropy are purely bosonic. In such cases, the index with microcanonical boundary conditions coincides with the absolute degeneracy of states~\cite{Sen:2009vz,Dabholkar:2010rm,Iliesiu:2022kny}. 
}
However, recent developments starting with~\cite{Cabo-Bizet:2018ehj} have clarified that, by turning on certain holonomies compatible with supersymmetry, the Euclidean gravitational path integral does compute a supersymmetric index, specifically, a refined version of the Witten index, providing a reliable gravitational counterpart for the microscopic computation.


\section{Microstate counting for AdS black holes}

The holographic principle offers a natural framework for counting black hole microstates, describing gravitational configurations as ensemble of quantum states in a dual quantum field theory. A concrete realization is the \emph{AdS/CFT correspondence}, conjectured by Maldacena~\cite{Maldacena:1997re}, which proposes a non-perturbative equivalence between quantum gravity in anti-de Sitter (AdS) spacetimes and a strongly coupled conformal field theory (CFT) in one dimension less. This conjecture has passed numerous non-trivial tests and is now a cornerstone of modern theoretical physics,  although a rigorous proof is still lacking. 
The first holographic account of black hole entropy in terms of states of a dual CFT was given in~\cite{Strominger:1997eq} for three-dimensional black holes with AdS asymptotics~\cite{Banados:1992wn} (these arise locally as the near-horizon regions of many black holes in string theory, including the D1-D5-P black hole reviewed above). In this context, the powerful intuition underlying the AdS/CFT correspondence was already anticipated by the work of Brown and Henneaux~\cite{Brown:1986nw}, who showed that the asymptotic symmetry group of AdS$_3$ gravity is generated by (two copies of) a Virasoro algebra. Consequently, any quantum excitation of AdS$_3$ must form a representation of this algebra, and hence be described by some two-dimensional CFT, whose central charge depends on the negative cosmological constant of the effective three-dimensional description. Building on this, in~\cite{Strominger:1997eq} a Cardy formula~\cite{Cardy:1986ie} (sensitive only to the central charge) was again used to extract the asymptotic growth of CFT states, thereby reproducing the Bekenstein-Hawking entropy of the three-dimensional black hole, and providing a bridge between Strominger-Vafa's original computation and holography. Maldacena’s formulation of the correspondence is, of course, far more powerful: it allows to obtain of the exact dual CFT rather than just its central charge, as well as it extends to higher-dimensional spacetimes. For a comprehensive introduction to AdS/CFT, see~\cite{Aharony:1999ti}.

In this section, we review recent progress in the holographic account of the entropy of black holes in AdS spacetimes (see~\cite{Zaffaroni:2019dhb} for a reference), focusing on the duality between quantum gravity on AdS$_5$ and four-dimensional ${\cal N}=1$ superconformal field theories (SCFT) as an example.

\paragraph{The original argument: decoupling limit.}

Consider a stack of $N$ parallel D$3$-branes separated by a transverse distance $r$. In the low-energy regime, ${\cal E} \ll \ell_{\rm s}^{-1}$, the massless open string modes on the worldvolume correspond to the fields of ${\cal N}=4$ super Yang-Mills (SYM). To isolate the theory, we keep the energy arbitrary, but take a \emph{decoupling limit} 
\begin{equation}\label{eq:AdS_CFT_decoupling}
\alpha' = \ell_{\rm s}^2 \rightarrow 0 \,,\qquad U = \frac{r}{\alpha'} = \text{fixed}\,,
\end{equation}
in which the dimensionless parameters $g_{\rm s}$ and $N$ are kept arbitrary. In this limit closed string modes as well as heavy vibrational states entirely decouple from the worldvolume theory. 
In the limit of $N$ coincident D$3$-branes ($r\rightarrow 0$) we are then left with a pure ${\cal N}=4$ $D=4$ SYM with gauge group SU$(N)$\footnote{To be precise, one would be left with a SYM theory with U$(N)$ gauge group. However a U$(1)$ factor decouples from the dynamics.}, which is known to be a supersymmetric CFT (SCFT).

Next, we consider the supergravity solution carrying D$3$-brane charges:
\begin{equation}
\begin{aligned}
\diff s^2 &= H_3^{-1/2} \diff s_{[\mathbb R^{1,3}]}^2 + H_3^{1/2} \left( \diff r^2 + r^2 \diff \Omega_5^2\right)\,,
\\
H_3 &= 1 + 4\pi \frac{g_{\rm s} N \alpha'{}^2}{r^4}\,,
\end{aligned}
\end{equation}
where $\diff \Omega_5^2$ is the metric on the unit $S^5$. Here, $N$ denotes the flux of the RR five-form that couples to the D$3$-branes (which we are not showing). After the decoupling limit \eqref{eq:AdS_CFT_decoupling}, the geometry reduces to
\begin{equation}
\diff s^2 = \alpha' \sqrt{4\pi g_{\rm s} N} \left[\left( \frac{U^2}{4\pi g_{\rm s} N}\diff s^2_{[\mathbb R^{1,3}]} + \frac{\diff U^2}{U^2}\right) +  \diff \Omega_5^2\right]\,.
\end{equation}
This shows that the near-horizon region ($r\rightarrow 0$) of a system of $N$ D$3$-branes describes a AdS$_5$ $\times$ $S^5$ spacetime. Note that the AdS$_5$ and the $S^5$ have the same radius, $\ell_{\rm AdS}^2 = \alpha' \sqrt{4\pi g_{\rm s} N}$. 
Due to the redshift, the energy of an excited string measured by an observer located at finite $r$ is given by ${\cal E}(r) = H_3^{1/4} {\cal E}$, being ${\cal E}$ the one measured by the asymptotic observer. In the decoupling limit, we then find ${\cal E} \sim U\,\sqrt{\alpha'}\,{\cal E}(r) \ll {\cal E}(r)$. This means that also in the supergravity description a decoupling occurs: excited string states with arbitrary energy in the near-horizon throat of the branes are completely decoupled from the physics of the asymptotic flat region of the spacetime, due to the very large redshift. Also, the isometry supergroup of AdS$_5$ $\times$ $S^5$ is $PSU(2,2|4)$, matching the global ${\cal N}=4$ superconformal symmetry group of the decoupled SYM theory on the worldvolume. 

These observations motivate the introduction of the following conjectured correspondence between the two decoupled descriptions: 
\begin{equation}
\text{Type IIB superstring theory on AdS$_5$ $\times$ $S^5$}\quad  \longleftrightarrow \quad  {\cal N}=4 \; \; D=4 \;\; {\rm SU}(N) \;\; \text{SYM}\,.
\end{equation}
%
The holographic dictionary between the dimensionless parameters of string ($g_{\rm s}$ and $\ell_{\rm AdS}/\ell_{\rm P}$) and field theory (the rank of the gauge group $N$ and the Yang-Mills coupling $g_{\rm YM}$) is soon found: 
\begin{equation}
g_{\rm YM}^2 = 4\pi g_{\rm s}\,,\qquad 4\pi N = \left( \frac{\ell_{\rm AdS}}{\ell_{\rm P}}\right)^4 \quad \implies \quad \lambda = g_{\rm YM}^2 N = \left( \frac{\ell_{\rm AdS}}{\ell_{\rm s}}\right)^4\,.
\end{equation}
Here, we also introduced the dimensionless 't Hooft coupling $\lambda$, that controls the perturbative expansion of the field theory.

Intuition for the duality came from considering the supergravity description of a system of $N$ D$3$-branes, which really can be trusted only in the low-curvature regime
\begin{equation}
N > g_{\rm s} N \gg 1 \quad \implies \quad \lambda \gg 1 \,.
\end{equation}
The supergravity regime, then, corresponds to a regime in which the dual SCFT is strongly coupled:
\begin{equation}
\text{Type IIB supergravity on AdS$_5$ $\times$ $S^5$}\quad  \longleftrightarrow \quad  \text{strongly coupled} \;\;{\cal N}=4 \;\; D=4\;\; \text{SYM}\,.
\end{equation}
This is the celebrated AdS/CFT correspondence.
This correspondence is a true duality: it relates a weakly coupled gravitational description to a strongly coupled quantum field theory. The converse also holds, but when the SCFT is weakly coupled, the dual gravitational description becomes strongly curved, and classical supergravity breaks down: in this regime, the duality involves the full Type IIB string theory on AdS$_5$ $\times$ $S^5$, which is considerably more difficult to study.

In the strong version of the correspondence, that has become widely accepted over the years, ${\cal N}=4$ SYM at strong coupling is supposed to provide an exact \emph{non-perturbative} definition of supergravity on AdS$_5$ $\times$ $S^5$, including quantum gravity effects, such as finite-$N$ or $1/\lambda$ corrections. More precisely, it defines a quantum gravity on spacetimes that are asymptotic to AdS$_5$ $\times$ $S^5$, since in the interior a wide range of dynamical processes is allowed by the presence of highly excited strings, leading for instance to black hole formation, graviton propagation etc.. 
%
In this context, precision tests of the duality are essential to probe the validity of the holographic principle beyond the supergravity approximation. In particular, subleading effects in the large-$N$ expansion play a central role in the recent developments explored in this Thesis. 
\begin{figure}[!htb]
	\centering
	\includegraphics[width=0.65\textwidth]{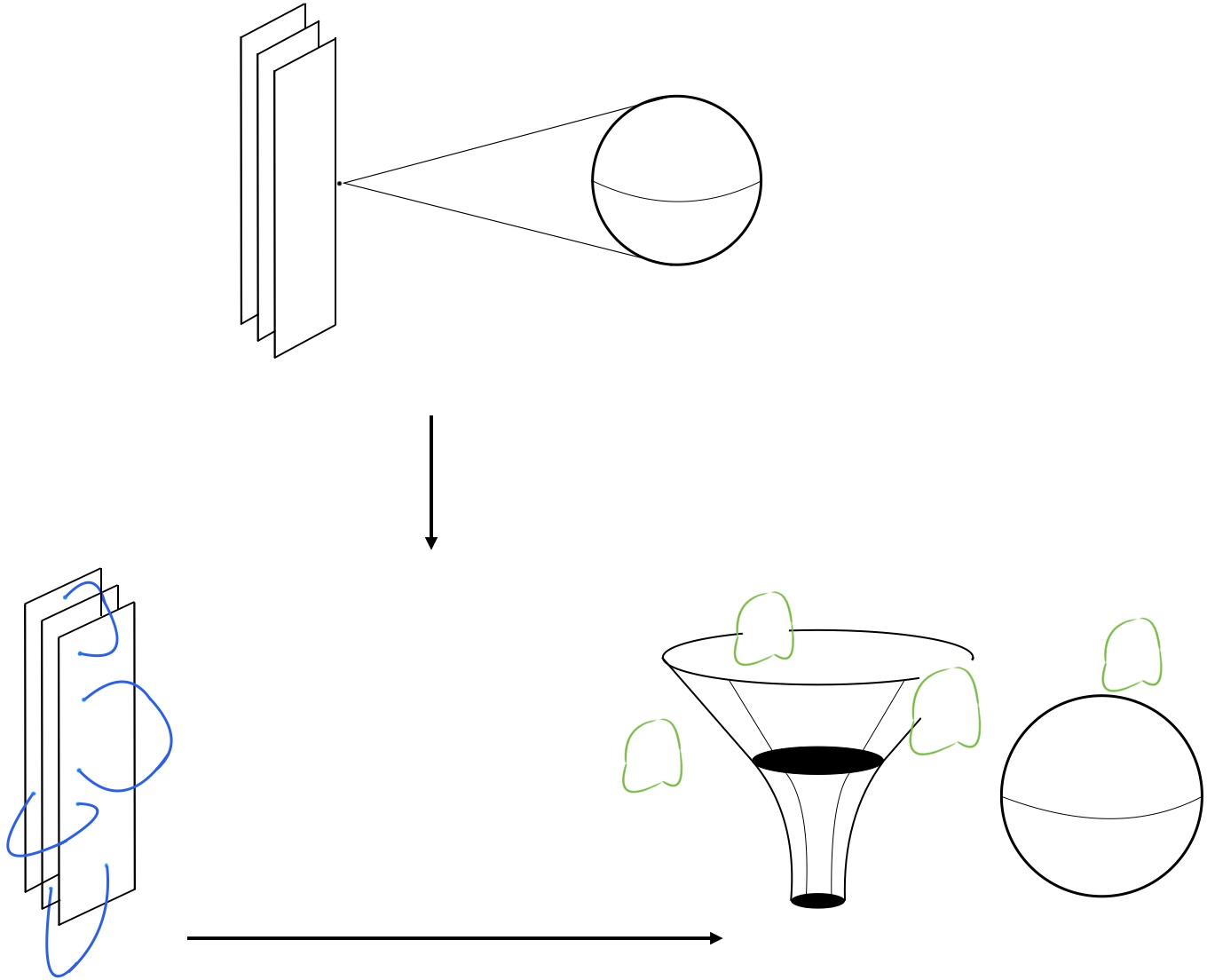}
	\caption{\it  AdS/CFT is a gauge/gravity duality. After a suitable decoupling limit, a stack of $N$ D$3$-branes admits two equivalent descriptions: Type IIB string theory on AdS$_5$ $\times$ $S^5$ and ${\cal N}=4$, $D=4$ SYM. When $g_{\rm s} N \ll 1$ (left) the SYM sitting on the brane worldvolume is weakly coupled. In the opposite regime, $g_{\rm s} N \gg 1$ (right), the SYM is strongly coupled, but the supergravity approximation is reliable.}
	\label{fig:ads_cft}
\end{figure} 

The holographic duality has been later extended beyond the original setup to encompass more general configurations.  In this Thesis, we will remain within the framework of five-dimensional supergravity, but we will focus on settings with reduced supersymmetry. In particular, we are interested in backgrounds of the form AdS$_5$ $\times$ $X_5$, where $X_5$ is a compact Einstein manifold satisfying $R^{[X_5]}_{mn} = 4 g_{mn}^{[X_5]}$. It can be shown that the ten-dimensional metric 
\begin{equation}
\diff s^2 = \left( 1 + \frac{\ell_{\rm AdS}^4}{r^4}\right)^{-1/2} \diff s_{[\mathbb R^{1,3}]}^2 +  \left( 1 + \frac{\ell_{\rm AdS}^4}{r^4}\right)^{1/2} \diff s_{Y_6}^2\,,\qquad \diff s_{Y_6}^2 = \diff r^2 + r^2 \diff s_{X_5}^2\,,
\end{equation}
where the AdS radius now depends on the volume of $X_5$,
\begin{equation}
\ell_{\rm AdS}^4 = \frac{\left(2\pi^2\alpha'\right)^2}{{\rm Vol}[X_5]}g_{\rm s} N\,,
\end{equation}
is still a solution of Type IIB supergravity provided we turn on $N$ units of five-form flux through $X_5$. Such geometries arise from placing $N$ D$3$-branes at the tip of the six-dimensional cone $Y_6$. 
To preserve some supersymmetry, the cone must be a \emph{Calabi-Yau} manifold. In this setup, in general only one quarter of the original 32 supercharges are preserved. Accounting for the presence of the D$3$-branes, which further break half of the supersymmetry, the resulting system retains 4 real supercharges. The base manifold $X_5$ is, then, called Sasaki-Einstein (but for special Sasaki-Einstein manifolds, such as the simplest case, i.e.~$S^5$, more supersymmetry is preserved). Therefore, for any Sasaki-Einstein manifold the near-horizon geometry is AdS$_5$ $\times$ $X_5$, and the dual theory is a four-dimensional ${\cal N}=1$ SCFT. The precise field theory depends on the specific choice of $X_5$, and we will discuss relevant examples later in this Thesis.

\paragraph{Master equation of AdS/CFT and AdS black holes.}

The correspondence can be formulated as the equivalence of the partition function of string theory on AdS$_5$ $\times$ $X_5$ and the generating functional of the dual ${\cal N}=1$ SCFT placed on the boundary of AdS$_5$~\cite{Witten:1998qj}. In this sense, the AdS/CFT correspondence realizes an \emph{holographic duality}.  This equivalence is compactly expressed by the relation:
\begin{equation}
\label{eq:master_eq_holography}
Z_{\rm string}\left[\phi_0 = \phi\Big|_{\partial {\rm AdS}_5}\right] = Z_{\rm CFT}\left[\phi_0\right]\,.
\end{equation}
On the right-hand side appears the generating functional for correlation functions in the SCFT,
\begin{equation}
Z_{\rm CFT}\left[ \phi_0\right] = \Big< {\rm e}^{\int_{\partial {\rm AdS}_5} \diff^4 x\,\phi_0(x) O(x)} \Big>\,,
\end{equation}
where $\phi_0$ is an arbitrary source function. The SCFT is defined on the conformal boundary of AdS$_5$, and correlation functions of the operator $O$ can be obtained by taking functional derivatives of $Z_{\rm CFT}[\phi_0]$ with respect to $\phi_0(x)$ and then setting $\phi_0 = 0$. 
The left-hand side of~\eqref{eq:master_eq_holography} represents the full string theory partition function on AdS$_5 \times X_5$. More precisely, one must consider bulk configurations that are \emph{asymptotically} AdS$_5 \times X_5$. In general, multiple backgrounds can contribute and the string partition function is expected to sum over all these contributions. The path integral is taken over all bulk fields $\phi$ with boundary conditions such that they approach $\phi_0$ at the conformal boundary of AdS$_5$. In this way, $\phi_0$ acts as a source for the dual operator $O$. 

In the limit of large $N$ and large 't Hooft coupling $\lambda$, the right-hand side of~\eqref{eq:master_eq_holography} provides a {\it non-perturbative definition of the gravitational path integral} for quantum gravity on AdS, with the given boundary conditions for the bulk fields:
\begin{equation}
\label{eq:mastereqadscft}
Z_{\rm grav}\left[ \phi_0 = \phi\Big|_{\partial{\rm AdS}_5} \right] =Z_{\rm CFT}\left[ \phi_0; N\gg 1 ,\,\lambda\gg 1\right]\,,
\end{equation}
Here, we have in mind a Euclidean setup in which the left-hand-side takes the form of a gravitational path integral \eqref{eq:grav_path_int}, where we path integrate over all bulk metric that induce at the boundary the same conformal structure as the background metric\footnote{This means that boundary conditions for the bulk metric amount to fixing only its conformal class at the boundary, i.e., metrics related by local overall rescalings are identified.} of the dual CFT. Since black holes appear as saddle points of the Euclidean gravitational path integral, we expect that the dual large-$N$ CFT captures the thermodynamics of these black holes, and that it can provide a description of their microstates.

\medskip

It is natural to expect that the entropy of AdS black holes can be reproduced by enumerating states in a dual CFT that carry the same conserved charges. Asymptotically AdS$_5$ (AAdS$_5$) black holes arise as solutions of five-dimensional gauged supergravity, possibly coupled to vector multiplets. A particularly relevant class of such solutions consists of supersymmetric, electrically charged, and rotating black holes, which uplift to Type IIB supergravity on AdS$_5 \times S^5$~\cite{Gutowski:2004yv,Kunduri:2006ek}. 
These black holes depend on two angular momenta $J_{1,2}$ corresponding to the Cartan generators of the AdS$_5$ isometry group, and three electric charges $Q_{1,2,3}$ corresponding to the Cartan of the $SO(6) \simeq SU(4)$ symmetry of $S^5$. They preserve only two real supercharges out of the 32 originally present in Type IIB supergravity, and are therefore $1/16$-BPS configurations. 
According to the AdS/CFT correspondence, these black holes should be holographically dual to ensembles of states in ${\cal N}=4$ SYM on a spatial $S^3$, carrying the same charges and preserving the same amount of supersymmetry. In particular, the $1/16$-BPS sector of the theory is characterized by the presence of two preserved supercharges, obeying the following (anti)commutation relations:
\begin{equation}
\label{eq:superalgebra_N4}
\begin{aligned}
\left\{{\cal Q},\,\bar {\cal Q}\right\} \,&=\, E - \Phi_*^I Q_I - \Omega_1^* J_1 - \Omega_2^* J_2\,, 
\\[1mm]
\left[J_{1,2},\,{\cal Q} \right] \,&=\, -\frac{1}{2}\left[ Q_I ,\,{\cal Q} \right] = \frac{1}{2}{\cal Q}\,,\qquad I = 1,2,3\,,
\end{aligned}
\end{equation}
where $E$ denotes the Hamiltonian, and $\Omega_{1,2}^*$ and $\Phi_*^I$ are constants that depend on the specific model (notation is chosen so as to match future conventions). Supersymmetric states are annihilated by both supercharges, thus they saturate the BPS bound $E = \Phi_*^I Q_I + \Omega_1^* J_1+ \Omega_2^* J_2$. Note that for AdS black holes the BPS bound involves angular momenta. To count states in the quantum field theory, one may consider two related but distinct quantities:
\begin{itemize}
\item The first is a (grand-canonical) partition function\footnote{
This can be understood as a standard thermal partition function ${\cal Z} = {\rm Tr} \left[ {\rm exp}\left(-\beta E + \beta\Omega_1 J_1+ \beta \Omega_2 J_2 +\beta\Phi^I Q_I\right)\right]$, with energy redefined according to \eqref{eq:superalgebra_N4}. Finally, one can redefine the chemical potentials, and introduce $\omega_{1,2} = \beta \left( \Omega_{1,2}- \Omega_{1,2}^*\right)$ and $\varphi^I = \beta \left( \Phi^I - \Phi^I_*\right)$. 
}
\begin{equation}
{\cal Z}\left( \varphi^I,\,\omega_{1,2}\right) = {\rm Tr}\left[ {\rm e}^{-\beta \{{\cal Q},\,\bar{\cal Q}\} + \omega_1 J_1 + \omega_2 J_2+ \varphi^I Q_I}\right]\,,
\end{equation} 
where the chemical potentials are introduced to ensure convergence of the trace in an appropriate domain. For generic values of the potentials, this function encodes the absolute degeneracy of states, which can be extracted via an inverse Laplace transform.
\item Alternatively, an index, specifically the \emph{superconformal index}~\cite{Romelsberger:2005eg,Kinney:2005ej}
\begin{equation}
\label{eq:SCI_basic}
{\cal I}\left(\varphi^{1,2},\,\omega_{1,2} \right) = {\rm Tr} \left[ \left( -\right)^{\rm F} {\rm e}^{-\beta \{ {\cal Q},\,\bar {\cal Q}\} } {\rm e}^{\omega_1 \left( J_1 + \frac{1}{2}Q_3\right) + \omega_2\left( J_2 + \frac{1}{2}Q_3\right) + \varphi^1 \left( Q_1 - Q_3\right) + \varphi^2 \left( Q_2 - Q_3\right)} \right]\,,
\end{equation}
which takes the form of a refined Witten index (with ${\rm F} = 2J_{1,2} = Q_{1,2,3}$), counting fermionic states with a minus and bosonic ones with a plus sign. As a consequence, it is independent of $\beta$. Note that, since the index can accommodate fugacities only for the conserved charges that commute with ${\cal Q}$, it contains one less chemical potential than the partition function. 
\end{itemize}
Both the grand-canonical partition function and the index can be computed by a Euclidean path integral $Z_{\rm CFT}$ on $S^1 \times S^3$, where $\beta$ becomes the radius of circle $S^1$.\footnote{To be precise, the superconformal index coincides with the Euclidean path integral up to a prefactor, $Z_{\rm CFT} = {\rm e}^{-\beta E_C}{\cal I}$. Here, $E_C$ denotes the Casimir energy (see, e.g.,~\cite{Assel:2014paa,Assel:2015nca}), that can be interpreted as the vacuum expectation value of the Hamiltonian.}  This can be interpreted as a finite temperature ($\beta^{-1}$) partition function with twisted boundary conditions along $S^1$ determined by the chemical potentials, which can be viewed as turning on angular velocities and electrostatic potentials for a thermal CFT. This resonates with the expectation that this thermal ensemble of states should reproduce the microstates of an electrically charged and rotating black hole.
The choice of spin structure on the $S^1$ circle distinguishes which of the two traces the Euclidean path integral is computing: if fermions acquire the same phase as their bosonic superpartners when twisted around the circle, supersymmetry is preserved and the resulting functional integral computes the superconformal index. 
On the other hand, if fermions and bosons do not have the same boundary conditions, the resulting path integral computes the grand-canonical partition function receiving contributions from non-BPS states.

Holography, then, instructs us to compute this Euclidean path integral in a strongly-coupled regime. This is, in general, an extremely challenging task. Again, supersymmetry offers a way forward. In fact, as in the case of asymptotically flat black holes discussed in the previous sections, also in the context of AdS black holes our focus will be on computing indices rather than absolute degeneracies. For instance, since the index is a protected observable, it turns out to be independent of the continuous 't Hooft coupling $\lambda$. As a consequence, it can be computed at weak coupling, where BPS states are more tractable, and the result can still be trusted at strong coupling. Additionally, supersymmetric gauge theories offer additional techniques for computing supersymmetric partition functions, including \emph{supersymmetric localization}. 

\paragraph{Localization.}

Localization is a powerful technique that allows to reduce complicated integrals (such as infinite-dimensional path integrals) to simpler ones (in the cases of interest to us, a finite-dimensional integral over a matrix model). The idea has a long history in physics, dating back to~\cite{Witten:1982im,Witten:1988ze}, and has found modern applications in the exact computation of supersymmetric partition functions starting with~\cite{Nekrasov:2002qd,Pestun:2007rz} (see~\cite{Pestun:2016zxk} for a review). While explicit computations can be highly non-trivial, the basic idea behind localization is quite simple. 
Let ${\cal Q}$ denote a fermionic symmetry in the theory (either nilpotent or squaring to a bosonic symmetry). Given a ${\cal Q}$-invariant path integral $Z$, we introduce a ${\cal Q}$-exact deformation controlled by the parameter $t$, such that $Z(t=0) = Z$:\footnote{For simplicity, we assume that the path integral measure is also ${\cal Q}$-invariant.}
\begin{equation}
Z = \int {\cal D}\left[\phi\right] \,{\rm exp}\left[ -I \right] \quad \rightarrow \quad Z(t) = \int {\cal D}\left[\phi\right] \,{\rm exp}\left[ -I + t{\cal Q} {\cal V} \right] \,,
\end{equation}
where $\phi$ denotes the collection of quantum fields in the theory. Since the deformation is ${\cal Q}$-exact, the deformed path integral is actually independent of the deformation parameter, $\frac{\diff}{\diff t}Z(t) = 0$. 
Thus, the original path integral can also be computed in the $t\rightarrow +\infty$ limit, where the integral simplifies dramatically. By choosing an appropriate ${\cal V}$,
this limit localizes the integral onto the critical set of ${\cal Q} {\cal V}$, which defines the localization manifold. Around this locus, one can expand the fields as $\phi = \phi_0 + t^{-1/2} \delta \phi$, where $\phi_0$ lies on the localization manifold, and $\delta \phi$ parametrizes arbitrary fluctuations about it. In the $t\rightarrow +\infty$ limit, only quadratic fluctuations survive, rendering the path integral one-loop exact. For a clever choice of ${\cal V}$, the final result takes the schematic form
\begin{equation}
Z = \int_{{\cal Q}{\cal V} =0} {\rm e}^{-I[\phi_0]} Z_{1-\text{loop}}\left[ {\cal Q} {\cal V}\right]\,,
\end{equation} 
where $Z_{1-\text{loop}}\left[ {\cal Q} {\cal V}\right]$ is computed by the ratio of the fermionic to bosonic determinants arising from the quadratic fluctuations of the operator ${\cal Q}{\cal V}$. 

Localization thus provides a systematic framework for evaluating supersymmetric path integrals exactly, reducing them to simpler expressions that can be computed explicitly in many examples.

\medskip

Progress in the statistical derivation of the entropy of AdS black holes has required the development of localization techniques, particularly on curved manifolds. For this reason, it was not until the last decade (starting from 2015), nearly twenty years after the Strominger-Vafa computation, that the first successful microscopic counting of AdS black holes in dimensions four or higher was achieved by Benini, Hristov, and Zaffaroni~\cite{Benini:2015eyy}. The setting they considered differs slightly from the one discussed in this Introduction, as it involves static, magnetically charged supersymmetric AdS$_4$ black holes. Their entropy was reproduced by computing the topologically twisted index of a three-dimensional SCFT (specifically, ABJM theory), using localization techniques~\cite{Benini:2015noa}. For a broader overview of these developments, we refer to the review~\cite{Zaffaroni:2019dhb}. 
Related progress was made a few years later, in 2018, for electrically charged and rotating AdS$_5$ black holes~\cite{Cabo-Bizet:2018ehj,Choi:2018hmj,Benini:2018ywd}. As we will discuss in more detail later, in this case the black hole entropy is reproduced by performing a constrained Legendre transform of the large-$N$ limit of the logarithm of the superconformal index, evaluated in a suitable regime of the chemical potentials (this key observation was first made in~\cite{Hosseini:2017mds}). A detailed discussion of the results of~\cite{Cabo-Bizet:2018ehj} will be the focus of the next chapter, providing the background for the developments that follow.

Before concluding this section, let us add a few relevant comments:
\begin{enumerate}
\item The CFT partition function at large $N$ provides a non-perturbative definition of the gravitational path integral with certain boundary conditions, via the master equation of the AdS/CFT correspondence \eqref{eq:mastereqadscft}. 
Then, a precise comparison between the two sides of the correspondence requires understanding how to compute the index from the gravitational side, via a Euclidean path integral over asymptotically AdS geometries. Of particular interest is identifying the precise contribution of the supersymmetric black hole to the index, from the perspective of both sides of the correspondence. Concretely, we aim to compare (at least in a semiclassical approximation) the on-shell action of the supersymmetric black hole with appropriate boundary conditions, evaluated directly in the bulk only using gravitational variables, with the large-$N$ limit of the logarithm of the superconformal index, in a suitable regime of the chemical potentials to be specified later.
\item In its strong form, the AdS/CFT correspondence is expected to hold at finite $N$, so the large-$N$ limit must really be understood as the leading term in a large-$N$ expansion. By focusing on a particular saddle of the index, the expansion of its logarithm predicts a series of subleading corrections: perturbative $1/N$ (power-law) terms, logarithmic corrections $\log N$, and non-perturbative (exponentially suppressed) corrections, e.g. ${\rm e}^{-N}$. Each of these terms provides a prediction for quantum effects in the quantum gravity path integral. Developing a gravitational understanding of these corrections is the goal of the program known as \emph{precision holography}. This program serves a dual purpose: on one hand, it offers concrete tests of the AdS/CFT correspondence beyond the classical $N\rightarrow +\infty$ limit, while on the other hand it provides insight into quantum aspects of supergravity and black hole physics, guided by exact results from the dual CFT.
Recent progress in this direction has been achieved in both four and five dimensions. Logarithmic corrections to black hole entropy are understood as one-loop fluctuations of light fields (including KK modes) in the gravitational path integral~\cite{Bobev:2023dwx}, while non-perturbative contributions have been interpreted as arising from wrapped Euclidean branes~\cite{Aharony:2021zkr}. Perturbative corrections, on the other hand, have been shown to correspond to higher-derivative terms in the supergravity effective action~\cite{Bobev:2020egg,Bobev:2021oku,Cassani:2022lrk,Bobev:2022bjm}. It is within this framework that the present Thesis is situated.
\end{enumerate}


\section{Gravitational path integral in Euclidean quantum gravity}
\label{sec:Grav_path_int}

In the previous sections we reviewed how string theory on asymptotically flat backgrounds and its holographic formulation in AdS spaces provide concrete frameworks for a statistical explanation of the Bekenstein-Hawking entropy of certain black holes. In both cases, the analysis begins with the computation of a supersymmetric index in the microscopic description:
\begin{itemize}
\item For asymptotically flat (AF) black holes in ungauged supergravity the relevant indices are helicity supertraces;
\item for asymptotically AdS (AAdS) black holes in gauged supergravity the relevant indices are either superconformal indices or topologically twisted indices.
\end{itemize}
Whenever these indices can be computed with accuracy, including subleading corrections in the asymptotic expansion, they provide quantitative predictions for quantum corrections to the black hole action (and entropy). Then, developing the gravitational counterpart offers valuable insight into the quantum physics of black holes. In both asymptotically flat and AdS cases, the focus is on protected quantities, that can be reliably evaluated in a parameter regime where the microscopic description (either in terms of bound states of fluctuating strings and branes or ensembles of quantum states of the dual CFT) is weakly coupled. Importantly, these results can be extrapolated to the gravitational regime, where one expects to be able to formulate a precise gravitational counterpart that encodes information about microstate counting.
Indeed, recent progress, starting with~\cite{Cabo-Bizet:2018ehj} (see~\cite{Iliesiu:2021are} for the first application to the AF case), has revealed that the Euclidean gravitational path integral \eqref{eq:grav_path_int}, when defined with appropriate boundary conditions compatible with supersymmetry, computes a supersymmetric index. This object, which we refer to as the \emph{gravitational index}, allows for direct and precise comparisons with microscopic computations.
The aim of this section is to review the formulation of the gravitational path integral as originally proposed by Gibbons and Hawking in~\cite{Gibbons:1976ue}, as a preparation for the subsequent discussion of the gravitational index. Here, we follow the review~\cite{Cassani:2025sim}.

\medskip

Since its introduction in~\cite{Gibbons:1976ue}, the gravitational path integral has played a central role in a wide range of developments, including the study of black hole thermodynamics, holography, and quantum cosmology (see, e.g.~\cite{PhysRevD.28.2960}), as well as more recent advances such as the derivation of the Page curve via replica tricks~\cite{Almheiri:2019qdq} and the analysis of quantum effects in near-extremal black holes~\cite{Iliesiu:2020qvm}, to name just a few.

However, the path integral \eqref{eq:grav_path_int}, formally introduced using only gravitational variables, is not well-defined in general. This is primarily (but not only) due to the non-renormalizable divergences that arise at some loop order in quantum gravity (see e.g. the classic reference~\cite{Goroff:1985th}), which indicate that gravity should be treated merely as a low-energy effective theory. Nonetheless, in certain controlled setups (as argued above, these involve cases with sufficient supersymmetry, a possible uplift to string theory, or a known CFT dual) we still expect to be able to reduce it to a well-defined quantity and study it more in detail, at least in some approximation. These are the cases we will focus on.

The path integral \eqref{eq:grav_path_int} is a function of the boundary conditions. The starting point to specify them is to identify the asymptotic symmetries of the gravitational theory, which determine a set of conserved global charges measured at infinity. We restrict to the case in which these include the energy $E$, some of angular momenta $J_i$, with $i =1,2,3,...$ depending on the spacetime dimensions, and a set of electric charges $Q_I$, $I=1,2,3,...$ associated with gauge symmetries. For concreteness, we will work in five-dimensional supergravity, where there are two independent angular momenta, $J_{1,2}$. 
We further assume that the spacetimes included in the path integral have an asymptotic circle $S^1$ of period $\beta$, which plays the role of the inverse temperature, parametrized by the Euclidean time direction. Then, one must determine how the other spacetime coordinates and all dynamical fields (both bosonic and fermionic) are identified upon being transported around the Euclidean time circle. This defines a set of chemical potentials conjugate to each conserved charge, 
specifying the set of boundary conditions relevant for the path integral. 

There are two equivalent ways to introduce the needed chemical potentials:
\begin{itemize}
\item \emph{Twist realization}: chemical potentials are implemented by imposing twisted identifications along $S^1$. 

Let $\phi_{1,2}$ be the $2\pi$-periodic angular coordinates advanced by the action of $J_{1,2}$, then, a revolution around the Euclidean time circle is accompanied by the following twisted coordinate identification:
\begin{equation}
\left( \tau,\,\phi_1,\,\phi_2\right) \sim \left( \tau + \beta ,\,\phi_1 - \ii \beta \Omega_1,\,\phi_2 -\ii\beta\Omega_2\right)\,,
\end{equation}
where $\Omega_{1,2}$ denote the angular velocities associated with $J_{1,2}$. 
Fields (including fermions), denoted in general by ${\cal X}$, are identified accordingly: 
\begin{equation}
{\cal X}\left( \tau+ \beta ,\,\phi_1 - \ii\beta \Omega_1,\,\phi_2 -\ii \beta \Omega_2\right) = \left( -\right)^{\rm F} {\rm e}^{\beta \Phi^I \hat q_I} {\cal X} \left( \tau ,\,\phi_1,\,\phi_2\right)\,,
\end{equation}
where ${\rm F}$ is the fermion number, $\hat q_I$ the charge of the field under $Q_I$, and $\Phi^I$ the corresponding electrostatic potential. 

With these identifications, the relevant spacetimes in the path integral are those with boundary geometry given by $S^1 \times S^3$ (AAdS$_5$ case) or that asymptote to $S^1 \times \mathbb R^4$ (AF$_5$), with vanishing gauge field. It is often convenient to switch to an equivalent description by performing a coordinate transformation $\phi_{1,2} \rightarrow \tilde \phi_{1,2} = \phi_{1,2} + \ii\Omega_{1,2} \tau$, along with a (large) gauge transformation $A \rightarrow A + \ii\Phi \diff \tau$, to eliminate the explicit twists, restoring untwisted identifications for all coordinates and fields.\footnote{Later, we will refer to asymptotically static coordinate systems as Boyer-Lindquist type coordinates, that typically satisfy twisted identifications. On the other hand, we define co-rotating coordinates those that identify an asymptotic frame that rotates, allowing us to realize angular velocities as asymptotic background holonomies.} Then, the chemical potentials will appear as certain components of the asymptotic metric and gauge fields, leading to the second possible realization of the potentials.
\item \emph{Background realization}: 
The fields now have untwisted identifications along $S^1$, but chemical potentials appear explicitly in the asymptotic gauge fields (including the non-diagonal components of the metric). 

AAdS$_5$ solutions: the boundary geometry involves a fibration of $S^3$ over the Euclidean time circle $S^1$, as well as some background gauge fields. Then, for an appropriate radial coordinate $r$, the spacetime asymptotes to
\begin{equation}
\begin{aligned}
\diff s^2& \rightarrow \ell_{\rm AdS}^2\frac{\diff r^2}{r^2} + r^2 \diff s^2_{\rm bdry}\,,
\\[1mm]
\diff s^2_{\rm bdry} &= \beta^2\diff \tilde \tau^2 + \diff \vartheta^2 + \sin^2\vartheta\left( \diff \tilde \phi_1 - \ii\beta\Omega_1 \diff\tilde \tau\right)^2 + \cos^2\vartheta \left( \diff \tilde \phi_2 - \ii\beta \Omega_2 \diff \tilde \tau \right)^2\,,
\\[1mm]
A^I &\rightarrow \ii \beta \Phi^I \diff\tilde \tau\,.
\end{aligned}
\end{equation}
where $\vartheta \in [0; \pi/2]$, $\tilde\tau \sim \tilde \tau +1$ and $\tilde\phi_{1,2} \sim \tilde\phi_{1,2} + 2\pi$. 

In holographic applications, the boundary geometry defines the background configuration for the dual four-dimensional CFT. 

AF$_5$ solutions: the spacetime asymptotes to $S^1 \times \mathbb R^4$, with appropriate background rotations and electrostatic potentials turned on:
\begin{equation}
\begin{aligned}
\diff s^2 &\rightarrow \beta^2 \diff \tilde \tau^2 +\diff r^2 + r^2\left[\diff \vartheta^2 + \sin^2\vartheta\left( \diff \tilde \phi_1 - \ii\beta\Omega_1 \diff\tilde \tau\right)^2 + \cos^2\vartheta \left( \diff \tilde \phi_2 - \ii\beta \Omega_2 \diff \tilde \tau \right)^2\right]
\\[1mm]
A^I &\rightarrow \ii \beta \Phi^I \diff\tilde \tau\,.
\end{aligned}
\end{equation}
\end{itemize}
The remarkable intuition by Gibbons and Hawking was to formally treat the gravitational path integral (defined with chemical potentials introduced as above and Dirichlet boundary conditions on all fields) as the partition function of an ordinary quantum system. Then, the Euclidean gravitational path integral with compactified imaginary time admits a quantum-statistical interpretation as grandcanonical partition function:
\begin{equation}
\label{eq:grav_part_func}
\begin{aligned}
Z_{\rm grav} \left( \beta,\Omega_{1,2},\Phi^I\right) &= {\rm Tr}\left[ {\rm e}^{-\beta\left( E - \Omega_1 J_1 -\Omega_2 J_2 - \Phi^I Q_I\right)} \right]
\\[1mm]
&= \sum_{E,\,J_{1,2},\,Q_I} d\left( E,J_{1,2},Q_I\right) {\rm e}^{-\beta \left( E - \Omega_1 J_1 -\Omega_2 J_2 - \Phi^I Q_I\right)}\,,
\end{aligned}
\end{equation}
where $d(E,J_{1,2},Q_I)$ denotes the degeneracy of states carrying the specified charges. Of course, in a theory of gravity, this trace is not well-defined since there is no rigorous understanding of the Hilbert space of quantum states forming the ensemble. The identification \eqref{eq:grav_part_func} is therefore meant to be purely formal. However, in certain controlled settings, such as those mentioned above, the trace is expected to acquire a more precise meaning.
For gauged supergravity with an holographic dual, the UV completion is provided by the dual CFT, then the trace can be understood as a partition function over the dual Hilbert space. For ungauged supergravity with a string theory uplift, the microscopic description is in terms of bound states of fluctuating strings and branes. 
%
\paragraph{Semiclassical approximation.} Before imposing supersymmetry to relate \eqref{eq:grav_part_func} to the \emph{gravitational index}, let us first clarify the regime in which we aim to evaluate the partition function. Because the gravitational path integral lacks a rigorous definition, the best controlled starting point is to work in a semiclassical approximation, which is expected to be well-defined~\cite{Gibbons:1976ue}. 
In this regime, the gravitational path integral reduces to a sum over saddle points (classical solutions of the Euclidean field equations) that satisfy the boundary conditions introduced earlier. The contribution of each saddle is given by the exponential of its classical on-shell action,\footnote{Evaluating the action $I$ on a given solution typically requires proper regularization and renormalization of divergences. For asymptotically AdS solutions a proper definition for the on-shell action requires regularization and the addition of counterterms to cancel divergences. This procedure is known as holographic renormalization. After renormalization, the resulting finite action depends on the boundary data of the bulk fields~\cite{Henningson:1998gx,deHaro:2000vlm,Bianchi:2001kw}.}
\begin{equation}\label{eq:sum_over_saddles}
Z_{\rm grav} \sim \sum_{\rm saddles} {\rm e}^{-I}\,.
\end{equation}
Finite temperature, electrically charged and rotating black hole solutions exists in five dimensional supergravity~\cite{Cvetic:1996xz,Chong:2005da,Chong:2005hr} and therefore their Wick-rotated version contribute as saddles in the semiclassical approximation. Importantly, one of the advantages of working in Euclidean signature is that black holes cap off smoothly at the event horizon, excising the curvature singularity. Many other solutions, most of which are probably still unknown, are also expected to contribute. Beyond the leading order, quantum corrections, such as higher-derivative terms or one-loop effects, can be systematically included around each saddle. 

The problem of computing the full path integral is therefore reduced to classifying all classical configurations compatible with the boundary conditions. In the Euclidean theory, these saddles are gravitational instantons with at least one isometry associated with the periodic imaginary time. Then, to determine the different phases of the theory, one must compare their Euclidean on-shell actions to identify the dominant contributions. 
To make this comparison efficient, it is desirable to compute the action using only minimal information about the saddle, such as topological data and boundary conditions. In this direction, Gibbons and Hawking showed that, at least for four‑dimensional gravitational instantons, the action can be reduced to contributions from certain fixed submanifolds of the isometry group~\cite{Gibbons:1979xm} (see also~\cite{BenettiGenolini:2019jdz} for a recent application in presence of supersymmetry). This result suggests the existence of a localization principle that could provide a systematic method for evaluating on‑shell actions, a point to which we will return later.

When the action is evaluated with Dirichlet boundary conditions for all fields, it becomes a function of the independent chemical potentials. As these parameters are varied, the dominant saddle can change, signalling a phase transition. 
Therefore, even in the semiclassical approximation, the path integral captures genuinely non-perturbative, quantum gravitational phenomena: it realizes a sum over distinct geometries (and topologies) and can exhibit non-trivial thermodynamic transitions between them. A paradigmatic example is the Hawking-Page transition~\cite{Hawking:1982dh}, which occurs in asymptotically AdS$_{d+1}$ spacetimes (with $d \geq 2$): at low temperatures, the dominant configuration is thermal AdS, whereas at high temperatures it is an AdS black hole. Using the master relation of AdS/CFT, this quantum gravity effect was related to the strong coupling phenomenon of the confinement/deconfinement transition in the dual gauge theory~\cite{Witten:1998zw}.

Once the leading saddle is identified, one can match the semiclassical expansion \eqref{eq:sum_over_saddles} to the expression \eqref{eq:grav_part_func} and extract a thermodynamic interpretation of the on-shell action:
\begin{equation}
\label{eq:general_qsr0}
I \left(\beta,\Omega_{1,2},\Phi^I \right) = - {\cal S}\left(E,J_{1,2},Q_I\right) + \beta E - \beta \Omega_1 J_1 -\beta \Omega_2 J_2 - \beta \Phi^I Q_I \,.
\end{equation}
This is the \emph{quantum statistical relation}, showing that $I/\beta$ corresponds to the Gibbs free energy, satisfying
\begin{equation}\label{charges_from_I}
J_{1,2} = -\frac{1}{\beta}\frac{\partial I}{\partial\Omega_{1,2}}\,,\qquad Q_I = -\frac{1}{\beta}\frac{\partial I}{\partial\Phi^I}\,,\qquad E = \frac{\partial I}{\partial\beta}\,,
\end{equation}
and that the entropy ${\cal S} = \log d\left( E,J_{1,2},Q_I\right)$ is related to the on-shell action by a Legendre transform.

\paragraph{Imposing supersymmetry.}
What we have discussed so far applies equally to gauged and ungauged supergravity. The next step is to understand how supersymmetry modifies the gravitational partition function \eqref{eq:grav_part_func}, turning it into a gravitational index. Since the preserved superalgebras differ in the two cases, we will not present the full technical derivation here. These details will be developed later in the Thesis, where the two setups will be carefully distinguished. For now, we focus on the general principles common to both.

Two related ingredients are involved:
\begin{enumerate}
\item In supergravity we consider a complex supercharge ${\cal Q}$ obeying the superalgebra
\begin{equation}
\label{eq:supergravity_algebra}
\begin{aligned}
\left\{ {\cal Q},\,\bar {\cal Q}\right\} = E - \Omega_1^* J_1 - \Omega_2^* J_2 - \Phi^I_* Q_I\,,
\\
\left[ J_{1,2} ,\,{\cal Q}\right] = \frac{1}{2}{\cal Q}\,,\qquad \left[ Q_I,\,{\cal Q}\right] = - r_I {\cal Q}\,,
\end{aligned}
\end{equation}
where the constants $\Omega_{1,2}^*$, $\Phi^I_*$, $r_I$ may in general vanish. When $r_I\neq 0$ for some $I$, 
the corresponding charge $Q_I$ rotates the supercharge ${\cal Q}$, and thus qualifies as an R-charge. We will normalize the charges such that $r_I$ is either $0$ or $1$, unless differently specified. In this sense, the angular momenta also act as R-charges in the algebra. This superalgebra can be used to eliminate the energy $E$ in the trace \eqref{eq:grav_part_func}. While doing so, it is convenient to redefine the chemical potentials via~\cite{Silva:2006xv}
\begin{equation}\label{eq:red_chem_pot}
\omega_{1,2} = \beta \left( \Omega_{1,2} - \Omega_{1,2}^*\right) \,,\qquad \varphi^I = \beta \left( \Phi^I - \Phi^I_*\right)\,.
\end{equation}
After this change of variables, the gravitational partition function becomes
\begin{equation}
Z_{\rm grav} \left( \beta,\,\omega_{1,2},\,\varphi^I\right) = {\rm Tr} \left[ {\rm e}^{-\beta \{ {\cal Q},\,\bar {\cal Q}\} + \omega_1 J_1 + \omega_2 J_2 + \varphi^I Q_I }\right]\,.
\end{equation}
At this stage, this expression is just a reparametrization of \eqref{eq:grav_part_func}, and the trace still retains explicit dependence on $\beta$, since we are not yet computing an index.

\item An index is defined by inserting $(-1)^{\rm F}$ in the trace. This insertion naturally arises by imposing a \emph{complex} linear constraint on the redefined chemical potentials associated with the charges that rotate the supercharge~\cite{Cabo-Bizet:2018ehj,Iliesiu:2021are}: 
\begin{equation}
\label{eq:constraint_gauged}
\omega_1 + \omega_2 - 2r_I \varphi^I = \pm 2\pi \ii\,, 
\end{equation}
where $r_I =1$ is $I$ is an R-symmetry, and $r_I = 0$ otherwise. This relation ensures that only the states annihilated by ${\cal Q}$ contribute to the trace, implementing the $\left(-\right)^{\rm F}$ insertion. It also admits a geometric interpretation: it constitutes a regularity conditions for a globally-defined Killing spinor in a disk topology, providing the correct spin structure to contribute to an index.
\end{enumerate}
We will return to these conditions later in the Thesis, where we examine in detail how they arise in both gauged and ungauged supergravity.
For now, our goal is simply to outline the general structure of the argument that turns \eqref{eq:grav_part_func} into an index.

The gravitational index is expected to provide a controlled framework for addressing questions that extend beyond the semiclassical approximation, including quantum corrections to the Bekenstein-Hawking area formula. These raise fundamental issues: how can such corrections be computed systematically? What is their physical origin? Can macroscopic results be compared with microscopic predictions with increasing precision, or even resummed into an exact expression? More broadly, studying the gravitational index may reveal general lessons about the role of the gravitational path integral in quantum gravity, insights that could remain valid even outside the supersymmetric context, which is the ultimate goal of this subject.

In gauged supergravity with a known CFT dual, the sharp predictions from the boundary theory, via the AdS/CFT correspondence, offer a natural guiding principle in this direction. In particular, over the past decade, major progress has been achieved by applying localization techniques in quantum field theories to compute exact path integrals, as reviewed earlier. One may wonder whether similar methods can be used to localize the supergravity path integral, at least in holographic settings, offering a way towards exact results in quantum gravity. This idea was explored starting with~\cite{Banerjee:2009af,Dabholkar:2010uh,Dabholkar:2011ec}, in setups with sufficient amount of supersymmetry and an uplift to string theory providing a UV completion (such as ${\cal N}=8$ four-dimensional supergravity). In such cases, one expects the gravitational path integral to localize over distinct supersymmetric configurations (i.e. geometries admitting a Killing spinor) that are compatible with fixed boundary conditions.\footnote{The fact that the path integral reduces to a discrete sum over supersymmetric configurations resonates with the Bethe Ansatz prescription on the dual CFT side (see for instance~\cite{Benini:2018ywd,Aharony:2021zkr}).} However, applying localization to supergravity involves major conceptual and technical challenges: for instance, it requires an off‑shell realization of supersymmetry,\footnote{In supergravity this is achieved using superconformal gravity after appropriate gauge fixing; see~\cite{Freedman:2012zz}. This formalism will be reviewed later in this Thesis in a different context, namely to construct higher‑derivative supergravity theories (see~\ref{part_two}).} as well as a consistent gauge‑fixing of local symmetries implemented through a deformed BRST algebra~\cite{deWit:2018dix,Jeon:2018kec}. Nonetheless, this framework has been successfully applied to compute the exact quantum entropy function, based on the AdS$_2$/CFT$_1$ correspondence~\cite{Sen:2008yk,Sen:2008vm}. In this case, the exact AdS$_2$ path integral with microcanonical boundary conditions computes the dimension of the putative Hilbert space of the extremal black holes in the theory, that have an AdS$_2$ near‑horizon geometry. Using this formalism, the computation can be carried out to remarkable precision~\cite{Iliesiu:2022kny}, and the resulting gravitational path integral reproduces the full expansion of the relevant microscopic index, hence the exact integer number of BPS states using only gravitational variables. Despite this success, many aspects of this application remain only partially understood~\cite{Sen:2023dps}, and extending these techniques to more general settings is still an open problem. 

In this Thesis we focus on the gravitational index in the semiclassical approximation, where it can be systematically expressed as a sum over classical saddles. Within this framework, several questions arise, some of which will be addressed in the following chapters. These include:
\begin{enumerate}
\item Which saddle, if any, captures the contribution of supersymmetric black holes to the index?
\item What other saddles compete with it, and which phase dominates for a given set of parameters?
\item What role do complex saddles play in Euclidean quantum gravity, and what are the rules to deal with them?
\item Is there a systematic way to compute on‑shell actions (and hence semiclassical contributions to the path integral) using only limited data such as topology and boundary conditions?
\item How can quantum corrections be computed systematically around each saddle, and to what extent can these macroscopic results be compared with microscopic predictions?
\end{enumerate}


\section{Outline and summary of results}

In the previous sections we elaborated on the historical developments of black hole microstate counting, moving from the earliest results to more recent insights. A central theme that has emerged is the role of supersymmetric indices as reliable tools for counting microstates, together with their gravitational counterparts: gravitational indices, computed from the gravitational path integral with suitable boundary conditions. The present Thesis focuses on recent advances in the study of the gravitational index from two complementary points. On one hand, in the holographic setting, where powerful predictions from dual conformal field theories provide benchmarks that guide the gravitational computation; on the other hand, in asymptotically flat spaces, where no dual holographic prediction is available. While the literature in this area has recently grown rapidly, this Thesis will concentrate on developments directly connected to the author’s contributions. We will focus on five-dimensional supergravity as a particularly rich arena in which to test predictions for quantum gravity. However, many of the methods and insights developed here admit natural generalizations to other spacetime dimensions.

The main objectives of the Thesis, addressing some of the fundamental questions raised earlier, are as follows:
\begin{description}
\item[Part \ref{part_two}] We investigate higher-derivative corrections to the entropy and on-shell action of AdS$_5$ black holes. These corrections are shown to reproduce precisely the subleading terms in the large-$N$ expansion of the superconformal index, in an appropriate regime of chemical potentials. This remarkable match offers a {\it precision test of the AdS/CFT correspondence}.
\item[Part \ref{part_three}] We construct and analyze new complex Euclidean saddles with flat asymptotics that provide competing contributions to the gravitational path integral that computes a supersymmetric index, shedding light on the intricate structure of the gravitational index. 
\end{description}
Pursuing these goals requires engaging with deep concepts and developing novel technical tools for the study of supergravity and quantum black holes. Along the way, we will review and highlight some of these methods, contributing to the expansion of the theoretical toolkit available for future investigations. 

\medskip 

This Thesis is organized into four parts. Part I is completed by chapter~\ref{sec:background}, which collects the background material necessary for the developments that follow. In this chapter we review recent results in the field-theoretic computation of the large-$N$ expansion of the relevant superconformal index, whose logarithm provides sharp predictions for the on-shell action of supersymmetric AdS$_5$ black holes, including subleading corrections in the large-$N$ expansion. We then begin to develop the gravitational counterpart, following~\cite{Cabo-Bizet:2018ehj}, by discussing the construction of the Euclidean saddle that captures the contribution of the supersymmetric AdS$_5$ black hole to the superconformal index. These saddles are supersymmetric, yet non-extremal (in a sense to be clarified later), rotating and complex solutions of the Euclidean field equations, with chemical potentials constrained by a specific complex linear relation of the form~\eqref{eq:constraint_gauged}. 
Parts~\ref{part_two} and~\ref{part_three} are each divided into three chapters. In both cases, the first chapter provides an introduction and a review of the technical tools required, while the subsequent chapters present original results based on~\cite{Cassani:2023vsa,Cassani:2024tvk,Cassani:2024kjn,Cassani:2025iix}. We draw our conclusions in part~\ref{part_four}. 

We provide here a summary of results for each chapter of parts~\ref{part_two} and~\ref{part_three}:

\paragraph{Part~\ref{part_two}:}

\begin{description}
\item[Chapter~\ref{chap:intro_2}, based on~\cite{Cassani:2022lrk}.] 
We study four-derivative corrections to five-dimensional minimal gauged supergravity. We explain how {\it off-shell superconformal techniques} can be used to discuss systematically higher-derivative corrections to the supergravity action. We focus on four-derivative terms and integrate out the auxiliary fields to obtain a consistent supergravity, exploiting the freedom to implement field redefinitions to recast this action in a much simpler form. We then evaluate the on-shell action of the AdS$_5$ black hole solution with two independent angular momenta and one electric charge at linear order in the corrections. After imposing supersymmetry, we are able to recast the action in terms of the supersymmetric chemical potentials and match the result obtained from the dual superconformal index in a suitable Cardy-like regime, that we specify. To achieve this, we use the fact that the two-derivative solution is enough. We use the on-shell action to determine the corrections to the black hole thermodynamics, including those to the entropy and the charges. We then specialize to the supersymmetric and extremal case and find a simple expression for the microcanonical entropy. 
%
\item[Chapter~\ref{sec:structure_2}, based on~\cite{Cassani:2023vsa}.]
We address some issues in higher-derivative gauged supergravity with Chern-Simons terms, focusing on the five-dimensional case. We discuss the variational problem with Dirichlet boundary conditions as well as holographic renormalization in asymptotically locally AdS spacetimes, and derive the corresponding boundary terms. We then employ Wald's formalism in order to define conserved charges associated to local symmetries (diffeomorphisms and U(1) gauge transformations), taking into account the effect of generic gauge Chern-Simons terms. 
These lead us to distinguish between Noether charges and Page (or Komar) charges which satisfy the Gauss law. We make use of the latter to compute corrections to the angular momentum and electric charge of the supersymmetric black hole in AdS$_5$ from its corrected near-horizon geometry, that we construct. This also allows us to derive the microcanonical form of the entropy, computed through Wald formula, as a function of the conserved charges relying entirely on the near-horizon geometry. 
%
\item[Chapter~\ref{chap:flavour}, based on~\cite{Cassani:2024tvk}.]
A Cardy-like regime of the four-dimensional superconformal index has been shown to be governed by 't Hooft anomalies and to single out a large-$N$ saddle carrying the Bekenstein-Hawking entropy of dual supersymmetric black holes in AdS$_5$. 
Here, we match the first subleading corrections to the saddle-point action with the four-derivative corrections to the black hole action including flavour symmetries. We consider five-dimensional gauged supergravity with vector multiplet and four-derivative couplings, and provide an effective theory reproducing the 't Hooft anomalies of the R- and flavour symmetries of generic holographic superconformal field theories at next-to-leading order in the large-$N$ expansion. Then we focus on a specific model dual to $\mathbb{C}^3/\mathbb{Z}_\nu$ quiver gauge theories, where the 't Hooft anomaly coefficients receive simple but sufficiently generic corrections. In this model, we evaluate the four-derivative corrections to the on-shell action of the supersymmetric multi-charge black hole, showing agreement with the flavoured Cardy-like formula from the index. We give a prediction for the corrected entropy of the supersymmetric black hole and discuss the general validity of our results. 
%
\end{description}

\paragraph{Part \ref{part_three}:}

\begin{description}
\item[Chapter~\ref{sec:intro3}, partially based on~\cite{Cassani:2024kjn}.]
We review recent developments in the application of {\it equivariant localization} in supergravity, and investigate localization of the gravitational on-shell action in odd dimensions, focusing on five-dimensional ungauged supergravity, with and without supersymmetry. We analyze the conditions for cancellation of boundary terms, so that the full action integral is given in terms of contributions from the fixed sub-manifolds of certain spacetime isometries, i.e. the odd-dimensional analog of the nuts and bolts of Gibbons-Hawking. 
\item[Chapter~\ref{chap:Black_hole}, based on~\cite{Cassani:2024kjn}.]
%
We construct asymptotically flat Euclidean supersymmetric non-extremal solutions with two independent rotations and an arbitrary number of electric charges, providing black hole saddles of the gravitational path integral that computes a supersymmetric index, and evaluate their action equivariantly. We find that these Euclidean saddles interpolate between supersymmetric extremal black holes and two-center horizonless microstate geometries. The interpolation involves dialing the temperature and implementing different analytic continuations. The corresponding on-shell action does not depend on temperature but is affected by the analytic continuations.
\item[Chapter~\ref{chap:bubbling}, based on~\cite{Cassani:2025iix}.]
We uncover a wealth of five-dimensional semiclassical saddles of the gravitational index with bubbling topology. These are complex finite-temperature  configurations asymptotic to $S^1\times\mathbb{R}^4$, solving the supersymmetry equations. We assume a ${\rm U}(1)^3$ symmetry given by the thermal isometry and two rotations, and present a general construction  based on a rod structure specifying the fixed loci of the ${\rm U}(1)$ isometries and their three-dimensional topology. These fixed loci may correspond to multiple horizons or three-dimensional bubbles, and they may have $S^3$, $S^2\times S^1$, or lens space topology. Allowing for conical singularities gives additional topologies involving spindles and branched spheres or lens spaces. As a particularly significant example, we analyze in detail the configurations with a horizon and a bubble just outside of it. We determine the possible saddle-point contribution of these configurations to the gravitational index by evaluating their on-shell action and the relevant thermodynamic relations. We also spell out two limits leading to well-definite Lorentzian solutions. The first is the extremal limit, which gives the BPS black ring and black lens solutions. The on-shell action and chemical potentials remain well-definite in this limit and should thus provide the contribution of the black ring and black lens to the gravitational index. The second is a limit leading to horizonless bubbling solutions, which have purely imaginary action. 
\end{description}


%% file: background.tex
\chapter{Background: Complex saddles and microstates of AdS$_5$ black holes}
\label{sec:background}

As we argued above, holography offers a framework to understand quantum gravity in asymptotically AdS spacetime via a dual CFT. In this section, our aim is to investigate the gravitational partition function with suitable supersymmetric boundary conditions (that is, a gravitational index), which, according to the master equation of AdS/CFT, is computed by the superconformal index of a dual four-dimensional SCFT. Within this setup, the CFT predictions provide valuable guidance for exploring quantum corrections to black hole entropy and the structure of the gravitational index.

In this and the following chapters we focus on AdS$_5$ black hole, which naturally lead us to consider a class of holographic $D=4$ ${\cal N}=1$ SCFTs. These include, in particular, the theories arising as the low energy limit of the worldvolume dynamics of D$3$-branes at the tip of the Calabi-Yau cone with Sasaki-Einstein base. Any such theory possesses a superconformal anomaly governed by the two central charges characterizing four-dimensional CFTs, $\aa,\cc$. As in two-dimensional CFTs, the central charges can be extracted by coupling the theory to a curved background metric. In this case, $\aa$ and $\cc$ appear in the trace of the energy-momentum tensor $T_{ij}$ and in the divergence of the R-current $J^i$, as~\cite{Anselmi:1997am,Cassani:2013dba}
\begin{align}\label{correcttrace}
 T_i{}^i  \, & =\, \,- \,
\frac{\aa}{16\pi^2}  \hat E + \frac{\cc}{16\pi^2}  \hat{C}^2    \, - \,  \frac{\cc}{6\pi^2}\,   \hat{F}^2  ~, \\[2mm]
\nabla_i  J^i \, & =\, \frac{\cc-\aa}{24\pi^2}\, \frac{1}{2}\,\epsilon^{ijkl}\hat{R}_{ijab} \hat{R}_{kl}{}^{ab}\, + \,\frac{5\aa-3\cc}{27\pi^2} \,\frac{1}{2}\,\epsilon^{ijkl}\hat{F}_{ij} \hat{F}_{kl}\,,
 \label{correctchiral}
 \end{align}
where $\hat C^2$ and $\hat E = \hat R_{ijkl}\hat R^{ijkl}-4 \hat R_{ij}\hat R^{ij}+\hat R^2$ denote the Weyl$^2$ and Euler invariants of the four-dimensional background geometry, respectively, while $\hat F_{ij}$ is the field strength of the background U(1) gauge field that canonically couples to the R-current.

The second equation shows that in four-dimensional SCFTs the central charges are directly related to the global anomaly coefficients of the R-symmetry. Indeed, comparing with the general expression for the anomaly of a U(1) current with cubic and linear coefficients denoted by ${\rm Tr}\mathcal{R}^3$ and ${\rm Tr}\mathcal{R}$, 
\be
\nabla_i  J^i  \, =\,-\frac{{\rm Tr}\mathcal{R}}{384\pi^2}  \, \frac{1}{2}\,\epsilon^{ijkl}\hat{R}_{ijab} \hat{R}_{kl}{}^{ab}  + \frac{ {\rm Tr}\mathcal{R}^3 }{48\pi^2}\, \,\frac{1}{2}\,\epsilon^{ijkl}\,\hat{F}_{ij} \hat{F}_{kl}\,,
\ee
one finds the relations
\be\label{relacTrR}
\aa = \frac{3}{32}(3\,{\rm Tr}\mathcal{R}^3 - {\rm Tr}\mathcal{R})\,, \qquad \cc = \frac{1}{32}(9\,{\rm Tr}\mathcal{R}^3 - 5\, {\rm Tr}\mathcal{R})\,.
\ee
While additional flavour symmetries may also be present (and can be anomalous), their properties depend on the specific theory at hand. By contrast, the relations above for the superconformal R-symmetry are universal.

\medskip

In recent years, significant progress has been made in defining and analyzing the SCFT partition function that counts the microstates of supersymmetric black holes in AdS spaces (see, e.g.~\cite{Closset:2013vra}). Our interest lies in cases where the microscopic index exhibits an exponential growth with some power of $N$, i.e. $\log {\cal I} \sim N^\gamma$ (with $\gamma>0$), which signals the presence of a black hole with macroscopic entropy in the gravitational dual. However, the original index~\eqref{eq:SCI_basic}, introduced in~\cite{Romelsberger:2005eg,Kinney:2005ej}, was found to behave as ${\cal O}(1)$ at large $N$, failing to reproduce the entropy of the supersymmetric AdS$_5$ black holes of~\cite{Gutowski:2004yv,Gutowski:2004ez,Chong:2005da,Chong:2005hr,Kunduri:2006ek}. However, this early result was obtained under the assumption of real fugacities, and the mismatch is resolved once complex fugacities are allowed.
Indeed, starting with~\cite{Cabo-Bizet:2018ehj,Choi:2018hmj,Benini:2018ywd}, 
a family of (complex) large-$N$ saddles of the superconformal index has been shown to correspond to supersymmetric black holes, establishing that the gravitational partition function matches the functional integral of the corresponding dual field theory at large $N$. 

%
In the regime of complex chemical potentials, the relevant SCFT generating function becomes a branched function of $\omega_1$ and $\omega_2$, with different branches related by the substitution $\omega_1 \rightarrow \omega_1 -2\pi \ii$. As reviewed below, isolating the black hole contribution requires moving to the \emph{second sheet} of this branched function and taking a Cardy-like limit of small chemical potentials $\omega_{1,2} \rightarrow 0$~\cite{Cassani:2021fyv}. Combined with the large-$N$ limit, in this regime the logarithm of the index yields a prediction for the black hole on-shell action, including subleading corrections in the large-$N$ expansion. 

\medskip

After briefly reviewing the relevant results on the gauge-theory side of the duality, we begin the study of the gravitational counterpart, following the analysis of~\cite{Cabo-Bizet:2018ehj}. Our goal is to understand how the Euclidean black hole saddle (with the appropriate boundary conditions), capturing the contribution of the general supersymmetric AdS$_5$ black hole to the gravitational index, can be constructed. We will show that the on-shell action of this saddle reproduces the leading contribution to the large-$N$ expansion of the logarithm of the superconformal index in the Cardy limit on the second sheet, and that the black hole entropy emerges from a constrained Legendre transform of the black hole action. This background analysis provides the foundation for the developments of the following chapters, on which this Thesis is built.

\medskip

In sec.~\ref{sec:field_theory_results}, we begin by reviewing the field-theoretic results that are essential for the subsequent gravitational analysis. We extend the known expressions to account for an arbitrary number of flavour symmetries, while incorporating finite-$N$ corrections. These generalizations are obtained through equivariant integration of the anomaly polynomial, with the details presented in app.~\ref{sec:field_th_section}. For the remainder of the chapter, however, we restrict to the leading large-$N$ term and switch off all flavour symmetries, thereby isolating the universal sector of holographic ${\cal N}=1$ SCFTs. We then turn to the gravitational counterpart of this analysis: in sec.~\ref{sec:grav_setup_0} we introduce the relevant gravitational index, while secs.~\ref{sec:Lor_susy_BH_5D} and~\ref{TwoDerReview} review the construction of the Euclidean saddle that captures the contribution of the supersymmetric AdS$_5$ black hole to this index.

\section{Summary of field theory results}
\label{sec:field_theory_results}

We start our analysis in field theory. We consider a four-dimensional $\mathcal{N}=1$ SCFT on the spatial manifold $S^3$.  We choose a supercharge $\mathcal{Q}$ satisfying the commutation relations 
\be\label{eq:comm_charges}
[J_{1},\mathcal{Q}]  = [J_{2},\mathcal{Q}] =  \frac12  \mathcal{Q}\,,\qquad [Q_I,\mathcal{Q}] = -r_I \mathcal{Q}\,,
\ee
where $J_1$, $J_2$ are the angular momenta generating the Cartan subalgebra of the ${\rm SO}(4)$ isometry of $S^3$, while $Q_I$, $I=1,\ldots,n+1$, are conserved charges made of linear combinations of the superconformal R-charge and the $n$ Abelian flavour charges that the theory may admit. In our conventions, the eigenvalues $r_I$ vanish if $Q_I$ is a flavour symmetry and  take the value $r_I = 1$ if it is a canonically normalized R-symmetry.
The superalgebra also involves the anticommutator
\begin{equation}
\label{eq:AdS_superalgebra}
\{{\cal Q},\,\bar {\cal Q}\} = E - \Omega_1^* J_1 - \Omega_2^* J_2 - \Phi^I_* Q_I\,,
\end{equation}
where $\Omega_{1,2}^*$, $\Phi^I_*$ are certain constant coefficients. 

The refined superconformal index can be defined as 
\be\label{our_flavored_index}
\mathcal{I} \,=\,   {\rm Tr}\,  \rme^{\pi i (1+n_0) {\rm F}} \, \rme^{-\beta \{\mathcal{Q},\bar{\mathcal{Q}}\}+
\omega_1  J_1 + \omega_2  J_2 + \varphi^{I} Q_{I} } \,,
\ee
with the constraint
\be\label{eq:linearconstraint_gen} 
\omega_1+ \omega_2  - 2 r_I \varphi^I \,=\, 2\pi i  n_0\,,\qquad n_0\in \mathbb Z\,,
\ee 
where the trace is taken over the Hilbert space of the theory on $S^3$, $\rm F$ is the fermion number, $n_0$ is an integer, and the \emph{complex} variables $\beta$, $\omega_1$, $\omega_2$, $\varphi^I$ are chemical potentials for the respective charges. The index $\mathcal{I}$ does not depend on $\beta$ and is a holomorphic function of $\omega_1,\omega_2,\varphi^I$, subject to the constraint \eqref{eq:linearconstraint_gen}. The latter is a supersymmetry condition, ensuring that the combination $\rme^{\pi i n_0 F} \, \rme^{\omega_1  J_1 + \omega_2  J_2 + \varphi^{I} Q_{I} }$ entering in \eqref{our_flavored_index} commutes with the supercharge. The integer $n_0$ was introduced in~\cite{Cabo-Bizet:2018ehj,Cabo-Bizet:2019osg}: although it can be removed by shifting e.g.\ $\omega_1\to\omega_1+2\pi i n_0$ so as to reach the standard definition of the index \eqref{eq:SCI_basic}, with a $(-1)^F$ insertion as originally formulated in~\cite{Romelsberger:2005eg,Kinney:2005ej} (which is valid for $n_0=0$, consistently with having real chemical potentials),  we find it convenient to keep it as, when set to $n_0=\pm 1$, it makes it manifest that the index can be obtained as a continuous limit of a non-supersymmetric, thermal partition function. This will be useful later when comparing with the gravitational partition function. 


At large-$N$, the superconformal index displays an intricate structure of complex saddles, including one that carries the entropy of the dual supersymmetric black hole in AdS$_5$, which may therefore be called the \emph{black hole saddle}. A convenient way to  isolate the contribution of the black hole saddle is to take a Cardy-like limit of small chemical potentials  $\omega_1,\omega_2 \to 0$, after setting $n_0=\pm 1$.  This Cardy-like limit is independent of the large-$N$ limit, therefore the expressions obtained for the index in this regime are valid at finite $N$ and allow for a large-$N$ expansion.
In microcanonical terms, the Cardy limit corresponds to a regime where the contributions to the index from dual states with sufficiently large charges become dominant and isolated. This Cardy-like limit can be viewed as the higher-dimensional analogue of the high-temperature Cardy limit originally formulated for two-dimensional CFTs. For two-dimensional theories, Cardy famously showed that the high-temperature asymptotics are governed by the two-dimensional trace anomaly~\cite{Cardy:1986ie}. 
It has been shown in a number of papers \cite{Choi:2018hmj,Honda:2019cio,ArabiArdehali:2019tdm,Kim:2019yrz,Amariti:2019mgp,Cabo-Bizet:2019osg,Gadde:2020bov,GonzalezLezcano:2020yeb,Goldstein:2020yvj,Amariti:2020jyx,Amariti:2021ubd,Cassani:2021fyv,ArabiArdehali:2021nsx} with progressively increasing accuracy, that also the index for the four-dimensional theory after the Cardy-like limit is controlled by the relevant anomalies, that are the cubic and linear 't Hooft anomalies of the SCFT conserved global currents. The asymptotic formula for the index in this regime is derived in appendix~\ref{sec:field_th_section} via \emph{equivariant integration} of the anomaly polynomial, following the approach of~\cite{Ohmori:2021dzb}.
By choosing $n_0=\pm 1$, corresponding to two equivalent saddles, in the constraint \eqref{eq:linearconstraint_gen}, 
\be\label{eq:linearconstraint} 
\omega_1+ \omega_2  - 2 r_I \varphi^I \,=\, \pm 2\pi i \,,
\ee  
the index in this regime is given by
\be\label{eq:logI}
\log \mathcal{I} \,=\, -I + \ldots\,,
\ee
\be\label{eq:index_asympt}
I \, =\, \frac{k_{IJK}\, \varphi^I\varphi^J\varphi^K}{6\,\omega_1\omega_2} - k_I \varphi^I \,\frac{\omega_1^2+\omega_2^2-4\pi^2}{24\,\omega_1\omega_2} \,,
\ee
where
\begin{equation}\label{def_anomaly_coeff}
k_{IJK}= {\rm{Tr}}\, (Q_I  Q_J Q_K)\,, \hspace{1cm} k_I={\rm Tr}\, Q_I\, 
\end{equation}
are the cubic and linear 't Hooft anomaly coefficients for the charges $Q_I$. This expression extends the flavoured formula given in~\cite{Kim:2019yrz} by including the non-divergent, polynomial terms in the regime of small $\omega_1,\omega_2$.

Eq.~\eqref{eq:index_asympt} should provide the exact action of the saddle of interest. The dots in \eqref{eq:logI} indicate that we are omitting terms that are exponentially suppressed in the limit $\omega_1,\omega_2 \to 0$, as well as a logarithmic term $\log |\mathcal{G}|$, where $|\mathcal{G}|$ is the order of a discrete one-form symmetry $\mathcal{G}$ that the theory may have, which is related to the degeneracy of saddles. The expression should apply both to Lagrangian and non-Lagrangian theories, not necessarily holographic, and is valid at finite $N$. In particular, it should be possible to extract higher-derivative corrections to the $N\rightarrow +\infty$ result by studying the large-$N$ expansion order-by-order. 

Explaining how the field theory result \eqref{eq:index_asympt} matches the gravitational on-shell action of the black hole saddle is one of the main focuses of this Thesis. For a standard holographic SCFT, in the large-$N$ expansion the 't Hooft anomaly coefficients can be expanded as
\begin{equation}
k_{IJK} = k_{IJK}^{(0)} + ... \,,\qquad k_I = ...\,,
\end{equation}
where $k_{IJK}^{(0)}$ denotes the leading-order anomaly coefficients,
while the ellipsis denote possible subleading terms in the large-$N$ expansion. The leading-order in this expansion is captured by the contribution to the gravitational path integral of the two-derivative on-shell action of gauged supergravity coupled to vector multiplets~\cite{Cassani:2019mms}. Indeed, according to the standard rules of AdS/CFT~\cite{Witten:1998qj}, global symmetries in the boundary theory correspond to gauge symmetries in the bulk theory. Thus, the flavoured index \eqref{our_flavored_index} captures the thermodynamic features of black holes carrying multiple electric charges, that are solutions to gauged supergravity coupled to vector multiplets. 
On top of that, the first subleading terms the large-$N$ expansion of \eqref{eq:index_asympt} are reproduced by four-derivative corrections to the gravitational on-shell action. This setup has been studied in~\cite{Cassani:2024tvk}, and will be the focus of part \ref{part_two}. 

\paragraph{Universal sector of ${\cal N}=1$, $D=4$ SCFT.} The result \eqref{eq:index_asympt} extends the Cardy-like formula obtained in~\cite{Cassani:2021fyv} to the case where flavour potentials are turned on. Turning them off corresponds to restricting the analysis to a \emph{universal sector} common to any holographic ${\cal N}=1$ SCFT, where only the R-charge and its corresponding chemical potential are present. The index in this case may be written as~\cite{Cabo-Bizet:2018ehj,Choi:2018hmj,Kim:2019yrz,Cabo-Bizet:2019osg}:
\begin{equation}\label{eq:R_index}
{\cal I} = {\rm Tr}\,{\rm e}^{-\ii\pi Q}{\rm e}^{-\beta\{{\cal Q},\,\bar{\cal Q}\}+ \omega_1\left( J_1 + \frac{1}{2}Q\right)+ \omega_2 \left( J_2 + \frac{1}{2}Q\right)} \,,
\end{equation}
where now $Q$ denotes the U$(1)$ R-charge. Indeed, \eqref{our_flavored_index} maps into \eqref{eq:R_index} by performing the replacement $\varphi^I Q_I \to \varphi Q $, where $\varphi = r_I \varphi^I$ is the R-symmetry chemical potential,\footnote{This replacement is derived by first going to a basis that isolates the R-symmetry from the flavour symmetries. Given any R-charge $Q = s^I Q_I$, with $r_Is^I=1$, we can apply the projectors \eqref{projectors_Rsymm} and  obtain the decomposition $Q_I = r_I Q + \widetilde{Q}_I$, where the $\widetilde{Q}_I = (\delta_I{}^J - r_I s^J)Q_J$ are flavour charges. Analogously, we can decompose the chemical potentials as $\varphi^I = s^I \varphi + \widetilde\varphi^I$, with  $\varphi = r_I \varphi^I$ and $\widetilde{\varphi}^I = (\delta^I{}_J - s^I r_J)\varphi^J$. It follows that $\varphi^I Q_I = \varphi Q + \widetilde \varphi^I \widetilde{Q}_I$. By turning off the flavour potentials, $\widetilde{\varphi}^I= 0$, we obtain the replacement above.\label{foot:change_basis}} together with the choice $n_0 = \pm 1$, and finally using the constraint \eqref{eq:linearconstraint_gen}, which in this case becomes
\begin{equation}
\omega_1 + \omega_2 - 2\varphi = \pm 2\pi \ii,,
\end{equation}
to eliminate $\varphi$. The combination of global charges that refine the index, $J_{1,2} + \frac{1}{2}Q$, commute with the supercharges, as seen from the superalgebra \eqref{eq:comm_charges}. Note that as a consequence of the linear constraint, the $\left(-\right)^{\rm F} = {\rm e}^{2\pi \ii J_1}$ factor, familiar from the Witten index \eqref{eq:Witten_index} by using the spin-statistics theorem, has been replaced by ${\rm e}^{-\ii\pi Q}$.\footnote{For this reason, the index was also dubbed  {\it R-charge index} or {\it index on the second sheet} in~\cite{Cassani:2021fyv}. As we mentioned above, we can go back to the standard Witten index by taking the transformation $\omega_1 \rightarrow \omega_1 + 2\pi \ii$.} Due to the superalgebra \eqref{eq:comm_charges}, bosonic and fermionic states in the same supermultiplet are still counted with opposite signs in the trace thanks to the phase ${\rm e}^{-\ii\pi Q}$, as required for an index. Using a three-dimensional field theory approach, equivalent to considering the Euclidean quantum field theory obtained after compactifying the Euclidean time circle in the $S^1\times S^3$ background, it was shown in~\cite{Cassani:2021fyv} that the superconformal index \eqref{eq:R_index} in the Cardy-like limit can be expressed as
\begin{equation}\label{eq:log_I}
\log {\cal I} = - {\rm Tr}{\cal R}^3 \frac{\varphi^3}{6\,\omega_1\omega_2} + {\rm Tr}{\cal R}\, \varphi\,\frac{\omega_1^2 + \omega_2^2 - 4\pi^2}{24\,\omega_1 \omega_2} + .\,.\,.\,,
\end{equation}  
where ${\rm Tr}{\cal R}^3= k_{RRR}$ and ${\rm Tr}{\cal R}= k_R$ denote the cubic and linear R-symmetry anomaly coefficients. Again, since this expression is valid at finite $N$, its large-$N$ expansion provides a prediction for higher-derivative (and possibly quantum, if the neglected terms were considered) corrections to the semiclassical black hole contribution to the gravitational partition function.  

An ensemble of states carrying only R-charge is \emph{universal}, in the sense that it can be found in any holographic ${\cal N}=1$, $D=4$ SCFT. Correspondingly, the needed ingredients to develop the gravitational counterpart of \eqref{eq:log_I} are all contained in five-dimensional minimal gauged supergravity (and its supersymmetric four-derivative corrections). Indeed, also black hole solutions to this minimal gauged supergravity are universal, since they can be embedded in any compactification admitting a supersymmetric AdS$_5$ vacuum~\cite{Gauntlett:2007ma,Cassani:2019vcl}. 

\medskip

In this chapter, we review how the Cardy limit of the logarithm of the superconformal index at leading order in the large-$N$ expansion is reproduced by the semiclassical contribution to the gravitational path integral of the supersymmetric, electrically charged, and rotating black hole solution to five-dimensional gauged supergravity. We focus on the universal sector captured by the index \eqref{eq:R_index}, following~\cite{Cabo-Bizet:2018ehj}. This simplified setup allows us to introduce the gravitational index relevant for computing the superconformal index and to analyze the associated saddle that capture the black hole contribution in a clearer and more tractable context, that is five-dimensional minimal gauged supergravity. This provides a more transparent starting point before moving on to the more intricate case of gauged supergravity coupled to vector multiplets. This, and the role of four-derivative corrections, which give rise to the first subleading terms in the large-$N$ expansion, will instead be addressed in part~\ref{part_two}.

\section{The gravitational setup}
\label{sec:grav_setup_0}

In this section, we introduce the necessary ingredients to develop the gravitational counterpart of the index \eqref{eq:R_index}.

We begin with the gravitational path integral \eqref{eq:grav_path_int}, which, under appropriate boundary conditions, admits the interpretation of a grand-canonical partition function:
\begin{equation}\label{eq:partition_function_gravity_5D}
Z_{\rm grav}\left( \beta,\,\Omega_{1,2},\,\Phi\right) = {\rm Tr}\left[{\rm e}^{-\beta \left( E -\Omega_1 J_1 - \Omega_2 J_2 - \Phi Q\right)} \right]\,.
\end{equation}
Here, the inverse temperature $\beta$, the angular velocities $\Omega_{1,2}$ and the electrostatic potential $\Phi$ are defined as in sec.~\ref{sec:Grav_path_int}. 

Our goal is to manipulate this expression to obtain a gravitational index, which, via the AdS/CFT correspondence, should reproduce the dual superconformal index. To this end, we consider a five-dimensional gauged supergravity theory endowed with a complex supercharge ${\cal Q}$ compatible with the superalgebra \eqref{eq:supergravity_algebra} by taking
\begin{equation}\label{eq:minimal_sugra_coeff}
\Omega_1^* = \Omega_2^* = g\,,\qquad \Phi_* = \frac{3}{2}g\,,\qquad r =1\,,
\end{equation}
where $g>0$ denotes the gauging parameter, that determines the radius of the supersymmetric AdS$_5$ vacuum admitted by the theory, $\ell_{\rm AdS} = g^{-1}$. 
Supersymmetric solutions are annihilated by this supercharge, and saturate the BPS bound:
\begin{equation}\label{eq:BPS_bound_5D_sugra}
E = g J_1 + g J_2 + \frac{3}{2}gQ\,.
\end{equation}

Following the steps highlighted in sec.~\ref{sec:Grav_path_int}, we introduce the redefined chemical potentials \eqref{eq:red_chem_pot}, constrained by\footnote{There is a slightly different way to impose supersymmetry that leads to the replacement $\ii\to -\ii$ in the linear constraint. The second case yields a distinct solution, but its analysis is completely analogous to the one we are presenting.}
\begin{equation}\label{eq:linear_constraint_minimal}
    \omega_1 + \omega_2 - 2\varphi = 2\pi\ii\,.
\end{equation}

Using this linear constraint to eliminate the dependence on $\varphi$, the partition function finally reproduces the superconformal index \eqref{eq:R_index},
 \begin{equation}\label{eq:gravitational_index_AdS}
    Z_{\rm grav} \left( \omega_{1,2}\right) ={\rm Tr}\,e^{-\ii\pi Q}e^{-\beta\{{\cal Q},\,\bar{\cal Q}\} + \omega_1 \left(J_1 + \frac{1}{2}Q\right) + \omega_2 \left(J_2 + \frac{1}{2}Q\right)}\,.
\end{equation}
The linear constraint \eqref{eq:linear_constraint_minimal} must be complex to produce the correct phase ${\rm e}^{-\ii\pi Q}$ in the trace, suggesting that the relevant gravitational saddles will involve complex configurations. 
Finally, the microscopic trace \eqref{eq:gravitational_index_AdS} is expected to be convergent provided we take
\begin{equation}
{\rm Re} \beta >0\,,\qquad {\rm Re}\,\omega_{1,2} <0\,.
\end{equation}

We conclude that with these supersymmetric boundary conditions, the gravitational path integral \eqref{eq:grav_path_int} computes a gravitational index.

To study this index, we work in the semiclassical approximation, where the path integral can be expanded a sum over saddle points. The problem, then, reduces to classifying all smooth Euclidean saddles compatible with the supersymmetric boundary conditions and comparing their on-shell actions. The relevant saddles are smooth, supersymmetric (i.e. admitting a globally-defined Killing spinor) Euclidean solutions to the supergravity equations of motion. Furthermore, the requirement that $Z_{\rm grav}$ is defined for arbitrary $\beta$ (and actually a finite $\beta$ is required for convergence of the trace) implies that the Euclidean saddles must possess a compactified Euclidean time with finite period. We refer to such solutions as non-extremal or finite-temperature, in analogy with the Lorentzian interpretation of $\beta$. Supersymmetric, yet non-extremal saddles can exist in Euclidean signature, provided one allows for complex configurations. Depending on the setup, this complexification may affect the metric, the conserved charges, or both, consistently with the fact that the chemical potentials are constrained by some complex relation \eqref{eq:linear_constraint_minimal}. As we will see, the on-shell action of such saddles, representing their semiclassical contribution to the gravitational path integral, is holomorphic in the complex chemical potentials and independent of $\beta$, as expected for a supersymmetric index.

Since the relevant saddles arise as solutions of five-dimensional minimal gauged supergravity~\cite{DAuria:1981yvr,Gunaydin:1984ak}, we provide a brief review of its main features.
\paragraph{${\cal N}=2$, $D=5$ minimal gauged supergravity.} 
Minimal ${\cal N}=2$ supergravity theory in five dimensions involves the supergravity multiplet, which contains the following fields:
\begin{itemize}
\item a f\"unfbein, $e^a_\mu$, with flat and curved spacetime indices $a=0,1,...,4$ and $\mu=0,1,...,4$, respectively;
\item an SU$(2)_R$ doublet of symplectic Majorana gravitini $\psi_\mu^i$, $i = 1,2$; 
\item a gravitphoton $A_\mu$, i.e. an Abelian gauge field.
\end{itemize}
To construct a gauged supergravity, a subgroup of the global R-symmetry is promoted to a local (gauge) symmetry. In the minimal five-dimensional theory, one typically gauges a U$(1)_R$ subgroup of the SU$(2)_R$ R-symmetry. This gauging is implemented by coupling the gravitini to the graviphoton, which serves as the gauge field for the U$(1)_R$ symmetry. 
The gauging procedure, then, replaces ordinary covariant derivatives acting on the gravitini with U$(1)_R$ covariant ones,
\begin{equation}
\nabla_\mu \psi_\nu^i \rightarrow D_\mu \psi_\nu^i = \nabla_\mu \psi_\nu^i - A_\mu \delta^{ij} \psi_\nu{}_i\,,
\end{equation}
The gravitini thus acquire charge under the Abelian gauge symmetry. Replacing the ordinary derivatives in the ungauged action by these covariant derivatives breaks its supersymmetry invariance, which is then restored by adding appropriate mass-like terms for the gravitini, a negative cosmological constant term, as well as by modifying the gravitini supersymmetry transformations.  In holography, this U$(1)_R$ gauge symmetry realizes in the bulk the R-symmetry of the dual SCFT.

This was the approach taken in~\cite{Gunaydin:1984ak} to construct the five-dimensional minimal gauged supergravity. An alternative and distinct construction employs conformal supergravity and superconformal tensor calculus~\cite{Bergshoeff:2004kh}, a framework that turned out to be particularly effective for systematically generating four-derivative corrections to the two-derivative action. For this reason this will be reviewed in sec.~\ref{sec:from_superconformal_to_supergravity}.

The bosonic action of the minimal gauged theory, in Euclidean signature, reads\footnote{
In this Chapter, we use an unconventional normalization for $A$, chosen so that the electric charge of this gauge symmetry matches the canonically normalized R-charge of the dual SCFT, facilitating comparisons with dual results. In these conventions, the limit to reach ungauged supergravity with canonical normalization must be taken with care: first one should rescale the gauge field $A \rightarrow \frac{\sqrt{3}g}{2}A$ (such that, $Q \rightarrow \frac{2}{\sqrt{3}g}Q$), and then send $g \rightarrow 0$.
} 
\begin{equation}\label{eq:action_minimal_gauged_sugra}
    I = -\frac{1}{16\pi G}\int_{\mathcal M_5}\left[\left(R + 12g^2\right)*1 - \frac{2}{3g^2}F \wedge * F + \frac{8\ii}{27g^3}F\wedge F \wedge A \right]\,,
\end{equation}
where $F = \diff A$ is the graviphoton field strength and $g>0$ is the cosmological constant parameter. The action necessarily includes a gauge Chern-Simons term, which plays a crucial role in holography: the global anomalies of the dual conformal field theory are captured by boundary terms arising from the variation of suitable Chern-Simons terms in the bulk supergravity action~\cite{Witten:1998qj}. This mechanism is fundamental for deriving the holographic dictionary, which relates dimensionless bulk parameters (such as $G g^3$ appearing in the Chern-Simons coefficient in the two-derivative theory) to the superconformal anomalies of the dual CFT. We provide a detailed discussion in appendix~\ref{app:B}.

Requiring vanishing gravitino variations in the gauged theory gives the Killing spinor equation:
\begin{equation}\label{eq:Killing_spinor_equations}
    \left[ \nabla_\mu - \frac{\ii}{12 g}\left(\gamma_\mu{}^{\nu\rho}-4\delta_\mu^\nu\,\gamma^\rho\right) F_{\nu\rho}- \frac{g}{2}\gamma^\mu - \ii A_\mu\right]\epsilon =0\,,
\end{equation}
where $\epsilon = \epsilon^1 + \ii \epsilon^2$ is a Dirac spinor and the gamma matrices $\gamma^\mu$ satisfy the Clifford algebra $\{ \gamma_\mu ,\,\gamma_\nu\} = 2g_{\mu\nu}$.\footnote{The charge conjugate spinor, $\tilde\epsilon$, which in the Euclidean theory should be treated as independent from $\epsilon$, satisfies an equation obtained from \eqref{eq:Killing_spinor_equations} by formally replacing $g\rightarrow -g$. Consequently, it we start from a Lorentzian supersymmetric solution, admitting Killing spinor $\epsilon$ with charge conjugate $\tilde \epsilon$, its analytic continuation to Euclidean signature yields a field configuration satisfying the doubled set of (complex) Killing spinor equations. This will be our working framework: we consider complex solutions obtained as analytic continuations of Lorentzian supersymmetric solutions. We note that such solutions in Lorentzian signature may be causally ill-behaved.}
This theory arises as a universal consistent truncation of ten- or eleven-dimensional supergravity on any internal geometry $M$ allowing for a supersymmetric AdS$_5\times M$ solution (where the product may be warped) \cite{Gauntlett:2007ma,Cassani:2019vcl}.  In particular, it is a consistent truncation of Type IIB supergravity on any Sasaki-Einstein five-manifold, including $S^5$, providing an explicit embedding into a UV-complete theory as the low-energy limit of type IIB string theory.

The minimal theory can couple to matter multiplets, such as vector and hypermultiplets. In particular, a vector multiplet contains a gauge field $A_\mu$, an SU$(2)$ doublet of gaugini $\lambda^i$, and a real scalar $\phi$. In this Thesis, we will later consider vector multiplet couplings, but not hypermultiplets. For $n$ vector multiplets, the $n$ scalars $\phi^x$, ($x = 1,2,...,n$) parametrize a scalar manifold, ${\cal M}_V$. As it turns out, the properties of ${\cal M}_V$ are uniquely determined by the constant, completely symmetric tensor $C_{IJK}$ appearing in the gauge Chern-Simons term, $C_{IJK} F^I \wedge F^J \wedge A^K$ (with $I = \{ 0,x\}$). More precisely, $C_{IJK}$ defines a cubic polynomial ${\cal C} = C_{IJK} X^I X^J X^K$ on an auxiliary $\mathbb R^{n+1}$ space with coordinates $X^I$ and the scalar manifold ${\cal M}_V$ is found by imposing ${\cal C}=1$, a structure known as \emph{very special geometry}~\cite{Gunaydin:1983bi} . The Lagrangian is entirely specified in terms of ${\cal C}$ and its derivatives. In chapter~\ref{chap:flavour}, we will revisit these features when considering the coupling to vector multiplets. In that context, although more general gaugings are possible, we will again restrict to the theory obtained by gauging a U$(1)_R$ subgroup of SU$(2)_R$, using a certain linear combination of the graviphoton and the other additional vector fields to implement the gauging. Unlike the minimal case, the holographic dictionary depends on the presence of flavour symmetries in the dual CFT and is, therefore, sensitive to the choice of internal manifold in the compactification.

\medskip

Returning to the minimal case, our first goal in this chapter will be to determine whether the earliest known supersymmetric AdS$_5$ black hole~\cite{Gutowski:2004ez} contributes to the gravitational index.


\section{Lorentzian supersymmetric black hole}
\label{sec:Lor_susy_BH_5D}

In this section, we review the supersymmetric AdS$_5$ black hole first constructed in~\cite{Gutowski:2004ez}. While this solution is extremal, admits a regular Lorentzian section, and preserves supersymmetry, we will argue that it does not possess the appropriate boundary conditions to serve as a valid saddle point for the gravitational index defined in the previous section. 

The Gutowski-Reall solution describes a one-parameter family\footnote{In particular, as we will see below, this family of solutions carries only a single independent rotation. While our main interest will later be in solutions with two independent rotations, in this section we focus on the singly rotating case, whose enhanced symmetry allows for a simpler treatment.}  of supersymmetric, asymptotically AdS$_5$ black holes. In Lorentzian signature, the metric and gauge field are given by
\begin{equation}\label{eq:gutowski_reall_bh}
\begin{aligned}
\diff s^2 &= - U(r)\Lambda(r)^{-1} \diff t^2 + \frac{\diff r^2}{U(r)} + \frac{r^2}{4}\Bigl[ \sigma_1^2 + \sigma_2^2 + \Lambda(r)\left(\sigma_3-W(r) \diff t\right)^2 \Bigr]\,,
\\[1mm]
A &= -\frac{3g}{2}\left[ \left( 1 - \frac{r_0^2}{r^2}- g^2\frac{r_0^4}{2r^2}\right) \diff t + g\frac{r_0^4}{4r^2}\sigma_3 \right]\,,
\end{aligned}
\end{equation}
where $\sigma_{1,2,3}$ are the left-invariant one-forms of SU$(2)$, expressed in terms of Euler angles $(\theta,\,\phi,\,\psi)$ as
\begin{equation}
        \sigma_1 = \cos\psi \diff \theta + \sin\psi \sin\theta \diff \phi\,,\qquad 
        \sigma_2 = -\sin\psi \diff \theta + \cos\psi \sin\theta \diff \phi \,,\qquad
        \sigma_3 = \diff \psi + \cos\theta \diff \phi\,,
\end{equation}
with $\theta \in [0;\pi]$, $\psi \sim \psi + 4\pi$ and $\phi \sim  \phi + 2\pi$. The metric functions are
\begin{equation}
    \begin{aligned}
        U(r) &= \left( 1- \frac{r_0^2}{r^2}\right)^2\left( 1 + 2g^2\,r_0^2 + g^2 r^2\right)\,,\qquad \Lambda(r) = 1 + g^2\frac{r_0^6}{r^4}- g^2\frac{r_0^8}{4r^6}\,,\\[1mm]
        W(r) &= \frac{2g}{\Lambda(r)}\left[\left( \frac{3}{2}+ g^2 r_0^2\right)\frac{r_0^4}{r^4}- \left(\frac{1}{2}+ g^2\frac{r_0^2}{4}\right)\frac{r_0^6}{r^6} \right]\,.
    \end{aligned}
\end{equation}
The function $U(r)$ exhibits a double zero at $r=r_0$, signalling the presence of an extremal event horizon, and the geometry is regular and causally well-behaved on and outside this horizon, for $r\geq r_0$. This is expected: supersymmetric solutions that admit a smooth Lorentzian section are necessarily extremal (see e.g.~\cite{Cvetic:2005zi}).

This solution carries equal angular momenta in the orthogonal two-planes, $J_1 = J_2 \equiv J$, given by
\begin{equation}
J = \frac{\pi g}{8G}r_0^4\left( 3 + 2g^2 r_0^2\right)\,,
\end{equation}
as well as an electric charge
\begin{equation}
    Q = \frac{\pi}{4gG}r_0^2\left( 2 + g^2 r_0^2\right)\,.
\end{equation}
Since there is only one independent parameter in this solution, the two charges are actually not independent, rather they satisfy a complicated non-linear relation of the form
\begin{equation}
Q^3 + \frac{2\pi}{Gg^3}J^2 = \left( 3Q + \frac{\pi}{2Gg^3}\right) \left( 3Q^2 - \frac{2\pi}{Gg^3}J\right)\,.
\end{equation}
The mass saturates the BPS bound \eqref{eq:BPS_bound_5D_sugra} derived from the superalgebra, with $J_1 = J_2= J$. Finally, the Bekenstein-Hawking entropy of the black hole is given by
\begin{equation}\label{eq:S_GR}
{\cal S} = \frac{\pi^2}{4G}r_0^3\sqrt{4 + 3 g^2 r_0^2} = \pi \sqrt{3Q^2 - \frac{2\pi}{Gg^3}J}\,.
\end{equation}

A main feature of supersymmetric spacetimes is the presence of a Killing spinor, solving the appropriate Killing spinor equations. The associated Killing vector, constructed as a bilinear of the Killing spinor, is called the \emph{supersymmetric Killing vector}, denoted by $V$~\cite{Gauntlett:2003fk}. For this solution, it reads
\begin{equation}\label{eq:GR_susy_KV}
    V = \partial_t + 2g\,\partial_\psi\,,
\end{equation}
and it coincides with the generator of the Killing horizon, as seen from the vanishing at the horizon of its norm
\begin{equation}
    |V|^2 = -U(r)\Lambda(r)^{-1}+ \frac{r^2}{4}\Lambda(r)\left(W(r) -2g\right)^2\,.
\end{equation}
This is a common feature of supersymmetric extremal black holes. 

Because of the double zero of the function $U(r)$, the geometry near the horizon develops an infinitely long AdS$_2$ throat. To analyze the near-horizon region, we define new coordinates:
\begin{equation}
    \tilde\psi = \psi - 2g \,t\,,\qquad \tilde t= \lambda t\,,\qquad \tilde r = \frac{8\left(1+3g^2r_0^2\right)}{r_0^2\sqrt{4+3g^2r_0^2}}\frac{r-r_0}{\lambda}\,,
\end{equation}
with the normalization chosen to simplify future expressions. Taking $\lambda \rightarrow 0$, while keeping the new coordinates fixed, allows us to probe the near-horizon region $r\sim r_0$. The angular shift $\psi \rightarrow \tilde \psi$ is required to obtain a finite metric in the limit, passing from Boyer-Lindquist type coordinates to a co-rotating frame, in which the generator of the event horizon becomes $V =  \partial_t$ .

The near-horizon geometry becomes
\begin{equation}
\label{eq:near_horizon_GR}
    \begin{aligned}
        \diff s^2 &= v_1 \left( - \tilde r^2 \diff \tilde t^2 + \frac{\diff \tilde r^2}{\tilde r^2}\right) + \frac{v_2}{4}\left[ \sigma_1^2 + \sigma_2^2 + v_3\left( \tilde\sigma_3 + w\,\tilde r\,\diff\tilde t\right)^2\right]\,,
        \\[1mm]
        A &= -e\,\tilde r\,\diff\tilde t-p\left( \tilde\sigma_3 + w\,\tilde r\,\diff \tilde t\right)\,,\qquad \qquad \tilde\sigma_3 = \diff \tilde \psi + \cos\theta \diff\phi\,,
    \end{aligned}
\end{equation}
with parameters 
\begin{equation}
\label{eq:GR_NH_coefficients}
\begin{aligned}
    v_1 &= \frac{r_0^2}{4\left(1 + 3g^2 r_0^2\right)}\,,\qquad v_2 = r_0^2\,,\qquad v_3 = 1+\frac{3}{4}g^2 r_0^2\,,
    \\[1mm]
w &=  \frac{3g\,r_0}{\left( 1 + 3g^2 r_0^2\right)\sqrt{4 + 3 g^2 r_0^2}}\,,\qquad e= \frac{w}{2}\,,\qquad p = \frac{3}{8}g^2\,r_0^2\,.
    \end{aligned}
\end{equation}
This describes a fibration of a squashed three-sphere (with U$(1) \times$ SU(2) isometry group) over AdS$_2$. The emergence of an infinitely long AdS$_2$ throat implies that, when analytically continuing this metric to Euclidean signature, the Euclidean time can only be compatified periodically with infinite period, $\beta \rightarrow +\infty$. Additionally, the on-shell action becomes IR divergent, making its semiclassical contribution to the gravitational path integral ill-defined. In contrast, recall that the gravitational index is well-defined only for finite $\beta$. 

A second independent obstruction arises from the behaviour of the chemical potentials. Passing to a co-rotating frame, one finds the asymptotic form of the metric and gauge field:
\begin{equation}
\begin{aligned}
\diff s^2 &\rightarrow \frac{1}{g^2}\frac{\diff r^2}{r^2} + g^2r^2\left[ -\diff t^2 + \frac{1}{4g^2}\left(\diff \theta ^2 + \sin^2\theta \diff \phi^2 + \left( \diff \tilde\psi + \cos \theta \diff \phi + \Omega^* \diff t\right)^2\right)\right]\,,
\\[1mm]
A &\rightarrow -\Phi_* \diff t\,,
\end{aligned}
\end{equation}
with $\Omega^* \equiv \Omega_1^* + \Omega_2^*= 2g$ and $\Phi_* = \frac{3}{2}g$, as in \eqref{eq:minimal_sugra_coeff}. From this expressions we identify the angular velocity and electrostatic potential of the solution:
\begin{equation}
\Omega \equiv \Omega_1 + \Omega_2 = \Omega^* \,,\qquad \Phi = \Phi_*\,.
\end{equation}
These values are frozen to the ones appearing in the superalgebra \eqref{eq:supergravity_algebra}. 

As a result, the redefined chemical potentials \eqref{eq:red_chem_pot} appearing in the gravitational index take the indeterminate form $\infty \cdot 0$ on this solution,  and there is no natural way to assign them complex values that satisfy the linear relation \eqref{eq:linear_constraint_minimal} responsible for the presence of the phase $e^{-\ii\pi Q}$ in the trace. Following the early analysis of~\cite{Silva:2006xv,Dias:2007dj,Morales:2006gm,Sen:2005wa}, one would be tempted to identify the relevant chemical potentials with the near-horizon coefficients $w$ and $e$ of \eqref{eq:GR_NH_coefficients},
\begin{equation}
    \omega_1 + \omega_2 \equiv \omega = 2\pi w\,,\qquad \varphi = 2\pi e\,,
\end{equation}
since they correspond to the non-normalizable modes for the AdS$_2 \times S^3$ geometry. However, these variables obey a \emph{real} linear constraint, $\omega - 2\varphi =0$. In other words, there is no reason to expect that the thermodynamics of the extremal and supersymmetric black hole involves the required insertion of the fermion number operator in the trace. 

\section{Euclidean black hole saddle}
\label{TwoDerReview}

In the previous section we argued that the supersymmetric black hole solution~\eqref{eq:gutowski_reall_bh} fails to satisfy the boundary conditions required to contribute to the gravitational index. The obstacles we identified can be overcome by introducing a complex deformation of the extremal solution, achieved by turning on a non-zero temperature, while preserving supersymmetry. The finite inverse-temperature $\beta$ can be viewed as an IR regulator that renders the on-shell action well-defined, with the extremal geometry recovered in the limit $\beta \to +\infty$.
Asymptotically AdS$_5$ saddles relevant for the gravitational index~\eqref{eq:gravitational_index_AdS} were first identified in~\cite{Cabo-Bizet:2018ehj}. Similar constructions have since been found in a wide range of asymptotically AdS spacetimes across different dimensions~\cite{Cassani:2019mms,Kantor:2019lfo,Bobev:2019zmz,Nian:2019pxj,Benini:2019dyp,Bobev:2020pjk,Larsen:2021wnu,Cassani:2021dwa,Boruch:2022tno,BenettiGenolini:2023ucp,Bobev:2023bxl,David:2025pby}, and, more recently, also in asymptotically flat spacetimes, starting with~\cite{Iliesiu:2021are}. We review the construction of~\cite{Cabo-Bizet:2018ehj} here, as it will be needed for later applications, and return to the asymptotically flat case in part~\ref{part_three}.

The supersymmetric saddles of~\cite{Cabo-Bizet:2018ehj} are obtained from the most general non-extremal, non-supersymmetric, asymptotically AdS$_5$ black hole solution of~\eqref{eq:action_minimal_gauged_sugra}, originally found in~\cite{Chong:2005hr} by Chong, Cvetic, Lu, Pope (CCLP), which contributes to a thermal partition function of the form~\eqref{eq:partition_function_gravity_5D}. One then takes a limit in which the charges saturate the BPS bound~\eqref{eq:BPS_bound_5D_sugra}, with chemical potentials constrained by~\eqref{eq:linear_constraint_minimal}, while keeping $\beta$ finite. This can be achieved by allowing for complex components in the metric. Therefore, the resulting geometry is ill-defined in Lorentzian signature but is allowed after Wick-rotation to Euclidean space.

\paragraph{The CCLP black hole.}

The most general asymptotically AdS$_5$ black hole solution to the theory \eqref{eq:action_minimal_gauged_sugra} can be expressed in the Boyer-Lindquist type coordinates $(\tau,\vartheta,\phi_1,\phi_2,r)$, after Wick-rotation $\tau = \ii t$ to Euclidean time~\cite{Chong:2005hr}:\footnote{We have renamed some of the quantities appearing in~\cite{Chong:2005hr}  as  $\phi_{\rm there}= \phi_1$, $\psi_{\rm there} = \phi_2$, $\nu_{\rm there} = \ii\,\nu_1$ and $\omega_{\rm there} = \ii\,\nu_2$.}  
\begin{eqnarray}
\diff s^2 &=& \frac{\Delta_\vartheta\, [(1+g^2 r^2)\rho^2 \diff \tau + 2 q \nu_1]
\, \diff \tau}{\Xi_a\, \Xi_b \, \rho^2} - \frac{2q\, \nu_1\nu_2}{\rho^2}
- \frac{f}{\rho^4}\Big(\frac{ \Delta_\vartheta \, \diff \tau}{\Xi_a\Xi_b} -\nu_2\Big)^2 + \frac{\rho^2 \diff r^2}{\Delta_r} +
\frac{\rho^2 \diff \vartheta^2}{\Delta_\vartheta} \nonumber\\
 &&+ \frac{r^2+a^2}{\Xi_a}\sin^2\vartheta\, \diff \phi_1^2 +
      \frac{r^2+b^2}{\Xi_b} \cos^2\vartheta\, \diff \phi_2^2\ ,\label{eq:CCLP_metric}\\[2mm]
A &=& -\frac{3\ii g q}{2\rho^2}\,
         \Big(\frac{\Delta_\vartheta\, \diff \tau}{\Xi_a\, \Xi_b}
       - \nu_2\Big) -\ii \zeta \, \diff \tau\ , \label{gaugepot}
\end{eqnarray}
where
\begin{align}\label{CCLPfunctions}
\nu_1 &= \ii b\sin^2\vartheta\, \diff \phi_1 + \ii a\cos^2\vartheta\, \diff \phi_2\,,\qquad\quad
\nu_2 = \ii a\sin^2\vartheta\, \frac{\diff \phi_1}{\Xi_a} +
              \ii b\cos^2\vartheta\, \frac{\diff \phi_2}{\Xi_b}\ ,\non\\
\Delta_r &= \frac{(r^2+a^2)(r^2+b^2)(1+g^2 r^2) + q^2 +2ab q}{r^2} - 2m
\ ,\non\\
\Delta_\vartheta &= 1 - a^2 g^2 \cos^2\vartheta -
b^2 g^2 \sin^2\vartheta\,,\qquad
\rho^2 = r^2 + a^2 \cos^2\vartheta + b^2 \sin^2\vartheta\,,\non\\
\Xi_a &=1-a^2 g^2\,,\quad \Xi_b = 1-b^2 g^2\ ,\ \ \qquad
f= 2 m \rho^2 - q^2 + 2 a b q g^2 \rho^2\ .
\end{align}
The angular coordinates $\phi_1,\phi_2$ are $2\pi$-periodic, while $\vartheta \in [0,\pi/2]$. 
 The constant $\zeta$ appearing in $A$ parameterizes a gauge choice that needs to be made so that the Euclidean section of the analytically continued solution is globally well-defined (more later).

The solution depends on the four parameters $a, b,m, q$, with  $a^2g^2<1, b^2g^2<1$. These control  four independent conserved charges: the energy $E$, the angular momenta $J_{1,2}$ associated with rotations in the $\phi_{1,2}$ directions, respectively, and the electric charge $Q$. Their expressions are:
\begin{align}\label{CCLPcharges}
E &= \frac{m\pi (2\Xi_a +2\Xi_b - \Xi_a\,\Xi_b) +2\pi qabg^2(\Xi_a+\Xi_b)}{4G\Xi_a^2\,\Xi_b^2}\ ,\qquad Q = \frac{\pi q}{2gG \Xi_a\, \Xi_b}\ ,\non\\[2mm]
&\ \  J_1 = \frac{\pi[2am + qb(1+a^2 g^2) ]}{4G \Xi_a^2\, \Xi_b}\ ,\qquad
J_2 = \frac{\pi[2bm + qa(1+b^2 g^2) ]}{4G \Xi_b^2\, \Xi_a}\ .
\end{align}

In the Lorentzian description the position $r=r_+$ of the outer event horizon would be given by the largest positive root of the equation $\Delta_r(r)=0$.\footnote{In practice, since solving for $r_+$ as a function of the parameters $a,b,m,q$ is very cumbersome, we rather solve $\Delta_r=0$ for $m$ in terms of $r_+$ and the other parameters. This gives
\be
m = \frac{(r_+^2+a^2)(r_+^2+b^2)(1+ g^2r_+^2) + q^2 +2ab q}{2r_+^2} \non \ .
\ee
 Hence the solution will be regarded as controlled by $a,b,r_+,q$.\label{foot:solm}}
In Euclidean signature, $r=r_+$ denotes the location where  the Killing vector 
\begin{equation}
W = \partial_\tau - \ii \Omega_1 \partial_{\phi_1} - \ii \Omega_2 \partial_{\phi_2}
\end{equation}
smoothly contracts (its analytic continuation $\ii W$ would be the generator of the event horizon in the Lorentzian solution) capping off the geometry, thereby excising the singularity. 
  
The parameter $r_+$ directly enters in the expressions for the thermodynamic potentials, namely the inverse Hawking temperature $\beta$, the angular velocities $\Omega_1,\Omega_2$, and the electrostatic potential $\Phi$~\cite{Chen:2005zj},
\be\label{temperature_CCLP}
T \equiv\beta^{-1} = \frac{r_+^4[(1+ g^2(2r_+^2 + a^2+b^2)] -(ab + q)^2}{2\pi\,
         r_+\, [(r_+^2+a^2)(r_+^2+b^2) + abq]}\ ,
\ee
\be\label{angular_velocities_CCLP}
\Omega_1 = \frac{a(r_+^2+ b^2)(1+g^2 r_+^2) + b q}{
               (r_+^2+a^2)(r_+^2+b^2)  + ab q}\ ,\qquad
\Omega_2 = \frac{b(r_+^2+ a^2)(1+g^2 r_+^2) + a q}{
               (r_+^2+a^2)(r_+^2+b^2)  + ab q}\ ,
\ee
\be\label{electrostatic_pot_CCLP}
\Phi = \frac{3g\, q \,r_+^2}{2\left[(r_+^2 + a^2)(r_+^2 + b^2)+abq\right]}\, .
\ee
Of course, $r_+$ also appears in the expression for the area of the horizon and thus in the Bekenstein-Hawking entropy,
\be\label{entropyCCLP}
{\cal S}= \frac{\rm Area}{4G} =\frac{\pi^2 [(r_+^2 +a^2)(r_+^2 + b^2) +a b q]}{2G\Xi_a \Xi_b r_+}
\ .
\ee
These quantities satisfy the first law of thermodynamics,
\be\label{firstlaw}
\diff E = T \diff {\cal S} + \Omega_1\, \diff J_1+ \Omega_2 \,\diff J_2 + \Phi\, \diff Q\ ,
\ee
One should fix the gauge so that the following regularity condition at the horizon is satisfied,
\begin{equation}
W^{\mu}A_{\mu}|_{r=r_{+}}=0\,,
\end{equation}
ensuring that the norm of the gauge field does not diverge there. 
The condition is satisfied by fixing the parameter $\zeta$ in \eqref{gaugepot} as
\be\label{regularxi}
\zeta = - \Phi \,,
\ee
implying that the electrostatic potential can be read as a boundary holonomy of the gauge field, $\Phi = -\ii W^\mu A_\mu|_{r\to\infty}$ (cf.~\ref{sec:Grav_path_int}). 

Near $r = r_+$, where the Euclidean geometry caps off, the absence of conical singularities imposes the twisted identifications
\begin{equation}
\left( \tau ,\,\phi_1,\,\phi_2 \right) \sim \left( \tau + \beta ,\,\phi_1 - \ii\beta \Omega_1,\,\phi_2 - \ii\beta\Omega_2\right)\,,
\end{equation}
when going around the Euclidean time direction once. These must be supplemented by the standard identifications ensuring a smooth asymptotic $S^3$:
\begin{equation}
\left( \tau ,\,\phi_1,\,\phi_2 \right) \sim \left( \tau ,\,\phi_1 + 2\pi,\,\phi_2 \right) \sim \left( \tau ,\,\phi_1 ,\,\phi_2 + 2\pi \right)\,.
\end{equation}
With these identifications, the geometry takes the form of a warped fibration of a deformed $S^3$ over a disk-like $\mathbb{R}^2$ factor. The U$(1)$ circle in the $\mathbb{R}^2$ factor involves, as usual, the compactified Euclidean time. Such disk-like topologies are a common feature of the Euclidean saddles of interest.


\paragraph{Imposing supersymmetry.}

The solution \eqref{eq:CCLP_metric} is supersymmetric if~\cite{Cvetic:2005zi}
\be\label{susyCCLP}
\begin{aligned}
q &= \frac{m}{1+ag+bg}\\
 &= -(a- \ii r_+)(b- \ii r_+)(1- \ii gr_+)\,,
\end{aligned}
\ee
where in the second line we have used the expression of $m$ in terms of $r_+$ and chosen one of the two roots of the resulting quadratic equation for $q$ (choosing the other root would just give expressions where $\ii$ is replaced by $-\ii$). 
This supersymmetry condition does not imply extremality, since the inverse temperature $\beta$ remains finite:
\begin{equation}
    \beta = -2\pi\frac{\left( a- \ii r_+ \right)\left( b - \ii r_+\right)\left( r_*^2 + \ii g^{-1} r_+\right)}{\left(r_*^2 - r_+^2\right)\left[2(1+ag+bg)r_+ + \ii g (r_*^2-3r_+^2)\right]}\,,
\end{equation}
where $r_*$ is given by\footnote{Following the notation of \cite{Cabo-Bizet:2018ehj}, a $*$ denotes quantities obtained by imposing both supersymmetry and extremality.
}
\be\label{r_BPS}
r_* =  \sqrt{\frac{1}{g}(a+b+abg)} \ .
\ee
Extremality follows only after additionally requiring that the metric function $\Delta_r$ has a double root, which is achieved by taking
 \be\label{BPSlimit}
 r_+\to r_*\,.
 \ee 
In this limit, the supersymmetry condition~\eqref{susyCCLP} fixes $q$ to the real value
 \begin{equation}
 q_*= g^{-1}\left( a+b\right)\left( 1 + a g\right)\left( 1 + b g\right)
 \end{equation}
 and, as a result, the Lorentzian section turns out to be real and causally well-behaved~\cite{Chong:2005hr}. In the supersymmetric and extremal limit, the chemical potentials approach the coefficients $\Omega_1^*,\Omega_2^*,\Phi_*$ appearing in the superalgebra \eqref{eq:supergravity_algebra}:
\be\label{leadingterms_chempot}
\beta \to \infty\ ,\qquad \Omega_1\to \Omega^*_1 = g\ ,\qquad \Omega_2\to \Omega^*_2 = g\ ,\qquad \Phi \to \Phi_* = \frac{3}{2}g .
\ee

In what follows, we impose supersymmetry while keeping $r_+$ generic. In this case, $q$ is complex according to~\eqref{susyCCLP}, and consequently the conserved charges~\eqref{CCLPcharges}, the chemical potentials~\eqref{temperature_CCLP},\eqref{angular_velocities_CCLP},\eqref{electrostatic_pot_CCLP}, and many components of the metric~\eqref{eq:CCLP_metric} become complex. Such a solution is inadmissible in Lorentzian signature but is perfectly acceptable after Wick rotation to Euclidean space.

In the supersymmetric solution, the complex conserved charges satisfy the linear BPS bound~\eqref{eq:BPS_bound_5D_sugra}. The associated complex redefined chemical potentials~\eqref{eq:red_chem_pot} are
\begin{equation}
\label{eq:susychemicalpotentials}
\begin{aligned}
\omega_1 \,&=\, \beta(\Omega_1-\Omega_1^*) \,=\, \frac{2\pi(ag-1)(b-\ii r_+)}{2(1+ag+bg)r_+ + \ii g (r_*^2-3r_+^2)}\,,\\[2mm]
\omega_2 \,&=\, \beta(\Omega_2-\Omega_2^*) \,=\,  \frac{2\pi(bg-1)(a-\ii r_+)}{2(1+ag+bg)r_+ + \ii g (r_*^2-3r_+^2)}\ ,\\[2mm]
\varphi \,&=\,\beta(\Phi-\Phi_*) \,=\, \frac{3g\pi(a-\ii r_+)(b-\ii r_+)}{2(1+ag+bg)r_+ + \ii g (r_*^2-3r_+^2)}\ .
\end{aligned}
\end{equation}
These satisfy the linear constraint~\eqref{eq:linear_constraint_minimal}, indicating that this supersymmetric but non-extremal configuration is a viable saddle for the index~\eqref{eq:gravitational_index_AdS}.

The presence of supersymmetry is confirmed by the existence of a Killing spinor supported by the solution, obtained as a solution to the Killing spinor equations for five-dimensional minimal gauged supergravity~\eqref{eq:Killing_spinor_equations}.  
An explicit solution in the background \eqref{eq:CCLP_metric}, after imposing the supersymmetric condition \eqref{susyCCLP}, has been constructed in~\cite{Cabo-Bizet:2018ehj}:
\begin{equation}
    \epsilon = e^{\frac{1}{2}\left[\left(g -2\Phi\right)\tau+ \ii\left(\phi_1 + \phi_2\right)\right]}\,f^{1/2}\,\epsilon_0\,,
\end{equation}
where $\epsilon_0$ is a constant spinor of unit norm, and
\begin{equation}
\begin{aligned}
    f &= \frac{g^{-1}\left( a + b + ab g\right) \left( 1 + \tilde m\right) + \tilde m \left( a^2 \cos^2\vartheta + b^2 \sin^2\vartheta \right) - r^2}{a^2 \cos^2\vartheta + b^2 \sin^2\vartheta + r^2}\,,
    \\[1mm]
\tilde m &= \frac{m\,g}{\left( a + b\right) \left( 1 + ag\right) \left( 1 + b g\right) \left( 1 + ag + bg \right)}-1\,. 
    \end{aligned}
\end{equation}
The Killing spinor bilinear $\bar\epsilon \gamma^\mu\epsilon \partial_\mu$ defines the supersymmetric Killing vector
\begin{equation}
    V \equiv -\ii \bar\epsilon \gamma^\mu\epsilon\, \partial_\mu = \partial_\tau -\ii \Omega_1^* \partial_{\phi_1} -\ii\Omega_2^*\partial_{\phi_2}\,.
\end{equation}

The linear constraint~\eqref{eq:linear_constraint_minimal} admits a geometric interpretation as a global regularity condition for this Killing spinor. In particular, it ensures that the spinor is antiperiodic around all contractible cycles, guranteeing that the geometry supports a well-defined spin structure.
To verify this, consider the spinorial Lie derivative
\begin{equation}
    {\cal L}_X\epsilon = X^\mu\nabla_\mu \epsilon - \frac{1}{4}\nabla_\mu X_\nu \gamma^{\mu\nu}\epsilon\,.
\end{equation}
For the relevant vectors, one finds
\begin{equation}
    {\cal L}_{\partial_\tau}\epsilon = \frac{1}{2}\left( g - 2\Phi\right) \epsilon\,,\qquad {\cal L}_{\partial_{\phi_1}} \epsilon = \frac{\ii}{2}\epsilon\,,\qquad {\cal L}_{\partial_{\phi_2}} = \frac{\ii}{2}\epsilon\,.
\end{equation}
Transporting the spinor along the orbits of $\partial_{\phi_{1,2}}$ yields
\begin{equation}
e^{2\pi{\cal L}_{\partial_{\phi_{1,2}}}}\epsilon = - \epsilon\,,
\end{equation}
showing that it is antiperiodic along these cycles, as required since they are the contractible circles of $S^3$.
A similar argument applies to the horizon generator $W$, which contracts at $r=r_+$. Acting on the spinor, it gives
\begin{equation}
    {\cal L}_W\epsilon = \frac{1}{2}\left(g-2\Phi + \Omega_1 + \Omega_2\right) \epsilon = \frac{1}{2\beta}\left(\omega_1 + \omega_2 - 2\varphi\right) \epsilon\,.
\end{equation}
When transported around the circle generated by $W$, the spinor then acquires the phase
\begin{equation}
    e^{\beta{\cal L}_W}\epsilon = e^{\frac{\omega_1 + \omega_2 - 2\varphi}{2}}\epsilon= -\epsilon\,,
\end{equation}
where the last equality follows from the constraint~\eqref{eq:linear_constraint_minimal}. Thus, the Killing spinor is also antiperiodic along the contractible circle at the horizon, confirming its global regularity.

\paragraph{Recovering the supersymmetric and extremal solution.} The supersymmetric extremal solution is obtained by \emph{first} imposing the supersymmetry condition~\eqref{susyCCLP} and then taking the extremal limit~\eqref{BPSlimit}. This procedure is completely smooth and yields a real Lorentzian solution upon Wick-rotating the Euclidean time back to Lorentzian signature~\cite{Chong:2005hr}. In this limit, the real charges take the form
\begin{equation}
\begin{aligned}
    Q^* &= \frac{\pi}{2g^2G}\frac{a+ b}{\left( 1- ag\right)\left( 1- bg\right)}\,,
    \\[1mm]
 J_1^* &= \frac{\pi}{4g G}\frac{\left( a+ b\right)\left( 2a + b + abg\right)}{\left(1- ag\right)^2 \left( 1-bg\right)}\,,\qquad J_2^* = \frac{\pi}{4g G}\frac{\left( a+ b\right)\left( a + 2b + abg\right)}{\left(1- ag\right) \left( 1-bg\right)^2}\,.
\end{aligned}
\end{equation}
Note that the adapted chemical potentials~\eqref{eq:susychemicalpotentials} remain complex even in the extremal limit. The extremal charges obey the non-linear relation
\be\label{nonlinear_rel_BPScharges}
 \left(3Q^*  + \frac{\pi}{2Gg^3} \right)\left( 3 \left(Q^*\right)^2 -  \frac{\pi}{Gg^3}(J^*_1+J^*_2)\right) \,=\, \left(Q^*\right)^3 + \frac{2\pi}{Gg^3}J^*_1 J^*_2\ .
\ee

The entropy of the supersymmetric extremal black hole is
\begin{equation}
\label{S_BPS}
\begin{aligned}
\mathcal{S}^* &= \frac{\pi^2 (a+b)r_*}{2Gg(1-ag)(1-bg)}
\\[2mm]
&=\pi \sqrt{3\left({ Q^*}\right)^2 -\frac{\pi}{G g^3} \big(J^*_1+J^*_2\big)}\ ,
\end{aligned}
\end{equation}
where in the second line we have expressed the result in terms of the charges~\cite{Kim:2006he}. 
As will be shown below, both~\eqref{S_BPS} and~\eqref{nonlinear_rel_BPScharges} follow directly from a constrained Legendre transform of the on-shell action~\cite{Hosseini:2017mds}.

When the rotation parameters are equal, $a = b$, the solution reduces to the one-parameter supersymmetric black hole reviewed in sec.~\ref{sec:Lor_susy_BH_5D},\footnote{The angular coordinate appearing in this section are related to the coordinates of section~\ref{sec:Lor_susy_BH_5D} by
$$ \vartheta = \frac{\theta}{2}\,,\qquad \qquad \phi_1 = \frac{\psi - \phi}{2}\,,\qquad \qquad \phi_2 = \frac{\psi + \phi}{2}\,.$$
} with
\begin{equation}\label{eq:GRtoCCLP}
    r_0 = \frac{1}{g}\sqrt{\frac{2ag}{1-ag}}\,.
\end{equation}

\section{Supersymmetric on-shell action and holography}
\label{sec:holographic_match}

In this section, we review the computation of the on-shell action for the supersymmetric saddle obtained by imposing \eqref{susyCCLP} in \eqref{eq:CCLP_metric}. We will show that the resulting on-shell action matches the field theory prediction for the logarithm of the superconformal index, \eqref{eq:log_I}. From this, the entropy \eqref{S_BPS} follows via a constrained Legendre transform, confirming that the saddle captures precisely the contribution of the physical supersymmetric black hole to the index, as anticipated. Our strategy is to first compute the on-shell action for the general, non-supersymmetric black hole, and only then impose the supersymmetry condition.

The action \eqref{eq:action_minimal_gauged_sugra}, evaluated on the solution \eqref{eq:CCLP_metric}, is divergent. These divergences can be removed either via the background subtraction method~\cite{Chen:2005zj}, or by holographic renormalization \cite{deHaro:2000vlm,Bianchi:2001kw}, followed by subtracting the on-shell action of the AdS$_5$ vacuum.\footnote{The reason for this subtraction is that we want to describe the gravity dual of the superconformal index $\mathcal{I}$ as opposed to the SCFT partition function $Z$ computed via the path integral. The two are related by the contribution of the vacuum in the path integral, that is $Z = \rme^{-I_{\rm AdS}}\, \mathcal{I} $, with $I_{\rm AdS}$ proportional to the Casimir energy~\cite{Assel:2014paa,Assel:2015nca,Martelli:2015kuk,BenettiGenolini:2016qwm}. In the saddle point approximation $Z\sim \rme^{-I_Z}$, then $\mathcal{I}\sim \rme^{-I}$ becomes  $ I= I_{Z} - I_{\rm AdS}$. Note that while $I_{Z}$ is sensitive to the scheme used, $I$ is not, since any finite counterterm that is added to our choice of boundary terms would contribute in the same way to the black hole action and the AdS action.} Both approaches yield the same expression~\cite{Cassani:2019mms}. Following this second preocedure, we define the \emph{renormalized} on-shell action as the sum
\begin{equation}
    I_{\rm ren} = I + I_{\rm bdry} - I_{\rm AdS}\,,
\end{equation}
where $I_{\rm bdry}$ collects boundary terms needed to render the Dirichlet variational problem well-posed, together with local counterterms to remove divergences. In our case,
\begin{equation}
    I_{\rm bdry} = -\frac{1}{8\pi G}\int_{\partial \mathcal M}\diff^4 x\sqrt{h}\left({\cal K} + 3g + \frac{1}{4g}\hat{\cal R}\right)\,,
\end{equation}
where $h$ denotes the determinant of the induced boundary metric $h_{ij}$, ${\cal K}$ is the trace of the boundary extrinsic curvature ${\cal K}_{\mu\nu} = \nabla_{(\mu}n_{\nu)}$ (here, $n_\mu$ is the outward-pointing unit normal to the boundary), and $\hat{\cal R}$ is the boundary Ricci scalar. Finally, $I_{\rm AdS}$ removes the non-zero contribution of the AdS vacuum
\begin{equation}
    I_{\rm AdS} = \frac{3\pi\beta}{32Gg^2}\,.
\end{equation}
The on-shell action, then, gives~\cite{Chen:2005zj} (see also~\cite{Kunduri:2005zg}): 
\be\label{OnShAction_CCLP}
I_{\rm ren} = \frac{\pi\beta}{4G\Xi_a\Xi_b}
\Big[m - g^2 (r_+^2 + a^2)(r_+^2 + b^2) -
\frac{q^2 r_+^2}{(r_+^2 + a^2)(r_+^2 + b^2)+abq}\Big]\ .
\ee
For brevity, we will often drop the subscript “${\rm ren}$” and reinstate it only when it is necessary to distinguish the bulk action \eqref{eq:action_minimal_gauged_sugra} from its renormalized version.

The expression \eqref{OnShAction_CCLP} satisfies the quantum statistical relation:
\be\label{QSR_CCLP}
I = \beta E - {\cal S} - \beta \Omega_1J_1 - \beta\Omega_2J_2 - \beta\Phi Q\ .
\ee
As discussed in~\ref{sec:Grav_path_int}, the on-shell action can be interpreted as the saddle-point value of the grand-canonical partition function, $\log Z_{\rm grav} \sim -I$.
In this framework, the action of an asymptotically (locally) AdS solution is a function of the boundary values of the bulk fields with Dirichlet boundary conditions; see, for instance, discussion in~\cite{Papadimitriou:2005ii}. Imposing regularity of the Euclidean geometry fixes the quantities $\beta, \Omega_1, \Omega_2, \Phi$, which thus appear explicitly in the boundary data. Consequently, the action can be regarded as a function of the chemical potentials, $I=I(\beta,\Omega_1,\Omega_2,\Phi)$. In contrast, the black hole entropy corresponds to the logarithm of a microcanonical partition function, obtained in the saddle-point approximation as the Legendre transform of $I$ with respect to all chemical potentials.

To obtain a supersymmetric version of the above thermodynamic relations, we use the BPS relation \eqref{eq:BPS_bound_5D_sugra} to eliminate $E$. Then, the quantum statistical relation \eqref{QSR_CCLP} and the first law \eqref{firstlaw} restricted to the supersymmetric ensemble become 
\be
I =  - S - \omega_1J_1 - \omega_2J_2 - \varphi\, Q\ ,
\ee
\be
\diff S + \omega_1\, \diff J_1+ \omega_2\, \diff J_2 +  \varphi\, \diff Q=0\,.
\ee
The action satisfying these supersymmetric relations can be obtained from \eqref{OnShAction_CCLP} by imposing the supersymmetry condition \eqref{susyCCLP}, yielding~\cite{Cabo-Bizet:2018ehj}:
\be\label{Isusy}
I = \frac{2\pi}{27G g^3}\,\frac{\varphi^3}{\omega_1\omega_2}\ ,
\ee
where $\varphi$ can be further eliminated via \eqref{eq:linear_constraint_minimal}. To compare this result to the superconformal index prediction, we apply the holographic dictionary summarized in appendix~\ref{sec:holographicdictionary}. At the two-derivative level, the map between the R-symmetry anomaly coefficients and the gravitational parameters is
\begin{equation}\label{eq:2d_anomaly_coeff}
    {\rm Tr}{\cal R}^3 = \frac{4\pi}{9Gg^3}\,,\qquad {\rm Tr}{\cal R} = 0\,,
\end{equation}
showing perfect consistency between \eqref{Isusy} and the large-$N$ limit of \eqref{eq:log_I},
\begin{equation}
I = - \log{\cal I} \left( \omega_1,\,\omega_2\right)\,.
\end{equation}

\paragraph{Constrained Legendre transform.}

It remains to explain how the supersymmetric and extremal entropy \eqref{S_BPS} can be recovered from the index formula \eqref{Isusy} (or equivalently from \eqref{eq:log_I} with the anomaly coefficients given in \eqref{eq:2d_anomaly_coeff}), thus providing an independent derivation directly from the field theory. Extensions of this procedure to include higher-derivative corrections and to apply to asymptotically flat saddles will be discussed in part~\ref{part_two} and part~\ref{part_three}, respectively. In the supersymmetric case, the entropy is obtained by extremizing the following function~\cite{Hosseini:2017mds,Cabo-Bizet:2018ehj}:
\be
\mathcal{S} = {\rm ext}_{\{\omega_1,\omega_2,\varphi,\Lambda\}} \left[ -I -\omega_1J_1-\omega_2J_2  -\varphi\, Q -\Lambda(\omega_1+\omega_2-2\varphi-2\pi i)\right]\,,
\ee
where the Lagrange multiplier $\Lambda$ implements the linear constraint \eqref{eq:linear_constraint_minimal}. The extremization equations are:
\be\label{derivativesI}
-\frac{\partial{I}}{\partial\omega_1} = J_1 + \Lambda  \,,\qquad
-\frac{\partial{I}}{\partial\omega_2} = J_2 + \Lambda  \,,\qquad
-\frac{\partial{I}}{\partial\varphi} = Q -2 \Lambda  \,,
\ee
together with \eqref{eq:linear_constraint_minimal}. A key property of the action \eqref{Isusy} is that it is homogeneous of degree one in the variables $(\omega_1,\omega_2,\varphi)$. By Euler’s theorem for homogeneous functions, it satisfies
\begin{equation}
I = \omega_1 \frac{\partial I}{\partial \omega_1} + \omega_2 \frac{\partial I}{\partial \omega_2} + \varphi \frac{\partial I}{\partial \varphi}\,.
\end{equation}
Substituting this relation into the expression above, one finds
\be
\begin{aligned}\label{SfromLegTransf}
\mathcal{S} \,&=\, {\rm ext}\left[ -I +\omega_1 \frac{\partial{I}}{\partial\omega_1}+\omega_2 \frac{\partial{I}}{\partial\omega_2}  +\varphi \frac{\partial{I}}{\partial\varphi} + 2\pi i\Lambda\right]\\
\,&=\, {\rm ext} \left[ 2\pi i\Lambda\right]\,,
\end{aligned}
\ee

The physical (real) entropy is obtained only if the solution for $\Lambda$ is purely imaginary. This requirement implies that the equation for $\Lambda$ must factorize in the form $(\Lambda^2+ {\cal X})({\rm rest})=0$ for some positive ${\cal X}$. The factorization condition turns out to be equivalent to the non-linear constraint among the charges, given in~\eqref{nonlinear_rel_BPScharges}. To determine the equation for $\Lambda$, we use the identity
\be
\frac{9}{2}\,{\rm Tr }\mathcal{R}^3 \, \frac{\partial{I}}{\partial\omega_1}\frac{\partial{I}}{\partial\omega_2} - \left(\frac{\partial{I}}{\partial\varphi}\right)^3 = 0\,,
\ee
which using \eqref{derivativesI} gives the cubic
\be
p_0 + p_1 \Lambda + p_2 \Lambda^2 + \Lambda^3 = 0\,,
\ee
with coefficients
\be
\begin{aligned}
p_0 &=  -\frac{1}{8} \left(Q^3 + \frac{9}{2} \,{\rm Tr}\mathcal{R}^3  J_1J_2 \right)\,,\\
p_1 &= \frac{1}{4}\left(3Q^2 - \frac{9}{4}\,{\rm Tr}\mathcal{R}^3  \, (J_1+J_2)\right)\,,\\
p_2 &= -\frac{3}{2}Q - \frac{9}{16}\,{\rm Tr}\mathcal{R}^3  \,.\\
\end{aligned}
\ee
Imposing the factorization condition
\be\label{nonl_constraint}
p_0=p_1p_2\,,
\ee 
reduces the cubic equation to
\be
( p_1 +\Lambda^2)( p_2 +\Lambda) = 0
\ee
which admits purely imaginary roots $\Lambda = \pm i \sqrt{p_1}$ for $p_1>0$. Choosing the sign that yields a positive entropy leads to the final expression
\be
\begin{aligned}
\mathcal{S} &=  2\pi \sqrt{p_1}\\[2mm]
&=\pi \sqrt{3Q^2 - \frac{9}{4}\,{\rm Tr}\mathcal{R}^3  \, (J_1+J_2)}\,,
\end{aligned}
\ee
in perfect agreement with \eqref{S_BPS}. Finally, note that the factorization condition~\eqref{nonl_constraint} is exactly equivalent to the non-linear relation among the charges in~\eqref{nonlinear_rel_BPScharges}.

%% file: research_one.tex
\part{Higher-derivative corrections and holography}
\label{part_two}

\chapter{Introduction to part II}
\label{chap:intro_2}

The second part of this Thesis is devoted to developing the gravitational counterpart of the full expression in~\eqref{eq:index_asympt}. This involves studying the first subleading corrections to black hole thermodynamics in five-dimensional supergravity with vector multiplets and higher-derivative couplings. String theory is expected to provide a well-behaved infinite series of higher-derivative corrections to two-derivative ten-dimensional supergravity, controlled for instance by the string tension~$\alpha'$. However very few terms are completely known, and their reduction to lower dimensions crucially depends on the details of the compactification, besides mixing with quantum effects in the Kaluza-Klein towers. Given these limitations, a more agnostic way to proceed is to adopt an effective field theory approach where all higher-derivative terms compatible with the symmetries of interest are included as corrections to the lower-dimensional two-derivative theory. Studying the self-consistency and consequences of the corrections, one can aim at extracting the imprint of a high-energy completion. However, handling higher-derivative terms in gravity or supergravity is a notoriously challenging task; yet, one can hope to obtain precise results in setups with enhanced control. These include theories that preserve sufficient supersymmetries and admit an off-shell formulation, as well as gauged supergravities with known holographic duals, where exact results from quantum field theory provide guidance in probing higher-derivative effects. This is the framework we consider. Nonetheless, achieving a consistent holographic match requires overcoming several technical issues. A substantial portion of this second part is, therefore, dedicated to developing techniques for dealing with higher-derivative supergravity, either by building on existing methods in the literature or by introducing new ones with the goal of reproducing~\eqref{eq:index_asympt} as a saddle-point contribution to the gravitational path integral. 

\medskip

The first question we must address is which supergravity action to consider. We focus on Lagrangians supplemented by four-derivative corrections:
\begin{equation}
{\cal L} = {\cal L}^{(2\partial)} + \alpha\, {\cal L}^{(4\partial)}  \,,
\end{equation}
where ${\cal L}^{(2\partial)}$ is the two-derivative bosonic Lagrangian of five-dimensional gauged supergravity, and $\alpha$ is some parameters with the dimension of length$^2$ which controls the higher-derivative terms. To determine the appropriate form of ${\cal L}^{(4\partial)}$ recall that the Cardy limit of the superconformal index is governed by the global anomalies of the dual CFT. Specifically, the expression~\eqref{eq:index_asympt} is controlled by the cubic and linear ’t Hooft anomaly coefficients, while the result for the universal sector~\eqref{eq:R_index} depends on the cubic and linear R-symmetry anomalies. A classic result in AdS/CFT~\cite{Witten:1998qj} is that field theory 't Hooft anomalies arise from the bulk supergravity action as boundary terms obtained by varying suitable \emph{Chern-Simons terms}, and that in holography this is precisely the mechanism by which gravity matches the dual field theory global anomalies. The relevant Chern-Simons terms that reproduce the cubic and linear 't Hooft anomalies are given by two- and four-derivative terms, respectively, and involve the bulk gauge fields associated with conserved currents. They take the form 
\be\label{CSterms_intro}
\ \frac{1}{24\pi^2}\left( k_{IJK} A^I \wedge \diff A ^J \wedge \diff A^K - \frac{1}{8}  k_I A^I \wedge R_{ab} \wedge R^{ab} \right)\,,
\ee
where $A^I$ are Abelian gauge fields and $R_{ab}$ is the Riemann curvature two-form. This suggests that the supersymmetrization of these Chern-Simons terms is precisely what is needed to match the first line of~\eqref{eq:index_asympt}. As we will see, this is indeed the case. The action obtained this way should be understood neither as a consistent truncation (at least not with the usual definition of consistent truncations as properties of the classical equations of motion, since the higher-derivative terms may be generated quantum mechanically in the compactification), nor as a low-energy effective action (since we are not including all massless modes), it is rather a supersymmetric effective action for the SCFT anomalies. 

The supersymmetrization of  the mixed gauge-gravitational Chern-Simons term, which involves four derivatives, is a challenging task. This is because in a four-derivative supergravity theory the supersymmetry transformations themselves receive ${\cal O}(\alpha)$ corrections, as the superalgebra only closes on-shell. In contrast, the construction of higher-derivative supersymmetric invariants is significantly simplified in theories that admit an \emph{off-shell} formulation, where the algebra closes independently of the equations of motion (i.e. off-shell)~\cite{Hanaki:2006pj, Bergshoeff:2011xn, Ozkan:2013uk, Ozkan:2013nwa,Ozkan:2024euj}. 
An off-shell formulation of ${\cal N}=2$, $D=5$ supergravity is provided by five-dimensional conformal supergravity~\cite{Bergshoeff:2001hc, Fujita:2001kv, Kugo:2002js, Bergshoeff:2002qk}. The larger symmetry group facilitates the classification of supermultiplets, and the \emph{superconformal tensor calculus} offers a systematic method for constructing invariant Lagrangians by providing embedding formulae for new multiplets into existing ones. In this framework, thanks to the presence of \emph{auxiliary fields} in the multiplets, the superalgebra closes off-shell, hence the supersymmetry transformations are fixed once and for all, as they do not depend on the specific Lagrangian. To obtain an \emph{off-shell} Poincaré supergravity, one then imposes suitable gauge-fixing conditions and assigns a non-zero VEV to a compensator field, thereby breaking the superconformal group down to the Poincaré subgroup. A concise review of this formalism is provided in appendix~\ref{chap:conformal_sugra}. 

To obtain an ${\cal N}=2$ supergravity theory with the correct propagating degrees of freedom, one must finally integrate out the auxiliary fields. This involves solving their equations of motion in terms of the dynamical fields and substituting the solutions back into the action. This procedure is valid as long as the auxiliary fields remain non-dynamical, that is, their equations of motion remain algebraic. However, once higher-derivative terms are introduced, auxiliary fields typically acquire kinetic terms and become dynamical. To preserve consistency, one is therefore forced to work at \emph{linear order} in the perturbative $\alpha$ expansion. At this order, it suffices to use the solutions to the two-derivative equations of motion for the auxiliary fields, since higher-derivative corrections would contribute only at ${\cal O}(\alpha^2)$, which are subleading in this approximation~\cite{Cremonini:2008tw}. Treating $\alpha$ as a small parameter is natural from the perspective of effective field theories, where higher-derivative terms arise as quantum corrections to the low-energy theory from UV physics. This includes, for example, $\alpha'$ corrections in string theory or subleading effects in the large-$N$ expansion of the dual CFT. 

Working perturbatively also allows for the use of field redefinitions to simplify the form of the supergravity action, leading to more manageable final expressions. Thanks to the power of field redefinitions, for instance, in minimal five-dimensional supergravity, the general corrected supersymmetric action (Eq.~\eqref{eq:4daction2}, following~\cite{Cassani:2022lrk}, which refines earlier results in~\cite{Hanaki:2006pj, Cremonini:2008tw,Baggio:2014hua, Bobev:2021qxx,Liu:2022sew}), involves only a handful of higher-derivative terms: the Gauss-Bonnet invariant, a contraction of the Weyl tensor with two graviphoton field strengths, a quartic graviphoton term, and the gauge-gravitational Chern-Simons term. Another powerful simplification occurs in four-dimensional minimal supergravity. Four-derivative corrections in this context have been constructed in~\cite{Bobev:2020egg,Bobev:2021oku}, where it was noted that all solutions of the two-derivative theory are also solutions after the corrections are included, allowing to obtain the holographic match without having to solve for the corrected geometries. This remarkable observation was later clarified in~\cite{Cassani:2023vsa}, where it was demonstrated that the supersymmetric terms with four derivatives can be reduced, via perturbative field redefinitions, to just the Gauss-Bonnet combination of curvature-squared terms, which is topological in four dimensions and does not affect the equations of motion.

Finally, the effective field theory treatment has the additional advantage of avoiding the typical pathologies of higher-derivative gravity theories, such as the emergence of unphysical propagating degrees of freedom, including ghosts or negative energy modes~\cite{Stelle:1977ry}.

\medskip

The next step is to evaluate the four-derivative action on an appropriate black hole background. Finding exact black hole solutions (electrically charged and rotating) in higher-derivative gravity theories is notoriously hard, even when working perturbatively in the small $\alpha$ expansion. A more tractable task involves computing the four-derivative corrections to the near-horizon geometry of the extremal, supersymmetric configuration. In fact, thanks to the enhanced isometries associated with the emergence of an AdS$_2$ factor, finding exact solutions to the equations of motion becomes simpler (for instance, they become algebraic for geometries with $J_1 = J_2$). While the corrected near-horizon geometry allows for the computation of extremal charges and entropy (see sec.~\ref{sec:corrections_GR}), it is insufficient for evaluating the full on-shell action (see discussion in sec.~\ref{sec:Lor_susy_BH_5D}).
Nevertheless, we are able to compute the corrected on-shell action without knowing the full corrected black hole solution. By adapting an argument from~\cite{Reall:2019sah} to asymptotically AdS spacetimes, we demonstrate that it suffices to evaluate the action on the uncorrected (i.e.~two-derivative) solution. A crucial ingredient for this to work with the same boundary conditions as the two-derivative theory (consistent with considering a grand-canonical ensemble, where the thermodynamic potentials are held fixed), as well as the perturbative treatment at linear order in $\alpha$.

In addition, computing the on-shell action requires a careful identification of the appropriate boundary terms. These include both the generalized Gibbons-Hawking-York terms, which ensure the differentiability of the action under Dirichlet boundary conditions, and the holographic counterterms necessary to cancel divergences and yield a finite, renormalized action. While a complete classification of higher-derivative boundary terms for theories involving the metric and gauge fields is not known, we show that the terms used e.g. in~\cite{Myers:1987yn,Cremonini:2009ih} are sufficient for our purposes. A detailed discussion of these boundary terms is provided in sec.~\ref{sec:holographicrenormalization}, with further details in secs.~\ref{sec:onshellact} and~\ref{sec:boundaryterms}.
With these ingredients in place, we proceed to compute the supersymmetric on-shell action and show that it correctly reproduces the Cardy limit of the logarithm of the dual superconformal index.

Our strategy in the second part of this thesis is to first address the technical challenges discussed above within the universal sector of the class of ${\cal N}=1$, $D=4$ SCFTs of interest, whose Cardy-limit prediction for the superconformal index was given in~\eqref{eq:log_I}. This is holographically reproduced by the on-shell action of minimal ${\cal N}=2$ five-dimensional gauged supergravity, which includes a single graviphoton and no vector multiplets. The analysis of four-derivative corrections to the AdS$_5$ black hole in this minimal setup, based on~\cite{Cassani:2022lrk,Cassani:2023vsa} (see also~\cite{Bobev:2022bjm} for related results), serves as a toy model that allows us to illustrate the relevant techniques in a simplified context, before tackling the more involved case of matter-coupled supergravity, which captures the effects of flavour symmetries in the dual CFT index. In section.~\ref{sec:from_superconformal_to_supergravity}, we detail the construction of the four-derivative effective action for minimal gauged supergravity, starting from the off-shell formulation provided by conformal supergravity. We then turn in sec.~\ref{sec:onshellact} to the computation of the on-shell action. The final expression, given in~\eqref{action_almost_done}, precisely matches the field theory prediction~\eqref{eq:log_I} upon using the holographic dictionary relating R-symmetry anomalies to the coefficients of the Chern-Simons terms, as derived in appendix~\ref{sec:holographicdictionary}. In sec.~\ref{sec:chargesandBPSlimit}, we compute the entropy of the supersymmetric and extremal black hole, comparing the result obtained from Euclidean methods to that derived from a (constrained) Legendre transform. Before proceeding to matter-coupled supergravity, chapter~\ref{sec:structure_2} provides a more detailed discussion of certain conceptual and technical issues that arise in higher-derivative supergravity. Focusing on minimal gauged supergravity, we examine the proper derivation of boundary terms for asymptotically locally AdS (AlAdS) solutions, the definition of conserved charges in the presence of higher-derivative and Chern-Simons terms, and how these charges and the entropy can be extracted from the corrected near-horizon geometry. We then return, in~\ref{chap:flavour}, to the holographic computation of \eqref{eq:index_asympt} in a five-dimensional gauged supergravity coupled to multiple vector multiplets (based on~\cite{Cassani:2024tvk}). Here, we encounter a surprising limitation: very few asymptotically AdS$_5$ black hole solutions carrying multiple electric charges and admitting uplift to string or M-theory are known, although more are expected to exist. In fact, only solutions with up to three independent electric charges have been explicitly constructed. These uplift to Type IIB supergravity on $S^5$ or its quotients $S^5/\Gamma$~\cite{Gutowski:2004yv,Cvetic:2004ny,Chong:2005da,Kunduri:2006ek,Mei:2007bn,Wu:2011gq}, and are dual to $\mathcal{N}=4$ SYM with gauge group ${\rm SU}(N)$ and the $\mathcal{N}=1$ $\mathbb{C}^3/\Gamma$ quiver gauge theories, namely the theories describing the low-energy dynamics of D3-branes probing a $\mathbb{C}^3/\Gamma$ singularity, where $\Gamma$ is a discrete subgroup of ${\rm SU}(3)$. As a result the holographic match we obtain is restricted to this subclass of solutions. More details on the specific setup are provided at the beginning of chapter~\ref{chap:flavour}. The final results for the on-shell action for this class of solutions are summarized in sec.~\ref{sec:resultsandmatch}.

\section{Preliminaries: Corrections to AdS$_5$ black hole thermodynamics from higher-derivative supergravity}

The present section, based on~\cite{Cassani:2022lrk}, is organized as follows. 
In section~\ref{FourDerAction} we work out the four-derivative effective action for ${\cal N}=2$ minimal gauged supergravity in five dimensions. In order to do so, we start from off-shell conformal supergravity in the standard Weyl formulation where curvature-squared invariants can be constructed systematically, gauge-fix the redundant symmetries to recover a consistent off-shell Poincar\'e supergravity, and finally integrate out the auxiliary fields at first order in the perturbative $\alpha$ expansion. Finally, the resulting action is simplified exploiting field redefinitions.\footnote{Some related computations have also appeared in~\cite{Cremonini:2008tw,Baggio:2014hua,Bobev:2021qxx,Liu:2022sew,Bobev:2022bjm}, while progress in four dimensions has been reported e.g. in~\cite{Bobev:2020egg,Bobev:2020zov,Bobev:2021oku,Genolini:2021urf}.} In section~\ref{sec:onshellact} we adapt to our setup the argument of~\cite{Reall:2019sah}, in order to evaluate the four-derivative action at linear order in $\alpha$ on the black hole solution. This allows us to match the field theory result \eqref{eq:log_I} after imposing supersymmetry. In section~\ref{sec:chargesandBPSlimit} we use the black hole thermodynamical relations to obtain the charges and the entropy, focusing on their extremal values. This leads us to the microcanonical form of the supersymmetric and extremal entropy \eqref{eq:four_d_susy_ent}.
Additionally, we further validate our results by directly performing the Legendre transform of the supersymmetric on-shell action, or equivalently of~\eqref{eq:log_I}, at linear order in ${\rm Tr}\,\mathcal{R}$. Besides representing a nice consistency check, this computation provides a very direct way to obtain the BPS entropy, extending to the higher-derivative setup the method already proven useful in~\cite{Cabo-Bizet:2018ehj,Cassani:2019mms}. In appendix~\ref{sec:holographicdictionary} we discuss the holographic dictionary between the gravitational couplings and the SCFT anomaly coefficients, in app.~\ref{app:eoms} we provide the higher-derivative equations of motion, while in appendix~\ref{app:field_red_minimal} field redefinitions are discussed. 


\subsection{Constructing the action: ${\cal N}=2$, $D=5$ supergravity from superconformal tensor calculus}
\label{sec:from_superconformal_to_supergravity}

\subsubsection{The two-derivative theory}

In this section we review the derivation of minimal gauged ${\cal N}=2$, $D=5$ supergravity starting from the off-shell formulation provided by conformal supergravity. The procedure follows three steps:

\begin{itemize}
\item Introduce the superconformal multiplets and construct invariant actions using superconformal methods;
\item Break the extra superconformal symmetries by imposing gauge-fixing conditions, reducing the symmetry to the super-Poincaré algebra;
\item Integrate out auxiliary fields introduced to achieve off-shell closure of the algebra.
\end{itemize}

The procedure to obtain a consistent supergravity is, however, not unique. To begin with one can make use of two inequivalent Weyl multiplets~\cite{Bergshoeff:2001hc}\footnote{
In five dimensions with ${\cal N}=2$ supersymmetries, conformal supergravity admits two inequivalent Weyl multiplets, known as the \emph{standard} and \emph{dilaton Weyl multiplets}, distinguished by their auxiliary matter content. In what follows we adopt the formulation based on the \emph{standard Weyl multiplet}; details on the dilaton Weyl multiplet may be found in appendix~\ref{chap:conformal_sugra} and the cited literature.}. Even when one of these has been chosen (in our case, we will consider the so-called \emph{standard Weyl multiplet}), one still has the freedom to choose which matter multiplets will act as \emph{compensators}, ensuring a consistent gauge-fixing of the redundant symmetries. Following~\cite{Ozkan:2013nwa}, we will consider a \emph{linear multiplet} and a \emph{vector multiplet} (instead of one hyper-multiplet, as in~\cite{Bergshoeff:2004kh}). Let us now introduce the multiplets relevant to our discussion (further details and transformation rules are relegated to appendix~\ref{chap:conformal_sugra}):

\paragraph{The standard Weyl multiplet.}

The standard Weyl multiplet contains the gauge fields associated with the generators of the $D=5$, ${\cal N}=2$ superconformal algebra (vielbein included), plus auxiliary matter fields needed for off-shell closure \cite{Bergshoeff:2001hc}. The independent components of the standard Weyl multiplet are:
\begin{equation}
\bigl(e^a{}_{\mu},\,\psi_{\mu}^i,\,b_{\mu},\,V_{\mu}^{ij},\,T_{ab},\,\chi^i,\,D\bigr)\,.
\end{equation}
Here $e^a{}_{\mu}$ is the vielbein and $\psi_{\mu}^i$ the gravitino, with $i=1,2$ being an SU$(2)_R$ index. The vectors $b_{\mu}$ and $V_{\mu}^{ij}$ are gauge fields for dilatations and SU$(2)_R$ rotations, respectively. The auxiliary fields consist of an antisymmetric tensor $T_{ab}$, a scalar $D$, and a symplectic Majorana spinor $\chi^i$. 

\paragraph{Matter multiplets.}
To build the gauged supergravity action we couple the Weyl multiplet to two types of matter multiplets:
\begin{itemize}
\item The \emph{abelian vector multiplet}:
\begin{equation}
\bigl(A_{\mu},\,\lambda^i,\,Y^{ij},\,X\bigr)\,,
\end{equation}
where $A_{\mu}$ is the gauge field, $\lambda^i$ the gaugino, $X$ a real scalar and $Y^{ij}=Y^{(ij)}$ an auxiliary SU$(2)_R$ triplet.

\item The \emph{linear multiplet} used as a compensator in the gauge-fixing:
\begin{equation}
\bigl(L^{ij},\,\varphi^i,\,E_a,\,N\bigr)\,,
\end{equation}
where $L^{ij}=L^{(ij)}$ is an SU$(2)_R$ triplet, $\varphi^i$ a fermionic doublet, $E_{a}$ a constrained vector (subject to a divergence-free condition), and $N$ a scalar.  
\end{itemize}

%
%

\paragraph{Superconformal actions and gauge-fixing.}

A consistent conformal supergravity that reduces to an \emph{off-shell} Poincar\'e minimal supergravity, after appropriate gauge-fixing of the redundant symmetries, can be constructed by combining the vector-linear \eqref{eq:L_VL}, linear \eqref{eq:L_linear} and vector \eqref{eq:L_vector2} superconformal actions. 

Specifically, for the case of a single vector multiplet one can construct the following Lagrangian:\footnote{In this Section, we take $n=0$, where $n+1$ denotes the number of vector multiplets. Also, our normalization of the gauge coupling constant differs from the one used in the appendix.}
\begin{equation}\label{eq:superc_conf_action_0}
    {\cal L}_{gR}^{\rm S} \,=\, - {\cal L}_{L}^{\rm S} - 3 {\cal L}_V^{\rm S} - \sqrt{6}\,g\, {\cal L}_{VL}\,.
\end{equation}
The action \eqref{eq:superc_conf_action_0} is invariant under the full superconformal group. To obtain an action invariant only under the super-Poincaré algebra, one must break the additional conformal symmetries by imposing gauge-fixing conditions. These conditions are implemented with the help of a compensator (in this case, the linear multiplet introduced above) to fix dilatations, special conformal boosts, and special supersymmetries, while reducing the SU$(2)_R$ symmetry to its U$(1)_R$ subgroup. Following~\cite{Ozkan:2013nwa}, we adopt the gauge-fixing conditions
\begin{equation}
\label{eq:gauge_fixing_cond}
    L_{ij} = \frac{1}{\sqrt{2}}\delta_{ij}\,,\qquad b_\mu =0\,,\qquad \varphi^i=0\,.
\end{equation}
Requiring invariance of these conditions under the full set of superconformal transformations (i.e. demanding $\delta b_\mu = \delta\varphi^i = \delta L_{ij} =0$, with transformation rules given in app.~\ref{chap:conformal_sugra}) uniquely fixes the parameters of dilatations, special conformal boosts, and special supersymmetry transformations, leaving only the super-Poincaré symmetries unbroken.

\paragraph{Poincar\'e supergravity: integrating out auxiliary fields.}
After gauge-fixing, \eqref{eq:superc_conf_action_0} gives the action for off-shell $\mathcal{N}=2$ Poincar\'e gauged supergravity~\cite{Ozkan:2013nwa}: 
\begin{equation}\label{eq:offshell2daction}
\begin{aligned}
S_{\rm{off\text{-}shell}}^{(2\partial)} &= \frac{1}{8\pi G}\int \diff^5 x\,e\,{\cal L}_{gR}^{\rm S}\\[1mm]
&=\frac{1}{16\pi G}\int \diff^5x \,e\left[\frac{1}{4}\left({\cal C}+3\right)R+\frac{1}{4}{\cal C}'' F^2+\frac{16}{3}\left(13{\cal C}-1\right)T_{\mu\nu}T^{\mu\nu}-8{\cal C}' F_{\mu\nu}T^{\mu\nu}\right.\\
&+8\left({\cal C}-1\right)D+2{\tilde V}_{\mu}{}^{ij}{\tilde V}^{\mu}{}_{ij}-2\sqrt{2}V_{\mu} E^{\mu}-2 E_{\mu}E^{\mu}-2N^2+\frac{1}{2}{\cal C}'' (\partial X)^2-{\cal C}'' Y^{ij}Y_{ij}\\
&\left.+\frac{1}{24}{\cal C}'''\epsilon^{\mu\nu\rho\sigma\lambda}F_{\mu\nu}F_{\rho\sigma} A_{\lambda}-2\sqrt{3}\,g Y_{ij}\delta^{ij}-2\sqrt{6}\,g A_{\mu}E^{\mu}-2\sqrt{6}\,g X N \right]\, ,
\end{aligned}
\end{equation}
where we introduced the cubic polynomial ${\cal C} \propto X^3$. To obtain the on-shell supergravity theory, we solve the auxiliary field equations of motion and then plug the solution back into the action. Let us first note that the scalar $X$ is fixed to an arbitrary constant by the equation of motion of $D$, which is satisfied by
\begin{equation}
{\cal C}=1\, , 
\end{equation}
and which further implies 
\begin{equation}
{\cal C}'=\frac{3}{X}\, , \hspace{1cm}{\cal C}''=\frac{6}{X^2}\, , \hspace{1cm} {\cal C}'''=\frac{6}{X^3}\,.
\end{equation}
The remaining equations of motion are solved by 
\begin{equation}\label{eq:auxfields}
\begin{aligned}
{\tilde V}_{\mu}^{ij}=\,&0\, , \qquad V_{\mu}=\,-\sqrt{3}g A_{\mu}\, , \qquad E_{\mu}=\,\,0\, ,\\[1mm]
N=\,&-\frac{\sqrt{3} gX}{\sqrt{2}}\, ,\qquad Y_{ij}=-\frac{gX^2}{2\sqrt{3}}\, \delta_{ij}\, ,\qquad T_{\mu\nu}=\,\frac{3}{16X}F_{\mu\nu}\, ,\\[1mm]
D=\,&-\frac{1}{32}\left(R-\frac{1}{4X^2}F^2+\frac{20}{3}g^2 X^2\right)\, .
\end{aligned}
\end{equation}
In this section, we prefer to choose the canonical normalization for the gauge field $A_{\mu}$\footnote{The gauge field appearing in \eqref{eq:action_minimal_gauged_sugra} is related to the one used in this section by $A^{({\rm here})} = \frac{2}{\sqrt{3}g}A^{({\rm there})}$.}, recovered once $X$ is fixed to
\begin{equation}
X=-\sqrt{3}\, .
\end{equation}
Using this value in \eqref{eq:auxfields}, and plugging this expressions back into \eqref{eq:offshell2daction}
we find\footnote{By $S$ we denote actions in Lorentzian signature, to distinguish them from their Euclidean counterpart, $I = -\ii S$.} 
\begin{equation}\label{eq:2_der_minimal_sugra_5D}
S^{(2\partial)}=\frac{1}{16\pi G}\int \diff^5x \, e \left[R-\frac{1}{4}F^2+12g^2 -\frac{1}{12\sqrt{3}}\epsilon^{\mu\nu\rho\sigma\lambda}F_{\mu\nu}F_{\rho\sigma} A_{\lambda}\right]\, ,
\end{equation}
which is nothing but the bosonic action of minimal gauged supergravity.

\subsubsection{The four-derivative effective action}\label{FourDerAction}

We now give the most general four-derivative corrections (up to field redefinitions) to minimal gauged supergravity which is compatible with its local symmetries, namely diffeomorphism, gauge invariance and supersymmetry. 

We start by supplementing the two-derivative action (\ref{eq:2_der_minimal_sugra_5D}) with four-derivative corrections, 
\begin{equation}\label{eq:4daction}
S=\frac{1}{16\pi G}\int \diff^5x \, e \left[d_{0}R+12g^2d_{1}-\frac{d_2}{4}F^2-\frac{d_3}{12\sqrt{3}}\epsilon^{\mu\nu\rho\sigma\lambda}F_{\mu\nu}F_{\rho\sigma} A_{\lambda}+\alpha \,{\mathcal L}_{4\partial}\right]\, ,
\end{equation}
where ${\mathcal L}_{4\partial}$ contains all the possible four-derivative terms constructed out of the low-energy degrees of freedom which are compatible with the local symmetries of the two-derivative theory. The parameter $\alpha$ has dimensions of length$^2$ and we assume that $\alpha {\cal R}\ll 1$, where $\cal R$ here denotes  the curvature scale of the solution. This guarantees us that terms with six or more derivatives can be safely neglected as they will be subleading with respect to the four-derivative terms that we have included. The most general four-derivative correction compatible with supersymmetry that we can write down is a linear combination of three independent supersymmetric invariants $\{{\cal L}_{i}^{(4\partial)}\}_{i=1,2,3}$\,,
\begin{equation}\label{eq:L4dSUSY}
{\cal L}_{4\partial}=\lambda_{1} \,{\cal L}_{1}^{(4\partial)}+\lambda_{2} \;{\cal L}_{2}^{(4\partial)}+ \lambda_{3} \,{\cal L}_{3}^{(4\partial)}\, ,
\end{equation}
where $\lambda_{1}$, $\lambda_{2}$ and $\lambda_{3}$ are dimensionless couplings.\footnote{The parameters $\lambda_{1,2,3}$ are not fixed in the effective theory and should in principle be determined from the UV completion.} These three invariants can be understood (for instance) as the supersymmetrizations of the $R_{\mu\nu\rho\sigma}R^{\mu\nu\rho\sigma}$, $R_{\mu\nu}R^{\mu\nu}$ and $R^2$ terms.
In addition to the four-derivative corrections, since $\alpha g^2$ is a dimensionless quantity, we can expect possible corrections to the two-derivative terms controlled by this parameter (hence, we are also assuming $\alpha g^2 \ll 1$). This kind of corrections appear through the dimensionless couplings $d_i$, which are of the form
\begin{equation}
d_{i}=1+\alpha g^2 \delta d_{i}\,,
\end{equation}
being $\delta d_{i}$ a linear combination of $\lambda_1$, $\lambda_2$ and $\lambda_3$, as we will see later.

The approach we follow to construct a basis of supersymmetric invariants $\{{\cal L}_{i}^{(4\partial)}\}_{i=1,2,3}$ involves the following steps:
\begin{itemize}
\item Construct the superconformal invariants containing curvature-square terms using techniques from superconformal tensor calculus (see e.g.~\cite{Hanaki:2006pj, Ozkan:2013nwa});
\item As for the two-derivative theory, four-derivative corrections to \emph{off-shell} Poincar\'e supergravity are obtained by imposing the gauge-fixing conditions \eqref{eq:gauge_fixing_cond};
\item Integrate out the auxiliary fields at linear order in $\alpha$;
\item Simplify the final form of the action by performing perturbative field redefinitions of the form
\begin{equation}\label{eq:fieldredefinition}
g_{\mu\nu}\rightarrow g_{\mu\nu}+\alpha\, \Delta_{\mu\nu}\, , \hspace{1cm} A_{\mu}\rightarrow A_{\mu}+\alpha\, \Delta_{\mu}\, ,
\end{equation}
which allow us to eliminate most of the four-derivative terms.
\end{itemize}

\paragraph{Summary of results.} We will show that the action (\ref{eq:4daction}) can be brought to the following form using field redefinitions,
\begin{equation}\label{eq:4daction1}
\begin{aligned}
&S=\frac{1}{16\pi G}\int \diff^5x \,e \left\{c_{0}R+12c_1g^2-\frac{c_2}{4}F^2-\frac{c_3}{12\sqrt{3}}\epsilon^{\mu\nu\rho\sigma\lambda}F_{\mu\nu}F_{\rho\sigma} A_{\lambda}\right.\\
&\left.+\,\lambda_1\alpha\left[R_{\mu\nu\rho\sigma}R^{\mu\nu\rho\sigma}-\frac{1}{2}R_{\mu\nu\rho\sigma}F^{\mu\nu}F^{\rho\sigma}+\frac{5}{36}\left(F^2\right)^2-\frac{13}{24}F^4-\frac{1}{2\sqrt{3}}\epsilon^{\mu\nu\rho\sigma\lambda}R_{\mu\nu\alpha\beta}R_{\rho\sigma}{}^{\alpha\beta} A_{\lambda}\right]\right\} ,
\end{aligned}
\end{equation}
where $F^4= F_{\mu\nu}F^{\nu\rho}F_{\rho\sigma}F^{\sigma\mu}$. The coefficients in front of the two-derivative terms are given by $c_i=1+\alpha g^2{\delta c}_{i}$, where
\begin{equation}
{\delta c}_{0}=4\lambda_2\, , \hspace{.5cm}\delta c_{1}=-\frac{10\lambda_1}{3}+4\lambda_2 \,, \hspace{.5cm} {\delta c}_{2}=\frac{32\lambda_1}{3}+4\lambda_2 \,, \hspace{.5cm}{\delta c}_{3}=-12\lambda_1+4\lambda_2 \,.
\end{equation}
Note that of the three coefficients in~\eqref{eq:L4dSUSY}, only $\lambda_1,\lambda_2$ appear in~\eqref{eq:4daction1}. In fact, below we will give an argument proving that it is possible to choose the basis of invariants such that the one controlled by $\lambda_3$ 
yields a vanishing contribution after implementing appropriate field redefinitions.  Alternatively, considering the same basis of four-derivative invariants as in Ref.~\cite{Liu:2022sew}, we obtain
\begin{equation}\label{eq:4daction2}
\begin{aligned}
S\,=&\,\frac{1}{16\pi G}\int \diff^5x \,e \left\{{\tilde c}_0 R+12{\tilde c}_1g^2-\frac{{\tilde c}_2}{4}F^2-\frac{{\tilde c}_3}{12\sqrt{3}}\epsilon^{\mu\nu\rho\sigma\lambda}F_{\mu\nu}F_{\rho\sigma} A_{\lambda}\right.\\[2mm]
&\left.\,+\,\lambda_1 \alpha \left[{\cal X}_{\text{GB}}-\frac{1}{2}C_{\mu\nu\rho\sigma}F^{\mu\nu}F^{\rho\sigma}+\frac{1}{8}F^4-\frac{1}{2\sqrt{3}}\epsilon^{\mu\nu\rho\sigma\lambda}R_{\mu\nu\alpha\beta}R_{\rho\sigma}{}^{\alpha\beta} A_{\lambda}\right]\right\}\, ,
\end{aligned}
\end{equation}
where ${\cal X}_{\text{GB}}=R_{\mu\nu\rho\sigma}R^{\mu\nu\rho\sigma}-4 R_{\mu\nu}R^{\mu\nu}+R^2$ is the Gauss-Bonnet invariant,  $C_{\mu\nu\rho\sigma}=R_{\mu\nu\rho\sigma}-\frac{2}{3}\left(R_{\mu[\rho}g_{\sigma]\nu}+R_{\nu[\sigma}g_{\rho]\mu}\right)+\frac{1}{6}Rg_{\mu[\rho}g_{\sigma]\nu}$ is the Weyl tensor and ${\tilde c}_{i}=1+\alpha g^2 {\delta \tilde c}_{i}$, with
\begin{equation}\label{ctildes}
{\delta \tilde c}_{0}=4\lambda_2\, , \hspace{.5cm}{\delta\tilde c}_{1}=-10\lambda_1+4\lambda_2 \,, \hspace{.5cm} {\delta \tilde c}_{2}=4\lambda_1+4\lambda_2 \,, \hspace{.5cm}{\delta 
\tilde c}_{3}=-12\lambda_1+4\lambda_2 \,.
\end{equation}
We observe that the corrections controlled by $\lambda_2$ are proportional to the two-derivative Lagrangian in both \eqref{eq:4daction1} and \eqref{eq:4daction2}. Thus, they can be simply interpreted as corrections to the Newton's constant $G$ (cf.~\eqref{eq:Geff}), which do not modify the solutions of the two-derivative theory. 

%
%

\subsubsection{Field redefinitions}\label{sec:fieldred}

Before introducing higher-derivative superconformal invariants, let us briefly discuss how certain field redefinitions can be used to simplify actions with four-derivative terms at linear order in the expansion parameter. 
Field redefinitions of the form~(\ref{eq:fieldredefinition}) induce the following terms into the action
\begin{equation}
S\rightarrow S-\frac{\alpha}{16\pi G}\int \diff^5x\,e\left[\left({\cal E}_{\mu\nu}-\frac{1}{2}g_{\mu\nu}\,{\cal E}\right)\Delta^{\mu\nu}-{\cal E}^{\mu}\Delta_{\mu}\right]+ {\cal O}(\alpha^2)\,,
\end{equation}
where ${\cal E}_{\mu\nu},\, {\cal E}^\mu$ are the two-derivative equations of motion,
\begin{equation}
\label{EoM2der}
\begin{aligned}
{\cal E}_{\mu\nu} &= R_{\mu\nu} + 4g^2 g_{\mu\nu} - \frac{1}{2}F_{\mu\rho}F_\nu{}^\rho + \frac{1}{12}g_{\mu\nu} F_{\rho\sigma}^2 =0\,,
\\[1mm]
{\cal E}^\mu &= \nabla_\nu F^{\nu\mu} - \frac{1}{4\sqrt{3}}\epsilon^{\mu\nu\rho\sigma\lambda} F_{\nu\rho} F_{\sigma\lambda} =0\,,
\end{aligned}
\end{equation}
and ${\cal E}=g^{\mu\nu}{\cal E}_{\mu\nu}$. The induced terms can be used to eliminate all four-derivative terms of the form ${\cal E}_{\mu\nu}K^{\mu\nu}$ and ${\cal E}_{\mu}L^{\mu}$ by choosing $\Delta_{\mu\nu}$ and $\Delta_{\mu}$ appropriately, which amounts to take
\begin{equation}
\Delta_{\mu\nu}=K_{\mu\nu}-\frac{1}{3}\, g_{\mu\nu} K\,, \hspace{1cm}\Delta_{\mu}=-L_{\mu}\,,
\end{equation}
where $K=g^{\mu\nu}K_{\mu\nu}$.
\noindent
In practice, this means that we can use the two-derivative equations of motion in the piece of the action of order $\alpha$ since all the terms that are proportional either to ${\cal E}_{\mu\nu}$ or ${\cal E}_\mu$ can be eliminated with a field redefinition, up to ${\cal O}(\alpha^2)$ terms which we are neglecting. Let us consider as an instance a term of the form $R_{\mu\nu}K^{\mu\nu}$. This term can be replaced in the action by 
\begin{equation}\label{eq:masterrule1}
R_{\mu\nu}K^{\mu\nu}\rightarrow \frac{1}{2}F_{\mu\rho}F_{\nu}{}^{\rho}K^{\mu\nu}-\left(4g^2+\frac{1}{12}F^2\right) K \, ,
\end{equation}
since the difference between the left and right-hand sides is ${\cal E}_{\mu\nu}K^{\mu\nu}$. For the same reason, a term $\nabla_{\rho}F^{\rho \mu}L_{\mu}$ in the action can be replaced by 
\begin{equation}\label{eq:masterrule2}
\nabla_{\rho}F^{\rho \mu}L_{\mu}\rightarrow \frac{1}{4\sqrt{3}}\epsilon^{\mu\nu\rho\sigma\lambda}F_{\nu\rho}F_{\sigma\lambda}L_{\mu}\, .
\end{equation}

A full set of replacement rules to implement field redefinitions in a convenient way are listed in appendix~\ref{app:field_red_minimal}. We emphasize that this applies in a general context, since we have not yet imposed supersymmetry of the action. This is what we will do next.

\subsubsection{Off-shell supersymmetric invariants}
\label{sec:susyinvariants}

In this section we give a more detailed proof that the action can be written as above. We consider the formulation of off-shell ${\cal N}=2$ Poincar\'e supergravity based on the standard Weyl multiplet~\cite{Bergshoeff:2004kh}, as reviewed above. Among the three independent supersymmetric invariants, we will focus on the two corresponding to the supersymmetric completions of the Weyl squared term $C_{\mu\nu\rho\sigma}^2$~\cite{Hanaki:2006pj} and of the Ricci-scalar squared term $R^2$~\cite{Ozkan:2013nwa}. They can also be interpreted as the supersymmetrizations of the $\epsilon^{\mu\nu\rho\sigma\lambda}R_{\mu\nu\alpha\beta}R_{\rho\sigma}{}^{\alpha\beta} A_{\lambda}$ and $\epsilon^{\mu\nu\rho\sigma\lambda}F_{\mu\nu}F_{\rho\sigma}A_{\lambda}$ Chern-Simons terms, which provide the correct contributions to reproduce the R-symmetry anomalies of the dual CFT. 
%
The third invariant, corresponding to the supersymmetric completion of the Ricci tensor squared term $R_{\mu\nu}^2$, was constructed in superspace in~\cite{Butter:2014xxa}, and has recently been studied in~\cite{Gold:2023dfe,Gold:2023ykx,Gold:2023ymc,Gold:2025ttt}.\footnote{In the context of matter-coupled supergravity theories, a complete basis for these invariants had also been constructed in the off-shell formulation based on the dilaton Weyl multiplet \cite{Ozkan:2013nwa}, but only in the ungauged limit.}
However, as argued in~\cite{Cassani:2022lrk}, there is no loss of generality in setting the third invariant to zero (i.e. $\lambda_3 = 0$). This can be made precise by means of field redefinitions: terms such as $R_{\mu\nu}^2$ can be eliminated using the substitution rules derived in app.~\ref{app:field_red_minimal}, and one generally expects such terms to generate only two-derivative corrections.\footnote{This has been explicitly verified for the full set of invariants constructed from the dilaton Weyl multiplet in minimal supergravity.} This expectation is also confirmed by the analysis in~\cite{Gold:2023ykx,Gold:2023ymc}, where, starting from a combination of the three independent invariants, a four-derivative effective action characterized by three parameters has been obtained: one controlling the genuine four-derivative corrections, and two affecting only the two-derivative sector. However, one of the two parameters affecting the two-derivative terms turns out to be unphysical, as it can be absorbed by a constant rescaling of the metric. After performing this redefinition, the resulting action precisely matches the one constructed here (up to a redefinition of the coefficients $\lambda_{1,2}$).

Let us then consider the following action, 
\begin{equation}
S_{\rm{off\text{-}shell}}=S_{\rm{off\text{-}shell}}^{(2\partial)}+\frac{\alpha}{16\pi G}\int \diff^5x\,e \left[\lambda_1\, {\cal L}^{\rm{off\text{-}shell}}_{C^2}+\lambda_{2}\,{\cal L}^{\rm{off\text{-}shell}}_{R^2}\right]\, ,
\end{equation}
where $S_{\rm{off\text{-}shell}}^{(2\partial)}$ denotes the off-shell Poincar\'e supergravity action given by \eqref{eq:offshell2daction}, and ${\cal L}^{\rm{off\text{-}shell}}_{C^2}$ and ${\cal L}^{\rm{off\text{-}shell}}_{R^2}$ are the off-shell four-derivative invariants constructed in \cite{Hanaki:2006pj, Ozkan:2013nwa}, and reviewed in app.~\ref{app:offshell_invariants}. 

In order to obtain an on-shell supergravity, we need to integrate out the auxiliary fields. It is important to note, however, that since we are working at first order in $\alpha$, we just need the solution to the two-derivative equations of motion of the auxiliary fields, Eq.~\eqref{eq:auxfields}. To understand why, let us denote by $\Phi_{\rm aux}$ the auxiliary fields, while $\Psi$ will denote the dynamical ones. In the four-derivative off-shell theory, the solution to the corrected equations of motion for the auxiliary fields schematically takes the form
\begin{equation}
\Phi_{\rm aux} (\Psi) = \Phi^{(2\partial)}_{\rm aux}(\Psi) + \alpha \Phi^{(4\partial)}_{\rm aux}(\Psi)\,.
\end{equation}
Now, an on-shell supergravity is obtained as
\begin{equation}\begin{aligned}
S_{\rm on\text{-}shell} &= S_{\rm{off\text{-}shell}}\Big|_{\Phi_{\rm aux}} = 
\\
&= S_{\rm{off\text{-}shell}}^{(2\partial)}\Big|_{\Phi^{(2\partial)}_{\rm aux}} + \alpha \Phi_{\rm aux}^{(4\partial)} \frac{\delta S_{\rm{off\text{-}shell}}^{(2\partial)}}{\delta \Phi_{\rm aux}}\Big|_{\Phi^{(2\partial)_{\rm aux}}} + \alpha S_{\rm{off\text{-}shell}}^{(4\partial)}\Big|_{\Phi^{(2\partial)}_{\rm aux}} + {\cal O}(\alpha^2)\,.
\end{aligned}
\end{equation}
All the terms that involve the corrections to the auxiliary fields are proportional to the two-derivative equations of motion of the latter, hence they do not contribute. Therefore, all we have to do in order to obtain the on-shell action is to substitute the values of the auxiliary fields in the four-derivative invariants. 

Let us first consider the Weyl squared invariant, which is given in \eqref{eq:offshell_Weyl_squared}.\footnote{In this section, we are considering the case in which $n =0$. Also, we have absorbed an overall factor of $-\sqrt{3}\lambda_I \bar X^I$ into $\lambda_{1}$ for convenience.} Substituting the value of the auxiliary fields, setting $X = -\sqrt{3}$, integrating by parts and making use of the Bianchi identities, we obtain
\begin{equation}\label{eq:Weyl20}
\begin{aligned}
&{\cal L}^{\rm{off\text{-}shell}}_{C^2}|_{\eqref{eq:auxfields}}=\,\,R_{\mu\nu\rho\sigma}R^{\mu\nu\rho\sigma}+\tfrac{1}{2}\,R_{\mu\nu\rho\sigma}F^{\mu\nu}F^{\rho\sigma}-\tfrac{5}{32} \,\left(F^2\right)^2+\tfrac{5}{8}\,F^4\\
&-\tfrac{1}{2\sqrt{3}}\, \epsilon^{\mu\nu\rho\sigma\lambda}R_{\mu\nu\alpha\beta}R_{\rho\sigma}{}^{\alpha\beta} A_{\lambda}
-\tfrac{4}{3}\,R_{\mu\nu}R^{\mu\nu}+\tfrac{1}{6}\,R^2+\tfrac{5}{12} \,RF^2-\tfrac{10}{3}\, R^{\mu\nu}F_{\mu\rho}F_{\nu}{}^{\rho}-\nabla_{\mu}F_{\nu\rho}\nabla^{\mu}F^{\nu\rho}\\
&+4\,\nabla_{\rho}F^{\rho\mu}\nabla_{\sigma}F^{\sigma}{}_{\mu}-\tfrac{\sqrt{3}}{4} \,\epsilon^{\mu\nu\rho\sigma\lambda}F_{\nu\rho}F_{\sigma\lambda}\nabla_{\delta}F^{\delta}{}_{\mu}-\tfrac{1}{2\sqrt{3}}\,\epsilon^{\mu\nu\rho\sigma\lambda}F_{\mu\nu}F_{\rho}{}^{\alpha}\nabla_{\sigma}F_{\lambda\alpha}-\tfrac{1}{6}{\cal E} F^2+\tfrac{1}{6}{\cal E}^2\\
&+\tfrac{g^2}{\sqrt{3}}\epsilon^{\mu\nu\rho\sigma\lambda}F_{\mu\nu}F_{\rho\sigma} A_{\lambda}\, .
\end{aligned}
\end{equation}
We observe that the elimination of the auxiliary fields in ${\cal L}^{\rm{off\text{-}shell}}_{C^2}$ gives rise not only to four-derivative corrections but also to ${\cal O}(\alpha g^2)$ (two-derivative) corrections. 

In turn, the evaluation of the $R^2$ term, which is given in \eqref{eq:offshell_R^2}, yields\footnote{Again, there is an overall factor of $\frac{9\sqrt{3}}{4}\sigma_I \bar X^I$ which we have absorbed in $\lambda_2$ for convenience.}
\begin{equation}\label{eq:R2}
{\cal L}^{\rm{off\text{-}shell}}_{R^2}|_{\eqref{eq:auxfields}}=\,g^2\left(R-12g^2-\tfrac{3}{4} F^2-\tfrac{\sqrt{3}}{9}\epsilon^{\mu\nu\rho\sigma\lambda}F_{\mu\nu}F_{\rho\sigma}A_{\lambda}+\tfrac{8}{3}{\cal E}\right) -\tfrac{1}{9}{\cal E}^2\, .
\end{equation}
As we can see, the contribution of this invariant to the four-derivative terms is trivial, as all of the four-derivative terms appear inside ${\cal E}^{2}$, which could have been dropped since it does not contribute to first order in $\alpha$. This means that, as we are going to discuss further, the contribution of this invariant boils down to a change in the coefficients of the two-derivative terms. 

At this stage we are ready to perform perturbative field redefinitions, following~\ref{app:field_red_minimal}. To do so, we first extract the coefficients $a_{i}$ and $b_{i}$ in \eqref{eq:genexpL4d} from the four-derivative terms in \eqref{eq:Weyl20}. We obtain the following values\footnote{We ignore the contributions from ${\cal E}F^2$ and ${\cal E}^2$ since they can be directly eliminated with a field redefinition without changing the rest of the action.}
\begin{equation}
\begin{aligned}
a_1=&\,\lambda_1\,, \hspace{0.5cm}a_2=\frac{\lambda_1}{2}\, ,\hspace{0.5cm}a_3=-\frac{5\lambda_1}{32}\, ,\hspace{0.5cm}a_4=\frac{5\lambda_1}{8}\,, \hspace{0.5cm}a_5=-\frac{\lambda_1}{2\sqrt{3}}\, ,\\[1mm]
b_1=&-\frac{4\lambda_1}{3}\,, \hspace{0.5cm}b_2=\frac{1}{6}\lambda_1\, ,\hspace{0.5cm}b_3=\frac{5\lambda_1}{12}\, ,\hspace{0.5cm}b_4=-\frac{10\lambda_1}{3}\,, \hspace{0.5cm}b_5=-\lambda_1\, ,\\[1mm]
b_6=&\,4\lambda_1\,, \hspace{0.5cm}b_7=0\, ,\hspace{0.5cm}b_8=-\frac{\sqrt{3}\lambda_1}{4}\, ,\hspace{0.5cm}b_9=-\frac{\lambda_1}{2\sqrt{3}}\,.
\end{aligned}
\end{equation}
In turn, from the coefficients in front of the two-derivative terms in \eqref{eq:Weyl20} and \eqref{eq:R2}, we  can read the $\delta d_{i}$ coefficients, getting
\begin{equation}
\delta d_0=\lambda_{2}\,, \hspace{0.5cm} \delta d_1=-\lambda_{2}\,,\hspace{0.5cm}\delta d_3=3\lambda_{2}\,, \hspace{0.5cm}\delta d_4=-12\lambda_1+4\lambda_{2}\,.
\end{equation}
Therefore, we are ready to use \eqref{eq:rulesa'}, \eqref{eq:rulesc}, \eqref{eq:rulesa''} and \eqref{eq:rulesctilde} in order to obtain 
\begin{equation}
\begin{aligned}
a'_{1}=&\lambda_1\, , \hspace{0.5cm} a'_{2}=-\frac{\lambda_1}{2}\, , \hspace{0.5cm}a'_{3}=\frac{5\lambda_1}{36}\, , \hspace{0.5cm} a'_{4}=-\frac{13\lambda_1}{64}\, ,\hspace{0.5cm} a'_5=-\frac{\lambda_1}{2\sqrt{3}}\, ,\\[1mm]
a''_{1}=&\lambda_1\, , \hspace{0.5cm} a''_{2}=-\frac{\lambda_1}{2}\, , \hspace{0.5cm}a''_{3}=0\, , \hspace{0.5cm} a''_{4}=\frac{\lambda_1}{8}\, ,\hspace{0.5cm} a''_5=-\frac{\lambda_1}{2\sqrt{3}}\, ,
\end{aligned}
\end{equation}
and, choosing ${\delta c}_0={\delta \tilde c}_0=4\lambda_2$,
\begin{equation}
\begin{aligned}
\delta c_{1}=&-\frac{10\lambda_1}{3}+4\lambda_2 \,, \hspace{1cm} {\delta c}_{2}=\frac{32\lambda_1}{3}+4\lambda_2 \,, \hspace{1cm}{\delta c}_{3}=-12\lambda_1+4\lambda_2 \,,\\[1mm]
{\delta\tilde c}_{1}=&-10 \lambda_1+4\lambda_2 \,, \hspace{1cm} {\delta\tilde c}_{2}=4\lambda_1+4\lambda_2 \,, \hspace{1cm}{\delta\tilde c}_{3}={\delta c}_3 \,,
\end{aligned}
\end{equation}
as anticipated in \eqref{eq:4daction} and \eqref{eq:4daction2}.


\subsection{Matching the supersymmetric black hole action with the dual index}\label{sec:onshellact}

In this section we review the computation of the on-shell action for the black hole of \cite{Chong:2005hr} at linear order in $\alpha$. Then we impose supersymmetry and match the dual index \eqref{eq:R_index} at the corresponding order in the large-$N$ expansion. 

\subsubsection{The two-derivative solution is enough for evaluating the action}

When setting up the computation of the on-shell action at linear order in $\alpha$, there are two crucial points that should be kept in mind. The first is that since we are working in the grand-canonical ensemble, the inverse temperature $\beta$ and the chemical potentials, $\Phi$, $\Omega_{1}$, $\Omega_{2}$, must be held fixed to their zeroth-order values given in sec.~\ref{TwoDerReview}. Another way to say this is that we hold fixed the values of the supergravity fields at the boundary of the asymptotically AdS solution, so as to maintain the original Dirichlet variational problem. On the other hand, the action, the entropy and the conserved charges are allowed to receive corrections. The second point is that the corrections to the bulk metric and gauge field are not needed in order to compute the $\Oa$ corrections to the thermodynamics, as recently argued in \cite{Reall:2019sah}. Following~\cite{Cassani:2022lrk}, we adapt this argument to the case at hands.

As reviewed in sec.~\ref{sec:holographic_match}, at zeroth-order in $\alpha$, the renormalized Euclidean action has three contributions:
the first is the bulk contribution,
the second is the Gibbons-Hawking boundary term that renders the Dirichlet variational problem for the metric well posed, 
finally, 
the boundary counterterms needed to remove the divergences due to the non-compactness of the space.\footnote{For the purposes of the present argument, the background subtraction to remove the contribution of the AdS vacuum plays no role.}
Taking now into account the four-derivative corrections, the full Euclidean action reads
\begin{equation}
I=I^{(2\partial)}+\alpha \,I^{(4\partial)}\,,
\end{equation}
where $I^{(4\partial)}$ will be again the sum of bulk and boundary contributions, as we will discuss momentarily. Let us now assume we have a solution to the corrected equations of motion. This must be of the form,
\begin{equation}\label{eq:perturbation}
g_{\mu\nu}=g^{(2\partial)}_{\mu\nu}+\alpha \,g^{(4\partial)}_{\mu\nu}\, , \hspace{1cm} A_{\mu}=A^{(2\partial)}_{\mu}+\alpha \,{A}^{(4\partial)}_{\mu}\,, 
\end{equation}
being $\left\{g^{(2\partial)}_{\mu\nu}, A^{(2\partial)}_{\mu}\right\}$ a solution of the zeroth-order equations of motion. Evaluating the action on the corrected solution and expanding in $\alpha$ yields
\begin{equation}
I=I^{(2\partial)}|_{\alpha=0}+\alpha\left(\partial_{\alpha} I^{(2\partial)}+I^{(4\partial)}\right)|_{\alpha=0}+{\cal O}\left(\alpha^2\right)\, .
\end{equation}
As we see, out of the three contributions that arise, the corrections to the metric and gauge field are only needed to compute $(\partial_{\alpha} I^{(2\partial)})|_{\alpha=0}$. However, it is possible to show that this contribution actually vanishes if one fixes the boundary conditions appropriately \cite{Reall:2019sah}. Although the authors of \cite{Reall:2019sah} focus on the asymptotically-flat case, they argue (see footnote~6) that the same will hold in asymptotically locally AdS spacetimes. Let us demonstrate that this is in fact the case.
Integrating by parts, one finds that 
\begin{equation}
\begin{aligned}
\delta I^{(2\partial)}=\,&-\frac{1}{16\pi G}\int_{\cal M} \diff^5x\, e\, \left[\left({\cal E}_{\mu\nu}-\frac{1}{2}g_{\mu\nu}\,{\cal E}\right)\delta g^{\mu\nu}+{\cal E}^{\mu} \,\delta A_{\mu} \right]\\[1mm]
&+\frac12\int_{\partial {\cal M}} \diff^{4}x \,\sqrt{h}\, T_{ij}\, \delta h^{ij}+\int_{\partial {\cal M}} \diff^{4}x \,\sqrt{h}\, j^{i}\, \delta A_{i}\,,
\end{aligned}
\end{equation}
where we have introduced the (zeroth-order) Brown-York energy momentum tensor \cite{Brown:1992br,Balasubramanian:1999re} $T_{ij}$ and the electric current $j^i$, which are defined as follows
\begin{equation}
T_{ij}=\frac{2}{\sqrt{h}}\frac{\delta I^{(2\partial)}}{\delta h^{ij}}\,, \hspace{1cm} j^i=\frac{1}{\sqrt{h}}\frac{\delta I^{(2\partial)}}{\delta A_{i}}\,.
\end{equation}
Regarding now the corrected solution \eqref{eq:perturbation} as a perturbation over the leading-order solution, we immediately see that the bulk term vanishes at order $\alpha$ because of the zeroth-order equations of motion. This is perhaps less obvious for the boundary terms, though.  In the case at hands, the boundary is the hypersurface $r=r_{\text{bdry}}$, where $r_{\text{bdry}}$ is a regulator that eventually we will send to infinity, $r_{\text{bdry}}\to \infty$. The behavior of $T_{ij}$ and $j^{i}$ for large $r_{\text{bdry}}$ in asymptotically locally AdS$_{5}$ solutions is \cite{deHaro:2000vlm,Bianchi:2001kw}
\begin{equation}
\sqrt{h}\, T_{ij}\sim {\cal O}\left(r_{\text{bdry}}^{2}\right)\,, \hspace{1cm} \sqrt{h}\, j^{i}\sim {\cal O}\left(r_{\text{bdry}}^{0}\right)\, ,
\end{equation}
implying that the boundary terms also vanish if one imposes the following asymptotic behavior on the corrections,
\begin{equation}\label{eq:boundaryconditions}
{h}^{(4\partial)}_{ij}={\cal O}\left(r_{\text{bdry}}^0\right)\,, \hspace{1cm}  A^{(4\partial)}_{i}={\cal O}{\left(r_{\text{bdry}}^{-2}\right)}\, .
\end{equation}
The first can always be achieved by a suitable rescaling of the Euclidean time coordinate $\tau=it$ \cite{Cremonini:2009ih}, namely 
\begin{equation}
\tau\to \frac{\tau}{g\ell}\,, 
\end{equation}
where $\ell$ is the corrected AdS radius. As for the asymptotic condition on the gauge field, there is always a gauge in which it is satisfied. Let us show that this is also the gauge in which the $A_{\mu}$ is regular at the horizon. Calling $r_++\alpha\,\delta r_+$ the corrected position of the horizon, we must verify that
\be
W^{\mu}A_{\mu}|_{r=r_{+}+\alpha \delta r_{+}}=0\,,
\ee
where $W^\mu$ is the Killing vector generating the horizon.
Since we are working in the grand-canonical ensemble, we need to keep the electric potential $\Phi = W^{\mu}A_{\mu}|_{r=r_{+}+\alpha \delta r_{+}} - W^{\mu}A_{\mu}|_{r=\infty}$ fixed, that is
\begin{equation}
\delta \Phi=\alpha\, W^{\mu}\Bigl( \left(\partial_r A^{(2\partial)}_\mu|_{r=r_+}\right)\delta r_+ +  A^{(4\partial)}_{\mu}|_{r=r_{+}}-A^{(4\partial)}|_{r=\infty}\Bigr)=0\, .
\end{equation}
 Together with \eqref{eq:boundaryconditions}, this leads to 
\begin{equation}
W^\mu\Bigl( \left(\partial_r A^{(2\partial)}_\mu|_{r=r_+}\right)\delta r_+ +  A^{(4\partial)}_{\mu}|_{r=r_{+}}\Bigr) =W^{\mu}A^{(4\partial)}_{\mu}|_{r=\infty}=0\, .
\end{equation}
It follows that
\begin{equation}
W^{\mu}A_{\mu}|_{r=r_{+}+\alpha \delta r_{+}}= W^{\mu}A^{(2\partial)}_{\mu}|_{r=r_{+}}\,,
\end{equation}
implying that the regularity condition is satisfied as it is assumed such in the uncorrected solution.
We indicate the behavior~\eqref{eq:boundaryconditions} by saying that the corrections preserve the asymptotic boundary conditions on the fields.

\subsubsection{Higher-derivative boundary terms}

Having established that the action can be evaluated on the uncorrected solution, it only remains to specify the boundary terms that supplement the bulk contribution in the $\Oa$ action $I^{(4\partial)}$.
 The boundary terms associated with a generic higher-derivative bulk action such as \eqref{eq:genexpL4d} are not known, although several effective prescriptions have been discussed in the literature before, see e.g.~\cite{Cremonini:2009ih, Bueno:2018xqc, Cano:2022ord} and references therein. In our simpler effective action \eqref{eq:4daction2}, the only four-derivative invariant for which the associated Gibbons-Hawking term $I^{(4\partial)}_{\rm{GH}}$ is known is the Gauss-Bonnet one, \cite{Teitelboim:1987zz, Myers:1987yn}. However this is enough for our purposes here, given the asymptotic behavior of the field strength in the solutions under study. As a matter of fact, only the Gauss-Bonnet term contains divergences, as previously noted also in \cite{Cremonini:2019wdk} in the static case. Therefore for $I^{(4\partial)}_{\rm{GH}}$ we take \cite{Teitelboim:1987zz, Myers:1987yn}
\begin{equation}\label{eq:GHGB}
\begin{aligned}
I^{(1)}_{\rm{GH}}=\,&\frac{\lambda_1}{8\pi G}\int_{\partial {\cal M}} \diff^{4}x\,\sqrt{h} \,\left[\frac{2}{3}{\cal K}^3-2{\cal K} {\cal K}_{ij}{\cal K}^{ij}+\frac{4}{3}{\cal K}_{ij}{\cal K}^{jk}{\cal K}_{k}{}^{i}+4  {\cal G}_{ij}{\cal K}^{ij}\right]\\[1mm]
&-\frac{\lambda_2g^2}{2\pi G}\int_{\partial {\cal M}}\diff^4x\sqrt{h}\, {\cal K}\,,
\end{aligned}
\end{equation}
where ${\cal G}_{ij}={\cal R}_{ij}-\frac{1}{2}h_{ij}{\cal R}$ is the Einstein tensor of the induced metric. 
Regarding the boundary counterterms, we follow the prescription of \cite{Cremonini:2009ih}, which amounts to shift the coefficients in front of the boundary counterterms already present at zeroth order in $\alpha$, namely
\begin{equation}
\label{eq:4d_I_count}
I^{(1)}_{\text{count}}=\frac{1}{8\pi G}\int_{\partial {\cal M}}\, \diff^4x \sqrt{h} \, \left(3\mu_1g +\frac{\mu_2}{4 g}\,{\cal R}\right)\, ,
\end{equation}
where $\mu_1$ and $\mu_2$ are chosen so as to cancel the $r^4$ and $r^2$ divergences. For the effective action \eqref{eq:4daction2}, we find the following values 
\begin{equation}\label{eq:mus}
\mu_1=\frac{4g^2}{3}\left(-4\lambda_1+3\lambda_{2}\right)\,, \hspace{1cm}\mu_2=4g^2\left(2\lambda_1+\lambda_{2}\right)\, .
\end{equation}
In chapter~\ref{sec:structure_2} we will provide a rigorous derivation for the higher-derivative counterterms employed in this section via the Hamilton-Jacobi method~\cite{deBoer:1999tgo,Fukuma:2000bz,Martelli:2002sp}.

\subsubsection{Results}

Given the setup above, it is technically demanding but otherwise straightforward to evaluate the action \eqref{eq:4daction2} on the two-derivative solution \eqref{eq:CCLP_metric}, \eqref{gaugepot}.\footnote{We also checked that the results of this section agree with the ones obtained from the action \eqref{eq:4daction1}, giving evidence of the claim regarding the invariance of the black hole thermodynamics under the class of field redefinitions we have used to simplify the four-derivative action. See~\cite{Cassani:2022lrk} for a complete discussion.} The next step is to impose supersymmetry. We do so by imposing the same condition as at the two-derivative level, that is~\eqref{susyCCLP}. One way to see that this is the correct condition even at order $\alpha$ is that the linear relation \eqref{eq:linear_constraint_minimal} between $\omega_1,\omega_2,\varphi$ must be satisfied, and none of these quantities depend on $\alpha$. 
  After imposing the supersymmetry condition, we find that remarkably the action only depends on the combinations of the parameters $a,b,r_+$ which enter in the supersymmetric chemical potentials $\omega_1,\omega_2$ and $\varphi=\frac{1}{2}(\omega_1+\omega_2 -2 \pi i)$ given in \eqref{eq:susychemicalpotentials}. Specifically, we obtain
\begin{equation}\label{action_almost_done}
I=\frac{2\pi}{27 G  g^3} \left(1- 4 ( 3 \lambda_1 - \lambda _2)\alpha  g^2\right) \frac{\varphi^3 }{\omega_1\omega_2}+\frac{2\pi  \alpha  \lambda _1}{3  G g} \frac{\varphi \left(\omega_1^2+\omega_2^2-4 \pi ^2\right)}{\omega_1\omega_2}\,.
\end{equation}
The fact that $\beta$ drops out of this expression indicates the validity of the supersymmetric thermodynamics reviewed in section~\ref{TwoDerReview} at linear order in the corrections.
We can now convert this result in field theory units. The dimensionless quantities $G g^3$, $\alpha g^2$ have a holographic counterpart in the dual SCFT central charges $\aa$, $\cc$, or equivalently in the R-symmetry anomaly coefficients ${\rm Tr}\mathcal{R}^3$, ${\rm Tr}\mathcal{R}$. In appendix~\ref{sec:holographicdictionary} we show that for the bulk action~\eqref{eq:4daction2} we have been studying, the dictionary between the gravitational  and the field theory coefficients is 
\be
\begin{aligned}
{\rm Tr}\,\mathcal{R}^3 \,&=\, \frac{16}{9}(5\aa-3\cc)\,=\, \frac{4\pi}{9G g^3} \left( 1 -4(3\lambda_1-\lambda_2) \alpha g^2  \right)\,, \\[1mm]
{\rm Tr}\,\mathcal{R} \,&=\, 16(\aa-\cc)\,=\, -\frac{16\pi \alpha\lambda_1}{G g}   \,.
\end{aligned}
\ee
Plugging this in~\eqref{action_almost_done} and eliminating $\varphi$ via \eqref{eq:linear_constraint_minimal} yields
\begin{equation}\label{susy_action_om1om2}
I\,=\, {\rm Tr}\,\mathcal{R}^3\, \frac{(\omega_1+\omega_2 -2 \pi i )^3 }{48\,\omega_1\omega_2}- {\rm Tr}\,\mathcal{R} \,\frac{(\omega_1 +\omega_2 -2 \pi i) \left(\omega_1^2+\omega_2^2-4 \pi ^2\right)}{48\,\omega_1\omega_2}\,,
\end{equation}
which precisely reproduces the prediction from the index on the second sheet reported in section~\ref{sec:field_theory_results} upon identifying $I = -\log\mathcal{I}$.


\subsection{Corrected entropy and constrained Legendre transform}
\label{sec:chargesandBPSlimit}

\subsubsection{Corrected entropy from the on-shell action}

As we will explicitly prove in chapter \ref{sec:structure_2}, 
the first law \eqref{firstlaw} and the quantum statistical relation \eqref{QSR_CCLP} remain valid at linear order in the four-derivative corrections, then, we can derive the non-supersymmetric charges and the entropy by varying the Euclidean on-shell action $I(\beta,\Omega_1,\Omega_2,\Phi)$ (or the Gibbs free energy $\mathcal{G}(T,\Omega_1,\Omega_2,\Phi)=I/\beta$) with respect to its arguments, as in \eqref{charges_from_I}. 
After imposing supersymmetry through~\eqref{susyCCLP}, 
all quantities have complex values.

In the rest of this section we focus on the supersymmetric and extremal 
limit of the conserved charges and entropy, which is obtained by imposing both the conditions \eqref{susyCCLP} and \eqref{r_BPS} on the parameters. 
The prescription to take the limit that we follow here is the same as in~\cite{Silva:2006xv}, namely we fix $q$ to its BPS value, that is $q = q^*=g^{-1}(a+b)(1+ag)(1+bg)$, and parametrize deviations with respect to the BPS locus by setting 
\begin{equation}
r_{+}=r_{*}+\epsilon\,, 
\end{equation}
where $r_{*}$ is given in \eqref{r_BPS} and $\epsilon$ is the expansion parameter. In this limit all charges are again real. 
Here we set $g=1$ for simplicity. The electric charge reads\footnote{Here, we are taking the same normalization for the electric charge as the one of sec.~\ref{TwoDerReview}, corresponding to the canonically normalized R-charge for the ease of comparison with field theory expressions.} 

\begin{equation}\label{eq:corrected_4d_charge}
Q^* = \frac{\pi(a+b)}{2G(1-a)(1-b)}\Big[1+ 4\lambda_1\alpha\,\Delta Q + 4\lambda_2\alpha\Big]\,\,,
\end{equation}
with
\[\small
\begin{aligned}
\Delta Q &= \frac{2}{3 \mathfrak{D}}\Big[ a^7 \left(5 b^2+4 b-1\right)-a^6 \left(7 b^3+34 b^2+57 b+26\right)
\\[1mm]
&-b (b+1)^2 \left(b^4+24 b^3+31 b^2+14 b+3\right)-a^5 \left(55 b^4+244 b^3+427 b^2+322 b+80\right)
\\[1mm]
&-a^4 \left(55 b^5+400 b^4+985 b^3+1142 b^2+586 b+100\right)\\[1mm]
&-a^3 \left(7 b^6+244 b^5+985 b^4+1688 b^3+1410 b^2+516 b+62\right)\\[1mm]
&+a^2 \left(5 b^7-34 b^6-427 b^5-1142 b^4-1410 b^3-844 b^2-220 b-20\right)\\[1mm]
&+a \left(4 b^7-57 b^6-322 b^5-586 b^4-516 b^3-220 b^2-44 b-3\right)
\Big]\,,
\end{aligned}\]
and 
\begin{equation}
\begin{aligned}
\mathfrak{D} &= (1+a)(1+b)(a+b)^2 [a^4-2 a^3 (b+1)-3 a^2 \left(3 b^2+8 b+3\right)
\\[1mm]
&-2 a \left(b^3+12 b^2+12 b+1\right) +b^4-2 b^3-9 b^2-2 b+1]\,\,.
\end{aligned}
\label{eq:denominator}\end{equation}
The angular momentum $J_1$ is given by
\begin{equation}\label{eq:corrected_4d_J1}
J_1^* = \frac{\pi(a+b)\left(b+2a + ab\right)}{4G(1-a)^2(1-b)}\Big[ 1+ 4\lambda_1\alpha\,\Delta J_1 + 4\lambda_2\alpha\Big]\,\,,
\end{equation}
where
\[\small
\begin{aligned}
\Delta &J_1 =\frac{-2}{ \mathfrak{D}(b+2a+ab)}\Big[a^8 \left(-3 b^2-2 b+1\right)+ a^7 \left(4 b^3+38 b^2+68 b+30\right)\\[1mm]
+& 2 a^6 \left(19 b^4+111 b^3+238 b^2+196 b+50\right)+ 2 a^5 \left(24 b^5+184 b^4+543 b^3+707 b^2+377 b+63\right)\\[1mm]
+&2 a^4 \left(6 b^6+109 b^5+509 b^4+1032 b^3+917 b^2+324 b+33\right)\\[1mm]
+&a^3 \left(-2 b^7+36 b^6+402 b^5+1398 b^4+2026 b^3+1234 b^2+294 b+24\right)\\[1mm]
+&a^2 \left(-b^8-2 b^7+58 b^6+428 b^5+1074 b^4+1076 b^3+442 b^2+76 b+5\right)\\[1mm]
+&2 a b \left(-b^7+23 b^5+118 b^4+204 b^3+131 b^2+32 b+3\right)-b^8+12 b^6+48 b^5+52 b^4+16 b^3+b^2
\Big].
\end{aligned}\]
The angular momentum $J_2$ is obtained from $J_1$ by exchanging the rotational parameters, $a\leftrightarrow b$. 
The mass $E^*$ satisfies the BPS bound \eqref{eq:BPS_bound_5D_sugra}, $E^* = g\left( J_1^* + J_2^*\right) + \frac{3}{2}g Q^*$, 
while the entropy is
\begin{equation}
\mathcal S^* = \frac{\pi^2(a+b)\sqrt{a+b+ab}}{2G(1-a)(1-b)}\Big[1+ 4\lambda_1\alpha \,\Delta\mathcal S + 4\lambda_2 \alpha \Big]\,\,,
\label{eq:entropy0}\end{equation}
with
\[\small
\begin{aligned}
\Delta\mathcal S& = \frac{2}{\mathfrak{D}}\Big[ a^7 (b+1)-a^6 (b+1)-a^5 \left(12 b^3+48 b^2+63 b+23\right)
\\[1mm]
&-a^4 \left(26 b^4+156 b^3+312 b^2+245 b+59\right)
\\[1mm]
&-a^3 \left(12 b^5+156 b^4+496 b^3+620 b^2+313 b+45\right)
\\[1mm]
&-a^2 \left(48 b^5+312 b^4+620 b^3+504 b^2+159 b+11\right) 
\\[1mm]
&+ a b \left(b^6-b^5-63 b^4-245 b^3-313 b^2-159 b-26\right)
\\[1mm]
&+b^2 \left(b^5-b^4-23 b^3-59 b^2-45 b-11\right)
\Big]\,\,.
\end{aligned}\]
These expressions are only valid at linear order in $\alpha$, therefore we are ignoring both explicit and implicit ${\cal O}{\left(\alpha^2\right)}$ corrections. At the zeroth-order in $\alpha$, these agree with the supersymmetric and extremal limit of the charges and the entropy given in section~\ref{TwoDerReview}, as they should. 

We find that the supersymmetric and extremal entropy can be written in terms of the above charges, using the holographic dictionary~\eqref{a_c_high_der}, as
\begin{equation}
\label{eq:four_d_susy_ent}
\mathcal S^* = \pi\sqrt{3\left(Q^*\right)^2 -8 \mathtt a (J_1^* + J_2^*) - 16\, \mathtt a (\mathtt{a}-\mathtt c) \frac{(J_1^*-J_2^*)^2}{\left(Q^*\right)^2-2\mathtt a(J_1^* + J_2^*)}}\,\,.
\end{equation}
Moreover, we find that the non-linear constraint on the charges now becomes
\be
\label{eq:four_d_susy_const}
\begin{aligned}
& \left[3 Q^*  + 4\left(2\,\aa-\cc\right) \right]\left[ 3 \left(Q^*\right)^2 -  8\cc\, (J_1^*+J_2^*)\right]   \\
&= \left(Q^*\right)^3 + 16 \left(3\cc-2\aa\right)J_1^*J_2^* +\,64\aa\, (\aa-\cc)\frac{(Q^*+\aa)(J_1^*-J_2^*)^2}{\left(Q^*\right)^2-2\aa (J_1^*+J_2^*)}\,.
\end{aligned}
\ee

\subsubsection{The constrained Legendre transform}\label{constrained_transform}

We now reproduce the expression for the supersymmetric entropy \eqref{eq:four_d_susy_ent} by directly evaluating the Legendre transform of the grand-canonical function~\eqref{susy_action_om1om2}, at linear order in ${\rm Tr}\,\mathcal{R}$. As previously discussed, this function admits two complementary interpretations: it represents either the supersymmetric black hole on-shell action at linear order in the four-derivative corrections, or the power-law contributions in $\omega_1,\omega_2$ to the logarithm of the superconformal index, cf.~Eq.~\eqref{eq:R_index}. We will explain how the constrained Legendre transform procedure, reviewed in sec.~\ref{sec:holographic_match} for the leading-order result, continues to apply to the corrected expression. 

We start by observing that~\eqref{susy_action_om1om2} can be rephrased as a homogeneous function of degree 1 in terms of the variables $\omega_1,\omega_2,\varphi$. This is done using the linear constraint \eqref{eq:linear_constraint_minimal} to eliminate the factors of $2\pi i$ (including the term $-4\pi^2 = (2\pi i)^2$),
so that
\be
\begin{aligned}\label{expressI}
I 
\,&=\, ({\rm Tr}\mathcal{R}^3-{\rm Tr}\mathcal{R})\, \frac{  \varphi^3}{6\,\omega_1\omega_2}  -  {\rm Tr}\mathcal{R}\,\frac{  \varphi  \left[  - 2\varphi(\omega_1+\omega_2)+ \omega_1^2 + \omega_2^2 +\omega_1\omega_2\right]}{12\,\omega_1\omega_2}\,.  
\end{aligned}
\ee
As above, the entropy is given by the following extremized function:
\be
\mathcal{S} = {\rm ext}_{\{\omega_1,\omega_2,\varphi,\Lambda\}} \left[ -I -\omega_1J_1-\omega_2J_2  -\varphi\, Q_R -\Lambda(\omega_1+\omega_2-2\varphi-2\pi i)\right]\,,
\ee
where the Lagrange multiplier $\Lambda$ implements the linear constraint.
Therefore, by applying Euler's theorem for homogeneous functions as in sec.~\ref{sec:holographic_match}, it follows that
\be
\begin{aligned}\label{SfromLegTransf}
\mathcal{S} \,&=\, {\rm ext}\left[ -I +\omega_1 \frac{\partial{I}}{\partial\omega_1}+\omega_2 \frac{\partial{I}}{\partial\omega_2}  +\varphi \frac{\partial{I}}{\partial\varphi} + 2\pi i\Lambda\right]\\
\,&=\, {\rm ext} \left[ 2\pi i\Lambda\right]\,,
\end{aligned}
\ee
A real entropy is only obtained if the solution for $\Lambda$ is purely imaginary. As we are going to discuss, this means that the equation for $\Lambda$ has to factorize as $(\Lambda^2+ {\cal X})({\rm rest})=0$ for some positive ${\cal X}$. The factorization condition turns out to be equivalent to the non-linear constraint among the charges.

Here, work at linear order in ${\rm Tr}\mathcal{R}\neq 0$. Equivalently, we can say we work at linear order in the four-derivative corrections in the gravitational solution.
We find that $I$ in \eqref{expressI} satisfies
\begin{equation}
\begin{aligned}\label{eq:constraint1}
&\frac{9}{2}\left(\text{Tr}\mathcal R^3-\text{Tr}\mathcal R \right) \frac{\partial I}{\partial\omega_1}\frac{\partial I}{\partial\omega_2}-\Big(\frac{\partial I}{\partial\varphi}\Big)^3-\frac{3}{2}\,\text{Tr}\mathcal R \left(\frac{\partial I}{\partial\omega_1} + \frac{\partial I}{\partial\omega_2} \right)\frac{\partial I}{\partial\varphi}-\frac{3}{4}\,\text{Tr}\mathcal R\,\Big( \frac{\partial I}{\partial \varphi}\Big)^2   \\[2mm]
&\,\simeq\,\frac{9}{8}\,\text{Tr}\mathcal R^3\,\text{Tr}\mathcal R\, \frac{\left(\frac{\partial I}{\partial\omega_1} - \frac{\partial I}{\partial \omega_2}\right)^2 }{\frac{\partial I}{\partial\varphi}} \,,
\end{aligned}
\end{equation}
where by the symbol $\simeq$ we indicate that the equality holds up to terms of order $({\rm Tr}\mathcal{R})^2$. The same notation is used in the rest of this section. Using~\eqref{derivativesI}, the equation above can be written as an equation for $\Lambda$,
\begin{equation}
  \frac{p_{-1}}{\Lambda-\frac{1}{2}Q} + p_0 + p_1 \Lambda  + p_2 \Lambda^2 + \Lambda^3  \simeq 0\,\,,
\label{eq:lambda1}\end{equation}
where the coefficients are given by
\be
\begin{aligned}
p_{-1} \,&=\, 4\aa \,(\aa-\cc)(J_1-J_2)^2\\[1mm]
p_0 &=  -\frac{1}{8} \left[Q^3  + 12 (\cc- \aa) Q (Q+2(J_1+J_2))+ 16 (3 \cc - 2 \aa) J_1J_2 \right]\,,\\[1mm]
p_1 &= \frac{3}{4}Q^2-2\aa\, (J_1+J_2)\,,\\[1mm]
p_2 &= -\frac{3}{2}Q-2\aa\,.
\end{aligned}
\ee
 Here, we used the dictionary \eqref{relacTrR} to express the R-symmetry anomaly coefficients $\text{Tr}\mathcal R^3,\,\text{Tr}\mathcal R$ in terms of the conformal anomalies $\aa,\,\cc$, since the resulting expressions are slightly more compact.

When $p_{-1}$ does not vanish and \eqref{eq:lambda1} is a quartic equation for $\Lambda$. In analogy with the discussion above, we choose the microcanonical charges real and require that there exist two purely imaginary roots of opposite sign, so that a real and positive entropy is obtained. This means that the equation has to factorize as
\begin{equation}
(\Lambda^2 + {\cal X}) (\Lambda^2 + \Upsilon \Lambda + {\cal Z}) \,\simeq\, 0\,. 
\label{eq:lambda2}\end{equation}
Comparing with (\ref{eq:lambda1}), we can read the coefficients
\begin{equation}
{\cal X} \,=\, \frac{p_0- \tfrac12 Q\, p_1}{p_2- \tfrac12 Q}\,\,,\qquad
\Upsilon \,=\, p_2 -\tfrac{1}{2}Q \,\,,\qquad
{\cal Z} \,=\, -\tfrac{1}{2}Q\,p_2 +\frac{p_1 p_2-p_0}{p_2-\frac{1}{2}Q}\,\,,
\label{eq:coefficients}\end{equation}
and find the factorization condition,
\begin{equation}
p_{-1} \left(p_2 - \tfrac12 Q \right) - (p_1 p_2-p_0)\left( p_1+ \tfrac14 Q^2 \right) + \frac{ (p_1p_2- p_0)^2}{p_2 - \tfrac12 Q}\,\simeq\, 0\,.
\label{eq:factorization}\end{equation}
Since $p_{-1} = \mathcal{O}(\aa-\cc)\sim {\rm Tr}\,\mathcal{R}$, we see that the equation is solved demanding that $p_0-p_1 p_2 = \mathcal{O}(\aa-\cc)$ too. Then the last term is higher-order and should be dropped, so at linear order the equation is solved by taking
\begin{equation}\label{nonl_constraint_general}
p_0 \,\simeq\, p_1 p_2 - \frac{p_{-1} \left(p_2 - \tfrac12 Q \right)}{p_1+ \tfrac14 Q^2} \,.
\end{equation}
Substituting the expressions for the $p$'s, this condition can be written as the constraint
\be
\label{eq:universal_constraint}
\begin{aligned}
& \left[3 Q  + 4\left(2\,\aa-\cc\right) \right]\left[ 3 Q^2 -  8\cc\, (J_1+J_2)\right]   \\
&\simeq Q^3 + 16 \left(3\cc-2\aa\right)J_1J_2 +\,64\aa\, (\aa-\cc)\frac{(Q+\aa)(J_1-J_2)^2}{Q^2-2\aa (J_1+J_2)}\,.
\end{aligned}
\ee
The entropy is then given by
\begin{equation}
\begin{aligned}
\mathcal S &= 2\pi\sqrt{{\cal X}} \,\simeq\, 2\pi\,\sqrt{p_1 -\frac{p_{-1}}{p_1+\frac14 Q^2}}\\[2mm]
&\simeq\pi\sqrt{3Q^2 -8 \mathtt a \left(J_1 + J_2\right) - 16\, \mathtt a \left(\mathtt{a}-\mathtt c\right) \frac{(J_1-J_2)^2}{Q^2-2\,\mathtt a\left(J_1 + J_2\right)}}\,\,,
\end{aligned}
\label{eq:entropy}\end{equation}
where in the second step we have used \eqref{nonl_constraint_general}.
Again we find perfect agreement with the expressions found by first varying the non-supersymmetric on-shell action and subsequently taking the supersymmetric and extremal limit, cf.~\eqref{eq:four_d_susy_ent} and \eqref{eq:four_d_susy_const}.

%% file: research_two.tex
\chapter{Boundary terms and conserved charges in higher-derivative gauged supergravity}

\label{sec:structure_2}

In this chapter, based on contribution~\cite{Cassani:2023vsa}, we clarify some issues that arise when dealing with
higher-derivative gauged supergravity in the context outlined above. 
We will adopt the four-derivative bosonic action whose construction has been reviewed in \ref{sec:from_superconformal_to_supergravity}.

The first question we address in section~\ref{sec:holographicrenormalization} concerns the boundary terms that need to be added to the five-dimensional gauged supergravity action. These are made by the generalized Gibbons-Hawking-York terms necessary to have a well-posed variational principle defining the equations of motion, and by the counterterms that remove the divergences appearing when the action is evaluated on an AlAdS solution. Generically these terms contain a finite piece and contribute to the final result for the on-shell action; it is thus important to take them into account when comparing the latter with a given field theory partition function.

We start by discussing the variational principle for the action, and show that despite the higher-derivative terms, it makes sense to just impose Dirichlet boundary conditions in an AlAdS spacetime.
Although our focus is on the supergravity action given in~\cite{Cassani:2022lrk}, we also make some more general considerations. In particular, we identify a four-derivative coupling of the metric and gauge field curvatures that yields two-derivative equations and thus admits a good Dirichlet variational principle, independently of the asymptotics of the spacetime and of whether it is regarded as a small correction to a two-derivative action.
We next implement holographic renormalization using the Hamilton-Jacobi method; this generalizes the work of \cite{Fukuma:2001uf,Liu:2008zf} as we include a gauge field, and of~\cite{Landsteiner:2011iq,Cremonini:2009ih} as we add the  (bosonic) supersymmetric completion of the higher-derivative terms considered there. 
The outcome of our analysis confirms that the boundary terms introduced in sec.~\ref{sec:from_superconformal_to_supergravity}, based on effective considerations and used to match the SCFT partition function, are indeed correct.

The second part of the chapter discusses conserved charges and black hole thermodynamics in presence of higher-derivative terms. Black hole thermodynamics still makes sense in the presence of higher-derivative corrections: for a generic diffeomorphism invariant higher-derivative action, the corrections to the two-derivative Bekenstein-Hawking area law for the entropy are computed by Wald's formula \cite{Wald:1993nt,Iyer:1994ys}. Moreover, for extremal black holes Sen's formalism based on the near-horizon geometry provides a convenient method to evaluate the Wald entropy and opens the way for the definition of the full quantum entropy \cite{Sen:2005wa,Sen:2007qy}. However, these formalisms typically require to consider gauge-invariant quantities, and need to be treated with care in presence of Chern-Simons terms. First, in section~\ref{sec:conservedchargesgeneral}, we revisit Wald's formalism~\cite{Wald:1993nt,Iyer:1994ys} to define the conserved charges associated to diffeomorphisms and U(1) gauge transformations, carefully keeping track of the contribution from gauge Chern-Simons terms.\footnote{The effect of purely gravitational Chern-Simons terms (which we are not considering) has been discussed before in e.g.~\cite{Tachikawa:2006sz, Elgood:2020nls}.} We emphasize that it is important to distinguish between the charges defined through the Noether procedure and those which satisfy the Gauss law, which are often referred to as Komar or Page charges. We provide explicit expressions for the latter, building on the generalized Noether theorem by Barnich, Brandt and Henneaux~\cite{Barnich:2000zw}. In particular, we present the formulae for angular momenta and electric charge associated with our five-dimensional higher-derivative supergravity action.
Then, in the appendix~\ref{sec:aladsthermodynamics}, we make use of these general results to demonstrate the first law of black hole mechanics as well as to prove the quantum statistical relation which identifies the Euclidean on-shell action (with Dirichlet boundary conditions) with the Legendre transform of the black hole entropy. This is done working in a gauge that is regular up to the horizon, so as to be able to use the Stokes' theorem to relate integrals  at the horizon and at the conformal boundary.\footnote{We note that our analysis of conserved charges does not rely on the linearization in the higher-derivative corrections, hence it would be applicable more generally if a gravitational action valid at non-linear order in the corrections, such as the Gauss-Bonnet action and its generalizations, is given.}
 
Our main motivation to work with charges that obey a Gauss law is that we can exploit them to compute the corrections to the angular momentum and the electric charge of the supersymmetric AdS$_5$ black hole whose two-derivative solution was given in~\cite{Gutowski:2004ez}: although the full corrected solution is not known and thus we cannot evaluate the charges by means of an asymptotic integral at the boundary, we can obtain them from our new formulae applied to the corrected near-horizon solution, which we also construct. As we will show, the results are in agreement with the expressions we derived in sec.~\ref{sec:chargesandBPSlimit} by varying the on-shell action, modulo a constant shift in the electric charge that we discuss. This computation also allows us to express the  black hole entropy as a function of the angular momentum and electric charge relying solely on the near-horizon solution.\footnote{We leave for future work a comparison of the charges obtained here via the Noether method with those given by the holographic energy-momentum tensor and electric current derived from the renormalized action. This would generalize the results of~\cite{Papadimitriou:2005ii} to theories including Chern-Simons terms and higher-derivative couplings.} Details of this computation can be found in section~\ref{sec:corrections_GR}.

\section{Boundary terms in five-dimensional higher-derivative supergravity}\label{sec:holographicrenormalization}

In this section, we work in five dimensions and consider the four-derivative theory for the metric and an Abelian gauge field, including Chern-Simons terms, given by \eqref{eq:4daction2}. However, we also make some general considerations. We discuss the boundary terms that are needed in order to obtain a well-defined variational problem and implement holographic renormalization in AlAdS solutions. In order to discuss such terms we need to first regulate the spacetime:
we assume this 
is foliated by hypersurfaces diffeomorphic to the conformal boundary, 
 and introduce adapted coordinates $x^\mu = \{x^i,z\}$, $i=0,1,2,3$, such that the hypersurfaces are labelled by $z$ and the conformal boundary is found at $z=0$. Adopting a Fefferman-Graham gauge, we can write the metric and gauge field in the form
\begin{equation}
\diff s^2\, =\, \ell^2\,\frac{\text dz^2}{z^2} + h_{ij}(x,z)\,\text d x^i \text dx^j\,\,, \qquad A= A_i(x,z)\, \diff x^i\,\,,
\label{eq:FGgauge}\end{equation}
where the parameter $\ell$ is identified with the radius of the AdS solution.
Then, $h_{ij}$ and $A_i$ are the induced fields on the hypersurface at fixed $z$,
and the outward-pointing normal vector with unit norm reads
\begin{equation}
n = -\frac{z}{\ell}\,\partial_z\,\,.
\label{eq:normalvector}\end{equation}
 We denote by $\mathcal{R}_{ij}$ and $\mathcal{R}$ the Ricci tensor and Ricci scalar of $h_{ij}$, respectively, while the extrinsic curvature ${\cal K}_{ij}$ of the hypersurface, its trace ${\cal K}$ and the normal derivative of the gauge field $E_i$ are given by
\be\label{eq:extrinsiccurv}
{\cal K}_{ij} = -\frac{z}{2\ell}\,\partial_z h_{ij}\,\,,\qquad {\cal K} = h^{ij} {\cal K}_{ij}\,\,,\qquad
E_i = n^\mu F_{\mu i} = -\frac{z}{\ell}\partial_z A_i\,\,.
\ee
We next assume the bulk spacetime terminates at the cutoff hypersurface $z=\epsilon$, and impose Dirichlet boundary conditions for the fields. In the gauge \eqref{eq:FGgauge}, this means that the field variations satisfy
 \be\label{eq:dirichlet}
\delta h_{ij}= \delta A_i=0 \quad\text{for}\ z=\epsilon\,. 
 \ee
In sec.~\ref{sec:onshellact}, we argued that the boundary terms to be added to the bulk action, which we repeat here for convenience, are:
\begin{equation}
\begin{aligned}\label{GHY_bdy_term}
S_{\text{GHY}}&=\frac{1+4\lambda_2\alpha g^2}{8\pi G}\int_{z=\epsilon}
 \diff^4x\sqrt{-h}\, {\cal K} \\[1mm]
& -\,   \frac{\alpha\lambda_1}{4\pi G}\int_{z=\epsilon}
 \diff^{4}x\,\sqrt{-h} \,\left[\frac{1}{3}{\cal K}^3-{\cal K} {\cal K}_{ij}{\cal K}^{ij}+\frac{2}{3}{\cal K}_{ij}{\cal K}^{jk}{\cal K}_{k}{}^{i}+ 2 \Big({\cal R}_{ij}-\frac{1}{2}h_{ij}{\cal R} \Big){\cal K}^{ij}\right],
\end{aligned}
\end{equation}
and
\begin{equation}\label{counterterms}
S_{\text{ct}}= -\frac{1}{8\pi G}\int_{z=\epsilon}
  \diff^4x \sqrt{-h} \, \left(3g \mu_1 +\frac{\mu_2}{4g}\,{\cal R}\right)\, ,
\end{equation}
where the coefficients $\mu_1$ and $\mu_2$ read
\begin{equation}\label{eq:mus2}
\mu_1\,=\, 1+ \left(-\frac{16}{3}\lambda_1+4\lambda_{2}\right)\alpha g^2\,, \hspace{1cm}\mu_2\,=\,  1+ \left(8\lambda_1+4\lambda_{2}\right)\alpha g^2\, .
\end{equation}
Here, $S_{\rm GHY}$ represents the generalized Gibbons-Hawking-York (GHY) boundary terms needed for the variational principle with Dirichlet boundary conditions to be well-posed asymptotically, while $S_{\text{ct}}$ denotes the local counterterms that remove the divergences from the action. Note that the counterterms take the same form as in the two-derivative theory, however their coefficients receive corrections at order $\alpha$. After having added the boundary terms, the cutoff is removed by taking $\epsilon \to 0$. The renormalized on-shell action $S_{\rm ren}$ is then defined as
\be
S_{\rm ren} \,=\, \lim_{\epsilon\to 0} \left(S_{\rm bulk} + S_{\rm GHY} + S_{\rm ct} \right)\,.
\ee
While the derivation given in~\ref{sec:onshellact} was based on the specific black hole solution studied there and was therefore only partial, below we present a more complete derivation of the boundary terms \eqref{GHY_bdy_term}, \eqref{counterterms}, valid for general AlAdS solutions, showing that they are indeed sufficient.

\subsection{The variational problem}\label{sec:variational_pb}

We justify here the GHY terms given in \eqref{GHY_bdy_term}. The first line is just the usual GHY term ensuring that the two-derivative part of the bulk action, i.e. the first line of~\eqref{eq:4daction2}, has a good Dirichlet variational problem. The second line of \eqref{GHY_bdy_term} is the boundary term introduced in~\cite{Myers:1987yn}, fixing the Dirichlet variational problem for the Gauss-Bonnet term appearing in our action. We now argue that this is all that we need as long as the variational problem is considered asymptotically in an AlAdS spacetime.

It is well-known that in higher-derivative gravity one cannot just impose Dirichlet boundary conditions, as the field equations are of order higher than two (the Gauss-Bonnet term is an exception as it generates second-order equations). 
 Identifying the additional boundary data to be specified is a non-trivial task, and different strategies have been proposed in the literature. In~\cite{Deruelle:2009zk}, and more recently~\cite{Erdmenger:2022nhz}, an approach involving auxiliary fields has been developed. The auxiliary fields are useful to identify the additional degrees of freedom that are propagated by the field equations, however this does not evade the need to fix more boundary data than the metric and (if present) the gauge field. Another approach consists of treating the higher-order terms as corrections to the two-derivative theory. Doing so one only has to study two-derivative equations, as at each order in the corrections the higher-derivative contributions are evaluated on a solution to the lower-order equations. In this perturbative approach, therefore, it should be possible to construct a generalized GHY boundary term order by order in the perturbative expansion.
 See e.g.~\cite{Smolic:2013gz} for a thorough discussion in the context of pure gravity.
 In the following we will show that when a gauge field is coupled to the metric curvature, the corresponding variational problems get combined, and this affects the GHY term. This issue was already encountered in~\cite{Cremonini:2009ih}. Instead of invoking the perturbative expansion in the higher-order corrections, we will prefer to follow a different strategy: we will assume the asymptotic behavior of the fields is the one taken in AlAdS spacetimes, and show that the terms that make the variational principle problematic are suppressed asymptotically.

Let us consider the term in the bulk action~\eqref{eq:4daction2} coupling the Weyl tensor to the gauge field strength, $C_{\mu\nu\rho\sigma}F^{\mu\nu}F^{\rho\sigma}$. We actually perform a slightly more general analysis and consider generic four-derivative couplings of the Riemann curvature with $F_{\mu\nu}$,
\begin{equation}
S_{4\partial,F} \,=\,  \int\text d^5 x\, e\left(u_1 R_{\mu\nu\rho\sigma} F^{\mu\nu} F^{\rho\sigma} + u_2 R_{\mu\nu}F^{\mu\rho}F_{\rho}{}^{\nu} + u_3R F^2\right)\,\,,
\label{eq:u123combination}\end{equation} 
where $u_1, u_2, u_3$ are arbitrary coefficients. The combination that gives the $C_{\mu\nu\rho\sigma}F^{\mu\nu}F^{\rho\sigma}$ term in our bulk action is
\be\label{values_u123}
u_2 = \frac{4}{3}u_1\,,\qquad u_3 = \frac{1}{6}u_1\,,\qquad \text{with}\ u_1= - \frac{\lambda_1\alpha}{32\pi G}\,\,.
\ee 
The metric variation of \eqref{eq:u123combination} reads:
\begin{equation}
\begin{aligned}
\delta S_{4\partial,F} \,=\,
\int\text d^5 x \, e \left(u_1 \delta R^\mu_{\,\,\,\nu\rho\sigma} F_\mu^{\,\,\,\nu} F^{\rho\sigma}+u_2\,\delta R_{\mu\nu}F^{\nu\rho}F_\rho^{\,\,\,\mu} + u_3\,g^{\mu\nu}\delta R_{\mu\nu} F^2\right)+ \ldots\,,
\end{aligned}
\label{eq:variationriemann}\end{equation}
with
\begin{equation}
\delta R^\mu_{\,\,\,\nu\rho\sigma} = 2\nabla_{[\rho}\delta\Gamma^\mu_{\,\,\,\sigma]\nu}\,\,,\qquad \delta\Gamma^\mu_{\sigma\nu} = \frac{1}{2}g^{\mu\rho}\left(2\nabla_{(\sigma}\delta g_{\nu)\rho} - \nabla_\rho\delta g_{\nu\sigma} \right)\,\,.
\label{eq:riemanntometric}\end{equation}
The ellipsis comprises terms that vanish upon imposing Dirichlet boundary conditions for the metric. In the next formulae, the ellipsis will continue to denote such terms, as well as all bulk terms that are not total derivatives and lead to the bulk equations of motion.
Plugging the first of (\ref{eq:riemanntometric}) into (\ref{eq:variationriemann}) 
and integrating by parts, we are left with the following surface terms
\begin{equation}
\begin{aligned}
\delta S_{4\partial,F} \,=\, \int\text d^5 x \,e\, \nabla_\rho\Big[ 2 u_1 \left(\delta\Gamma^\mu_{\,\,\,\nu\lambda} F_\mu^{\,\,\,\nu} F^{\rho\lambda}\right)+\delta\Gamma^\rho_{\,\,\,\mu\nu}\left(u_2\,F^{\nu\lambda}F_\lambda^{\,\,\,\mu}+ u_3\,g^{\mu\nu}F^2\right) - \\
-\delta\Gamma^\nu_{\,\,\,\nu\mu} \left(u_2\,F^{\rho\lambda}F_\lambda^{\,\,\,\mu} + u_3\,g^{\mu\rho} F^2\right)\Big]+\ldots\,\,.
\end{aligned}
\label{eq:variationriemann3}\end{equation}
Passing to the boundary integral on the cutoff hypersurface and using the second of \eqref{eq:riemanntometric} we obtain
\begin{equation}
\begin{aligned}
\delta S_{4\partial,F}  = \!\int
 \text d^4 x\sqrt{-h}\,n_\mu\Big[2u_1 \nabla_\nu\delta g_{\lambda\rho} F^{\rho\nu}F^{\mu\lambda} -\frac{1}{2}g^{\nu\sigma}\nabla_\rho\delta g_{\nu\sigma} \left(u_2\, F^{\mu\lambda}F_\lambda{}^{\rho}+ u_3\,g^{\mu\rho} F^2 \right)
 \\[1mm]
\ + \Big(g^{\mu\rho}\nabla_\sigma\delta g_{\nu\rho}-\frac{1}{2}\nabla^\mu\delta g_{\sigma\nu}\Big)\Big(u_2 \,F^{\nu\lambda}F_\lambda^{\,\,\,\sigma} + u_3\, g^{\sigma\nu}F^2 \Big) \Big] + \ldots \,,
\end{aligned}
\label{eq:variationriemann4}\end{equation}
where we recall that $n^\mu$ is the  unit normal vector.
We now assume the Fefferman-Graham gauge \eqref{eq:FGgauge} and recall the definitions \eqref{eq:extrinsiccurv} for the normal derivatives of the fields.
It is then straightforward to show that \eqref{eq:variationriemann4} becomes
\begin{equation}
\begin{aligned}
\delta S_{4\partial,F} \,=\, 
  \int
  \text d^4x\sqrt{-h}&\Big[
-\left(4u_1 -u_2\right)  \delta {\cal K}_{ij} E^{i} E^{j} + \left(u_2-4u_3\right) h^{kl} \delta {\cal K}_{kl}  E^i E_i \\[1mm]
&\,\ -u_2 \,\delta {\cal K}_{ij} F^{ik} F_k^{\,\,\,j} - 2u_3\, h^{kl} \delta {\cal K}_{kl} F^{ij}F_{ij}\Big]+ \ldots\,\,,
 \end{aligned}
\label{eq:GHY0}\end{equation}
where we have used the relation between variations of the induced metric and variations of the extrinsic curvature,
\begin{equation}
n^\mu \nabla_\mu \delta h_{ij} \,=\,2\delta {\cal K}_{ij} \,\,.
\end{equation}
It follows that the generalized GHY term yielding a good Dirichlet variational problem for the metric with the action \eqref{eq:u123combination} is 
\begin{equation}
S_{{\rm GHY},F}= \int \diff^4x\sqrt{-h}\Big[ (4u_1 -u_2) {\cal K}_{ij}E^i E^j - (u_2-4u_3) {\cal K} E^{i} E_i + u_2 {\cal K}_{ij}F^{ik}F_k^{\,\,\,j}  + 2u_3 {\cal K} F^{ij}F_{ij}\Big].
\label{eq:GHY1}
\end{equation}
In particular, using the values \eqref{values_u123} for the coefficients, we obtain the GHY term for the $C_{\mu\nu\rho\lambda}F^{\mu\nu}F^{\rho\lambda}$ term appearing in our bulk supergravity action, 
\begin{equation}\label{eq:GHY_Weyl}
S_{\rm GHY,Weyl}= -\frac{\lambda_1\alpha}{48\pi G}\int \text d^4x\sqrt{-h}\left[4{\cal K}_{ij}E^i E^j -  {\cal K} E^{i} E_i  + 2{\cal K}_{ij}F^{ik}F_k{}^{j} +\frac{1}{2} {\cal K} F^{ij}F_{ij}\right]\,\,.\end{equation}

While this fixes the Dirichlet variational principle for the metric, it spoils the 
one for the gauge field, which was so far well-posed. Indeed, \eqref{eq:GHY1} and  \eqref{eq:GHY_Weyl} involve the normal derivatives $E_i$ of the gauge field, that are not fixed by Dirichlet boundary conditions.
 As anticipated, we circumvent this issue by assuming that the fields respect the conditions for an AlAdS solution.
 In the coordinates specified by \eqref{eq:FGgauge}, this means that the induced fields admit the following small-$z$ expansion:
\begin{equation}
\begin{aligned}
h_{ij}(x,z) &= \frac{1}{z^2}\left[h_{ij}^{(0)}(x) + z^2h_{ij}^{(2)}(x) + z^4\left( h_{ij}^{(4)}(x) + \tilde h_{ij}^{(4)}(x)\,\log z\right) + \mathcal O(z^5)\right] \,\,,\\[1mm]
A_i(x,z) &= A_i^{(0)}(x) + z^2 \left(A_i^{(2)}(x) + \tilde A_i^{(2)}(x)\,\log z\right) + \mathcal O(z^3)\,\,. 
\end{aligned}
\label{eq:expansion}\end{equation}
This implies the following leading behaviour of the intrinsic curvatures of  boundary tensors,
\be
F_{ij} = \mathcal O(1)\,\,,\ \quad\nabla_i= \mathcal O(1)\,\,,\ \quad
\mathcal R^i_{\,\,\,jkl} =\mathcal O(1)\,\,,\ \quad \mathcal R_{ij} =\mathcal O(1)\,\,,\ \quad \mathcal R = \mathcal O(z^2)\,\,,
\label{eq:intrinsictensors}
\ee
while the extrinsic curvatures behave as
\begin{equation}
\begin{aligned}
& {\cal K}_{ij} = \frac{1}{z^2}h_{ij}^{(0)} + \mathcal O(1)\,\,,\qquad
\nabla_i {\cal K}_{jk} = \mathcal O(1)\,\,,\qquad {\cal K} = 4+ \mathcal{O}(z^2)\,\,,\\[1mm]
&E_i =  -\frac{z^2}{\ell} \left(2A_i^{(2)}+ \tilde A_i^{(2)}\,(1+2\log z) \right) +\mathcal{O}(z^3) \,\,.
\label{eq:extrinsictensors}
\end{aligned}
\end{equation}
Using these expansions, it is easy to see that the terms in \eqref{eq:GHY1} (and hence \eqref{eq:GHY_Weyl}) involving $E_i$ are suppressed in the limit $\epsilon \to 0$ that removes the radial regulator. Therefore, even though the Dirichlet problem for the gauge field is ill-defined for finite values of the cutoff, the issue disappears in the limit where the latter is removed. We conclude that the Dirichlet problem for the action $S_{4\partial,F}+ S_{{\rm GHY},F}$ given in  \eqref{eq:u123combination}, \eqref{eq:GHY1} is well-posed {\it asymptotically} in AlAdS spacetimes.

We further observe that while the terms involving $E_i$ are suppressed, the last two terms in \eqref{eq:GHY1} are finite in the asymptotic expansion, hence generically they contribute to the renormalized action. However, for the specific case of the $C_{\mu\nu\rho\sigma}F^{\mu\nu}F^{\rho\sigma}$ coupling appearing in our bulk supergravity action, the whole GHY term is suppressed asymptotically since the finite contributions from the last two terms in \eqref{eq:GHY_Weyl} add up to zero.
For this reason, we omitted the GHY term associated with $C_{\mu\nu\rho\sigma}F^{\mu\nu}F^{\rho\sigma}$ in our boundary terms \eqref{GHY_bdy_term}.

An analogous argument was previously used in~\cite{Landsteiner:2011iq} 
to treat the gauge-gravitational Chern-Simons term, $\epsilon^{\mu\nu\rho\sigma\lambda}R_{\mu\nu\alpha\beta}R_{\rho\sigma}{}^{\alpha\beta} A_{\lambda}$, which also appears in our bulk action~\eqref{eq:4daction2}. By requiring that the action reproduces the gauge-gravitational anomaly on a general hypersurface a boundary term was derived in~\cite{Landsteiner:2011iq}, however it contains normal derivatives of the extrinsic curvature and does not fix the Dirichlet variational problem for the metric.\footnote{Nevertheless, it simplifies the expression for the action when this is expressed in ADM variables by cancelling a bulk total derivative. In sec.~\ref{sec:HJmethod} we show that this is also the case for our boundary term~\eqref{eq:GHY1}.} This issue is avoided by noting that the terms not fixed by the Dirichlet boundary conditions are suppressed asymptotically for field configurations respecting the expansion~\eqref{eq:expansion}. 
While we refer to~\cite{Landsteiner:2011iq} for details,\footnote{See also~\cite{Grumiller:2008ie} for a study in four dimensions, where the gauge field is replaced by an axion-like scalar.} here we simply conclude that it is not necessary to associate a boundary term to the gauge-gravitational Chern-Simons term for our purposes.

Noting that the Dirichlet problem for $F^4$ is well-defined with no boundary term added, the analysis of the variational problem for the bulk action~\eqref{eq:4daction2} is then complete. 

This concludes our argument for the GHY boundary term given in  \eqref{GHY_bdy_term}. In section~\ref{sec:HJmethod} we will derive the boundary counterterms \eqref{counterterms} that implement holographic renormalization by removing the divergences from the on-shell action. 

\subsubsection{A fully well-defined Dirichlet problem}
As an aside to our main discussion, we observe that there exists a special way to couple the Riemann curvature with the gauge field strength $F_{\mu\nu}$ such that the Dirichlet problem is well-defined both for the metric and the gauge field at finite values of the radial cutoff $\epsilon$. This corresponds to choosing the coefficients in \eqref{eq:u123combination} as
\be
u_2 = 4u_1\,,\qquad  u_3 = u_1\,,
\ee
so that the GHY term \eqref{eq:GHY1} does not contain normal derivatives of the gauge field, $E_i$.
The corresponding combination of bulk terms is
\begin{equation}
\begin{aligned}
S_{4\partial,F} \,&=\, u_1\int \diff^5x \,e \left(R_{\mu\nu\rho\sigma} F^{\mu\nu} F^{\rho\sigma} + 4 R_{\mu\nu}F^{\mu\rho}F_{\rho}{}^\nu + R F^2\right)\\
\,&=\, -  u_1\int \diff^5x \,e \,\epsilon^{\lambda\mu\nu\rho\sigma} R_{\mu\nu}{}^{\mu'\nu'}\epsilon_{\lambda\mu'\nu'\rho'\sigma'} F_{\rho\sigma}F^{\rho'\sigma'}\,,
\label{eq:specialcombination}
\end{aligned}
\end{equation} 
and the GHY term reads
\begin{equation}
S_{{\rm GHY},F} \,=\, 2u_1\int\text d^4x\sqrt{-h}\left( 2{\cal K}_{ij}F^{ik}F_k^{\,\,\,j} +{\cal K}F_{ij}F^{ij} \right)\,\,.
\label{eq:genGHY_F}
\end{equation}
In this case, the variational problems for the metric and the gauge field are both well-posed. Indeed, variations of $S_{4\partial,F} + S_{{\rm GHY},F}$ with respect to the gauge field take the schematic form
\begin{equation}
\int\text d^5 x\, e\left(\text{vector EoM} \right)^i\delta A_i + \int\text d^4x\sqrt{-h} \left(\text{bdy terms}\right)^i\delta A_i 
+\int\text d^4x\sqrt{-h}\, \frac{\delta S_{\rm GHY}}{\delta F_{ij}}2\nabla_i\delta A_j\,\,.
\label{eq:variationgaugefield}\end{equation}
The second piece denotes boundary terms emerging from the bulk action when integrating by parts so as to obtain the vector equations of motion (EoM). Since the bulk action does not involve terms with derivatives of the field strength, these boundary terms are simply proportional to $\delta A_i$. Moreover, since the GHY term (\ref{eq:genGHY_F}) does not contain normal components of the field strength, the term in (\ref{eq:variationgaugefield}) involving its variation vanishes identically when we impose the Dirichlet boundary conditions \eqref{eq:dirichlet}.

One can also check that both the metric and gauge field equations generated by this action are  of second-order, consistently with the fact that we have found a GHY term. 

We stress that the present proof does not require the fields to satisfy the AlAdS expansion~\eqref{eq:expansion}, nor it needs $\alpha$ to be small. Moreover, although we have given it for five bulk dimensions, it works in the same way in arbitrary dimension. So the sum of \eqref{eq:specialcombination} and \eqref{eq:genGHY_F} provides an analog of the Gauss-Bonnet term and its corresponding GHY term for the couplings of the Riemann curvature with a gauge field.

\subsection{Holographic renormalization}\label{sec:HJmethod}

\subsubsection{Brief review of the Hamilton-Jacobi method}\label{sec:hamiltonjacobigeneral}

After discussing the variational principle, we can turn to the cancellation of the divergences in the on-shell action by means of holographic renormalization. We will adopt the approach based on the Hamilton-Jacobi method~\cite{deBoer:1999tgo,Fukuma:2000bz,Martelli:2002sp} -- see also~\cite{Papadimitriou:2004ap} for a related method, and~\cite{Elvang:2016tzz,Papadimitriou:2016yit} for nice recent discussions. This systematizes the analysis of counterterms and appears particularly convenient in the presence of higher-derivative terms. Compared to existing analyses using the Hamilton-Jacobi method in higher-derivative gravity, such as~\cite{Fukuma:2001uf,Liu:2008zf}, here we include couplings to the gauge field $A$.

The idea is to solve perturbatively the Hamilton-Jacobi equation
\begin{equation}\label{eq:hamiltonjacobi}
  {\cal H}\left[\frac{16\pi G_{\rm eff}}{\sqrt{-h}}\frac{\delta S_{\text{reg}}}{\delta h_{ij}}\,,\ \frac{16\pi G_{\rm eff}}{\sqrt{-h}}\frac{\delta S_{\text{reg}}}{\delta A_{i}}\ ;\ h_{ij}\,,\ A_i\right]=0\,\,,
\end{equation}
where the regulated action 
\be
S_{\text{reg}}=S_{\rm bulk}+S_{\rm GHY}
\ee
 is the action evaluated on a solution in the spacetime ending at $z=\epsilon$. Here, $\mathcal{H}$ is the Hamiltonian density of the theory, where the role of time is being played by the radial coordinate $z$. Eq.~\eqref{eq:hamiltonjacobi} arises as the Hamiltonian constraint associated with reparametrization invariance. 
In terms of $S_{\text{reg}}$, the renormalized on-shell action $S_{\rm ren}$ is given by
\begin{equation}
S_{\text{ren}} \,=\, S_{\text{reg}} + S_{\rm ct}\,\,,
\label{def:onshellrenormalizedaction}\end{equation}
where it is understood that in the right hand side we neglect all terms that are suppressed in the limit $\epsilon\to 0$.
By construction this is finite, or at most logarithmically divergent in the case where a holographic Weyl anomaly is present. 
 Then Eq.~\eqref{eq:hamiltonjacobi} should be regarded as a first-order partial differential equation for $S_{\text{ren}}$ and $S_{\rm ct}$. Here we are only interested in solving the leading orders of \eqref{eq:hamiltonjacobi} near the boundary, so as to determine the local counterterms that remove the power-law divergences from the on-shell action, as well as the holographic Weyl anomaly.
 
 In order to do that we first need to identify the divergent sector of the on-shell action. It is convenient to rewrite the Lagrangian density $\mathcal{L}_{\rm reg}$, that for a $(d+1)$-dimensional spacetime is related to the regulated action as $S_{\rm reg} = \frac{1}{16\pi G_{\rm eff}}\int\text d^{d+1}x\,e\,\mathcal L_{\rm reg}$, in terms of the fields $h_{ij}$ and $A_i$ induced on the hypersurface at fixed $z$, making use of Gauss-Codazzi relations. After doing that, the Lagrangian density can be expanded as
\begin{equation}
\mathcal L_{\rm reg} = \mathcal L^{(0)} + \mathcal L^{(2)} + ... + \mathcal L^{(d)} + ...\,\,,
\end{equation}
where every term $\mathcal L^{(2k)}$ comprises terms whose leading-order behavior near the boundary is $\mathcal{O}(z^{2k})$ when the asymptotic expansion of the fields is substituted
 (in the case of pure gravity this expansion is equivalent to a derivative expansion).
The leading terms in the expansion, starting from $\mathcal L^{(0)}$ up to $\mathcal L^{(d-2)}$, when integrated, yield power-law divergences in $S_{\rm reg}$. These divergences should be cancelled by the counterterms $S_{\rm ct}$, as requested by \eqref{def:onshellrenormalizedaction}. For even $d$ there can also be a term $\mathcal L^{(d)}$ in the Lagrangian, which gives a logarithmically divergent contribution to the action. The higher-order terms in the expansion of $\mathcal L_{\rm reg}$ are sufficiently suppressed asymptotically and do not affect the divergent sector, hence can be neglected for the purposes of the present discussion. 

Implementing the Hamilton-Jacobi method asymptotically then consists of the following steps.

\paragraph{Introducing the Hamiltonian.} First, we need to determine the Hamiltonian density $\mathcal H$, after rewriting $\mathcal L_{\rm reg}$ using the Gauss-Codazzi relations. 
 This is defined as
\begin{equation}\label{def:hamiltonian}
\mathcal H \,=\,   \frac{16\pi G_{\rm eff}}{\sqrt{-g}}\left(\frac{\delta S_{\rm reg}}{\delta \dot h_{ij}}\,\dot h_{ij} + \frac{\delta S_{\rm reg}}{\delta \dot A_i}\,\dot A_i\right) -\mathcal L_{\rm reg}\,\,,
\end{equation}
where the dot means a derivative with respect to $z$.
The canonical momenta are given by
\begin{equation}
\pi^{ij} \,\equiv\,  -\frac{16\pi G_{\rm eff}}{\sqrt{-h}}\,\frac{\delta S_{\rm reg}}{\delta \dot h_{ij}} =\frac{16\pi G_{\rm eff}}{2\sqrt{-g}}\frac{\delta S_{\rm reg}}{\delta {\cal K}_{ij}}
\,\,,\qquad  \
\pi^i \,\equiv\,  -\frac{16\pi G_{\rm eff}}{\sqrt{-h}}\frac{\delta S_{\rm reg}}{\delta \dot A_i} =\frac{16\pi G_{\rm eff}}{\sqrt{-g}}\frac{\delta S_{\rm reg}}{\delta E_i}
\,\,,
\label{def:conjugatemomenta}\end{equation}
hence, recalling \eqref{eq:extrinsiccurv}, the Hamiltonian density becomes 
\begin{equation}
\mathcal H \,=\, 2\pi^{ij} {\cal K}_{ij} + \pi^i E_i-\mathcal L_{\rm reg}\,\,.
\label{eq:hamiltonian}\end{equation}
The momenta may be expanded as
\begin{equation}
\pi^{ij} = \pi^{ij}_{(2)} + \pi^{ij}_{(4)} + ... + \pi^{ij}_{(2+d)} + ... \,\,,
\end{equation}
and similarly for $\pi^i$. Here, each term is associated to a specific term in the expansion of $\mathcal L_{\rm reg}$. In particular, for even $d$ there is a term $\pi^{ij}_{(2+d)}$ coming from variations of $\mathcal L^{(d)}$.
It follows that the Hamiltonian density can be expanded as
\begin{equation}
\mathcal H \,=\, \mathcal H^{(0)} + \mathcal H^{(2)} + ... + \mathcal H^{(d)} + ... \,\,.
\end{equation}

\paragraph{Hamiltonian in terms of momenta.} The definition \eqref{def:conjugatemomenta} for the momenta can be inverted in order to express the extrinsic tensors ${\cal K}_{ij}$ and $E_i$ as functions of the momenta $\pi^{ij}$ and $\pi^i$. In this way we obtain the Hamiltonian \eqref{eq:hamiltonian} as a function of the momenta, $\mathcal H = \mathcal H(\pi^{ij}, \pi^i; h_{ij} , A_i)$.

We can discuss the dependence of the different terms in the asymptotic expansion of the Hamiltonian on the different terms in the momenta. 
We notice that $\mathcal H^{(0)}$ can only depend on the momentum conjugate to the metric via the trace of $\pi^{ij}_{(2)}$, since it is the only possible combination with the correct order in the asymptotic expansion. Similarly, $\mathcal H^{(2k)}$ will depend on the trace of $\pi^{ij}_{(2+2k)}$ and some combinations of the momenta $\pi^{ij}_{(n)}$ with $n<2k+2$. 
Finally, $\mathcal H^{(d)}$ depends on the trace of $\pi^{ij}_{(2+d)}$ and on the lower-order terms.

\paragraph{Counterterms and Weyl anomaly.} The Hamilton-Jacobi method now prescribes to replace the momenta appearing in the Hamiltonian \eqref{eq:hamiltonian} with the functional derivatives,
\begin{equation}
\pi^{ij} = \frac{16\pi G_{\rm eff}}{\sqrt{-h}}\left(-\frac{\delta S_{\rm ren}}{\delta h_{ij}} + \frac{\delta S_{\rm ct}}{\delta h_{ij}}\right)\,,\qquad\quad
\pi^{i} = \frac{16\pi G_{\rm eff}}{\sqrt{-h}}\left(-\frac{\delta S_{\rm ren}}{\delta A_i} + \frac{\delta S_{\rm ct}}{\delta A_i}\right)\,.
\label{eq:momentahj}\end{equation}
This yields the Hamilton-Jacobi equation \eqref{eq:hamiltonjacobi}, that can be solved order by order in the expansion of $\mathcal{H}$. 
The momenta can be split in two contributions
\begin{equation}
\pi = \pi^{(\rm ren)} + \pi^{\rm (ct)} \,\,,
\end{equation}
where $\pi^{(\rm ren)}$ contains the terms contributing to the renormalized action, while $\pi^{\rm (ct)}$ contributes to power-law divergences.
 We can now introduce an ansatz for the counterterm action $S_{\rm ct}$ as a linear combination of all local covariant boundary terms that diverge asymptotically, with coefficients to be determined. Expanding the Hamilton-Jacobi equation \eqref{eq:hamiltonjacobi} order by order gives a set of equations $\mathcal{H}^{(n)}=0$. Those constraining the power-law divergent sector of $S_{\rm reg}$ are the leading-order ones up to $\mathcal H^{(n)}$ with $n<d$. These just depend on $\pi^{\rm (ct)}$ and allow us to extract the coefficients fixing $S_{\rm ct}$. When $d$ is even we should also consider  $\mathcal H^{(d)}=0$: from this equation we can extract the holographic Weyl anomaly, since it depends on $h_{ij}\frac{1}{\sqrt{-h}}\frac{\delta S_{\rm ren}}{\delta h_{ij}}$ through the trace of $\pi^{ij}_{(2+d)}$ \cite{Martelli:2002sp,Fukuma:2001uf}.

\subsubsection{Implementing the procedure for our action}
\label{sec:hamiltonjacobi5D}

We now apply the procedure reviewed above to five-dimensional minimal gauged supergravity with four-derivative couplings.

\paragraph{Divergent sector.} In order to obtain the Hamiltonian $\mathcal H$ as in \eqref{eq:hamiltonian}, we first express the Lagrangian density $\mathcal L_{\rm reg} = \mathcal L_{\rm bulk} + \mathcal L_{\rm GHY}$ using the Gauss-Codazzi relations. 
In the gauge specified by \eqref{eq:FGgauge}, the non-vanishing components of the Christoffel symbol read
\begin{equation}
\begin{aligned}
&\qquad\quad\Gamma^i_{\,\,\,jk} = \gamma^i_{\,\,\,jk}\,\,,\qquad \Gamma^z_{\,\,\,ij} = -\frac{1}{2}g^{zz}\partial_z h_{ij}=\frac{z}{\ell} {\cal K}_{ij}\,\,,\\[1mm]
&\Gamma^z_{\,\,\,zz}=\frac{1}{2}g^{zz}\partial_z g_{zz} = -z^{-1}\,\,,\quad \,\,\, \Gamma^i_{\,\,\,jz} = \frac{1}{2}h^{ik}\partial_z h_{jk}=-\frac{\ell}{z}{\cal K}^i_{\,\,\,j} \,\,,
\end{aligned}
\label{eq:gamma}\end{equation}
where $\gamma$ denotes the Christoffel symbol defined from $h_{ij}$.
The non-vanishing components of the Riemann tensor are
\begin{equation}
\begin{aligned}
&R^i_{\,\,\,jkl} = \mathcal R^i_{\,\,\,jkl}- {\cal K}^i_{\,\,\,k}{\cal K}_{jl}+ {\cal K}^i_{\,\,\,l}{\cal K}_{jk}\,\,,\\[1mm] 
&R^z_{\,\,\,ijk} = \frac{z}{\ell}\left(\nabla_j {\cal K}_{ik} - \nabla_k {\cal K}_{ij}\right)\,\,, \qquad R^z_{\,\,\,izj} = \frac{z}{\ell}\partial_z{\cal K}_{ij} + {\cal K}_{ik} {\cal K}^k_{\,\,\,j}\,\,,
\end{aligned}
\label{eq:riemann1}\end{equation}
where $\mathcal R_{ijkl}$ and $\nabla_i$ are the Riemann tensor and the Levi-Civita connection built out of $h_{ij}$.
The components of the Ricci tensor can be written as
\begin{equation}
\begin{aligned}
R_{ij} &= \mathcal R_{ij} - {\cal K} {\cal K}_{ij} +   \frac{z}{\ell}\partial_z {\cal K}_{ij} + 2{\cal K}_{ik} {\cal K}^k_{\,\,\,j}\,\,,\\[1mm]
R_{zi} &= 
  -\frac{\ell}{z}\left( \nabla_j {\cal K}^j_{\,\,\,i} - \nabla_i {\cal K}\right)\,\,,\\[1mm]
R_{zz} 
 &= \frac{\ell^2}{z^2}\left(\frac{z}{\ell}\partial_z {\cal K} - {\cal K}_{ij}{\cal K}^{ij}\right)\,\,,
\end{aligned}
\label{eq:ricci1}\end{equation}
while the Ricci scalar reads
\begin{equation}
\begin{aligned}
R &= \mathcal R + 2\frac{z}{\ell}\partial_z {\cal K} -{\cal K}_{ij}{\cal K}^{ij}-{\cal K}^2\\[1mm]
&= \mathcal R -{\cal K}_{ij}^2 + {\cal K}^2 + 2e^{-1}\,\partial_z\left(\sqrt{-h}\,{\cal K}\right)\,\,.
\end{aligned}
\label{eq:ricciscalar}\end{equation}

We next use the Gauss-Codazzi relations above to identify the divergent part of the action $S_{\text{reg}}=S_{\rm bulk}+S_{\rm GHY}$, where $S_{\rm bulk}$ and $S_{\rm GHY}$ were the first is the Lorentzian counterpart of \eqref{eq:4daction2} and the second was given in \eqref{GHY_bdy_term}.
For the sector made of the two-derivative action, the Gauss-Bonnet term, and the respective GHY terms, we can follow the analysis of~\cite{Liu:2008zf} -- with the  difference that we also include the gauge kinetic term $F_{\mu\nu}F^{\mu\nu}$. Using the Gauss-Codazzi relations, one finds that the whole generalized GHY term is 
cancelled by an opposite contribution coming from the bulk action, leaving us with
\begin{equation}
\begin{aligned}
 S_{\rm reg} & \supset\ \frac{1}{16\pi G_{\rm eff}} \int\text d^5 x \, e\left(\overline R + 12\hat c_1g^2 - \frac{1}{4}\hat c_2 F_{ij}^2 -\frac{1}{2}\hat c_2E_i^2\right) \\[1mm]
& +\alpha\lambda_1\Big\{\left(\overline R^2_{ijkl} - 4\overline R^2_{ij} + \overline R^2 \right)  + 4\Big[\mathcal R {\cal K}^2 - \mathcal R {\cal K}_{ij}^2 -4 \mathcal R_{ij} {\cal K}^{ij} {\cal K} +4 \mathcal R_{ij} {\cal K}^{ik}{\cal K}_k^{\,\,\,j} \\[1mm]
& +2 \mathcal R_{ijkl} {\cal K}^{ik} {\cal K}^{jl}  + 2 {\cal K}^2 {\cal K}_{ij}^2 -  ({\cal K}_{ij}^2)^2 +2 {\cal K}_{ij}{\cal K}^{jk}{\cal K}_{kl}{\cal K}^{li}-\frac{1}{3}{\cal K}^4 -\frac{8}{3}{\cal K} {\cal K}^i_{\,\,\,j}{\cal K}^{jk}{\cal K}_{ki}\Big]
\Big\},
\end{aligned}
\label{eq:gb4}\end{equation}
where the coefficients 
\begin{equation}
\hat c_1 = 1-10\lambda_1\alpha g^2\,\,, \quad\qquad \hat c_2 = 1+ 4\lambda_1\alpha g^2
\end{equation}
correspond with the coefficients $\tilde c_1$, $\tilde c_2$ given in~\eqref{eq:4daction2},\eqref{ctildes} with $\lambda_2 =0$, since the effect of this parameter has been absorbed in the redefinition \eqref{eq:Geff} of the gravitational coupling constant, $G \to G_{\rm eff}$. Furthermore, we introduced
\begin{equation}
\begin{aligned}
\overline{R}_{ijkl} &= \mathcal R_{ijkl} - {\cal K}_{ik}{\cal K}_{jl} + {\cal K}_{il} {\cal K}_{jk}\,\,,\\[1mm]
\overline{R}_{ij} &= \mathcal R_{ij} - {\cal K} {\cal K}_{ij} + {\cal K}_{ik}{\cal K}^k{}_j\,\,,\\[1mm]
\overline{R} &= \mathcal R - {\cal K}^2 + {\cal K}_{ij}{\cal K}^{ij}\,\,.
\end{aligned}
\label{eq:overliner}\end{equation}

We now argue that the remaining part of $S_{\rm reg}$ does not lead to divergences. The only non-obvious terms are $C_{\mu\nu\rho\sigma}F^{\mu\nu} F^{\rho\sigma}$ and the gauge-gravitational Chern-Simons term in  \eqref{eq:4daction2}. The latter was discussed in~\cite{Landsteiner:2011iq}: as we already noticed in section~\ref{sec:variational_pb}, a boundary term was introduced there, but it is suppressed asymptotically in an AlAdS background and can therefore be neglected for our purposes. 
The reasoning for $C_{\mu\nu\rho\sigma}F^{\mu\nu} F^{\rho\sigma}$ is similar, as we already discussed in section~\ref{sec:variational_pb}. Let us see this from the perspective of the Hamilton-Jacobi method. In order to do so, we  come back to the action \eqref{eq:u123combination} with generic coefficients $u_1,u_2,u_3$, for which we identified the GHY term \eqref{eq:GHY1}. Using (\ref{eq:riemann1})--(\ref{eq:ricciscalar}) and (\ref{eq:intrinsictensors}), (\ref{eq:extrinsictensors}), one can show that
\begin{equation}
\begin{aligned}
&u_1\,R_{\mu\nu\rho\lambda} F^{\mu\nu} F^{\rho\lambda} + u_2\,R_{\mu\nu}F^{\nu\rho} F_\rho^{\,\,\,\mu} + u_3\, RF^2 =\\[1mm]
& -2\left(u_1- u_2\right){\cal K}_{ik} {\cal K}_{jl} F^{ij} F^{kl}- 2u_2\,{\cal K}_{ij}{\cal K}^{jk}F_{kl}F^{li} +u_3\left(8 {\cal K} {\cal K}_{ij}F^{ik}F_k^{\,\,\,j} - {\cal K}_{ij}^2 F_{kl}^2 + {\cal K}^2 F_{ij}^2 \right)\\[1mm]
&+e^{-1}\,\partial_z\left(\sqrt{-h}\left(u_2 {\cal K}_{ij} F^{jk}F_{k}^{\,\,\,i} + 2u_3\,{\cal K} F_{ij}^2 \right)\right) + \mathcal O(z^5)\,\,. 
\end{aligned}
\label{eq:weyltermgeneral}\end{equation}
The total derivative is exactly cancelled by the boundary terms in \eqref{eq:GHY1} that are not suppressed asymptotically. 
 Plugging the asymptotic behavior of the extrinsic curvature given in  (\ref{eq:extrinsictensors}), we obtain
\begin{equation}
\begin{aligned}
&u_1\,R_{\mu\nu\rho\lambda} F^{\mu\nu} F^{\rho\lambda} + u_2\,R_{\mu\nu}F^{\nu\rho} F_\rho^{\,\,\,\mu} + u_3\, RF^2 =\\
&\left(-2u_1 + 4u_2 -20 u_3\right) F_{ij}F^{ij} + e^{-1}\,\partial_z\left(\sqrt{-h}\left(-u_2 + 8u_3\right)F_{ij}F^{ij}\right) + \mathcal O(z^5)\,\,.
\end{aligned}
\label{eq:weyltermgeneral2}\end{equation}
These terms do not lead to power-law divergences in the on-shell action, but a priori the first one  yields a $\log \epsilon$ divergence when integrated in the bulk and thus contributes to the holographic Weyl anomaly, while the boundary term gives a finite contribution.

However, when the Weyl combination $C_{\mu\nu\rho\sigma}F^{\mu\nu} F^{\rho\sigma}$ is considered, corresponding to the choice $u_2 = \frac{4}{3}u_1$ and $u_3 = \frac{1}{6}u_1$,
Eq.~\eqref{eq:weyltermgeneral2} simplifies and the whole expression is further suppressed asymptotically. 

We conclude that the divergences of the regularized on-shell action are captured entirely by expression \eqref{eq:gb4}, as anticipated.\footnote{The $E_iE^i$ term appearing in \eqref{eq:gb4} does not lead to divergences either.}
 The Hamilton-Jacobi equations we have to solve, therefore, have the same form as the Einstein-Gauss-Bonnet problem discussed in~\cite{Liu:2008zf}, plus a contribution from the Maxwell kinetic term that only affects the logarithmic divergence.

\paragraph{The Hamiltonian.} Computing the momenta \eqref{def:conjugatemomenta} and evaluating the Hamiltonian density (\ref{eq:hamiltonian}), we arrive at
\begin{equation}
\begin{aligned}
\mathcal H = -\Big( \overline R + 12\hat c_1g^2 - \frac{1}{4}\hat c_2 F_{ij}^2 \Big) -\alpha\lambda_1\left( \overline R_{ijkl}^2 -4\overline R_{ij}^2 + \overline R^2\right) + ... \,\,.
\end{aligned}
\label{eq:hamiltoniandensity}\end{equation}
Here, the ellipsis comprise all the subleading contributions to the Hamiltonian, that correspond to finite contributions to the regulated action $S_{\rm reg}$, that requires no renormalization, and just affect the definition of the finite part of the renormalized action $S_{\rm ren}$. Recall that we just want to solve the Hamilton-Jacobi constraint \eqref{eq:hamiltonjacobi} asymptotically, in order to extract the counterterms action $S_{\rm ct}$ and the Weyl anomaly. 

By inverting the relation between the momenta, defined according to \eqref{def:conjugatemomenta}, and the extrinsic curvature $K_{ij}$, we obtain the following expression for the Hamiltonian~\cite{Liu:2008zf},
\begin{equation}
\begin{aligned}
\mathcal H(\pi^{ij}; h_{ij}, A_i) =&  -\Big[ \mathcal R + 12\hat c_1 g^2 -\frac{1}{2}\hat c_2 F_{ij}^2 + \pi_{ij}^2 - \frac{1}{3}\pi^2 + \alpha\lambda_1\Big( \mathcal R^2 - 4 \mathcal R_{ij}^2 + \mathcal R_{ijkl}^2 \\[1mm]
&-\frac{16}{3}\mathcal R_{ij}\pi^{ij} \pi + \frac{10}{9}\mathcal R \pi^2 - 2\mathcal R \pi_{ij}^2 + 8\mathcal R_{ij}\pi^{jk}\pi_k^{\,\,\,i} + 4\mathcal R_{ijkl} \pi^{ik}\pi^{jl} \\[1mm]
&+ 2\pi^i_{\,\,\,j}\pi^j_{\,\,\,k}\pi^k_{\,\,\,l}\pi^l_{\,\,\,i} - (\pi_{ij}^2)^2 -\frac{16}{9}\pi\pi^i_{\,\,\,j}\pi^j_{\,\,\,k}\pi^k_{\,\,\,i} + \frac{10}{9}\pi^2 \pi_{ij}^2 - \frac{11}{81}\pi^4
\Big) \Big]+\text{...}\,\,,
\end{aligned}
\label{eq:hamiltonianliuesabra}\end{equation}
where $\pi\equiv \pi^{ij}h_{ij}$, and we neglected $\mathcal O(\alpha^2)$ terms.

\paragraph{The counterterms.} The ansatz for the possible counterterms is   
\begin{equation}
S_{\rm ct} = -\frac{1}{16\pi G_{\rm eff}}\int\text d^4x\sqrt{-h}\left( W + C \mathcal R\right)\,\,,
\label{eq:countertermsactionansatz}\end{equation}
 where $W \propto g$ and  $C\propto g^{-1}$ by dimensional analysis.
 All other local, covariant terms built out of intrinsic boundary fields are either constant at the boundary, hence depend on the choice of renormalization scheme, or suppressed. 

As explained in sec.~\ref{sec:hamiltonjacobigeneral}, the power-law divergent sector of the Hamiltonian 
 just depends on 
\begin{equation}
\frac{16\pi G_{\rm eff}}{\sqrt{-h}}\,\frac{\delta S_{\rm ct}}{\delta h_{ij}} \,=\, -\frac{W}{2}\,h^{ij} + C\left( \mathcal R^{ij} - \frac{1}{2}\mathcal R h^{ij}\right)\,\,.
\label{eq:momentahjct}\end{equation}
After substituting \eqref{eq:momentahjct} into \eqref{eq:hamiltonianliuesabra}, the first two orders of the Hamilton-Jacobi equations give algebraic conditions for determining the coefficients $W$ and $C$ introduced in \eqref{eq:countertermsactionansatz}~\cite{Liu:2008zf},
\begin{equation}
\begin{aligned}
\mathcal H^{(0)}= 12\hat c_1 g^2- \frac{1}{3}W^2  - \frac{\alpha\lambda_1}{162}W^4=0\,\,,\\
\mathcal H^{(2)}=\mathcal R\left(1- \frac{1}{3}WC + \alpha\lambda_1 \frac{1}{9}W^2\left( 1- \frac{1}{9}WC\right) \right)=0\,\,.
\end{aligned}
\label{eq:hamiltonjacobi5D}\end{equation}
These two equations are solved by 
\begin{equation}
W= 6g\left(1-\frac{16}{3}\lambda_1\alpha g^2\right)\,, \hspace{1cm} C=\frac{1}{2g}\left(1+8\lambda_1\alpha g^2\right)\,.
\label{eq:countertermscoefficients}
\end{equation}
As anticipated, these coefficients are in agreement with the boundary terms we gave in  \eqref{counterterms} \eqref{eq:mus}. 
 The theory is indeed renormalized by the same counterterms as the two-derivative action, just the respective coefficients get corrected.

\paragraph{The holographic Weyl anomaly.} The only remaining divergence in the on-shell action is the logarithmic one, which leads to the holographic Weyl anomaly. In the following we check that the Hamilton-Jacobi equation  \eqref{eq:hamiltonjacobi} gives the result expected from \cite{Fukuma:2001uf},\footnote{See also~\cite{Nojiri:1999mh,Blau:1999vz}, where the holographic Weyl anomaly for higher-derivative gravity was calculated using the original method of~\cite{Henningson:1998gx}.} with an additional contribution from the gauge field strength $F_{ij}$, which was not considered in that reference.

As we argued in sec.~\ref{sec:hamiltonjacobigeneral}, we should consider the equation $\mathcal H^{(4)}=0$ as it depends on $h_{ij}\frac{\delta S_{\rm ren}}{\delta h_{ij}}$, that gives the trace of the holographic energy-momentum tensor. Substituting the definition \eqref{eq:momentahj} for the momenta as functional derivatives into \eqref{eq:hamiltonianliuesabra}, where now the values of $C$ and $W$ are given by \eqref{eq:countertermscoefficients}, we find that $\mathcal H^{(4)}=0$ yields
\begin{equation}
\begin{aligned}
32\pi G_{\rm eff}\left(1-4\alpha g^2\lambda_1\right) g\,{\frac{h_{ij}}{\sqrt{-h}}\frac{\delta S_{\rm ren}}{\delta h_{ij}}}=&-\frac{1}{4g^2}\left({\cal R}_{ij}{\cal R}^{ij}-\frac{1}{3}{\cal R}^2-g^2 F_{ij}F^{ij}\right)\\[1mm]
&-\lambda_1\alpha\left(C_{ijkl}C^{ijkl}-{\cal R}_{ij}{\cal R}^{ij}+\frac{1}{3}{\cal R}^2-g^2 F_{ij}F^{ij}\right)\,,
\end{aligned}
\end{equation}
where $C_{ijkl}C^{ijkl}={\cal R}_{ijkl}{\cal R}^{ijkl}-2 {\cal R}_{ij}{\cal R}^{ij}+\frac{1}{3}{\cal R}^2$ is the square of the Weyl tensor of $h_{ij}$.
In this equation, the left-hand side contains precisely the trace of the quasi-local energy-momentum tensor, $h_{ij}\frac{\delta S_{\rm ren}}{\delta h_{ij}} = -\frac{1}{2}\sqrt{-h}\, T^i{}_i$. Therefore we have found
\begin{equation}
T^i{}_i \,=\,\frac{1+4\lambda_2 \alpha g^2}{64\pi G g^3}\left({\cal R}_{ij}{\cal R}^{ij}-\frac{1}{3}{\cal R}^2-g^2 F_{ij}F^{ij}\right) +\frac{\lambda_1\alpha}{16\pi G g}\left(C_{ijkl}C^{ijkl}-2g^2 F_{ij}F^{ij}\right)\,,
\end{equation}
where we also used \eqref{eq:Geff}. Pushing $\sqrt{-h}\, T^i{}_i$ to the conformal boundary gives a finite quantity, that is the trace of the holographic energy-momentum tensor.
This is to be compared with the general expression of the Weyl anomaly of ${\cal N}=1$ SCFTs,
\begin{equation}
T^{i}{}_i = -\frac{\aa}{16\pi^2} \hat{\cal X}_{\rm GB}+\frac{\cc}{16\pi^2}\left(\hat{C}_{ijkl}\hat{C}^{ijkl}-\frac{8}{3}{\hat F}_{ij}{\hat F}^{ij}\right)\,,
\end{equation}
where here $\hat{\cal X}_{\rm GB}$, $\hat{C}_{ijkl}\hat{C}^{ijkl}$ and ${\hat F}_{ij}{\hat F}^{ij}$ denote the Gauss-Bonnet, Weyl-squared and Maxwell terms constructed using the metric on the conformal boundary, and $\hat{F}_{ij}$ is the field strength of the gauge field $\hat A_i=\frac{\sqrt{3} g}{2} A_i$  coupling canonically to the R-current of the dual SCFT. The two expressions match upon using the dictionary \eqref{a_c_high_der} for the anomaly coefficients, that is what we wanted to verify.

\section{Conserved charges}\label{sec:conservedchargesgeneral}

The aim of this section is to derive conserved charges associated to the gauge symmetries of the theory: diffeomorphisms and U$(1)$ gauge transformations. Before entering into the technical details of the derivation, let us briefly discuss the complications that one encounters when doing so. The main issue is that Noether currents associated to local symmetries, here denoted as $J^\mu$, are trivial, meaning that on-shell they can always be written as
\begin{equation}
J^{\mu}=\nabla_{\nu}k^{\mu\nu}\, ,
\end{equation}
where $k^{\mu\nu}=-k^{\nu\mu}$ so that the current is conserved, $\nabla_\mu J^\mu = 0$. The main consequence of this is that the associated Noether charge is arbitrary. Let ${\cal C}$ be a Cauchy slice and integrate the Noether current on it to define a conserved charge:
\begin{equation}
Q =\int_{\cal C}{\boldsymbol \epsilon}_{\mu} J^{\mu}=\frac{1}{2}\int_{\partial {\cal C}}\,{\boldsymbol \epsilon}_{\mu\nu}\, k^{\mu\nu}\, ,
\end{equation}
where we have made use of the notation
\begin{equation}
{\boldsymbol\epsilon}_{\mu_1 \dots \mu_n}\equiv \frac{1}{\left(D-n\right)!}\epsilon_{\mu_1 \dots \mu_n \nu_1 \dots \nu_{D-n}}\, \diff x^{\nu_{1}}\wedge \dots \diff x^{\nu_{D-n}}\, ,
\label{def:boldsymbolepsilon}
\end{equation}
being $D$ the spacetime dimension. The charge defined in this way, though a conserved one, can take any value since at this stage $k^{\mu\nu}$ is arbitrary. Indeed, we could have defined a second current,
\begin{equation}
{J'}^{\mu}=J^{\mu}+\nabla_{\nu}\left(\Delta k\right)^{\mu\nu}=\nabla_{\nu} \left(k^{\mu\nu}+\left(\Delta k\right)^{\mu\nu}\right)\,,
\end{equation}
which is also conserved and is therefore equivalent to $J^{\mu}$ if we do not ask for additional requirements. However, the associated charges do not coincide in general since
\begin{equation}
Q'-Q = \frac{1}{2}\int_{\partial{\cal C}}{\boldsymbol\epsilon}_{\mu\nu}\left(\Delta k\right)^{\mu\nu}\,.
\end{equation}

It turns out that the additional requirement that we have to impose in order to define the charge unambiguously is that the associated current vanishes on-shell, so that the $(D-2)$-form ${\bf k}=\frac{1}{2}{\boldsymbol \epsilon}_{\mu\nu}k^{\mu\nu}$ is conserved. 
This follows from the generalized Noether's theorem of Barnich, Brandt and Henneaux \cite{Barnich:2000zw}, which in simple terms states that there is a one-to-one correspondence between the equivalence classes of gauge parameters $\xi$ which are field symmetries (i.e., which leave  the fields invariant: $\delta_{\xi}\Psi=0$) and that of $(D-2)$-forms ${\mathbf k}_{\xi}$ which are conserved on-shell, $\diff{\mathbf k}_{\xi}=0$. We say that two gauge parameters are equivalent if their difference vanishes on-shell, whereas two conserved $(D-2)$-forms, ${\bf k}_{\xi}, {\bf k}'_{\xi}$ are equivalent if they differ by an exact form, ${\mathbf k'}_{\xi}\sim {\mathbf k}_{\xi}$ if ${\mathbf k'}_{\xi}={\mathbf k}_{\xi}+\diff{\mathbf l}_{\xi}$. 

\noindent
The charge is then given by 
\begin{equation}
Q_{\xi}=\int_{\partial {\cal C}} {\mathbf k}_{\xi}\,,
\end{equation}
and clearly does not depend on the representative of the equivalence class that is chosen to compute it since the integral of $\diff {\bf l}_{\xi}$ vanishes by Stokes theorem. For further details and for a pedagogic introduction to this topic we refer to~\cite{Compere:2006my, Compere:2018aar}.

\subsection{Wald formalism in the presence of Chern-Simons terms}\label{sec:Waldformalism}

In order to apply these ideas to the gauge symmetries of five-dimensional minimal gauged supergravity, we shall make use of the formalism developed by Wald in \cite{Wald:1993nt, Iyer:1994ys}, slightly modifying it to account for the effect of Chern-Simons terms. In particular, we will not demand the Lagrangian to be invariant under U$(1)$ gauge transformations, as it transforms by a total derivative when these terms are present -- see~\eqref{eq:delta_xiL2} below. As we will see, this crucially affects the definition and properties of the charges.

 Before getting started, let us introduce some notation that will be useful in the remaining of the paper. Following \cite{Wald:1993nt, Iyer:1994ys}, we define the $D$-form Lagrangian $\mathbf L$ as\footnote{In this section we are setting $16\pi G=1$ for convenience. It will be reinstated later.} 
\begin{equation}
\mathbf L={\boldsymbol\epsilon}\, {\mathcal L}\,, \hspace{1cm} S=\int {\bf L}\, .
\end{equation}
Under a generic variation of the fields, which we schematically denote by $\Psi=\{g_{\mu\nu}, A_{\mu}\}$, we have
\begin{equation}\label{eq:generaldL}
\delta {\mathbf L}={\mathbf E}_{\Psi}\delta \Psi+ \diff {\boldsymbol\Theta}(\Psi, \delta \Psi)\,,
\end{equation}
where ${\mathbf E}_{\Psi}=0$ are the equations of motion of the theory and $\boldsymbol\Theta(\Psi, \delta \Psi)$ is the term which is generated through integration by parts.

We now specify \eqref{eq:generaldL} to the most general gauge transformation given our field content. This consists of a diffeomorphism generated by a vector field $\xi^{\mu}$ plus a U$(1)$ gauge transformation parametrized by a function $\chi$, 
\begin{equation}\label{eq:deltafields}
\delta_{\xi}g_{\mu\nu}={\cal L}_{\xi}\, g_{\mu\nu}\,, \hspace{1cm} \delta_{\xi, \chi}A_{\mu}={\cal L}_{\xi} A_{\mu}+\nabla_{\mu}\chi\,,
\end{equation}
As in \cite{Elgood:2020svt}, we find it useful to introduce the so-called momentum map, $P_{\xi, \chi}$, which is defined as follows, 
\begin{equation}\label{eq:defP}
P_{\xi, \chi}=\chi+\iota_{\xi}A\, .
\end{equation}
Using it, we can write the transformation of the gauge field as
\begin{equation}\label{eq:delta_xiA}
\delta_{\xi, \chi} A=\iota_{\xi} F+ \diff P_{\xi, \chi}\,.
\end{equation}
When restricting to gauge parameters which are field symmetries, that is $\delta_{\xi}g=0$, $\delta_{\xi, \chi}A=0$, we have that $\xi$ becomes a Killing vector of the metric, as usual, while $P_{\xi, \chi}$ satisfies the momentum map equation
\begin{equation}\label{eq:momentummap}
\diff P_{\xi, \chi}=-\iota_{\xi}F\, ,
\end{equation}
which further implies $\delta_{\xi}F=0$.

The transformation \eqref{eq:deltafields} acts on the $D$-form Lagrangian $\bf L$ as follows
\begin{equation}\label{eq:delta_xiL1}
\begin{aligned}
\delta_{\xi, \chi}{\bf L}=\,&{\boldsymbol\epsilon}\left({\cal E}_{\mu\nu}\,\delta_{\xi}g^{\mu\nu}+{\cal E}^{\mu}\,\delta_{\xi, \chi}A_{\mu}\right)+\diff {\boldsymbol \Theta}_{\xi, \chi}\\[1mm]
=\,&{\boldsymbol\epsilon}\left[\nabla_{\mu}\left(-2\,\xi^{\nu}\,{\cal E}^{\mu}{}_{\nu}\,+P_{\xi, \chi}\,{\cal E}^{\mu}\right)\right]+\diff {\boldsymbol \Theta}_{\xi, \chi}\\[1mm]
=\,& \diff\left({\bf S}_{\xi, \chi}+{\boldsymbol\Theta}_{\xi, \chi}\right)\,,
\end{aligned}
\end{equation}
where we have defined 
\begin{equation}
{\boldsymbol\Theta}_{\xi, \chi}={\boldsymbol \Theta}\left(\Psi, \delta_{\xi, \chi}\Psi\right)\,, \hspace{1cm} {\bf S}_{\xi, \chi}={\boldsymbol\epsilon}_{\mu}\left(-2\,\xi^{\nu}\,{\cal E}^{\mu}{}_{\nu}\,+P_{\xi, \chi}\,{\cal E}^{\mu}\right)\,,
\end{equation}
and
\begin{equation}\label{def:eom}
{\cal E}_{\mu\nu}=\frac{\delta S}{e\, \delta g^{\mu\nu}}\,, \hspace{1cm} {\cal E}^{\mu}=\frac{\delta S}{e\, \delta A_{\mu}}\,.
\end{equation}
Moreover, we have used that (as a consequence of gauge invariance) ${\cal E}_{\mu\nu}$ and ${\cal E}^{\mu}$ satisfy the following off-shell identities,
\begin{equation}
2\nabla^{\mu}{\cal E}_{\mu\nu}=F_{\mu\nu}\,{\cal E}^{\mu}\, , \hspace{1cm} \nabla_{\mu}{\cal E}^{\mu}=0\,.
\end{equation}
In addition to \eqref{eq:delta_xiL1}, we assume that $\bf L$ transforms as follows,
\begin{equation}\label{eq:delta_xiL2}
\begin{aligned}
\delta_{\xi, \chi}{\bf L}\,=\,&{\cal L}_{\xi}{\bf L}+\diff\left(\chi{\boldsymbol \Lambda}\right)\,=\,\diff\left(\iota_{\xi}{\bf L}+\chi{\boldsymbol \Lambda}\right)\,,
\end{aligned}
\end{equation}
where $\boldsymbol\Lambda$ is a closed $(D-1)$-form that determines the transformation of ${\bf L}$ under gauge transformations. Hence, the results that we derive in this and the next section apply to theories whose Lagrangians satisfy Eq.~\eqref{eq:delta_xiL2}. A particular case is the Lagrangian of five-dimensional minimal gauged supergravity with four-derivative corrections presented in \eqref{eq:4daction2}, which will be analyzed in detail in the next subsection. 
Putting together \eqref{eq:delta_xiL1} and \eqref{eq:delta_xiL2}, we obtain the conservation of the Noether current defined as
\begin{equation}\label{eq:Noethercurrent}
{\bf J}_{\xi, \chi}={\bf S}_{\xi, \chi}+{\boldsymbol \Theta}_{\xi, \chi}-\iota_\xi {\bf L}-\chi{\boldsymbol\Lambda}  \, ,
\end{equation}
which means that, at least locally,
\begin{equation}
{\bf J}_{\xi, \chi}=\diff {\bf Q}_{\xi, \chi}\,,
\end{equation}
being ${\bf Q}_{\xi, \chi}$ the Noether surface charge. Note that this is not yet the conserved $(D-2)$-form of the generalized Noether's theorem \cite{Barnich:2000zw}, as ${\bf J}_{\xi, \chi}$ does not vanish on-shell when $\xi, \chi$ are field symmetries. However, we can always improve it (adding a total derivative) in a way such that it does. Following the approach of \cite{Bazanski:1990qd, Kastor:2008xb, Ortin:2021ade}, we find that the improved current that vanishes on-shell is,
\begin{equation}\label{eq:improvedJ}
{\bf J}_{\xi, \chi}\to {\bf J}_{\xi, \chi}+\diff {\boldsymbol\Xi}_{\xi, \chi}\,, 
\end{equation}
where ${\boldsymbol\Xi}_{\xi, \chi}$ is a local $(D-2)$-form such that $\diff{\boldsymbol\Xi}_{\xi, \chi}=\iota_\xi {\bf L}+\chi{\boldsymbol\Lambda}$.\footnote{Notice that the integrability condition of ${\boldsymbol\Xi}_{\xi, \chi}$ is automatically satisfied if $\xi, \chi$ are field symmetries.} The surface charge ${\bf k}_{\xi, \chi}$ associated to this improved current  (and that is therefore conserved on-shell) is given by
\begin{equation}\label{eq:improvedcurrent}
{\bf k}_{\xi, \chi}={\bf Q}_{\xi, \chi}+{\boldsymbol\Xi}_{\xi, \chi}\,, \hspace{1cm} \diff{\bf k}_{\xi, \chi}={\bf J}_{\xi, \chi}+\diff {\boldsymbol\Xi}_{\xi, \chi}={\bf S}_{\xi, \chi}+{\boldsymbol \Theta}_{\xi, \chi}\, .
\end{equation}

Eventually, the charges that we obtain upon integration are,
\begin{equation}\label{eq:generalcharge}
{Q}_{\xi, \chi}=\int_{\partial\cal C}{\bf k}_{\xi, \chi}=\int_{\partial\cal C}\left({\bf Q}_{\xi, \chi}+{\boldsymbol\Xi}_{\xi, \chi}\right)\, .
\end{equation}
In the purely gravitational case, these charges are referred to as generalized Komar integrals, see e.g.~\cite{Komar:1958wp,Kastor:2009wy, Bazanski:1990qd, Kastor:2008xb, Ortin:2021ade} and references therein, while for a pure gauge symmetry they are called Page charges~\cite{Page:1983mke,Marolf:2000cb}. Their most interesting property  for the purposes of this work is that they obey the Gauss law. Let us assume that the spacetime can be foliated with a family of spacelike hypersurfaces $\Sigma_{z}$ diffeomorphic to the (conformal) boundary, labelled by a radial coordinate $z$. If regularity is assumed, then we have that
\begin{equation}\label{eq:Pageness}
\int_{\Sigma_{z}\, \cap \, {\cal C}}{\bf k}_{\xi, \chi}=\int_{\Sigma_{z'}\, \cap \, {\cal C}}{\bf k}_{\xi, \chi}\,,
\end{equation}
as a consequence of Stokes theorem. We will make use of this property to derive the quantum statistical relation in appendix~\ref{sec:aladsthermodynamics} and to compute the corrected charges (electric charge and angular momentum) for the supersymmetric black hole of~\cite{Gutowski:2004ez} from the corrected near-horizon solution found in \cite{Cassani:2022lrk}. Let us then define the electric charge and angular momenta as particular cases of \eqref{eq:generalcharge}:

\paragraph{Electric charge.} In order to define the electric charge, we restrict ourselves to a pure gauge transformation, i.e.~$\xi=0$. In this case, $\chi$ must be constant if it corresponds to a field symmetry and the improvement term ${\boldsymbol\Xi}_{0, \chi}$ simply reads
\begin{equation}\label{eq:chargexi}
{\boldsymbol\Xi}_{0, \chi}=\chi {\boldsymbol\Delta}\,, \hspace{1cm}\diff {\boldsymbol\Delta}={\boldsymbol\Lambda}\,,
\end{equation}
so that\footnote{We fix $\chi=1$ without loss of generality.} 
\begin{equation}\label{eq:pagecharge}
Q\equiv Q_{0, 1}=\int_{\partial {\cal C}}{\bf k}_{0, 1}=\int_{\partial {\cal C}}\left({\bf Q}_{0, 1}+{\boldsymbol\Delta} \right)\, .
\end{equation}
This corresponds with the notion of Page charge \cite{Page:1983mke,Marolf:2000cb}.

\paragraph{Angular momenta.} Angular momenta can be defined as the Komar integral associated to the angular Killing vectors $\xi^{\mu}=\phi_{k}^{\mu}$, $k=1,2,\ldots$,
\begin{equation}\label{eq:defJi}
J_{k}=-\int_{\partial {\cal C}}{\bf k}_{\phi_k, 0}=-\int_{\partial {\cal C}}{\bf Q}_{\phi_k, 0}\, ,
\end{equation}
where we have used that the integral of the improvement term ${\boldsymbol\Xi}_{\xi, 0}$ vanishes when $\xi^{t}=0$.

\subsection{Specifying to five-dimensional higher-derivative supergravity}\label{sec:conservedchargesourtheory}

Let us now specify to the four-derivative theory of interest to us \eqref{eq:4daction2} and derive the corresponding expressions for the Noether current \eqref{eq:Noethercurrent} and surface charge \eqref{eq:generalcharge}. Let us start defining the following auxiliary tensors,\footnote{We note that  ${\cal P}^{\mu\nu\rho\sigma}={P}^{\mu\nu\rho\sigma}+{\Pi}^{\mu\nu\rho\sigma}+ \text{(anti)symmetrizations}$, where $\displaystyle P^{\mu\nu\rho\sigma}$ is the tensor introduced in \eqref{eq:Pmunurhosigma} in order to write down the equations of motion. Its explicit expression, as well as the one for $\displaystyle \frac{\partial {\cal L}'}{\partial F_{\mu\nu}}$, for the action of interest can be found in \eqref{Ptensor_explicit}.}
\begin{equation}\label{eq:auxiliarytensors}
{\cal P}^{\mu\nu\rho\sigma} = \frac{\partial\mathcal L}{\partial R_{\mu\nu\rho\sigma}}\,\,,\hspace{1cm} {\mathcal F}^{\mu\nu} =-2\,\frac{\partial\mathcal L'}{\partial F_{\mu\nu}}\,\,, \hspace{1cm}\Pi^{\mu\nu\rho\sigma} = -\frac{\lambda_1\alpha}{\sqrt{3}}\epsilon^{\mu\nu\alpha\beta\gamma}R_{\alpha\beta}{}^{\rho \sigma}A_{\gamma}\,\,,\hspace{1cm}
\end{equation}
where ${\mathcal L'}={\cal L}-{\cal L}_{\rm{CS}}$ is the Lagrangian density without the contribution of the Chern-Simons terms, which are treated separately. It is assumed (and there is no loss of generality in doing so) that the above auxilary tensors inherit the algebraic symmetries of the associated curvature tensors, namely
\begin{equation}
\mathcal P^{(\mu\nu)\rho\sigma} =\, 0 \,\,, \hspace{7.5mm} {\mathcal P}^{\mu\nu\rho\sigma}={\mathcal P}^{\rho\sigma\mu\nu}\,\,,\hspace{7.5mm}{\mathcal P}^{\mu[\nu\rho\sigma]}=0\,\,,\hspace{7.5mm}{\mathcal F}^{(\mu\nu)} = 0\,\,.
\label{eq:propertiesauxiliarytensors}
\end{equation}
According to \eqref{eq:Noethercurrent}, the information we need in order to compute the Noether current ${\bf J}_{\xi, \chi}$ consists of the tensors ${\cal E}_{\mu\nu}$, ${\cal E}^{\mu}$, the boundary term ${\boldsymbol\Theta}={\boldsymbol \epsilon}_{\mu}\Theta^{\mu}$ and $\boldsymbol \Lambda={\boldsymbol \epsilon}_{\mu}\Lambda^{\mu}$. Making use of \eqref{EinsteinEqCorrected} and \eqref{MaxwellEqCorrected}, we have that
\begin{equation}\label{eq:eoms2}
\begin{aligned}
\mathcal E_{\mu\nu}=\,&\mathcal P_{(\mu}{}^{\rho\sigma\lambda} R_{\nu)\rho\sigma\lambda} - 2\nabla^\rho\nabla^\sigma\mathcal P_{\rho(\mu\nu)\sigma} +\frac{1}{2} {\mathcal F}_{\rho(\mu}F_{\nu)}^{\,\,\,\,\rho} -\frac{1}{2}g_{\mu\nu} \mathcal L' -\Pi_{(\mu}^{\,\,\,\,\,\rho\sigma\lambda} R_{\nu)\rho\sigma\lambda}\,\,,\\[1mm]
\mathcal E^\mu =\,& \nabla_\nu\mathcal F^{\nu\mu} - \frac{\tilde c_3}{4\sqrt{3}}\epsilon^{\mu\nu\rho\sigma\lambda} F_{\nu\rho} F_{\sigma\lambda} - \frac{\lambda_1\alpha}{2\sqrt{3}}\epsilon^{\mu\nu\rho\sigma\lambda} R^{\delta\gamma}_{\,\,\,\,\,\,\,\nu\rho} R_{\delta\gamma\sigma\lambda}\,\,,
\end{aligned}
\end{equation}
and, using \eqref{bdry_term_metric_var}, \eqref{Theta0}, \eqref{Theta1}, \eqref{Theta2}, \eqref{Theta_final},
\begin{equation}\label{eq:theta}
{\Theta}^{\mu} = 2\mathcal P^{\rho\sigma\mu\nu}\nabla_\sigma\delta g_{\nu\rho} - 2\nabla_\sigma\mathcal P^{\mu\nu\rho\sigma} \delta g_{\nu\rho} -{\mathcal F}^{\mu\nu} \delta A_\nu - \frac{\tilde c_3}{3\sqrt{3}}\epsilon^{\mu\nu\rho\sigma\lambda} \delta A_\nu A_\rho F_{\sigma\lambda}\,\,.
\end{equation}
It will be useful later to split the boundary term ${\boldsymbol\Theta}$ into its ``gravitational'' and ``electromagnetic'' contributions, ${\boldsymbol\Theta}={\boldsymbol\Theta}^{\rm g}+{\boldsymbol\Theta}^{\rm {em}}$, where
\begin{equation}\label{eq:split_theta}
\begin{aligned}
{\boldsymbol\Theta}^{\rm g}=\,&{\boldsymbol \epsilon}_{\mu}\left(2\mathcal P^{\rho\sigma\mu\nu}\nabla_\sigma\delta g_{\nu\rho} - 2\nabla_\sigma\mathcal P^{\mu\nu\rho\sigma} \delta g_{\nu\rho}\right)\,,\\[1mm]
{\boldsymbol\Theta}^{\rm em}=\,&{\boldsymbol \epsilon}_{\mu}\left( -{\mathcal F}^{\mu\nu} \delta A_\nu - \frac{\tilde c_3}{3\sqrt{3}}\epsilon^{\mu\nu\rho\sigma\lambda} \delta A_\nu A_\rho F_{\sigma\lambda}\right)\,.
\end{aligned}
\end{equation}
Finally, the transformation of the action under U$(1)$ gauge transformations is characterized by 
\begin{equation}
{\Lambda}^\mu =\,-\epsilon^{\mu\nu\rho\sigma\lambda} \left(\frac{\tilde c_3}{12\sqrt{3}} F_{\nu\rho} F_{\sigma\lambda} + \frac{\lambda_1\alpha}{2\sqrt{3}}R_{\nu\rho\delta\gamma} R_{\sigma\lambda}{}^{\delta\gamma}\right)\,, \hspace{1cm} \Lambda^{\mu}=\nabla_{\nu}\Delta^{\mu\nu}\,,
\label{eq:lambda}
\end{equation}
where
\begin{equation}\label{eq:Deltamunu}
\Delta^{\mu\nu} = -\epsilon^{\mu\nu\rho\sigma\lambda}\left[\frac{\tilde c_3}{6\sqrt{3}}A_\rho F_{\sigma\lambda} +\frac{\lambda_1\alpha}{3\sqrt{3}}\Omega_{\rm{CS}}{}_{\rho\sigma\lambda}\right]\,\,,
\end{equation}
and $\Omega_{\rm{CS}}$ is the Lorentz-Chern-Simons three-form constructed out of the gravitational spin connection $\omega^{ab}=\,\omega_{\mu}^{ab}\,\diff x^{\mu}$,
\begin{equation}\label{eq:lorentzchernsimons}
\Omega_{\rm{CS}}=\diff \omega^{ab}\wedge \omega_{ab}-\frac{2}{3}\, \omega^{a}{}_{b}\wedge \omega^{b}{}_{c}\wedge \omega^{c}{}_a\,,
\end{equation}
where $a, b, \dots$ are flat indices.

Putting these results together, we find the following expression for the Noether current ${\bf J}_{\xi, \chi}={\boldsymbol \epsilon}_{\mu}J^\mu_{\xi,\chi}$, 
\begin{equation}
\begin{aligned}
J^\mu_{\xi,\chi}=\, &\left(-2\mathcal P^{(\mu|\rho\sigma\lambda} R^{|\nu)}_{\,\,\,\,\,\rho\sigma\lambda} +4\nabla_\rho\nabla_\sigma\mathcal P^{\rho(\mu\nu)\sigma} - \mathcal F^{\rho(\mu}F^{\nu)}_{\,\,\,\,\rho}\right)\xi_\nu +P_{\xi,\chi}\nabla_\nu\mathcal F^{\nu\mu}\\[1mm]
+& 2\mathcal P^{\rho\sigma\mu\nu}\nabla_\sigma\delta_{\xi} g_{\nu\rho} - 2\nabla_\sigma\mathcal P^{\mu\nu\rho\sigma} \delta_{\xi} g_{\nu\rho} -\mathcal F^{\mu\nu} (\delta_{\xi,\chi} A)_\nu - \xi^\mu \mathcal L_{\rm{CS}}\\[1mm]
-& P_{\xi,\chi}\epsilon^{\mu\nu\rho\sigma\lambda}\left(\frac{\tilde c_3}{4\sqrt{3}} F_{\nu\rho} F_{\sigma\lambda} +\frac{\alpha\lambda_1}{2\sqrt{3}} R^{\delta\gamma}{}_{\nu\rho} R_{\sigma\lambda\delta\gamma}\right) - \frac{\tilde c_3}{3\sqrt{3}}\epsilon^{\mu\nu\rho\sigma\lambda} (\delta_{\xi,\chi} A)_\nu A_\rho F_{\sigma\lambda}-\\[1mm]
&-\frac{2\lambda_1\alpha}{\sqrt{3}}\epsilon^{(\mu|\rho\alpha\beta\gamma}A_\alpha R_{\beta\gamma}{}^{\sigma\lambda}R^{|\nu)}{}_{\rho\sigma\lambda} \xi_\nu +\chi\epsilon^{\mu\nu\rho\sigma\lambda}  \left(\frac{\tilde c_3}{12\sqrt{3}}F_{\nu\rho} F_{\sigma\lambda} +\frac{\alpha\lambda_1}{2\sqrt{3}}R_{\nu\rho}{}^{\delta\gamma} R_{\sigma\lambda\delta\gamma}\right).
\end{aligned}
\label{eq:oneformnoethercurrent}\end{equation}
Using the transformation rules \eqref{eq:deltafields} and the fact that
\begin{equation}
\begin{aligned}
-\xi^\mu\mathcal L_{\rm{CS}} &\,=\,\epsilon^{\mu\nu\rho\sigma\lambda}\left[\iota_\xi A  \left(\frac{\tilde c_3}{12\sqrt{3}}F_{\nu\rho} F_{\sigma\lambda} +\frac{\alpha\lambda_1}{2\sqrt{3}}R^{\delta\gamma}_{\,\,\,\,\,\,\,\nu\rho} R_{\delta\gamma\sigma\lambda}\right) \right.\\[1mm]
&\left.\qquad\quad\  +\,\xi^\tau A_\rho\left(\frac{\tilde c_3}{3\sqrt{3}}F_{\tau\nu} F_{\sigma\lambda}
 +\frac{2\alpha\lambda_1}{\sqrt{3}}R_{\tau\nu\delta\gamma} R_{\sigma\lambda}^{\,\,\,\,\,\,\,\delta\gamma} \right)\right],
\end{aligned}
\label{eq:chernsimonsidentity}\end{equation}
it is possible to manipulate \eqref{eq:oneformnoethercurrent}, extending the calculation in Appendix A of~\cite{Ortin:2021ade} to our more general setup, to obtain that
\begin{equation}\label{eq:oneformnoethercurrent2}
{\bf J}_{\xi,\chi} =\, {\boldsymbol \epsilon}_{\mu}\nabla_\nu\left[-4\nabla_\sigma \mathcal P^{\nu\mu\sigma\rho}\xi_\rho +2 \mathcal P^{\nu\mu\sigma\rho}\nabla_\sigma\xi_\rho +P_{\xi,\chi}\left(\mathcal F^{\nu\mu} +\frac{\tilde c_3}{3\sqrt{3}}\epsilon^{\nu\mu\rho\sigma\lambda}A_\rho F_{\sigma\lambda} \right)\right]\,.
\end{equation}
The associated Noether surface charge ${\bf Q}_{\xi, \chi}$ is therefore given by \begin{equation}\label{eq:noethersurfacechargecomponents}
{\bf Q}_{\xi,\chi} =\,\frac{1}{2}\boldsymbol {\epsilon}_{\mu\nu}\left[4\nabla_\sigma \mathcal P^{\mu\nu\sigma\rho}\xi_\rho - 2 \mathcal P^{\mu\nu\sigma\rho}\nabla_\sigma\xi_\rho - P_{\xi,\chi}\left( \mathcal F^{\mu\nu} +\frac{\tilde c_3}{3\sqrt{3}}\epsilon^{\mu\nu\rho\sigma\lambda}A_\rho F_{\sigma\lambda} \right)\right]\,,
\end{equation}
and we observe that it can be written as the sum of two contributions,
\begin{equation}\label{eq:surfacecharge2}
\mathbf Q_{\xi,\chi} = \mathbf Q_\xi^{\rm g} +P_{\xi,\chi} \,\mathbf Q_{0,1}\,\,,
\end{equation}
where ${\bf Q}^{\rm g}_{\xi}$ is the generalized Komar $(D-2)$-form \cite{Compere:2006my},
\begin{equation}
{\bf Q}^{\rm g}_{\xi} = \,\frac{1}{2}\boldsymbol {\epsilon}_{\mu\nu}\left(4\nabla_\sigma \mathcal P^{\mu\nu\sigma\rho}\xi_\rho - 2 \mathcal P^{\mu\nu\sigma\rho}\nabla_\sigma\xi_\rho\right)\,\,, 
\label{eq:defQg}
\end{equation}
and 
\begin{equation}\label{eq:Q_{0,1}}
{\bf Q}_{0,1} = -\frac{1}{2}\boldsymbol {\epsilon}_{\mu\nu}\left(\mathcal F^{\mu\nu} +\frac{\tilde c_3}{3\sqrt{3}}\epsilon^{\mu\nu\rho\sigma\lambda}A_\rho F_{\sigma\lambda} \right)\,\,.
\end{equation}

Let us provide an explicit expression for the electric charge defined in \eqref{eq:pagecharge}. Taking into account \eqref{eq:Q_{0,1}} and the expression for ${\boldsymbol\Delta}$ given in \eqref{eq:Deltamunu}, we find that 
\begin{equation}\label{eq:k_{0,1}}
\begin{aligned}
{\bf k}_{0,1} =\,& -\frac{1}{2}\boldsymbol \epsilon_{\mu\nu}  \left[ \mathcal F^{\mu\nu} +\epsilon^{\mu\nu\rho\sigma\lambda}\left(\frac{\tilde c_3}{2\sqrt{3}}A_\rho F_{\sigma\lambda} + \frac{\lambda_1\alpha}{3\sqrt{3}}\Omega_{\rm{CS}}{}_{\rho\sigma\lambda}\right)\right]\,\,,
\end{aligned}
\end{equation}
and therefore the electric charge reduces to
\begin{equation}
\label{eq:electric_charge_page}
\begin{aligned}
Q=-\int_{\partial {\cal C}}\left(\star {\cal F}-\frac{\tilde c_3}{\sqrt{3}}F\wedge A-\frac{2\lambda_1\alpha}{\sqrt{3}}\Omega_{\rm{CS}}\right)\,,
\end{aligned}
\end{equation}
which coincides with the notion of Page charge, as anticipated.

Finally, we can give the expression for the angular momentum \eqref{eq:defJi}. Considering the surface charge \eqref{eq:noethersurfacechargecomponents} for an angular Killing vector $\phi_k$ and taking $\chi=0$ gives
\begin{equation}\label{eq:surfacechargeangmom}
 J_{k} = -\frac{1}{2}\int_{\partial {\cal C}}\!\boldsymbol {\epsilon}_{\mu\nu}\left[4\nabla_\sigma \mathcal P^{\mu\nu\sigma\rho}(\phi_k)_\rho - 2 \mathcal P^{\mu\nu\sigma\rho}\nabla_\sigma(\phi_k)_\rho - \left(\iota_{\phi_k} A\right)\left( \mathcal F^{\mu\nu} +\frac{\tilde c_3}{3\sqrt{3}}\epsilon^{\mu\nu\rho\sigma\lambda}A_\rho F_{\sigma\lambda} \right)\right].
\end{equation}

In particular, we can apply the above integral to evaluate the angular momentum for the two-derivative minimal gauged supergravity, \eqref{eq:2_der_minimal_sugra_5D}. In this case, the explicit expressions for the tensors \eqref{eq:auxiliarytensors} are particularly simple,
\begin{equation}
\mathcal P^{\mu\nu\rho\sigma} = g^{\mu[\rho}g^{\sigma]\nu}\,\,,\quad\quad \mathcal F^{\mu\nu} = F^{\mu\nu}\,\,,
\end{equation}
and the angular momentum \eqref{eq:surfacechargeangmom} reproduces the one given in~\cite{Barnich:2005kq,Suryanarayana:2007rk,Hanaki:2007mb,Cassani:2018mlh}.

\section{Corrected charges of BPS AdS$_5$ black hole from near-horizon geometry}\label{sec:corrections_GR}

In this section we apply the formalism above to the supersymmetric black hole solution to five-dimensional minimal gauged supergravity. In particular, we want to reproduce the corrected thermodynamical charges directly from the near-horizon geometry. We expect to be able to do that since in section \ref{sec:conservedchargesgeneral} we have constructed charges satisfying a Gauss law, that can be equivalently evaluated on any radial slice of the spacetime at fixed time.\footnote{The same approach has been followed recently in \cite{Cano:2023dyg} to study corrections to the extremal Kerr entropy in the context of the heterotic string and cubic gravities. For an analysis in the ungauged case see~\cite{deWit:2009de}.} 
Reproducing the black hole charges and its entropy from the near-horizon geometry is particularly useful in higher-derivative gravity, where full corrected black hole solutions are particularly difficult to obtain. Instead, the corrections to the near-horizon geometry of extremal solutions are more easily found thanks to its enhanced symmetry.

\subsection{Supersymmetric black hole and near-horizon geometry}

The most general known black hole solution to two-derivative minimal gauged supergravity in five dimensions  was given in~\cite{Chong:2005hr}, and reviewed in sec.~\ref{TwoDerReview}. Here we will restrict to the supersymmetric and extremal solution with equal angular momenta, $J_1=J_2\equiv J$, that is the solution first given in~\cite{Gutowski:2004ez}, whose features have been summarized in sec.~\ref{sec:Lor_susy_BH_5D}. 
The near-horizon geometry, \eqref{eq:near_horizon_GR}, is a fibration over AdS$_2$ of a compact space with ${\rm SU}(2) \times {\rm U}(1)$ isometry, compatible with the geometry of a three-sphere squashed by the rotation along an axis. 
Higher-derivative corrections to the near-horizon geometry have been obtained in~\cite{Cassani:2022lrk} by solving the equations of motion from the action~\eqref{eq:4daction2}, which we derive in appendix~\ref{app:eoms}.\footnote{More recently, a similar analysis has been extended to the case of two independent angular momenta in~\cite{Cano:2024tcr}.} The metric can be parametrized by $(t,r,\psi,\theta,\phi)$ coordinates as
\begin{equation}
\begin{aligned}
\text ds^2 &= v_1\left[-r^2\text dt^2 + \frac{\text dr^2}{r^2}\right] + \frac{v_2}{4}\Bigl[\sigma_1^2 + \sigma_2^2+ v_3 \left(\sigma_3 + w\, r\text dt\right)^2\Bigr]\,\,,\\
A &=- e\,r\,\text dt - p\left(\sigma_3 + w\,r\text dt\right)\,\,,
\end{aligned}
\label{def:ads2timess3}\end{equation}
where $\sigma$'s are the left-invariant Maurer-Cartan one-forms of SU$(2)$, given in terms of Euler angles on $S^3$ $(\psi,\theta,\phi)$ as,
\begin{equation}
\sigma_1 = \cos\psi \text d\theta + \sin\psi \sin\theta \text d\phi \,\,,\,\,\sigma_2 = -\sin\psi\text d\theta + \cos\psi \sin\theta \text d\phi \,\,,\,\,\, \sigma_3 = \text d\psi + \cos\theta \text d\phi\,\,,
\label{def:su2invariantforms}\end{equation}
and the coefficients are fixed to
\begin{equation}\label{eq:nearhorizoncoefficients2d}
\begin{aligned}
&v_1=\,\frac{\chi ^2}{4g^2 \left(1+3 \chi ^2\right)}+\alpha \,\delta v_1\, , \hspace{0.5cm} v_2=\frac{\chi ^2}{g^2}+\alpha \,\delta v_2\, , \hspace{0.5cm}v_3=1+\frac{3 \chi ^2}{4}+\alpha \,\delta v_3\, , \\[1mm]
&p=\frac{\sqrt{3} \chi ^2}{4g}+\alpha \,\delta p\, , \hspace{0.25cm} w=\, \frac{3 \chi }{\left(1+3 \chi ^2\right)\sqrt{4+3 \chi ^2}}+\alpha \,\delta w\, , \hspace{0.25cm}
e=\frac{\sqrt{3} \chi}{g \left(1+3 \chi ^2\right)\sqrt{4+3 \chi ^2}}+\alpha\, \delta e\, ,
\end{aligned}
\end{equation}
contain the deviations from the $\alpha=0$ solution. Here, the (dimensionless) parameter  $\chi$ is related to the only parameter $r_0$ of \eqref{eq:near_horizon_GR}-\eqref{eq:GR_NH_coefficients} by $\chi=g\,r_0$.\footnote{To simplify the notation, we have dropped the '$\sim$' appearing in \eqref{eq:near_horizon_GR}.}
Working perturbatively in $\alpha$, the problem of solving the equations of motion reduces to a linear system of algebraic equations,
\begin{equation}
{\cal M}{\cal X}={\cal N}\, ,
\end{equation}
where ${\cal M}={\cal M}(\chi; g)$ is a degenerate $6\times 6$ matrix, ${\cal X}=(\delta v_1, \delta v_2, \delta v_3, \delta p, \delta e, \delta w)^T$ and ${\mathcal N}={\mathcal N}(\chi; g)$ is a vector which encodes the contribution to the equations of motion of the corrections. The general solution to this equation is
\begin{equation}
{\cal X}={\cal X}^{H}+{\cal X}^{P}\,,
\end{equation}
where ${\cal X}^{H}$ is the homogeneous solution, ${\mathcal M}{\cal X}^{H}=0$, and ${\cal X}^{P}$ is a particular solution. The latter carries information about the new physics while the homogeneous solution (which is non-trivial since the matrix ${\cal M}$ is not invertible) parametrizes the freedom that we still have to fix the electrostatic potential and angular velocity at the horizon for the extremal solution. 
Solving the homogeneous system in terms of $\delta e$ and $\delta w$ yields the following solution
\begin{equation}
\begin{aligned}
{\delta v}^{H}_1=\,&\frac{\chi  \left(9 \chi ^6-12 \chi ^4-68 \chi ^2-48\right)\delta e}{2 g \sqrt{9 \chi ^2+12} \left(27 \chi
   ^8+162 \chi ^6+144 \chi ^4+4 \chi ^2-12\right)}\\[1mm]
&-\frac{\chi ^3 \left(9 \chi ^6+78 \chi ^4+154 \chi ^2+88\right)\delta w}{4 g^2 \sqrt{3 \chi ^2+4} \left(27 \chi
   ^8+162 \chi ^6+144 \chi ^4+4 \chi ^2-12\right)}\,,\\[1mm]
{\delta v}^{H}_2=\,&4\left(1+3\chi^2\right)^2\delta v_1^{H}\,,\\[1mm]
{\delta v}^{H}_3=\,&-\frac{\sqrt{3}  g \chi ^3 \left(3 \chi ^2+4\right)^{3/2} \left(45 \chi ^4+66 \chi ^2+17\right)\delta e}{2
   \left(3 \chi ^4+16 \chi ^2+6\right) \left(9 \chi ^4+6 \chi ^2-2\right)}\\[1mm]
&-\frac{3 \chi  \left(3 \chi ^2+1\right) \left(3 \chi ^2+4\right)^{3/2} \left(3 \chi ^6+7 \chi
   ^4+2\right)\delta w}{4 \left(3 \chi ^4+16 \chi ^2+6\right) \left(9 \chi ^4+6 \chi ^2-2\right)}\,,\\[1mm]
\delta p^{H}=\,&\frac{\chi  \left(3 \chi ^2+1\right) \sqrt{3 \chi ^2+4} \left(9 \chi ^6-48 \chi ^4-62 \chi
   ^2-6\right)\delta e}{2 \left(3 \chi ^4+16 \chi ^2+6\right) \left(9 \chi ^4+6 \chi ^2-2\right)}\\[1mm]
&-\frac{\chi  \left(3 \chi ^2+1\right) \sqrt{9 \chi ^2+12} \left(9 \chi ^8+69 \chi ^6+70 \chi ^4+10
   \chi ^2+4\right)\delta w}{4 g \left(3 \chi ^4+16 \chi ^2+6\right) \left(9 \chi ^4+6 \chi ^2-2\right)}\,,\\[1mm]
{\delta e}^{H}=\;\,&\delta e\, ,\qquad {\delta w}^{H}=\,\,\delta w\, ,
\end{aligned}
\end{equation}
while a particular solution is given by 
\begin{equation}\label{eq:particularsol}
\begin{aligned}
{\delta v}^{P}_1=\,&\frac{\lambda _1 \left(18 \chi ^6+21 \chi ^4+14 \chi
   ^2+2\right)}{27 \chi ^6+27 \chi ^4-2}\,,\qquad {\delta v}^{P}_2=\,\frac{4 \lambda _1 \left(36 \chi ^6+51 \chi ^4+10 \chi ^2+6\right)}{9 \chi ^4+6 \chi
   ^2-2}\,,\\[1mm]
{\delta v}^{P}_3=\,&-\frac{\lambda _1 g^2  \left(3 \chi ^2+4\right) \left(63 \chi ^4+78 \chi ^2-38\right)}{2 \left(9 \chi ^4+6 \chi
   ^2-2\right)}\,,\\[1mm]
   \delta p^{P}=\,&\frac{\sqrt{3} \lambda _1 g  \left(3 \chi ^2+4\right) \left(27 \chi ^4-6 \chi ^2+10\right)}{4 \left(9 \chi ^4+6 \chi
   ^2-2\right)}\, ,\qquad \delta e^{P}=\,0\, , \qquad \delta w^{P}=\,0\, .
\end{aligned}
\end{equation}
As discussed in \cite{Morales:2006gm, Dias:2007dj} building on the formalism of~\cite{Sen:2005wa,Sen:2008vm}, the variables $e$ and $w$, that corresponds to the non-normalizable modes of the AdS$_2 \times S^3$ solution, are identified with the thermodynamical variables conjugate to the electric charge and angular momentum. Thus, if we want to obtain the corrections keeping them fixed, we must impose the following choice of boundary conditions,
\begin{equation}
\delta e=0\, , \hspace{1cm} \delta w=0\, ,
\end{equation}
in which case the homogeneous part of the solution vanishes and the solution is simply equal to the particular one in \eqref{eq:particularsol}.


\subsection{Entropy and charges from the near-horizon geometry}\label{sec:chargefromhorizon}

Having found the corrected near-horizon geometry, we can now compute the entropy of the black hole. As shown in the appendix~\ref{sec:aladsthermodynamics}, the black-hole entropy ${\cal S}$ satisfying the correct first law of thermodynamics in our higher-derivative gravity can still be computed by means of the Wald formula \eqref{eq:Waldentropy}. Note that the entropy obtained using this formula would not be gauge invariant in general.\footnote{This issue has been recently addressed in \cite{Elgood:2020xwu, Elgood:2020nls} in the context of the heterotic superstring theory, where a gauge-invariant generalization of the Wald formula has been obtained. One way to preserve gauge invariance, at the expense of breaking diffeomorphisms, would be to start from a five-dimensional action where the mixed gauge-gravitational Chern-Simons term has been integrated by parts, and use the prescription of \cite{Tachikawa:2006sz} for the entropy. We do not do this, as it would correspond to a scheme where the mixed gauge-gravitational anomaly manifests itself in the non-conservation of the holographic energy-momentum tensor, rather than of the R-current. In any case, we have checked that the prescription of \cite{Tachikawa:2006sz} yields the same result as ours, since the total derivative does not contribute in the present case.
} In the case at hands, there is only one gauge transformation which affects the entropy, which is $A\to A+\diff \psi$. However, this gauge transformation breaks the ${\rm SU}(2)\times {\rm U}(1)$ symmetry of the near-horizon solution and in fact is not globally well defined. Hence, it is reasonable to expect that the direct application of the Wald formula in the gauge we are using should produce the right result. This is exactly the case, as we are going to show next.

The final expression for the Wald entropy is,
\begin{equation}
\begin{aligned}
{\cal S} &=-2\pi\int_{\cal H} \diff^3x\, \sqrt{\gamma}\,{\cal P}^{\mu\nu\rho\sigma}\, n_{\mu\nu}\, n_{\rho\sigma}
\\[1mm]
{\cal S}& =\frac{\pi ^2\chi ^3\sqrt{4+3 \chi ^2}}{4 Gg^3}\left[1+4 \lambda _2\alpha  g^2 +24 \lambda _1\alpha  g^2 \frac{9 \chi ^4+18 \chi ^2+8}{9 \chi ^4+6 \chi ^2-2}\right]\, ,
\end{aligned}
\end{equation}
which in terms of the parameter $a$ (see Eq.~\eqref{eq:GRtoCCLP}) reads
\begin{equation}\label{eq:entropy_near_horizon}
{\cal S}=\frac{\pi ^2 ag \sqrt{ag (a g+2)}}{Gg^{3}\left(1-ag\right)^2}\left[1+4 \lambda _2\alpha  g^2 +48\lambda _1 \alpha  g^2 \frac{2 a^2 g^2+5 a g+2}{11 a^2 g^2+8 a g-1}\right]\,,
\end{equation}
 nicely matching the expression obtained from the on-shell action, \eqref{eq:entropy0} with $b=a$. This is a very robust check of the validity of our results. 
 
In section \ref{sec:conservedchargesourtheory} we have derived the expression for the electric charge $Q$ as a  Page charge. We report here the integral we have to perform on the horizon geometry,
\begin{equation}
Q = -\frac{1}{32\pi G_{\rm eff}}\int_{\mathcal H}\sqrt{\gamma}\,n_{\mu\nu}\left[\mathcal F^{\mu\nu} +\epsilon^{\mu\nu\rho\sigma\lambda}\left( \frac{\hat c_3}{2\sqrt{3}}A_\rho F_{\lambda\sigma} + \frac{\lambda_1 \alpha}{3\sqrt{3}}\,\Omega_{\rm CS\,\rho\lambda\sigma}\right) \right]\,\,,
\label{eq:electricchargeathorizon}\end{equation}
where $\mathcal F^{\mu\nu}$ has been introduced in \eqref{eq:auxiliarytensors}.
Notice that the presence of the Lorentz-Chern-Simons term \eqref{eq:lorentzchernsimons} in this formula implies that the value of the charge is frame-dependent. The possible shifts due to a frame rotation are parametrized by an integer $k\in \mathbb Z$ as~\cite{Witten:2007kt}
\begin{equation}\label{eq:csshift}
Q \rightarrow Q + \frac{\pi\lambda_1\alpha}{4\sqrt{3}G_{\rm eff}} k\,\,.
\end{equation}
Choosing the vielbein basis
\begin{equation}
e_0 = \sqrt{v_1} r\text dt\,\,,\,\,\,\, e_1 = \sqrt{v_1}\frac{\text dr}{r} \,\,,\,\,\,\, e_2=\frac{\sqrt{v_2}}{2}\sigma_1 \,\,,\,\,\,\, e_3= \frac{\sqrt{v_2}}{2}\sigma_2\,\,,\,\,\,\, e_4 = \frac{\sqrt{v_2 v_3}}{2}\left( \sigma_3 + w\, r\text dt \right)\,\,
\end{equation}
 and evaluating \eqref{eq:electricchargeathorizon} on the near-horizon solution \eqref{def:ads2timess3}, 
\eqref{eq:nearhorizoncoefficients2d}, \eqref{eq:particularsol}, we obtain\footnote{Due to the normalization for the gauge field chosen in this chapter, the electric charge \eqref{eq:bpselectriccharge} does not correspond to the canonically normalized R-charge, used in~\eqref{eq:corrected_4d_charge}. The two charges are related by $Q_{\rm here} = \frac{\sqrt{3} g}{2} Q_{\rm there}$.} 
\begin{equation}
Q^* = \frac{\sqrt{3}\pi a}{2g(1-a g)^2 G_{\rm eff}}\left[1 + 4\lambda_1 \alpha g^2\frac{2\left(2 + 7 a g + 57 a^2 g^2 + 73 a^3 g^3 + 23 a^4 g^4 \right)}{3ag(11 a^2g^2 + 8 ag -1)}  \right]\,\,.
\label{eq:bpselectriccharge}\end{equation}
This can be compared with the supersymmetric and extremal electric charge computed in~\eqref{eq:corrected_4d_charge} by varying the black hole on-shell action and imposing the supersymmetric and extremal limit on the parameters. 
If we denote by $Q^*_{\rm there}$ the charge given there, we have
\begin{equation}
\frac{\sqrt{3}g}{2}Q^*_{\rm there} = Q^*+ \frac{2\pi}{\sqrt{3}G_{\rm eff}}\lambda_1 \alpha\,\,,
\label{eq:comparingelectriccharges}\end{equation}
hence the two charges differ by a contribution that does not depend on the parameters of the solution (and therefore does not affect the first law of thermodynamics). We expect that this discrepancy should be due to a rotation of the vielbein causing the shift \eqref{eq:csshift}, with $k=8$.

Similarly, we can evaluate the angular momentum using the formula \eqref{eq:surfacechargeangmom} for the corresponding Page charge, that is
\begin{equation}
J = -\frac{1}{32\pi G_{\rm eff}}\int_{\mathcal H}\sqrt{\gamma}\,n_{\mu\nu} \!\left[4\nabla_\sigma \mathcal P^{\mu\nu\sigma\rho}\eta_\rho - 2 \mathcal P^{\mu\nu\sigma\rho}\nabla_\sigma\eta_\rho - \iota_\eta A\left(\mathcal F^{\mu\nu} + \frac{\hat c_3}{3\sqrt{3}}\epsilon^{\mu\nu\rho\sigma\lambda}A_\rho F_{\sigma\lambda} \right) \right],
\label{eq:angularmomentumathorizon}\end{equation}
where $\eta = 2\partial_\psi$,
 Evaluating this formula on the near-horizon 
solution gives
\begin{equation}
J^* = \frac{a^2\left(3+ag\right)\pi}{2g\left(1-ag\right)^3 G_{\rm eff}}\left[1 + 24\lambda_1 \alpha g^2\frac{8a^4g^4 + 25 a^3g^3 + 29 a^2 g^2 + 9 ag +1}{ag(11 a^2g^2 + 8 ag -1)} \right] \,\,,
\label{eq:bpsangularmomentum}\end{equation}
which precisely matches the expression obtained from thermodynamical considerations in \eqref{eq:corrected_4d_J1} when $b=a$ (note that there is no frame ambiguity in this formula).

The entropy \eqref{eq:entropy_near_horizon} can then be expressed (neglecting $\mathcal O(\alpha^2)$ terms) as a function of the extremal charges derived above as
\begin{equation}\label{eq:bpsentropy}
\mathcal S^* = \pi\sqrt{ \frac{4}{g^2} (Q^*)^2 -\frac{2\pi}{g^3 G_{\rm eff}} J^*}\,\,.
\end{equation}
This form of the microcanonical entropy is in agreement with the one found in~\cite{Cassani:2022lrk,Bobev:2022bjm}, capturing four-derivative corrections to the leading-order result given in~\eqref{eq:S_GR}. Here we have derived it by relying just on the near-horizon solution. 

We can also check that a \emph{near-horizon version of the first law} is satisfied, where variations of the BPS entropy are related to variations of the charges from the near-horizon solution,
\begin{equation}\label{nh1stlaw}
\delta\mathcal S^* = 2\pi \left(w\,\delta J^* +e\,\delta Q^*\right)\,,
\end{equation}
where $e$ and $w$ are the coefficients determined in \eqref{eq:nearhorizoncoefficients2d}. The absence of the mass in \eqref{nh1stlaw} is in line with the fact that in the supersymmetric regime this is no more an independent quantity, as it is fixed by a linear combination of the angular momentum and electric charge.

%% file: research_three.tex
\chapter{Higher-derivative corrections to flavoured BPS black hole thermodynamics and holography}
\label{chap:flavour}

In this chapter, based on contribution~\cite{Cassani:2024tvk}, we derive a precise holographic match between the SCFT results \eqref{eq:index_asympt} and the on-shell action of the dual supersymmetric black hole, providing a first-principle microscopic explanation of its entropy, which is the final goal of the second part of this Thesis. In order to do so, we will employ the techniques developed and reviewed in the previous chapters. 

Our first step is the evaluation of the Legendre transform of~\eqref{eq:index_asympt} at first subleading order in the large-$N$ limit for a class of holographic theories whose 't Hooft anomaly coefficients satisfy certain requirements that we specify, obtaining in this way a prediction for the corrected entropy of the dual supersymmetric black hole as a function of the electric charges and angular momenta. Among the field theories falling in the class we consider, there are $\mathcal{N}=4$ SYM with gauge group ${\rm SU}(N)$ and the $\mathcal{N}=1$ $\mathbb{C}^3/\Gamma$ quiver gauge theories, namely the theories describing the low-energy dynamics of D3-branes probing a $\mathbb{C}^3/\Gamma$ singularity, where $\Gamma$ is a discrete subgroup of ${\rm SU}(3)$. For these theories the corrections to the large-$N$ results are suppressed by a factor of $1/N^2$. 

For $\mathcal{N}=4$ SYM the cancellations due to maximal supersymmetry give the following linear and cubic anomaly coefficients
\begin{equation}
\text{${\cal N} =4$ SYM}\qquad  \longleftrightarrow \qquad k_I = 0\,,\quad  k_{IJK}= \frac{N^2-1}{2}|\epsilon_{IJK}|\,,
\end{equation}
where $I,\,J,\,K = 1,2,3$ and where we have assumed a basis such that all charges are R-symmetries. 
This makes it straightforward to obtain the corrected entropy from \eqref{eq:index_asympt}. However, in the case of $\mathbb{C}^3/\Gamma$ theories both 't Hooft anomaly coefficients $k_{IJK}$ and $k_I$ receive simple but sufficiently generic corrections to the leading-order result, that make these theories an interesting class to study. Focussing on the case $\Gamma=\mathbb{Z}_\nu$, we obtain the following anomaly coefficients
\begin{equation}
\mathbb{C}^3/\mathbb Z_{\nu} \,\,\text{quiver theories} \qquad \longleftrightarrow \qquad k_I = - \nu \,r_I \,,\quad k_{IJK} = \frac{\nu N^2}{2}|\epsilon_{IJK}| + k_{(I} r_{J} r_{K)} \,,
\end{equation}
with $r_I = 1/2$ for all charges. We give our prediction for the corrected entropy of a dual supersymmetric black hole in section~\ref{subsec:entropy_orbifolds}. 

A further motivation for considering the orbifold theories has to do with the gravity side of the correspondence. There, we face the issue that, perhaps surprisingly, very few asymptotically AdS$_5$ black hole solutions carrying multiple electric charges and admitting an uplift to string theory or M-theory are explicitly known, though more are expected to exist. In fact, the only  examples are solutions carrying at most three independent electric charges, uplifting to Type IIB supergravity on $S^5$ or its quotients $S^5/\Gamma$, which are dual to $\mathcal{N}=4$ SYM or the $\mathbb{C}^3/\Gamma$ theories we consider~\cite{Gutowski:2004yv,Cvetic:2004ny,Chong:2005da,Kunduri:2006ek,Mei:2007bn,Wu:2011gq}. 

Following the approach we already adopted in the minimal case~\cite{Cassani:2022lrk}, we start from off-shell supergravity including four-derivative invariants and work at linear order in the couplings governing the latter. This makes it possible to easily eliminate the auxiliary fields by solving algebraic equations of motion, 
and to perform field redefinitions that simplify the resulting Lagrangian. Clearly, in the present case the computations are more complicated than in~\cite{Cassani:2022lrk} due to the many terms involving the scalar fields belonging to the vector multiplets.

Besides these technical complications, we encounter a more fundamental issue: we find that the four-derivative off-shell invariants available in the literature~\cite{Hanaki:2006pj,Ozkan:2013nwa,Ozkan:2016csy} are in general not sufficient to achieve a perfect match with the corrections to the dual 't Hooft anomaly coefficients.
For instance, the issue arises when considering the five-dimensional gravity dual of $\mathcal{N}=4$ SYM, by which here we mean a five-dimensional supergravity theory that at the two-derivative level uplifts to Type IIB supergravity on $S^5$ and that reproduces the full cubic 't Hooft anomalies of ${\rm SU}(N)$ $\mathcal{N}=4$ SYM. As already noted, in our basis the corrections   just shift the overall factor in the cubic coefficient $k_{IJK}$ as $N^2\to N^2-1$, however we have found no way to reproduce this starting from the known off-shell invariants.
We overcome this difficulty by proposing a simple modification of the two-derivative theory that does the job (and also reproduces the corrections obtained starting from the known invariants and going on-shell at linear order). 
The supergravity model capturing the anomalies of the orbifold theories, on the other hand, requires two different sets of corrections, which involve genuinely four-derivative terms, i.e. the supersymmetric completion of the mixed gauge-gravitational Chern-Simons term.

Our results for the bosonic sector of the four-derivative corrections to gauged supergravity coupled to an arbitrary number of vector multiplets is given in  section~\ref{sec:final_Lagr}. Of course, it contains much more than Chern-Simons terms and, as anticipated, it is considerably more involved than the minimal gauged supergravity case we studied previously. We show that it is sufficient for reproducing any 't Hooft anomaly coefficient of a dual SCFT at next-to-leading order in the large-$N$ expansion.  This also allows us to complete the discussion of  \cite{Tachikawa:2005tq,Hanaki:2006pj} for the gravity dual of $a$-maximization by including general next-to-leading order corrections.

Next we specialize to the supergravity model reproducing the\ 't Hooft anomalies of the $\mathbb{C}^3/\Gamma$ quiver theories. 
We compute the corrected supersymmetric black hole on-shell action within this model 
and match it  with the SCFT formula~\eqref{eq:index_asympt}. Due to the intrinsic complication of the calculation, in order to check this equivalence we partially resort to numerics and assume equality of the two a priori independent angular velocities, $\omega_1=\omega_2$. 
The result justifies why the entropy obtained in section~\ref{subsec:entropy_orbifolds} from the SCFT formula is a prediction for the corrected black hole entropy.
 Our result generalizes the match obtained at the two-derivative level in \cite{Cassani:2019mms}, as well as the universal four-derivative result of~\cite{Cassani:2022lrk,Bobev:2022bjm}.

The rest of the chapter is organized as follows. In section~\ref{sec:Legendre_transf_gen} we evaluate the Legendre transform of \eqref{eq:index_asympt} at first subleading order in the large-$N$ expansion, generalizing \ref{constrained_transform}, and obtain our prediction for the corrected black hole entropy with flavour charges turned on. In section~\ref{sec:orbifold_section} we apply our results to $\mathcal{N}=1$ orbifold theories. In section~\ref{sec:fourder_action} we present our (bosonic)  Lagrangian for supergravity including vector multiplet and four-derivative couplings. In section~\ref{sec:onshellaction} we specify the supergravity model dual to the orbifold SCFT's and evaluate the corrected  on-shell action of supersymmetric black hole solutions,  matching \eqref{eq:index_asympt}. 
We conclude in section~\ref{sec:Conclusions}. In appendix~\ref{sec:amaxim_section} we discuss $a$-maximization and  its gravity dual, including the corrections and completing the analysis of \cite{Tachikawa:2005tq,Hanaki:2006pj} at the four-derivative level. In appendices~\ref{app:offshell_invariants} and~\ref{app:field_redefinitions} we give the off-shell supersymmetry invariants and some field redefinitions needed in section~\ref{sec:fourder_action}. 

\section{Corrected entropy via Legendre transform}\label{sec:Legendre_transf_gen}

In this section, we obtain a prediction for the corrected entropy of supersymmetric multi-charge AdS${}_5$ black holes by taking the Legendre transform of the formula for $I$ given in \eqref{eq:index_asympt}. Doing this in full generality is a difficult task, hence we make some convenient assumptions on the 't Hooft anomaly coefficients that we specify next.

\subsection{Assumptions}\label{sec:assumptions}

The formula \eqref{eq:index_asympt}  holds at finite $N$ and independently of whether the SCFT is holographic. However, for a holographic theory we can study it in the large-$N$ expansion. In this chapter we will focus on (the leading and) the next-to-leading terms in the large-$N$ expansion, assuming the theory has a weakly-coupled holographic dual. Then we can write 
\be\label{eq:expansion_k's}
k_{IJK} = k^{(0)}_{IJK}+k^{(1)}_{IJK} + \ldots\,, \qquad\ k_I = 0 + k^{(1)}_I +\ldots\,, 
\ee
where the dots denote possible higher-order terms.
 At leading-order, eq.~\eqref{eq:index_asympt} reduces to
\be\label{eq:index_asympt_leading}
I^{(0)} \, =\, \frac{k^{(0)}_{IJK}\, \varphi^I\varphi^J\varphi^K}{6\,\omega_1\omega_2} \,,
\ee
where $k^{(0)}_{IJK}$ denotes the leading-order cubic 't Hooft anomaly. This agrees with a number of existing leading-order results, starting from the conjectured formula in Appendix~A of~\cite{Hosseini:2018dob}.

Our first assumption, that has already been used in~\cite{Cassani:2019mms}, regards the leading-order cubic anomaly $k^{(0)}_{IJK}$. We assume the existence of a constant fully-symmetric tensor $k^{(0)}{}^{IJK}$ such that
\begin{equation}\label{eq:cubic_relation}
k^{(0)}{}^{IJK}k^{(0)}_{J(LM}k^{(0)}_{NP)K}=\gamma \, \delta^{I}_{(L}k^{(0)}_{MNP)}\,,
\end{equation}
where $\gamma$ is some coefficient. Without loss of generality, we can fix the convenient normalization
\be\label{eq:normalization_k}
8k^{(0)}{}^{IJK}r_Ir_J r_K =1\,.
\ee
Then one can prove that\footnote{The proof goes as follows. Contracting~\eqref{eq:cubic_relation} with $r_I\bar s^L\bar s^M\bar s^N\bar s^P $ and decomposing the indices along the R-symmetry $Q_R = \bar s^I Q_I$ and the flavour symmetries by means of the projections~\eqref{projectors_Rsymm}, one obtains
$$
\gamma \,=\, \frac{1}{8}\, k^{(0)}_{RRR} + 2\, r_I r_J\, k^{(0)IJ L} \,\widetilde k^{(0)}_{L RR}  +  \frac{r_I \,k^{(0)I  J L} \,\widetilde k^{(0)}_{ J RR}\, \widetilde k^{(0)}_{ L RR}}{k^{(0)}_{RRR}}\,,
$$
where the index $R$ denotes the projection along the R-symmetry while $\widetilde k_{IRR}$  has the  $I$ index projected on the flavour directions, see appendix~\ref{amax_at_largeN} for details. Around Eq.~\eqref{eq:extremization_a_leading_order} we also show that $ \widetilde k^{(0)}_{I RR}  = 0$ for holographic theories with a weakly-coupled gravity dual, which leads us to~\eqref{expr_gamma}.} 
\be\label{expr_gamma}
\gamma\,=\, \frac{1}{8}k^{(0)}_{RRR} \,=\, \frac{4}{9}\, \aa^{(0)}\,,
\ee
where $\aa^{(0)}$ is the leading-order term of the Weyl anomaly coefficient $\aa$.
 We should note that property \eqref{eq:cubic_relation} is non-generic, for instance the cubic 't Hooft anomalies for the four global symmetries of the conifold theory do not satisfy it~\cite{Amariti:2019mgp}. For the five-dimensional supergravity which matches these global anomalies holographically, \eqref{eq:cubic_relation} holds when the scalar manifold is a symmetric space~\cite{Gunaydin:1983bi}.

Our second assumption regards the corrections: we assume the following relation between the cubic and linear coefficients,
\be\label{relation_k3_k1}
k_{IJK} \,=\,   k^{(0)}_{IJK}  + k_{(I} r_J r_{K)}  \,.
\ee
This condition implies a relation between the first-order corrections to the cubic and linear 't Hooft anomaly coefficients for the superconformal R-symmetry, $k^{(1)}_{RRR}=k^{(1)}_R$ (in order to see this one has to use $a$-maximization, see appendix~\ref{amax_at_largeN}).
 The condition is satisfied quite generally by the four-dimensional $\mathcal{N}=1$ quiver gauge theories which describe D3-branes probing the tip of a Calabi-Yau conical singularity, whose gravity dual is given by type IIB string theory on the Sasaki-Einstein base of the Calabi-Yau cone.  
 These are $\mathcal{N}=1$ quiver gauge theories made of $\nu$ SU$(N)$ nodes connected by chiral superfields, and \eqref{relation_k3_k1} holds as long as there are bifundamental chiral fields but {\it no} adjoint ones.  
 In order to see this, we note that for these theories the 't Hooft anomaly coefficients \eqref{def_anomaly_coeff} are made of a $\mathcal{O}(N^2)$ term and a $\mathcal{O}(1)$ term (that is, a term independent of $N$). Given the dimensions of the respective representations,  bifundamental fermions contribute with $N^2$ to the anomaly while adjoint fermions contribute with $N^2-1$. If no matter fields transform in the adjoint, then only the gaugini contribute to the $\mathcal{O}(1)$ correction. Note that the gaugino has charge $r_I$ under $Q_I$, that is the same charge as the supersymmetry parameter.
Then the form of the 't Hooft anomaly coefficients for these theories is:
\be\label{eq:exp_anomaly_coeff}
k_{IJK}  \,=\,  k^{(0)}_{IJK}  - \nu \, r_I r_J r_K \,,\qquad
k_I  \,=\,  - \nu\, r_I\,,
\ee
where the explicit expression of the leading $\mathcal{O}(N^2)$ cubic coefficient $k^{(0)}_{IJK}$ depends on the details of the quiver.
 On the other hand, it is a general fact that for quiver gauge theories of the type considered, $k_I$ contains no $\mathcal{O}(N^2)$ term provided the symmetry is non-anomalous, namely the $Q_I$-gauge-gauge anomaly vanishes as we assume here.\footnote{The argument is reviewed e.g.\ in Appendix B of~\cite{Cabo-Bizet:2020nkr}; there it is given for an R-symmetry but extension to flavour symmetries is straightforward. 
}
Clearly, \eqref{eq:exp_anomaly_coeff} implies \eqref{relation_k3_k1}.

Projecting onto the R-symmetry as discussed in appendix~\ref{amax_at_largeN}, we obtain the corrections to the R-symmetry anomaly coefficients, $k_{RRR} = k_{RRR}^{(0)}-\nu$, $k_R = -\nu$. Then, recalling the relation between the R-symmetry anomaly coefficients and the Weyl anomaly coefficients $\aa$, $\cc$ given in~\eqref{relacTrR}, we find the corrections to the latter:
\be\label{Weyl_coeffs_gen_quivers}
{\mathtt a}\,=\,\aa^{(0)} -\frac{3\nu}{16}\,,  \qquad  {\mathtt c}\,=\,\aa^{(0)} -\frac{\nu}{8}\,,
\ee
where by $\aa^{(0)} = \cc^{(0)}$ we denote the leading-order term in the large-$N$ expansion.

In the following we keep calling $\nu$ the parameter controlling the corrections. For  the class of quivers specified above, $\nu$ denotes the number of gauge groups. More generally, we impose \eqref{relation_k3_k1} and denote $\nu=-k^{(1)}_{R}=-k^{(1)}_{RRR}\,$.

\subsection{Legendre transform}

The Legendre transform consists of the extremization principle
\begin{equation}\label{eq:Legendretransform}
\mathcal{S}= {\rm{ext}}_{\left\{\varphi^I,\, \omega_{1}, \,\omega_{2},\, \Lambda\right\}}\left[-I-\omega_1 J_1-\omega_2 J_2-\varphi^I Q_I -\Lambda \left(\omega_1+\omega_2-2r_I \varphi^I- 2\pi i\right)\right]\, ,
\end{equation}
which gives the microcanonical form of the entropy (namely, the entropy as a function of the charges and angular momenta). This will be evaluated by extending to the present higher-derivative case the method already introduced in sections~\ref{sec:holographic_match} and \ref{constrained_transform}. 
The extremization equations are
\begin{equation}\label{eq:extremization_eqs}
-\frac{\partial I}{\partial \varphi^I}=Q_I-2 r_I \Lambda\,, \hspace{1cm} -\frac{\partial I}{\partial \omega_{1}}=J_{1}+\Lambda\,, \hspace{1cm} -\frac{\partial I}{\partial \omega_{2}}=J_{2}+\Lambda\,, 
\end{equation}
together with the linear constraint~\eqref{eq:linearconstraint} which follows from the variation with respect to the Lagrange multiplier $\Lambda$. For definiteness we have made the upper sign choice in \eqref{eq:linearconstraint}; making the other choice leads essentially to the same computations, in particular it gives the same reality condition for the entropy and the same final expression for it.
It is convenient for our purposes to rewrite the expression \eqref{eq:index_asympt} for $I$ using the linear constraint \eqref{eq:linearconstraint} in such a way that it reads:
\begin{equation}\label{eq:rewriting_index}
I= \frac{\left(k_{IJK}-k_Ir_Jr_K\right)\varphi^I \varphi^J \varphi^K} {6\, \omega_1 \omega_2}-\frac{k_I\varphi^I}{12}\left(\frac{\omega_1}{\omega_2}+\frac{\omega_2}{\omega_1}+1\right) \,+\frac{k_Ir_J\varphi^I\varphi^J}{6}\left(\frac{1}{\omega_1}+\frac{1}{\omega_2}\right) ,
\end{equation}
since now it is manifestly a homogeneous function of degree one with respect to $\varphi^I$, $\omega_1$, $\omega_2$. Euler's theorem then implies that the entropy is simply given by the extremum value of the Lagrange multiplier,
\begin{equation}\label{eq:Legendre}
\mathcal{S}=2\pi i \Lambda|_{\rm{ext}}\, .
\end{equation}

Next we use our assumption \eqref{relation_k3_k1}.  
 Remarkably, this implies that the ${\cal O}(1)$ corrections in the first term of \eqref{eq:rewriting_index} cancel.
Then the effective action becomes:
\begin{equation}\label{eq:index_quiver}
I= \frac{k^{(0)}_{IJK}\varphi^I \varphi^J \varphi^K} {6 \,\omega_1 \omega_2}-\frac{k_I\varphi^I}{12}\left(\frac{\omega_1}{\omega_2}+\frac{\omega_2}{\omega_1}+1\right) \,+\frac{k_Ir_J\varphi^I\varphi^J}{6}\left(\frac{1}{\omega_1}+\frac{1}{\omega_2}\right) \,.
\end{equation}

\paragraph{Leading contribution to the entropy.}
As a useful warm-up, we start by recalling how the Legendre transform  is implemented at leading-order in the large-$N$ expansion~\cite{Cassani:2019mms}. We then consider  \eqref{eq:index_asympt_leading}.
  Using our assumption \eqref{eq:cubic_relation} on the leading-order coefficients, it is not hard to see that it satisfies
\begin{equation}\label{eq:I}
k^{(0)}{}^{IJK}\frac{\partial I^{(0)} }{\partial \varphi^I}\frac{\partial I^{(0)} }{\partial \varphi^J}\frac{\partial I^{(0)} }{\partial \varphi^K}- 2\aa^{(0)}  \frac{\partial I^{(0)} }{\partial \omega_1}\frac{\partial I^{(0)} }{\partial \omega_2}=0\, .
\end{equation}
After using the extremization equations \eqref{eq:extremization_eqs}, this becomes a polynomial equation for $\Lambda$:
\begin{equation}\label{eq:Lambda}
\Lambda^3+p_2 \Lambda^2+p_1 \Lambda+p_0=0\, ,
\end{equation}
where 
\begin{equation}\label{eq:p_coeffs}
\begin{aligned}
p_2\,&=\, -12 k^{(0)}{}^{IJK}r_Ir_J Q_K- 2\aa^{(0)} \, ,\\[1mm]
p_1\,&=\, 6 k^{(0)}{}^{IJK}r_IQ_J Q_K-    2\aa^{(0)}  \left(J_1+J_2\right)\, ,\\[1mm]
p_0\,&=\, -k^{(0)}{}^{IJK}Q_IQ_J Q_K- 2\aa^{(0)} J_1J_2\, .
\end{aligned}
\end{equation}
From \eqref{eq:Legendre} we see that for the entropy to be real, $\Lambda$ must be a purely imaginary number. This implies a condition on the charges and angular momenta. In terms of the coefficients of the polynomial equation, such condition reads
\begin{equation}
 p_0 = p_1 p_2\,.
\end{equation}
Then \eqref{eq:Lambda} factorizes as
\begin{equation}
 \left(\Lambda^2+p_1\right)\left(\Lambda+ p_2\right)=0\,.
\end{equation}
Taking the purely imaginary root $\Lambda=-i \sqrt{p_1}$ (assuming $p_1>0$), we find that the supersymmetric extremal entropy is given by
\begin{equation}\label{eq:BPS_entropy_0thorder}
\mathcal{S}^{(0)}\,=\, 2\pi \sqrt{6 k^{(0)}{}^{IJK}r_IQ_J Q_K- 2\aa^{(0)}\left(J_1+J_2\right)}\, .
\end{equation}

\paragraph{First-order corrections.}
Now we perform the Legendre transform keeping the corrections in \eqref{eq:index_quiver}. We work at linear order in $k_I$. In this approximation, it is possible to check that the corrections to \eqref{eq:I} are the following,
\begin{equation}\label{eq:I_corrected}
\begin{aligned}
&\left(k^{(0)}{}^{IJK}\frac{\partial I}{\partial \varphi^I}\frac{\partial I}{\partial \varphi^J}\frac{\partial I}{\partial \varphi^K}- 2 \aa^{(0)} \frac{\partial I}{\partial \omega_1}\frac{\partial I}{\partial \omega_2}\right)\frac{\partial I}{\partial \omega_1}\frac{\partial I}{\partial \omega_2} + \frac{1}{4}k^{(0)}{}^{IJK}\frac{\partial I}{\partial \varphi^I}\frac{\partial I}{\partial \varphi^J}k_K\left[3\frac{\partial I}{\partial \omega_1}\frac{\partial I}{\partial \omega_2}+\right.\\[1mm]
&+\,\left(\frac{\partial I}{\partial \omega_1}-\frac{\partial I}{\partial \omega_2}\right)^2+\frac{3}{\aa^{(0)}   }k^{(0)}{}^{LMN}\frac{\partial I}{\partial \varphi^L}\frac{\partial I}{\partial \varphi^M}r_N \left(\frac{\partial I}{\partial \omega_1}+\frac{\partial I}{\partial \omega_2}\right)\bigg]=0\, .
\end{aligned}
\end{equation}
As before it boils down to a polynomial equation for $\Lambda$, now of order five:
\begin{equation}\label{eq:Lambda2}
\begin{aligned}
{\cal P}_{5}\left(\Lambda\right)\equiv&\left(\Lambda+J_1\right)\left(\Lambda+J_2\right)\left(\Lambda^3+p_2 \Lambda^2+p_1 \Lambda+p_0\right)+  \frac{1}{4} {\cal P}_2\left(\Lambda; k_I\right)\Big[3 \left(\Lambda+J_1\right)\left(\Lambda+J_2\right)\\[1mm]
& + \left(J_1-J_2\right)^2-\frac{3}{\aa^{(0)}}{\cal P}_2\left(\Lambda; r_I\right)\left(2\Lambda+J_1+J_2\right)\Big]=0\,,
\end{aligned}
\end{equation}
where $p_0, p_1, p_2$ are still given by \eqref{eq:p_coeffs} and we have introduced
\begin{equation}
{\cal P}_2\left(\Lambda; v_I\right)=k^{(0)}{}^{IJK}  v_I\left(2r_J\Lambda -Q_J\right)\left(2r_K\Lambda -Q_K\right)\,.
\end{equation}
From now on we specify the $k_I$ as in the second of \eqref{eq:exp_anomaly_coeff}. Though this is not really necessary in order to work out the Legendre transform, it makes the final expressions slightly simpler.  Then ${\cal P}_5\left(\Lambda\right)$ in \eqref{eq:Lambda2} becomes
\begin{equation}\label{eq:Lambda2}
\begin{aligned}
{\cal P}_{5}\left(\Lambda\right)\equiv&\left(\Lambda+J_1\right)\left(\Lambda+J_2\right)\left(\Lambda^3+p_2 \Lambda^2+p_1 \Lambda+p_0\right)- \frac{\nu}{4}\,{\cal P}_2\left(\Lambda; r_I\right)\left[3 \left(\Lambda+J_1\right)\left(\Lambda+J_2\right)\right.\\[1mm]
&\left.+ \left(J_1-J_2\right)^2-\frac{3}{\aa^{(0)}}{\cal P}_2\left(\Lambda; r_I\right)\left(2\Lambda+J_1+J_2\right)\right]=0\,.
\end{aligned}
\end{equation}
We also note that
\begin{equation}
{\cal P}_2\left(\Lambda; r_I\right)=\frac{1}{2}\Lambda^2+ \frac{1}{3}\left(p_2 + 2\aa^{(0)}\right)\Lambda+\frac{p_1}{6}+\frac{1}{3}\aa^{(0)} \left(J_1+J_2\right)\, .
\end{equation}
As before, in order to obtain a real entropy, we must impose the factorization of the polynomial, which in this case takes the form
\begin{equation}\label{eq:factorization}
{\cal P}_{5}\left(\Lambda\right)=\left(\Lambda^2+{\cal X}\right)\left(\Upsilon_0+\Upsilon_1 \Lambda +\Upsilon_2 \Lambda^2+\Upsilon_3\Lambda^3\right)\, ,
\end{equation}
where ${\cal X}, \Upsilon_{0}, \Upsilon_1, \Upsilon_2$ and $\Upsilon_3$ are just coefficients. This factorization translates into a condition on the coefficients of ${\cal P}_{5}\left(\Lambda\right)$, and eventually on the charges $Q_I$ and angular momenta $J_1$,$J_2$.

\paragraph{Solution in the $J_1=J_2$ case.} Let us illustrate in detail the case where there is only one independent angular momentum, $J_1=J_2\equiv J$, as this will be the case for which we will actually calculate the on-shell action on the gravity side. We note that in this case the polynomial \eqref{eq:Lambda} factorizes as ${\cal P}_5(\Lambda)=\left(\Lambda+J\right){\cal P}_{4}(\Lambda)$, where
\begin{equation}
{\cal P}_4\left(\Lambda\right)=\left(\Lambda+J\right)\left(\Lambda^3+p_2 \Lambda^2+p_1 \Lambda+p_0\right)-\nu\, {\cal P}_2\left(\Lambda; r_I\right)\left[\frac{3}{4}\left(\Lambda+J\right)-\frac{3}{2\aa^{(0)}}{\cal P}_2\left(\Lambda; r_J\right)\right]\,.
\end{equation}
So the factorization condition can be written as
\begin{equation}
{\cal P}_{4}\left(\Lambda\right)=\left(\Lambda^2+{\cal X}\right)\left(\Upsilon_0+\Upsilon_1 \Lambda +\Upsilon_2 \Lambda^2\right)\, .
\end{equation}
Comparing the last two expressions and working perturbatively, one finds the solution 
\be
\begin{aligned}
\Upsilon_2\,&=\, 1+\frac{3\nu }{8 \aa^{(0)}}\,,\\[1mm]
\Upsilon_1\, &=\, p_2+ J  +\frac{\nu}{2} \left(\frac{5}{4}+\frac{p_2}{\aa^{(0)}}\right)\,,\\[1mm]
\Upsilon_0 \, &=\, J p_2+   \frac{\nu}{8}   \left[ 5 J +\frac{p_1^2+4 J^2 p_2^2+J p_1\left(8p_2-3J  \right)}{3 \aa^{(0)} \left(p_1+J^2  \right)}\right]\,,\\[1mm]
\label{eq:X_J1=J2}
{\cal X}\,&=\, p_1 + \frac{\nu}{6} \left[\aa^{(0)} + \frac{5}{2} p_2 -\frac{ p_1\left(p_1 -p_2^2+2 J p_2 \right)}{ \aa^{(0)}   \left(p_1+J^2  \right)}\right]\,,
\end{aligned}
\ee
and the factorization condition leading to the non-linear constraint among the charges reads
\begin{equation}\label{eq:nonlinearconstraint_J1=J2}
p_0 -p_1p_2\,=\,\frac{\nu}{6} \left[\frac{5}{2}\left(p_1 +p_2^2\right)+ \aa^{(0)}\left(p_2-J  \right)+\frac{ p_1\left(p_2-J  \right)\left(p_1 + p_2^2  \right)}{\aa^{(0)}  \left(p_1+J^2  \right)}\right]\,.
\end{equation}
From the expression for ${\cal X}$ we immediately obtain the entropy:
\be\label{eq:entropy_equalJ}
\mathcal{S}\,=\, 2\pi \sqrt{p_1 + \frac{\nu}{12} \left[2 \aa^{(0)}+ 5 p_2 -\frac{2 p_1\left(p_1 -p_2^2+2 J p_2 \right)}{\aa^{(0)}   \left(p_1+J^2  \right)}\right]}\,.
\ee
We emphasize that this expression can only be trusted at linear order in the correction, even if we have not explicitly linearized the square root (the reason for not doing so being that the way it is derived suggests that the form $\mathcal{S} = 2\pi i\Lambda = 2\pi \sqrt{{\cal X}}$ of the entropy may hold beyond linear order).

\paragraph{General solution $J_1\neq J_2$.} In the case of two unequal angular momenta, the expression for the entropy receives additional corrections and reads
\begin{equation}\label{eq:entropy_J1neqJ2}
\begin{aligned}
\mathcal{S}&=\,2\pi\sqrt{p_1}\left\{1+\frac{\nu}{24}\left[\frac{\left(p_{1}+J_{+}^2\right)\left(\aa^{(0)} \left(p_{1}+J_{+}^2\right)\left(2\aa^{(0)}+5 p_2\right)-2p_1\left(p_1-p_2^2+2 J_+ p_2\right)\right)}{{\aa^{(0)}}p_1\left[p_1+\left(J_++J_-\right)^2\right]\left[p_1+\left(J_+-J_-\right)^2\right]}\right.\right.\\[1mm]
&\left.\left.+\,\frac{J_-^2\left[\left(p_2+2\aa^{(0)}\right)\left(2p_1\left(p_2+\aa^{(0)}+2J_+\right)+\aa^{(0)} J_-^2\right)-2\aa^{(0)} J_+^2\left(3p_2+ 2\aa^{(0)}\right)-2p_1^2\right]}{{\aa^{(0)}}p_1\left[p_1+\left(J_++J_-\right)^2\right]\left[p_1+\left(J_+-J_-\right)^2\right]}\right]\right\},
\end{aligned}
\end{equation}
where $J_{\pm}=\frac{J_1\pm J_2}{2}$. In turn, the non-linear constraint reads
\begin{equation}\label{eq:nonlinearconstraint_J1neqJ2}
\begin{aligned}
p_0-p_1p_2=\,&\frac{\nu}{6}\left\{\aa^{(0)}\left(p_2-J_+\right)+\frac{\left(p_1+p_2^2\right)\left(p_1+J_+^2\right)\left[5 \aa^{(0)}\left(p_1+J_+^2\right)+2p_1\left(p_2-J_+\right)\right]}{2\, \aa^{(0)} \left[p_1+\left(J_++J_-\right)^2\right]\left[p_1+\left(J_+-J_-\right)^2\right]}\right.\\[1mm]
&\quad\left.+\,\frac{\left(p_1+p_2^2\right)J_-^2\left[ \aa^{(0)} \left(6p_1-6J_+^2+J_-^2\right)+2p_1(p_2+J_+)\right]}{2\, \aa^{(0)} \left[p_1+\left(J_++J_-\right)^2\right]\left[p_1+\left(J_+-J_-\right)^2\right]}\right\}\,.
\end{aligned}
\end{equation}

\paragraph{Recovering the universal case.} 

As a sanity check, we verify that the results above are in agreement with those obtained in \cite{Cassani:2022lrk,Bobev:2022bjm} for the universal case where all flavour charges are turned off and one is left only with the R-charge $Q_R$ and  the angular momenta $J_1,J_2$. 
This case is reached by setting $Q_I = r_I Q_R$ for all $I$'s, see footnote~\ref{foot:change_basis}.
Then, recalling \eqref{eq:normalization_k},
the coefficients \eqref{eq:p_coeffs} reduce to
\begin{equation}\label{eq:coeffs_minimalcase}
\begin{aligned}
p_2\,&=\, -\frac{3}{2}Q_R-   2\aa^{(0)} \,, \\[1mm]
p_1\,&=\, \frac{3}{4}Q_R^2 - 2\aa^{(0)} \left(J_1+J_2\right)\,, \\[1mm]
p_0\,&=\, -\frac{1}{8}Q_R^3-  2\aa^{(0)} J_1 J_2\,. \\[1mm]
\end{aligned}
\end{equation}
Plugging these into \eqref{eq:entropy_J1neqJ2} and using the non-linear constraint \eqref{eq:nonlinearconstraint_J1neqJ2}, one can verify that the expression for the entropy reduces to
\begin{equation}
\begin{aligned}
\mathcal{S}\,
&=\, \pi \sqrt{3Q_R^2-8\aa \left(J_1+J_2\right)-16\aa^{(0)} \left(\aa-\cc\right)\frac{\left(J_1-J_2\right)^2}{Q_R^2-2\aa^{(0)}\left(J_1+J_2\right)}}\,,
\end{aligned}
\end{equation}
where we have used \eqref{Weyl_coeffs_gen_quivers} to introduce the corrected Weyl anomaly coefficients.
This agrees with our previous results \eqref{eq:entropy}. The non-linear constraint \eqref{eq:nonlinearconstraint_J1=J2} also reduces correctly to \eqref{eq:universal_constraint}.

\section{Application to $\mathcal{N}=1$ orbifold theories}\label{sec:orbifold_section}

Given the general field theory results discussed in the previous section, we would now like to write down explicit expressions in some concrete examples, to be then studied on the gravity side. 
A limitation with matching these results with a dual gravitational computation is represented by the restricted number of known asymptotically AdS$_5$ multi-charge black hole solutions uplifting to ten- or eleven-dimensional supergravity. Indeed, besides the universal case that involves just the R-charge and applies to any compactification admitting a supersymmetric AdS$_5\times M$ solution,
the only known such solution has three independent electric charges and uplifts to type IIB supergravity on $S^5$ or the $S^5/\Gamma$ orbifold, dual to ${\rm SU}(N)$ $\mathcal{N}=4$ SYM or the $\mathbb{C}^3/\Gamma$ orbifold theories. In the case of $\mathcal{N}=4$ SYM, the Cardy asymptotics of the index involve extra simplifications due to the underlying maximal supersymmetry,  specifically one has the exact expressions $k_{IJK} =\frac{N^2-1}{2}|\epsilon_{IJK}|$ and $k_I=0$. The orbifold theories, instead, have a more interesting set of corrections, which as such offer a 
more ``realistic'' view on the corrections of generic $\mathcal{N}=1$ SCFT's. This provides our main motivation for considering such theories here. We start by briefly recalling the features that will be relevant for us and obtain the anomaly coefficients.

\subsection{Anomaly coefficients}

The structure of the orbifold theories describing the low-energy limit of a stack of D3-branes probing a $\mathbb{C}^3/\Gamma$ singularity has been discussed long ago~\cite{Douglas:1996sw,Douglas:1997de,Kachru:1998ys,Lawrence:1998ja}.\footnote{The superconformal index of orbifold theories has been studied in~\cite{Nakayama:2005mf,Arai:2019wgv}.}
 In order to preserve $\mathcal{N}=1$ supersymmetry, we require  the finite group $\Gamma$ to be a subgroup of $ {\rm SU}(3)\subset {\rm SU}(4)$, with ${\rm SU}(4)\simeq{\rm SO}(6)$ corresponding to rotations in the $\mathbb{R}^6\simeq\mathbb{C}^3$ space transverse to the branes.
 
\begin{figure}
	\centering
	\includegraphics[width=0.7\textwidth]{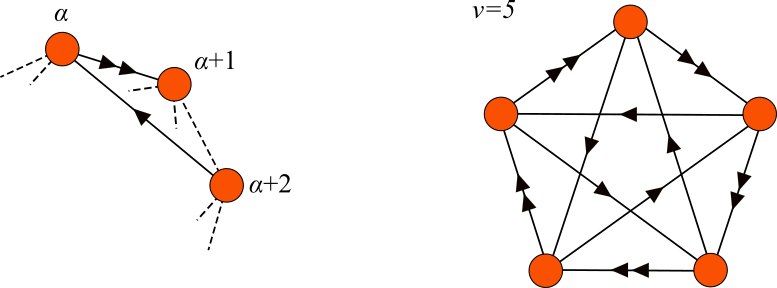}
	\caption{\it The quivers describing the low-energy limit of D3-branes probing a $\mathbb{C}^3/\mathbb{Z}_\nu$ singularity.  Left: a generic ${\rm SU}(N)$ node $\alpha$ is connected to the node $(\alpha+1)$ by a doublet of outgoing arrows  
	and to the node $(\alpha+2)$ by an incoming arrow 
	 (an arrow pointing from node $\alpha$ to node $\beta$ denotes a chiral superfield in the $({\bf N}, {\bf \bar N})$ bifundamental representation of ${\rm SU}(N)_{\alpha}\times {\rm SU}(N)_{\beta}$). Right: the case $\nu = 5$.}
	\label{fig:quivers}
\end{figure}

  We will focus on the $\Gamma=\mathbb{Z}_\nu$ orbifolds whose action on the $(z_1,z_2,z_3)$ coordinates of $\mathbb{C}^3$ is generated by the element 
  \be
 \Theta ={\rm diag} (\rme^{\frac{2\pi i}{\nu}},\, \rme^{\frac{2\pi i}{\nu}},\,\rme^{-\frac{4\pi i}{\nu}})\,.
  \ee 
 We take $\nu\geq 3$ so that the quotient preserves exactly $\mathcal{N}=1$ supersymmetry. For $\nu=2$ one has $\Gamma \subset {\rm SU}(2)$, so the quotient preserves $\mathcal{N}=2$ supersymmetry and, when described in $\mathcal{N}=1$ language, involves chiral superfields in the adjoint representation (from the decomposition of the $\mathcal{N}=2$ vector multiplets into $\mathcal{N}=1$  multiplets), a case falling out of our assumptions in section~\ref{sec:assumptions} (though it would not be hard to study it separately).\footnote{We could also consider other $\mathbb{C}^3/\Gamma$ orbifolds with $\Gamma\subset{\rm SU}(3)$, including the more general Abelian case $\Gamma=\mathbb{Z}_{\nu_1}\times \mathbb{Z}_{\nu_2}$  
  as well as non-Abelian cases, see e.g.~\cite{Hanany:1998sd}. 
  We expect that the study of these more complicated examples does not involve qualitatively new features.
}
Note that $\Gamma$ commutes with the ${\rm SU}(2)$ acting on $(z_1,z_2)$, so a ${\rm U}(1)$ global symmetry enhances to ${\rm SU}(2)$.
The resulting theories 
 are quiver gauge theories of the type discussed in the previous section, namely they contain $\nu$ ${\rm SU}(N)$ nodes, connected by bifundamental chiral superfields. Each node $\alpha$ is connected to the node $(\alpha+1)$ by a doublet of chiral fields transforming in the $({\bf N}, {\bf \bar N})$ bifundamental representation of ${\rm SU}(N)_\alpha\times {\rm SU}(N)_{\alpha+1}$ and to the node $(\alpha+2)$ by a chiral field transforming in the  $({\bf \bar N}, {\bf N})$ representation of ${\rm SU}(N)_{\alpha}\times {\rm SU}(N)_{\alpha+2}$. In figure~\ref{fig:quivers} we show the generic structure at a node (to be repeated for all nodes) and the quiver for $\Gamma = \mathbb{Z}_5$ as an example.

For odd $\nu$ the orbifold action only has the origin of $\mathbb{C}^3$ as its fixed point, hence the base space $S^5/\mathbb{Z}_\nu$ is smooth and the low-energy spectrum of Type IIB string theory on this space is simply given by the orbifold projection of the supergravity modes on $S^5$. On the other hand, for even $\nu$ there is a $\mathbb{Z}_2\subset\mathbb{Z}_\nu$ subgroup generated by the element $\Theta^{\nu/2}$ which leaves the complex line in $\mathbb{C}^3$ parameterized by $z_3$ invariant. This translates in an invariant circle in $S^5$, implying that the resulting orbifold space $S^5/\mathbb{Z}_\nu$ is singular and leads to a light twisted sector for the string modes localized at the invariant circle. 

The Abelian global symmetries are the R-symmetry and two ${\rm U}(1)$ flavour symmetries, whose generators span the Cartan subalgebra of ${\rm SO}(6)$. When $\nu$ is even there is also a non-anomalous ${\rm U}(1)$ baryonic symmetry;  on the gravity side this  acts in the twisted sector and is not visible at the level of Type IIB supergravity. Because of this we will switch off the baryonic charge. 

We choose a basis where the global charges $Q_I$ are all R-charges with $r_I=\frac{1}{2}$, $I=1,2,3$  (since this is the basis that is naturally obtained when reducing Type IIB supergravity on $S^5$). It follows that the fermion $\psi$ in a chiral multiplet with charge $q_I$ has charge $q_I - \frac{1}{2}$, while the gaugino has charge $+\frac{1}{2}$ under all $Q_I$'s. The charge assignement for the fermions in the theory are given in table~\ref{table:charges_fermions}.
\begin{table}
\centering
\begin{tabular}{|c|c|c|c|c|c|}
 \hline
 Field & multiplicity & $Q_1$ & $Q_2$ & $Q_3$ \\
 \hline
$\phantom{\Big[} \!\!\!\psi_1$  & $ \nu N^2$  & $ \frac{1}{2}$ &  $-\frac{1}{2}$  &  $-\frac{1}{2}$    \\[2mm]
$\psi_2$ & $ \nu N^2$  &  $-\frac{1}{2}$  &  $\frac{1}{2}$  &  $-\frac{1}{2}$   \\[2mm]   
$\psi_3$   &  $\nu N^2$  & $-\frac{1}{2}$  &  $-\frac{1}{2}$  &  $\frac{1}{2}$   \\[2mm]  
gaugini   &  $\nu (N^2-1)$ &  $\frac{1}{2}$  & $\frac{1}{2}$  & $\frac{1}{2}$   \\[1mm]  
 \hline
\end{tabular}
\caption{Multiplicities and charge assignments for the fermion fields in the $\mathbb{C}^3/\mathbb{Z}_\nu$ quiver theories. $\psi_1,\psi_2,\psi_3$ are the fermion fields belonging to the $\nu$ triplets of bifundamental chiral multiplets.}
\label{table:charges_fermions}
\end{table}
 
Evaluating the 't Hooft anomaly coefficients using their definition \eqref{def_anomaly_coeff} one finds:
\be
\begin{aligned}\label{tHooft_coeff_orbifolds}
k_{IJK}   \, &=\, k^{(0)}_{IJK } + k^{(1)}_{IJK}\,,\quad\ \text{with}\quad\ 
k^{(0)}_{IJK }  \, =\, \frac{\nu N^2}{2} |\epsilon_{IJK}|\,,\quad\ k^{(1)}_{IJK} = -\frac{\nu}{8}\,,\\[1mm]
k_I  \,&=\, - \frac{\nu}{2}\,.
\end{aligned}
\ee
 These satisfy relation \eqref{relation_k3_k1} since $r_I=\frac{1}{2}$ for all charges.
The superconformal R-symmetry is given by the exact relation
\be
\mathcal{R}\,=\, \frac{2}{3}\left(Q_1+Q_2+Q_3\right)\,.
\ee
It follows from~\eqref{relacTrR} that the (exact) 
 Weyl anomaly coefficients read
\begin{equation}\label{a_c_orbifolds}
{\mathtt a}\,=\,\frac{\nu N^2}{4}-\frac{3\nu}{16}\,,\qquad\quad {\mathtt c}\,=\,\frac{\nu N^2}{4}-\frac{\nu}{8}\, .
\end{equation}
The value of $\cc-\aa= \frac{\nu}{16}$ has been matched with a supergravity (string theory) computation  in~\cite{ArabiArdehali:2013jiu}.

\subsection{Corrected entropy}\label{subsec:entropy_orbifolds}

 For the  tensor $k^{(0)}{}^{IJK}$ we take
\begin{equation}\label{eq:specifyingkIJK}
 k^{(0)}{}^{IJK}=\frac{1}{6}|\epsilon^{IJK}|\,,
\end{equation}
where $\epsilon_{IJK}$ is the Levi-Civita symbol and $\epsilon^{IJK}=\delta^{II'}\delta^{JJ'}\delta^{KK'}\epsilon_{I'J'K'}$. Note that $ k^{(0)}{}^{IJK}$ satisfies the normalization condition~\eqref{eq:normalization_k} as well as the cubic relation \eqref{eq:cubic_relation}, with $\gamma = \frac{4}{9}\,{\mathtt a}^{(0)} = \frac{\nu N^2}{9}$.
 Since all assumptions are satisfied, we can use the general results of section~\ref{sec:Legendre_transf_gen}.
The coefficients \eqref{eq:p_coeffs} read
\be
\begin{aligned}
p_2 &= -(Q_1+Q_2+Q_3)-2 \aa^{(0)} \,,\\[1mm]
p_1 &= Q_1Q_2+Q_2Q_3+Q_3Q_1 - 2 \aa^{(0)}(J_1+J_2) \,,\\[1mm]
p_0 &= -Q_1Q_2Q_3 -2 \aa^{(0)}J_1J_2\,.
\end{aligned}
\ee
Then from \eqref{eq:BPS_entropy_0thorder} one finds that the expression for the leading-order entropy reads:
\begin{equation}\label{2der_S_explicit_model}
\mathcal{S}^{(0)}\,=\,  2\pi \sqrt{Q_1 Q_2+Q_2 Q_3 +Q_1 Q_3-  2 \, {\mathtt a}^{(0)}\left(J_1+J_2\right)}\, .
\end{equation}
Apart from the multiplicative factor of $\nu$ hidden in the anomaly coefficient ${\mathtt a}^{(0)}$, this expression is the same as the one that is obtained for ${\rm SU}(N)$ $\mathcal{N}=4$ SYM. 
However, when we include the corrections things become more interesting: while for $\mathcal{N}=4$ SYM the replacement $\aa^{(0)}= \frac{N^2}{4}\to \aa = \frac{N^2-1}{4}$ in \eqref{2der_S_explicit_model} accounts for all $1/N^2$ corrections to the black hole entropy (in the Cardy limit),
 for the orbifold theories we obtain a more complicated expression. We provide the explicit form of the entropy for the slightly simpler case of $J_1=J_2\equiv J$:
\begin{equation}\label{eq:entropy_orbifolds_eqJ}
\mathcal{S}= 2\pi \sqrt{Q_1 Q_2+Q_2 Q_3 +Q_1 Q_3- 4 \, {\mathtt a}J+\frac{2({\mathtt c}-{\mathtt a})}{3{\mathtt a}}\frac{  \mathcal{U}(1,2,3) + \mathcal{U}(2,3,1)+ \mathcal{U}(3,1,2)}{Q_1 Q_2+Q_2 Q_3 +Q_1 Q_3-4 \, {\mathtt a}J+J^2}}\,,
\end{equation}
where
\be
\mathcal{U}(1,2,3) \,=\, \left[ Q_1 Q_2 -J(Q_3+2\aa) \right](Q_1-Q_2)^2 \,,
\ee
and $\aa$, $\cc$ have been given in eq.~\eqref{a_c_orbifolds}. It is understood that the result is only valid at first order in the $1/N^2$ corrections. 
 The constraint \eqref{eq:nonlinearconstraint_J1=J2} can be written as
\be
\begin{aligned}
&\left[Q_1+Q_2+Q_3 +2(2\aa-\cc)\right] (Q_1Q_2+Q_2Q_3+Q_3Q_1-4\cc J) - Q_1Q_2Q_3 - 2(3\cc-2\aa)J^2\\[1mm]
&+ \frac{2(\cc-\aa)}{3\aa}\,\frac{ \mathcal{T}(1,2,3) + \mathcal{T}(2,3,1)  + \mathcal{T}(3,1,2) }{{Q_1 Q_2+Q_2 Q_3 +Q_1 Q_3-4 {\mathtt a}J+ J^2}}  \,=\,0\,,
\end{aligned}
\ee
with
\be\label{eq:Tfunction_eqJ}
\mathcal{T}(1,2,3)  \,=\, \left[(3 Q_1 + 3 Q_2 - 2 Q_3 - 2 J) Q_3 - 6 \aa J \right] (Q_1+Q_2+2\aa) (Q_1-Q_2)^2.
\ee

\section{Four-derivative ${\cal N}=2$ U(1)$_{R}$-gauged supergravity in five dimensions}\label{sec:fourder_action}

The details of the field theory enter in the formula \eqref{eq:index_asympt} only through the anomaly coefficients. This suggests that in order to reproduce such formula via a holographic computation it is enough to consider a matter-coupled five-dimensional supergravity which reproduces the anomalies. The goal of this section is to construct such theory. More specifically, given the applications that we have in mind, we shall consider a four-derivative extension of ${\cal N}=2$ five-dimensional gauged supergravity coupled to an arbitrary number $n$ of abelian vector multiplets, with gauge group consisting of a U(1)$_{R}$ subgroup of the SU(2) R-symmetry group and $n$ additional U(1) isometries of the scalar manifold. We shall not consider the coupling to hyper- or tensor multiplets.

Our starting point in this section will be an off-shell formulation of ${\cal N}=2$ supergravity that includes the relevant supersymmetric four-derivative invariants. Eventually we will integrate out the auxiliary fields (working at linear order in the corrections) to obtain a four-derivative effective action for the propagating degrees of freedom, further exploiting the possibility of performing perturbative field redefinitions to reduce the number of independent terms in the action. The procedure mimics the one we carefully explained in \ref{sec:from_superconformal_to_supergravity} in the case of minimal supergravity.

The plan of this section is the following. We start in section~\ref{sec:2der_action} introducing the basics of off-shell five-dimensional ${\cal N}=2$ supergravity; in particular reviewing how the on-shell theory is recovered once the auxiliary fields have been integrated out. Then in section~\ref{sec:4der_action} we repeat the same process including the relevant four-derivative off-shell invariants, treating them as a perturbation. We conclude in section~\ref{sec:final_Lagr} summarizing the final form of the Lagrangian. The reader can safely skip the first two parts if not interested in the derivation of the results.

\subsection{Two-derivative ${\cal N}=2$ gauged supergravity in five dimensions}\label{sec:2der_action}

${\cal N}=2$, $D=5$ off-shell Poincar\'e supergravity can be obtained from superconformal methods after fixing the redundant gauge symmetries. 
Here, we follow the procedure outlined in sec.~\ref{sec:from_superconformal_to_supergravity} based on the standard Weyl formulation and using as compensators a vector and a linear multiplet. After fixing the gauge redundancies, one gets an off-shell supergravity theory whose two-derivative bosonic Lagrangian is given by~\eqref{eq:off_shell_poinc_general}.
Here, the totally symmetric constant tensor $C_{IJK}$ will specify the \emph{very special geometry} of the scalar manifold. The gauging parameters $g_{I}$ select the linear combination of the vector fields $A^I_{\mu}$ that gauges the U(1) R-symmetry. 

Let us now integrate out the auxiliary fields in order to obtain a Lagrangian for the propagating degrees of freedom. This amounts to solving their equations of motion and plugging the solution back into~\eqref{eq:off_shell_poinc_general}. The solution to the equations of motion of the auxiliary fields $P_{\mu}, V^{ij}_{\mu}, T_{\mu\nu}, N$ and $Y^I_{ij}$ is
\begin{equation}\label{eq:solauxfields}
\begin{aligned}
P_{\mu}=\,&0\,, \hspace{5mm} {V'}_{\mu}^{ij}=\,0\,,  \hspace{5mm} V_{\mu}=\,-\frac{3}{\sqrt{2}}g_{I}A^{I}_{\mu}\, , \hspace{5mm} T_{\mu\nu}=\,\frac{3}{16}\,{X}_I F^{I}_{\mu\nu}\,,\\[1mm]
N=\,& -\frac{3}{2}\,g_{I}X^I\,, \hspace{5mm} Y^{I}_{ij}=\, -\frac{3}{\sqrt{2}}\,{\cal C}^{IJ}g_{J}\,\delta_{ij}\,,
\end{aligned}
\end{equation}
where ${\cal C}^{IJ}$ denotes the inverse of ${\cal C}_{IJ}$. In addition to this, the auxiliary field $D$ (which plays the role of a Lagrange multiplier) imposes a constraint on the scalars $X^I$,
\begin{equation}\label{eq:cubicconstraint}
{\cal C} = C_{IJK}\, X^I X^J X^K=1\,,
\end{equation}
which implies that there are only $n$ independent scalars $\phi^x$, $x=1, \dots, n$. We can thus regard the $X^I$ as functions of the physical scalars, $X^I=X^I(\phi^x)$.  The expression for $D$, which will be needed when studying the higher-derivative theory, is found from the following combination of the equations of motion of the scalars,
\begin{equation}\label{eq:D}
X^I \frac{\delta \mathcal L_{2\partial}^{\rm off-shell}}{\delta X^I}=0\,.
\end{equation}
Making use of some of the expressions in \eqref{eq:solauxfields}, one finds
\begin{equation}\label{eq:D}
D = -\frac{1}{32}\left[R +\left(\frac{1}{3}\,\mathcal C_{IJ}- \frac{9}{4}X_I X_J\right) F_{\mu\nu}^I F^J{}^{\mu\nu} +2 \mathcal C_{IJ} \partial_\mu X^I \partial^\mu X^J-12\mathcal C^{IJ} g_I g_J +12 (g_I X^I)^2
\right]\,.
\end{equation}
 Finally, we substitute the expressions \eqref{eq:solauxfields} into~\eqref{eq:off_shell_poinc_general} to recover the well-known bosonic supergravity Lagrangian for the propagating degrees of freedom \cite{Gunaydin:1984ak} (see also \cite{Gunaydin:1999zx, Ceresole:2000jd, Bergshoeff:2004kh}):\footnote{The complete theory can be found in  \cite{Bergshoeff:2004kh}. The dictionary between the fields and couplings here and in that reference is the following:
$$C_{IJK}^{\rm here} = \mathcal{C}_{IJK}^{\rm there}\,,\qquad X^I_{\rm here} = h^I_{\rm there}\,,\qquad a^{\rm here}_{IJ} = a^{\rm there}_{IJ}\,, \qquad
 A^I_{\rm here} = \sqrt{\tfrac{2}{3}} A^I_{\rm there}\,,\qquad   \tfrac{\sqrt3}{2} g_I^{\rm here} = (g\xi_I)^{\rm there} \,.
$$}
\begin{equation}\label{eq:2dSUGRA}
{\cal L}_{2\partial}=R-2{\cal V}-\frac{3}{2}\, a_{IJ}\, \partial_{\mu} X^I \partial^{\mu} X^J-\frac{3}{4}\, a_{IJ} F^I_{\mu\nu} F^{J}{}^{\mu\nu}
+\frac{1}{4}C_{IJK}\epsilon^{\mu\nu\rho\sigma\lambda}F^{I}_{\mu\nu} F^J_{\rho\sigma} A^{K}_{\lambda} \, ,
\end{equation}
where we have defined
\begin{equation}
a_{IJ}=3\,{X}_I \,{X}_{J}-\frac{1}{3}\,{\cal C}_{IJ}\,,
\end{equation}
and where the scalar potential ${\cal V}$ is given by
\begin{equation}
{\cal V}=\, -\frac{9}{4}\, (g_I X^I)^2-\frac{9}{2}\,{\cal C}^{IJ}g_I g_J\, .
\end{equation}
Defining the metric of the scalar manifold as
\begin{equation}\label{eq:scmetric}
{\mathfrak g}_{xy}=\frac{3}{2}\, a_{IJ}\partial_x X^I \partial_yX^J\, ,
\end{equation}
we can rewrite the two-derivative Lagrangian as
\begin{equation}\label{eq:action2dSUGRA} 
{\cal L}_{2\partial}=R-2{\cal V}-{\mathfrak g}_{xy}\, \partial_{\mu} \phi^x  \partial^{\mu} \phi^y-\frac{3}{4}\, a_{IJ}\,F^I_{\mu\nu} F^J{}^{\mu\nu}+\frac{1}{4}\,C_{IJK}\epsilon^{\mu\nu\rho\sigma\lambda}F^{I}_{\mu\nu} F^{J}_{\rho\sigma} A^{K}_{\lambda} \, .
\end{equation}
The tensors which are set to zero by the two-derivative equations of motion are:
\begin{eqnarray}\label{eq:2dEOMEinst}
{\cal E}_{\mu\nu}&=&R_{\mu\nu}-\frac{1}{2}\,g_{\mu\nu}\, \left(R-2{\cal V}\right) -T^{\rm{vectors}}_{\mu\nu}-T^{\rm{scalars}}_{\mu\nu}\,,\\[1mm]
\label{eq:2dEOMvec}
{\cal E}^{\mu}_{I}&=&\nabla_{\nu}\left(3\,a_{IJ}{F^J}^{\nu\mu}\right)+\frac{3}{4}\,C_{IJK}\epsilon^{\mu\nu\rho\sigma\lambda}F^{J}_{\nu\rho}F^{K}_{\sigma\lambda}\,, \\[1mm]
\label{eq:2dEOMsc}
{\cal E}_{x}&=&\nabla_{\mu}\left(2\,{\mathfrak g}_{xy}\partial^{\mu}\phi^y\right)-2\,\partial_{x}{\cal V} -\partial_{x}{\mathfrak g}_{yz} \partial_{\mu} \phi^y \partial^{\mu}\phi^z-\frac{3}{4}\,\partial_{x}a_{IJ} F^I_{\mu\nu}F^{J}{}^{\mu\nu}\, ,
\end{eqnarray}
where 
\begin{equation}
\begin{aligned}
T^{\rm{vectors}}_{\mu\nu}=\,&\frac{3}{2}\,a_{IJ}\left(F^{I}_{\mu\rho}\,{F^J}_{\nu}{}^{\rho}-\frac{1}{4}g_{\mu\nu}\,F^I_{\rho\sigma} F^J{}^{\rho\sigma}\right)\,,\\[1mm]
T^{\rm{scalars}}_{\mu\nu}=\,&\frac{3}{2}\, a_{IJ}\,\left(\partial_\mu X^I \partial_\nu X^J-\frac{1}{2}g_{\mu\nu}\,\partial_{\rho} X^I \partial^{\rho} X^J\right)\, .
\end{aligned}
\end{equation}

\subsection{Four-derivative corrections}
\label{sec:4der_action}

Our goal now is to obtain a four-derivative extension of ${\cal N}=2$, $D=5$ U(1)$_{R}$-gauged supergravity coupled to an arbitrary number of vector multiplets. To this aim, we modify the procedure followed in the two-derivative case adding the relevant four-derivative supersymmetric invariants. The off-shell Lagrangian will then contain two pieces,
\begin{equation}\label{eq:offshellLagrangian4der}
{\cal L}^{\rm{off-shell}}={\cal L}^{\rm{off-shell}}_{2\partial}+\alpha \, {\mathcal L}^{\rm{off-shell}}_{4\partial}\,,
\end{equation}
Before specifying the form of ${\mathcal L}^{\rm{off-shell}}_{4\partial}$, let us explain the general procedure we are going to follow to integrate out the auxiliary fields at linear order in $\alpha$. 

Let us denote by $\Phi_{\rm{aux}}$ all the auxiliary fields except for the combination of the scalars $X^I$ that is not dynamical, which is treated separately for the sake of clarity. The solution to the corrected equations of motion for the auxiliary fields derived from \eqref{eq:offshellLagrangian4der} is in general of the form,
\begin{equation}\label{eq:Psi}
\Phi_{\rm{aux}}\left(\Psi\right)=\Phi^{(0)}_{\rm{aux}}\left(\Psi\right)+\alpha \,\Phi^{(1)}_{\rm{aux}}\left(\Psi\right)\,,
\end{equation}
where $\Psi\equiv \{g_{\mu\nu}, A^I, \phi^x\}$ denotes the dynamical fields. If  ${\mathcal L}^{\rm{off-shell}}_{4\partial}$ depends on $D$, the cubic constraint of the very special geometry \eqref{eq:cubicconstraint} will receive corrections. Let us assume a generic modification
\begin{equation}
\left(C_{IJK}+\alpha\, C^{(1)}_{IJK}\right){X}^I {X}^J {X}^K=1\, ,
\end{equation}
where $C^{(1)}_{IJK}$ is an arbitrary symmetric tensor. Denoting by $X^{(0)}{}^{I}$ the scalars satisfying the original constraint $C_{IJK}X^{(0)}{}^{I}X^{(0)}{}^{J}X^{(0)}{}^{K}=1$, we have that
\begin{equation}
X^{I}=X^{(0)}{}^{I}+\alpha \,X^{(1)}{}^I\, , 
\end{equation}
with $X^{(1)}{}^I$ obeying the constraint
\begin{equation}
C_{IJK} X^{(0)}{}^{I}X^{(0)}{}^{J}X^{(1)}{}^{K}=-\frac{1}{3}\,C^{(1)}_{IJK}X^{(0)}{}^{I}X^{(0)}{}^{J}X^{(0)}{}^{K}\equiv -\frac{1}{3}\, {\cal C}^{(1)}\, ,
\end{equation}
whose solution is 
\begin{equation}\label{eq:sol_cc}
X^{(1)}{}^{I}=-\frac{1}{3}\, {\cal C}^{(1)} \,X^{(0)}{}^{I}\, , \hspace{1cm} X^I=X^{(0)}{}^{I}\left(1-\frac{\alpha}{3}\, {\cal C}^{(1)}\right)\, .
\end{equation}
Substituting the expressions for the auxiliary fields \eqref{eq:Psi} and the solution to the cubic constraint \eqref{eq:sol_cc} into the two-derivative off-shell Lagrangian, we get 
\begin{equation}
\begin{aligned}
{\cal L}^{\rm{off-shell}}_{2\partial}\Big|_{\{X^I(\Psi), \Phi_{\rm{aux}}(\Psi)\}}=\,&{\cal L}^{\rm{off-shell}}_{2\partial}\Big|_{(0)}+\alpha\, \Phi^{(1)}_{\rm{aux}}\frac{\delta{\cal L}^{\rm{off-shell}}_{2\partial}}{\delta \Phi_{\rm{aux}}}\Bigg|_{(0)}-\frac{\alpha}{3}\,{\cal C}^{(1)}\, X^{(0)}{}^{I}\, \frac{\delta{\cal L}^{\rm{off-shell}}_{2\partial}}{\delta X^I}\Bigg|_{(0)}\\[1mm]
=\,&{\cal L}^{\rm{off-shell}}_{2\partial}\Big|_{(0)}\,,
\end{aligned}
\end{equation}
up to boundary and ${\cal O}(\alpha^2)$ terms. The subscript $(0)$ in the above equation means evaluation using the zeroth-order expressions for the auxiliary fields $\Phi_{\rm{aux}}\to \Phi^{(0)}_{\rm{aux}}$ and for the scalars $X\to X^{(0)}{}$. Therefore, the first term yields the same result as before: the two-derivative Lagrangian \eqref{eq:2dSUGRA}. Instead, the second and  third term vanish, as they contain the two-derivative equations of motion for the auxiliary fields. Let us remark in particular that the combination of the scalar equations that appears in the third term is precisely the one that yields the equation of motion for the Lagrange multiplier $D$, \eqref{eq:D}. This is a consequence of the fact that the solution to the modified cubic constraint is in general of the form \eqref{eq:sol_cc}. 

We have just justified that we can make use of the zeroth-order expressions for the auxiliary fields and for the cubic constraint in the two-derivative off-shell Lagrangian. Thus, the final four-derivative on-shell Lagrangian ${\cal L}$ will be given by
\begin{equation}
{\cal L}\equiv {\cal L}^{\rm{off-shell}}\Big|_{\{X^I(\Psi), \Phi_{\rm{aux}}(\Psi)\}}=\left({\cal L}^{\rm{off-shell}}_{2\partial}+\alpha \,{\cal L}^{\rm{off-shell}}_{4\partial}\right)\Big|_{(0)}\,.
\end{equation}
We emphasize that when following the procedure just outlined we will be writing the resulting Lagrangian in terms of the constrained scalars $X^{(0)}{}^{I}$, which differ from those appearing in the parent off-shell theory, \eqref{eq:sol_cc}. 

 In what follows we apply this procedure including a specific combination of four-derivative off-shell invariants to be specified along the way. Then we will argue that this choice of  invariants suffices to obtain the most general four-derivative effective action, at least for our present purposes. To avoid the clutter, we will not write the superscript $(0)$ in the scalars $X^{(0)}{}^{I}$; we will denote them simply as $X^I$, keeping in mind that they satisfy the same constraint as in the original two-derivative theory. In addition, we will ignore terms involving the two-derivative equations of motion, as those can be removed with perturbative field redefinitions without affecting the rest of the terms. 
 
We find instructive to first consider the ungauged limit. As a matter of fact, the gauging does not enter into the four-derivative part of the Lagrangian. In turn, it gives rise to a set of two-derivative corrections.

\subsubsection{Ungauged limit}

In principle, given our purposes in this section, we must now add the most general linear combination of off-shell four-derivative invariants.
On general grounds we expect three of them, but based on our discussion of sec.~\ref{sec:from_superconformal_to_supergravity}, we expect that only two of them will play a non-trivial role. Furthermore, we can argue that in the ungauged case ($g_I = 0$) all the supersymmetric invariants based on Ricci curvature can be eliminated with perturbative field redefinitions.

Let us then specify ${\cal L}^{\rm{off-shell}}_{4\partial}$ to be a linear combination of the Weyl-squared invariant ${\cal L}^{\rm{off-shell}}_{C^2}$ and the Ricci-scalar-squared invariant ${\cal L}^{\rm{off-shell}}_{R^2}$, whose explicit expressions are provided in Eqs.~\eqref{eq:offshell_Weyl_squared} and \eqref{eq:offshell_R^2}, respectively.
\begin{equation}
{\cal L}^{\rm{off-shell}}_{4\partial}={\cal L}^{\rm{off-shell}}_{C^2}+{\cal L}^{\rm{off-shell}}_{R^2}\, .
\end{equation}
Next we show that the $R^2$ invariant yields a trivial contribution, as argued above. After using the expressions for the auxiliary fields \eqref{eq:solauxfields} derived in the previous section and the expression for $D$ given in \eqref{eq:D}, we find that it reduces to
\begin{equation}
{\mathcal L}^{\rm{off-shell}}_{R^2}\Big|_{(0)}\,=\,{\sigma}_I X^I \left(-\frac{3}{8}\,R+\frac{8}{3}\, T^2+4 D\right)^2\Big|_{(0)}\,=\,\frac{1}{9}{\sigma}_I X^I {\cal E}^2\, ,
\end{equation}
where ${\cal E}={\cal E}_{\mu}{}^{\mu}$ is the trace of Einstein equations. Therefore, this term can be directly ignored since it can be removed by means of field redefinitions, up to ${\cal O}(\alpha^2)$ corrections. 
This provides explicit evidence in favour of our previous claim, according to which the most general four-derivative effective Lagrangian can be obtained by just considering the Weyl-squared invariant, or equivalently any other supersymmetric invariant containing $R_{\mu\nu\rho\sigma}R^{\mu\nu\rho\sigma}$. Therefore, 
\begin{equation}
{\cal L}_{4\partial}\,=\,{\mathcal L}^{\rm{off-shell}}_{4\partial}\Big|_{(0)}\,=\,{\mathcal L}^{\rm{off-shell}}_{C^2}\Big|_{(0)}\, .
\end{equation}
After integration by parts, use of Ricci identities and ignoring terms which can be removed with field redefinitions (without affecting the rest), we get
\begin{equation}\label{eq:Weyl2}
\begin{aligned}
 {\cal L}_{4\partial} &=\lambda_{M}X^{M} C_{\mu\nu\rho\sigma}C^{\mu\nu\rho\sigma}+{\mathtt D}_{IJ}\,C_{\mu\nu\rho\sigma}{F^I}^{\mu\nu}{F^J}^{\rho\sigma}+{\mathtt E}_{IJKL} \, F^I_{\mu\nu} F^J{}^{\mu\nu} \,F^K_{\rho\sigma} F^L{}^{\rho\sigma} \\[1mm]
&\!\!\!+ \widetilde{\mathtt E}_{IJKL} \, F^{I}_{\mu\nu} F^{J}{}^{\nu\rho} \,F^{K}_{\rho\sigma} F^{L}{}^{\sigma\mu}+\frac{27}{8}\lambda_{M}X^{M} \left(\partial_{\mu}X_I\, \partial^{\mu}X^I\right)^2+{\mathtt G}_{IJKL}\,\partial_{\mu} X^I \partial^{\mu} X^J \, F^K_{\rho\sigma} F^L{}^{\rho\sigma}\\[1mm]
&\!\!\!+12\lambda_M X^M\partial_{\mu}X_I \partial_{\nu}X_J  F^{J}{}^{\mu\rho}F^{I}{}^{\nu}{}_{\rho}-\frac{2}{3} \lambda_I {\mathcal F}^{\mu\alpha}{\mathcal F}^{\nu}{}_{\alpha}\nabla_{\nu}\nabla_{\mu}X^{I}+\frac{1}{4}\lambda_I  X_J\, \epsilon^{\mu\nu\rho\sigma\lambda}\nabla_{\alpha}F^{[I}_{\mu\nu}F^{J]}_{\rho\sigma}{\mathcal F}_{\lambda}{}^{\alpha}\\[1mm]
&\!\!\!+{\mathtt H}_{IJKL}\, \epsilon^{\mu\nu\rho\sigma\lambda}{F^I}_{\mu\nu} {F^{J}}_{\rho}{}^{\alpha} {F^K}_{\sigma\alpha} \partial_{\lambda} X^L+\frac{1}{2}\lambda_I \epsilon^{\mu\nu\rho\sigma\lambda} R_{\mu\nu\alpha\beta}R_{\rho\sigma}{}^{\alpha\beta}{A}^I_{\lambda} \\[1mm]
&\!\!\!-\lambda_M X^M\left(\frac{2}{3}R_{\mu\nu}{\mathcal F}^{\mu\alpha}{\mathcal F}^{\nu}{}_{\alpha}-\frac{1}{6}R\, {\mathcal F}^2\right)+\frac{2}{3}\lambda_M X^M \nabla_{\mu}{\mathcal F}^{\mu\nu}\nabla_{\rho}{\mathcal F}^{\rho}{}_{\nu}\\[1mm]
&\!\!\!+{\mathtt W}_{IJ}\,\epsilon^{\mu\nu\rho\sigma\lambda}F^I_{\mu\nu}F^J_{\rho\sigma}\nabla^{\alpha}{\mathcal F}_{\alpha\lambda}\,,
\end{aligned}
\end{equation}
where $\lambda_I$ is a dimensionless coupling, ${\cal F}$ is defined as $\mathcal {F}=3X_I F^I $, and the different couplings appearing in the Lagrangian read
\begin{equation}\label{eq:couplings}
\begin{aligned}
{\mathtt D}_{IJ}=\,&3\lambda_{I} X_{J}-\frac{9}{2}\lambda_{M}X^{M}X_IX_J\,,\\[1mm]
{\mathtt E}_{IJKL}=\,& \frac{3}{16}\lambda_{M}X^{M}\left(\frac{1}{2}a_{IJ}a_{KL}+3a_{IJ}X_K X_L- 9X_I X_J X_K X_L\right)-\frac{3}{8}a_{IJ}X_K\lambda_L\,,\\[1mm]
\widetilde{\mathtt E}_{IJKL}=\,&\frac{81}{8}\lambda_{M}X^{M}X_I X_JX_K X_L-\frac{9}{2}X_I X_JX_K \lambda_L\,,\\[1mm]
{\mathtt G}_{IJKL}=\,&\frac{9}{8}\lambda_{M}X^{M}a_{IJ}\left(a_{KL}+3X_K X_L\right)-\frac{9}{4}a_{IJ}X_K\lambda_L-6\lambda_{M}X^{M}{a}_{IK}{a}_{JL}\,,\\[1mm]
{\mathtt H}_{IJKL}=\,& -\frac{21}{4}\lambda_J \,a_{IL} \,X_K-\frac{3}{4}\lambda_I\, a_{JL}\, X_K\,,\\[1mm]
{\mathtt W}_{IJ}=\,&-\frac{1}{8}\lambda_I X_J+\frac{3}{4}\lambda_M X^M X_I X_J\,.
\end{aligned}
\end{equation}

\subsubsection{Including the gauging}

The gauged case is more subtle, as it contains an additional length scale set by the effective cosmological constant or, equivalently, the gauging parameters $g_I$. This is precisely what allows for corrections to the two-derivative terms, since $\alpha g_I g_J$ is dimensionless. These will play indeed a crucial role in this story, as they will eventually account for the corrections to the cubic `t Hooft anomaly coefficients, $k^{(1)}_{IJK}$. 

A main consequence of these two-derivative corrections is that the reasoning used in the ungauged case to argue that supersymmetric invariants just containing Ricci curvature yield a trivial contribution (namely, removable with suitable field redefinitions) does not  work anymore. However, it should be true that their contributions reduce to corrections to the two-derivative terms. This is exactly the logic that was used in the minimal supergravity case to argue that the effective action presented  in \cite{Cassani:2022lrk} was the most general one, even if we did not use the complete basis of off-shell four-derivative invariants. 
%

Given this, we consider the following off-shell Lagrangian,
\begin{equation}\label{eq:finaloffshellLagrangian}
{\cal L}^{\rm{off-shell}}\,=\,{\cal L}_{2\partial}^{\rm{off-shell}}\Big|_{C_{IJK} \to C_{IJK}+\alpha {\widetilde \lambda}_{IJK}}+\alpha \,{\cal L}^{\rm{off-shell}}_{C^2}\,,
\end{equation}
where ${\widetilde \lambda}_{IJK}$ is an arbitrary symmetric constant tensor of mass dimension 2. Splitting the first term in \eqref{eq:finaloffshellLagrangian} into its zeroth- and first-order contributions, we have:
\begin{equation}
{\cal L}_{2\partial}^{\rm{off-shell}}\Big|_{C_{IJK} \to C_{IJK}+\alpha {\widetilde \lambda}_{IJK}}= {\cal L}_{2\partial}^{\rm{off-shell}}+\alpha \,\Delta{\cal L}^{\rm{off-shell}}_{2\partial}\,,
\end{equation}
where ${\cal L}_{2\partial}^{\rm{off-shell}}$ is the one in~\eqref{eq:off_shell_poinc_general} and 
\begin{equation}\label{eq:2ndinvariantoffshell}
\begin{aligned}
\Delta{{\cal L}}^{\rm{off-shell}}_{2\partial}=\,&8\, {\widetilde \lambda}\left(\frac{1}{32} R+D+\frac{26}{3}T_{\mu\nu}T^{\mu\nu}\right)+\frac{1}{4} \,{\widetilde \lambda}_{IJ} F^I_{\mu\nu}F^J{}^{\mu\nu}+\frac{1}{2}\,{\widetilde \lambda}_{IJ}\,{\partial}_{\mu}X^I {\partial}^{\mu}X^J\\[1mm]
&-{\widetilde \lambda}_{IJ}\, Y^I{}^{ij}Y^J{}_{ij}-8 {\widetilde \lambda_I}  F^I_{\mu\nu}T^{\mu\nu}+\frac{1}{4}{\widetilde \lambda}_{IJK}\epsilon^{\mu\nu\rho\sigma\lambda}F^I_{\mu\nu} F^J_{\rho\sigma}A_\lambda^K\, ,
\end{aligned}
\end{equation}
where we have defined
\begin{equation}\label{eq:tildel_contracted}
{\widetilde\lambda}_{IJ}=6\,{\widetilde\lambda}_{IJK}X^K\,, \hspace{1cm} {\widetilde\lambda}_{I}=3\,{\widetilde\lambda}_{IJK}X^JX^K\,, \hspace{1cm}  {\widetilde\lambda}={\widetilde\lambda}_{IJK}X^I X^JX^K\,.
\end{equation}

Let us integrate out the auxiliary fields.  The Weyl-squared invariant now gives additional two-derivative corrections with respect to the ungauged case:
\begin{equation}
{\cal L}^{\rm{off-shell}}_{C^2}\Big|_{(0)}=\,{\cal L}_{4\partial}+\Delta {\cal L}^{C^2}_{2\partial}\,,
\end{equation}
where ${\cal L}_{4\partial}$ is still given by \eqref{eq:Weyl2} and
\begin{equation}\label{eq:DeltaL2dC2} 
\begin{aligned}
\Delta {\cal L}^{C^2}_{2\partial}=\,\,&6 \,\lambda_M X^M \left[{\cal V} + 3\left (g_M X^M\right)^2\right]^2- 9\lambda_M X^M \left[{\cal V}+3\left (g_M X^M\right)^2\right] a_{IJ} \,\partial_{\mu} X^I \partial^{\mu} X^J  \\[1mm]
& +f_{IJ} \,F^I_{\mu\nu}F^J{}^{\mu\nu} - \frac{3}{2}\lambda_{(I}g_Jg_{K)}\,\epsilon^{\mu\nu\rho\sigma\lambda}F^I_{\mu\nu} F^J_{\rho\sigma}A_\lambda^K\,,
\end{aligned}
\end{equation}
where
\begin{equation}
\begin{aligned}
f_{IJ}=\,&-\frac{3}{2}\left[{\cal V}+3\left ( g_M X^M\right)^2\right]\left[\lambda_M X^M \left(3X_I X_J + a_{IJ}\right) - 2\lambda_{(I}X_{J)} \right]+ 36 \lambda_K \,{\cal C}^{KL}\,g_L g_{(I} X_{J)}\\[1mm]
&  - 6 \lambda_M X^M\, g_I g_J\,.
 \end{aligned}
\end{equation}
In turn, the contribution from the second invariant $\Delta{{\cal L}}^{\rm{off-shell}}_{2\partial}$ is
\begin{equation}\label{eq:2ndinvariant}
\begin{aligned}
\Delta{\widetilde{\cal L}}_{2\partial}\equiv&\ \Delta{{\cal L}}^{\rm{off-shell}}_{2\partial}\Big|_{(0)}=\,\frac{{\widetilde \lambda}}{4} \,\left[R-6{\cal V}-18 \left(g_I X^I\right)^2\right]-9 \,{\widetilde \lambda}_{IJ} \,{\cal C}^{IK}{\cal C}^{JL}g_Kg_L\\[1mm]
&+\frac{1}{4}\left[{\widetilde \lambda}_{IJ} + 3{\widetilde \lambda} \left(3X_I X_J +\frac{1}{4}a_{IJ}\right)-6 {\widetilde \lambda}_{I} X_J\right]F^I_{\mu\nu} F^J{}^{\mu\nu}\\[1mm]
&+\frac{1}{2}\left({\widetilde\lambda}_{IJ}+\frac{9}{4} \,{\widetilde \lambda} \,a_{IJ}\right)\partial_{\mu} X^I \partial^{\mu} X^{J}+\frac{1}{4}{\widetilde \lambda}_{IJK}\epsilon^{\mu\nu\rho\sigma\lambda} F^I_{\mu\nu} F^J_{\rho\sigma}A^K_{\lambda}\,.
\end{aligned}
\end{equation}
Thus, the complete Lagrangian is
\begin{equation}\label{eq:4Der_Lagr}
\begin{aligned}
 {\cal L}=\,&\,R-2{\cal V}-\frac{3}{2}\, a_{IJ}\, \partial_{\mu} X^I \partial^{\mu} X^J-\frac{3}{4}\, a_{IJ}\,F^I_{\mu\nu} F^J{}^{\mu\nu}
+\frac{1}{4}C_{IJK}\epsilon^{\mu\nu\rho\sigma\lambda}F^{I}_{\mu\nu} F^{J}_{\rho\sigma} A^{K}_{\lambda} \\
&+\alpha \left({\cal L}_{4\partial} +\Delta {\cal L}^{C^2}_{2\partial}+\Delta {\widetilde{\cal L}}_{2\partial}\right) \,,
\end{aligned}
\end{equation}
where ${\cal L}_{4\partial}$, $\Delta {\cal L}^{C^2}_{2\partial}$ and $\Delta {\widetilde{\cal L}}_{2\partial}$ are given in eqs.~\eqref{eq:Weyl2}, \eqref{eq:DeltaL2dC2} and \eqref{eq:2ndinvariant}, respectively.

Some comments are in order. First, we expect that this effective Lagrangian captures, for particular choices of ${\widetilde\lambda}_{IJK}$, the corrections that one would obtain when considering any other basis of off-shell invariants. In particular we have checked this for the $R^2$ invariant of \cite{Ozkan:2013nwa}, as well as for the ``off-diagonal'' invariants constructed in \cite{Ozkan:2016csy}. However, let us emphasize that \eqref{eq:2ndinvariant} generalizes all of them, as the correction to the gauge Chern-Simons term is controlled by a (symmetric) tensor ${\widetilde \lambda}_{IJK}$, which is the most general possibility.  This is crucial as it will allow us to match any correction to the cubic anomalies of the dual field theories. 

Second, we note that when ${\widetilde \lambda}_{IJK}\propto C_{IJK}$, the correction from the second invariant \eqref{eq:2ndinvariant} reduces to a correction of Newton's constant, after a suitable constant rescaling of the metric is performed. This is exactly what happens for the supergravity theory dual to ${\cal N}=4$ SYM. In addition, for this theory the corrections coming from the Weyl-squared invariant also trivialize, as the anomaly matching imposes $\lambda_I=0$, as a consequence of the fact that $k_I=0$ at all orders in the large-$N$ expansion.

Finally, we note that the Lagrangian \eqref{eq:4Der_Lagr} can be further simplified using perturbative field redefinitions. In particular we can use them to remove the last four terms in \eqref{eq:Weyl2}, as well as all the two-derivative corrections to the Ricci scalar term. When doing this, however, we will modify some of the couplings to the remaining terms. We relegate the details of this procedure to appendix~\ref{app:field_redefinitions} and simply present the final form of the action (with the couplings updated) in the next subsection.

\subsection{Final form of the four-derivative effective Lagrangian}
\label{sec:final_Lagr} 

After implementing suitable perturbative field redefinitions (see appendix~\ref{app:field_redefinitions}) to reduce the number of terms in \eqref{eq:4Der_Lagr}, we arrive to the following final Lagrangian:
\begin{equation}\label{eq:finalL}
\begin{aligned}
{\cal L}=\,&\, R-2{\cal V}-\frac{3}{2}\, {a}_{IJ}\, \partial_{\mu} {X}^I \partial^{\mu} {X}^J-\frac{3}{4}\, {a}_{IJ}\,F^I_{\mu\nu} F^J{}^{\mu\nu}
+\frac{1}{4}{ C}_{IJK}\epsilon^{\mu\nu\rho\sigma\lambda}F^I_{\mu\nu} F^J_{\rho\sigma} A^K_{\lambda}\\
&+ \alpha \left({\cal L}_{4\partial}+{\Delta{\cal L}}_{2\partial}\right)\,,
\end{aligned}
\end{equation}
where
\begin{equation}\label{eq:4_der_terms_final}
\begin{aligned}
\mathcal L_{4\partial} &= \, \lambda_M X^M \,{\cal X}_{\rm{GB}} +D_{IJ} \,C_{\mu\nu\rho\sigma}\, F^{I\mu\nu} F^{J\rho\sigma} + E_{IJKL}\,F^I_{\mu\nu} F^J{}^{\mu\nu}\, F^K_{\rho\sigma} F^L{}^{\rho\sigma}
\\[1mm]
+&\,{\widetilde E}_{IJKL}\, F^I_{\mu\nu}F^{J\,\nu\rho} \,F^K_{\rho\sigma} F^{L\,\sigma\mu} + I_{IJKL}\,\partial_{\mu} X^I \partial^{\mu} X^J\, \partial_{\nu} X^K \partial^{\nu} X^L+ H_{IJKL}\, \partial_{\mu} X^I  \partial^{\mu} X^J\, F^K_{\rho\sigma}F^L{}^{\rho\sigma}
\\[1mm]
+&\,{\widetilde H}_{IJKL}\, \partial_\mu X^I \partial^\nu X^J \, F^{K\,\mu\rho}F^L_{\nu\rho}-\,6X_I X_J\lambda_K \,  F^{I}{}^{\mu\alpha} F^{J}{}^\nu{}_\alpha\, \nabla_\nu\partial_\mu X^K 
\\[1mm]
+&\,\frac{3}{4}\lambda_{[I}X_{J]}X_K\,\epsilon^{\mu\nu\rho\sigma\lambda}\nabla_\alpha F^I_{\mu\nu}F^J_{\rho\sigma} F^K_\lambda{}^\alpha + W_{IJKL}\, \epsilon^{\mu\nu\rho\sigma\lambda} F^I_{\mu\nu} F^J_\rho{}^\alpha F^K_{\sigma\alpha} \,\partial_\lambda X^L 
\\[1mm]
+&\,\frac{1}{2}\lambda_I\,\epsilon^{\mu\nu\rho\sigma\lambda} R_{\mu\nu\alpha\beta} \,R_{\rho\sigma}{}^{\alpha\beta}A_\lambda^I\,,
\end{aligned}
\end{equation}
being ${\cal X}_{\rm{GB}} = R_{\mu\nu\rho\sigma}^2 - 4 R_{\mu\nu}^2 + R^2$ is the Gauss-Bonnet combination. The four-derivative couplings are given by
\begin{equation}
\begin{aligned}
D_{IJ}=& \,3\lambda_{I} X_{J}-\frac{9}{2}\lambda_{M}X^{M}X_IX_J \,,
\\[1mm]
E_{IJKL} =&\,\lambda_M X^M \left(-\frac{27}{16}X_I X_J X_K X_L - \frac{9}{8}a_{IJ} a_{KL} + \frac{39}{16}a_{IJ} X_K X_L - \frac{3}{4}a_{IK} a_{JL} + \frac{9}{4}a_{IK} X_J X_L \right) \\[1mm]
&- \frac{3}{4}a_{J[I}\lambda_{L]}X_K -\frac{9}{8}\lambda_I X_J X_K X_L 
\,,\\[1mm]
{\widetilde E}_{IJKL}=&\, \lambda_M X^M \left( \frac{81}{8}X_I X_J X_K X_L + 6a_{IJ}a_{KL} + \frac{3}{2}a_{IK} a_{JL} -9X_I X_J a_{KL} - \frac{9}{2}X_I X_K a_{JL}\right)-\\[1mm]
&- \frac{9}{4}X_I X_J X_K \lambda_L -\frac{3}{4}a_{IK} X_J \lambda_L
\,,\\[1mm]
I_{IJKL}=&\, \lambda_M X^M \left( \frac{3}{2}a_{IJ}a_{KL} + 6a_{K(I}a_{J)L}\right)
\,,\\[1mm]
H_{IJKL}=& \,-\frac{3}{2}\lambda_M X^M a_{IJ} a_{KL} - 6\lambda_M X^M a_{IK} a_{JL} + \frac{45}{8}\lambda_M X^M a_{IJ} X_K X_L - \frac{9}{4}a_{IJ} X_K \lambda_L 
\,,\\[1mm]
{\widetilde H}_{IJKL}=&\,\lambda_M X^M\left(12a_{IL}a_{JK} + 6a_{IK}a_{JL}+ 12a_{IJ}a_{KL} - 9a_{IJ}X_K X_L\right)\,\,, \\[1mm]
W_{IJKL}=& -\frac{21}{4}\lambda_J a_{IL} X_K - \frac{3}{4}\lambda_J X_I a_{KL} + 3\lambda_M X^M \left(2a_{IJ} a_{KL} -3X_I X_J a_{KL} \right)
\,.
\end{aligned}
\end{equation}
Finally, $\Delta {\cal L}_{2\partial}$ contains all the two-derivative corrections: 
\begin{equation}
\begin{aligned}
{\Delta {\cal L}}_{2\partial}=\,&-2\Delta {\cal V}+\left\{\frac{1}{2}\left({\widetilde\lambda}_{IJ}+3{\widetilde \lambda} \,a_{IJ}\right) - 3\lambda_M X^M \left[4{\cal V}+9\left (g_M X^M\right)^2\right]a_{IJ} \right\}\,\partial_{\mu} X^I  \partial^{\mu} X^J \\[1mm]
& -\frac{3}{4}\Delta a_{IJ} F^I_{\mu\nu} F^J{}^{\mu\nu}+\frac{1}{4}\left({\widetilde \lambda}_{IJK}- 6\lambda_{(I}g_Jg_{K)}\right)\epsilon^{\mu\nu\rho\sigma\lambda} \,F^I_{\mu\nu} F^J_{\rho\sigma}A_\lambda^K\,,
\end{aligned}
\end{equation}
where ${\Delta}{\cal V}$ is the correction to the scalar potential, whose explicit expression reads
\begin{equation}
\begin{aligned}
\Delta {\cal V}=\,&-\lambda_M X^M \left[\frac{4}{3}\,{\cal V}^2 +18\left (g_M X^M\right)^2{\cal V}+27\left(g_M X^M\right)^4\right]+\frac{1}{3}\,{\widetilde \lambda}\,{\cal V}+\frac{9}{4}\,{\widetilde \lambda}\left(g_I X^I\right)^2\\[1mm]
&+\frac{9}{2} \,{\widetilde \lambda}_{IJ} \,{\cal C}^{IK}{\cal C}^{JL}g_Kg_L\,,
\end{aligned}
\end{equation}
and
\begin{equation}
\begin{aligned}
\Delta a_{IJ}=\,&2\left[{\cal V}+ 3\left (g_M X^M\right)^2\right]\left[\lambda_M X^M \left(3X_I X_J + a_{IJ}\right) - 2\lambda_{(I}X_{J)} \right]+8\lambda_M X^M\, g_I g_J\\[1mm]
& -48 \lambda_K \,{\cal C}^{KL}\,g_L g_{(I} X_{J)}+\frac{2}{3}\lambda_M X^M {\cal V} \left(a_{IJ}-2X_IX_J\right)\\[1mm]
&-\frac{1}{3}{\widetilde \lambda}_{IJ} - {\widetilde \lambda} \left(3X_I X_J +\frac{1}{3}\,a_{IJ}\right)+2{\widetilde \lambda}_{I} X_J\,.
\end{aligned}
\end{equation}

We conclude observing that the contribution from the invariant controlled by the coupling ${\widetilde \lambda}_{IJK}$ can be cast as the original two-derivative Lagrangian,
\begin{equation}\label{eq:tilded_Lagr}
{\cal L}|_{\lambda_I=0}= R-2{\widetilde{\cal V}}-\frac{3}{2}\,  {\widetilde a}_{IJ}\, \partial_{\mu} {\widetilde X}^I \partial^{\mu} {\widetilde X}^J-\frac{3}{4}\, {\widetilde a}_{IJ}\,F^I_{\mu\nu} F^J{}^{\mu\nu}
+\frac{1}{4}{\widetilde C}_{IJK}\epsilon^{\mu\nu\rho\sigma\lambda}F^{I}_{\mu\nu} F^{J}_{\rho\sigma} A^{K}_{\lambda}\, ,
\end{equation}
with a shifted Chern-Simons coupling
\begin{equation}\label{eq:subtildeaction1}
{\widetilde C}_{IJK}\,=\, C_{IJK}+\alpha \,{\widetilde \lambda}_{IJK}\,.
\end{equation}
In \eqref{eq:tilded_Lagr}, the tilded quantities are defined as in the two-derivative theory just using ${\widetilde C}_{IJK}$ instead of $C_{IJK}$. In particular the scalars ${\widetilde X}$ satisfy the cubic constraint with ${\widetilde C}_{IJK}$, which means that they are given in terms of $X^I$ by
\begin{equation}\label{eq:subtildeaction2}
{\widetilde X}^I=X^I\left(1-\frac{\alpha}{3}\,{\widetilde \lambda}\right)\, .
\end{equation}

\section{Supersymmetric black hole action and holographic match}\label{sec:onshellaction}

Given a holographic $\mathcal{N}=1$ SCFT$_4$, one may expect that the higher-derivative five-dimensional gauged supergravity reproducing its 't Hooft anomalies admits an asymptotically AdS$_5$ supersymmetric black hole solution whose on-shell action matches the large-$N$ expansion of the formula~\eqref{eq:index_asympt}. In this section we prove the validity of such expectation at linear order in the four-derivative corrections for the specific model dual to the $\mathbb{C}^3/\mathbb{Z}_\nu$ quiver theories, in the case of equal angular velocities, $\omega_1=\omega_2$.

We begin  in section \ref{sec:holographicdictionary2} by providing the dictionary between the SCFT 't Hooft anomaly coefficients and the supergravity couplings. We will also specify the choices of couplings $\lambda_I$ and $\widetilde\lambda_{IJK}$ that we will consider. This extends app.~\ref{sec:holographicdictionary} to the case of many vector multiplets. 
In section \ref{sec:blackhole} we review the black hole solution of the two-derivative theory and its thermodynamics; then we take the supersymmetric (and extremal) limit. We next turn to the four-derivative theory: in section \ref{sec:boundaryterms} we discuss the boundary terms to be included in order to holographically renormalize the theory, and in section \ref{sec:setupaction} we revisit the argument showing that we do not need the corrected solution to evaluate the four-derivative action at linear order in the corrections. In section~\ref{sec:resultsandmatch} we present our final result for the supersymmetric on-shell action, matching the large-$N$ expansion of~\eqref{eq:index_asympt}. 

\subsection{Holographic dictionary for anomaly coefficients}\label{sec:holographicdictionary2}

The precise holographic dictionary between the field theory 't Hooft anomaly coefficients and the supergravity Chern-Simons couplings can be obtained by equating the anomalous variation of the respective partition functions under a transformation induced by the charges $Q_I$ \cite{Witten:1998qj}.\footnote{Equivalently, we could take the formal exterior derivative of the supergravity Chern-Simons terms and  compare the resulting six-form with the SCFT anomaly polynomial~\eqref{anomaly_poly}. }

We will use a hat to indicate the field theory background fields on $\mathcal{M}_4$ and $i,j,k$ for the spacetime indices on $\mathcal{M}_4$.\footnote{These should not be confused with the SU(2) indices $i=1,2$ used in section~\ref{sec:fourder_action}, which however will not appear in the present section.} We denote by $J^k_I$ the SCFT current associated with $Q_I$, by $\hat A^I_k$ the background gauge field that canonically couples to it, transforming as
\be
\label{gauge_var_SCFT}
\delta_{\Lambda} \hat{A}^I_k = \partial_k\Lambda^I\,,
\ee 
and by $\hat F^I_{jk}= 2\partial_{[j}\hat A^I_{k]}$ its field strength. Then the variation of the SCFT partition function reads (in Lorentzian signature),
\begin{equation}\label{variation_ZCFT}
\begin{aligned}
\delta_{\Lambda} \log Z_{\rm CFT} \,&=\,  -i \int_{{ \cal M}_4}\diff^4 x\, \hat e \, \Lambda^I \,\nabla_k  J^k_I \\[1mm]
\,&=\, - \frac{i}{{96\pi^2}} \int_{{ \cal M}_4}\diff^4 x\, \hat e \, \Lambda^I \left( k_{IJK} \,\hat\epsilon^{\,ijkl}\hat {F}^J_{ij} \hat {F}^K_{kl} \, - \,  \frac{1}{8}\,k_I\,\hat\epsilon^{\,ijkl}\hat {R}_{ijab} \hat {R}_{kl}{}^{ab}\right)\,.
\end{aligned}
\end{equation}

This is to be compared with the corresponding variation of the gravitational partition function.
 In the classical saddle-point approximation, the gravitational partition function is given by the renormalized supergravity action evaluated on-shell. For the action given in section~\ref{sec:final_Lagr}, the non-invariant sector made of the Chern-Simons terms yields under the variation \eqref{gauge_var_SCFT},\footnote{When we apply the Stokes theorem and pass from the bulk to the boundary, we introduce an unusual minus sign. This is because the orientation we use for $\mathcal{M}_4 = \partial \mathcal{M}$ is opposite to the orientation induced from the bulk by contracting the bulk volume form with the vector $\frac{\partial}{\partial r}$ normal to the boundary. Concretely, in this section the positive orientation in the bulk is given by $\epsilon_{tr123}>0$, while the positive orientation in the boundary is given by $\epsilon_{t 123}>0$, where $1,2,3$ label the coordinates for the spatial slices of the boundary.}
\begin{equation}\label{variation_Zgrav}
\begin{aligned}
\delta_{\Lambda} \log Z_{\rm grav} &\simeq i\, \delta_{\Lambda} S\\[1mm]
\,=&\, -\frac{i}{16 \pi G g^3}\int_{\partial {\cal M}} \diff^4x \,\hat{e}\, \Lambda^I \left( \frac{1}{4}\, C^{\rm (\alpha)}_{IJK} \,\hat{\epsilon}^{\,ijkl}\hat{F}^J_{ij}\hat{F}^K_{kl} \,+\, \frac{\alpha\lambda_I g^2}{2}\,\hat{\epsilon}^{\,ijkl}\hat{R}_{ijab}\hat{R}_{kl}{}^{ab}\right)\,,
\end{aligned}
\end{equation}
where we have identified the supergravity gauge potentials $A^I$ restricted to the conformal boundary $\partial \mathcal{M}$ with the field theory background gauge potentials $\hat A^I$ on $\mathcal{M}_4 =\partial\mathcal{M}$ via 
\be
A^I|_{\partial \mathcal{M}} \,=\, \frac{\hat A^I}{g} \,,
\ee
 $g$ being a parameter with the dimensions of a mass, which will later appear as the inverse radius of the two-derivative AdS solution. Here it is needed in order to make the mass dimensions consistent.\footnote{According to the conventions used in this section, the gravitational electric charges and the respective electrostatic potentials will also carry an extra factor of $g$ compared to the corresponding field theory quantities. 
 }
We have also introduced the corrected gauge Chern-Simons coupling,
\be\label{correctedCIJK}
C^{\rm (\alpha)}_{IJK} \,=\,  C_{IJK} -6 \alpha \lambda_{(I} g_J g_{K)} + \alpha \widetilde\lambda_{IJK} \,.
\ee

Comparing \eqref{variation_ZCFT} and \eqref{variation_Zgrav}, we obtain the desired holographic dictionary:
\begin{align}
k_{IJK} \,&=\,  \frac{ 3\pi }{2Gg^3} \,C^{(\alpha)}_{IJK}  \,, \label{dict_cubic_anom} \\[1mm]
k_I \,&=\, -\frac{24\pi}{Gg}\, \alpha\lambda_I\,. \label{dict_linear_anom}
\end{align}
The first relation can be split into leading-order and correction terms as:
\be
k^{(0)}_{IJK} \,=\,  \frac{ 3\pi }{2Gg^3} \,C_{IJK} \,,\qquad   k^{(1)}_{IJK} \,=\,   \frac{ 3\pi\alpha }{2Gg^3} \,  \big(-6  \lambda_{(I} g_J g_{K)} +  \widetilde\lambda_{IJK}\big)\,.
\ee

Using the dictionary above, we can rephrase the SCFT formula \eqref{eq:index_asympt} in gravitational variables as
\be\label{eq:gravityactionprediction}
I \, =\,    \frac{ \pi }{4G} \, \frac{C^{(\alpha)}_{IJK} \varphi^I_{\rm gr}\varphi^J_{\rm gr}\varphi^K_{\rm gr}}{\omega_1\omega_2} + \frac{\pi}{G}\, \alpha\lambda_I \varphi_{\rm gr}^I \,\frac{\omega_1^2+\omega_2^2-4\pi^2}{\omega_1\omega_2} \,,
\ee
with linear constraint given by
\be\label{eq:constraint_gravity}
\omega_1+ \omega_2  -  \frac{3}{\sqrt2} \, g_I  \varphi_{\rm gr}^I \,=\, \pm 2\pi i \,,
\ee
where $\varphi^I_{\rm gr}=g^{-1}\varphi^I$ and we have used the identification 
\be\label{eq:rel_rI_gI}
 r_I \,= \, \frac{3}{2\sqrt2} \frac{g_I}{g}  \,,
 \ee
 which is derived in appendix~\ref{app:sugradual}. The precise way the  supersymmetric chemical potentials $\varphi_{\rm gr}^I$ and $\omega$ should be evaluated on the gravity side will be specified below.
In the following we check the expectation that there exists a corrected supersymmetric AdS$_5$ black hole solution whose on-shell action matches \eqref{eq:gravityactionprediction} in the model dual to the $\mathbb{C}^3/\mathbb{Z}_\nu$ quiver theories of section~\ref{sec:orbifold_section}, for the case of equal angular velocities.

\subsection{Specialization to the orbifold theories}\label{sec:special_relations}
 
Since we eventually want to match the gravitational action with the $\mathbb{C}^3/\mathbb{Z}_\nu$ quiver theories  in section~\ref{sec:orbifold_section} for generic $\nu$, we will take the supergravity Lagrangian \eqref{eq:finaloffshellLagrangian} with $n =2$ vector multiplets (thus ignoring the baryonic symmetry available for even $\nu$). This contains Abelian vector fields $A_\mu^I$, $I=1,2,3$, coupling to the three global symmetries of the dual quantum field theories, and is known as U(1)$^3$ model. 

As a first thing, we observe that the general assumption we have made in~\eqref{relation_k3_k1}, which is in particular satisfied by the orbifold theories discussed in section~\ref{sec:orbifold_section}, translates via the dictionary above into the following relation,
\be
C^{(\alpha)}_{IJK} \,=\,   C_{IJK}  - 18 \alpha  \lambda_{(I} g_J g_{K)}  \,,
\ee
Comparing with \eqref{correctedCIJK}, this means that we need to choose
\be\label{choice_tildelambda}
\widetilde\lambda_{IJK} \,=\,   - 12   \lambda_{(I} g_J g_{K)} \,,
\ee
demonstrating that in order to discuss this  class of theories both couplings $\lambda_{I}$ and $\widetilde\lambda_{IJK}$ are needed. Notice that with this choice the prediction \eqref{eq:gravityactionprediction} for the on-shell action can be rewritten in the simpler form
\be\label{eq:gravityactionpredictionBis}
I \, =\,    \frac{ \pi }{4G} \, \frac{C_{IJK} \varphi^I_{\rm gr}\varphi^J_{\rm gr}\varphi^K_{\rm gr}}{\omega_1\omega_2} - \frac{2\pi}{G}\, \alpha\lambda_I \varphi_{\rm gr}^I \,\left[ 1 \mp 2\pi i \left(\frac{1}{\omega_1}+\frac{1}{\omega_2}\right) \right]\,.
\ee

We now focus more specifically on the gravity dual of the $\mathbb{C}^3/\mathbb{Z}_\nu$ quiver theories of section~\ref{sec:orbifold_section} for generic $\nu$. Using the dictionary above as well as information from appendix~\ref{sec:amaxim_section}, the SCFT anomaly coefficients \eqref{tHooft_coeff_orbifolds} translate into the following supergravity couplings,
\be
C_{IJK}  =  \frac{1}{6}|\epsilon_{IJK}|\,, \qquad\ \widetilde\lambda_{IJK} =   - 12   \lambda_{(I} g_J g_{K)} \,,\qquad\ \lambda_I = \frac{ g_I}{8\sqrt2 \,g}\,, \qquad   I=1,2,3\,,
\ee
with 
\be
g_I = \frac{\sqrt{2}}{3}\,g \,,
\ee 
and
\be
\frac{\pi}{2Gg^3}= \nu N^2\,,\qquad\quad \alpha g^2 = \frac{1}{4N^2}\,.
\ee
Also, the Weyl anomaly coefficients  \eqref{a_c_orbifolds}  can be expressed in gravitational units as
\be
\aa  \,=\,   \frac{\pi}{8Gg^3} \,  \left(1  -3\alpha g^2   \right)  \,,\qquad\quad
\cc \,=\, \frac{\pi}{8Gg^3} \,  \left(1  - 2 \alpha g^2 \right) \,.
\ee

\subsection{The two-derivative black hole solution}\label{sec:blackhole}

A general asymptotically AdS black hole solution of the model specified above has three independent electric charges $Q^{\rm gr}_I$ and two angular momenta $J_1$, $J_2$. Here we will focus on the case of equal angular momenta $J_1 = J_2\equiv J$. At the two-derivative level, the supersymmetric and extremal solution in this regime has been derived in~\cite{Gutowski:2004yv}, while the non-supersymmetric, thermal solution has been found in~\cite{Cvetic:2004ny}. The corrections to the solution introduced by the four-derivative terms are not known, however as we will see this still allows us to obtain the on-shell action at first order in the corrections.

The action of the two-derivative ${\rm U}(1)^3$ model reads
\begin{equation}\label{eq:u1cubemodelaction}
\begin{aligned}
S\, =\, \frac{1}{16\pi G}\int \diff^5x\, e\,& \bigg[  R + 4g^2\sum_{I=1}^3\left( X^I\right)^{-1}-\frac{1}{2}\partial\vec\phi^{\,2} -\frac{1}{2}\sum_{I=1}^3\left(X^I\right)^{-2}\,F^I_{\mu\nu} F^{I\,\mu\nu}\\
&\ +\frac{1}{24}\epsilon^{\mu\nu\rho\sigma\lambda}\, |\epsilon_{IJK}\, |  F^I_{\mu\nu} F^J_{\rho\sigma} A^K_\lambda
\bigg]\,,
\end{aligned}
\end{equation} 
where $A^I\,,\,I=1,2,3$, are Abelian gauge fields, $\vec \phi = \left(\phi_1,\phi_2\right)$ are two real scalar fields and
\begin{equation}
X^1 = {\rm e}^{-\frac{1}{\sqrt{6}}\phi_1 - \frac{1}{\sqrt{2}}\phi_2}\,,\quad\quad X^2 =  {\rm e}^{-\frac{1}{\sqrt{6}}\phi_1 +\frac{1}{\sqrt{2}}\phi_2}\,,\quad\quad X^3= {\rm e}^{\frac{2}{\sqrt{6}}\phi_1}\,,
\end{equation}
satisfy the constraint $X^1 X^2 X^3=1$.
This model is a consistent truncation of Type IIB supergravity on $S^5/\mathbb{Z}_\nu$, where the Abelian gauge fields arise as KK vectors gauging the  U(1)$^3\subset$ SO(6) isometries of $S^5$. The action \eqref{eq:u1cubemodelaction} follows from the general expression \eqref{eq:2dSUGRA}  by taking
\begin{equation}\label{eq:conventionsveryspecialgeometry}
C_{IJK} = \frac{1}{6}|\epsilon_{IJK}|\,,\qquad\quad a_{IJ} =\frac{1}{3\left(X^I\right)^2}\delta_{IJ}\,,\qquad\quad g_I = \frac{\sqrt{2}}{3}\,g\,,\qquad\quad I,J,K=1,2,3\,.
\end{equation}
The scalar potential $\mathcal V = -2g^2\sum_{I=1}^3\left(X^I\right)^{-1}$ is extremized for
\be
\bar X^I=1  \quad \Rightarrow\quad   \bar X_I = C_{IJK}\bar X^J\bar X^K = \frac{1}{3}   \,,\qquad\quad I=1,2,3\,.
\ee 
It follows that the corresponding AdS$_5$ solution has radius $g^{-1}$. The gauging parameters can be expressed in terms of the vacuum value of the scalars as
\begin{equation}\label{eq:gaugingparameters}
g_I = \sqrt{2}\,g\,\bar X_I\,.
\end{equation}
This relation shows that the AdS$_5$ vacuum solution is supersymmetric, see the derivation of~\eqref{scalars_AdS_sol} in the appendix.\footnote{Throughout this section, we denote by a bar the AdS$_5$ value of the scalar fields and functions thereof.}

The metric, gauge, and scalar fields for the asymptotically AdS$_5$ black hole solution can be expressed in a non-rotating frame at infinity using the coordinates $\left(t,r,\theta,\phi,\psi\right)$ as
\begin{equation}\label{eq:metric5d}
\text ds^2 = \left(H_1 H_2 H_3\right)^{1/3}\left[ -\frac{r^2 Y}{f_1}\text dt^2+ \frac{r^4}{Y}\text dr^2+\frac{r^2}{4}\left(\sigma_1^2 + \sigma_2^2\right)+\frac{f_1}{4r^4H_1H_2H_3}\left(\sigma_3 -\frac{2f_2}{f_1}\text dt\right)^2
\right]\,,
\end{equation}
\begin{equation}\label{eq:scalars5d}
A^I = \left(\frac{2m}{r^2H_I}s_I\,c_I + z_I\right)\text dt + \left(\frac{m\,a}{r^2 H_I}\left(c_I\, s_J\,s_K-s_I\,c_J\,c_K\right)\right)\sigma_3\,,\quad\quad X^I = \frac{\left(H_1 H_2 H_3\right)^{1/3}}{H_I}\,,
\end{equation}
where the indices $I,J,K$ in $A^I$ are never equal, and the constant parameters  $z_I$ will be fixed later as  gauge choices. The solution is given in terms of the SU(2) left-invariant one-forms parametrized by the Euler angles on $S^3$, $\theta \in [0,\pi]$, $\phi\in[0,2\pi]$ and $\psi\in[0,4\pi]$,
\begin{equation}
\sigma_1 = \cos\psi \, \text d\theta + \sin\psi\sin\theta \, \text d\phi \,,\quad \sigma_2 = -\sin\psi \, \text d\theta + \cos\psi \sin\theta \, \text d\phi\,,\quad \sigma_3 = \text d\psi + \cos\theta \, \text d\phi\,.
\end{equation}
We have also introduced the following radial functions,
\begin{align}
H_I(r) &= 1+\frac{2m \,s_I^2}{r^2}\,, \non \\[1mm] 
f_1(r)&= 4 a^2 m^2 \left[2\,s_1 s_2 s_3 \left(c_1c_2c_3-s_1s_2s_3\right)-s_1^2 s_2^2-s_1^2 s_3^2-s_2^2 s_3^2\right]+2 a^2 m\, r^2+H_1H_2 H_3\,  r^6, \non \\[1mm]
f_2(r) &= 2m\,a\left(c_1c_2c_3-s_1s_2s_3\right)r^2+ 4m^2as_1s_2s_3\,,\non\\[1mm]
f_3(r)&= 4 g^2a^2 m^2 \left[2\, s_1 s_2 s_3 \left(c_1c_2c_3-s_1s_2s_3\right)-s_1^2 s_2^2-s_1^2 s_3^2-s_2^2 s_3^2\right]+2 a^2 m \left(g^2 r^2+1\right)\,, \non \\[1mm]
Y(r) &= f_3(r) + g^2 r^6H_1H_2H_3 + r^4 - 2m\, r^2\,.
\end{align}
The parameters $s_I$ and $c_I$ are shorthand notations for $s_I = \sinh \delta_I$ and $c_I =\cosh\delta_I$. Therefore, the black hole depends on five independent parameters $(m,\delta_{1},\delta_2,\delta_3,a)$, roughly corresponding to the five independent conserved charges (mass, three electric charges and angular momentum). The black hole has a Killing horizon, whose location, denoted by $r_+$, is given by the largest positive root of $Y(r)$.

The thermodynamical chemical potentials of the non-extremal black hole are the angular velocity $\Omega$, the electrostatic potentials $\Phi^I$ and the inverse Hawking temperature $\beta$. The angular velocity is read from the condition that the Killing vector $\xi = \partial_t + \Omega\,\partial_\psi$ becomes null at the Killing horizon. The electrostatic potentials are defined by the gauge invariant combination $\Phi^I = \iota_\xi A^I|_{r_+}-\iota_\xi A^I|_{+\infty}$. Their expressions are
\begin{equation}\label{eq:thermalchemicalpotentials2d}
\Omega = 2\frac{f_2(r_+)}{f_1(r_+)}\,,\quad\quad \Phi^I = \frac{2m}{r_+^2 H_I(r_+)}\left(s_I c_I + \frac{1}{2}a\,\Omega\left(c_I \,s_J \,s_K - s_I\, c_J \,c_K\right)\right)\,,\quad I\neq J\neq K\,.
\end{equation}
The inverse Hawking temperature is identified with the  period $\beta$ of the compactified Euclidean time $\tau = it$; this is fixed by regularity of the Euclideanized solution to
\begin{equation}\label{eq:inversetemperature2d}
\beta = 4\pi\,r_+ \sqrt{f_1(r_+)}\left(\frac{\text d Y}{\text dr}\,(r_+)\right)^{-1}\,.
\end{equation}
Regularity of the Euclidean solution at $r_+$ also fixes the gauge constants $z_I$ introduced in \eqref{eq:scalars5d} through the requirement that the gauge field has no component along the shrinking direction identified by the Killing vector at the horizon, that is
\begin{equation}
\iota_\xi A^I|_{r_+}=0 \quad\implies\quad z_I = -\Phi^I\,.
\end{equation}
As a consequence, the electrostatic potentials can be read off from the asymptotic boundary value of the gauge fields. The same holds for the angular velocity upon changing the angular coordinate $\psi$ so that the connection encoded in the $g_{t\psi}$ component of the metric vanishes at $r_+$.

The conserved charges read~\cite{Cvetic:2005zi}
\begin{equation}\label{eq:conservedcharges2d}
\begin{aligned}
E \,&=\, \frac{\pi}{4G}\,m\left(3 + g^2 a^2+ 2 s_1^2 + 2s_2^2 + 2s_3^2\right)\,,\\
J \,&=\, \frac{\pi}{2G}\,m\,a\left(c_1c_2c_3-s_1s_2s_3\right)\,,\\
Q^{\rm gr}_I \,&=\, \frac{\pi}{2G}\,m\,s_I\,c_I\,,
\end{aligned}
\end{equation}
while the Bekenstein-Hawking entropy is given by
\begin{equation}\label{eq:entropy2d}
\mathcal S \,=\, \frac{\pi^2}{2G}\sqrt{f_1(r_+)}\,.
\end{equation}

The expression for the Euclidean on-shell action of the two-derivative theory \eqref{eq:u1cubemodelaction} for the black hole solution \eqref{eq:metric5d}, \eqref{eq:scalars5d} can be found in~\cite{Cassani:2019mms}. We will distinguish between the action obtained there using holographic renormalization with a minimal subtraction scheme, that we denote by $I_{Z}$, and the action $I$ of interest here. These are related as $I = I_{Z}-I_{\rm AdS}$, where $I_{\rm AdS}$ denotes the contribution of the AdS vacuum, which in the scheme used in~\cite{Cassani:2019mms}  reads
\begin{equation}
I_{\rm AdS} = \beta E_{\rm AdS}\,\,,\quad\quad E_{\rm AdS} = \frac{3\pi}{32g^2\,G}\,,
\end{equation}
where $E_{\rm AdS}$ is the energy of the AdS$_5$ vacuum solution, dual to the field theory Casimir energy.  

It can be shown that the following quantum statistical relation holds, 
\begin{equation}\label{eq:qsr}
I= \beta E -\mathcal S - \beta\Omega J -\beta\Phi^I Q^{\rm gr}_I\,,
\end{equation} 
showing that $I$ is the Legendre transform of the entropy and therefore, since the latter is a  function of the conserved quantities $(E,J,Q_I)$, should be seen as a function of the  chemical potentials, $I=I\left(\beta,\Omega,\Phi^I\right)$. This leads to the interpretation of the Euclidean solution with action $I$ as a saddle of a grand-canonical partition function, $Z_{\rm grand}\simeq \rme^{-I}$.
The interpretation is in agreement with the fact that the on-shell action with Dirichlet boundary conditions is a function of the conformal boundary values of the bulk fields~\cite{Papadimitriou:2005ii}, and that these just encode the chemical potentials once regularity of the Euclidean solution is imposed.

\subsection{Supersymmetric thermodynamics}\label{sec:BPSlimit}

The supersymmetric and extremal black hole~\cite{Gutowski:2004yv} develops and infinitely long AdS$_2$ throat in its near-horizon geometry, for this reason in order to obtain a supersymmetric 
 black hole thermodynamics it is convenient to turn on some temperature as a regulator. We will follow the method of~\cite{Cabo-Bizet:2018ehj} and take the limit in two steps: first we impose the supersymmetry condition (but not the extremality one), evaluating the black hole on-shell action together with all related thermodynamic quantities in this regime, and only at the end we take the extremal limit. The advantage of this method is that the supersymmetric action turns out to be independent of the temperature, hence the extremal limit is smooth. Moreover, the supersymmetric non-extremal solutions appear to precisely capture the saddles of the dual superconformal index. The price to pay is that these configurations correspond to a complexified section of the original solution. 
At the two-derivative level, the supersymmetric limit of black hole thermodynamics for the solution under study has been discussed in~\cite{Cassani:2019mms}. We review here the main steps as a preparation to the four-derivative case, which will follow the same pattern.

The black hole is supersymmetric when the parameter $a$ satisfies the condition~\cite{Cvetic:2005zi}
\begin{equation}\label{eq:susycondition}
a\,=\,\frac{1}{g\,\text t_1 \text t_2\text t_3}\,,
\end{equation}
where we have defined $\text t_I \equiv \rme^{\delta_I}$.
As a consequence of this condition, the black hole charges \eqref{eq:conservedcharges2d} satisfy the linear relation
\begin{equation}
E \,=\, 2g\,J + \bar X^I Q^{\rm gr}_I\,.
\end{equation}
This relation does not automatically imply extremality, namely the vanishing of Hawking temperature $\beta^{-1}$. Following~\cite{Cassani:2019mms}, to study the extremal limit $m\rightarrow m^*$, it is more convenient to trade the parameter of the (Euclidean) black hole $m$ for the position of the outer horizon $r_+$, by solving $Y(r_+)=0$. Since this is a third order equation in $m$, its solutions are better manageable after a suitable change of coordinate, such that
\begin{equation}\label{eq:bpsradialcoord}
r^2\,=\, \frac{\zeta^2}{g^2}-2m\,s_1^2\,.
\end{equation}
In terms of the new coordinate $\zeta$, $Y(\zeta_+)=0$ becomes of second order in $m$ and can be easily solved as
\begin{equation}\label{eq:defmplus}
\begin{aligned}
&\frac{m_\pm}{m_*} =\\
=&\frac{\zeta_+^4\left( 2\left(\text t_1^4+1\right)\text t_2^2\,\text t_3^2-\text t_1^2\left(1+\text t_2^2\,\text t_3^2\right)\left(\text t_2^2+\text t_3^2\right)\right)+2\zeta_+^2\left(-2 + \text t_2^2\,\text t_3^2\left(1 + \text t_1^4\right)\right)-4\pm\left(\zeta_*^2-\zeta_+^2\right)\Upsilon}{2\zeta_*^2\left(-1-2\text t_2^2\,\text t_3^2+\text t_1^2\left(\text t_1^2\,\text t_2^2\,\text t_3^2+\text t_2^2+\text t_3^2\right) + \zeta_+^2\left(\text t_1^2-\text t_2^2\right)\left(\text t_1^2-\text t_3^2\right)\text t_2^2\,\text t_3^2\right)}
\end{aligned}
\end{equation}
where
\begin{equation}
\Upsilon= \frac{2}{\zeta_*^2}\sqrt{4+4\zeta_+^2\left( 1-\text t_1^4\,\text t_2^2\,\text t_3^2\right)+\zeta_+^4\,\text t_1^4\left(\text t_2^2-\text t_3^2\right)^2}\,,
\end{equation}
and
\begin{equation}\label{eq:extremalcondition}
\zeta_*^2 = \frac{2}{-1+\text t_2^2\,\text t_3^2}\,\,,\quad\quad m_* = \frac{4\text t_1^2\,\text t_2^2\,\text t_3^2}{g^2\left(-1+\text t_1^2\text t_2^2\right)\left(-1+\text t_1^2\text t_3^2\right)\left(-1+\text t_2^2\text t_3^2\right)}\,.
\end{equation}
Notice that as $\zeta_+$ becomes sufficiently close to $\zeta_*$, $m_\pm$ become imaginary due to the square-root in $\Upsilon$. Therefore, for general $\zeta_+$ we have identified a family of supersymmetric, complexified and non-extremal Euclidean solutions. In the extremal limit, that is obtained by sending $\zeta_+\rightarrow \zeta_*$, the solutions $m_\pm$ become real and equal to each other, and coincide with the BPS value $m_*$ in \eqref{eq:extremalcondition}. 
The supersymmetric non-extremal Lorentzian solution has in general closed time-like curves~\cite{Cvetic:2005zi}, but it turns out that the condition for avoiding them is to take the parameter $m$ to its extremal value \eqref{eq:extremalcondition}. Therefore, the extremal solution is regular also in Lorentzian signature.\footnote{The supersymmetric and extremal solution was first derived in~\cite{Gutowski:2004yv}. The latter depends on three independent parameters $\mu_I$ , $I=1,2,3$, related to the $\delta_I$ as
$
\mu_I = \frac{2}{\rme^{2(\delta_J + \delta_K)}-1}\,,
$
where $I,J,K$ can only assume different values. The extremal metric is regular if 
$
\mu_I >0\,,$ and $ 4\mu_1\mu_2\mu_3\left( 1+ \mu_1 + \mu_2 + \mu_3\right) > \left( \mu_1 \mu_2 + \mu_1 \mu_3 + \mu_2\mu_3\right)^2\,.
$
} 
The position of the BPS horizon can be expressed in terms of the original $r$ coordinate as
\begin{equation}\label{eq:rstar}
r_*^2 =\frac{2}{g^2} \frac{1-\text t_1^2\,\text t_2^2-\text t_2^2\,\text t_3^2-\text t_1^2\,\text t_3^2+2\text t_1^2\,\text t_2^2\,\text t_3^2}{\left(-1+\text t_1^2\text t_2^2\right)\left(-1+\text t_1^2\text t_3^2\right)\left(-1+\text t_2^2\text t_3^2\right)}\,.
\end{equation}
In the supersymmetric and extremal limit the chemical potentials \eqref{eq:thermalchemicalpotentials2d}, \eqref{eq:inversetemperature2d} become 
\begin{equation}
\begin{aligned}
&\beta^* = +\infty\,,\quad\quad \Omega^* = 2g\,\,,\quad\quad \Phi^*{}^I  = 1\,.
\end{aligned}
\end{equation}

After imposing the supersymmetry condition \eqref{eq:susycondition}, the quantum statistical relation \eqref{eq:qsr} becomes
\begin{equation}\label{eq:susyqsr}
I  \,=\, -\mathcal S - \omega J - \varphi_{\rm gr}^I Q^{\rm gr}_I\,,
\end{equation} 
where we have introduced the refined chemical potentials 
\begin{equation}\label{eq:susychempot}
\omega= \beta\left( \Omega-\Omega^*\right)\,,\quad\quad \varphi^I_{\rm gr} = \beta\left(\Phi^I-\Phi^*{}^I\right)\,.
\end{equation}
Using  \eqref{eq:susycondition}, one can check that these satisfy the linear constraint
\begin{equation}\label{eq:linearconstraint_grav}
\omega- g\, (\varphi^1_{\rm gr} + \varphi^2_{\rm gr} + \varphi^3_{\rm gr})\,=\, \pm 2\pi i\,,
\end{equation}
which ensures the correct (anti)periodicity of the Killing spinor. The sign choice  corresponds to the two branches for the supersymmetric solution given by sign choice in \eqref{eq:defmplus}. The supersymmetric on-shell action takes the simple form
\begin{equation}
I\,=\, \frac{\pi}{G}\frac{\varphi^1_{\rm gr}\varphi^2_{\rm gr}\varphi^3_{\rm gr}}{\omega^2}\,,
\end{equation}
which is independent of the inverse temperature $\beta$. Recalling \eqref{eq:conventionsveryspecialgeometry}, these findings agree with~\eqref{eq:gravityactionprediction}, \eqref{eq:constraint_gravity}, setting $\alpha=0$ and $\omega_1=\omega_2=\frac{\omega}{2}$ there.

\subsection{Corrected boundary terms}\label{sec:boundaryterms}

We now turn to the evaluation of the four-derivative corrections to the Euclidean on-shell action.
This is a sum of three different contributions
\begin{equation}
I_{ Z} \,=\, I_{\rm bulk} + I_{\rm GH} + I_{\rm count}\,,
\end{equation}
where each term has an expansion in $\alpha$. The bulk action is the Euclidean version of the one given in section \ref{sec:final_Lagr}. We denote by $I_{\rm GH}$ (from Gibbons-Hawking) the terms required to guarantee that the variational problem with Dirichlet boundary conditions on all fields is well posed, while $I_{\rm count}$ is a set of covariant boundary counterterms removing the divergences of the action.

Regarding $I_{\rm GH}$, the only terms that we need are the standard Gibbons-Hawking term for the two-derivative action, and the one associated to the Gauss-Bonnet combination of four-derivative  curvature terms~\cite{Myers:1987yn}. Putting these together, $I_{\rm GH}$ reads
\begin{equation}\label{eq:Igh}
\begin{aligned}
I_{\rm GH}=& -\frac{1}{8\pi G}\int_{\partial\mathcal M} \text d^4 x\sqrt{h}\,{\cal K} + \\
&+\frac{\alpha}{4\pi G}\int_{\partial\mathcal M} \text d^4 x\sqrt{h}\, \left(\lambda_I X^I\right) \left(\frac{1}{3}{\cal K}^3- {\cal  K} {\cal K}^{ij}{\cal K}_{ij} + \frac{2}{3}{\cal K}_{ij}{\cal K}^{jk}{\cal K}_k{}^i+2 \mathcal G_{ij}{\cal K}^{ij} \right)\,,
\end{aligned}
\end{equation}
where ${\cal K}_{ij}$ is the extrinsic curvature of the boundary, $\mathcal G_{ij} = \mathcal R_{ij}- \frac{1}{2}h_{ij} \mathcal R$ is the Einstein tensor of the induced boundary metric $h_{ij}$, with $\mathcal R_{ij}$ built out of $h_{ij}$. Additional boundary terms that are produced by studying the variational problem of the bulk action are either vanishing under Dirichlet boundary conditions or sufficiently suppressed in asymptotically locally AdS spacetimes 
(cf.~\ref{sec:structure_2}). We emphasize that an advantage of having performed the field redefinitions described in appendix~\ref{app:field_redefinitions}, is that we can use a basis of curvature-square invariants comprising the Gauss-Bonnet combination, for which the appropriate boundary term is known.

We next discuss the counterterm action $I_{\rm count}$. Schematically, this is a sum of three contributions,
\begin{equation}
I_{\rm count} \,=\, I^{(0)}_{\rm count} + I^{(1)}_{\rm count} + \widetilde I^{\,(1)}_{\rm count}\,,
\end{equation}
where $I_{\rm count}^{(0)}$ comprises the counterterms for the $\alpha=0$ theory. If we restrict  to  solutions with no non-normalizable modes for the scalar fields, such as the solution \eqref{eq:metric5d}, \eqref{eq:scalars5d}, then for $I_{\rm count}^{(0)}$ it is sufficient to consider~\cite{Cassani:2019mms}
\begin{equation}\label{eq:Ict0}
I^{(0)}_{\rm count} =- \frac{1}{16\pi G}\int \text d^4x \sqrt{h}\,\left( \mathcal W + \Xi \, \mathcal R\right)\,,
\end{equation}
where 
\begin{equation}\label{eq:countertermsfunction}
\mathcal W = -6g\, \bar X_I X^I\,\,, \quad\quad\quad \Xi = -\frac{1}{2g}\,\bar X^I X_I\,.
\end{equation}
Turning to the corrections, $I^{(1)}_{\rm count}$ defines the counterterms needed to renormalize the higher-derivative terms coming from the terms that couple to $\lambda_I$, whereas $\widetilde I^{\,(1)}_{\rm count}$ is associated to the terms that couple to $\widetilde\lambda_{IJK}$. 
For what concerns $I^{(1)}_{\rm count}$, following the prescription in~\cite{Cremonini:2009ih,Cassani:2022lrk}, we expect that the counterterms are obtained through a shift of order $\alpha$ in the coefficients of the counterterms that are already present in the two-derivative theory. For the minimal supergravity theory, this has been demonstrated by applying the Hamilton-Jacobi method, see~\ref{sec:holographicrenormalization}. We expect that the same prescription is valid here as well, since the divergences of the on-shell action come from $I_{\rm GH}$ rather than from the bulk integral, analogously to the case analysed in the mentioned references. Assuming this ansatz and demanding cancellation of the divergences we arrived at the following counterterm,
\begin{equation}\label{eq:Ict1}
\begin{aligned}
I_{\rm count}^{(0)} + I_{\rm count}^{(1)}=-\frac{1}{16\pi G}&\int_{\partial\mathcal M}\text d^4 x\sqrt{h}\left [\left(1-\frac{16}{3}g^2\alpha\,\lambda_I X^I\left(\bar X_K X^K\right)^2 \right)\mathcal W  \right.\\[1mm]
&\qquad\quad\ \qquad +\left( 1+ 8g^2\alpha\,\lambda_I X^I\left(\bar X_K X^K\right)^2\right)\Xi\,\mathcal R
\bigg ]\,.
\end{aligned}
\end{equation}
In order to obtain $\widetilde I_{\rm count}^{(1)}$, we recall that the theory including the corrections proportional to $\widetilde\lambda_{IJK}$ is equivalent to the two-derivative theory \eqref{eq:tilded_Lagr} together with the substitution rules \eqref{eq:subtildeaction1}, \eqref{eq:subtildeaction2}. Also, considering the supersymmetric AdS$_5$ condition \eqref{scalars_AdS_sol} in the theory \eqref{eq:tilded_Lagr}, we see that  the corresponding inverse AdS radius $\widetilde{g} = \frac{1}{\sqrt2} \,g_I\overline{\widetilde X}{}^I$  is mapped into
\begin{equation}\label{eq:tildeg}
\widetilde g \,=\, g\left( 1- \frac{\alpha}{3}\,\widetilde\lambda_{IJK}\bar X^I \bar X^J \bar X^K  \right)\,.
\end{equation}
The counterterms can, then, be derived starting from the two-derivative expression \eqref{eq:countertermsfunction}, now for the theory with tilded variables, and applying these substitution rules.
We find
\begin{equation}\label{eq:Ict2}
\begin{aligned}
I_{\rm count}^{(0)}+ \widetilde I_{\rm count}^{\,(1)} =&\, -\frac{1}{16\pi G}\int_{\partial\mathcal M} \text d^4 x\sqrt{h}\ \bigg[ \left(1-\frac{\alpha}{3}\,\widetilde\lambda -\alpha\,\widetilde\lambda_{IJK}\bar X^I \bar X^J\bar X^K \right)\mathcal W \\[1mm]
&\,-6g\,\alpha\,\widetilde\lambda_{IJK} X^I\bar X^J\bar X^K+ \left(1-\frac{2}{3}\alpha\,\widetilde\lambda\right)\Xi\,\mathcal R - \frac{\alpha}{6g}\,\bar X^I\,\widetilde\lambda_I\,\mathcal R
\bigg]\,,
\end{aligned}
\end{equation}
where $\widetilde\lambda_I$ and $\widetilde\lambda$ were defined in \eqref{eq:tildel_contracted}.
 
Putting everything together, we can give a set of counterterms for the bulk action \eqref{eq:finalL} that removes all $\sim r^4$ and $\sim r^2$ divergences from the on-shell action of asymptotically AdS solutions,
\begin{equation}\label{eq:generalcounterterms}
\begin{aligned}
I_{\rm count}
&= -\frac{1}{16\pi G}\!\int_{\partial\mathcal M} \!\!\text d^4 x\sqrt{h}\bigg\{\!\!\left[1-\alpha\left(\frac{16}{3}g^2\lambda_I X^I\left(\bar X_J X^J\right)^2+\frac{1}{3}\widetilde\lambda+\,\widetilde\lambda_{IJK}\bar X^I \bar X^J\bar X^K\right)\right] \mathcal W\quad\,\\[1mm]
& -6g\,\alpha\,\widetilde\lambda_{IJK} X^I\bar X^J\bar X^K+\left[\left(1+\alpha\,\Big(8g^2\lambda_I X^I\left(\bar X_J X^J\right)^2-\frac{2}{3}\,\widetilde\lambda\Big)\right)\Xi -\frac{\alpha}{6g}\bar X^I \widetilde\lambda_I\right]\mathcal R
\bigg\}\,.
\end{aligned}
\end{equation}

A rigorous derivation of these counterterms should be possible by adapting the arguments of~\cite{Batrachenko:2004fd,Cassani:2023vsa,Cremonini:2009ih} to the more involved case under study here. 
 However, the present effective prescription appears to give the correct set of counterterms for computing the supersymmetric action we are interested in, as we will discuss later.

\subsection{Setting up the computation}\label{sec:setupaction}

We want to evaluate the on-shell action at linear order in the four-derivative parameter $\alpha$. In order to do this, it is in fact not necessary to know the $\mathcal O(\alpha)$ corrected solution. The argument proving this claim can be found in~\cite{Reall:2019sah} for asymptotically flat solutions, and was adapted to asymptotically AdS spacetimes without scalars in section~\ref{sec:onshellact}, based on~\cite{Cassani:2022lrk}. Here we revisit it specifying its conditions of validity in the presence of scalar fields. 

Schematically, the on-shell action has the form $I(\Psi)\equiv I^{(0)}(\Psi) + \alpha \, I^{(1)}(\Psi)$, where by $\Psi$ we denote the set of all fields, in this case $\Psi = (g_{\mu\nu} \,,\, A_\mu^I \,,\, X^I(\phi^x))$. When we include $\mathcal O(\alpha)$ corrections, a generic solution of the equations of motion is expressed as $\Psi = \Psi^{(0)} + \alpha\, \Psi^{(1)}$. We can, then, expand the action at first order in $\alpha$ around the leading-order solution,
\begin{equation}
I(\Psi) \,=\, I^{(0)}(\Psi^{(0)})+ \alpha\left( \partial_\alpha I^{(0)}(\Psi^{(0)}, \Psi^{(1)})+ I^{(1)}(\Psi^{(0)}) \right) + \mathcal O(\alpha^2)\,.
\end{equation}
As shown in~\cite{Cassani:2022lrk}, the only term that depends on the corrected solution, that is $\partial_\alpha I^{(0)}$, indeed vanishes when the two-derivative equations of motion are satisfied, up to a boundary term, that has the schematic form
\begin{equation}
\begin{aligned}\label{eq:bdry_term_corrections}
\partial_\alpha I^{(0)} \,\sim\, \int_{\partial\mathcal M_{r_{\rm bdry}}} \!\!\!\text d^4x&\,\bigg [\,\frac{\delta I^{(0)}}{\delta h^{ij}} \bigg |_{\Psi^{(0)}}\,h^{(1)}{}^{ij}+ \frac{\delta I^{(0)}}{\delta A^I_i}\bigg |_{\Psi^{(0)}}\, A_i^{(1)}{}^I+\\
&+\left(\mathtt n^\mu \frac{\delta I^{(0)}_{\rm bulk}}{\delta(\partial^\mu X^I)}+
\frac{\delta I^{(0)}_{\rm count}}{\delta X^I}\right)_{\Psi^{(0)}}\,\left(\delta^I{}_J- X^I X_J\right)X^{(1)}{}^J
\bigg]\,,
\end{aligned}
\end{equation}
where $r_{\rm bdry}$ is a radial cut-off that we introduce to regulate the divergences, such that the boundary $\partial\mathcal M_{r_{\rm bdry}}$ is located at $r=r_{\rm bdry}$ and $\mathtt n$ is the outward pointing normal vector. Eventually, the cutoff is removed by taking $r_{\rm bdry}\rightarrow +\infty$. The projector $\left(\delta^I{}_J- X^I X_J\right)$ ensures that only corrections satisfying the condition $X_I X^{(1)I}=0$ play a role (recall that  in our conventions the scalars satisfy the constraint $X_I X^I =1$ both at the two- and at the four-derivative level). 

The boundary term \eqref{eq:bdry_term_corrections} vanishes if we impose boundary conditions such that the $\mathcal O(\alpha)$ corrections  to the solution do not modify the leading-order asymptotic behaviour of the fields, so as not to modify the original Dirichlet boundary conditions. When scalar fields are present, we should distinguish whether their non-normalizable modes are turned on in the two-derivative solution or not, due to the different asymptotic behaviour in these two cases. We will focus on the  scalars of interest for this paper, which have squared mass $m^2 = -4g^2 $ and couple to operators with conformal dimension $\Delta=2$; their asymtptotic behavior can be found e.g.\ in~\cite{Bianchi:2001kw}, and is $\mathcal O(r_{\rm bdry}^{-2} \,\log r_{\rm bdry}^2)$ for the non-normalizable modes and $\mathcal O(r_{\rm bdry}^{-2})$ for the normalizable modes.

For solutions where the scalars have non-normalizable modes, the asymptotic behaviour of the fields is such that\footnote{In this case we need to use the general set of counterterms given in section 4 of~\cite{Cassani:2018mlh}.} 
\begin{equation}\label{eq:asymptoticboundarybehaviour}
\begin{aligned}
&\frac{\delta I^{(0)}}{\delta h^{ij}}\sim \mathcal O \left(r_{\rm bdry}^2\right)\,,\quad\quad \frac{\delta I^{(0)}}{\delta A_i}\sim \mathcal O \left(r_{\rm bdry}^0\right)\,, \\[1mm]
&\left(\mathtt n^\mu \frac{\delta I^{(0)}_{\rm bulk}}{\delta(\partial^\mu X^I)}+
\frac{\delta I^{(0)}_{\rm count}}{\delta X^I}\right)\left(\delta^I{}_J- X^I X_J\right)\ \sim\ \mathcal O\left(\frac{r_{\rm bdry}^2}{\log r_{\rm bdry}^2}\right)\,\,.
\end{aligned}
\end{equation}
As a consequence of \eqref{eq:asymptoticboundarybehaviour}, the boundary term vanishes if the corrected solution has the asymptotic behavior
\begin{equation}
h_{ij}^{(1)}\,=\, \mathcal O\big(r_{\rm bdry}^0\big) \,,\quad\quad A_i^{(1)}{}^I\,=\,\mathcal O\big(r_{\rm bdry}^{-2}\big)\,,\quad\quad X^{(1)}{}^I\,=\, \mathcal O\big(r_{\rm bdry}^{-2}\big)\,.
\end{equation}
As shown in~\cite{Cassani:2022lrk}, for the metric and the gauge fields this condition can always be met through a choice of gauge, and requires that the induced fields at the conformal boundary are not modified by $\mathcal O(\alpha)$ corrections. This is in harmony with the fact that the Dirichlet boundary conditions define a grand-canonical ensemble, where the chemical potentials, that can be read from the boundary values of the bulk metric and gauge fields after imposing regularity of the Euclidean solution, are kept fixed at their two-derivative expressions when including the corrections. 

However, in this work we are interested in two-derivative solutions where the scalars only have normalizable modes as in  \eqref{eq:scalars5d}, in which case one has the regular behaviour,
\begin{equation}
\left(\mathtt n^\mu \frac{\delta I^{(0)}_{\rm bulk}}{\delta(\partial^\mu X^I)}+
\frac{\delta I^{(0)}_{\rm count}}{\delta X^I}\right)\left(\delta^I{}_J- X^I X_J\right) \ \sim\ \mathcal O\left(r_{\rm bdry}^0\right)\,.
\end{equation}
Consequently, for the last term in \eqref{eq:bdry_term_corrections} to vanish it is sufficient that the vacuum value of the corrected scalars, which is of order $\sim r_{\rm bdry}^0$, is not modified, namely
\begin{equation}\label{eq:asymptbehaviour}
X^{(1)}{}^I= \mathcal O\left(r_{\rm bdry}^{-2} \log r_{\rm bdry}^2\right)\,.
\end{equation}

In order to understand if this condition is satisfied, we need to determine the $\mathcal{O}(\alpha)$ corrections to the value of the scalars in the supersymmetric AdS$_5$ solution. We thus set $A_\mu^I=0$, choose the AdS$_5$ metric, and take constant scalars $X^I =  \bar X^I + \alpha \, \bar X^{(1)}{}^I$. 
By studying the scalar field equations it is possible to show that the terms in the Lagrangian proportional to $\lambda_I$ do not correct the vacuum value of the scalars, whereas the terms depending on $\widetilde\lambda_{IJK}$ imply the following relation\footnote{We interpret this as a corrected supersymmetric AdS$_5$ vacuum condition. Indeed one obtains the same relation by starting from the two-derivative theory with tilded variables in \eqref{eq:tilded_Lagr}, writing the supersymmetric vacuum condition \eqref{scalars_AdS_sol} in this theory, that is $g_I = \sqrt 2\, \widetilde g\, \widetilde{C}_{IJK}\widetilde X^J\widetilde X^K$, and transforming the latter to the present variables using \eqref{eq:subtildeaction1}, \eqref{eq:subtildeaction2} and~\eqref{eq:tildeg}.}
\begin{equation}
g_I = \sqrt{2}\,g\left[\,\bar a_{IJ}\left(\bar X^J + \alpha \, \bar X^{(1)}{}^J\right)\left(1-\alpha \,\widetilde\lambda_{KLM}\bar X^K \bar X^L\bar X^M\right) +\alpha\,\widetilde\lambda_{ILM}\bar X^L \bar X^M \right]\,.
\end{equation}
Therefore, the corrections to the scalars in the supersymmetric vacuum have the form
\begin{equation}\label{corr_to_scalars}
\bar X^{(1)}{}^I= -\bar a^{IJ}\left(\delta_J{}^K-\bar X_J  \bar X^K \right)\widetilde\lambda_{KLM}\bar X^L \bar X^M\,.
\end{equation}
Hence, our condition \eqref{eq:asymptbehaviour} is satisfied only for a choice of coupling $\widetilde\lambda_{IJK}$ such that 
\begin{equation}\label{eq:secondinvariantcoefficient}
\left(\delta_I{}^J -  \bar X_I \bar X^J  \right)\widetilde\lambda_{JKL}\bar X^K \bar X^L=0\,.
\end{equation}

In fact the corrections we study, specified in \eqref{choice_tildelambda}, do satisfy \eqref{eq:secondinvariantcoefficient}. This proves that it is legitimate to use the two-derivative solution to evaluate the corrected action.

We conclude this section commenting on the field theory interpretation of \eqref{corr_to_scalars}, \eqref{eq:secondinvariantcoefficient}. Let us first consider just the effect of the invariant controlled by ${\widetilde \lambda}_{IJK}$. Using the holographic dictionary above, this means the anomaly coefficients $k_{I}$ vanish, while $k^{(1)}_{IJK}$ are left arbitrary, proportional to ${\widetilde \lambda}_{IJK}$. Then, one can show that $a$-maximization at next-to-leading order in the large-$N$ expansion (see appendix~\ref{amax_at_largeN}) imposes
\be
-2k^{(0)}_{IJK} {\bar s}^J \delta {\bar s}^{K}=(\delta_{I}{}^{J}-r_{I}{\bar s}^J)k^{(1)}_{JKL}{\bar s}^K{\bar s}^L\,.
\ee
This is precisely the field theory dual of \eqref{corr_to_scalars}, and from the above equation we see that \eqref{eq:secondinvariantcoefficient} corresponds to demanding that the coefficients $\bar s^I$ that determine the superconformal R-symmetry at leading-order do not receive corrections, $\bar s^I +\delta\bar s^I =\bar s^I$. In this case \eqref{eq:secondinvariantcoefficient} states that the coupling $\widetilde \lambda_{IJK}$ with one index projected on the flavour symmetries and the remaining two projected on the superconformal R-symmetry vanishes. 

Turning on the invariant controlled by $\lambda_I$ does not modify this discussion as its contribution to $k^{(1)}_{IJK}$ and $k_I$ automatically cancel each other when implementing $a$-maximization. This is reflected in supergravity by the fact that $\lambda_I$ does not correct the value of the scalars in the supersymmetric AdS$_5$ vacuum.

\subsection{Results for the supersymmetric on-shell action}\label{sec:resultsandmatch}

Despite the great simplification following from evaluating the corrected action on the uncorrected solution, the evaluation of the higher-derivative terms on the fields \eqref{eq:metric5d}, \eqref{eq:scalars5d} remains a technically challenging problem. We have not been able to perform a fully analytic computation of the four-derivative on-shell action in the most general case, so we have resorted to a mix of numerical and analytical checks of the result. 
This was done by assigning many different numerical values to the black hole parameters $\delta_I$, while keeping the remaining parameters $a$, $m$ (and $g$) analytic.
However,  in some limits we could carry out the full computation analytically: this was possible
 in the ungauged limit $g=0$ 
 keeping all electric charges independent, and in the case where $\lambda_I =0$ and two of the three charges are set equal.

A non-trivial consistency check we have made on our result has been to verify that the Gibbs free energy $\mathcal G = I/\beta$, vanishes in our supersymmetric and extremal limit. This means that the action $I$ can (and will) be finite. An extremal but non-supersymmetric expression for $I$ would instead give a finite $\mathcal G$ as the $\mathcal{O}(\beta)$ terms on the right hand side of \eqref{eq:qsr} would fail to cancel between them. Here, it is important to recall from section~\ref{sec:blackhole} that our definition of $I$ is such that $I = I_{Z} - I_{\rm AdS}$, and that the AdS action to be subtracted receives $\mathcal{O}(\alpha)$ corrections.
 For the case at hands we find
\begin{equation}
I_{\rm AdS} \,=\, \frac{3\pi\beta}{32Gg^2}\left( 1+ \alpha\, \widetilde\lambda_{IJK}\bar X^I\bar X^J\bar X^K\right) \,=\, \frac{3}{4}g\beta\, \mathtt a\,.
\end{equation}
In the last equality we used~\eqref{a_holo_corrected} to rewrite the expression in terms of the corrected Weyl anomaly coefficient  $\mathtt a$ of the dual SCFT and the ratio $g\beta$ of the $S^1$ and $S^3$ radii on $\partial\mathcal{M}$. 

After this consistency check, in order to obtain the supersymmetric action, following the procedure highlighted in section \ref{sec:BPSlimit}, we substitute $a$ with its supersymmetric value \eqref{eq:susycondition}, and define the supersymmetric chemical potentials as \eqref{eq:susychempot}. Then, we change coordinate using \eqref{eq:bpsradialcoord}, and trade the dependence on $m$ with that on $\zeta_+$ via \eqref{eq:defmplus}.
Again, we find that the supersymmetric on-shell action in this regime has a complicated dependence on the parameters of the solution, however when written in terms of the supersymmetric chemical potentials \eqref{eq:susychempot} it takes a simple form. This is independent of the value of $\beta$, namely it is valid both before and after taking the limit $\zeta_+ \rightarrow \zeta_*$.  It reads 
\begin{equation}
\begin{aligned}
I 
\,&=\, \frac{\pi}{G}   \frac{\varphi_{\rm gr}^{1} \varphi_{\rm gr}^{2} \varphi_{\rm gr}^{3}}{\omega^{2}}     -\frac{2\pi}{G}\alpha\lambda_I \varphi_{\rm gr}^{I} \,\frac{\omega  \mp 8\pi i }{\omega}\,,
\end{aligned}
\end{equation} 
which perfectly agrees with the field theory prediction~\eqref{eq:gravityactionprediction} (in its simplified form~\eqref{eq:gravityactionpredictionBis}) after setting $\omega_1=\omega_2=\frac{\omega}{2}$ and using the identifications in section~\ref{sec:special_relations}.
It follows that the corrected  entropy for the supersymmetric extremal black hole under study is given by the Legendre transform of this expression, given in \eqref{eq:entropy_orbifolds_eqJ}.

\paragraph{More general validity of our results.}
We have also computed the supersymmetric on-shell action for a more general choice of higher-derivative couplings $\lambda_I$, $\widetilde \lambda_{IJK}$, consisting of those which do not require knowledge of the corrected solution to compute the action at $\mathcal O(\alpha)$. As explained in section \ref{sec:setupaction}, these are the corrections that satisfy \eqref{eq:secondinvariantcoefficient}. In these computations, we employ the same set of counterterms \eqref{eq:generalcounterterms} to remove the divergences.
Again, we have performed extensive numerical checks of the final form of the on-shell action.\footnote{Specifically, we define $\lambda_I = \alpha_1\,A_I$, $\widetilde\lambda_{IJK} = \alpha_2\, B_{IJK}$, for some constant parameters $\alpha_{1,2}$ and where the entries of $A_I$ and $B_{IJK}$ are randomly generated integers. The black hole parameters $\text t_I$ are chosen as randomly generated primes. The computation is still analytic with respect to the remaining black hole parameters and $g$.}  
For all choices that satisfy \eqref{eq:secondinvariantcoefficient}, we have obtained that the supersymmetric action can be written as
\begin{equation}
I \,=\, \frac{\pi}{G} \left(\frac16|\epsilon_{IJK}| - 6\alpha\,\lambda_{I}g_{J}g_{K} +\alpha\,\widetilde\lambda_{IJK} \right) \frac{\varphi_{\rm gr}^{I} \varphi_{\rm gr}^{J} \varphi_{\rm gr}^{K}}{\omega^{2}}  +\frac{2\pi}{G}\alpha\lambda_I \varphi_{\rm gr}^{I} \,\frac{\omega^{2}-8\pi^2}{\omega^{2}}\,.
\end{equation}  
This supports the expectation that the holographic match we have been discussing in this work should hold beyond the specific example considered.

\section{Discussion and open problems}\label{sec:Conclusions}

%
In the second part of this Thesis, we have studied four-derivative corrections to the thermodynamics of asymptotically AdS black holes in five-dimensional gauged supergravity. Our analysis covered both solutions of minimal gauged supergravity and certain multi-charge black holes.  At the two-derivative level, these belong to a ${\rm U}(1)^3$ gauged supergravity, which arises as a consistent truncation of Type IIB supergravity on $S^5$ or the orbifold $S^5/\Gamma$, dual to ${\rm SU}(N)$ $\mathcal{N}=4$ SYM or the $\mathbb{C}^3/\Gamma$ quiver theories. In these setups, the corrections to the action were determined by requiring that the Chern-Simons terms reproduce the $1/N^2$ corrections to the SCFT 't Hooft anomaly coefficients, and then supersymmetrizing these terms. In $\mathcal{N}=4$ SYM the effect of $1/N^2$ corrections is especially simple, amounting to the exact replacement $N^2 \to N^2-1$ in the cubic anomaly coefficient. We therefore focused more on the orbifold quiver theories, where the corrections exhibit a richer structure.

We have computed the on-shell action at linear order in the corrections. In the minimal case, we also verified that the counterterms we used to renormalize the action are consistent with the prescriptions of holographic renormalization. In both minimal and matter-coupled supergravity, the action can be expressed in terms of the supersymmetric chemical potentials $\omega_1,\omega_2$ and matches the prediction from the Cardy-like limit of the dual (flavoured) superconformal index. For the cases of interest to us, explicit knowledge of the corrected solution was not needed. We have specified the condition for this property to hold in the presence of scalar fields, and interpreted it holographically as the superconformal R-symmetry not receiving corrections at next-to-leading order in the large-$N$ expansion. For multi-charge black holes the match was established only in the simplifying case of equal angular momenta, due to the intrinsic difficulty of dealing with higher-derivative terms in geometries with many parameters and reduced symmetry. Extending the analysis to unequal angular momenta (starting from the two-derivative solution given in~\cite{Wu:2011gq}) should be conceptually straightforward, though computationally even more demanding.

We have also derived the corrected supersymmetric and extremal black hole entropy as a function of the conserved charges by taking the Legendre transform of the action. 
In the minimal theory, and in the limit $J_1=J_2$, we confirmed this result independently by computing the corrected near-horizon geometry of the BPS black hole of~\cite{Gutowski:2004ez} and evaluating Wald's formula. A similar direct check for multi-charge black holes remains out of reach, as it requires the explicit corrected solution, or at least its extremal near-horizon limit. In that case the entropy could still be obtained from Wald’s formula, while the charges could be evaluated by extending the methods of chapter~\ref{sec:structure_2} to include vector multiplet couplings. It should then be possible to express the entropy as a function of the charges just using near-horizon data and match the prediction we obtained in these chapters.

\medskip

It is natural to ask how our effective field theory arises from top-down constructions accounting for different stringy and/or quantum effects. In particular, one can focus on obtaining the Chern-Simons terms, which directly capture the central charges $\aa,\cc$ of the dual field theory. The higher-dimensional origin of the $\epsilon^{\mu\nu\rho\sigma\lambda} {\rm} R_{\mu\nu}{}^{\alpha\beta} R_{\rho\sigma\alpha\beta}A_\lambda$ term in Type IIB compactifications on certain Sasaki-Einstein manifolds was studied e.g.\ in~\cite{Liu:2010gz,ArabiArdehali:2013jiu,ArabiArdehali:2013vyp}. For these terms, one-loop effects from the tower of supergravity Kaluza-Klein modes are sufficient: the R-symmetry linear anomaly coefficient $k_R \sim (\aa-\cc)$ was shown to be correctly reproduced this way in~\cite{ArabiArdehali:2013jiu}. Even when the coefficient $\lambda_1\sim (\aa-\cc)$ vanishes exactly (as for Type IIB on $S^5$) there are still corrections to $\epsilon^{\mu\nu\rho\sigma\lambda} F_{\mu\nu}  F_{\rho\sigma}A_\lambda$ controlled by $\lambda_2$,  the other parameter  in our effective action. These also arise from quantum effects in the Kaluza-Klein towers. In the case of Type IIB on $S^5$, gauge modes running in the loop are responsible for the shift $N^2\to N^2-1$ in $\aa=\cc$\ \cite{Bilal:1999ph} (see also the discussion in~\cite{Beccaria:2014xda}).  It would be interesting to study these effects more systematically, possibly in more general compactifications.

A different type of corrections, arising from $\alpha'{}^3$ eight-derivative terms in Type IIB on $S^5$, has been analyzed in~\cite{Melo:2020amq}. These do not generate the terms in our effective action. They encode corrections in the 't Hooft coupling and thus cannot contribute to the holographic description of the SCFT index, which is independent of continuous parameters. In fact, the authors of~\cite{Melo:2020amq} evaluated these corrections on the black hole solution of~\cite{Chong:2005hr}, finding that they yield a vanishing contribution to the on-shell action when supersymmetry is imposed.

It would also be intriguing to study how the four-derivative corrections we consider affect the near-BPS thermodynamics that has recently been studied, e.g. in~\cite{Boruch:2022tno}. In particular, one could extend the Schwarzian theory of \cite{Boruch:2022tno} and determine the contributions of our $\lambda_1,\lambda_2$ terms to the mass gap found there.

Finally, our results are expected to hold beyond the specific AdS/CFT dual pairs discussed here, and we have given some evidence in this sense. It will be intriguing to explore further the expectation that, given a holographic $\mathcal{N}=1$ SCFT$_4$,  the higher-derivative five-dimensional gauged supergravity reproducing its 't Hooft anomalies admits an asymptotically AdS$_5$ supersymmetric black hole solution whose on-shell action matches the SCFT formula~\eqref{eq:index_asympt}. 
It is likely that {\it equivariant localization}, whose role in supergravity has been recently emphasized starting with~\cite{BenettiGenolini:2023kxp,Martelli:2023oqk} and that will be extensively reviewed in the following chapters, could be helpful in this direction. Indeed, the localization theorem in equivariant cohomology allows us to reduce complicated integrals, such as on-shell actions, to simpler ones localized at the fixed points of certain spacetime isometries, showing that the final result can be expressed in terms of a limited set of data, such as topology and boundary conditions, without the need to know the full explicit geometry, but only assuming the existence of a solution with the prescribed properties.

%% file: research_four.tex
\part{Asymptotically flat saddles for the gravitational index}
\label{part_three}

\chapter{Introduction to part III}
\label{sec:intro3}

In the third part of this Thesis we investigate how the general ideas of the gravitational index and its supersymmetric finite-temperature saddles apply when asymptotically flat boundary conditions are imposed. As discussed above, in gauged supergravity theories with a CFT dual these ideas can be made precise thanks to the guidance of the dual description, allowing for detailed comparisons between macroscopic and microscopic predictions. Our aim is to transfer the lessons learned by studying AdS black holes to the case of asymptotically flat spacetimes, where no dual SCFT is available for comparison. Indeed, although progress on Euclidean saddles of the gravitational index was initially driven by AdS black holes, the asymptotically flat case provides a simpler setting in which to construct supersymmetric solutions and study their contributions to the index, for instance by building on the general classification of~\cite{Gauntlett:2002nw}. We therefore expect this setup to yield valuable insights into the structure of supersymmetric indices.

Recently, it has been understood that supersymmetric asymptotically flat black holes contribute to the gravitational grand-canonical partition function with supersymmetric boundary conditions, which takes the form of a refined Witten index. The situation is analogous to the AdS case discussed in chapter~\ref{sec:background}: while in Lorentzian signature supersymmetric black holes are necessarily extremal and therefore develop an infinitely long throat that makes a direct evaluation of the action ill-defined, in the Euclidean regime one can consider analytic continuations of the black hole that are supersymmetric but non-extremal. These contribute with a finite action that is independent of the inverse temperature $\beta$, that in turn reproduces also the extremal action upon sending $\beta\to \infty$. This provides an effective prescription to assign a finite on-shell action and chemical potentials to the extremal supersymmetric black holes that have a well-behaved Lorentzian section. Progress in this direction has included studies of supersymmetric indices in four dimensions~\cite{Iliesiu:2021are,Hristov:2022pmo,H:2023qko,Boruch:2023gfn,Hegde:2023jmp,Boruch:2025biv}, in five dimensions~\cite{Anupam:2023yns,Cassani:2024kjn,Adhikari:2024zif,Boruch:2025qdq,Cassani:2025iix,Boruch:2025sie}, as well as for small black holes~\cite{Chowdhury:2024ngg,Chen:2024gmc,Bandyopadhyay:2025jbc}, making this an established and promising arena for future developments.

\medskip

In this Thesis, we work in five-dimensional supergravity and impose supersymmetric boundary conditions defining a grand-canonical supersymmetric index. We will mostly discuss pure supergravity, but also consider multi-charge black hole saddles in presence of couplings to vector multiplets. 

Let us set the stage by briefly recalling how one can define a supergravity path integral calculating the supersymmetric index of interest (focusing on the minimal theory for simplicity of notation), following the steps highlighted in sec.~\ref{sec:Grav_path_int}:
\begin{enumerate}
\item We consider a five-dimensional asymptotically flat spacetime where one can define the energy $E$, two independent angular momenta $J_1,J_2$,  
 and an electric charge $Q$. These conserved quantities generate evolution along Lorentzian time $t$, translations along $2\pi$-periodic angular coordinates $\phi_1,\phi_2$, and gauge transformations of the supergravity abelian vector field, respectively. Now we Wick-rotate $t=-\ii \tau$ and compactify the Euclidean time $\tau$ to a circle $S^1$ of period $\beta$, which we interpret as an inverse temperature. We also consider a supercharge ${\cal Q}$. In ungauged supergravity the electric charges commute with the chosen supercharge ($r_I =0$ in~\eqref{eq:supergravity_algebra}), and we choose conventions such that the coefficients appearing in~\eqref{eq:supergravity_algebra} take the values
 \begin{equation}
 \Omega_1^* = \Omega_2^* = 0\,,\qquad \Phi_* = \sqrt{3}\,,
 \end{equation}
 implying that the BPS bound reads
 \begin{equation}
 \label{eq:superalgebraaf}
 E -\sqrt{3} Q =0\,.
 \end{equation}
\item We specify the angular coordinate identifications, including those for a revolution around the thermal $S^1$, as
\begin{equation}
\label{eq:globalidentifications0}
\left( \tau,\,\phi_1,\,\phi_2\right) \sim \left( \tau + \beta,\,\phi_1 - \ii  \omega_1,\,\phi_2 - \ii\omega_2\right) \sim \left( \tau ,\,\phi_1 + 2\pi,\,\phi_2\right) \sim \left( \tau ,\,\phi_1,\,\phi_2+ 2\pi\right)\,,
\end{equation}
where $\omega_i$, $i=1,2$, are refined angular velocities, related to the usual angular velocities $\Omega_i$ appearing in black hole thermodynamics by \eqref{eq:red_chem_pot}. 
 We also introduce an electrostatic potential $\Phi$ via the holonomy of the supergravity abelian gauge field around the asymptotic circle, $\beta\Phi \,=\, -\ii \int_{S^1_\infty} A$. 
With these boundary conditions, together with suitable Dirichlet-like boundary conditions for the fields in the theory, the gravitational path integral \eqref{eq:grav_path_int} can be seen as a thermal partition function in the grand-canonical ensemble, depending on the potentials. When the theory is embedded in a UV-complete theory such as string theory, this partition function may be interpreted as a trace over microstates of the form
\begin{equation}\label{eq:ZasTrace}
Z_{\rm grav}(\beta,\omega_1,\omega_2,\Phi)\, =\, {\rm Tr} \,\rme^{-\beta\left( E -\Phi Q\right) + \omega_1 J_1 + \omega_2 J_2 }\,.
\end{equation}
\item We choose supersymmetric boundary conditions, so that the partition function above reduces to a refined Witten index~\cite{Witten:1982df}, receiving contributions from supersymmetric states only. 
To do so, we need the argument of the trace \eqref{eq:ZasTrace} to anticommute with the supercharge, so that pairs of bosonic and fermionic states related to each other by the action of $\mathcal{Q}$ give opposite contributions and cancel out. This is obtained by demanding that the refined chemical potentials satisfy a relation of the form \eqref{eq:constraint_gauged}, which in the present setup becomes
\be
\label{eq:susyconstraint}
\omega_1+\omega_2 \,=\,  2\pi \ii \qquad (\text{mod}\ 4\pi \ii )\,.
\ee
 Solving \eqref{eq:susyconstraint} for $\omega_1$ and noting that $\rme^{\pm2\pi \ii  J_1} = (-1)^{\rm F}$, we can write
$\rme^{\omega_1 J_1+\omega_2J_2} = (-1)^{\rm F} \rme^{\omega_2 (J_2-J_1)}\,$. Also using the expression for $E$ from the first in~\eqref{eq:superalgebraaf}, and redefining the electrostatic potential as in \eqref{eq:red_chem_pot}, the trace~\eqref{eq:ZasTrace} eventually takes the form of a refined Witten index,
\begin{equation}
\label{eq:microscopicindex}
Z_{\rm grav}(\omega_2, \varphi)\, =\, {\rm Tr}\, \left(-1\right)^{\rm F} {\rm e}^{-\beta\{\mathcal Q,\,\bar{\mathcal Q} \}} {\rm e}^{\omega_2\left( J_2-J_1\right)+\varphi Q}\,,
\end{equation}
that only receives contributions from states preserved by the action of $\mathcal Q$, $\bar{\mathcal Q}$ and is independent of $\beta$.
We will denote the gravitational path integral with chemical potentials $\omega_1,\omega_2,\varphi$ satisfying the constraint \eqref{eq:susyconstraint} the {\it gravitational index}. The trace is expected to be convergent if
\begin{equation}
{\rm Re}\beta >0\,,\qquad \qquad {\rm Re}\,\omega_2 <0\,,\qquad \qquad {\rm Re}\,\varphi <0\,.
\end{equation}
The generalization to the multi-charge case, relevant when coupling pure supergravity to $n$ vector multiplets, is straightforward and follows from the simple substitution $\varphi Q \to \varphi^I Q_I$, with $I = 0,1,...,n$.
\end{enumerate}
Following the approach introduced in section \eqref{sec:Grav_path_int}, we will thus study this gravitational partition function in the semiclassical approximation, where it is expected to be dominated by saddles given by regular solutions to the classical equations of motion of five-dimensional ungauged supergravity satisfying the assigned  boundary conditions.

\medskip

Since we eventually want to make the connection with supersymmetric solutions in Lorentzian signature, it may be useful to briefly summarize the state of the art in the Lorentzian setup.
In five dimensions, the celebrated uniqueness theorems that hold in four dimensions do not apply. This means that
 specifying the conserved quantities measured at infinity (mass, angular momenta, electric charge) is not enough for determining an asymptotically flat solution with an event horizon. Relatedly, horizons can have more general topology than in four dimensions. In fact, the first example of a regular solution violating black hole uniqueness was provided by a black ring, namely a solution with an  $S^2\times S^1$ horizon topology, which can have the same conserved quantities measured at infinity as a black hole with $S^3$ horizon topology~\cite{Emparan:2001wn}. 
 
Let us focus on minimal five-dimensional supergravity, and consider asymptotically flat field configurations having two independent axial symmetries (leading to the angular momenta $J_1,J_2$) in addition to time translation invariance, so that the symmetry group includes $\mathbb{R}\times {\rm U}(1)\times {\rm U}(1)$.
With these assumptions, it has been shown that the cross-section of supersymmetric horizons can have the topology of $S^3$, $S^2\times S^1$, or the lens space $L(n,1)\simeq S^3/\mathbb{Z}_n$ \cite{Breunholder:2017ubu}. Explicit supersymmetric solutions with these horizon topologies have been constructed: the $S^3$ horizon is realized by the BMPV black hole~\cite{Breckenridge:1996is} (see also~\cite{Cvetic:1996xz,Chamseddine:1998yv}), the $S^2\times S^1$ topology is realized by the black ring of~\cite{Elvang:2004rt}, while the $L(n,1)$ topology is realized by the black lens solutions of\cite{Kunduri:2014kja,Tomizawa:2016kjh}.

There are also regular supersymmetric solutions with non-trivial topology and charges despite the absence of a horizon, see e.g.~\cite{Bena:2005va,Berglund:2005vb,Bena:2007kg,Gibbons:2013tqa}. These are called topological solitons, or ``bubbling'' solutions, due to the presence of a ``foam'' of two-cycles threaded by flux of the supergravity gauge field, and play a central role in the fuzzball and microstate geometry programs~\cite{Mathur:2005zp,Skenderis:2008qn,Bena:2022rna}.
Horizons can also coexist with bubbling topology in the domain of outer communication; this was first emphasized in~\cite{Kunduri:2014iga}, where a solution with a bubble outside a black hole horizon was presented. More generally, it is possible to add a black hole horizon to a background topological soliton containing many bubbles. Additionally, one can construct
 solutions with multiple disconnected horizons, the first examples being the concentric black rings of~\cite{Gauntlett:2004wh}.
This variety of combinations is enabled by the linearity of the supersymmetry equations, which allows one to sum up the functions that determine the different horizons and bubbles. 

Besides the conserved quantities at infinity, the information needed to specify a solution is effectively encoded in the {\it rod structure}, based on a formalism originally developed for pure gravity in~\cite{Emparan:2001wk,Harmark:2004rm,Hollands:2007aj}. 
Given the two-dimensional orbit space obtained by modding out the five-dimensional spacetime by the action of the $\mathbb{R}\times {\rm U}(1)\times {\rm U}(1)$ isometries, the rod structure consists of specifying the isometries which degenerate and their fixed loci in the orbit space. These fixed loci can be arranged into consecutive segments, the rods, forming an infinite line. 
If it is a spacelike ${\rm U(1)}$ that degenerates, then the corresponding rod together with the orbits of the surviving ${\rm U}(1)$ forms a fixed two-cycle in the spatial section of the geometry.
 Depending on whether the latter ${\rm U}(1)$ also degenerates at the endpoints of the rod or not, the cycle can have the topology of a two-sphere, a disc, or a tube.\footnote{When the solutions are embedded in string theory, one may also allow for orbifolds of the sphere and disc topologies, which yield conical singularities.} If, instead, the  Killing vector becoming null is timelike, then the corresponding rod together with the axial ${\rm U}(1)\times {\rm U}(1)$ forms an event horizon, whose topology depends on which of these axial ${\rm U}(1)$'s shrinks at the rod endpoints.
Extremal horizons (including all supersymmetric horizons in Lorentzian signature) correspond to the limiting case where the horizon rod has vanishing length, namely they are given by special isolated points on the rod axis. In order to specify the solution, one should supplement the rod structure with the magnetic flux of the supergravity gauge field through the non-contractible fixed loci, or alternatively provide the conjugate potentials.
A classification of asymptotically flat, biaxisymmetric supersymmetric solutions to minimal supergravity using the rod structure formalism was given in~\cite{Breunholder:2017ubu}.\footnote{Supersymmetric solutions with just one axial symmetry have been studied in \cite{Katona:2022bjp}. Although we will not consider such solutions in this paper, we observe that they share many features with the biaxisymmetric ones, so we anticipate that at least part of our analysis will go through in that case as well.}
These magnetic fluxes and their conjugate potentials have been argued to also play a role in black hole thermodynamics, although they are not associated to an asymptotic symmetry: a generalization of the first law of black hole mechanics in a background topological soliton has been proven in~\cite{Copsey:2005se,Kunduri:2013vka}. 

\medskip 

The existence of this rich variety of supersymmetric solutions makes five-dimensional ungauged supergravity a particularly interesting arena to investigate the structure of the index. In fact, the aim of the next chapters is to explore how the supersymmetric solutions reviewed above contributes to the gravitational index, by constructing the corresponding non-extremal Euclidean saddles. This naturally leads us to a systematic analysis of supersymmetric solutions. Our general strategy is to start from the classification of Lorentzian timelike supersymmetric solutions of~\cite{Gauntlett:2002nw}, obtained by directly solving the supersymmetry equations, and then study the regularity conditions for the Euclidean sections, obtained after Wick-rotation. In this way, we identify a new infinite class of smooth Euclidean geometries with arbitrary horizon topologies and an arbitrary number of bubbles outside the horizons that can serve as saddles of the index. Extremal solutions with a well-defined Lorentzian section, including those reviewed above, are then recovered in the limit $\beta \to \infty$, showing that our constructions provide finite-temperature deformations of the previously known extremal black holes.

This analysis also reveals an intriguing connection between supersymmetric extremal black holes (endowed with an event horizon in their Lorentzian description) and the horizonless topological solitons of~\cite{Bena:2005va,Berglund:2005vb,Bena:2007kg,Gibbons:2013tqa}. Starting from a finite-temperature supersymmetric saddle with arbitrary topology connected by the limit $\beta \to \infty$ to an extremal black hole with horizon and finite entropy, the opposite limit, $\beta \to 0$, (followed by analytic continuation back to Lorentzian signature) yields a horizonless configuration with vanishing entropy. Since the supersymmetric on-shell action is independent of $\beta$, both limiting solutions can be assigned a finite action. 
This observation suggests that the latter, usually studied in the context of the microstate geometry program, may also admit a grand-canonical interpretation as saddles of a gravitational partition function.
However, the two limiting solutions require different analytic continuations of the parameters to make sense in Lorentzian signature. Thus, although they seem to related to the same Euclidean saddle, extremal black holes and horizonless solitons can be distinguished by examining the on-shell action after the appropriate analytic continuations. In particular, for extremal solutions one typically finds regions of parameter space with ${\rm Re}\,I > 0$, as required for convergence of the path integral, while for horizonless solitons the on-shell action turns out to be purely imaginary. Although we have explicitly verified this only in a limited number of examples, we expect it to be a generic feature of such horizonless configurations. This raises doubts about whether these solitons can genuinely be regarded as acceptable saddles of the gravitational index, a question that requires further investigation.

\medskip

In order to determine the different phases of the theory arising when dialling the chemical potentials and discuss their thermodynamics, one should in principle know all classical solutions satisfying the assigned boundary conditions and compare their Euclidean action in order to identify the dominant ones. Given the difficulty of dealing with the new solutions we construct, it is desirable to develop techniques allowing to obtain the on-shell action just using minimal information such as topological data and boundary conditions, that might work even when the solution is not explicitly known. In this way one could learn a priori what is the contribution to the gravitational partition function of the saddles under consideration. The work of Gibbons-Hawking \cite{Gibbons:1979xm}, giving the on-shell action of four-dimensional gravitational instantons in terms of contributions from the fixed-submanifolds under a U$(1)$ action of their isometry group, i.e. zero-dimensional {\it nuts} and two-dimensional {\it bolts}, is a pioneering development in this direction.

Recently, starting with~\cite{BenettiGenolini:2023kxp,Martelli:2023oqk}, it has been realized that the mathematical tool of equivariant localization plays a role in supergravity and is useful for computing on-shell actions (related earlier ideas can be found e.g.\ in~\cite{BenettiGenolini:2019jdz,Hosseini:2019iad}).
This technique, based on the Berline-Vergne-Atiyah-Bott (BVAB) formula in equivariant cohomology~\cite{BerlineVergne,Atiyah:1984px}, allows to reduce the integral over a manifold to the fixed-point locus of a symmetry. The integration space may be either the external spacetime manifold, or the compactification manifold of higher-dimensional supergravity (or a submanifold thereof). By now, equivariant localization has been applied to the computation of a variety of gravitational observables, including black hole entropies, holographic central charges, free energies and scaling dimensions of operators.
Investigations so far have mostly focused on even-dimensional spaces, where Killing vectors can have isolated fixed points allowing for a simpler analysis~\cite{BenettiGenolini:2023yfe,BenettiGenolini:2023ndb,Colombo:2023fhu,BenettiGenolini:2024kyy,Suh:2024asy,Couzens:2024vbn,Hristov:2024cgj,BenettiGenolini:2024xeo,BenettiGenolini:2024hyd,Crisafio:2024fyc,BenettiGenolini:2024lbj}. More recently, equivariant localization has also been employed to compute the on-shell action in five-dimensional supergravity (both gauged and ungauged)~\cite{Cassani:2024kjn,Colombo:2025ihp,BenettiGenolini:2025icr,Colombo:2025yqy,Park:2025fon}, although a definitive framework in this setting has yet to be established.

Here, we discuss equivariant localization of the supergravity action in odd-dimensional spaces following~\cite{Cassani:2024kjn}, focusing on ungauged five-dimensional supergravity. The fixed locus of a Killing vector always has even codimension and hence, in five dimensions, it must have dimension at least one. It is therefore natural to adopt the terminology of {\it nuts} and {\it bolts} introduced in~\cite{Gibbons:1979xm}: we will use {\it nuts} for fixed loci of maximal codimension (i.e. one-dimensional submanifolds), and {\it bolts} for those of dimension three. Within this framework, we provide a localization formula for the on-shell action of ungauged five-dimensional supergravity, expressed in terms of nut and bolt contributions together with a set of boundary terms. We also show that there is a specific choice of localization scheme such that the boundary terms cancel out, and the on-shell action is entirely given by the fixed-point contributions. In particular, the nut contributions are determined by the holonomy of certain one-form potentials $\nu$ constructed from the gauge fields. We first apply this formula to recover the on-shell action of the simplest saddle, corresponding to the supersymmetric black hole with spherical horizon, studied in chapter~\ref{chap:Black_hole}.
For more general saddles, however, several complications, that we discuss below, arise. 

\medskip

The remainder of part~\ref{part_three} is organized as follows. In section~\ref{sec:equiv_5d} we review the BVAB localization theorem and its recent applications in the context of supergravity, extending its original formulation to the case of manifolds with boundary, and explain how the on-shell action of five-dimensional supergravity can be computed via equivariant localization. We show that the action localizes onto contributions from the fixed submanifolds of a U$(1)$ isometry, together with a set of boundary terms. As an illustration, we apply the theorem to a non-supersymmetric background of interest and discuss the scheme choice that ensures the cancellation of boundary terms.
In chapter~\ref{chap:Black_hole} we turn to supersymmetric solutions. We present a general construction of asymptotically flat Euclidean saddles and we first analyze in detail the simplest class of two-center configurations. This class includes the so-called {\it black hole saddle}, namely the saddle encoding the contribution to the index of the supersymmetric black hole with spherical horizon. Then we apply our localization formula to evaluate their on-shell action, connecting to both black holes and topological solitons. We extend the analysis to black hole saddles relevant for the multi-charge gravitational index, arising from five-dimensional supergravity coupled to vector multiplets.
Finally, in chapter~\ref{chap:bubbling} we go back to the minimal theory and study the rod structure of multi-center solutions, deriving the regularity conditions required to obtain smooth saddles of the index, aiming towards a general classification. Particular attention is devoted to the thermodynamics and general features of three-center solutions, which contain both a horizon rod and a non-trivial bubble. As we show, these saddles realize the non-extremal counterparts of the well-known supersymmetric black ring and black lens, and allow us, for the first time, to assign chemical potentials and a finite on-shell action to these solutions.

\section{Localization of the 5D supergravity action}
\label{sec:equiv_5d}

As discussed above, localization has provided a powerful tool for studying supersymmetric quantum field theories and, more recently, for investigating precision holography, since it enables the exact computation of certain observables at strong coupling~\cite{Pestun:2016zxk,Zaffaroni:2019dhb}. In these contexts, localization applies to the supersymmetric infinite-dimensional path integral, which, under favorable circumstances, becomes one-loop exact and reduces to a finite-dimensional integral. Early attempts to extend this framework to the supergravity path integral were made starting with~\cite{Banerjee:2009af,Dabholkar:2010uh,Dabholkar:2011ec}, but much remains to be understood. Nonetheless, recent developments of~\cite{BenettiGenolini:2023kxp,Martelli:2023oqk} have clarified that a finite-dimensional version of localization, applying in spacetime instead of in field space, plays a role in (super)gravity. More precisely, many observables, including the renormalized action, localize onto contributions from fixed-submanifolds of a suitable isometry, with the derivation relying on the Berline-Vergne-Atiyah-Bott (BVAB) fixed-point theorem of equivariant cohomology~\cite{BerlineVergne,Atiyah:1984px} (see~\cite{Pestun:2016qko,BerlineEtAlBook,Alekseev:2000fe,Cremonesi:2013twh} for a more extensive review of finite-dimensional localization techniques). In this sense, the semiclassical gravitational path integral in its saddle-point approximation, which is the regime considered in this Thesis, localizes in the gravitational description. In supersymmetric gauge theories both versions of localization, infinite and finite-dimensional, indeed play a role, as first highlighted in the seminal work of~\cite{Nekrasov:2002qd}. It is therefore natural to ask whether an analogous interplay holds in (super)gravity, as recent developments seem to suggest, and how the equivariant cohomology used to localize the action relates to that used to localize the full path integral. Addressing this question will likely require going beyond supergravity, incorporating auxiliary fields to obtain an off-shell formulation (as offered, e.g. by superconformal gravity after appropriate gauge fixing, see discussion in part \ref{part_two}) as well as (super)ghosts to implement consistent gauge-fixing of local symmetries (a framework to do so has been proposed in~\cite{deWit:2018dix}). The goal of this section, however, is not to elaborate further on these directions, but rather to review recent progress in applying equivariant localization to the supergravity action.

\medskip

The first example of localization of the supergravity action was offered in the context of minimal ${\cal N}=2$ gauged supergravity in four dimensions~\cite{BenettiGenolini:2019jdz,BenettiGenolini:2023kxp}. In this setting, every supersymmetric Euclidean AlAdS$_4$ solution admits a canonical supersymmetric Killing vector $\xi$, constructed as an appropriate bilinear in the Killing spinors. The fixed-point set of $\xi$ lies in the bulk of the spacetime and consists of (usually pairs of) isolated fixed points, known as {\it nuts}, and fixed two-dimensional sub-manifolds, called {\it bolts}. For simplicity, we restrict the discussion to the nut contributions.
At a nut, the tangent space splits as $\mathbb R^2 \oplus \mathbb R^2$, and the Killing vector can always be expressed as
\begin{equation}
\xi = \epsilon_1 \partial_{\varphi_1} + \epsilon_2 \partial_{\varphi_2}\,,
\end{equation}
where $\varphi_1$ and $\varphi_2$ are the polar angle coordinates on each copy of $\mathbb R^2$, and $\epsilon_1$, $\epsilon_2$ are the {\it equivariant parameters} measuring the weight of the linearized action of $\xi$ at the nut. At such fixed points the Killing spinor necessarily becomes chiral or antichiral, and nuts are accordingly labeled by $\pm$.
With these ingredients, the on-shell action takes the universal form
\begin{equation}\label{eq:loc_action_4D}
I = \frac{\pi}{2G_4}\sum_{{\rm nuts}_\mp} \pm \frac{\left(\epsilon_1 \pm \epsilon_2\right)^2}{\epsilon_1\,\epsilon_2} + \,\,\text{bolt contributions}\,.
\end{equation}  
Thus, the result is completely determined by topological data, namely, the weights of the supersymmetric Killing vector at its fixed points. By analyzing the global properties of the Euclidean saddles, these weights can be related to the chemical potentials defining the boundary conditions.\footnote{Indications that (supersymmetric) gravitational partition functions should have a very constrained dependence on the chemical potentials come from holography: it is known that supersymmetric field theory partition functions in curved space have a very constrained dependence on the geometry and other background fields~\cite{Closset:2013vra}, hence the same should be true in the dual gravitational partition function, see e.g.~\cite{BenettiGenolini:2016tsn} for a discussion.} 

A systematic derivation of \eqref{eq:loc_action_4D} follows from the BVAB localization theorem. A detailed review of the theorem is given in section~\ref{sec:BVAB}; here we only recall the ingredients needed to understand how the above result arises and to highlight the differences with the five-dimensional case. The starting point is the existence of a Killing vector, possibly with fixed points. One then introduces the equivariant differential
\begin{equation}\label{eq:diff_equiv} 
\diff_\xi = \diff - \iota_\xi\,,
\end{equation}
and looks for polyforms $\Psi$ satisfying $\diff_\xi \Psi = 0$, to which the theorem can be applied. Such {\it equivariantly closed polyforms} should provide the equivariant completion of the integral under consideration (in this case, the four-dimensional supergravity action).
The key intuition of~\cite{BenettiGenolini:2023kxp} is that in supergravity many such polyforms can be constructed as polynomials of Killing spinor bilinears, obtained by contracting the preserved Killing spinor (and its conjugate) with gamma matrices generating the Clifford algebra. The Killing spinor equations can then be manipulated into a set of differential and algebraic relations satisfied by these bilinears, which in turn can be used to show that certain polyforms built out of them are indeed equivariantly closed. Importantly, this closure requires only a subset of the supersymmetry conditions and equations of motion, and the relevant polyforms exist partially off-shell. As a consequence, the localization result holds for all supersymmetric solutions of the theory (when they exist), depending only on topological data and boundary conditions, without the need for the full explicit metric.
The strength of localization in four dimensions thus follows from the fact that the supersymmetric Killing vector admits fixed points, so that the relevant equivariant cohomology reduces to the cohomology defined by the associated equivariant differential.

\medskip

A natural question is whether analogous techniques can be employed to show that the five-dimensional action also reduces to a similarly constrained, universal form. However, the situation in five dimensions is different. In the supersymmetric non-extremal saddles we consider the supersymmetric Killing vector has no fixed points. To obtain a Killing vector with fixed points in these geometries, one must instead combine the supersymmetric Killing vector with an additional isometry commuting with the preserved supercharges. Crucially, this extra isometry cannot be expressed as a bilinear in the Killing spinor. As a result, applying equivariant localization to the five-dimensional supergravity action requires a reformulation of the formalism of~\cite{BenettiGenolini:2023kxp}, one that does not rely exclusively on Killing spinor bilinears.
In this section, based on contribution~\cite{Cassani:2024kjn}, we develop such a framework. We present a localization formula for the on-shell action of ungauged five-dimensional supergravity, depending on nut and bolt contributions together with a set of boundary terms, and which remains valid even in the absence of supersymmetry. Furthermore, we show that with a suitable choice of localization scheme the boundary terms can be cancelled, so that the on-shell action is determined solely by the fixed-point contributions. This corresponds to selecting an appropriate representative within the cohomology class defined by the closure condition.

As a first application we indeed discuss a non-supersymmetric example: we consider the Euclidean continuation of the general charged and doubly-rotating black hole solution. The ${\rm U}(1)^3$ symmetry of the solution and the $S^3$ topology of the horizon lead us to identify {\it two} Killing vectors, each with {\it one} (one-dimensional) nut. Each of these vectors is a combination of two ${\rm U}(1)$ isometries, with the third ${\rm U}(1)$ generating the fixed circle at the nut. We show that the corresponding localization formula reproduces the correct expression for the on-shell action. 
In order to reach the final expression, we need some information about the solution close to the nut and at infinity, analogously to the four-dimensional analysis of~\cite{Gibbons:1979xm}. The supersymmetric limit of this solution gives smooth non-extremal geometries representing the black hole contribution to the Euclidean gravitational path integral computing a supersymmetric index. In the presence of supersymmetry, the equivariant integral is determined by fewer data than in the general case, and the final result is entirely specified by the boundary conditions, like in the four-dimensional analysis of~\cite{BenettiGenolini:2019jdz}. A more detailed discussion of the supersymmetric case is presented in chapter~\ref{chap:Black_hole}.

We encounter, however, a crucial limitation of the technique: the polyform to which the localization theorem is applied must be globally well-defined. In odd dimensions this requirement is more restrictive than in four dimensions, primarily because a nut corresponds to a circle that may shrink to zero size elsewhere in the space, imposing non-trivial regularity conditions on the one-form bottom component of the polyform. Moreover, five-dimensional supergravity includes a gauge Chern-Simons term, which introduces additional gauge ambiguities in the definition of the polyform. In particular, when the gauge fields themselves are not globally well-defined forms, as is typically the case in the presence of non-trivial topologies, the construction of a polyform that remains everywhere regular requires a careful patch-wise formulation. Additionally, supersymmetric saddles with more general topologies often fail to admit a real positive-definite metric, introducing an additional challenge, since the standard BVAB theorem assumes a Riemannian manifold.

\medskip

Beyond the approach of~\cite{Cassani:2024kjn}, two alternative methods for computing the odd-dimensional action equivariantly have been proposed. The first, explored in~\cite{Cassani:2024kjn,BenettiGenolini:2025icr}, consists of performing a dimensional reduction along a compact direction distinct from the supersymmetric Killing vector (that is, a compact direction that cannot be expressed as a Killing spinor bilinear). This reduction allows one to apply equivariant localization in the resulting four-dimensional theory, where the supersymmetric Killing vector can have isolated fixed points, a setting that is far better understood~\cite{BenettiGenolini:2023kxp}. However, as emphasized in~\cite{BenettiGenolini:2025icr}, the dimensional reduction does not introduce singularities only when the compact U$(1)$ acts freely, which is typically not the case. Furthermore, the reduced action decomposes into three parts: contributions from the fixed points of the supersymmetric Killing vector, boundary terms, as well as an additional term that may not be globally well-defined when the gauge fields are not global forms. This technique must therefore be revisited when attempting to compute the action of Euclidean saddles with arbitrary topology.
A second approach has been put forward in~\cite{Colombo:2025ihp}, based on a localization theorem alternative to BVAB, valid for odd-dimensional manifolds admitting a foliation~\cite{goertsches2015localizationchernsimonstypeinvariants}. Applied to the supersymmetric geometries relevant here, where a natural foliation is provided by the orbits of the supersymmetric Killing vector itself, this method can be viewed as a dimensional reduction along these orbits (which has no fixed points in these geometries), followed by localization in four dimensions. The final result is nevertheless independent of the choice of Killing vector in the four-dimensional description, while it depends on the one determining the foliation. Also, very recently it has been extended through a patch-wise formulation to the case of non globally-defined gauge fields~\cite{Colombo:2025yqy}.

\medskip

The significance of global regularity constraints in odd-dimensional applications, where fixed points have at least dimension one, already appears in the simple case of the equivariant computation of the volume of $S^3$, which we study in section~\ref{sec:volumeS3}. This example also provides a natural setup for comparison with alternative approaches to equivariant localization in odd-dimensional spaces~\cite{Colombo:2025ihp,BenettiGenolini:2025icr}. Although we will not pursue these alternative routes further (nor show how the five-dimensional black hole action could be recovered through them) we highlight their key differences with respect to~\cite{Cassani:2024kjn} in the context of the $S^3$ volume computation. 

\medskip

In section~\ref{sec:BVAB} we review the BVAB fixed-point formula in the presence of boundaries. section~\ref{sec:volumeS3} is devoted to the equivariant computation of the volume of the three-sphere, which we illustrate using three complementary techniques. In section~\ref{sec:equiv5d} we then derive the localized expression for the five-dimensional on-shell action, and in section~\ref{sec:nonsusy_ex} we apply it to the general non-supersymmetric black hole solution.

\subsection{BVAB formula in spaces with  boundary}
\label{sec:BVAB}

We start with a brief review of the Berline-Vergne-Atiyah-Bott (BVAB)  equivariant localization formula \cite{BerlineVergne,Atiyah:1984px} for a torus symmetry. 
While the original BVAB formula is valid in compact spaces, here we will emphasize the modifications occurring for manifolds with a boundary, which will be needed in order to apply the formula to the evaluation of the gravitational action. 

 Let $(\mathcal{M},g)$ be a Riemannian manifold of dimension $D$, admitting a torus\footnote{Namely a product of U$(1)$'s, where the emphasis is on the fact that the group is compact.} isometry group. Let $\xi$ be a Killing vector field, generating an infinitesimal isometry of ${\cal M}$. We assume for the moment that $\mathcal{M}$ is compact with no boundary, we will add  boundary contributions at the end.  
We consider polyforms on $\mathcal{M}$, namely formal sums $\poly= \sum_n \poly_{(n)},$ where $\poly_{(n)}$ are forms of fixed degree $n=0,\ldots,D$. In particular, odd/even polyforms are made just of forms of odd/even degree.
The $\xi$-equivariant differential \eqref{eq:diff_equiv} 
sends odd/even polyforms into even/odd ones and squares to minus the Lie derivative along $\xi$, that is $\text d_\xi^2 = -\cal{L}_\xi$. 
Then, if we restrict to 
 polyforms satisfying  ${\cal L}_\xi \poly=0$,
the $\xi$-equivariant differential  can be used to define an equivariant cohomology.
A polyform $\poly$ is equivariantly closed if $\text d_\xi \poly=0$, while it
 is equivariantly exact if $\poly= \diff_\xi \Upsilon $ for some well-defined polyform $\Upsilon$. The equivariant cohomology  then consists of equivariantly closed modulo equivariantly exact polyforms.  This gives odd/even cohomology classes.
Note that the condition $\text d_\xi \poly=0$ leads to a set of recursive relations between forms of different degree (but same parity),  of the type
\be\label{descenteqs}
\iota_\xi \poly_{(n)} = \text d\poly_{(n-2)} \,.
\ee

We now consider the integral of a $\xi$-equivariantly closed polyform over a compact manifold $\mathcal{M}$ and sketch the localization argument showing that this only receives contributions from the fixed-point locus of the symmetry generated by $\xi$,  which we denote by ${\cal M}_0$.
First, the integral of a polyform over a $D$-dimensional manifold $\mathcal{M}$ picks by definition the component with the right degree to be integrated over,
$\int_{\cal M} \poly\, =\,  \int_{\cal M} \poly_{(D)}\,.
$
Hence by the ordinary Stokes' theorem  the integral of an equivariantly exact polyform over a compact manifold vanishes,
\begin{equation}
\int_{\cal M}\diff_\xi \Upsilon\, =\, \int_{\cal M} \text d\Upsilon_{(D-1)}\,= \,0\,,
\end{equation}
implying that integrals of equivariantly closed polyforms only depend on the equivariant cohomology class of the integrand. 

Next, we show that equivariantly closed polyforms on ${\cal M}$ are equivariantly exact on ${\cal M}\setminus {\cal M}_0$. Given the one-form dual to the vector $\xi$,  $\xi^{\flat} = \xi^\mu g_{\mu\nu}\text dx^\nu$, its equivariant differential  $\diff_\xi \xi^{\flat} = \diff \xi^{\flat} - |\xi|^2$ is invertible on ${\cal M}\setminus {\cal M}_0$, with inverse
\begin{equation}
\left( \diff_\xi \xi^{\flat}\right)^{-1} =-\frac{1}{|\xi|^2}\sum_{j=0}\left( \frac{\diff\xi^{\flat}}{|\xi|^2}\right)^j\,.
\end{equation}
It is easy to see that $\left(\diff_{\xi} \xi^{\flat}\right)^{-1}$ is equivariantly closed. It is convenient to consider the regulated space ${\cal M}_\epsilon$ obtained by removing infinitesimal tubular neighbourhoods about each connected component of ${\cal M}_0$, such that the original space is recovered in the $\epsilon\rightarrow 0$ limit. On the regulated space we can define the equivariant polyform
\begin{equation}
\Theta_{\xi} = \xi^{\flat} \wedge \left(\diff_{\xi} \xi^{\flat}\right)^{-1}\,,
\end{equation}
that by construction satisfies $\diff_\xi \Theta_{\xi} = 1$. An equivariantly closed polyform can now be written as
\begin{equation}\label{closed_is_exact}
\poly \,= \,\left(\text d_\xi \Theta_{\xi}\right)\poly \,=\, \text d_\xi \left( \Theta_{\xi} \poly \right)\,,
\end{equation}
showing that equivariantly closed polyforms on ${\cal M}$ are exact on ${\cal M}_\epsilon$. Then by the Stokes' theorem the integral of $\poly$ only receives contributions from $\partial {\cal M}_\epsilon$, that is the boundary of the infinitesimal neighbourhood about ${\cal M}_0$.  The BVAB  formula specifies what is such contribution in the $\epsilon\rightarrow 0$ limit: for any smooth $\poly$, it gives~\cite{BerlineVergne,Atiyah:1984px}\footnote{There is a simple refinement of the formula for the case where $\mathcal{M}$ has orbifold singularities, see e.g.~\cite{BenettiGenolini:2023kxp,Martelli:2023oqk,BenettiGenolini:2023ndb} for applications to supergravity. We do not consider it here.}
\begin{equation}
\label{eq:BVABformula}
\int_{\cal M} \poly \,=\, \int_{{\cal M}_0}\frac{\iota^*\poly}{ e_{\xi}({\cal N})}\,,
\end{equation}
where one should sum over the connected components of ${\cal M}_0$. On general grounds, each connected component must be of even codimension $2k$. The term $\iota^*\poly$ denotes the pullback over ${\cal M}_0$ of $\poly$ via the embedding $\iota: {\cal M}_0 \hookrightarrow {\cal M}$, while $e_{\xi}({\cal N})$ is the equivariant Euler form of the normal bundle ${\cal N} $ to ${\cal M}_0$ in ${\cal M}$. 
Quite generally, ${\cal N}$ splits into a direct sum of $k$ line bundles $L_i$, $i=1,\ldots,k$, which locally looks like $ \left(\mathbb R^2\right)^{k}$,
where each orthogonal $\mathbb R_i^2\subset \left(\mathbb R^2\right)^{k}$ has a corresponding angular coordinate $\phi_i$ with period $2\pi$, such that $\xi = \sum_{i=1}^{k} \epsilon_i\, \partial_{\phi_i}$. Then, the inverse of the Euler form has the expansion (see e.g.\ \cite{BenettiGenolini:2023ndb})
\be\label{exp_inv_Euler}
\frac{1}{e_\xi({\cal N})} \,=\, \frac{(2\pi)^k}{\Pi_{i=1}^k \epsilon_i } \left[ 1 - \sum_{1\leq i\leq k} \frac{2\pi}{\epsilon_i}\,c_1(L_i) + \sum_{1\leq i\leq j \leq k} \frac{(2\pi)^2}{\epsilon_i\epsilon_j} c_1(L_i)\wedge c_1(L_j)+ \ldots \right] \,,
\ee
 where $c_1(L_i)$ is the first Chern class of $L_i$.

Given the steps above, it is straightforward to adapt the BVAB formula to the case where the manifold ${\cal M}$ has a boundary, $\partial {\cal M}$ \cite{Couzens:2024vbn}. We assume that the latter does not contain fixed points of $\xi$.
The boundary $\partial {{\cal M}}_\epsilon$ introduced above is now given by the union of the boundaries of the infinitesimal neighbourhoods about the connected components of the fixed point locus, and the original boundary $\partial {\cal M}$. It follows that when integrating \eqref{closed_is_exact} and applying Stokes' theorem we need to include the boundary term on $\partial {\cal M}$, which reads
\be
 \int_{\partial {\cal M}} \Theta_{\xi} \poly \,=\, -\sum_{j=0}^{\left[\frac{D-2}{2}\right]} \int_{\partial {\cal M}} \eta \wedge \left(\diff \eta \right)^j\wedge \poly_{(D-2-2j)}\,,
\ee
where  we introduced the one-form
\be\label{eq:defeta}
\eta=\frac{\xi^{\flat}}{|\xi|^2}\,,
\ee
which is well defined on $\partial{{\cal M}}$ since by assumption $\xi$ does not vanish there.
Adding this boundary contribution to the BVAB formula \eqref{eq:BVABformula}, we arrive at the final expression
\begin{equation}\label{eq:BVAB_bdry}
\int_{\cal M} \poly \,=\, \int_{{\cal M}_0}\frac{\iota^*\poly}{ e_{\xi}({\cal N})}\,-\,\sum_{j=0}^{\left[{\frac{D-2}{2}}\right]} \int_{\partial {\cal M}} \eta \wedge \left(\diff \eta \right)^j\wedge \poly_{(D-2-2j)}\,.
\end{equation}

In the following we will be mostly interested in applying the equivariant localization formula to evaluate the gravitational action in odd-dimensional spaces. In particular, we will discuss in detail the five-dimensional case. In this case the connected components of the fixed locus of a Killing vector are either one-dimensional, or three-dimensional. By analogy with the four-dimensional terminology of Gibbons-Hawking~\cite{Gibbons:1979xm}, we will call `nuts' the former and `bolts' the latter.

\subsection{A simple example: the volume of $S^3$}\label{sec:volumeS3}

We now discuss a simple example (in compact space) that will be instructive in view of the applications to gravitational theories in the next sections. We consider a unit three-sphere $S^3$ and compute its volume, that is ${\rm Vol}_{S^3}=2\pi^2$, via equivariant localization. The metric is taken as
\be
\label{eq:metricS3}
\diff s^2 \,=\, \frac{1}{4}\diff \theta^2 + \sin^2\frac{\theta}{2}\, \diff \phi_1^2 + \cos^2\frac{\theta}{2}\, \diff \phi_2^2\,,
\ee
where $\theta\in [0,\pi]$, and $\phi_{1},\phi_2$  parameterize two circles $C_1$, $C_2$ of length $2\pi$. Here 
$S^3$ is described as a torus $C_1\times C_2$ foliated  over the interval parameterized by $\theta$: at the extremum  of the interval in  $\theta=0$ the circle $C_1$ shrinks to zero size, while $C_2$ remains finite. Conversely, at  $\theta=\pi$ it is $C_2$  that collapses while $C_1$ remains finite. 

We can localize with respect to the action of either one of the Killing vectors
\be
\xi_1 \,=\, \epsilon_1 \partial_{\phi_1}\,,\qquad\qquad \xi_2\, =\, \epsilon_2 \partial_{\phi_2}\,,
\ee with the result being independent of the choice made. Let us pick $\xi_1$: its fixed locus is the circle $C_2$ in $\theta=0$.
The following polyform is $\xi_1$-equivariantly closed and has the $S^3$ volume form as its top component:
\be
\Psi \,=\, \frac{1}{4} \sin\theta\, \diff\theta\wedge\diff \phi_1\wedge \diff\phi_2 \, + \,  \frac{\epsilon_1}{4}  \left(\cos\theta + \mu \right)\diff\phi_2\,,
\ee
where $\mu$ is an integration constant of the equivariant closure condition \eqref{descenteqs}; however this is not free, as regularity of the one-form  at $\theta=\pi$ (where $\xi_2$ vanishes) imposes $\mu=1$. The equivariant localization formula gives for the volume integral:
\be
{\rm Vol}_{S^3} \,=\, \int_{S^3} \Psi \,=\,  \frac{2\pi}{\epsilon_1}  \int_{C_{2,\theta=0}} \Psi_{(1)} \,=\,    \left(1 + \mu \right)\pi^2 \,,
\ee
which for $\mu=1$ is the correct result. 
In the computation we have kept $\mu$ generic so as to emphasize the importance of checking regularity everywhere on the manifold, not only close to the fixed locus of the vector chosen to localize. 

\paragraph{A second route: dimensional reduction on $S^1$. }Another remark is that we could follow an alternate route, namely reduce to two dimensions and use localization there. In order to do so, we introduce new angular coordinates $\phi = \phi_1+\phi_2$, $\psi = \phi_2 - \phi_1$,
 so that the metric is expressed as
\be
\diff s^2 \,=\, \frac{1}{4}\left[ \diff \theta^2 + \sin^2\theta\, \diff\phi^2 + (\diff \psi + \cos\theta\, \diff \phi)^2 \right]\,,
\ee
while our Killing vector becomes
\be
\xi_1 \,=\, \epsilon_1 (\partial_\phi - \partial_\psi)\,.
\ee
Here $S^3$ is seen as the Hopf fibration $S^1\hookrightarrow S^3\to S^2$ of the circle of length $4\pi$ parameterized by $\psi$ over the unit two-sphere $S^2$ parameterized by $\theta,\phi$. The vector $\partial_\psi$ has constant norm and we can straightforwardly reduce along its orbit down to the two-sphere. The reduced volume integral is
\be\label{eq:volS3split}
{\rm Vol}_{S^3} \,=\,  \int_0^{4\pi} \!\diff\psi    \int_{S^2}  \frac{1}{8} \sin\theta \,\diff\theta\wedge \diff \phi 
\,,
\ee
while the vector reduces to
\be
\xi_1^{\rm 2D} \,=\,  \epsilon_1 \partial_\phi\,.
\ee
We can now apply equivariant localization in 2D using the equivariantly closed polyform
\be
\Psi^{\rm 2D} \,=\, \iota_{\partial_\psi}\Psi \,=\,  \frac18\sin\theta \,\diff\theta\wedge\diff \phi + \frac{\epsilon_1}{8} \,(\cos\theta + \mu)\,,
\ee
where now the integration constant $\mu$ is not constrained by any regularity condition as it appears in a zero-form.
On $S^2$, the vector $\xi^{\rm 2D}$ has isolated fixed points at both poles $\theta= 0$, $\theta = \pi$. So  \eqref{eq:volS3split} is given by a sum over two contributions, 
\be
{\rm Vol}_{S^3} \,=\,\int_0^{4\pi} \!\diff\psi   \int_{S^2} \Psi^{\rm 2D} \,=\,   4\pi \cdot\frac{2\pi}{\epsilon_1} \cdot \frac{\epsilon_1}{8} \left[   (1 + \mu)  -  (-1 + \mu) \right]  \,=\, 2\pi^2\,,    
\ee
with the integration constant trivially canceling out.

Comparing the two procedures above, we see that while the  two-dimensional integral is computed by a sum over both poles of the sphere, in the former three-dimensional computation one pole gives the fixed circle $C_2$ entering in the localization formula, while the other pole contributes indirectly, through the regularity condition for the pullback of the local one-form $\Psi_{(1)}$ on $C_2$.

\subsubsection{An alternative formula: Localization for odd-dimensional foliations} 

Here, we consider an alternative to the BVAB localization formula, proposed in~\cite{goertsches2015localizationchernsimonstypeinvariants}. This formula applies to odd-dimensional manifolds with a torus action that degenerates to an $S^1$ action on a finite number of loci, which we denote by $L_a$. Each such locus then comprises a nut, diffeomorphic to $S^1$. Additionally, the manifold must be equipped with a vector $\kappa$ that defines a foliation. The localization formula derived in~\cite{goertsches2015localizationchernsimonstypeinvariants} can be applied to evaluate integrals of the form 
\begin{equation}\label{eq:int_foliation}
\int_{{\cal M}_D} \eta \wedge \Psi^{D-1}\,,
\end{equation}
as a sum of contributions from the degeneration loci. Here, $\eta$ is any one-form satisfying $\iota_\kappa \diff \eta =0$ and $\Psi^{D-1}$ is a even-degree basic polyform, meaning it satisfies 
\begin{equation}
\iota_\kappa \Psi^{D-1} = {\cal L}_\kappa \Psi^{D-1} =0\,.
\end{equation} 
Furthermore, $\Psi^{D-1}$ must be equivariantly closed with respect to another vector field, $t$ (non-collinear to $\kappa$ in ${\cal M}_D \setminus \cup_a L_a$): $\diff_t \Psi^{D-1} \,=\,0\,=\, {\cal L}_t \Psi^{D-1}$. The localization formula rewrites the above integral as
\begin{equation}
\label{eq:alt_BBVA}
\int_{{\cal M}_D}\eta \wedge \Psi^{D-1} = \left( - 2\pi \right)^{\frac{D-1}{2}} \sum_a \frac{\Psi_{(0)}^{D-1}\Big|_{L_a} \,\int_{L_a} \eta}{\prod_{i=1}^{\frac{D-1}{2}}\left( \frac{t_a^0}{\kappa_a^0}\kappa_a^i-t_a^i\right)}\,.
\end{equation}
where $t_a^0$ and $\kappa_a^0$ (provided they are non-zero) denote to the components tangent the direction generating the fixed circle at $L_a$ of $t$ and $\kappa$, respectively, while $t^i_a$ and $\kappa_a^i$ are the components along the transverse degenerating directions. Each individual term in the sum depends on the choice of $t$, but once all the contributions are combined, the $t$-dependence drops out. This formula has recently been revisited in~\cite{Colombo:2025ihp}, where it was extended to manifolds with boundary and applied to the computation of the on-shell action for the supersymmetric AdS$_5$ black hole. In this chapter, we only review how the formula can be used in a simple example (without boundary), the computation of the volume of the three-sphere, to give intuition about how this additional technique works. For further details we refer to~\cite{Colombo:2025ihp}. 

On $S^3$ with metric \eqref{eq:metricS3}, we take $\kappa = \partial_{\phi_2}- \partial_{\phi_1}$ and rewrite the volume as
\begin{equation}
{\rm Vol}_{S^3} = -\frac{1}{2}\int_{S^3} \eta \wedge \diff \eta \,,
\end{equation}
for a smooth one-form
\begin{equation}
 \eta = \kappa^\flat = \cos^2\frac{\theta}{2}\diff\phi_2 - \sin^2\frac{\theta}{2}\diff \phi_1\,.
\end{equation}
To compute this integral using \eqref{eq:alt_BBVA}, we introduce its equivariant completion
\begin{equation}
{\rm Vol}_{S^3} = -\frac{1}{2}\int_{S^3} \eta \wedge \diff \eta =-\frac{1}{2}\int_{S^3} \eta \wedge \diff_t \eta\,,\qquad \diff_t \eta = \diff \eta + \left( t^1 \sin^2\frac{\theta}{2} - t^2\cos^2\frac{\theta}{2}\right)\,, 
\end{equation}
where $t = t^1 \partial_{\phi_1} + t^2 \partial_{\phi_2}$. We have obtained an integral of the form \eqref{eq:int_foliation} with $\Psi^2 = \diff_t \eta$. As discussed above, the three-sphere has two localization loci, $L_{1}$ and $L_2$. At $L_1$ $(\theta = 0)$ we have
\begin{equation}
\Psi^2_{(0)}\Big|_{L_1} = - t^2\,,\qquad \int_{L_1} \eta = 2\pi\,,\qquad t^0_1 = t^2\,,\qquad t^1_1= t^1\,,\qquad \kappa_1^0 = 1 = - \kappa_1^1\,,
\end{equation}
while at $L_2$ ($\theta = \pi$) we find
\begin{equation}
\Psi^2_{(0)}\Big|_{L_2} = t^1\,,\qquad \int_{L_2} \eta = -2\pi\,,\qquad t^0_2 = t^1\,,\qquad t^1_2= t^2\,,\qquad \kappa_2^0 =- 1 = - \kappa_2^1\,.
\end{equation}
Combining the two contributions, formula~\eqref{eq:alt_BBVA} gives
\begin{equation}
{\rm Vol}_{S^3} =\pi \left( \frac{2\pi t^2}{t^1 + t^2} + \frac{2\pi t^1}{t^1 + t^2}\right) = 2\pi^2\,,
\end{equation}
which reproduces the correct result, independent of $t$. In some sense, this technique can be viewed as dimensionally reducing along $\kappa$ and then performing localization in one lower dimension, similarly to what we did above. Its main advantage, however, is that the final result depends only on the vector field $\kappa$, which specifies the foliation, and is independent of the choice of the auxiliary localizing vector $t$.


\subsection{5D supergravity on-shell action as an equivariant integral}\label{sec:equiv5d}

We now turn back to the BVAB localization formula and apply it to the evaluation of the action of minimal five-dimensional supergravity on a (bosonic) solution. In Euclidean signature, the action reads
\footnote{This action can be obtained from \eqref{eq:action_minimal_gauged_sugra} by rescaling the gauge field $A \rightarrow \frac{\sqrt{3}}{2}gA$ and then taking the ungauged limit $g\rightarrow 0$. In this chapter, we also set the gravitational constant to $G_5 = 1$.}

\begin{equation}\label{eq:actionminimal5d}
I_{\rm bulk} \,=\, - \frac{1}{16\pi }\int_{\cal M} \left( R\, \star_5 \!1 - \frac{1}{2}F\wedge \star_5 F + \frac{\ii}{3\sqrt{3}}A\wedge F \wedge F\right)\,,
\end{equation}
where $A$ is an abelian gauge field, and $F=\text dA$. It will also be useful to introduce the local three-form
\be\label{def_G_dualF}
G = \star_5 F - \frac{\ii}{\sqrt{3}}A\wedge F\,.
\ee
  The equations of motion read
\begin{equation}\label{eq:eqsofmotion5d}
\begin{aligned}
&R_{\mu\nu} - \frac{1}{2}F_{\mu\rho}F_\nu{}^\rho + \frac{1}{12}g_{\mu\nu} F_{\rho\sigma}F^{\rho\sigma}=0\,\,,\\
&\text dG=0\,.
\end{aligned}
\end{equation}

We consider a solution admitting a torus symmetry and we choose a vector field $\xi$ such that
\be
{\cal L}_\xi g \,=\, {\cal L}_\xi F \,=\, 0\,.
\ee
We will also pick a gauge such that 
\be
{\cal L}_\xi A \,=\, 0\,.
\ee

We now show that the on-shell action can be written as the integral of an equivariantly-closed polyform. Using the trace of the Einstein equation, \eqref{eq:actionminimal5d} can be written as
\begin{equation}\label{eq:action_after_einstein}
I_{\rm bulk} \,=\, \frac{1}{48\pi}\int_{\cal M} F\wedge G\,.
\end{equation}
The integrand $F\wedge G$ is the top-form $\Psi_{(5)}$ of an odd equivariantly closed polyform $\Psi = \Psi_{(5)}+ \Psi_{(3)} + \Psi_{(1)}$,  whose other components can be taken as
\begin{equation}\label{eq:polyform5d2}
\begin{aligned}
\Psi_{(3)} =&\, -\left(\iota_\xi A+\crho\right) G + \nu\wedge F \,,\\
\Psi_{(1)} =&\, -\left(\iota_\xi A+\crho\right)\nu + \iota_\xi \nu \,A\,,
\end{aligned}
\end{equation}
where the one-form $\nu$ is determined by the conditions 
\begin{equation}\label{eq:definitionnu5d_original}
\text d\nu = \iota_\xi G\,,\qquad\  \diff \iota_\xi \nu = 0\,,\qquad\ \crho\, \iota_\xi \nu = 0\,,
\end{equation}
and $\crho$ is a constant. The integrability condition $\diff \iota_\xi G =0$ for the first equation is satisfied using ${\cal L}_\xi G\equiv (\diff \iota_\xi + \iota_\xi \diff)\, G=0$ together with the Maxwell equation $\diff G=0$. Note that one-form potential $\nu$ is essentially the magnetic dual of the KK reduction of the gauge field $A$ along the orbits of $\xi$.

Whenever the vector $\xi$ vanishes somewhere, the contraction $\iota_\xi \nu$ -- which must be constant by the second in \eqref{eq:definitionnu5d_original} -- should be set to zero, so as to have a polyform $\poly$ that is regular along the shrinking direction. This  also solves the third in \eqref{eq:definitionnu5d_original}. Hence whenever $\xi$ has zeros, $\nu$ is defined by 
\be\label{eq:definitionnu5d}
\text d\nu = \iota_\xi G\,,\qquad\quad \iota_\xi \nu = 0\,.
\ee 
In this case the constant $\crho$ remains arbitrary. We will illustrate the role of $\crho$ in the example below, where it will be clear that it parameterizes different possible `schemes'.
We also stress that $\Psi$ should be everywhere regular, not just at the fixed locus of the vector that is used to localize, as made clear by the simple $S^3$ example of section~\ref{sec:volumeS3}.
 
Having constructed an equivariantly closed polyform whose top-component is the integrand of the on-shell bulk action, we can evaluate the latter by means of the BVAB localization formula \eqref{eq:BVAB_bdry}.
Expanding the formula, we find that the bulk integral can be expressed as
\be
\begin{aligned}\label{eq:localized_action_5d}
 I_{\rm bulk} \,&=\, \frac{1}{48\pi} \,\bigg[ \,\frac{(2\pi)^2}{ \epsilon_1 \epsilon_2 } \int_{{\rm nuts}} \iota^*\poly_{(1)}  +
\frac{2\pi}{\epsilon} \int_{{\rm bolts}}  \Big( \iota^*\poly_{(3)} - \frac{2\pi}{\epsilon}\, \iota^*\poly_{(1)}\wedge  c_1(L)  \Big) 
  \\[2mm]
&  \hspace{1.8cm}  -   \int_{\partial {\cal M}} \left(  \eta \wedge \poly_{(3)}+  \eta \wedge \diff \eta \wedge \poly_{(1)}  \right) \,\bigg] \,.
\end{aligned}
\ee
To this we must add the renormalized GHY boundary term,
\begin{equation}\label{eq:ghy_def}
I_{\rm GHY} \,=\, -\frac{1}{8\pi }\int_{\partial {\cal M}} \text d^4 x\left( \sqrt{h}\,{\cal K}- \sqrt{h_{\rm bkg}}\,{\cal K}_{\rm bkg}\right),
\end{equation}
where $h$ is the determinant of the induced metric and ${\cal K} $ is the extrinsic curvature of the boundary, while $h_{\rm bkg}$ and ${\cal K}_{\rm bkg}$ denote the corresponding quantities for a flat background with the same asymptotics as the solution under consideration, which cancels the divergences in the former term.
The full action then is 
\be
I = I_{\rm bulk} + I_{\rm GHY}\,.
\ee

Whenever the Killing vector has no fixed points and the regularity assumptions are satisfied, the action reduces to a boundary term on $\partial{\cal M}$. In this case one may have $\iota_\xi \nu\neq 0$, and the last in \eqref{eq:definitionnu5d} is solved by taking $\crho=0$. However, this would be essentially equivalent to using the Maxwell equation to reduce the action \eqref{eq:action_after_einstein} to the boundary term $I_{\rm bulk} =\frac{1}{48\pi}\int_{\partial\cal M} A\wedge \star_5 F \,,$ and then evaluating this term directly. 
 Therefore, here we will be interested in pursuing a different route: we will  choose a vector with zeros and find a choice of the constant $\crho$ such that the entire action is given by the contribution at the fixed locus.
In this case, since regularity at a nut requires $\iota_\xi A|_{\rm nut} = 0$, the nut contribution in \eqref{eq:localized_action_5d} can be expressed a little more explicitly as
 \be\label{eq:specify_Psi1}
 \int_{{\rm nuts}}\iota^*\poly_{(1)} \,=\, - \crho\, \int_{{\rm nuts}}\iota^* \nu\,,
 \ee
 so it is determined by the holonomies of the potential $\nu$ at the nuts.
 
 In the following, we illustrate the localization formula with a particularly relevant example which admits a supersymmetric limit. This will also be at the basis of the further developments in the next sections, where we will study more general supersymmetric solutions.
We will also provide the generalization to the case where the supergravity action includes an arbitrary number of vector multiplets, see section~\ref{sec:onshellaction_multicharge}.

\subsection{A non-supersymmetric example}\label{sec:nonsusy_ex}

We consider a general (non-supersymmetric and non-extremal) asymptotically flat black hole solution to the equations \eqref{eq:eqsofmotion5d}, which is included in the family first found in~\cite{Cvetic:1996xz}. We are going to compute the Euclidean on-shell action by means of the localization formula, showing in a concrete example how the localization argument works for odd-dimensional spaces, even in the absence of supersymmetry. 

\paragraph{The asymptotically flat black hole.}

The solution carries energy $E$, electric charge $Q$, and two independent angular momenta $J_1$, $J_2$. A convenient parametrization can be obtained by taking the ungauged limit  of the solution of minimal five-dimensional gauged supergravity given in~\ref{TwoDerReview}. We focus on a real Euclidean section.  In coordinates $(r,\tau,\vartheta,\phi_1,\phi_2)$, the solution is given by
\begin{equation}\label{eq:cclpsolution}
\begin{aligned}
\text ds^2 \,&=\, \left(\text d\tau -\frac{2q\,\nu_1}{\rho^2}\right)\,\diff\tau - 2 q\,\frac{\nu_1\,\nu_2}{\rho^2}-\frac{2m\rho^2-q^2}{\rho^4}\left(\text d\tau + \nu_2\right)^2+\rho^2\left(\frac{\text dr^2}{\Delta_r}+ \text d\vartheta^2\right)\\[1mm]
\, &\quad \,+ \left(r^2-\aa^2\right)\sin^2\vartheta\,\text d\phi_1^2+\left(r^2-\bb^2\right)\cos^2\vartheta\, \text d\phi_2^2\,,
\\[1mm]
A\,&=\,  -\ii\, \frac{\sqrt{3}\,q}{\rho^2}\left( \text d\tau + \nu_2\right) + \ii \,\Phi\,\text d\tau\,,
\end{aligned}
\end{equation}
where 
\begin{equation}\label{eq:bh_functions}
\begin{aligned}
\nu_1\,&=\, \bb \sin^2\vartheta\, \text d\phi_1 + \aa\cos^2\vartheta \,\text d\phi_2\,\,,\quad\quad \nu_2 = \aa \sin^2\vartheta\, \text d\phi_1 + \bb \cos^2\vartheta\, \text d\phi_2\,\,,\\[1mm]
\Delta_r \,&=\, \frac{\left(r^2-\aa^2\right)\left(r^2-\bb^2\right)+q^2-2 \aa\bb q}{r^2}-2m\,,\\[1mm]
\rho^2\,&= \, r^2 -  \aa^2\cos^2\vartheta -  \bb^2 \sin^2\vartheta\,\,.
\end{aligned}
\end{equation}
The angular coordinates $\phi_{1},\phi_2$ are $2\pi$-periodic, while $\vartheta\in [0,\pi/2]$.
In order to obtain a real metric after Wick-rotating to Euclidean time, we have analytically continued the rotational parameters $a,b$ appearing in the original Lorentzian solution as $a= \ii \,\aa$ and $b= \ii\, \bb$. This is useful in order to study regularity of the Euclidean section, where we will assume $\aa,\bb$ real, as well as in order to satisfy the assumptions behind the equivariant localization argument. Note that the gauge field is then purely imaginary.\footnote{Although we will not do so here, one may find it convenient to also analytically continue the gauge field to a real field,
$A \to \ii A$, which would reabsorb the factor of $\ii$ appearing in front of the Chern-Simons term in \eqref{eq:actionminimal5d}. This would give a solution to the Euclidean five-dimensional supergravity discussed in~\cite{Sabra:2016abd}.}
The conserved charges are given in terms of the parameters $(m,q,\aa, \bb)$ by 
\begin{equation}\label{eq:cclpcharges}
E = \frac{3\pi}{4}m\,,\quad\quad Q = \frac{\sqrt{3}\pi}{4}q\,,\quad\quad J_1 = \frac{\pi \ii}{4}\left(2 \aa m + \bb q\right)\,,\quad\quad J_2 =\frac{\pi \ii}{4}\left(2\bb m + \aa q\right)\,,
\end{equation}
where $J_1$ and $J_2$ are associated with rotations generated by the Killing vectors $\partial_{\phi_1}$ and $\partial_{\phi_2}$, respectively.
For any conserved charge we can define a conjugate chemical potential, namely the inverse Hawking temperature $\beta$, the angular velocities $\Omega_{1}, \Omega_2$ and the electrostatic potential $\Phi$.
These can be computed in the standard way and read
\begin{equation}\label{eq:betacclp}
\beta = \frac{2\pi\,r_+\left[\left( r_+^2 -\aa^2 \right)\left( r_+^2 - \bb^2\right) - \aa\bb q\right]}{r_+^4 -\left( q- \aa\bb \right)^2}\,,
\end{equation}
\begin{equation}\label{ang_vel_nonsusy}
\Omega_1 = \ii \frac{\aa\left( r_+^2 - \bb^2\right)+ \bb q}{\left( r_+^2 - \aa^2\right)\left( r_+^2 - \bb^2\right) - \aa\bb q}\,\,,\quad\quad \Omega_2 = \ii \frac{\bb\left( r_+^2 - \aa^2\right)+ \aa q}{\left( r_+^2 - \aa^2\right)\left( r_+^2 - \bb^2\right) - \aa\bb q}\,\,,
\end{equation}
\begin{equation}\label{elec_pot_nonsusy}
\Phi = \frac{\sqrt{3}\, q\,r_+^2}{\left( r_+^2 -\aa^2 \right)\left( r_+^2 - \bb^2\right) - \aa\bb q}\,,
\end{equation}
where $r_+$ is determined by solving the equation $\Delta_r(r_+) =0$. In the Lorentzian solution, $r=r_+$ gives the event horizon, while in Euclidean signature it is the position where the solution caps off.\footnote{Note that the condition $\rho^2>0$, necessary for positive-definiteness of the metric, is non-trivial in the Euclidean solution:  we demand that this holds as long as $r> r_+$, namely we demand $r_+^2>  {\rm max}\,(\aa^2,\bb^2) $. In the supersymmetric case where $m=q$, this condition is  satisfied as long as $m>0$.
} Given the reality conditions chosen for the parameters, we see that in this Euclidean section $\beta$ and $\Phi$ are real while $\Omega_1,\Omega_2$ are purely imaginary.
 Finally,  the Bekenstein-Hawking entropy, computed as 1/4 the area of the three-sphere at $r=r_+$, is given by
\be\label{entropyE_minimal5d}
{\cal S} \,=\, \pi^2\,\frac{\left( r_+^2 -\aa^2 \right)\left( r_+^2 - \bb^2\right) - \aa\bb q}{2r_+}\,.
\ee

The angular velocities are such that the Killing vector 
\begin{equation}
W \,=\, \partial_\tau - \ii \Omega_1 \partial_{\phi_1} - \ii \Omega_2 \partial_{\phi_2}\,
\end{equation}
satisfies $W_\mu W^\mu |_{r_+} =0$ and thus vanishes at  $r=r_+$ (its analytic continuation $\ii W$ being the generator of the event horizon in Lorentzian signature). Regularity of the solution at $r_+$ requires  the Euclidean time to be compactified with period $\beta$, with the following twisted identification on the coordinates,
\begin{equation}
\left( \tau\,,\,\phi_1\,,\,\phi_2\right) \sim \left( \tau + \beta\,,\,\phi_1 - \ii \Omega_1 \beta\,,\,\phi_2 - \ii \Omega_2\beta\right)\,,
\end{equation}
in addition to the standard identifications
\begin{equation}
\left( \tau\,,\,\phi_1\,,\,\phi_2\right)\, \sim\, \left( \tau \,,\,\phi_1 +2\pi \,,\,\phi_2 \right)\, \sim\, \left( \tau \,,\,\phi_1  \,,\,\phi_2+2\pi \right)\,.
\end{equation}
Also,  in the expression for $A$  in~\eqref{eq:cclpsolution} we have fixed the gauge such that the field is regular at $r=r_+$, namely it satisfies
$
W^\mu A_\mu\big|_{r_+} =0\,.
$

With these identifications, the Euclidean solution we are considering topologically is $\mathbb R^2\times S^3$, with the first factor being parameterized by the  radial coordinate $r$ and  polar coordinate $\tau$, and the $S^3$ being described as already seen in section~\ref{sec:volumeS3}. 

\paragraph{Localization of the action.}

 The isometry group is U$(1)^3$, generated by the Killing vectors $W$, generating rotations of period $\beta$ around the Euclidean thermal circle in $\mathbb R^2$, and $\partial_{\phi_{1}}$, $\partial_{\phi_{2}}$ in $S^3$.\footnote{One can introduce coordinates $\tilde\tau=\tau$, $\tilde\phi_1 = \phi_1+\ii \Omega_1\tau$, $\tilde\phi_2 = \phi_2 + \ii \Omega_2\tau$ which satisfy the untwisted identifications $$(\tilde\tau\,,\,\tilde\phi_1\,,\,\tilde\phi_2)\, \sim\, (\tilde\tau+\beta\,,\,\tilde\phi_1\,,\,\tilde\phi_2)\, \sim\, (\tilde\tau\,,\,\tilde\phi_1+2\pi\,,\,\tilde\phi_2)\, \sim\, (\tilde\tau\,,\,\tilde\phi_1\,,\,\tilde\phi_2+2\pi)\,.$$ In these coordinates, the vectors generating the U$(1)^3$ symmetry read  more simply as $W = \partial_{\tilde\tau}$, $\partial_{\tilde\phi_1}$, $\partial_{\tilde\phi_2}$.} 
As we have seen, each of these vectors vanishes on a three-dimensional hypersurface: at $r=r_+$, at $\vartheta =0$, and at $\vartheta = \pi/2$, respectively, hence these loci are bolts of the Euclidean geometry. 

We instead choose either one of the following Killing vectors for localizing the action:
\begin{equation}
\begin{aligned}
\label{eq:definexicclp}
\xi_N \,&=\, W +\ii\, (\Omega_1+\Omega_2)\, \partial_{\phi_1}\\[1mm]
\,&=\, \partial_\tau + \ii \,\Omega_2\, (\partial_{\phi_1}-\partial_{\phi_2})\,,
\end{aligned}
\end{equation}
which vanishes at the one-dimensional `north pole'  $\mn:  \{r=r_+\,,\ \vartheta = 0\}$, hence according to our terminology it has a nut at  $\mn$, or
\begin{equation}
\begin{aligned}
\xi_S \,&=\, W +\ii \left(\Omega_1+\Omega_2\right)\partial_{\phi_2} \\[1mm]
\,&=\, \partial_\tau - \ii \Omega_1 (\partial_{\phi_1}-\partial_{\phi_2})\,,
\end{aligned}
\end{equation}
which has a nut at  the `south pole' $\ms : \{ r=r_+\,\ \vartheta = \pi/2\}$. The result for the action integral must be independent of whether we choose to run the localization argument using $\xi_N$ or $\xi_S$. We are going to discuss in detail the case where we pick $\xi_N$, then we will comment on using $\xi_S$.
From the first line in~\eqref{eq:definexicclp}, we read that the equivariant parameters specifying the weights of the action of $\xi_N$ at the nut $\mn$ are 
\begin{equation}\label{eq:equivparamscclp}
\epsilon_1 = \frac{2\pi}{\beta}\,\,,\quad\quad \epsilon_2 = \ii \left(\Omega_1+\Omega_2\right)\,. 
\end{equation}

The localization argument states that the on-shell action reduces to a one-dimensional integral over $\mn$, up to a set of boundary terms. The first step in order to apply the localization formula \eqref{eq:localized_action_5d} is to compute the one-form potential $\nu$ by solving \eqref{eq:definitionnu5d}. We find it convenient to split the vector $\xi_N$ into the sum
\begin{equation}
\xi_N = V -   2\ii\Omega_2 \,U\,\,,\qquad\text{with}\qquad V =  \partial_\tau \,,\quad\quad U =  -\frac{1}{2} \left(\partial_{\phi_1} - \partial_{\phi_2}\right)\,,
\end{equation}
and consequently to split the one-form potential $\nu_N$ associated to $\xi_N$ as  $\nu_N = \nu_V -  2\ii\Omega_2\, \nu_U$, with 
\begin{equation}\label{eq:definenuVU}
\text d\nu_V = \iota_V G\,\,,\quad\qquad \text d\nu_U = \iota_U G\,,\qquad\quad \iota_{\xi_N} \nu_N = 0\,.
\end{equation}
This will allow us to easily compare with localization with respect to the other vector, $\xi_S$, which also is a combination of   $V$ and $U$.
We find the expressions
\begin{equation}
\label{eq:nucclp}
\begin{aligned}
\nu_V = &-\,\frac{\sqrt{3}\,\ii\,q}{\rho^2}\left[ \Omega_1 \Omega_2\,\rho^2 \,\text d\tau -\ii \Omega_2  \left(r_+^2 - \bb^2\right) \sin^2\vartheta\,\text d\phi_1 -\ii\Omega_1  \left( r_+^2 - \aa^2\right)\cos^2\vartheta\,\text d\phi_2  \right]\,,
\\[2mm]
\nu_U =&\,\ \frac{\sqrt{3}\, \ii \,q}{2\rho^2}\,  \bigg[\left(\ii(\Omega_1- \Omega_2)\,\rho^2 + \aa\cos^2\vartheta - \bb \sin^2 \vartheta + \frac{\Phi}{\sqrt{3}}\left(\bb\cos^2\vartheta - \aa \sin^2 \vartheta\right)\right)\text d\tau \\
&\qquad\qquad + \left(r^2-\bb^2\right) \sin^2\vartheta\, \text d\phi_1- \left( r^2 - \aa^2\right)\cos^2\vartheta\, \text d\phi_{2} \bigg]\,.
\end{aligned}
\end{equation} 
We emphasize that these are everywhere regular; in particular, they are regular at the nut $\mn$ as they satisfy $\iota_{\xi_N} \nu_V |_{\mn}=\iota_{\xi_N} \nu_U |_{\mn}=0\,$. 

We can now construct the three- and one-form in \eqref{eq:polyform5d2}, as well as the one-form $\eta$ in \eqref{eq:defeta}, and compute the different terms appearing in the localization formula \eqref{eq:localized_action_5d}. In order to compute the boundary terms, we will need the asymptotic behaviours for $r\to \infty$,
\be\label{eq:asympt_behav}
\eta \ \to\ \frac{1}{\ii\Omega_2} \left(\sin^2\vartheta\,\diff\phi_1 - \cos^2\vartheta\,\diff\phi_2 \right)\,,\qquad\qquad
\iota_{\xi_N} A \ \to\ \ii\, \Phi\,.
\ee
We find that the boundary term involving $\Psi_{(3)}$ is suppressed asymptotically.
On the other hand, since we have fixed a regular gauge, at the nut we have $\iota_{\xi_N} A |_{\mn}=0$.
Then the formula for the on-shell action reduces to a sum of three contributions,
\be\label{eq:local_action_bh}
 I \,=\, \frac{-\crho}{48\pi}\,\frac{(2\pi)^2}{ \epsilon_1 \epsilon_2 } \int_{\mn}\!\iota^*\nu_N   \, \,  +\,\, \frac{\ii\, \Phi+\crho}{48\pi} \int_{\partial {\cal M}}   \eta \wedge \diff \eta \wedge \nu_N   \, \, +\, \,I_{\rm GHY} \,.
\ee
Evaluating the boundary integrals using \eqref{eq:asympt_behav} we find
\begin{equation}\label{eq:bdrytermcclp}
 \frac{\ii\, \Phi+\crho}{48\pi}\int_{\partial{\cal M}} \eta \wedge \text d\eta \wedge\nu_N \,=\,-  \frac{\beta}{3}\left(\Phi - \ii \crho\right)Q\,,
\end{equation} 
\begin{equation}\label{eq:ghycclp}
I_{\rm GHY}\, =\, \frac{\beta}{3}\,E\,.
\end{equation}
The contribution from the fixed locus at $\mn$ can again be divided into the one coming from $\nu_V$ and the one from $\nu_U$. 
By pulling back on the nut we find
\begin{equation}\label{nutcclp_split}
\int_{\mn}\iota^*\nu_V\,=\, -8\Omega_1Q\,, \qquad\qquad
-2\ii \Omega_2\int_{\mn}\iota^*\nu_U\,=\, -8\Omega_2Q\,.
\end{equation}
Recalling the equivariant parameters given in \eqref{eq:equivparamscclp}, the full nut contribution to the action reads
\begin{equation}\label{eq:nutcclp}
\frac{- \crho}{48\pi} \frac{(2\pi)^2}{\epsilon_1\epsilon_2}\int_{\mn}\! \iota^*\nu_N\,=\, -\frac{\ii \crho}{3}\,\beta\,Q \,.
\end{equation}

Summing everything up we find the result
\begin{equation}\label{eq:result_action}
I = \frac{\beta}{3}\left(E-\Phi Q\right)\,,
\end{equation}
which is independent of the arbitrary coefficient $\crho$ introduced in~\eqref{eq:polyform5d2}, confirming that the latter represents a scheme choice.
This expression can be recast in a more familiar form by recalling the Smarr relation satisfied by the quantities given in \eqref{eq:cclpcharges}--\eqref{entropyE_minimal5d}, which reads
\begin{equation}
\beta E - \frac{3}{2}\,\mathcal S - \frac{3}{2}\left(\beta\Omega_1J_1 + \beta\Omega_2J_2\right) - \beta \Phi\,Q \,=\,0\,.
\end{equation}
Using this relation,~\eqref{eq:result_action} can be expressed as
\begin{equation}\label{QSE_general}
I = - \mathcal S +\beta\left(E -\Omega_1J_1 - \Omega_2J_2-\Phi\,Q\right)\,,
\end{equation}
that is the usual quantum statistical relation  identifying the on-shell action as a grand-canonical thermodynamic potential (see~\eqref{eq:general_qsr0}).

Besides recovering the on-shell action via the localization formula, that is without explicitly evaluating the bulk integral, our main point here is that since the coefficient $\crho$ is arbitrary, we can choose it in such a way that the contribution from the asymptotic boundary terms, that is the sum of  \eqref{eq:bdrytermcclp} and \eqref{eq:ghycclp}, vanishes. This is achieved by taking
\begin{equation}\label{eq:cvarrho}
\ii \crho =   \Phi - \frac{E}{Q}\,.
\end{equation}
 Then, the full on-shell action arises as a nut contribution from $\mn$. 
 
 The lesson learned in this example, that there is a `localization scheme' such that the boundary terms cancel and the action arises solely from the fixed point set of the Killing vector, appears to be a general fact. Generically, in order to fix the constant $\crho$ to the desired value one should know some data of the solution that do not correspond merely to boundary conditions, such as the mass-to-charge ratio $E/Q$ in our example. 
 However, in the supersymmetric case this ratio is fixed by the supersymmetry algebra, so for supersymmetric solutions the constant $\crho$ is actually fixed by the choice of the potential $\Phi$, that is by boundary conditions only. Below we discuss the supersymmetric limit of the present solution. In the next sections we will present a  more systematic supersymmetric analysis. 
 
Let us also comment on working equivariantly with respect to $\xi_S$ instead of $\xi_N$. In terms of the vectors $V$ and $U$, we have $\xi_S = V +2 \ii\Omega_1 U$. The associated one-form potential is $\nu_S = \nu_V +2\ii \Omega_1 \nu_U$, with $\nu_V$ and $\nu_U$ being still given by \eqref{eq:nucclp}, and one can check that $\iota_{\xi_S}\nu_S =0$ as required by the equivariant closure condition. Fixing $\crho$ as in \eqref{eq:cvarrho} so that the boundary terms cancel (this is independent of whether we are using $\xi_N$ or $\xi_S$), the entire action is given by the nut contribution. The result must be the same as the one obtained earlier working equivariantly with respect to $\xi_N$. Since the equivariant parameters are also the same, we deduce that the following identity must hold
\be
\int_{{\cal M}_N}\iota^* \nu_N \,=\,  \int_{{\cal M}_S}\iota^* \nu_S \,,
\ee
Expressing $\nu_N$ and $\nu_S$ in terms of $\nu_V$ and $\nu_U$, this can also be written as
\be
\int_{{\cal M}_N}\iota^* \nu_V -2 \ii \Omega_2 \int_{{\cal M}_N}\iota^* \nu_U \,=\,  \int_{{\cal M}_S}\iota^* \nu_V +2 \ii \Omega_1 \int_{{\cal M}_S}\iota^* \nu_U \,.
\ee
Actually, a direct computation shows that these integrals over ${\cal M}_N$ and ${\cal M}_S$ are pairwise equal:
\begin{equation}
\int_{\mn}\iota^*\nu_V\,=\, -8\Omega_1Q \,=\,  2 \ii \Omega_1 \int_{{\cal M}_S}\iota^* \nu_U\,, \quad\quad
-2\ii \Omega_2\int_{\mn}\iota^*\nu_U\,=\, -8\Omega_2Q =  \int_{{\cal M}_S}\iota^* \nu_V\,.
\end{equation}
This implies that using the Killing vector $V$ and the associated potential $\nu_V$ is enough for obtaining the full action integral provided we integrate $\nu_V$ both on $\cal{M}_N$ and $\cal{M}_S$.

\subsubsection{Supersymmetric limit}\label{sec:susylimit_BH}
  
 The metric and gauge field \eqref{eq:cclpsolution} allow for a solution to the Killing spinor equation if $m=q$. Imposing this condition implies that the conserved charges \eqref{eq:cclpcharges} satisfy the linear relation \eqref{eq:superalgebraaf}. However, it does not imply that the solution becomes extremal, as the inverse Hawking temperature \eqref{eq:betacclp} remains finite~\cite{Cabo-Bizet:2018ehj}. While the Riemannian metric is  regular, this non-extremal configuration has no well-definite Lorentzian counterpart for generic values of the parameter.

It is convenient to introduce the supersymmetric chemical potentials\footnote{These are just the ungauged versions of those defined in sec.~\ref{TwoDerReview}. As in that case, there is actually a second branch of expressions for the chemical potentials associated with the supersymmetric solution (arising because the equation $\Delta_r(r_+)=0$ has two solutions, see \eqref{eq:solq} below). Starting from the expressions in~\eqref{eq:cclpsusychempot}, the expressions for the other branch are obtained by sending $r_+ \to -r_+$  and flipping the overall signs. Since these other expressions lead to similar results, we will not explicitly display them in this paper.}
\begin{equation}
\label{eq:cclpsusychempot}
\begin{aligned}
\varphi \,=&\,\, \beta\big(\, \Phi - \sqrt{3}\,\big) \,=\, -2\sqrt{3}\pi\,\frac{\left(r_+ - \aa\right)\left(r_+ -\bb\right)}{2r_+ - \aa - \bb} \,,\\
\omega_1\, =&\,\, \beta\Omega_1\,=\, 2\pi\ii\,\frac{r_+ -\bb}{2r_+ - \aa -\bb} \,\,,\quad\quad \omega_2 \,=\, \beta\Omega_2\,=\, 2\pi\ii\, \frac{r_+ -\aa}{2r_+ - \aa -\bb}\,,
\end{aligned}
\end{equation} 
which satisfy
\be\label{eq:susyconstraint2}
 \omega_1+\omega_2 \,=\, 2\pi \ii\,.
 \ee
 The quantum statistical relation~\eqref{QSE_general} then reads
 \begin{equation}
I = - \mathcal S -\omega_1J_1 -\omega_2J_2-\varphi\,Q\,.
\end{equation}

We then turn to the localization argument. We will work in the scheme where the boundary terms cancel out, given by the choice~\eqref{eq:cvarrho}. In the supersymmetric case, this reads
\be\label{eq:susy_c}
\ii \crho \,=\,   \Phi - \sqrt{3}\,,
\ee
hence it is fixed by boundary conditions only, as anticipated.\footnote{In four-dimensional supergravity, this amounts to a gauge choice, and corresponds to the `supersymmetric gauge' taken in~\cite{BenettiGenolini:2019jdz}.} 
The vector $V$ is the supersymmetric Killing vector arising as a bilinear of the Killing spinor, while $U$ generates an additional symmetry that preserves the supercharge; this will be discussed in detail in section~\ref{sec:susyaction}.
 The on-shell action is given by the sum $I = \mathcal I_N + \mathcal I_S$, where 
\begin{equation}\label{eq:cclpsusygravblocks}
\mathcal I_N = -\frac{\ii}{24\sqrt{3}}\,\frac{\varphi^3}{\omega_2}\,\,,\qquad\qquad \mathcal I_S = -\frac{\ii}{24\sqrt{3}}\,\frac{\varphi^3}{\omega_1}\,,
\end{equation}
 are the supersymmetric versions of \eqref{nutcclp_split}, \eqref{eq:nutcclp}, which are now shown to just depend on the boundary conditions $\varphi$, $\omega_1$ and $\omega_2$.
The on-shell action can then be expressed as
\begin{equation}\label{eq:cclpsusyaction}
I \,=\, \frac{\pi}{12\sqrt{3}}\frac{\varphi^3}{\omega_1\,\omega_2}\,.
\end{equation}
This is also the ungauged ($g\to 0$) version of the result we reviewed around~\eqref{Isusy}, here recovered via equivariant localization.


\chapter{
Black hole saddles for the index
}
\label{chap:Black_hole}

In this chapter, based on contribution~\cite{Cassani:2024kjn}, we turn to a systematic analysis of finite-temperature supersymmetric solutions. Our primary goal is to identify the “black hole saddle” that accounts for the contribution to the index from the supersymmetric Lorentzian black holes with spherical ($S^3$) horizon~\cite{Breckenridge:1996is,Cvetic:1996xz,Chamseddine:1998yv}. To this end, we will consider their Euclidean, supersymmetric, non-extremal counterparts and show how their on-shell action can be computed through equivariant localization. We also show how the action is determined in larger classes of supersymmetric solutions. 

The black hole saddles were originally constructed by taking suitable supersymmetric yet non-extremal limits of the corresponding non-supersymmetric solutions. This method was used, for example, in section~\ref{TwoDerReview}, following the original study of the supersymmetric AdS$_5$ black hole in~\cite{Cabo-Bizet:2018ehj}, and similarly in~\cite{Iliesiu:2021are} for four-dimensional black holes with flat asymptotics. In the same spirit, the supersymmetric limit of the general doubly-rotating asymptotically flat five-dimensional black hole, as discussed in section~\ref{sec:susylimit_BH}, yields the black hole saddle, i.e. a smooth non-extremal solution representing the black hole contribution to the Euclidean gravitational path integral that computes a supersymmetric index. However, this construction is only available when the corresponding non-supersymmetric thermal solution is known, which is often not the case.

A more systematic approach to constructing these solutions is to solve the supersymmetry equations directly. This yields a general classification of their local form, after which one can impose regularity conditions. This method, developed in~\cite{Gauntlett:2002nw}, led to a comprehensive classification of Lorentzian supersymmetric solutions in five-dimensional minimal supergravity, providing a fertile framework for discovering new supersymmetric configurations whose non-supersymmetric counterparts may not be known. In this chapter we adopt this general strategy, starting by reviewing the classification of~\cite{Gauntlett:2002nw}, after Wick-rotation to Euclidean signature. In this setting, the local structure of the general solution takes the form of a U$(1)$ fibration over a four-dimensional base space, with the U$(1)$ direction parametrized by the coordinate associated with the supersymmetric Killing vector, constructed as a Killing spinor bilinear. Our main interest lies in recovering the black hole saddles, while leaving the study of more general classes of solutions to the next chapter. As we will see, constructing the black hole saddles requires geometries with an additional U$(1)$ isometry commuting with the supercharges. This is realized by choosing the four-dimensional base space to be of Gibbons-Hawking type, still within the classification of~\cite{Gauntlett:2002nw}. A distinctive feature of these supersymmetric solutions is that they are completely specified by a set of harmonic functions, which must be chosen appropriately so as to yield smooth Euclidean geometries satisfying the correct boundary conditions. The dependence on harmonic functions makes the classification of ungauged supergravity solutions particularly tractable, providing a simpler framework for systematically constructing supersymmetric configurations and analyzing their contributions to the index, as opposed to the classification of solutions in gauged supergravity, first obtained in~\cite{Gauntlett:2003fk}, which relies on a considerably more intricate formalism.

\medskip

A second main goal of this chapter is to compute the on-shell action of the black hole saddles, providing an application of the BVAB theorem in odd dimensions and in a supersymmetric setting, where it is expected to be most effective. For the time being, then, we will focus on supersymmetric Euclidean saddles which admit a real metric (as required by the BVAB theorem), which also simplify the regularity analysis. 

In the presence of supersymmetry, the equivariant integral is fixed by fewer data than in the general case, similarly to what has been found in even dimension. However, we face an issue that makes the five-dimensional analysis different from the even-dimensional one: the supersymmetric Killing vector is nowhere vanishing in  the Euclidean solutions of interest.
 This should be contrasted with the situation in even dimension, where the supersymmetric Killing vector can be used to localize the action to fixed-point contributions; for instance, in the supersymmetric non-extremal limit of the Kerr-Newman black hole this vector has a nut at the poles of the horizon two-sphere~\cite{Whitt:1984wk} (see e.g.~\cite{BenettiGenolini:2016tsn,BenettiGenolini:2019jdz,Bobev:2020pjk,Cassani:2021dwa,BenettiGenolini:2023kxp,BenettiGenolini:2024xeo,BenettiGenolini:2024kyy,BenettiGenolini:2024hyd,BenettiGenolini:2024lbj,Crisafio:2024fyc} for more general 4D solutions with a cosmological constant where the action is computed in terms of fixed points of the supersymmetric Killing vector). This leads us to consider a different vector for localizing: we assume a Gibbons-Hawking base-space, possessing an additional ${\rm U}(1)$ isometry that preserves the supercharge and use it to construct two Killing vectors with one nut.\footnote{For black hole saddles, these vectors are just the supersymmetric limit of the ones identified in the non-supersymmetric case mentioned above.} The supersymmetric solution is then characterized by harmonic functions on three-dimensional flat space. 
 
 As we showed in the previous chapter, by choosing a gauge in which the gauge field is smooth in the Euclidean section, the localized on-shell action depends on the nut holonomy of a one-form potential, denoted by $\nu$. We provide a general expression for the local potential $\nu$ under the assumption that the base-space is of Gibbons-Hawking type, independently of the number of sources for the harmonic functions. However, as we discussed in section~\ref{sec:intro3}, regularity constraints in odd-dimensional applications of the BVAB theorem are more restrictive, preventing us from writing a general globally-defined potential $\nu$. To make progress, we analyze in detail only the two-center case where the solution naturally exhibits an enhanced ${\rm U}(1)^3$ symmetry. This class of solutions includes the black hole saddles we are interested in. In this setup, we are able to complete the derivation and demonstrate that the action integral is entirely determined by the coefficients of the harmonic functions at their centers, while remaining independent of the distance between the two centers. Crucially, this implies that the final result continues to hold in the limit where the two centers coalesce, which corresponds to the extremal limit of this family of non-extremal geometries. We also observe that the resulting formula for the action can be written in the form of a sum over the centers. Finally, we identify the scheme choice that ensures the cancellation of boundary terms. This choice is fully characterized by the supersymmetry algebra and the boundary conditions, providing a general prescription.
 
The setup above gives us the opportunity to discuss an alternative approach to the evaluation of the five-dimensional action: at least when the additional ${\rm U}(1)$ isometry is freely acting, we can reduce to four dimensions and then apply equivariant localization there. In this way, we are able to match our results with those obtained by localizing with respect to the supersymmetric Killing vector in four dimensions. In app.~\ref{sec:exampleaction4d}, we explicitly show this in the example of  the Kerr-Newman solution and its uplift to five dimensions, which is also captured by harmonic functions with two centers. We find that the sum over the two centers in the five-dimensional formula is mapped into a sum over the nuts in four dimensions. A similar approach, applying the BVAB theorem after dimensional reduction, was later explored for five-dimensional solutions of gauged supergravity in~\cite{BenettiGenolini:2025icr}.

\medskip

Finally, we extend our analysis to five-dimensional supergravity coupled to an arbitrary number of vector multiplets. In this case, the supersymmetric non-extremal black hole saddles are not already available for arbitrary rotation and electric charges, so we construct them by directly solving the supersymmetry equations. Then we evaluate their action via equivariant localization and analyze their thermodynamic properties by taking suitable analytic continuations. 

We elaborate on an intriguing connection between the supersymmetric extremal black hole and horizonless topological solitons. Namely, we find that the supersymmetric black hole saddles with finite inverse temperature $\beta$ interpolate between the supersymmetric extremal black holes of~\cite{Breckenridge:1996is,Cvetic:1996xz,Chamseddine:1998yv},  which are obtained  taking $\beta \to \infty$ and continuing to Lorentzian signature,  and two-center horizonless solutions of~\cite{Giusto:2004id,Bena:2005va, Berglund:2005vb}, that we reproduce by taking the $\beta\to 0$ limit and turning to Lorentzian signature. 
 Although the supersymmetric on-shell action does not depend on $\beta$ and has fixed functional dependence on the remaining chemical potentials, the two limiting solutions require different analytic continuations of the parameters to make sense in Lorentzian signature. This implies that they arise for different values of the chemical potentials and carry a different action. The conserved charges also take different values.

The rest of the chapter is organized as follows. In section~\ref{sec:gen_ssol} we introduce some general features of supersymmetric solutions to pure five-dimensional supergravity having the assumed symmetry. In section~\ref{sec:susyaction}, we exploit the extra mileage of supersymmetry to evaluate the on-shell action equivariantly in a class of solutions with ${\rm U}(1)^3$ symmetry, including black hole saddles. In section~\ref{sec:multicharge} we extend our results to matter-coupled supergravity: we construct new supersymmetric non-extremal black hole saddles carrying multiple electric charges, compute their on-shell action via localization, discuss their thermodynamic properties, and show that they interpolate between the extremal black hole and the horizonless Lorentzian solutions. 
Appendix~\ref{sec:4Dexample} computes the action of the Kerr-Newman solution via equivariant localization,  as needed in section~\ref{sec:exampleaction4d}, where we compare our approach with localization in four dimensions. Appendix~\ref{sec:DiracMisner} contains a complementary analysis of the regularity of multi-charge saddles.

 
%

\section{Supersymmetric solutions with biaxial symmetry}
\label{sec:gen_ssol}

Bosonic supersymmetric solutions of the field equations \eqref{eq:eqsofmotion5d} admit a super-covariantly constant spinor $\epsilon$, called Killing spinor, satisfying a differential equation which arises from setting to zero the supersymmetry variation of the gravitino.
The local form of the bosonic supersymmetric solution is determined by analysing the differential forms that can be constructed as bilinears of the Killing spinor~\cite{Gauntlett:2002nw}. 

\paragraph{Lorentzian solutions.} One can construct a supersymmetric Killing vector field $\tilde V$ as a one-form spinor bilinear. Introducing local coordinates such that $\tilde V = \partial_t$, the Lorentzian metric locally takes the form
\begin{equation}
\text ds^2 = - f^2\left( \text dt +\omega\right)^2 + f^{-1}\diff\hat s^2\,\,,
\end{equation}
where $\diff\hat s^2$ is a four-dimensional hyper-K\"ahler metric, and $\omega $ is a local one-form on the hyper-K\"ahler base. We are interested in considering supersymmetric solutions in the case in which $\tilde V$ is a timelike Killing vector, with $f>0$. 

The gauge field, and the associated field strength, can be expressed as
\begin{equation}\label{eq:susysolminimal}
\begin{aligned}
A \,=&\,\, \sqrt{3}\left[-f\left( \text dt + \omega\right) + \hat A \right]\,\,,\\
F \,=&\,\, \text dA \,=\, \sqrt{3}\left[-\text df \wedge \left(\text dt + \omega \right)- f\text d\omega + \text d\hat A \right]\,,
\end{aligned}
\end{equation}
where $\hat A$ is a gauge field on the hyper-K\"ahler base. 

We will now assume that the four-dimensional hyper-K\"ahler manifold admits an additional U$(1)$ isometry whose Killing vector field, denoted as $\partial_\psi$, preserves its triholomorphic structure. This case was also analyzed in~\cite{Gauntlett:2002nw}. One can show that then the hyper-K\"ahler metric takes the Gibbons-Hawking form
\begin{equation}\label{eq:susysolminimalghbase1}
\text d\hat s^2\,= \, H^{-1} \left( \text d\psi + \chi\right)^2 + H\,\diff s^2_{\mathbb R^3}\,,
\end{equation}
where $\chi$ is a one-form on a three-dimensional flat base-space $\mathbb R^3$ with coordinates $\bold x = \left(x,y,z\right)$, determined by the equation
\begin{equation}\label{eq:defchi}
\star_3 \text d\chi\, =\, \text dH\,,
\end{equation}
where $H$ is a harmonic function on $\mathbb R^3$ and $\star_3$ the Hodge dual in $\mathbb R^3$. Assuming further that the Killing vector $\partial_\psi$ generates a symmetry for the full five-dimensional spacetime, one can decompose the one-form appearing in the solution as
\begin{equation}\label{eq:susysolminimalghbase2}
\begin{aligned}
\omega \,=\,\, \omega_\psi \left( \text d\psi +\chi\right) + \breve\omega\,,
\qquad\qquad 
\hat A \,=\,\, \frac{K}{H}\left( \text d\psi + \chi\right) + \breve A
\,.
\end{aligned}
\end{equation}
All the functions and one-forms introduced above are determined in terms of a set of four harmonic functions $(H,K,L,M)$ with sources on $\mathbb R^3$. Specifically, the functions $\omega_\psi$ and $f$ are given by
\begin{equation}\label{eq:susysolminimalghbase3}
\omega_\psi \,=\,\frac{K^3}{H^2}+\frac{3}{2}\frac{K\,L}{H}+M \,\,,\qquad\qquad f^{-1} \,=\, \frac{K^2}{H}+L\,,
\end{equation}
while $\breve\omega$ and $\breve A$ are local one-forms on $\mathbb R^3$ satisfying the equations
\begin{equation}\label{eq:susysolminimalghbase3_bis}
\star_3 \text d\breve\omega\,=\, H\text dM - M \text dH +\frac{3}{2}\left(K\text dL - L\text dK \right)\,\,,\qquad\qquad \star_3\text d\breve A \,=\, -\text dK\,.
\end{equation}
%

\paragraph{Euclidean solutions.}

We start by outlining our strategy for obtaining supersymmetric solutions to Euclidean (complexified) supergravity.
In suitable conventions, the Killing spinor equation reads
\be\label{KillingSpEqDirac}
\left[ \nabla_\mu - \frac{\ii}{8\sqrt 3}\left( \gamma_{\mu}{}^{\nu\rho} - 4\delta^\nu_\mu\gamma^\rho\right) F_{\nu\rho} \right]\epsilon  \,=\, 0\,, 
\ee
where $\epsilon$ is a Dirac spinor.\footnote{Eq.~\eqref{KillingSpEqDirac} corresponds to the ungauged version of \eqref{eq:Killing_spinor_equations}.}  
In the general framework of Euclidean supergravity, all degrees of freedom should be doubled, i.e.\ the metric and the gauge field become complex, while the gravitino and its Lorentzian charge-conjugate  should be seen as independent. It follows that the supersymmetry spinor parameter and its Lorentzian charge-conjugate should also be taken as independent in Euclidean signature. We denote by $\epsilon$ and $\tilde\epsilon$ these two a priori independent Dirac spinors. It turns out that the Lorentzian charge-conjugate equation of the Killing spinor equation \eqref{KillingSpEqDirac} takes exactly the same form, so $\epsilon $ and $\tilde\epsilon$ need to satisfy the very same Killing spinor equation.
 This implies that if we have a Lorentzian supersymmetric solution depending on some parameters, any analytic continuation of those parameters to the complex domain will give a field configuration solving the doubled Killing spinor equations, as well as the equations of motion following from \eqref{eq:actionminimal5d}, where the fields are assumed complex. This will be our working framework, namely we will consider complex solutions which arise as analytic continuations of Lorentzian supersymmetric solutions.\footnote{This applies to the local supersymmetry conditions. As we will see, global regularity in Euclidean signature will lead us to supersymmetric non-extremal solutions that do not have a \hbox{Lorentzian counterpart.}} We will not impose any global a priori condition on these Lorentzian supersymmetric solutions, these will be studied after the analytic continuation.

According to the classification of~\cite{Gauntlett:2002nw}, supersymmetric timelike solutions with an extra symmetry commuting with the supercharges take the form
\begin{equation}
\label{eq:susy_conf}
\begin{aligned}
\diff s^2 \,&=\, f^2 \left( \diff \tau + \ii \omega_\psi \left( \diff \psi + \chi\right) + \ii \breve\omega \right)^2 + f^{-1}\left[ H^{-1}\left( \diff \psi + \chi \right)^2 + H \, \diff s^2_{\,\mathbb{R}^3}\right]\,,\\[1mm]
\frac{\ii }{\sqrt{3}}\,  A \,&=\, -f \left( \diff \tau + \ii \omega_\psi \left( \diff\psi + \chi \right) + \ii \breve\omega\right) + \ii H^{-1}\,K \left( \diff \psi + \chi \right) + \ii\breve A 
 + \diff\alpha\,,
\end{aligned}
\end{equation}
where $\tau = \ii t$ is the Euclidean time parameterizing the orbits of the supersymmetric Killing vector $V = \partial_\tau$, while $\psi$ parameterizes the orbits of the additional isometry preserving the Killing spinor. Also, $\alpha$ is a function to be specified later parametrizing the freedom to perform a gauge transformation. 
The functions $\omega_\psi$ and $f$, as well as the one-forms $\chi$, $\breve\omega$ and $\breve A$ are determined according to \eqref{eq:defchi}, \eqref{eq:susysolminimalghbase3} and \eqref{eq:susysolminimalghbase3_bis}.

As customary, we shall assume a multi-center ansatz for the harmonic functions,
\begin{equation}
\label{eq:multicenter_ansatz}
H=h_0+\sum_{a=1}^{s}\frac{h_a}{r_a}\,,\hspace{5mm} \ii K=\kk_0+\sum_{a=1}^{s}\frac{\kk_a}{r_a}\,,\hspace{5mm} L=\ell_0+\sum_{a=1}^{s}\frac{\ell_a}{r_a}\, ,  \hspace{5mm}\ii M=\mm_0 +\sum_{a=1}^{s}\frac{\mm_a}{r_a}\,,\quad
\end{equation}
where $r_a = |\bold x - {\bold x}_a|$, with ${\bold x}_a$ denoting the location of the $a$-th source,  and $s$ being the number of such sources (usually called `centers' in this context) on the $\mathbb{R}^3$ base-space. In Lorentzian signature, all  harmonic functions would be taken real; in particular $K$ and $M$ would be expressed in terms of coefficients $\kk_0 =\ii k_0$, $\kk_a =\ii k_a$, $\mm_0=\ii m_0$, $\mm_a=\ii m_a$, with real $k_0$, $k_a$, $m_0$, $m_a$. Here, we have chosen to work with the analytically continued parameters $\kk_0$, $\kk_a$, $\mm_0$, $\mm_a$,  which satisfy the condition that if all of them are real, then the Euclidean metric in \eqref{eq:susy_conf} is real (however, below we will also allow for complex choices of these parameters). Then the solution depends on $4(s+1)$ parameters, which are the arbitrary coefficients of the four harmonic functions. Some of these parameters will be fixed by imposing asymptotic conditions on the solution, or by requiring that the geometry is regular in the interior.  

In this paper we shall restrict to solutions admitting an additional isometry, rotating a plane in the $\mathbb{R}^3$ base. This implies that all the centers are placed on the rotation axis, which we identify with the $z$-axis. The location of the $a$-th center is thus specified by its position $z=z_a$ on this axis. To describe these solutions a convenient set of coordinates are cylindrical coordinates on $\mathbb{R}^3$ given by $(\rho,\phi,z)$, where $(\rho,\phi)$ are polar coordinates in the $\mathbb{R}^2$-plane orthogonal to the $z$-axis,
\begin{equation}
x = \rho \cos\phi \,,\qquad \qquad y = \rho \sin \phi \,,
\end{equation}
In addition 
we will also find it useful to introduce sets of spherical coordinates $(r_a,\theta_a,\phi)$ with origin at each of the Gibbons-Hawking centers,
\be 
\label{eq:coordinatesystem2}
\rho = r_a\sin\theta_a\,,\quad z = z_a+ r_a \cos\theta_a\,\, \leftrightarrow\,\,  r_a = \sqrt{\rho^2 + (z-z_a)^2}\,,\quad
\cos\theta_a = \frac{z-z_a}{\sqrt{\rho^2 + (z-z_a)^2}}\,.
\ee
Finally, we will need a system of polar coordinates centered on the origin of $\mathbb R^3$, 
\begin{equation}\label{eq:coordinatesystem1}
x = r\sin\theta\cos\phi\,\,,\quad\quad y = r\sin\theta\sin\phi\,\,,\quad\quad z = r\cos\theta\,,
\end{equation}
with $\theta \in [0;\pi]$ and $\phi \sim \phi + 2\pi$. 

\medskip 

In terms of the coordinates of \eqref{eq:coordinatesystem2}, one readily finds that the following expressions solve the equations for  $\chi$ and $\breve A$:
\begin{equation}\label{eq:explicit_oneform1}
\chi\,=\,\sum_{a=1}^{s}h_a \cos \theta_a\, \diff \phi\,,\hspace{1cm}\quad
\ii {\breve A}\,=\,-\sum_{a=1}^{s}\kk_a \cos \theta_a\, \diff \phi\,.
\end{equation}
Solving the equation for $\breve \omega$ requires a bit more effort. We relegate the details to appendix~\ref{app:omega_eq}, and present here just the final result, which is: 
\begin{equation}
\label{eq:omega_final}
\ii\breve \omega=\sum_a \ii\ww_a \cos\theta_a \diff \phi +\sum_{a}\sum_{b>a}\frac{C_{ab}}{\delta_{ab}}\left(1+\cos\theta_a\right)\left(1-\frac{r_a+\delta_{ab}}{r_b}\right)\diff \phi\,, 
\end{equation}
where
\begin{equation}
\label{eq:omega_a_gen}
\ii\ww_{a}=h_0\mm_a-\mm_0 h_a+\frac{3}{2}\left(\kk_0 \ell_a-\ell_0\kk_a\right)-\sum_{b\neq a}\frac{C_{ab}}{|\delta_{ab}|}\, ,
\end{equation}
\begin{equation}
\label{eq:C_ab_gen}
C_{ab}=h_a \mm_b- h_b \mm_a+ \frac{3}{2}\left(\kk_a \ell_b-\kk_b \ell_a\right)\, ,
\end{equation}
and $\delta_{ab}\equiv z_a-z_b$. In deriving \eqref{eq:omega_final}, it has been assumed that $\delta_{ab}>0$ if $a<b$. Thus, only the absolute value of $\delta_{ab}$ appears in \eqref{eq:omega_final}. We have also fixed an integration constant such that 
\be
\sum_a \ww_a =0\,.
\ee

\paragraph{Asympotic flatness condition.} We shall consider solutions that asymptote to ${S}^1\times {\mathbb R}^4$, where $S^1$ is parameterized by the $\tau$ coordinate. This implies, in particular, that the Gibbons-Hawking base space must be asymptotically $\mathbb R^4$. The asymptotics of the multi-centered Gibbons-Hawking base-space are controlled by the function $H$~\cite{Gibbons:1979xm,Bena:2007kg,Gibbons:1978tef,Gibbons:1987sp,Gibbons:2013tqa}. Asymptotic flatness is indeed achieved imposing
\begin{equation}
h_0\,=\,0 \,, \hspace{1cm} \sum_a h_a\,=\,1\, ,
\end{equation}
and that the metric functions $f$ and $\omega_{\psi}$ tend to constant values at infinity ($1$ and $0$ respectively). The latter conditions fix the constant terms of the remaining harmonic functions as
\begin{equation}\label{eq:cond_par_0}
\kk_0\,=\,0\,, \hspace{1cm} \ell_0\,=\,1\,, \hspace{1cm}\mm_0 \,=\, -\frac{3}{2}\sum_a\kk_a\, .
\end{equation}
 It is convenient to recall that the metric and gauge field strength are invariant under the following shifts of the harmonic functions,
\be
\begin{aligned}
&K \ \to K + \mu H\,,\\[1mm]
&L \ \to \ L-2\mu K - \mu^2 H\,,\\[1mm]
& M \ \to \ M - \frac{3}{2}\mu L + \frac32 \mu^2K + \frac12\mu^3H\,,
\end{aligned}
\ee
where $\mu$ is a constant. We fix this redundancy by imposing 
\begin{equation}
\sum_{a} \kk_a\,=\,0\, .
\end{equation}

\paragraph{Cap condition.} One of the main advantages of working in  Euclidean signature is that the solution one obtains after Wick-rotating a thermal black hole smoothly caps off at the position of the horizon, thereby excising the singularity. Since the goal of this chapter is to construct gravitational solitons of this kind, we have to demand that the solution has a cap, instead of the infinite AdS$_{2}$ throat characteristic of the near-horizon of supersymmetric and extremal black holes. As it turns out, this is equivalent to demanding 
\begin{equation}
\lim_{r_a\to 0} f^{-1}\,=\, {\cal O}(r_a^0)\, , \hspace{1cm}\lim_{r_a\to 0} \omega_{\psi}\,=\, {\cal O}(r_a^0)\, ,
\end{equation}
which in turn implies that the coefficients of the harmonic functions $L$ and $M$ are determined in terms of those of $H$ and $K$ by
\begin{equation}
\label{eq:cap_cond}
\ell_a\,=\, \frac{\kk_a^2}{h_a}\,, \hspace{1cm} \mm_a\,=\, -\frac{\kk^3_a}{2h_a^2}\, .
\end{equation}
Indeed, if $\lim_{r_a\rightarrow 0} f =0$ the solution develops an infinitely long AdS$_2$ throat near the center, then $f^2$ has a double pole and the solution is necessarily extremal (cf.~\cite{Boruch:2023gfn}). 

Not surprisingly, one can verify that these conditions are also satisfied by the smooth horizonless ``microstate'' geometries constructed in \cite{Bena:2005va, Berglund:2005vb}. Indeed, a central feature of these solutions is the absence of an AdS$_{2}$ throat. We further discuss the relation between our solutions and those of \cite{Bena:2005va, Berglund:2005vb} in some explicit examples presented later in this Chapter.

Additionally, in this class of solutions the following property holds
\begin{equation}\label{eq:smooth_center_condition}
   - h_a \,\lim_{r_a\rightarrow 0}\omega_\psi \,=\, \ww_a\,,
\end{equation}
which will be relevant when analyzing regularity of the metric around the centers in the next section.

\paragraph{The solution so far.}
Imposing the above conditions, the harmonic functions specialize to
\begin{equation}
\label{eq:harm_fct_sol_so_far}
H=\sum_{a}\frac{h_a}{r_a}\,,\quad\ \ \ii K= \sum_{a}\frac{\kk_a}{r_a}\,,\quad\ \ L=1+ \sum_{a} \frac{\kk_a^2}{h_a r_a}  \, , \quad\ \ \ii M= -\frac{1}{2}\sum_{a}  \frac{\kk^3_a}{h_a^2 r_a}\,,\quad
\end{equation}
with $\sum_a h_a=1$ and $\sum_{a} \kk_a=0
$. The solution then depends on $2s-2$ harmonic function coefficients, as well as $s-1$ distances between the centers. 
The coefficients \eqref{eq:omega_a_gen}, \eqref{eq:C_ab_gen} determining the local one-form $\breve \omega$ given in~\eqref{eq:omega_final} boil down to
\begin{equation}
\label{eq:breve_omega_a}
\ii \ww_a\,=\,-\frac{3}{2}\,\kk_a-\sum_{b\neq a} \frac{C_{ab}}{|\delta_{ab}|}\,, \,
\end{equation}
and
\begin{equation}
C_{ab}\,=\, \frac{h_a h_b}{2}\left(\frac{\kk_a}{h_a}-\frac{\kk_b}{h_b}\right)^3\, .
\end{equation}
The coefficients $\ww_a$ satisfy $\sum_a w_a = 0$ and also determine  the function $\omega_\psi$ near to the $a$-th center via~\eqref{eq:smooth_center_condition}.

\paragraph{Asymptotic charges.} The mass and angular momenta, associated with the symmetries generated by the Killing vectors $\partial_t $, $\partial_\phi$ and $\partial_\psi$, can be read from the asymptotic expansion of the Lorentzian metric. To this aim, we introduce the coordinates
\begin{equation}\label{eq:phi1phi2coords}
{\tilde r}^2\,=\, 4r\,, \hspace{1cm}\phi_1\,=\,\frac{\phi-\psi}{2}\,, \hspace{1cm} \phi_{2}\,=\, \frac{\phi+\psi}{2}\, ,
\end{equation}
and use the  Lorentzian time $t\,=\,-\ii \tau$. The
 asymptotic expansion of the relevant components of the metric then reads
\begin{equation}
\begin{aligned}
g_{tt}\,=\,&-1+\frac{8\sum_a\ell_a}{{\tilde r}^2}+\dots\,, \\[1mm]
g_{t\phi_1}\,=\,&\frac{4\ii \sum_a\left(-2\mm_a+ 3\kk_a z_a\right)\sin^2 \frac{\theta}{2}}{{\tilde r}^2}+\dots\,, \\[1mm]
g_{t\phi_2}\,=\,&\frac{4\ii \sum_a\left(2\mm_a+ 3\kk_a z_a\right)\cos^2 \frac{\theta}{2}}{{\tilde r}^2}+\dots\,, \\[1mm]
\end{aligned}
\end{equation}
from which we extract that the mass $E$ and the angular momenta $J_{\pm}\,=\, \frac{J_1\pm J_2}{2}\,$:
\begin{equation}
\label{eq:mass+ang_momenta}
E\,=\, 3\pi\sum_a \frac{\kk_a^2}{h_a}\, , \hspace{1cm}
J_+\,=\,-3\pi\ii \sum_a\kk_a z_a \,, \hspace{1cm} J_{-}\, =\, -\pi\ii \sum_a \frac{\kk_a^3}{h_a^2}\, .
\end{equation}
 Since the solutions are supersymmetric, according to \eqref{eq:superalgebraaf} the electric charge $Q$ must be related to the mass by
\begin{equation}
E\,=\,\sqrt{3}\,Q\, .
\end{equation}
We can verify this by considering the asymptotic expansion of the field strength,
\begin{equation}
F_{t{\tilde r}}\,=\, \sqrt{3}\,\partial_{{\tilde r}}f\,=\, \frac{8\sqrt{3}}{{\tilde r}^3}\sum_a \frac{\kk_a^2}{h_a}+\dots\, ,
\end{equation}
from which it follows that the electric charge is
\begin{equation}
\label{eq:electric_charge}
Q\,=\, \sqrt{3}\,\pi \sum_a \frac{\kk_a^2}{h_a}\, .
\end{equation}


\subsection{Global identifications and asymptotic boundary conditions}
\label{sec:global_id}
Using the coordinates introduced above, the asymptotic metric and gauge field read
\begin{equation}\label{eq:asymptotics}
\begin{aligned}
\diff s^2\ &\longrightarrow\ \diff \tau^2 +\diff {\tilde r}^2 + {\tilde r}^2 \diff s^2_{S^3}\,,\\
A\ &\longrightarrow\  \sqrt{3}\,\ii \left(\diff\tau - \diff{\alpha}_\infty\right) \,, 
\end{aligned}
\end{equation}
 where again  ${\tilde r}^2\,=\, 4r$, while $\alpha_\infty$ is defined as the asymptotic value of $\alpha$, and
\be\label{eq:S3metric_2center}
\diff s^2_{\,S^3} \,=\, \frac{1}{4} \left[  \diff\theta ^2 + \sin^2 \theta \diff \phi^2 +  \left( \diff \psi + \cos\theta \diff\phi\right)^2 \right]\,.
\ee
For this to describe the unit metric on a smooth  $S^3$ parameterized by $(\theta,\phi,\psi)$, we must take 
 $\theta\in [0,\pi]$ and demand regularity at the poles in $\theta=0$ and $\theta =\pi$,  which leads to the identifications 
\begin{equation}\label{S3_angles_identifications}
\left( \phi ,\,\psi \right)  \,\sim\, \left( \phi + 2\pi ,\,\psi - 2\pi \right) \,\sim\, \left( \,\phi ,\,\psi + 4\pi \right)\,.
\end{equation}
Crucially, since we are compactifying the Euclidean time we must  also specify how the angular coordinates are identified when going around the thermal $S^1$. This can be expressed as
\begin{equation}\label{beta_identifications}
\left( \tau ,\,\phi ,\,\psi \right) \,\sim\, \left( \tau + \beta ,\,\phi -  \ii \omega_+,\,\psi + \ii \omega_-\right) \,,
\end{equation}
where $\beta$ is interpreted as inverse temperature, while $\omega_\pm$ play the role of angular velocities (at least when these are real quantities).  
Overall, the angular coordinates satisfy the identifications,\footnote{\label{foot:phi1phi2basis}Making the transformation $\phi = \phi_1+\phi_2$, $\psi = \phi_2-\phi_1$, we obtain standard $2\pi$-periodic coordinates $\phi_1,\phi_2$ satisfying the more symmetric identifications \eqref{eq:globalidentifications0}
while the metric \eqref{eq:S3metric_2center} becomes
$$\diff s^2_{S^3}\,=\, \diff\vartheta^2 + \sin^2\vartheta \,\diff\phi_1^2+\cos^2\vartheta \,\diff\phi_2^2\,,\qquad \text{with}\ \vartheta = \theta/2\,.$$
To the vectors $\partial_{\phi_1}=\partial_{\phi}-\partial_{\psi}$ and $\partial_{\phi_2}=\partial_{\phi}+\partial_{\psi}$ we associate the angular momenta $J_1,J_2$ and the angular velocities $\omega_1,\omega_2$, which are those appearing in section \ref{sec:Grav_path_int}. These are related to the quantities appearing in this chapter as 
$$\omega_\pm \,=\, \omega_1 \pm \omega_2\,,\qquad\qquad J_\pm \,=\, \frac{1}{2}(J_1\pm J_2)\,,$$ which explains the origin of the $\pm$ labels used in the main text. 
Note that in our conventions, $J_+$ advances $\phi$, while $J_-$ advances $-\psi$.}
\begin{equation}\label{eq:coord_identif}
\left( \tau ,\,\phi ,\,\psi \right) \,\sim\, \left( \tau + \beta ,\,\phi - \ii \omega_+,\,\psi + \ii \omega_-\right) \,\sim\, \left( \tau ,\,\phi + 2\pi ,\,\psi - 2\pi \right) \,\sim\, \left( \tau ,\,\phi ,\,\psi + 4\pi \right)\,.
\end{equation}
A basis of independent Killing vectors having closed $2\pi$-periodic orbits under the identifications above is given by
\be\label{eq:U(1)isometries}
 \frac{1}{2\pi}\left(\beta\partial_\tau -\ii \omega_+\partial_\phi +\ii \omega_-\partial_\psi\right)
\,,\qquad\quad \partial_\phi - \partial_\psi \,,\qquad\quad  \partial_\phi + \partial_\psi\,.
\ee
These generate the ${\rm U}(1)^3$ isometry which will play a central role in the following. We will denote the first as the {\it thermal isometry}, and the other two as axial isometries.
The orbits of $\partial_\phi \pm \partial_\psi$ are contractible, since they collapse to zero size  at either one of the poles of the asymptotic $S^3$.  
Then fermion fields must be antiperiodic when going around such orbits.
Recalling \eqref{eq:coord_identif}, we infer that shifts of the form $\omega_\pm \to \omega_\pm + 4\pi \ii n_\pm$, with $n_\pm\in\mathbb Z$, are an invariance of the boundary conditions for both bosonic and fermionic fields. As far as the boundary conditions are concerned,  
we can thus take 
\be\label{eq:restriction_bdry_cond}
{\rm Im\, \omega_\pm} \in [0,4\pi)
\ee
with no loss of generality. 
Regarding the thermal isometry, we assume that the theory admits a supersymmetric non-extremal black hole solution having precisely inverse temperature $\beta$ and angular velocities $\omega_\pm$ (this will be discussed later in this chapter). Then the first vector in~\eqref{eq:U(1)isometries} is proportional to the generator of the horizon, and its orbits are contractible in the bulk of this solution.
Well-definiteness of the spinorial supersymmetry parameter requires it to be antiperiodic when completing one revolution   around the contractible orbits, which imposes
\be\label{eq:choice_om+}
 \omega_+ \,=\, 
  2\pi \ii\,.
 \ee
In order to see this, note that the supersymmetry parameter is neutral under $\partial_\tau$, while (in our conventions) it has charge $+1/2$ under both $\partial_\phi \pm \partial_\psi$, namely it has charge $1/2$ under $\partial_\phi$ and charge 0 under $\partial_\psi$; hence the choice \eqref{eq:choice_om+} is needed for the spinor to acquire the correct phase $\rme^{ 2\pi \ii \cdot \frac12}=-1$. 

The boundary conditions are completed by specifying the holonomy of the supergravity gauge field $A$ around the asymptotic $S^1$, which measures an electrostatic potential. We define it as
\begin{equation}\label{eq:def_varphi}
    \varphi \,=\, -\ii \int_{S^1_\infty} \big(A - \sqrt 3\, \ii \,\diff\tau\big) \,=\, - \sqrt 3 \int_{S^1_\infty} \diff\alpha_\infty  \,,
\end{equation}
where in the second equality we have used the asymptotic form of the gauge field given in~\eqref{eq:asymptotics}.

Regularity in the bulk provides a map between  $\beta,\omega_\pm,\varphi$ and the parameters of the solution introduced above. This depends on which combinations of the ${\rm U}(1)$ isometries degenerate.

\subsection{Two-center solutions}\label{sec:2c_ssol}

We begin with two-center solutions, described by harmonic functions of the form
\begin{equation}
\label{eq:harmonicfunctiongeneral2c}
\begin{aligned}
H \,=&\,\, h_0+\frac{h_N}{r_N}+ \frac{h_S}{r_S}\,\,,\qquad\qquad\qquad\qquad \ii K \,=\, \mathsf k\left( \frac{1}{r_N}-\frac{1}{r_S}\right)\,,\\[1mm]
L\, =&\,\, 1+ \mathsf k^2\left( \frac{1}{h_N\,r_N}+\frac{1}{h_S\,r_S}\right)\,\,,\quad
\ii M\,=\, -\frac{\mathsf k^3}{2}\left(\frac{1}{h_N^2\,r_N}-\frac{1}{h_S^2\,r_S}\right)\,,
\end{aligned}
\end{equation}
where $r_{N,S}$ denote the distances in $\mathbb R^3$ from the two centers. 

In this chapter we work with a real metric (while the gauge field becomes purely imaginary). This choice has two advantages: on the one hand, it gives us greater control over the regularity conditions; on the other, it remains compatible with the assumptions of the localization theorem. To implement this setup, we have performed an analytic continuation of the coefficients of the harmonic functions $K$ and $M$ (setting $k = - \ii \mathsf{k}$, where $k$ are the Lorentzian coefficients). As we will see, this requirement will be relaxed in the next chapter.

\paragraph{ALF spaces.}

Here, we also consider a slightly more general ansatz for the harmonic functions. In particular, asymptotically flat geometries arise by setting $h_0 = 0$ and $h_+ \equiv h_N + h_S = 1$ in \eqref{eq:harmonicfunctiongeneral2c}. For our purposes, however, we also allow more general values of the coefficients $h_0$ and $h_+$ in the harmonic function $H$, corresponding to configurations that are not asymptotically flat, that are required for the discussion in appendix~\ref{app:4Dexample}.

Indeed, when $h_0\neq 0$, the harmonic function tends to $H \to h_0$ at large $r$, and the four-dimensional hyper-Kähler base is asymptotic to the twisted product $S^1 \times \mathbb R^3$. Such spacetimes are referred to as asymptotically locally flat (ALF) (see e.g.~\cite{Gibbons:2013tqa}). This behaviour requires
\begin{equation}\label{eq:harmonicfunctionsalf}
H = h_0 + \mathcal O(r^{-1})\,\,,\quad\quad K = \mathcal O(r^{-2})\,\,,\quad\quad M = \mathcal O(r^{-2}) \,\,,\quad\quad L = 1+ \mathcal O(r^{-1})\,,
\end{equation}
which in turn implies $\ell_0=1$, $\sum_a \kk_a = 0$, and $\sum_a \mm_a=0$. Using \eqref{eq:cap_cond}, one then finds $h_N=h_S$. In this case, the periodicity $\Delta_\psi$ of the $S^1$ can be determined by studying the global properties of the ALF solution.\footnote{If we take $\Delta_\psi = 4\pi$ together with $h_0= 0$, then we must take $h_+ \in \mathbb Z$ to have a smooth geometry realizing a straightforward generalization of the AF solutions, by allowing for a simple asymptotic orbifold. Indeed, in this case the hyper-K\"ahler base is an asymptotically locally Euclidean (ALE) space, describing an asymptotic $\mathbb R^4/\mathbb Z_{|h_+|}$.\label{foot:ALE}} 
To distinguish between AF and ALF solutions, we will keep the parameters $h_{0}$, $h_{N,S}$, and $\Delta_\psi$ arbitrary throughout this chapter, specifying their values when required.


\medskip 

Two-center solutions always possess an additional U$(1)$ isometry generated by $\partial_\phi$, corresponding to rotations about the axis connecting the two centers. Without loss of generality, we place the centers on the $z$-axis at $z_{N,S}=(0,0,\pm \delta/2)$, where $\delta$ is their separation. In this coordinate system, they coincide with the north and south poles of the two-sphere parametrized by the coordinates $(\theta,\phi)$ of \eqref{eq:coordinatesystem1}, and therefore we will refer to them as $\mathcal M_N$ and $\mathcal M_S$, respectively.

\subsubsection{Regularity in the bulk and parameters of the solution}\label{sec:onshellactionpreliminaries}

In this section we are interested in studying the two-center solutions and compute their on-shell action as an application of the BVAB theorem reviewed above. 
In order to apply the localization formula we need a Killing vector with fixed points. One may a priori consider the supersymmetric Killing vector $V =\partial_\tau$, whose squared norm is given by  $g_{\tau\tau} = f^2$. In fact, for supersymmetric and extremal black holes the horizon three-sphere is a bolt for this vector. 
However, in the present chapter we are mostly interested in non-extremal solutions, with a thermal circle of finite length $\beta$.
Indeed, to ensure the absence of AdS$_2$ throats characteristic of extremal solutions, we required in \eqref{eq:cap_cond} that $f$ is a non-vanishing function on the base-space. As a result, then the value of $f$ at the north center is
\be
f_N\,\equiv \, f\big|_{{\cal M}_N} \,=\,  \frac{\delta}{\delta\left( 1+ \frac{h_0\kk^2}{h_N^2}\right)+ \frac{\kk^2h_+^2}{h_N^2h_S}}\,,
\ee
while the value $f_S$ at the south center is obtained from this expression by exchanging $h_N$ and $h_S$. The extremal limit is obtained by sending $\delta\to 0$ (it will be clear later that this implies $\beta\to\infty$), so that the two centers coalesce; we see that in this limit we recover the vanishing value $f_N=f_S=0$.
 
 We therefore consider a different vector for localizing.\footnote{Of course, one can still apply the BVAB formula \eqref{eq:localized_action_5d} using the vector $V=\partial_\tau$. As already noted in section~\ref{sec:equiv5d}, if all regularity assumptions are satisfied this results in the on-shell action being expressed as a combination of boundary terms.}   
 Using the additional ${\rm U}(1)$ symmetry, we can combine the two commuting isometries $\partial_\tau$ and $\partial_\psi$ and consider Killing vectors of the type
\begin{equation}\label{eq:kvsusy}
\xi_a = V + \Omega_a\, U\,, \qquad\text{with}\qquad V=  \partial_\tau \,\,,\quad U = \partial_\psi\,,
\end{equation}
where $\Omega_a$ is a parameter, and $a= N,S$ labels the possible choices. This choice is very natural for solutions with a Gibbons-Hawking base-space, since it generates an isometry that preserves the hyper-K\"ahler complex structures. 
The norm is given by
\begin{equation}
|\xi_a|^2\,=\, f^2\left( 1+ \Omega_a\,\ii \omega_\psi\right)^2 + \frac{\Omega_a^2}{f\,H}\,.
\end{equation}
This vanishes at the centers ${\mathcal M_a}$, where the harmonic function $H$ diverges, if we choose
\begin{equation}\label{eq:vanishingnorm}
\begin{aligned}
\Omega_a &= -\frac{1}{\ii \omega_\psi}\bigg|_{\mathcal M_a} \,\,,\quad\quad a= N,S\,,
\\
&= \frac{h_a}{\ii w_a}\,,
\end{aligned}
\end{equation}
provided $ \omega_\psi$ remains finite at the centers, as ensured by \eqref{eq:cap_cond}. To obtain the second line, we used \eqref{eq:smooth_center_condition}.

Therefore, the Euclidean geometries of interest are characterized by four key quantities, which will play a role in the localization argument. Firstly, the inverse-Hawking temperature, $\beta$, and the electrostatic potential, $\Phi$. Additionally, these solutions possess a pair of Killing vectors of the form \eqref{eq:kvsusy}, each with fixed points at one of the two centers. These vectors are determined by $\Omega_N$ and $\Omega_S$, which encode the relevant local data around the fixed-point sets.
Below, by studying the regularity conditions of the solution about the centers we are going to relate these four quantities to the  parameters of the solution
\begin{equation}
\left( \beta\,,\Phi\,,\,\Omega_N\,,\,\Omega_S\right)\ \longleftrightarrow\ \left( \delta\,,\,\mathsf k\,,\,h_N\,,\,h_S\right)\,.
\end{equation}
On top of that, by studying the global properties of the saddle, one can relate $\Omega_{N,S}$ to the chemical potentials $\omega_{1,2}$.

\paragraph{Regularity of the metric at the centers.}
In order to study the regularity of the metric around the centers of the harmonic functions it is convenient to introduce two sets of spherical coordinates on the three-dimensional base-space $\mathbb R^3$, as in \eqref{eq:coordinatesystem2}, originating from either one of the two centers.
Using these coordinates in \eqref{eq:omega_final} and \eqref{eq:explicit_oneform1} we determined the one-forms $\chi$ and $\breve\omega$ by solving eqs.~\eqref{eq:defchi}, \eqref{eq:susysolminimalghbase3}. We specialize here the above expressions to the two-center case:
\begin{equation}\begin{aligned}
\chi =&\, \left( h_N \cos\theta_N + h_S \cos\theta_S\right) \diff \phi\,,\\[1mm]
\breve\omega =&\, \left( w_N \cos\theta_N + w_S \cos\theta_S\right) \diff\phi + w_{\rm reg}\,,
\end{aligned}
\end{equation}
with
\begin{equation}\label{eq:omegaterms}
\begin{aligned}
\ii w_{N} \,= &\,\,-\frac{h_0\kk^3}{2h_N^2} -\frac{3}{2}\mathsf k -\frac{\kk^3 h^3_+}{2h_N^2 h_S^2\delta}\,,\hspace{1cm}\ii w_{S} \,=\, \,\frac{h_0\kk^3}{2h_S^2} +\frac{3}{2}\mathsf k +\frac{\kk^3 h^3_+}{2h_N^2 h_S^2\delta}\,,\\[1mm]
\ii w_{\rm reg} \,=&\,\, \frac{\kk^3 h^3_+}{2h_N^2 h_S^2\delta}\left(\cos\theta_N +1\right) \left( 1-\frac{r_N + \delta}{r_S}\right)\diff\phi\,,
\end{aligned}
\end{equation}
where $w_{\rm reg}$ denotes the regular part of $\breve\omega$, that vanishes on the entire $z$-axis. 
We notice that for both AF and ALF spacetimes one finds $w_N = -w_S$.
We rewrite the metric \eqref{eq:susy_conf} isolating the coordinate $\psi$, obtaining
\begin{equation}
\diff s^2 =Y\left( \diff\psi + \chi + \frac{\ii \omega_\psi f^2}{Y}\left( \diff\tau + \ii \breve\omega\right)\right)^2 + \frac{f}{HY}\left( \diff\tau + \ii \breve\omega\right)^2 + \frac{H}{f}\diff s^2_{\mathbb R^3}\,,
\end{equation}
where 
\begin{equation}
Y = f^2\,(\ii\omega_\psi)^2 + \frac{1}{f\,H}\,.
\end{equation}
Next, we expand the metric at leading order around the north center. To do so, we introduce a radial coordinate $\tilde r_N = 2 r^{1/2}_N$ and consider the limit $\tilde r_N\rightarrow 0$. We obtain
\begin{equation}\label{eq:metricaroundnut}
\begin{aligned}
\diff s^2 \,& \underset{\tilde r_N\to 0}{\longrightarrow} \ \frac{f_N^2}{\Omega_N^2}\Big( \diff \psi + \left( h_N \cos\theta_N + h_S\right) \diff\phi - \Omega_N \left( \diff\tau + \ii w_N\left( \cos\theta_N -1\right) \diff\phi\right)\Big)^2 \\[1mm]
&\ +\frac{h_N}{f_N}\left[\diff \tilde r_N^2 + \frac{\tilde r_N^2}{4}\left(\frac{\Omega_N^2}{h_N^2}\left(\diff \tau + \ii w_N\left( \cos\theta_N -1\right)\diff \phi \right)^2 + \left( \diff\theta_N^2 + \sin^2\theta_N \diff\phi^2\right) \right)\right]\,.
\end{aligned}
\end{equation}
Regularity of this metric is guaranteed by \eqref{eq:vanishingnorm}, 
which further implies that the center is the smooth origin of an $\mathbb R^4$ factor, with $(\theta_N,\phi,\tau)$ describing a round $S^3$ shrinking along the $\tilde r_N$ radial direction,  provided $\beta$ satisfies 
\begin{equation}\label{eq:inversehawkingtemperaturesusy}
\beta = 4\pi  \ii w_N\,,
\end{equation} 
and the coordinates satisfy suitable identifications.\footnote{This can be extended by allowing for orbifolds of the type $\left( S^1 \times \mathbb R^4\right) / \mathbb Z_{|m|}$ (which can be freely acting), for a certain integer $m$, with $|m|>1$, at the centers. Such orbifolded saddles will be considered in the next chapter.} 
In order to show that the full five-dimensional metric around the center is smooth, we introduce the new coordinates
\begin{equation}\label{eq:regularnutcoordinate}
\tilde\tau =\frac{\Omega_N}{2h_N}\tau \,\,,\quad\quad \phi_N = \phi-\frac{\Omega_N}{2h_N}\tau\,\,,\quad\quad \psi_N = \psi + h_+ \phi - \Omega_N \tau\,,
\end{equation}
with
\begin{equation}
\tilde r_N = \sqrt{y_1^2 + y_2^2} \,\,,\quad\quad \theta_N = 2\,{\rm arctan}\left( \frac{y_1}{y_2}\right)\,,
\end{equation}
which give\footnote{We note that when writing \eqref{eq:spaceatnut} we are ignoring mixed components of the metric,  $g_{\psi_N \phi_N}$ and $g_{\psi_N {\tilde\tau}}$, which are of order ${\cal O}\left({\tilde r}_N^2\right)$.} 
\begin{equation}\label{eq:spaceatnut}
\diff s^2  \, \underset{\tilde r_N\to 0}{\longrightarrow} \  \frac{f_N^2}{\Omega_N^2}\,\diff\psi_N^2+ \frac{h_N}{f_N}\left( \diff y_1^2 + y_1^2\,\diff\phi_N^2+ \diff y_2^2 + y_2^2\,\diff\tilde\tau^2\right)\,,
\end{equation}
where the compact coordinates $\tilde\tau$ and $\phi_N$ are $2\pi$-periodic. On the other hand, the compact coordinate $\psi_N$ is identified with period $\Delta_\psi$, which can be determined by studying the global properties of the specific solution. 

Using the coordinates introduced in \eqref{eq:regularnutcoordinate}, one can write the Killing vector that vanishes at the north pole as
\begin{equation}\label{eq:kvnorth}
\xi_N = \epsilon_1^N \partial_{\tilde\tau} + \epsilon_2^N \partial_{\phi_N}\,,
\end{equation}
where $\tilde\tau$ and $\phi_N$ are standard polar coordinates of period $2\pi$ that rotate each orthogonal copy of $\mathbb R^2$, and $\epsilon_{1,2}^N$ are the weights of such rotations. Namely, $\epsilon_{1,2}^{N}$ is the set of equivariant parameters that specifies the action of the Killing vector $\xi_{N}$ at the nut. As is evident from \eqref{eq:spaceatnut}, the tangent space to the nut comprises an $\mathbb R^2\oplus \mathbb R^2$ factor, and each point in $\mathbb R^2\oplus \mathbb R^2$ is associated with an $S^1$ parameterized by $\psi_N$ that remains invariant under the action of the Killing vector \eqref{eq:kvnorth}. 
The equivariant parameters can be explicitly computed by starting with the Killing vector given in \eqref{eq:kvsusy}. After performing the change of coordinate \eqref{eq:regularnutcoordinate}, we express the vector as shown in \eqref{eq:kvnorth}. Then, they are given by
\begin{equation}\label{eq:equivparams}
\epsilon_1^N = \frac{2\pi}{\beta}= - \epsilon_2^N\,.
\end{equation}
A similar expansion around the south pole gives 
the equivariant parameters
\begin{equation}\label{eq:equivparams2}
\epsilon_1^S = \frac{2\pi}{\beta}= \epsilon_2^S\,.
\end{equation}

\paragraph{Regularity of the one-forms.} We require that the gauge field $A$ be regular at the nuts. To achieve that, it is sufficient to choose $\diff\alpha = \alpha^{(\tau)}\diff \tau$,
and impose the conditions:
\begin{equation}\label{eq:reggaugevector}
\iota_{\xi_a}A\big|_{\mathcal M_a} \,=\, 0 \quad \Leftrightarrow \quad  \left[\alpha_\tau -  f\left( 1+ \Omega_a\,\ii\omega_\psi \right) +  \Omega_a\frac{\ii K}{H}\right]_{\mathcal M_a}\!\!=
 0\,,\quad\ a = N,S.
\end{equation}
These two separate conditions should be met for a given value of $\alpha^{(\tau)}$. Using \eqref{eq:vanishingnorm}, the equations \eqref{eq:reggaugevector} have a unique solution 
\begin{equation}\label{eq:gaugechoice}
\alpha^{(\tau)} \,=\, -\frac{4\pi}{\beta}\kk \,.
\end{equation}
In this gauge, the electrostatic potential is determined from the asymptotic behaviour of the gauge field, which approaches $A \rightarrow \ii\sqrt{3} \left(1 -\alpha^{(\tau)}\right) \diff \tau$. Consequently, we find
\begin{equation}
\Phi = \sqrt{3}\left( 1- \alpha^{(\tau)}\right) = \sqrt{3}\left( 1 + \frac{4\pi}{\beta}\kk\right)\,.
\end{equation}

\paragraph{In summary.} The regularity analysis establishes the following map between the coefficients of the harmonic functions and the thermodynamic quantities, that serve as independent variables of the solution:
\begin{equation}\label{eq:mapcoeffchempot}
\mathsf k = \frac{\varphi}{4\sqrt{3}\pi}\,\,,\quad\quad h_N = \frac{\beta \Omega_N}{4\pi}\,\,,\quad\quad h_S = - \frac{\beta\Omega_S}{4\pi}\,,
\end{equation}
where we used $\varphi = \beta \left(\Phi - \sqrt{3}\right)$. Here, $\beta$ is related to the parameters of the geometry through the first relation in \eqref{eq:inversehawkingtemperaturesusy}, together with \eqref{eq:omegaterms}. The same regularity conditions also yield an expression for the equivariant parameters in terms of the boundary data, as given in \eqref{eq:equivparams}.


 \section{On-shell action of two-center solutions\label{sec:susyaction}}


In this section we compute the on-shell action equivariantly for the class of solutions described above. While some of the steps involved apply more generally to the full family of multi-center solutions with Gibbons-Hawking base-space, regularity conditions in odd-dimensional cases play a crucial role and must be addressed case-by-case. For concreteness, we will restrict our discussion to two-center solutions, leaving for future work the extension to more intricate setups, such as harmonic functions with additional centers.

We start by constructing the equivariantly closed form \eqref{eq:polyform5d2} needed for the localizing the action. We are free to work equivariantly either with respect to $\xi_N$ or $\xi_S$. Let us consider $\xi_N$. The only missing ingredient is the evaluation of the one-form potential $\nu$ according to its definition \eqref{eq:definitionnu5d}.\footnote{From now on we drop the `$N$' from $\nu_N$ so as to simplify the notation.} Note that $\nu$ is defined locally, i.e.\ up to an exact term. This should be chosen so that $\nu$ is globally well-defined.
A convenient way to determine $\nu$ is to split it into the sum
\begin{equation}
\nu = \nu_V +\Omega_N\, \nu_{U}\,\,,\quad\quad \diff \nu_V= \iota_V G\,\,,\quad\quad \diff \nu_{U} = \iota_{U} G\,,
\end{equation}
and use the relations given by the supersymmetry equations for the general solution with Gibbons-Hawking base-space \eqref{eq:susy_conf}. 
A lengthy, but straightforward, computation gives
\begin{equation}\label{eq:nuV}
\nu_V = \sqrt{3}\,\ii V^\flat - f\,A- \alpha^{(\tau)} \,A\,,
\end{equation}
where $V^\flat = g_{\mu\nu}V^\nu\diff x^\mu$, and
\begin{equation}\label{eq:nuU}
\nu_{U} = -\sqrt{3}\,\ii \left[\varepsilon + \ii \left(\frac{K}{H}-f\omega_\psi \right)\left(\ii \breve A + \alpha^{(\tau)} \diff\tau \right) + f\frac{\ii K}{H}\left( \diff \tau + \ii \breve \omega\right)+\mu\right]\,,
\end{equation}
with
\begin{equation}
\diff \varepsilon = \star_3 \diff L\,,
\end{equation}
In \eqref{eq:nuU} we used the freedom to add an exact term $\mu\equiv \mu_\tau \diff \tau + \mu_\psi \diff \psi$, for some constant coefficients $\mu_\tau$ and $\mu_\psi$. These are determined by solving the conditions
\begin{equation}\label{eq:regularitynususy}
\iota_{\xi_N} \nu = 0\,\,,\quad\quad \iota_{\xi_S} \nu\big|_S = 0\,,
\end{equation} 
where the first follows from regularity together with equivariant closure as in \eqref{eq:definitionnu5d}, while the second is required by regularity.
Using \eqref{eq:gaugechoice}, \eqref{eq:vanishingnorm} and \eqref{eq:gaugechoice}, these give 
\begin{equation}
\begin{aligned}
\iota_{\xi_N}\nu\,=\, \frac{\ii}{\sqrt{3}}\left(\left(\Phi- \sqrt{3}\right)^2 - 3\Omega_N \left( \mu_\tau + \Omega_N \mu_\psi\right)\right)=0\,\,,\\[1mm]
\iota_{\xi_S}\nu\big|_S\,=\, \frac{\ii}{\sqrt{3}}\,\frac{\Omega_N}{\Omega_S}\left( \left(\Phi- \sqrt{3}\right)^2 - 3\Omega_S (\mu_\tau + \Omega_S \mu_\psi )\right) =0\,,
\end{aligned}
\end{equation}
which are solved by
 \begin{equation}
\mu_\tau = \frac{\Omega_N + \Omega_S}{3\,\Omega_N\,\Omega_S}\left( \Phi- \sqrt{3}\right)^2\,\,,\quad\quad \mu_\psi = -\frac{\left( \Phi- \sqrt{3}\right)^2}{3\,\Omega_N\,\Omega_S}\,. 
\end{equation}

Let us observe that the expressions for $\nu_V$ \eqref{eq:nuV} and $\nu_U$ \eqref{eq:nuU} are valid for any five-dimensional supersymmetric solution with Gibbons-Hawking base-space of the type reviewed in \eqref{eq:susy_conf}, regardless of the number of centers. However, the regularity conditions for $\nu$ must be analyzed on a case by case basis. Here, using \eqref{eq:regularitynususy}, we have determined the explicit value of $\mu$ valid for the class of two-center solutions under study.


\subsection{Boundary terms}\label{sec:actionboundaryterms}

We are finally ready to evaluate the on-shell action. We first calculate the boundary terms, consisting of those from the localization argument in the second line of \eqref{eq:localized_action_5d} and the renormalized GHY boundary term~\eqref{eq:ghy_def}.  The computation in this section will confirm that the value of the constant $\crho$ that cancels the total boundary contribution in supersymmetric solutions is the one given in \eqref{eq:susy_c}, even for more general asymptotic behaviours than the asymptotically flat one discussed in section~\ref{sec:nonsusy_ex}. 

The boundary terms from the second line of \eqref{eq:localized_action_5d} are all finite, so we can directly evaluate them at $r\to \infty$. 
It follows that the one-form $\eta = \xi_N^\flat/|\xi_N|^2$ asymptotically for $r\to\infty$ goes as
\begin{equation}
\eta\, \rightarrow\, \frac{h_0\, \diff\tau + \Omega_N \left( \diff\psi + h_+ \cos\theta\diff\phi\right)}{h_0 + \Omega_N^2}\,.
\end{equation}
Asymptotically, the non-suppressed contribution of $\Psi_{(3)}$ is dominated by the term proportional to $\star_5 F$, which goes as
\begin{equation}
\star_5 F \rightarrow - \sqrt{3}\,\ii \left(\diff\psi + h_+ \cos\theta\diff\phi\right)\wedge \star_3 \diff f^{-1}  \,\to\,  \sqrt{3}\,\ii \frac{\kk^2 h_+}{h_Nh_S} \sin\theta\, \diff \theta\wedge \diff\phi\wedge\diff\psi \,,
\end{equation}
where in the second step we used \eqref{eq:cap_cond}.
 This gives
\begin{equation}\label{eq:bdrytermalf1}
\frac{1}{48\pi}\int_{\partial \mathcal M}\left(\iota_{\xi_N} A + \crho\right) \eta\wedge \star_5 F\, =\, \frac{\Upsilon}{48\sqrt{3}\pi}\frac{h_0}{h_0+\Omega_N^2}\frac{\left( \Phi- \sqrt{3}\right)^2}{\Omega_N \,\Omega_S}\left(-\Phi + \ii \crho\right)\,,
\end{equation}
where $\Upsilon \equiv \frac{(h_0+\Omega_N^2)^{2}}{\Omega_N^2} \int_{\partial \mathcal M} \diff\tau \wedge \eta \wedge \diff\eta= - h_+  \int_{\partial \mathcal M} \sin\theta\,\diff\tau \wedge \diff \theta\wedge \diff\phi \wedge \diff\psi$.
Therefore, this contribution vanishes for AF spacetimes, which are recovered by taking $h_0= 0$.

The boundary term involving $\Psi_{(1)}$ evaluates to
\begin{equation}\label{eq:bdrytermalf2}
\frac{1}{48\pi}\int_{\partial \mathcal M}\left(\iota_{\xi_N}A + \crho\right)\eta \wedge \diff \eta \wedge \nu \,=\,  \frac{\Upsilon}{48\sqrt{3}\pi}\frac{\Omega_N^2}{h_0 + \Omega_N^2}\frac{\left( \Phi- \sqrt{3}\right)^2}{\Omega_N \,\Omega_S}\left( -\Phi + \ii \crho\right)\,.
\end{equation}
Note that the sum \eqref{eq:bdrytermalf1}+\eqref{eq:bdrytermalf2} does not depend on $h_0$.

The renormalized GHY boundary term \eqref{eq:ghy_def} should be evaluated on a hypersurface at fixed $r$, with the limit  $r\to \infty$ being taken at the end of the computation. It can be written as
\begin{equation}
I_{\rm GHY} \,=\, -\frac{1}{8\pi}\int_{\partial \mathcal M}\diff^4x\left(n^\mu\partial_\mu\sqrt{h} - n_{\rm bkg}^\mu\partial_\mu\sqrt{h_{\rm bkg}}\right)\,,
\end{equation}
where the square root of the determinant of the induced metric is  $\sqrt{h}= f^{-1/2}H^{1/2}r^2\sin\theta$, the unit outward-pointing normal vector reads $n = f^{1/2}H^{-1/2}\partial_r$, while $\sqrt{h_{\rm bkg}}$ and $n_{\rm bkg}$ are the corresponding quantities for the reference background spacetime. 
 The result is
\begin{equation}\label{eq:ghyale}
\begin{aligned}
I_{\rm GHY}\,=\, -\frac{1}{8\pi}\lim_{r\rightarrow  \infty}\int_{\partial \mathcal M}\diff ^4x\, \frac{r^2}{2}\partial_r f^{-1} \sin\theta+ \mathcal O(r^{-1})\,=\, \frac{\Upsilon}{48\pi}\frac{\left( \Phi- \sqrt{3}\right)^2}{\Omega_N\,\Omega_S}
\,,
\end{aligned}
\end{equation}
which is independent of $h_0$.

It is now clear that the total boundary contribution, which is the sum of \eqref{eq:bdrytermalf1}, \eqref{eq:bdrytermalf2} and \eqref{eq:ghyale}, vanishes if we set $\crho$ as in \eqref{eq:susy_c} for both AF and ALF spacetimes.

\subsection{Fixed-point contribution}\label{sec:actionnut}

Since the final result for the action is independent of $\crho$, we can assume the choice in  \eqref{eq:susy_c}, which as we have just demonstrated makes the total boundary contribution vanish for the entire class of solutions considered in this section. Consequently, from the first line of \eqref{eq:localized_action_5d} and \eqref{eq:specify_Psi1}, we have that the full on-shell action is given by a bulk contribution localized at the one-dimensional nut,
\begin{equation}\label{eq:actionfromnut1}
I 
 \,=\, -\frac{\left(2\pi\right)^2}{48\pi}\, \crho \int_{\mathcal M_N} \frac{\iota^* \nu}{\epsilon_1^N\,\epsilon_2^N}  \,.
\end{equation}

 The pullback of $\nu_V$ \eqref{eq:nuV} and $\nu_{U}$ \eqref{eq:nuU} on the north pole, using \eqref{eq:vanishingnorm} and \eqref{eq:gaugechoice}, reads
\begin{equation}\label{eq:pullbacknususy}
\begin{aligned}
\int_{\mathcal M_N}\iota^*\nu_V \,=&\,\, \Delta_\psi \left( \Phi- \sqrt{3}\right)\frac{K}{H}\,\Big|_{\mathcal M_N}\,,
\\[1mm]
\Omega_N\int_{\mathcal M_N} \iota^*\nu_{U} \,=&\,\,- \Delta_\psi \left( \Phi- \sqrt{3}\right)\frac{K}{H}\,\Big|_{\mathcal M_S}\,.
\end{aligned}
\end{equation} 
Note that the pullback of $\nu_U$ is given by a quantity evaluated at the south pole ${\cal M}_S$ and arises entirely from the exact form $\mu$ that we introduced around \eqref{eq:nuU} to ensure regularity of $\nu$. This highlights, once again, the central role of regularity conditions in the localization argument in odd dimensions. These conditions connect local information from the south pole to the integral evaluated at the north pole. 

As a consequence of \eqref{eq:pullbacknususy}, Eq.~\eqref{eq:actionfromnut1} naturally splits into two separate blocks \\ $I = \mathcal I_N + \mathcal I_S$, where
\begin{equation}\label{eq:gravblocks}
\begin{aligned}
\mathcal I_N \,\equiv&\,\, -\frac{\left(2\pi\right)^2}{48\pi}\crho \int_{\mathcal M_N} \frac{\iota^*\nu_V}{\epsilon_1^N\,\epsilon_2^N} \,=\, \frac{\pi\Delta_\psi}{12\sqrt{3}} \frac{\left( \Phi- \sqrt{3}\right)^3}{\epsilon_1^N\,\epsilon_2^N\,\Omega_N}\,,\\[1mm]
\mathcal I_S \,\equiv&\,\, -\frac{\left(2\pi\right)^2}{48\pi}\crho\, \Omega_N \int_{\mathcal M_N} \frac{\iota^*\nu_{U}}{\epsilon_1^N\,\epsilon_2^N}\,=\, -\frac{\pi\Delta_\psi}{12\sqrt{3}} \frac{\left( \Phi- \sqrt{3}\right)^3}{\epsilon_1^N\,\epsilon_2^N\,\Omega_S}\,,
\end{aligned}
\end{equation}
with each block depending on $\Omega_N$ or $\Omega_S$, respectively. 

Finally, using \eqref{eq:mapcoeffchempot}, together with the value \eqref{eq:equivparams} of the equivariant parameters, we obtain the result for the on-shell action. We can express the action in terms of the coefficients of the harmonic functions, as 
\begin{equation}\label{eq:onshellactionfinal}
I \,=\,\mathcal I_N + \mathcal I_S=\,-\pi \Delta_\psi\frac{h_N + h_S}{h_N \,h_S}\,\mathsf k^3\,.
\end{equation}
%

\paragraph{The action as a sum of gravitational blocks.}

Note that $\mathcal I_S$ depends on the ratio $K/H$ computed at the south pole. It is then natural to ask whether we could compute $\mathcal I_S$ from a contribution that entirely comes from the south pole. This is indeed the case, as it is straightforward to show that
\begin{equation}
\mathcal I_S = -\frac{\left(2\pi\right)^2}{48\pi}\crho\int_{\mathcal M_S} \frac{\iota^*\nu_V}{\epsilon_1^S\epsilon_2^S}\,.
\end{equation}
Then, the full on-shell action can be expressed as a sum of terms that depend solely on $\nu_V$, computed both at the north and south pole,
\begin{equation}
I \,=\, \mathcal I_N + \mathcal I_S \,=\,  -\frac{\left(2\pi\right)^2}{48\pi}\crho\left( \int_{\mathcal M_N} \frac{\iota^*\nu_V}{\epsilon_1^N\epsilon_2^N} + \int_{\mathcal M_S} \frac{\iota^*\nu_V}{\epsilon_1^S\epsilon_2^S}\right)\,,
\end{equation}
which reminds of a sum over gravitational blocks~\cite{Hosseini:2019iad}. Recall that $V$ is the Killing vector obtained as a bilinear of the Killing spinor of the supersymmetric solution. This decomposition of the five-dimensional action extends what found in section \ref{sec:nonsusy_ex} for the black hole saddles and aligns with previous applications of equivariant localization in different contexts~\cite{Martelli:2023oqk,BenettiGenolini:2023yfe,Colombo:2023fhu,BenettiGenolini:2024kyy}, where gravitational blocks have been related to fixed-point contributions.


\subsection{Recovering the supersymmetric black hole saddles}\label{sec:exampleaction5d}

As a first application of the general formalism developed above, we reconsider the supersymmetric black hole saddle of section \ref{sec:susylimit_BH}, reviewing how it fits in the class of supersymmetric solutions with Gibbons-Hawking base-space and recovering the action \eqref{eq:cclpsusyaction}.
This will provide us with a useful guide in order to construct our more general multi-charge black hole saddles in section~\ref{sec:multicharge}. 

Following~\cite{Hegde:2023jmp}, we show that the solution \eqref{eq:cclpsolution}, after imposing supersymmetry as done in section~\ref{sec:susylimit_BH}, can be expressed in the form of a supersymmetric solution with a Gibbons-Hawking base-space \eqref{eq:susy_conf} with two centers. First, the map 
\begin{equation}
\phi = \phi_1 + \phi_2\,\,,\quad\quad \psi = \phi_2 - \phi_1\,\,,\quad\quad \theta = 2\vartheta
\end{equation}
 relates the coordinates $(\vartheta,\phi_1,\phi_2)$ appearing in \eqref{eq:cclpsolution} with those appearing in the Gibbons-Hawking ansatz, where $\theta,\phi$ belong to the three-dimensional flat base space $\mathbb R^3$ of \eqref{eq:susy_conf}. Starting from the global identifications of $\phi_1$ and $\phi_2$, we deduce that the angular coordinates $(\phi\,,\psi)$ have twisted identifications
\begin{equation}
\left( \phi\,,\,\psi\right) \,\sim\, \left( \phi + 2\pi\,,\psi+ 2\pi\right)\, \sim\, \left(\phi\,,\, \psi + 4\pi\right)\,,
\end{equation}
which fixes the period of $\psi$ as $\Delta_\psi = 4\pi$. 
For completeness, we report the full map between the Cartesian coordinate on $\mathbb R^3$, $(x,y,z)$ and the coordinates of the solution \eqref{eq:cclpsolution} after the supersymmetric limit, $(r,\theta = 2\vartheta,\phi = \phi_1 +\phi_2)$, that is
\begin{equation}
\begin{aligned}
x \,=&\,\, \frac{1}{4}\sqrt{\left(r^2-r_+^2\right)\left(r^2-\left(-r_+ + \aa + \bb\right)^2\right)}\,\sin\theta\cos\phi\,,\\
y\,=&\,\, \frac{1}{4}\sqrt{\left(r^2-r_+^2\right)\left(r^2-\left(-r_+ + \aa + \bb\right)^2\right)}\,\sin\theta\sin\phi\,,\\
z\,=&\,\, \frac{1}{8}\left(2r^2-r_+^2 -\left(-r_++\aa+\bb\right)^2 \right)\,\cos\theta\,.
\end{aligned}
\end{equation} 
We see that the `horizon' at $r=r_+$ is mapped into a segment belonging to the $z$-axis, whose extrema coincide with the north and south pole of the two-sphere parametrized by $(\theta,\phi)$. These are the loci where the Killing vectors $\xi_N$, $\xi_S$ used in the localization argument vanish, and the centers of the harmonic functions are located. As above, we denote the north pole as $\mathcal M_N :\{r=r_+\,,\vartheta=0\}$, and the south pole as $\mathcal M_S : \{r=r_+,\,\vartheta=\pi/2\}$. The distance between the two centers is found to be 
\be
\delta \,=\, \frac{1}{4}\left( \aa + \bb\right)\left(2r_+ -\aa -\bb\right)\,.
\ee 
The full set of harmonic functions that specify the solution is given by \eqref{eq:harmonicfunctiongeneral2c}, upon setting $h_0=0$. We repeat them here for convenience
\begin{equation}
\label{eq:harmonicfuncblackholesaddle}
\begin{aligned}
H &= \frac{h_N}{r_N} + \frac{h_S}{r_S}\,,\qquad \qquad h_N + h_S =1\,,\\[1mm]
\ii K &= \kk\left( \frac{1}{r_N}- \frac{1}{r_S}\right) \,,\qquad \qquad \ii M = -\frac{\kk^3}{2}\left( \frac{1}{h_N^2\,r_N}- \frac{1}{h_S^2\,r_S}\right)\,,\\[1mm]
L &= 1 + \kk^2\left( \frac{1}{h_N\,r_N} + \frac{1}{h_S\,r_S}\right)\,.
\end{aligned}
\end{equation}
Here, the parameters $h_{N,S}$ and $\kk$ are related to the supersymmetric chemical potentials \eqref{eq:cclpsusychempot} by 
\begin{equation}
\begin{aligned}
h_N & \,=\, \frac{\omega_2}{2\pi\ii} \,=\,  \frac{r_+ -\aa}{2r_+ - \aa -\bb}\,\,,
\\ 
h_S &\,=\, \frac{\omega_1}{2\pi\ii} \,=\,\frac{r_+ -\bb}{2r_+ - \aa -\bb}\,\,,\\
\mathsf k &\,=\, \frac{\varphi}{4\sqrt{3}\,\pi} \,=\,  -\frac{\left(r_+ - \aa\right)\left(r_+ -\bb\right)}{2\left(2r_+ - \aa - \bb\right)}\,.
\end{aligned}
\end{equation}
It then follows that 
%
 the two Killing vectors \eqref{eq:kvsusy} with nuts at the north or south center are specified by
\begin{equation}
\beta \Omega_N \,= \,4\pi h_N \,=\, -2\ii\omega_2\,\,,\quad\quad \beta\Omega_S \,= \, -4\pi h_S= 2\ii\omega_1\,.
\end{equation}
Remarkably, for these asymptotically flat solutions the supersymmetric linear constraint $\omega_+ = \omega_1 + \omega_2 =  2\pi \ii$
translates into the asymptotic flatness condition $h_N + h_S \equiv h_+ =1$.\footnote{A generalization of these solutions is obtained by allowing for a simple orbifold, by taking $h_N+h_S\equiv h_+ \in \mathbb{Z}$. Recall the discussion in footnote \ref{foot:ALE}. The constraint between the angular velocities then becomes $\omega_1+\omega_2 = 2\pi \ii |h_+|$. Both the action and the Bekenstein-Hawking entropy of the ALE solutions are divided by $h_+$.}

We have shown that the asymptotically flat supersymmetric black hole of section~\ref{sec:susylimit_BH} can be reinterpreted as a two-center supersymmetric geometry with a Gibbons-Hawking base. Its on-shell action, given in \eqref{eq:onshellactionfinal} and obtained via the localization theorem, exactly reproduces the result of \eqref{eq:cclpsusyaction}. This confirms that by directly solving the supersymmetry equations we have indeed constructed the Euclidean black hole saddle.

 \subsubsection{Contribution to the supersymmetric index and physical solutions}\label{contribindex}

The asymptotically flat Euclidean saddle we just constructed possesses the correct boundary conditions to be contributing to a gravitational path integral computing a supersymmetric index counting microstates. Having evaluated the action on the Euclidean solutions as above, we can now ask whether its real section represents a genuine saddle of the partition function~\eqref{eq:microscopicindex}. In this section we will use both equivalent parametrizations of this solution. The first, introduced in section~\ref{sec:susylimit_BH}, takes $(\aa,\bb,r_+)$ as independent parameters, with $m=q$ determined by the condition $\Delta_r(r_+)=0$ \eqref{eq:bh_functions}, which gives
\begin{equation}
q = r_+^2 + \aa \bb \pm r_+ \left( \aa + \bb \right)\,.
\end{equation}
Alternatively, we may employ the parametrization of section~\ref{sec:2c_ssol}, where the independent parameters are $(h_N,\kk,\delta)$. The relation between these two parametrizations was discussed in section~\ref{sec:exampleaction5d}.

 As a first thing, we observe that for the Euclidean solutions with real metric discussed above, we find
 \begin{equation}
 \aa,\,\bb,\,r_+ \,\in\,\mathbb R \quad \leftrightarrow \quad h_N,\,\kk,\,\delta \,\in\,\mathbb R \quad \leftrightarrow \quad \omega_1,\,\omega_2 \,\in\,\ii \,\mathbb R\,.
 \end{equation}
In the interpretation where the partition function  \eqref{eq:microscopicindex} is given by a sum over microstates, a generic purely imaginary value of the chemical potential $\omega_2$ may be problematic as the sum may not converge. For this reason, it is not clear to us if the Euclidean solutions with real $\aa,\bb, r_+$ (and $q$) should be regarded as saddles of \eqref{eq:microscopicindex}. Nevertheless we note that $\varphi$ and $I$ are real in this case, and one can arrange for $\varphi<0$, $I>0$ (this is obtained by taking $r_+>\aa$, $r_+>\bb$, or $\kk>0$).

We then implement suitable analytic continuations of the parameters and ask what sections of the solution is connected to a physical (i.e. well-behaved in Lorentzian signature) configuration.\footnote{When discussing saddles of a gravitational partition function and their thermodynamic interpretation, indeed, one may depart from the real section of the solution, see e.g.~\cite{Brown:1990fk,Witten:2021nzp} for discussions a priori of supersymmetry.} Consider supersymmetric solutions with parameters $a=\ii \aa$, $b = \ii \bb$, $r_+$, with $q$ determined by solving $\Delta_r(r_+) = 0$, and now conveniently expressed as
 \be\label{eq:solq}
 q = r_+^2 -ab \pm \sqrt{ -r_+^2(a+b)^2} \,.
 \ee
 We see from this formula that in general some of the parameters should be taken complex. Let us in particular discuss the cases directly related to physical solutions, namely the cases where the conserved charges \eqref{eq:cclpcharges} and the entropy~\eqref{entropyE_minimal5d} are real. Reality of the conserved quantities implies that the parameters $q$, $a$, $b$ should all be taken real. From \eqref{eq:solq} we see that there are two possibilities: either $b = -a$, or $r_+^2<0$. 
 
 \medskip
 
Taking $b = -a$ (and Wick-rotating back to real time $t=-\ii\tau$) we recover the extremal black hole, where we should also demand $q>a^2$ so that $r_+ = r_- = \sqrt{q-a^2}$ is real. From \eqref{eq:betacclp}--\eqref{elec_pot_nonsusy} it follows that $\beta = \infty$, $\Omega_1=\Omega_2 = \Phi - \sqrt 3 =0 $, while $J_+ = \frac{J_1 + J_2}{2}=0$. Taking this parametrization we obtain
\begin{equation}
b= -a,\,\,a,\,r_+ \,\in\,\mathbb R\quad \leftrightarrow \quad \delta=0\,,\,\,h_N = h_S^*\,,\, \,\kk \in \mathbb R \quad \leftrightarrow \quad \omega_1 = -\omega_2^*\,,\,\,\varphi \in \mathbb R\,,
\end{equation}
with ${\rm Re}\,h_N = {\rm Re}\, h_S = 1/2$. 
Moreover both ${\rm Re}\, \omega_2 = -2\pi {\rm Im} \,h_N$ and $\varphi$ can be taken negative (by taking $a,r_+ >0$), while $I$ is real and positive. We conclude that the extremal and supersymmetric black hole can arise as a suitable limit ($\delta \to 0$) of the index saddle with complex $h_N$, and $\kk$ real. 
  
On the other hand, considering the  case where $r_+^2<0$ we observe that the entropy \eqref{entropyE_minimal5d} would be purely imaginary, so it must vanish for the solution to be physical; this means that we must take $\left( r_+^2 +a^2 \right)\left( r_+^2 + b^2\right) +ab q =0$. This condition, together with $\Delta_r(r_+)=0$, gives\footnote{The other possible solutions are such that either $r_+$ or $q$ vanish.}
\be
q\,=\, -(2a+b)(a+2b)\,,\qquad\qquad r_+^2\,=\,-(a+b)^2\,.
\ee
  It also follows that $\beta=0$ while $\Omega_1,\Omega_2,\Phi$ diverge. The corresponding Lorentzian solution turns out to be the supersymmetric topological soliton discussed in~\cite{Chong:2005hr}, in the ungauged limit $g\to 0$. In this solution, the orbits of the angular Killing vector $\partial_{\phi_1}$ shrink to zero size as $\tilde r \equiv r^2- r_+^2\to 0$ and the geometry caps off with no horizon. In this regime we find
  \begin{equation}
  \begin{aligned}
  h_N = \frac{2a + b}{3\left( a+b\right)} \,,\qquad h_S = \frac{a + 2b}{3\left( a+b\right)}\,, \qquad \kk = -\ii \frac{\left( 2a + b\right)\left( a + 2b\right)}{6\left( a+b\right)} \,, \qquad \delta = \frac{3}{4}\left( a+ b\right)^2 >0 ,
  \end{aligned}
  \end{equation}
  and consequently $\omega_1,\omega_2,\varphi \,\in\,\ii\,\mathbb R$. It follows that the on-shell action \eqref{eq:cclpsusyaction} is also purely imaginary.
  
 On top of that, it turns out that in the present ungauged setup one cannot avoid a conical singularity at the tip of the local geometry with polar coordinates $(\tilde r,\phi_1)$ and still have a non-trivial asymptotically flat solution. Allowing for a conical singularity introduces the quantization condition
  \be
 h_S = \frac{a+2b}{3(a+b)} = p \,\in\, \mathbb{Z} \,,\qquad \text{with $|p|>1$\,.}
  \ee
Then one can check that the four-dimensional hypersurface at fixed time has an $\mathbb{R}^4/\mathbb{Z}_{|p|}$ orbifold singularity of the ALE type located at $(r^2 = r_+^2$, $\theta = 0)$ as well as an $\mathbb{R}^4/\mathbb{Z}_{|1-p|}$ singularity at $(r^2 = r_+^2$, $\theta = \pi)$. The singularity analysis is similar to the one for the solutions of~\cite{Giusto:2004kj}, of which the present solution is a five-dimensional avatar.
Relying on our formulae despite the singularity, we find that the chemical potentials $\omega_1, \omega_2$, are finite and purely imaginary and quantized, their expressions being
 \be
 \omega_1 = 2\pi \ii\,p \,\,, \qquad  \omega_2 = 2\pi \ii \left(1-p\right) \,\,.
 \ee
The microscopic sum defined by \eqref{eq:microscopicindex} is, then, blind to $J_1-J_2=2J_-$, since this takes integer eigenvalues both for fermionic and bosonic states and $\omega_2$ is a multiple of $2\pi \ii$. We also observe that in the limit $\beta \to 0$ the global identifications \eqref{eq:coord_identif} reduce to
\begin{equation}
\left( \tau,\,\phi,\,\psi \right) \sim \left( \tau,\,\phi + 2\pi ,\,\psi + \ii \omega_- \right) \sim \left( \tau,\,\phi + 2\pi ,\,\psi - 2\pi \right) \sim \left( \tau,\,\phi,\,\psi + 4\pi \right)\,,
\end{equation}
involving only $\phi$ and $\psi$, while $\tau$ acts a mere spectator. The mutual compatibility between these identifications is guaranteed by the above quantization condition, that implies $\omega_- = \omega_1 - \omega_2 = 2\pi \ii\left( 1 + 2p\right)$. It follows that the physical horizonless soliton also arises as a suitable limit ($\beta \to 0$) of the index saddle with certain parameters complexified, namely $\kk \in \ii \, \mathbb R$, while $h_N,h_S \in \mathbb Z$. We will have more to say about this horizonless solution and its generalizations in section~\ref{sec:physical_prop}.


\section{Multi-charge supersymmetric saddles}\label{sec:multicharge}

This section is devoted to the construction and analysis of new supersymmetric non-extremal Euclidean black hole solutions carrying multiple electric charges $Q_I$, which provide saddles of the gravitational path integral that computes a supersymmetric index refined by the conjugate electric potentials $\varphi^I$.
The solutions will be obtained by solving the supersymmetry equations directly, rather than taking a supersymmetric limit of the corresponding non-supersymmetric and non-extremal black hole, as we did in section~\ref{sec:susylimit_BH} for the minimal  setup. The main reason is that non-extremal multi-charge solutions are only known for specific supergravity models~\cite{Cvetic:1996xz} or in absence of rotation \cite{Meessen:2011bd, Meessen:2012su}, while we intend to perform a general analysis.  

The section is organized as follows. We start recalling the main features of ${\cal N}=2$ supergravity coupled to vector multiplets in section~\ref{sec:matter_coupled_sugra}. Then in section \ref{sec:susy_sol_matter_coupled_sugra} we review the general classification of timelike supersymmetric solutions with an additional U$(1)$, closely following \cite{Gutowski:2004yv, Gutowski:2004bj, Gauntlett:2004qy, Bellorin:2006yr}. In section~\ref{sec:gen_saddles} we construct our novel supersymmetric non-extremal solutions. Finally, in section~\ref{sec:physical_prop} we study their thermodynamic properties. In particular, we evaluate their on-shell action, further extending the equivariant technology of the previous sections to the matter-coupled case.

\subsection{5D supergravity coupled to vector multiplets}
\label{sec:matter_coupled_sugra}

The bosonic field content of five-dimensional $\mathcal N=2$ supergravity coupled to $n$ vector multiplets consists of the metric $g_{\mu\nu}$, $n+1$ vector fields $A^I$ and $n+1$ scalars $X^I$. The latter obey the following cubic constraint,
\begin{equation}
C_{IJK}X^I X^J X^K=1\,, 
\end{equation}
where $C_{IJK}$ is a fully-symmetric and constant tensor which determines the couplings of the bosonic action. The latter is given by\footnote{This is the ungauged ($g\to 0$) version of the two-derivative matter-coupled action \eqref{eq:2dSUGRA}.}
\begin{equation}\label{eq:action_sugra_matter-coupled}
S=\frac{1}{16\pi}\int \left(R\star_5 1-\frac{3}{2}\, a_{IJ}\,\diff X^I \wedge \star_5 \diff X^J-\frac{3}{2}\, a_{IJ} F^I\wedge \star_5 F^{J}
+C_{IJK}A^I\wedge F^{J}\wedge  F^K\right)\, ,
\end{equation}
where
\begin{equation}
\label{eq:defa_{IJ}&X_I}
a_{IJ}=3{X}_I {X}_{J}-2\,{C}_{IJK}X^K\,,  \hspace{1cm} X_I=C_{IJK}X^J X^K\,.
\end{equation}

In what follows we shall restrict our attention to supergravity models for which the scalar manifold is a symmetric space \cite{Gunaydin:1983bi}. This implies the existence of a fully-symmetric and constant tensor $C^{IJK}$ satisfying
\begin{equation}
C^{IJK}C_{J(LM}C_{NP)K}=\frac{1}{27} \, \delta^{I}_{(L}C_{MNP)}\, .
\end{equation}
Contracting the above equation with $X^L X^M X^N X^P$, one can invert the second of \eqref{eq:defa_{IJ}&X_I}:
\begin{equation}
X^I=27 C^{IJK}X_J X_K\, .
\end{equation}
Further contracting with $X_I$ yields
\begin{equation}
C^{IJK}X_I X_J X_K=\frac{1}{27}\, .
\end{equation}


\subsection{Timelike supersymmetric solutions with a U$(1)$ symmetry}
\label{sec:susy_sol_matter_coupled_sugra}

Let us review the general form of supersymmetric solutions in five-dimensional $\mathcal N=2$ ungauged supergravity coupled to vector multiplets. We shall assume that the supersymmetric Killing vector is timelike. In such case, using coordinates adapted to the isometry, the metric and vector fields are given by
\begin{eqnarray}
\label{eq:susy_metric}
\diff s^2&\,=\,&-f^2\left(\diff t+\omega\right)^2 + f^{-1}\diff {\hat s}^2\,, \\[1mm]
\label{eq:susy_vector}
A^{I}&\,=\,& -X^{I} f \left(\diff t+\omega\right)+ {\hat A}^{I}+{\alpha_\tau^I}\diff t\,,
\end{eqnarray}
where $\alpha_\tau^I$ is a constant parametrizing a possible gauge choice. The metric function $f$, the scalars $X^I$, and the one-forms $\omega$ and ${\hat A}^I$ are defined on a four-dimensional hyper-Kahler (base) space whose line element is denoted by $\diff{\hat s}^2$. Supersymmetry imposes the following equations
\begin{eqnarray}
\label{eq:SUSY1}
{\hat F}^I&\,=\,&\hat{\star}\,{\hat F}^I\,, \\ [1mm]
\label{eq:SUSY2}
\diff \,{\hat\star}\, \diff \left(f^{-1}X_I\right)&\,=\,& C_{IJK} {\hat F}^{J}\wedge {\hat F}^{K}\,, \\[1mm]
\label{eq:SUSY3}
f\left(\diff \omega + {\hat \star}\,\diff \omega \right)&\,=\,& 3X_I {\hat F}^I\,, 
\end{eqnarray}
where $\hat \star$ denotes the Hodge star operator with respect to the hyper-Kahler metric.

Following the steps of section~\ref{sec:gen_ssol}, let us assume the hyper-Kahler metric has an additional U$(1)$ isometry respecting the triholomorphic structure. Then it must take the Gibbons-Hawking form \eqref{eq:susysolminimalghbase1}, that we repeat below for convenience,
\begin{equation}
\label{eq:basespace}
\diff {\hat s}^2=H^{-1}\left(\diff \psi +\chi\right)^2 + H \diff s^2_{\mathbb R^3} \,.
\end{equation}
We recall that the function $H$ and the local one-form $\chi$ are related by  $\diff H\,=\, \star_{3} \diff \chi\,.$
Under the additional isometry, the local one-forms $\omega$, ${\hat A}^{I}$ naturally split as
\begin{eqnarray}
\omega&\,=\,&\omega_{\psi}\left(\diff \psi +\chi\right)+{\breve \omega}\,,\\[1mm]
{\hat A}^{I}&\,=\,&\frac{K^I}{H}\left(\diff \psi +\chi\right)+ {\breve A}^I\,, 
\end{eqnarray}
where again all the quantities appearing ($\omega_\psi, K^I, {\breve \omega},{\breve A}^I$) are defined on ${\mathbb R}^3$. Introducing the functions $L_I$ and $M$ as follows 
\begin{eqnarray}
f^{-1}X_I&\,=\,&L_I + C_{IJK}K^JK^K H^{-1}\,, \\[1mm]
\omega_{\psi}&\,=\,& M + \frac{3}{2} L_I K^I H^{-1}+C_{IJK} K^I K^J K^K H^{-2}\,,
\end{eqnarray}
it can then be shown that the supersymmetry equations \eqref{eq:SUSY1}--\eqref{eq:SUSY3}  boil down to the following equations
\begin{eqnarray}
\diff \star_3 \diff L_{I}&\,=\,&0\,, \hspace{1cm} \diff \star_3 \diff M\,=\,0\,\,,\\[1mm]
\label{eq:monopole}
\star_3 \diff K^{I}&\,=\,&-{\breve F}^{I}\,,\\[1mm]
\label{eq:omega3d}
\star_3\diff {\breve \omega}&\,=\,&H\diff M-M\diff H +\frac{3}{2}\left(K^I\diff L_I-L_I \diff K^I\right)\,\, .
\end{eqnarray}
Thus, in summary, timelike supersymmetric solutions with a U$(1)$$_{\psi}$ isometry are  determined by $2(n+2)$ harmonic functions on ${\mathbb R}^3$:
\begin{equation}
\left\{H, M, K^I, L_I\right\}\, .
\end{equation}


\subsection{Saddles for supersymmetric multi-charge black holes}
\label{sec:gen_saddles}

The aim of this section is to construct  supersymmetric solutions providing saddles of the Euclidean gravitational path integral associated to asymptotically-flat BPS (supersymmetric and extremal) black holes with electric charges $Q_I$ and angular momentum $J_{-}=\tfrac{J_1-J_2}{2}$.\footnote{The linear combination $J_{+}=\tfrac{J_1+J_2}{2}$ will vanish as a consequence of extremality.} We shall then look for Euclidean solutions ($\tau=\ii t$) with a U$(1)$$_{\psi} \times$ U$(1)$$_{\phi}$ isometry that satisfy the following asymptotic boundary conditions for $\tilde r\rightarrow+\infty$,
\begin{eqnarray}
\diff s^2&\,\rightarrow\,&\diff \tau^2+ \diff\tilde r^2 +\frac{\tilde r^2}{4} \left[\diff\theta^2+\sin^2\theta \,\diff \phi^2 + \left(\diff \psi+\cos\theta \diff\phi\right)^2\right]\,,\\[1mm]
\label{eq:bdry_cond_A^I}
A^{I}&\,\rightarrow\,& \ii \Phi^I \diff \tau\,,\\[1mm]
X^{I}&\,\rightarrow\,&\Phi^I_{*}\,,
\end{eqnarray}
along with the  twisted identifications
\begin{equation}
\label{eq:twisted_bdry_conditions}
\begin{aligned}
\left(\tau, \psi, \phi\right)\ \sim&\ \left(\tau+\beta, \psi+\beta \ii \Omega_{-}, \phi-\beta \ii \Omega_{+}\right)\,\\[1mm] 
\ \sim&\ \left(\tau, \psi+2\pi, \phi+2\pi\right)
\sim\ \left(\tau, \psi-2\pi, \phi+2\pi\right)\,\, .
\end{aligned}
\end{equation}
The last two, together with the fact that $\theta \in [0, \pi]$, imply that the metric asymptotes to $S^{1}_{\beta} \times {\mathbb R}^{4}$. The asymptotic behaviour of matter-coupled supersymmetric saddles we are constructing, then, falls into a subset of the ones considered around \ref{sec:global_id}.

All in all, the solutions will be specified by 
\begin{equation}
\beta, \Omega_{-}, \Phi^I\hspace{1cm} \text{or, equivalently,} \hspace{1cm} Q_I, J_{+}, J_{-}\, ,
\end{equation}
and the (constrained) moduli  $\Phi^I_{*}$.  The mass $E$ of these solutions, being supersymmetric, is fixed by 
\begin{equation}
E=\Phi^I_{*} Q_I\, , 
\end{equation}
which has been shown to be equivalent to \cite{Cabo-Bizet:2018ehj, Iliesiu:2021are}
\begin{equation}
\beta \Omega_{+}=2\pi {\ii}\, .
\end{equation}

In order to construct a solution with such properties, we consider the following multi-center harmonic functions,
\begin{equation}
H=h_0 +\sum_{a=N,S} \frac{h_a}{r_a}\,, \hspace{3mm}\ii M=\mm_0 +\sum_{a=N,S} \frac{\mm_a}{r_a}\,, \hspace{3mm} \ii K^I=\kk^I_0 +\sum_{a=N,S} \frac{\kk^I_a}{r_a}\,, \hspace{3mm}L_I={\ell}_{I, 0} +\sum_{a=N,S} \frac{{\ell}_{I, a}}{r_a}\,,   
\end{equation}
where the notation is the same as in section~\ref{sec:gen_ssol}.

Having specified the choice of harmonic functions, we can solve the equations for the one-forms $\chi$, ${\breve A}^{I}$ and ${\breve \omega}$ using two sets of spherical coordinates in $\mathbb R^3$ centered at the poles of the harmonic functions, as introduced in \eqref{eq:coordinatesystem2}. Locally, these are given by
\begin{eqnarray}
\label{eq:chi}
\chi&\,=\,&\left(h_N\cos \theta_N+h_S \cos \theta_S\right) \diff \phi\,, \\[1mm]
\label{eq:breve_AI}
\ii {\breve A}^{I}&\,=\,&-\left(\kk^I_N\cos \theta_N+\kk^I_S \cos \theta_S\right) \diff \phi\,, \\[1mm]
\label{eq:breve_omega}
\breve \omega&\,=\,&\left(w_N\cos \theta_N+w_{S}\cos \theta_S\right)\diff \phi +w_{\rm {reg}}\,, 
\end{eqnarray}
where 
\begin{equation}
\label{eq:omegaNS}
\begin{aligned}
\ii w_N\,=\,&h_0 \mm_N-\mm_0 h_N+\frac{3}{2}\left(\kk^I_0 \ell_{I, N}-\ell_{I, 0} \kk^I_{ N}\right)-\frac{2\left(h_N\mm_S-h_S\mm_N\right)+3\left(\kk^I_N\ell_{I, S}-\kk^I_S\ell_{I, N}\right)}{2\delta}\, ,\\[1mm]
\ii w_S\,=\,&h_0 \mm_S-\mm_0 h_S+\frac{3}{2}\left(\kk^I_0 \ell_{I, S}-\ell_{I, 0} \kk^I_{S}\right)+\frac{2\left(h_N\mm_S-h_S\mm_N\right)+3\left(\kk^I_N\ell_{I, S}-\kk^I_S\ell_{I, N}\right)}{2\delta}\, ,
\end{aligned}
\end{equation}
and 
\begin{equation}
\ii w_{\rm {reg}}=\frac{2\left(h_N\mm_S-h_S\mm_N\right)+3\left(\kk^I_N\ell_{I, S}-\kk^I_S\ell_{I, N}\right)}{2\delta}\left(\cos\theta_N+1\right)\left(1-\frac{r_N+\delta}{r_S}\right)\diff \phi\, .
\end{equation}
The next step is to fix the parameters of the solution in terms of $\beta, \Phi^I, \Omega_{-}$ and the moduli by demanding asymptotic flatness and regularity. Before that, as in section~\ref{sec:gen_ssol}, we make use of the freedom to shift $K^I\to K^{I}+ \lambda^I H$ to set
\begin{equation}
\label{eq:cond_k^I}
\kk^I_N=-\kk^{I}_S\, .
\end{equation}

\paragraph{Asymptotic flatness.} We start demanding that the metric of the base space \eqref{eq:basespace} asymptotes to ${\mathbb R}^{4}$ for large $r$, where $r$ is the radial direction of a set of coordinates centered at the origin, as given by \eqref{eq:coordinatesystem1}.\footnote{Here, $r$ is related to $\tilde r$ appearing in \eqref{eq:bdry_cond_A^I} by $\tilde r^2=4r$.} This amounts to imposing 
\begin{equation}
h_0=0\, , \hspace{1cm} h_N+h_S=1\,.
\end{equation}
Next, we impose that
\begin{equation}
\lim_{r\to \infty} f=1\,, \hspace{1cm} \lim_{r\to \infty}\omega_{\psi}=0\,,
\end{equation}
which yields the conditions 
\begin{equation}
\label{eq:asymptotic_flatness}
\kk^I_{0}=0\,, \hspace{1cm} C^{IJK}{\ell}_{I, 0}{\ell}_{J, 0}{\ell}_{K, 0}=\frac{1}{27}\,, \hspace{1cm} \mm_0=-\frac{3}{2}{\ell}_{I, 0}\left(\kk^I_N+\kk^I_S\right)=0\, ,
\end{equation}
where in the last equation we used~\eqref{eq:cond_k^I}.
\paragraph{Non-extremality and regularity conditions.} In analogy with section~\ref{sec:susyaction}, we demand a non-extremal behaviour at the centers for the metric functions $f$ and $\omega_{\psi}$. Concretely, we impose the conditions
\begin{equation}
\lim_{r_a\to 0}f^{-1}X_I= {\cal O}\left(r_a^0\right)\, , \hspace{1cm} \lim_{r_a\to 0}\omega_{\psi}= {\cal O}\left(r_a^0\right)\, ,
\end{equation}
which fix the coefficients ${\ell}_{I, a}$, $\mm_a$ in terms of $\kk^I_{a}$ and $h_a$ as
\begin{equation}
\ell_{I, a}=C_{IJK}\frac{\kk^J_a \kk^K_a}{h_a}\,, \hspace{1cm} \mm_a=-\frac{1}{2}C_{IJK}\frac{\kk^I_a \kk^J_a \kk^K_a}{h^2_a}\, .
\end{equation}
Therefore, the solution is fully determined in terms of $k^{I}_{N}, h_N$, the distance between the centers $\delta$ and ${\ell}_{I, 0}$. The latter are in correspondence with the moduli of the solution
\begin{equation}
\lim_{r\to \infty} X^{I}\,=\,\Phi^{I}_{*}\,=\, 27 \,C^{IJK}\ell_{J, 0} \ell_{K, 0}\, .
\end{equation}
Consequently, the remaining parameters must be in correspondence with the temperature and chemical potentials:
\begin{equation}
\left(\kk^{I}_{N}, h_N, \delta\right)\hspace{5mm}\leftrightarrow \hspace{5mm}\left(\beta, \Phi^I, \Omega_{-}\right)\, .
\end{equation}
The simplest way of obtaining the precise relation is by demanding regularity of the metric and gauge fields near the centers. The analysis is essentially the same as the one made in section~\ref{sec:susyaction}. The metric near the north pole has been given in \eqref{eq:metricaroundnut}. A similar expression holds around the south pole. Given these expressions, it is clear that regularity of the metric around the north and south poles imposes the conditions
\begin{equation}
\label{eq:OmegaNS2}
\Omega_{a} \,\equiv\, -\frac{1}{\ii\omega_\psi}\Big|_{\mathcal M_a} \,=\, \frac{h_a}{\ii w_a}\, , \quad a = N,S\,,
\end{equation}
which are compatible with \eqref{eq:smooth_center_condition}. Now, the following change of coordinates
\begin{equation}
\psi_N=\psi+\phi -\Omega_N \tau\,, \hspace{1cm} \phi_N=\phi-\frac{\Omega_N}{2h_N}\tau\,, \hspace{1cm} {\tilde \tau}=\frac{\Omega_N}{2h_N}\tau\,, 
\end{equation}
brings the metric around the north pole \eqref{eq:metricaroundnut} to the form
\begin{equation}
\diff s^2   \, \underset{\tilde r_N\to 0}{\longrightarrow} \  \frac{f_N^2}{\Omega_N^2} \diff \psi_N^2 +\frac{h_N}{f_N}\left[d{\tilde r}_N^2+{\tilde r}_N^2\left(\frac{1}{4}\diff\theta_N^2+\cos^2\frac{\theta_N}{2}\diff {\tilde\tau}^2+\sin^2\frac{\theta_N}{2}\diff {\phi}^2_N\right)\right]\, .
\end{equation}
Thus, as discussed in section~\ref{sec:susyaction}, this is equivalent to the product ${S}^1\times {\mathbb R^4}$, provided $\tilde\tau$ and $\phi_N$ are periodically identified, with period equal to $2\pi$ each. Consistency with the global identifications \eqref{eq:twisted_bdry_conditions} demands $\psi_N\sim \psi_N+4\pi$ together with
\begin{equation}
\Omega_N=-\ii \left(\Omega_{+}-\Omega_{-}\right)\,, \hspace{1cm} \Omega_S=\ii \left(\Omega_{+}+\Omega_{-}\right)\, ,\hspace{1cm} \beta=4\pi \ii w_{N}\, ,
\end{equation}
which are equivalent to
\begin{equation}
\label{eq:DiracMisner}
\beta \Omega_{+}=2\pi \ii\,, \hspace{1cm} \beta \Omega_{-}=2\pi \ii\left(h_S-h_N\right)\,, \hspace{1cm} \beta=4\pi \ii w_{N}\, .
\end{equation}
An alternative way of arriving to the same result would be to study the conditions for the cancellation of Dirac-Misner singularities. This is discussed in appendix~\ref{sec:DiracMisner}.

Finally, we obtain the relation between the parameters $\kk^I $ and the chemical potentials $\Phi^I$. To this aim, we first demand regularity of the gauge fields at the north and south poles. Namely, we impose $\iota_{\xi_N}A^I\big|_{{\cal M}_N}=\iota_{\xi_S}A^I\big|_{{\cal M}_S}=0$. Both yield the same condition, which is the following
\begin{equation}
\label{eq:reg_AI}
\beta \alpha_\tau^{I}=-4\pi \kk_N^I\, .
\end{equation}
Now we make use of the boundary conditions \eqref{eq:bdry_cond_A^I} to obtain $-\alpha_\tau^I+\Phi^I_{*}=\Phi^I$. This allows us to rewrite the above regularity condition \eqref{eq:reg_AI} as follows
\begin{equation}\label{eq:PhiI~kI}
\beta(\Phi^I-\Phi^I_{*})\,=\,4\pi \kk_N^I\, .
\end{equation}


\subsection{Physical properties of the solutions}\label{sec:physical_prop}

In this section we first make use of the equivariant approach developed in section \ref{sec:equiv5d} to compute the on-shell action of the Euclidean solutions derived in section \ref{sec:gen_saddles}. 
We then perform its Legendre transform to obtain the entropy, allowing at the same time for more general values of the parameters through analytic continuations. Even assuming the charges are real, the entropy we obtain through this procedure is in general complex. Demanding that the imaginary part vanishes yields a constraint on the charges of the solution. We find that there are two possibilities, depending on the assumptions  on the charges. The first possibility, which gives a finite entropy, corresponds to the extremal limit $\beta\to \infty$ of our configurations. We explicitly show that this extremal limit gives known BPS black hole solutions \cite{Breckenridge:1996is, Cvetic:1996xz, Chamseddine:1998yv, Gutowski:2004bj, Ortin:2015hya} after Wick-rotating to Lorentzian time, analogously to other similar cases discussed in previous work.  The second possibility is a new observation: we find that it corresponds to the limit $\beta \to 0$ and yields a supersymmetric topological soliton with vanishing entropy. In this case, we show that the corresponding Lorentzian solution is a two-center horizonless microstate geometry~\cite{Giusto:2004id,Bena:2005va, Berglund:2005vb}. Therefore, both the BPS black hole and the horizonless geometry arise as limits of our supersymmetric, finite $\beta$ solution.\footnote{Although we illustrate these results  in the present multi-charge setup, we remark that they also apply to the pure supergravity case of section~\ref{sec:susyaction}, consistently with the discussion in section~\ref{contribindex}.}


\subsubsection{Euclidean on-shell action via equivariant integration}
\label{sec:onshellaction_multicharge}

After using the trace of Einstein equations, we can recast the bulk contribution to the on-shell action as 
\begin{equation}
I_{\rm{bulk}}=\frac{1}{16\pi}\int F^I\wedge G_I\,,
\end{equation}
where 
\begin{equation}
G_{I}=a_{IJ}\star_5 F^J-\ii \, C_{IJK}F^J\wedge A^K\, .
\end{equation}
We emphasize that $\star_5$ is the Hodge star operator with respect to the Euclidean metric, hence the appearance of the imaginary unit $\ii$. The matter-coupled generalization of the equivariantly-closed polyform in \eqref{eq:polyform5d2} is
\begin{equation}
\Psi=F^I\wedge G_I + \left[-\left(\iota_{\xi}A^I+\crho^I\right)G_I+\nu_{I}\wedge F^I \right]+ \left[-\left(\iota_{\xi}A^I+\crho^I\right) \nu_I\right]\,, 
\end{equation}
where $\crho^I$ are for the time being arbitrary constants and $\nu_I$ is defined as follows
\begin{equation}
\diff {\nu}_{I}=\iota_{\xi}G_I\,, \hspace{1.5cm} \iota_{\xi}\nu_I=0\,.
\end{equation} 

As in section~\ref{sec:susyaction}, we restrict to supersymmetric solutions with an extra U$(1)$$_{\psi}$ isometry, and consider two linear combinations of the supersymmetric Killing vector and the generator of the extra U$(1)$ isometry, namely
\begin{equation}
\xi_{a}=\partial_{\tau}+ \Omega_{a} \, \partial_{\psi}\, ,
\end{equation}
where $\Omega_a$ is given by \eqref{eq:OmegaNS2}.
We choose to localize with respect to $\xi_N$. The determination of the equivariant parameters follows exactly the same steps as in section~\ref{sec:susyaction}. Therefore, we just provide the final result
 \begin{equation}
\label{eq:equivariant_parameters}
\epsilon^{N}_1=\frac{2\pi}{\beta}=-\epsilon^{N}_2\,.
\end{equation}

The one-form $\nu_I$ associated to $\xi_{N}$ is given by 
\begin{equation}
\begin{aligned}
\nu_I\,&=\, \ii C_{IJK}\left[\left(f X^J+\alpha_\tau^J\right)\ii {\hat A}^{K}-fX^J \alpha_\tau^{K} \ii \omega\right] - \ii \Omega_{N} \Big[\varepsilon_{I}+C_{IJK}fX^JH^{-1}\ii K^K \left(\diff\tau+\ii {\breve \omega}\right) \\[1mm]
&\quad\ \left.+\,C_{IJK}\left(H^{-1}\ii K^J-f X^J\ii\omega_{\psi}\right)\left(\ii {\breve A}^{K}+\alpha_\tau^K \diff \tau\right)\right]+ \ii\mu_{I}\, ,
\end{aligned}
\end{equation}
where the one-form $\varepsilon_I$ is related to the function $L_I$ through $\diff \varepsilon_I=\star_3 \diff L_I$, and $\mu_I$ is a closed one-form to be determined demanding regularity of $\nu_I$. Let us emphasize that except for this regularity analysis, which must be done case by case, the above expression for $\nu_I$ applies to any timelike supersymmetric solution with an extra U$(1)$${}_{\psi}$. As we saw in the previous sections, there is a scheme, which here corresponds to choosing
\begin{equation}\label{eq:choice_cI}
\ii \crho^{I}\,=\, \Phi^I-\Phi^I_{*}\, ,
\end{equation}
 such that the boundary terms cancel and the whole on-shell action comes from the integral of $\nu_I$ at the nut.

In what follows we specialize to the solutions studied in section \ref{sec:gen_saddles}. Demanding regularity, we obtain that 
\begin{equation}
\mu_I=C_{IJK}\,\alpha_\tau^{J}\,\frac{\kk^K_N}{h_S}\left(\diff \psi-\Omega_N \diff \tau\right)\, ,
\end{equation}
which already allows us to evaluate the nut contribution to the on-shell action. We will also assume the choice \eqref{eq:choice_cI}. Therefore the on-shell action is given by
\begin{equation}
\label{eq:onshellaction_multicharge}
I\,=\,\frac{1}{16\pi}\int_{{\cal M}_{N}} \frac{\iota^* \Psi_{(1)}}{\frac{\epsilon^N_1}{2\pi} \frac{\epsilon^N_2}{2\pi}}\,=\,-4\pi^2\frac{C_{IJK}\kk^I_N\kk^J_N\kk^K_N}{h_N h_S}\,=\,\pi \,\frac{C_{IJK}\varphi^I \varphi^J\varphi^K}{\omega_{+}^2-\omega_{-}^2}\, ,
\end{equation}
where we have introduced the quantities $\varphi^I, \omega_{+}, \omega_{-}$, defined as  
\begin{equation}
\varphi^I=\beta\left(\Phi^I-\Phi^I_{*}\right)\,, \hspace{1cm} \omega_{+}=\beta \Omega_{+}\,, \hspace{1cm} \omega_{-}=\beta \Omega_{-}\, ,
\end{equation}
which in terms of the parameters of the solutions read
\begin{equation}
\label{eq:chemical_pot_multicharge}
\varphi^I=4\pi \kk^I_N\,, \hspace{1cm} \omega_{+}=2\pi \ii \hspace{1cm} \omega_{-}=2\pi \ii \left(h_S-h_N\right)\, .
\end{equation}


\subsubsection{Entropy and charges from the on-shell action} 

The extremization principle introduced in sec.~\ref{sec:holographic_match} for AdS black holes can, with straightforward modifications, also be used to derive the entropy from the grand-canonical action in the case of asymptotically flat black holes. In particular, the entropy ${\cal S}$ of the saddles is obtained from:
\begin{eqnarray}
\label{eq:S}
{\cal S}&=&{\rm{ext}}_{\{\varphi^I, \omega_{\pm}, \Lambda\}}\left[-I-\varphi^I Q_I -\omega_{-} J_{-}-\omega_{+} J_{+}-\Lambda\left(\omega_{+}-2\pi \ii\right)\right]\, ,\\[1mm]
Q_I&=& -\frac{\partial I}{\partial \varphi^I}\,, \hspace{1cm} J_{-}=-\frac{\partial I}{\partial \omega_{-}}\,, \hspace{1cm} J_{+}=-\frac{\partial I}{\partial \omega_{+}}-\Lambda\,, \hspace{1cm} \omega_{+}=2\pi \ii\, ,\quad
\end{eqnarray}
which gives rise to the following expression 
\be\label{eq:entropy_multicharge}
{\cal S}\,=\,4\sqrt{\pi}\sqrt{C^{IJK}Q_I Q_J Q_K-\frac{\pi}{4}J_{-}^2}-2\pi \ii J_{+} \,.
\ee
The expressions for the charges $Q_I$ and the angular momentum $J_{-}$ as a function of the parameters of the solution can be obtained using the above extremization equations, together with \eqref{eq:onshellaction_multicharge} and \eqref{eq:chemical_pot_multicharge}:
\begin{equation}
\label{eq:charges}
Q_{I}=3\pi\frac{C_{IJK}\kk^J_N \kk^K_N}{h_N h_S}\,, \hspace{1cm}  \ii  J_{-}=\pi \frac{h_S-h_N}{h_N^2 h_S^2}C_{IJK}\kk^I_{N}\kk^J_N \kk^K_N\, .
\end{equation}
In turn, a direct calculation shows that $J_{+}$ is  given by
\begin{equation}
\label{eq:Jphi}
\ii J_{+}\,=\, 3\pi \ell_{I, 0}\kk^I_N \delta\, .
\end{equation}

Now we would like to interpret the entropy \eqref{eq:entropy_multicharge} and the charges \eqref{eq:charges} and \eqref{eq:Jphi} as those corresponding to a physical (Lorentzian) solution. To this aim, we assume first that there exists an analytic continuation of the parameters such that the charges and angular momenta become real.  Even after assuming this, the entropy \eqref{eq:entropy_multicharge} remains complex unless a constraint is imposed on the charges.
The constraint that one must impose clearly depends on the sign of the combination of the charges appearing in the argument of the square root, 
\be\label{eq:argument_squareroot}
C^{IJK}Q_I Q_J Q_K-\frac{\pi}{4}J_{-}^2\, .
\ee
When this is positive, the entropy is real only when $J_{+}=0$. It is well known that this is the constraint satisfied by a supersymmetric and extremal black hole with these charges \cite{Breckenridge:1996is, Cvetic:1996xz, Chamseddine:1998yv, Gutowski:2004bj, Ortin:2015hya}. Therefore, $J_{+}=0$ must follow from extremality $\beta\to \infty$. It is instructive to explicitly verify this by expressing $\beta$ as a function of the charges. Using the above expressions for the charges and angular momenta in terms of the parameters, alongside with \eqref{eq:DiracMisner} and \eqref{eq:omegaNS}, one finds that 
\begin{equation}
\label{eq:beta(Q)}
\beta=-18\,\frac{C^{IJK}{\bar X}_I Q_J Q_K}{{\cal S}+2\pi \ii J_{+}}\frac{\cal S}{\ii J_{+}}\, ,
\end{equation}
where ${\bar X}_{I}\equiv C_{IJK}\Phi^{J}_{*}\Phi^K_{*}$. Thus, we conclude that an extremal solution with finite entropy must necessarily have vanishing $J_{+}$. The Lorentzian black hole solution with these properties is studied in section \ref{label:extBHs}, where we show that its entropy agrees with \eqref{eq:entropy_multicharge}.

The second possibility is that \eqref{eq:argument_squareroot} is non-positive.\footnote{The limit in which \eqref{eq:argument_squareroot} vanishes can also be reached from the extremal black hole, but it gives rise to a singular solution, as it corresponds to a black hole with vanishing horizon area. This is the reason why we shall not discuss this possibility.} In such case the expression \eqref{eq:entropy_multicharge} is purely imaginary, implying that it should vanish. This is equivalent to the following relation among the charges,
\begin{equation}
\label{eq:constraint_horizonless_sol}
C^{IJK}Q_I Q_J Q_K-\frac{\pi}{4}\left(J_{-}^2- J^2_{+}\right)=0\,,
\end{equation}
which is then expected to be realized in a horizonless solution. From \eqref{eq:beta(Q)} it is further deduced that such solution must arise in the $\beta \to 0$ limit, provided the charges are not vanishing. Interestingly, we find that the Lorentzian solution that realizes these properties is the two-center horizonless geometry of \cite{Bena:2005va, Berglund:2005vb}. This is further discussed in section \ref{sec:fuzz}. 


\subsubsection{$\beta\to \infty$ limit: extremal black holes}
 \label{label:extBHs}
From the last of \eqref{eq:DiracMisner} and the expression for $w_{N}$ provided in \eqref{eq:omegaNS}, we see that the $\beta\to\infty$ limit corresponds to the limit in which the distance between the centers goes to zero. Taking the $\delta\to 0$ limit in the harmonic functions yields
\begin{equation}
H \to \frac{1}{r}\,, \hspace{1cm} K^{I}\to 0\,, \hspace{1cm} L_I \to \ell_{I, 0}+\frac{\ell_I}{r}\,, \hspace{1cm} M \to \frac{m}{r}\,,
\end{equation}
where
\begin{equation}
\ell_I=C_{IJK}\frac{\kk^J_N \kk^{K}_N}{h_Nh_S}\,, \hspace{1cm} \ii m=\frac{1}{2}C_{IJK}\kk^{I}_N \kk^J_N \kk^{K}_N \frac{ \left(h_N-h_S\right)}{h_N^2h_S^2}\, .
\end{equation}
After both a Wick rotation and an analytic continuation of $h_N-h_S$,
\begin{equation}
\label{eq:Wick}
\tau\to \ii t\,, \hspace{1cm} h_N\to\frac{1}{2}\left(1+\ii \Delta_h\right)\,, \hspace{1cm} h_S\to\frac{1}{2}\left(1-\ii \Delta_h\right)\,,
\end{equation}
one obtains a supersymmetric and extremal black hole with charges $Q_I$ and angular momentum $J_{-}$ given by
\begin{equation}
\label{eq:charges_extremal}
Q_I=3\pi \ell_I=6\pi \,\frac{C_{IJK}\kk^J_N \kk^K_N}{1+\Delta^2_h}\,,  \hspace{1cm} J_{-}=-2\pi m=-4\pi \frac{\Delta_h\, C_{IJK}\kk^I_N \kk^J_N \kk^K_N}{\left(1+\Delta^2_h\right)^2} \, ,
\end{equation}
while $J_{+}$ vanishes. The Bekenstein-Hawking entropy, computed as one quarter of the horizon area, is
\begin{equation}
\label{eq:Sextremal}
{\cal S}=4\sqrt{\pi}\sqrt{C^{IJK}Q_I Q_J Q_K-\frac{\pi}{4}J_{-}^2}\, ,
\end{equation}
in precise agreement with the Legendre transform of the saddle on-shell action \eqref{eq:S}, supplemented with the reality condition $J_{+}=0$.\footnote{In the case of two vector multiplets ($n=2$), and for a specific choice of $C_{IJK}$ corresponding, e.g., to an embedding into a toroidally compactified string theory, the saddle we constructed reduces to the solution analyzed in~\cite{Anupam:2023yns}. In this setup, the extremal entropy is that of the BMPV black hole, reproduced by the logarithm of the helicity supertrace index in the large-charge limit, as reviewed around~\eqref{eq:BMPV_ent}. More precisely, the gravitational index we compute corresponds (up to a Laplace transform) to the helicity supertrace index, once the appropriate number of fermionic zero modes have been reabsorbed. For further details, we refer the reader to~\cite{Anupam:2023yns}.}

\subsubsection{$\beta\to 0$ limit: two-center microstate geometries}
\label{sec:fuzz}
Finally we discuss the $\beta\to 0$ limit. Using again \eqref{eq:DiracMisner} and \eqref{eq:omegaNS}, we get that the condition $\beta \to 0$ fixes the distance between the centers in terms of the remaining parameters,
\begin{equation}
\label{eq:bubble_eq}
\delta= - \frac{C_{IJK}\kk^I_N \kk^J_N \kk^K_N}{3 \ell_{I, 0} \kk^I_N h_N^2 h_S^2}\, .
\end{equation}
In this limit, a real Lorentzian solution is obtained by the analytic continuation 
\begin{equation}
\label{eq:analytic_cont_microstate_geom}
\tau\to \ii t\,, \hspace{1cm} \kk^I_{N}\to \ii k^I_N\,.
\end{equation}
In particular one can verify that the charges $Q_{I}$ and both angular momenta $J_{\pm}$ become real, satisfying \eqref{eq:constraint_horizonless_sol}, which implies a vanishing entropy. The solution obtained following this procedure corresponds to a horizonless two-center microstate geometry \cite{Bena:2005va, Berglund:2005vb}. Indeed, the relations satisfied by the coefficients of the harmonic functions in microstate geometries are identical to those imposed for the saddles in section~\ref{sec:gen_saddles}. The main difference between these two solutions lies on global aspects. More concretely, the conditions arising from the removal of Dirac-Misner singularities are different, ultimately due to the fact that in our saddles the time coordinate is periodically identified. For microstate geometries, the absence of Dirac-Misner strings imposes two conditions \cite{Bena:2005va, Berglund:2005vb}: the so-called `bubble equations' (thoroughly studied in \cite{Avila:2017pwi}) and that $h_N$ is an integer. The bubble equation\footnote{There are $s-1$ bubble equations, being $s$ the number of centers. } is precisely recovered upon taking the $\beta\to 0$ limit of the last of \eqref{eq:DiracMisner}. As a matter of fact, one can check that the analytic continuation of \eqref{eq:bubble_eq} precisely corresponds to the solution to the bubble equation. In turn, the condition $h_N\in {\mathbb Z}$ follows from demanding consistency of the twisted identifications \eqref{eq:twisted_bdry_conditions} when $\beta$ is set to zero. Indeed, using \eqref{eq:DiracMisner} we see that they imply $\left(\psi, \phi\right)\sim \left(\psi +2\pi \left(h_N-h_S\right), \phi+2\pi\right)\sim \left(\psi +2\pi, \phi+2\pi\right)\sim \left(\psi -2\pi, \phi+2\pi\right)$, which only makes sense when $h_N$ (and therefore $h_S=1-h_N$) is an integer.

%

%% file: research_five.tex
\chapter{Bubbling saddles of the gravitational index}
\label{chap:bubbling}


In this chapter, based on contribution~\cite{Cassani:2025iix}, we investigate how the cornucopia of supersymmetric Lorentzian solutions reviewed in chapter \ref{sec:intro3} relates to saddles of the gravitational index. 
 We are thus led to study finite-temperature deformations of the extremal solutions, which we assume to have ${\rm U}(1)^3$ symmetry. As we explained in chap.~\ref{sec:intro3}, the main features of the solutions are encoded in the rod structure, which specifies the isometries that degenerate and their fixed loci. In this context, studying finite-temperature geometries means considering supersymmetric horizons containing a rod of finite size, and studying the associated regularity conditions.\footnote{In the context of minimal five-dimensional supergravity, the rod formalism including non-supersymmetric, non-extremal horizons is discussed e.g.\ in~\cite{Armas:2009dd,Armas:2014gga}.} This is a non-trivial task since the solutions have a complex metric, however we will still be able to address their topology.  
The extremal limit, leading to solutions that have a Lorentzian counterpart, can then be understood as a limit where this horizon rod contracts to a point. 

Our study shows that the set of saddles of the gravitational index  with fixed $\beta$, $\omega_1$, $\omega_2$, $\varphi$ is much richer than the class identified as ``black hole saddles'' discussed in chapter \ref{chap:Black_hole}.
By combining different types of rods, namely {\it horizon rods} and {\it bubbling rods}, we provide a general  framework for constructing solutions with multiple disconnected horizons of different non-trivial topologies and an arbitrary number of bubbles outside the horizons.\footnote{For parallel progress in four-dimensional setup see~\cite{Boruch:2025biv}. The 4D/5D uplift of such four-dimensional saddles gives five-dimensional solutions closely related to the one discussed in this chapter~\cite{Boruch:2025sie}.} 
These are distinguished by the number of times the orbits of the associated rod vector wind around the three basis ${\rm U}(1)$'s in the geometry, ~\eqref{eq:U(1)isometries}. The  winding numbers needed to specify this information characterize the topology of the fixed loci. In particular, if the orbits of the rod vector wind around the thermal circle, then the fixed locus is a horizon, while if they do not, then it is a three-dimensional bubble. 
In this framework, the black hole saddles of~\cite{Anupam:2023yns,Hegde:2023jmp,Cassani:2024kjn,Adhikari:2024zif,Boruch:2025qdq} correspond to the case where there is a unique compact rod, and the orbits of the rod vector wind just once around the thermal circle. A simple  generalization of this case is one where the orbits wind multiple times around the thermal circle as well as around an axial circle; this corresponds to a supersymmetric orbifold of the black hole solution, and provides a saddle which only makes sense in the complexified setup. Similar orbifolded solutions have been shown to play an interesting role in the study of asymptotically AdS$_5 \times S^5$ black holes in Type IIB supergravity, where they appear as competing contributions to the superconformal index alongside the supersymmetric AdS$_5$ black hole (reviewed in sec.~\ref{TwoDerReview})~\cite{Aharony:2021zkr}.\footnote{Related phenomena in the AdS$_4$ setting have been discussed in~\cite{BenettiGenolini:2023rkq}.} In particular, it was demonstrated that the on-shell action of these orbifolded black holes reproduces the contribution of known Bethe roots of the ${\cal N}=4$ SYM index. The presence of such saddles in the grand-canonical index gives rise to a rich phase structure in the space of chemical potentials~\cite{Cabo-Bizet:2019eaf}. By contrast, their role in ungauged supergravity has not been previously explored. Similarly to the AdS case, we expect that their contribution to the gravitational index can lead to a markedly erratic phase diagram.

We work out the topology of the rod vector fixed loci within the five dimensional geometry, including the Euclidean time circle fibration in the description. We find that in general the topology is the one of an $L(\mathtt p,\mathtt q)$ lens space, and allows for $S^3$ and $S^2\times S^1$ as special cases. The integers $\mathtt p$, $\mathtt q$ characterizing the lens space are determined by the winding numbers associated with the rod under study and the two adjacent ones, in a way that we discuss in detail. One can also consider non-freely-acting orbifolds of the above topologies, which produce horizons and bubbles with conical singularities, namely branched spheres, branched lens spaces and products of $S^1$ and a spindle.\footnote{Starting with~\cite{Ferrero:2020laf,Ferrero:2020twa}, geometries based on spindles have provided a new direction of investigation in the context of supergravity and supersymmetric holography. 
}

The characterization of the solutions also requires to specify the electric flux through the three-dimensional bubbles or, alternatively, certain conjugate potentials that we define and compute via equivariant localization. When the abelian gauge group is compact, namely ${\rm U}(1)$, this leads us to introduce an additional integer for each compact rod. For bubbling rods, the potential conjugate to the electric flux is now quantized and directly specified by the additional integer. For horizon rods, the integer enters in the linear map relating the horizon electrostatic potential and the chemical potential $\varphi$ appearing in the index \eqref{eq:microscopicindex}.

We then focus on the possible saddle-point contribution of our solutions to the gravitational index. This requires computing the supergravity on-shell action and expressing it in terms of the variables appearing in the index.
For the case with just one horizon but arbitrarily many bubbles, we infer a formula for the on-shell action. 
Relatedly, we study the thermodynamics of our solutions in the presence of non-trivial topology outside the horizon, focussing on the contribution of the electric flux through the three-dimensional bubbles.

As particularly significant examples of our general construction, we study in detail the class of solutions with two compact rods, one being a horizon rod and the other being a bubbling rod. This includes the supersymmetric non-extremal versions of the BPS black ring and black lens of~\cite{Elvang:2004rt,Kunduri:2014kja}. 
By a direct computation we evaluate the on-shell action of these solutions. This requires some care, since the gauge Chern-Simons term of five-dimensional supergravity has to be evaluated patchwise. The expression we obtain for the on-shell action reads
\begin{equation}
\label{eq:I_gen_intro}
I\,=\,\frac{\pi}{12\sqrt{3}}\left[\frac{\varphi^3}{\omega_1\omega_2}-\frac{\left(\pp_1\varphi- \omega_2\, \Phi^{\BR}\right)^3}{\pp_1^2\,\omega_2\left(\pp_1\omega_1 + \left(\pp_1-1\right)\omega_2\right)}\right]\, ,
\end{equation}
where $\pp_1\in \mathbb{Z}$ and $\Phi^{I_1}$ is the potential conjugate to the electric flux through the bubble.
We verify that by taking the extremal $\beta\to \infty$ limit of the solutions thus constructed and demanding regularity outside the horizon we land indeed on the BPS black ring and black lens. These correspond to the cases $\pp_1=1$ and $\pp_1=-1$ in~\eqref{eq:I_gen_intro}, respectively. The on-shell action remains well-definite in the limit and may thus be seen as the saddle-point contribution of these BPS solutions to the index.
As a check of our on-shell action formula, we recover the correct Bekenstein-Hawking entropy of the black ring and black lens by a Legendre transformation.

In addition to the extremal $\beta\to\infty$ limit, we consider a different limit leading to Lorentzian solutions, which involves sending $\beta\to 0$. This limit transforms horizon rods into bubbling rods, and requires that $\omega_2$, $\varphi$ satisfy suitable quantization conditions. We thus obtain horizonless bubbling solutions, belonging to the family found in~\cite{Bena:2005va,Berglund:2005vb}.
 In this way, we can assign chemical potentials and an on-shell action to these solutions as well. However, these quantities are purely imaginary, raising the question of whether the gravitational index can be extended to this regime and the horizonless solutions be regarded as acceptable saddles. 

\medskip

The rest of the chapter is organized as follows. 
In section~\ref{sec:rodstr} we illustrate the rod structure and analyze the topology of the rod vector fixed loci, distinguishing between horizon rods and bubbling rods. A summary of the general classification which follows from this analysis is given in section~\ref{sec:rodstrt}. We also illustrate the global properties of the gauge field and its role in the thermodynamics. In section~\ref{sec:2centersol} we revisit the two-center black hole solution emphasizing the need to complexify it, and discussing its shifts and orbifolds satisfying the same boundary conditions. In section~\ref{sec:3centersol} we discuss in detail the three-center solution, providing the on-shell action and specifying the Lorentzian limits leading to the black ring, the black lens and a horizonless topological soliton. We draw our conclusions in section~\ref{sec:conclusions}.
Two appendices contain a brief account on lens spaces, and the explicit evaluation of the on-shell action for the three-center solution.

\section{Rod structure and topology of fixed loci}\label{sec:rodstr}

Building on the general construction of supersymmetric solutions with ${\rm U}(1)^3$ isometry presented in section~\ref{sec:gen_ssol}, we now examine their global properties, with the aim of determining the additional conditions required for these solutions to serve as asymptotically flat saddles contributing to the supersymmetric index.

\subsection{The rod structure}\label{sec:rodstrt}

The global properties of the solutions are characterized by a rod structure. This is closely related to the one introduced for Lorentzian solutions~\cite{Harmark:2004rm,Hollands:2007aj,Breunholder:2017ubu}, however we will include the Euclidean time circle in the description.
In order to illustrate the rod structure in our setup, recall that the solutions we consider have a ${\rm U}(1)^3$ isometry generated by linear combinations of the supersymmetric Killing vector $\partial_\tau$, the Gibbons-Hawking fibre $\partial_\psi$ and the vector $\partial_\phi$ generating rotations around the $z$-axis in the $\mathbb R^3$ base of the Gibbons-Hawking space, as seen in the previous section. Denoting by $\cal M$ the five-dimensional spacetime manifold, we can thus introduce the orbit space ${\cal M}/{\rm U}(1)^3$.
Parameterizing the $\mathbb R^3$ base of the Gibbons-Hawking space with cylindrical coordinates $(\rho\geq 0,\phi,z)$, the orbit space is the half-plane spanned by $(\rho,z)$, with boundary the infinite $z$-axis.
The action of the isometries generates a three-torus at each interior point of the orbit space, while it is degenerate at the boundary. The boundary line is divided into segments -- the rods -- characterized by the specific {\rm U}(1) isometry which degenerates there. At the intersection of two adjacent rods, a ${\rm U}(1)\times {\rm U}(1)$ isometry degenerates, hence intersection points are corners of the orbit space. In our supersymmetric setup, these points coincide with the centers $z_a$ of the harmonic functions introduced in section~\ref{sec:gen_ssol}.
 We call {\it rod vectors} the Killing vectors whose closed orbits collapse at a rod.
 The fixed locus of a rod vector is the three-dimensional space made by the rod and the two-torus foliated over it. 
 At the intersection of two rods, a combination of the adjacent rod vectors degenerates; the corresponding fixed locus is a one-dimensional circle.
 Borrowing the terminology of~\cite{Gibbons:1979xm}, we will call {\it bolts} and {\it nuts} these co-dimension two and co-dimension four loci, respectively.
 
\medskip

\begin{figure}[h!]
    \centering
    \includegraphics[width=1\linewidth]{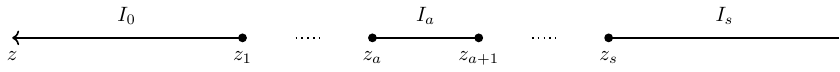}    
    \caption{\it Generic representation of the rod structure. The compact rod $I_a$ joins the Gibbons-Hawking centers placed at $z_a$ and $z_{a+1}$, while the dots represents any possible arrangement of compact rods. The first and last ones are semi-infinite segments originating from $z_1$ and $z_s$.}
    \label{fig:rod_notation}
\end{figure}

We label by $I_a= [z_a,\,z_{a+1}]$ the rod joining the centers at $z_a$ and $z_{a+1}$, with $a= 0,1,\ldots,s$, as illustrated in figure~\ref{fig:rod_notation}. Our convention is that  $z_0=+\infty$ and $z_{s+1}=-\infty$; hence $I_0$ and $I_s$ denote the semi-infinite rods $(+\infty,z_1]$ and $[z_{s},-\infty)$, respectively.
After fixing an arbitrary overall constant, the Killing vector 
degenerating at the rod $I_a$ is 
\begin{equation}
\label{eq:xi_A}
\xi_{I_a}\,=\, \partial_{\phi}- \ii {\breve \omega}_{I_a} \, \partial_{\tau} -\chi_{I_a} \,\partial_{\psi} \,, 
\end{equation}
where ${\breve \omega}_{I_a}$ and $\chi_{I_a}$ denote the $\phi$-component of the one-forms $\breve \omega$ and $\chi$ evaluated at the rod $I_a$. These are constant and read
\begin{equation}
\label{eq:formsatA}
{\breve \omega}_{I_a}\,=\,\sum_{b>a}\ww_{b}-\sum_{b\le a} \ww_{b} \,=\,-2\sum_{b\leq a}\ww_b\,, \hspace{1.2cm} \chi_{I_a}\,=\,\sum_{b> a}h_{b}-\sum_{b\le a}h_{b}\,=\,1-2\sum_{b\leq a}h_b\, .
\end{equation}
Notice that the rod vectors associated with the two semi-infinite rods $I_0$ and $I_s$ are always $\partial_{\phi}- \partial_{\psi}$ and $\partial_{\phi}+ \partial_{\psi}$, respectively (indeed, these are precisely the vectors having fixed points at infinity).

The rod structure of a supersymmetric finite-temperature solution is fixed once certain integers are specified.
Indeed, regularity requires that the orbits of $\xi_{I_a}$ close, hence \eqref{eq:xi_A} must be a linear combination of the three ${\rm U}(1)$ generators~\eqref{eq:U(1)isometries} with integer coefficients, where we should recall that we have fixed $\omega_+=2\pi\ii$. This combination can be parameterized as\footnote{This is obtained as follows. In terms of the basis vectors \eqref{eq:U(1)isometries} generating the 
independent $2\pi$-periodic ${\rm U(1)}$'s, each rod vector $\xi_I$ must be of the form
$$
\xi_{I}\,=\,  N_0\left(\frac{\beta}{2\pi}\partial_\tau +\partial_\phi + \frac{\ii \omega_-}{2\pi} \partial_\psi\right) + N_1 \left(\partial_{\phi}-\partial_{\psi}\right) + \,N_2 \left(\partial_{\phi}+\partial_{\psi}\right)\,,
$$
with integer $N_0,N_1,N_2$ satisfying $N_0+N_1+N_2=1$, so that the coefficient of $\partial_\phi$ is 1. In \eqref{eq:combinationU1s}, we have used $N_1=1-N_0-N_2$ and renamed $N_0\equiv\nn$, $N_2\equiv\pp$. 
}
\be\label{eq:combinationU1s}
\xi_{I_a}\,=\,  \nn_a \left(\frac{\beta}{2\pi}\partial_\tau +\partial_\phi + \frac{\ii \omega_-}{2\pi} \partial_\psi\right) + (1-\nn_a) \,\partial_\phi + (2\pp_a+\nn_a-1)\,\partial_\psi\,,\qquad\
 \nn_a, \,\pp_a \in \mathbb{Z}\,.\qquad
\ee
Comparing with \eqref{eq:xi_A} and using \eqref{eq:formsatA}, we find that for each rod $I_a$ we have
\be\label{map_from_rod_vec}
 \sum_{b\leq a}\ii\ww_b \,=\, \nn_a\frac{\beta}{4\pi} \,,\qquad\qquad  \sum_{b\leq a}h_b \,=\, \pp_a +\frac{\nn_a}{2}\left(1 +  \frac{\ii \omega_-}{2\pi}\right)\,.
\ee
These relations provide a map between the solution parameters (on the left hand side) and a choice of integers $(\nn_a,\,\pp_a)$ for each rod, together with the boundary conditions $\beta,\omega_-$ which are fixed at infinity and are independent of the rod considered (on the right hand side).

In sections~\ref{sec:corner_classification}, \ref{sec:bolt_topology} we perform a detailed regularity analysis and show how it constrains the rod structure data and determines the topology of the fixed loci associated with the rod vectors. Before diving into this,  for the reader's convenience we provide a summary of the outcome of the analysis.

We will distinguish between {\it horizon rods} and {\it bubbling rods}, depending on the value of $\nn_a$.  Let us discuss them in turn.

\paragraph{Bubbling rods.} Bubbling rods have $\nn_a=0$, hence the rod vector does not have a component along the thermal direction and reduces to
\be
\label{eq:KV_bubbling_rod}
\xi_{I_a}\,=\,  \,\partial_\phi + (2\pp_a-1)\,\partial_\psi\,.
\ee
From \eqref{map_from_rod_vec} it follows that
\be
\label{eq:bubbling_rods_properties}
\sum_{b \leq a}\ww_b \,=\, 0\,,\qquad\qquad
\sum_{b\leq a}h_b \,=\, \pp_a \in \mathbb{Z}\,.
\ee
Note that the definition includes the semi-infinite rods $I_0$ with $\pp_0=0$, and $I_{s}$ with $\pp_{s}=1$.
Moreover, if $I_{a-1}$ and $I_{a}$ are adjacent bubbling rods, the relations above imply 
\be\label{eq:breveom=0}
\ww_a \,=\, 0\,,
\ee
which is known as {\it bubble equation}~\cite{Bena:2007kg}, as well as 
\be
\label{eq:ha_bubbling_rod}
 h_a \,=\,  \pp_a-\pp_{a-1}   \in \mathbb{Z}\,.
\ee
Studying the solution near the intersection point at $z=z_a$, one can see that there is a $\mathbb C^2/\mathbb Z_{|h_a|}$ orbifold singularity, unless one chooses $h_a=\pm1$, in which case the metric is smooth around that point. We will allow for such orbifold singularities in our general discussion, since they may make sense when the solution is uplifted to string theory. 
The conditions for bubbling rods summarized here are equivalent to those known from previous analyses in Lorentzian signature, see e.g.~\cite{Bena:2007kg,Hollands:2007aj}.

\paragraph{Horizon rods.} Horizon rods have $\nn_a \neq 0$, hence their rod vector contains the generator of the thermal circle and the corresponding bolt is an Euclidean horizon.
We find that if $I_a$ is a horizon rod, then the two adjacent rods $I_{a-1}$ and $I_{a+1}$ must be bubbling rods.
 From~\eqref{map_from_rod_vec} it follows that the solution parameters associated to the horizon rod endpoints $z_a$ and $z_{a+1}$ satisfy 
\be\label{eq:rel_beta_breveom}
\ww_a  \,=\, - \ww_{a+1} \,=\, \nn_a \frac{\beta}{4\pi\ii}\,,
\ee
\be
\label{eq:ha_horizon_rod}
 h_a \,=\,   \pp_a -  \pp_{a-1} + \frac{\nn_a}{2}\left(1+ \frac{\ii \omega_-}{2\pi}\right)\,,\qquad\quad   h_{a+1} \,=\,  \pp_{a+1} - \pp_a - \frac{\nn_a}{2}\left(1+ \frac{\ii \omega_-}{2\pi}\right)\,,
\ee
implying
\be\label{eq:sum_h's}
h_a +h_{a+1} \,=\, \pp_{a+1} -  \pp_{a-1}\,\in\,\mathbb{Z}\,.
\ee
We emphasize that these relations must be satisfied by each horizon rod $I_a$ with the same $\beta$, $\omega_-$, since these are boundary conditions fixed at infinity.

We define the angular velocities of the horizon associated with $I_a$ as the constant coefficients $\omega_\pm^{I_a}$ that can be read from the rod vector expressed as\footnote{These definitions assume that $n_a$ is positive. In general, one should define the horizon angular velocities as $\xi_{I_a} \,=\, \frac{{\rm sign} (n_a)}{2\pi}\left(|\nn_a|\beta\partial_\tau - \ii\omega_+^{I_a}\partial_\phi + \ii\omega_-^{I_a}\partial_\psi\right)$, since the vector in parenthesis now generates evolution along the thermal circle in the same direction as the first of the basis vectors in \eqref{eq:U(1)isometries}. Accordingly, the expressions for the angular velocities become 
$$
\omega^{I_a}_+\,=\, 2\pi \ii \,{\rm sign} \left(n_a\right)\,, \hspace{1cm} \omega_{-}^{I_a}\,=\,|n_a|\,\omega_-+2\pi\ii\left[\,{\rm sign}(n_a)\left(1-2\pp_a\right)-|n_a|\,\right]\,. 
$$
Therefore, by choosing the sign of $n_a$, we can engineer different horizons in a multi-horizon solution to realize both allowed possibilities for the angular velocity: $\omega_+^{I_a} = \pm 2\pi\ii$. To avoid cluttering the notation, we focus in the main text on the case $n_a>0$, although all of the expressions we derive can be readily extended to the general case. Note that configurations such that $\omega_+^{I_a}$ changes sign while $\omega_-^{I_a}$ remains invariant are obtained by sending $n_a \to - n_a$ and $p_a\to 1-p_a$. Making this transformation for all rods $I_a$, together with $z_a \to -z_a$, sends the solution into itself, since it is equivalent to performing the $\pi$-rotation in $\mathbb{R}^3$ which sends  $z\to -z$ and $\phi \to -\phi$.
} 
\be\label{eq:velocities_in_rod_vec}
\xi_{I_a} \,=\, \frac{1}{2\pi}\left(\nn_a\beta\partial_\tau - \ii\omega_+^{I_a}\partial_\phi + \ii\omega_-^{I_a}\partial_\psi\right)\,.
\end{equation}
These are given by 
\begin{equation}\label{eq:ang_vel_hor}
    \omega_+^{I_a} \,=\, 2\pi \ii\,,\qquad \quad\omega_-^{I_a} \,=\,  \nn_{a}\omega_- + 2\pi \ii \left(1- \nn_a -2\pp_a\right) \,. 
\end{equation}
Hence the horizon angular velocity $\omega_+^{I_a}$ is the same as the chemical potential $\omega_+=2\pi\ii$, while $\omega_-^{I_a}$ is related to the chemical potential $\omega_-$ by a linear combination controlled by the integers $\nn_a$, $\pp_a$.

\medskip

As we will illustrate in the examples, the information above gives a systematic way to construct solutions by introducing one rod after the other, starting from the semi-infinite rod $I_0$, assigning integers $\nn_a$ and $p_a$ to each new rod $I_a$, and ending with the other semi-infinite rod, $I_s$. We emphasize that in order to conclude that a solution with a given rod structure  actually exists, one has to solve the algebraic equations~\eqref{eq:breveom=0}, \eqref{eq:rel_beta_breveom}, where the expression for $\ww_a$ was given in~\eqref{eq:breve_omega_a}. These equations fix the distances  between the centers in terms of the other parameters and, for horizon rods, in terms of the chosen $\beta$. When there are many centers, the equations become hard to solve (a method has been proposed in \cite{Avila:2017pwi}). A complete analysis of these equations in the complexified setup goes beyond the scope of the present work; in particular, one should check if they are solved by real or complex values of $\delta_{ab}$. We will comment more on this issue while discussing some explicit examples in the next sections.

\paragraph{Topology of bolts.} For each rod vector, we can discuss the topology of the associated fixed locus in the five-dimensional solution. This is a three-dimensional bolt, made by a foliation along the rod of the two circles that are non-degenerate in the interior of that rod. We will distinguish  between {\it bubbling bolts} and {\it horizon bolts}, associated with bubbling rods and horizon rods, respectively. 

We have mentioned that horizons can only sit between two bubbling bolts. On the other hand, next to a bubbling rod $I_a$ one can have either a bubbling rod or a horizon rod, and the topology of the bolt associated with $I_a$ is sensitive to the neighbours.
Indeed, the torus foliation along $I_a$ making the bubbling bolt involves the closed orbits of a Killing vector  involving $\partial_\tau$; these collapse at one of the two rod endpoints if the adjacent rod is a horizon rod, while they remain of finite size if the adjacent rod is a bubbling one. We should therefore distinguish between the following cases, depending on the nature of the adjacent bolts.

\begin{figure}[h!]
    \centering
    \includegraphics[width=\linewidth]{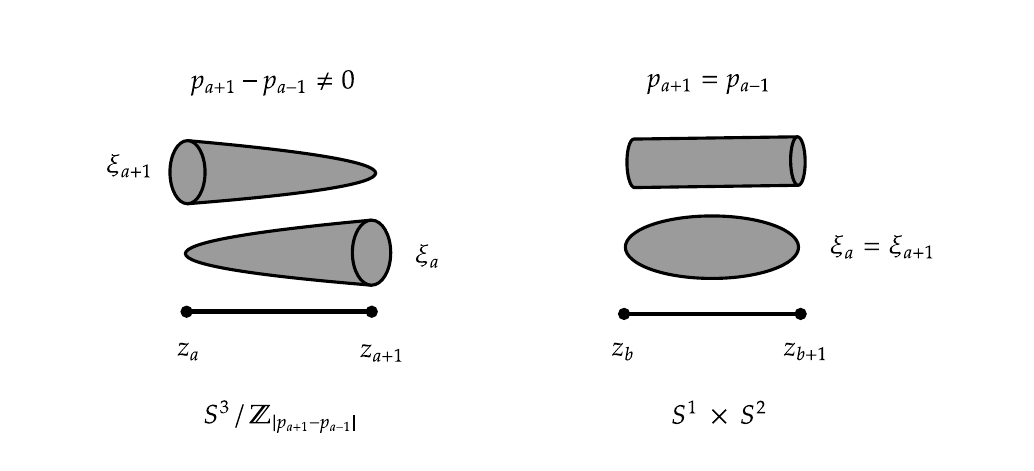}
    \caption{\it For $|n_a|=1$, bolts associated to horizon rods have $L(|p_{a+1}-p_{a-1}|, 1)$ topology. In the case $p_{a+1}-p_{a-1}\neq 0$, each of the two axial ${\rm U}(1)$ isometries shrinks at one rod endpoint, realizing the lens space topology $S^3/{\mathbb Z}_{|p_{a+1}-p_{a-1}|}$. When $p_{a+1}=p_{a-1}$, instead, the same ${\rm U}(1)$ shrinks at both endpoints, while the other never does; then, the horizon has $S^1\times S^2$ topology.}
    \label{fig:horizontopology}
\end{figure}

\begin{itemize}
\item \textbf{bubble$_{a-1}$--HORIZON$_{a}$--bubble$_{a+1}$.} We find that the bolt associated with a horizon rod $I_a$ has lens space topology
\begin{equation}
   L\Bigl(|\nn_a\left(\pp_{a+1} - \pp_{a-1}\right)|\,, \ 1 + \mathtt a\,\left(\pp_{a+1} - \pp_{a-1}\right)\Bigr) \, \simeq \, S^3/{\mathbb Z}_{|\nn_a\left(\pp_{a+1} - \pp_{a-1}\right)|}\,,
\end{equation}
where $\mathtt a$ is an integer defined via the equation
\begin{equation}
    \mathtt a \left(p_a - p_{a+1}\right) + \mathtt b \,\nn_a\left(\pp_{a-1} - \pp_{a+1}\right)\, =\,1\,,\qquad \mathtt a,\,\mathtt b \in \mathbb Z\,.
\end{equation}
In the special case $|n_a\left(\pp_{a-1} - \pp_{a+1}\right)| =1$, the horizon has ${S}^3$ topology, while for $p_{a-1}=p_{a+1}$ the topology is that of ${S}^1 \times {S}^2$. For $|n_a|=1$, this reduces to the  classification of (non-extremal) horizon topologies appearing in the Lorentzian analysis of~\cite{Hollands:2007aj}, as depicted in figure~\ref{fig:horizontopology}.

\item \textbf{horizon$_{a-1}$--BUBBLE$_{a}$--horizon$_{a+1}$.} 
The bolt associated to the bubbling rod $I_a$ sitting between two horizons also has lens space topology
\begin{equation}
    L\Bigl(|\nn_{a-1}\left( \pp_a - \pp_{a+1}\right)-\nn_{a+1}\left( \pp_a - \pp_{a-1}\right)|,\ 1 + \mathtt a\, \left( \pp_{a+1} - \pp_{a-1}\right)\Bigr)\,,
\end{equation}
with
\begin{equation}
    \mathtt a\left(\pp_a - \pp_{a+1}\right) + \mathtt b \left[\nn_{a-1}\left( \pp_a - \pp_{a+1}\right)-\nn_{a+1}\left( \pp_a - \pp_{a-1}\right)\right] = 1\,,\qquad \mathtt a,\,\mathtt b \in \mathbb Z\,.
\end{equation}
The circles that collapse smoothly at the rod endpoints intersecting  the two neighbouring horizons are both generated by vectors involving $\partial_\tau$. This tells us that if we consider the hypersurface at constant $\tau$ then we find a 2-tube~\cite{Breunholder:2017ubu}.

\item \textbf{horizon$_{a-1}$--BUBBLE$_{a}$--bubble$_{a+1}$}, or \textbf{bubble$_{a-1}$--BUBBLE$_{a}$--horizon$_{a+1}$}. 
The bolts have topology
\begin{equation}
L\Bigl(|n_{a-1}|,\, p_{a-1}-p_a\Bigr)\, \hspace{5mm}\text{or}\hspace{5mm} L\Bigl(|n_{a+1}|,\, p_{a+1}-p_a\Bigr)\,,
\end{equation}
respectively. This assumes $|h_{a+1}|=1$ (in the first case) or $|h_{a}|=1$ (in the second case), so that we have a smooth topology. If these conditions are not met, then there is a conical singularity $\mathbb{C}/\mathbb{Z}_{|h_{a+1}|}$ (in the first case), or $\mathbb{C}/\mathbb{Z}_{|h_{a}|}$ (in the second case) at the bubbling centers, and the lens space is branched. For $|n_{a\mp1}|=1$ this reduces to a branched $S^3$.  
The circle that smoothly collapses at the rod endpoint touching  the  horizon is generated by a vector involving $\partial_\tau$, while the one at the other endpoint does not. This implies that the hypersurface at constant $\tau$ has disc topology, possibly with a conical singularity at the tip.

\item \textbf{bubble$_{a-1}$--BUBBLE$_{a}$--bubble$_{a+1}$.}
The bolt associated with the rod $I_a$ generically has topology
\begin{equation}
    {S}^1\times \Sigma_{[h_a,h_{a+1}]}\,,
\end{equation}
 where $\Sigma_{[h_a,h_{a+1}]}=\mathbb{WCP}^1_{[h_a,h_{a+1}]}$ denotes a complex weighted projective space, also known as \emph{spindle}. This  is topologically a sphere with  
  axial symmetry and orbifold singularities at the two poles, of the type $\mathbb C/\mathbb Z_{|h_a|}$ and $\mathbb C/\mathbb Z_{|h_{a+1}|}$, leading to conical deficit angles $2\pi(1-1/|h_{a}|)$ and $2\pi(1-1/|h_{a+1}|)$, respectively. If $|h_a|=1$ and $|h_{a+1}|\neq1$ (or vice-versa) then we have a teardrop. If $|h_a|=|h_{a+1}|=1$ then we have a smooth $S^2$.
\end{itemize}

\subsection{Regularity of the metric}\label{sec:corner_classification}

We next come to our detailed regularity analysis, proving the claims summarized above. This consists of checking absence of Dirac-Misner strings as well as smoothness of the geometry at the  rod endpoints -- up to orbifold singularities. We stress that while these regularity conditions are standard when the metric and the gauge field are real, they become much less obvious for a complexified solution. However, we explicitly show how implementing suitable {\it complex} changes of coordinates one can reduce to the usual analysis in a real space near to the potentially singular points. We will comment further on the need to consider complex solutions in section~\ref{sec:2centersol}.

In order to study regularity of the geometry, it is convenient to choose coordinates adapted to the rod vectors $\xi_{I_a}$ by making the transformation
\begin{equation}\label{eq:rod_adapt_coords}
\tau_{a}\,=\, \frac{2\pi}{\beta}\left(\tau + \ii {\breve \omega}_{I_a} \,\phi\right)\, , \hspace{1cm} \psi_a\,=\, \psi+\chi_{I_a} \,\phi + c_a\left(\tau + \ii {\breve \omega}_{I_a}\, \phi\right)\,, \hspace{1cm} \phi_a\,=\, \phi\, ,
\end{equation}
where $c_a$ is a constant that will be fixed momentarily. Note that this is a linear transformation with a priori complex coefficients.
This change of coordinates leads to $\xi_{I_a}\,=\, \partial_{\phi_a}$,
 and is related to cancellation of Dirac-Misner string singularities.
 The latter arise when the connection one-forms $\chi$ and $\breve\omega$ have a non-vanishing $\phi$-component on the $z$-axis, where $\diff \phi$ is not well-defined. A string singularity on the $z$-axis is just a coordinate singularity if there exists a coordinate transformation that removes it, and the transformation \eqref{eq:rod_adapt_coords} indeed does the job. However, 
 we still have to show compatibility of the new coordinates  in the overlaps between the different regions containing the different rods $I_a$. Let us prove this. 
 We start by specifying the condition for the new coordinates  to obey untwisted periodic identifications. Recalling \eqref{eq:coord_identif}, we deduce the identifications 
\begin{equation}
\begin{aligned}
\left(\tau_a,\, \psi_a,\, \phi_a\right)\ \sim&\  \left(\tau_a+2\pi,\, \psi_a+\ii\omega_- +2\pi+\beta c_a,\, \phi_a\right)\ \sim\ \left(\tau_a,\, \psi_a+4\pi, \,\phi_a\right)\\[1mm]
\ \sim &\ \left(\tau_a+\frac{4\pi^2 \,\ii\, \breve \omega_{I_a}}{\beta},\, \psi_a+ 2\pi\left(\chi_{I_a}-1+\ii \breve \omega_{I_a} \,c_a\right), \,\phi_a+2\pi \right)\, .
\end{aligned}
\end{equation}
These are untwisted  provided
\begin{equation}\label{eq:DMgeneral0}
\ii\omega_-+2\pi+\beta c_a\,=\, 4\pi {\tilde p}_a\,, \hspace{1cm}\frac{2\pi\ii {\breve \omega}_{I_a}}{\beta} \,=\, -\nn_a\,, \hspace{1cm}  \chi_{I_a}- 1+\ii {\breve \omega}_{I_a}\, c_a \,=\, -2 \pp_a\,,
\end{equation}
where $\nn_a$, $\pp_a$, $\tilde \pp_a$ are a set of convenient integers. Indeed in this way we have
\begin{equation}
\left(\tau_a,\, \psi_a,\, \phi_a\right)\ \sim\  \left(\tau_a+2\pi,\, \psi_a,\, \phi_a\right)\ \sim\ \left(\tau_a,\, \psi_a+4\pi,\, \phi_a\right)
\ \sim \ \left(\tau_a ,\, \psi_a  ,\,\phi_a+2\pi\right)\, .
\end{equation}
With the conditions \eqref{eq:DMgeneral0} imposed, the Killing vector $\xi_{I_a}$  contracts smoothly on the rod $I_a$, so that the normal space close to the bolt forms a smooth $\mathbb R^2$:
\begin{equation}\label{eq:normal_space_rod}
    \diff s^2 \,=\, \ldots \,+\, f^{-1}H\left(\diff \rho^2 + \rho^2 \diff\phi_a^2\right)\,.
\end{equation}
Solving the first in \eqref{eq:DMgeneral0} for $c_a$, 
\begin{equation}
\label{eq:c^A_tau}
\beta c_a\,=\, 4\pi {\tilde p}_a-2\pi -\ii\omega_-\, ,
\end{equation}
we end up with
\begin{equation}
\label{eq:DiracMisnergeneral}
\frac{2\pi \ii {\breve \omega}_{I_a}}{\beta}\,=\, -\nn_a\, ,\hspace{1cm}
\chi_{I_a}\,=\,1-2\pp_a- \nn_a\left(1+\frac{\ii\omega_-}{2\pi}-2 {\tilde p}_a\right)\, .
\end{equation}
We note that we can set ${\tilde p}_a=0$ without loss of generality, since it can always be reabsorbed in a redefinition of $\pp_a$. We will do so in the following. Eqs.~\eqref{eq:DiracMisnergeneral} imply that on the overlap between two charts containing the rod $I_a$ and the rod $I_{b}$, respectively, one has
\begin{equation}\label{eq:change_patch}
\tau_{b}\,=\, \tau_a- \left(\nn_{b}-\nn_{a}\right)\phi_a\, ,\qquad \qquad\phi_{b}\,=\,\phi_{a} \,, \qquad\qquad
\psi_{b}\,=\,\psi_a-2\left(p_{b}-p_{a}\right) \phi_a \, ,
\end{equation}
which ensures compatibility of these coordinate patches, as we wanted to show.  

Let us comment on the meaning of discussing regularity conditions for a complexified metric. We note that the original coordinates $(\tau, \phi,\psi)$ should take complex values, since they satisfy the generically complex identifications~\eqref{eq:coord_identif}. However,  we can assume that the new coordinates $(\tau_a, \phi_a,\psi_a)$ introduced by the complex transformation~\eqref{eq:rod_adapt_coords} are real, since they have real period; this can be done in every patch, since the map~\eqref{eq:change_patch} between $(\tau_b, \phi_b,\psi_b)$ and $(\tau_a, \phi_a,\psi_a)$ is real. We conclude that if we describe the solution using the real coordinates $(\tau_a, \phi_a,\psi_a)$ in every patch, then the complexification is pushed into the metric (and gauge field) components, and a regularity statement such as the one associated with \eqref{eq:normal_space_rod},  where we check that the orbits of the rod vector $\xi_{I_a}\,=\, \partial_{\phi_a}$  smoothly shrink to zero size, appears to make sense.

Using \eqref{eq:formsatA} and setting $\tilde p_a=0$, Eqs.~\eqref{eq:DiracMisnergeneral} reduce to \eqref{map_from_rod_vec}, showing that the condition for absence of Dirac-Misner string singularities leads us to introduce a pair of  independent integers $(\nn_a,\,\pp_a)$ for each rod $I_a$, which  determine the rod vectors as in \eqref{eq:combinationU1s}. In general, Eqs.~\eqref{eq:DiracMisnergeneral} form a set of $2\left(s-1\right)$ non-trivial equations, the ones associated to the first ($I_0$) and last ($I_s$) rods being trivially solved as they do not involve the parameters of the solution.\footnote{It is sufficient to take $\pp_0= \nn_0 = \nn_s= 0$ and $\pp_s = 1$, since $\breve\omega_{I_0}= \breve\omega_{I_s} =0$ and $\chi_{I_0} = -\chi_{I_s} = 1$.} The remaining $2(s-1)$ equations fall in two categories. The   first set in \eqref{eq:DiracMisnergeneral} can be solved for the $h_a$, providing an expression of the latter in terms of $\omega_{-}$ and the integers $\nn_a$, $\pp_a$ (or viceversa). The second set of equations corresponds to a finite-$\beta$ version of the \emph{bubble equations} arising in the context of horizonless topological solitons \cite{Bena:2005va, Berglund:2005vb, Gibbons:2013tqa}.

\medskip

We now study the metric near the Gibbons-Hawking centers. At this stage, it is useful to distinguish two types of centers, {\it horizon poles} if $\ww_a \neq 0$, and {\it bubbling centers} if $ {\ww_a}=0$.

\paragraph{Smoothness at bubbling centers ($\ww_a =0$).} We start by analyzing the geometry near a bubbling center. To this aim, we introduce the new radial coordinate $\tilde r_a = 2\sqrt{r_a}$. Keeping only the relevant terms in the ${\tilde r}_a \to 0$ limit, one finds that
\begin{equation}
\diff s^2\,=\, f_a^2 \,\diff {\tau'_a}^2+ \frac{h_a}{f_a}\left\{\diff {\tilde r}_a^2+\frac{{\tilde r}_a^2}{4}\left[\left(\diff\psi'_a+\cos\theta_a \diff \phi'_a\right)^2+\diff \theta_a^2+\sin^2\theta_a\, \diff {\phi'_a}^2\right]\right\}\,+\dots\,, 
\end{equation}
where, using \eqref{map_from_rod_vec}, we have introduced the coordinates
\begin{equation}
\tau'_{a}\,=\, \tau-\frac{\beta\nn_{a-1}}{2\pi} \,\phi\,, \hspace{1cm} \psi'_a\,=\, \frac{\psi+\left(1-2Y_a-h_a\right)\phi -\frac{\ii\omega_- + 2\pi}{\beta}\tau}{h_a}\,, \hspace{1cm} \phi'_a\,=\,\phi\, ,
\end{equation}
and
\begin{equation}
Y_a \,=\, \sum_{b<a}h_b\,.
\end{equation}
The periodic identifications of the new coordinates are inherited from \eqref{S3_angles_identifications} and \eqref{beta_identifications}, namely: 
\begin{equation}
\label{eq:ident_aux_center}
\begin{aligned}
\left(\tau'_{a},\, \psi'_{a},\, \phi'_{a}\right) \sim &\left(\tau'_{a}+\beta,\, \psi'_{a},\, \phi'_{a}\right)\sim  \left(\tau'_{a}, \,\psi'_{a}+\frac{4\pi}{h_a},\, \phi'_{a}\right)\\[1mm]
\sim& \left (\tau'_a-\beta\nn_{a-1},\, \psi'_{a}-\frac{2\pi\left(2Y_a+h_a\right)}{h_a},\, \phi'_{a}+2\pi\right)\, .
\end{aligned}
\end{equation}
Demanding regularity of the metric at $\theta_{a}=0,\, \pi$ yields the following conditions
\begin{equation}
\label{eq:regularity_aux_center}
\nn_{a-1}= \frac{4\pi}{\beta}\sum_{b<a}\ii\ww_b=\,0 \,, \hspace{1cm}Y_a= p_{a-1} \in \mathbb Z \,,\hspace{1cm} h_a\,=\,\pm 1\,,
\end{equation}
which imply that the metric near a bubbling center is that of  $S^1\times {\mathbb R}^4$. The last condition in \eqref{eq:regularity_aux_center} can be relaxed to allow for integer values of $h_a$, if one admits discrete orbifold singularities. In such case, the periodic identifications \eqref{eq:ident_aux_center} become
\begin{equation}
\left(\tau'_{a},\, \psi'_{a},\, \phi'_{a}\right)\ \sim \ \left(\tau'_{a}+\beta,\, \psi'_{a},\, \phi'_{a}\right)\ \sim \ \left(\tau'_{a}, \,\psi'_{a}+\frac{4\pi}{h_a},\, \phi'_{a}\right)
\ \sim\ \left (\tau'_a, \,\psi'_{a}+2\pi, \,\phi'_{a}+2\pi\right)\, ,
\end{equation}
and the geometry near the center is that of $S^1\times {\mathbb R}^4/{\mathbb Z}_{|h_a|}$, where $h_a \in \mathbb Z$ and the ${\mathbb Z}_{|h_a|}$ quotient acts in the same way on the polar angles of the two orthogonal $\mathbb{R}^2$ planes in $\mathbb{R}^4$. 

The first condition in \eqref{eq:regularity_aux_center}, when combined with \eqref{eq:DiracMisnergeneral} and with the requirement $\ww_a =0$ characterizing bubbling centers, imposes that both the integers $\nn_{a}$ and $\nn_{a-1}$, associated to the two rods adjacent to the center, must vanish: 
\begin{equation}\label{aux_center_conditions}
\nn_{a-1}\,=\,\nn_{a}\,=\, 0\, .
\end{equation}
This implies that no bubbling center can be placed inside a horizon. Finally, let us further note that  the integers $p_a$ can be expressed in terms of $h_a$ and $p_{a-1}$, by the relation
\begin{equation}\label{eq:quantization_bubbling_center}
    p_a = h_a +p_{a-1} = \sum_{b\leq a}h_b \in \mathbb Z\,,
\end{equation}
as it  follows from \eqref{eq:regularity_aux_center}. Therefore, the local analysis near a bubbling center gives the conditions listed around \eqref{eq:bubbling_rods_properties}.

\paragraph{Smoothness at horizon poles ($\ww_a \neq 0$).} The metric near a horizon pole is given by 
\begin{equation}
\diff s^2 \,=\, \left(\frac{\ii {\ww}_a f_a}{h_a}\right)^2 \diff\psi_a'^2+ \frac{h_a}{f_a}\left[\diff {\tilde r}^2_a+{\tilde r}_a^2 \left(\frac{\diff \theta^2_a}{4}+ \sin^2\frac{\theta_a}{2}\,\diff \phi'^2_a + \cos^2 \frac{\theta_a}{2}\,\diff \tau_a'^2\right)\right]+\dots\,, 
\end{equation}
where we have introduced coordinates such that
\begin{equation}
\label{eq:coordinates_horizon_poles}
\begin{aligned}
\psi_a'\,=\,& \psi+\left(1-2Y_a -2 h_a X_a\right)\phi - \frac{h_a \tau}{\ii \ww_a}\,,\\[1mm] 
\tau_a' \,=\,& \frac{\tau}{2\ii \ww_a}+ X_a \phi\,, \hspace{1cm} \phi_a'\,=\, \left(1-X_a\right)\phi - \frac{\tau}{2\ii \ww_a}\,,
\end{aligned}
\end{equation}
with
\begin{equation}
\label{eq:def_X_a}
X_a\,=\, -\sum_{b<a} \frac{\ww_b}{\ww_a}\,, \hspace{1cm} Y_a\,=\, \sum_{b<a} h_b\, .
\end{equation}
The periodic identifications of the new coordinates, inherited from \eqref{S3_angles_identifications}, \eqref{beta_identifications}, are given by
\begin{equation}
\label{eq:identifications_horizon_poles}
\begin{aligned}
\left(\psi_a', \tau_a',\, \phi_a'\right)\sim&\,\left(\psi_a'+ 4\pi, \tau_a',\, \phi_a'\right)\,\sim\,  \left(\psi_a' -4\pi \left(Y_a+h_a X_a\right),\, \tau_a'+2\pi X_a,\, \phi_a'+2\pi \left(1-X_a\right)\right)\\[1mm]
\sim &\, \left(\psi_a'+\ii \omega_-+2\pi-\frac{\beta h_a}{\ii\ww_a},\, \tau_a' +\frac{\beta}{2\ii {\ww}_a},\, \phi_a'-\frac{\beta}{2\ii{\ww}_a}\right)\, .
\end{aligned}
\end{equation}
 Demanding regularity of the metric at $\theta_{a}=0, \pi$  boils down to the condition
\begin{equation}
X_a\in \{0,1\}\, .
\end{equation}
The choice determines whether the horizon pole is a north or a south pole. 
\begin{itemize}
\item \textbf{Horizon north pole.} A horizon north pole is defined by the condition $X_a=0$, which implies
\begin{equation}
 \nn_{a-1}\,=\,0\,, \hspace{1cm} \ww_a\,=\, \nn_a \frac{\beta}{4\pi \ii}\,.
\end{equation}
These follow after using the definition of $X_a$ in \eqref{eq:def_X_a}, together with \eqref{eq:DiracMisnergeneral} and \eqref{eq:formsatA}, and simply tell us that $I_{a-1}$ must be a bubbling rod, while $I_a$ is a horizon rod. Moreover, from the second of \eqref{eq:DiracMisnergeneral}, one deduces
\begin{equation}
Y_a\,=\,\sum_{b\le a} h_a\,=\, p_{a-1}\, , \hspace{1cm} h_a\,=\,p_{a}-p_{a-1}+\frac{\nn_a}{2}\left(1+\frac{\ii\omega_-}{2\pi}\right)\, .
\end{equation}
Putting this information together, the above identifications \eqref{eq:identifications_horizon_poles} reduce to 
\begin{equation}
\label{eq:identifications_horizon_NP}
\begin{aligned}
\left(\psi_a', \tau_a',\, \phi_a'\right)\sim&\,\left(\psi_a'+ 4\pi, \tau_a',\, \phi_a'\right)\,\sim\,  \left(\psi_a'\, ,\tau_a',\, \phi_a'+2\pi\right)\\[1mm]
\sim &\, \left(\psi_a'+\frac{4\pi\left(p_{a-1}-p_a\right)}{\nn_a},\, \tau_a' +\frac{2\pi}{\nn_a},\, \phi_a'-\frac{2\pi}{\nn_a}\right)\, ,
\end{aligned}
\end{equation}
which inform us that the space near a horizon north pole is $\left(S^1\times {\mathbb R}^4\right)/{\mathbb Z}_{|\nn_a|}$. The orbifold action is free as long as $p_{a-1}\neq p_{a}$.
\item \textbf{Horizon south pole.} A horizon south pole $z_a$ is in turn defined by the condition $X_{a}=1$, which implies 
\begin{equation}
\ww_{a}\,=\, -\nn_{a-1} \frac{\beta}{4\pi \ii}\, , \hspace{1cm} \nn_{a}\,=\, \sum_{b<a}\ww_b\,=\,0 \, .
\end{equation}
Then, $I_{a-1}$ and $I_{a}$ must be horizon and bubbling rods respectively. This observation also implies that the center $a-1$ must necessarily be a horizon north pole. Namely, horizon poles are always paired. On top of these conditions, one further deduces from \eqref{eq:DiracMisnergeneral} that 
\begin{equation}
    Y_{a}+h_{a}X_{a}\,=\,\sum_{b\le {a}}h_b\,=\,p_{a} \,, \hspace{1cm} h_{a}\,=\,p_{a}-p_{a-1}-\frac{\nn_{a-1}}{2}\left(1+\frac{\ii\omega_-}{2\pi}\right)\, .
\end{equation}
Thus, the periodic identifications of the coordinates read
\begin{equation}
\label{eq:identifications_horizon_SP}
\begin{aligned}
\left(\psi_{a}', \tau_{a}',\, \phi_{a}'\right)\sim&\,\left(\psi_{a}'+ 4\pi, \tau_{a}',\, \phi_{a}'\right)\,\sim\,  \left(\psi_{a}'\, ,\tau_{a}',\, \phi_{a}'+2\pi\right)\\[1mm]
\sim &\, \left(\psi_{a}'+\frac{4\pi\left(p_{a}-p_{a-1}\right)}{\nn_{a-1}},\, \tau_{a}' -\frac{2\pi}{\nn_{a-1}},\, \phi_{a}'+\frac{2\pi}{\nn_{a-1}}\right)\, .
\end{aligned}
\end{equation}
Then the space near the south pole is  $(S^1\times {\mathbb R}^4)/{\mathbb Z}_{|\nn_{a-1}|}$, acting freely provided $p_{a-1}\neq p_a$.
\end{itemize}
Thus, the summary of our findings (see also table~\ref{tab:summary_poles}) is that a regularity analysis does not allow horizon rods to be adjacent to each other: there must be bubbling rods right before and right after. 
Moreover, if $|\nn_a|\neq 1$, there will appear orbifold singularities at a pole of the horizon if either $p_{a+1}$ or $p_{a-1}$ are equal to $p_a$. If $|\nn_a|=1$, there is no such regularity condition on the $p$'s of the adjacent rods. 

\begin{table}[h!]
    \centering
    \begin{tabular}{c|ccccc}
         $z_a$\;& \;  $\ww_a$\;& \; $\nn_{a-1}$\;& \; $I_{a-1}$\;& \; $\nn_a$ \;& \;$I_a$\\
         \hline
 Horizon north pole\;& \; $\neq 0$\;& \; $= 0$\;& \; bubble \;& \; $\neq 0$ \;& \; horizon\\
         Horizon south pole\;& \; 
     $\neq 0$\;& \; $\neq0$\;& \; horizon \;& \; $=0$ \;& \; bubble\\
 Bubbling center\;& \; $=0$\;& \; $=0$\;& \; bubble \;& \; $=0$ \;& \; bubble\\\end{tabular}
    \caption{\it Classification of centers and their adjacent rods. Centers are distinguished by which rods meet there. This is reflected in the allowed values for the coefficients $w_a$ of the one-form $\breve\omega$ and the integer $\nn_a$, as shown by the regularity analysis.}
    \label{tab:summary_poles}
\end{table}

It can readily checked be that the expressions found here directly imply eqs.~\eqref{eq:rel_beta_breveom}, \eqref{eq:ha_horizon_rod}, \eqref{eq:sum_h's}. This  shows the consistency of our previous analysis with the local study around the horizon poles. 

 \subsection{Classification of bolt topologies}\label{sec:bolt_topology}

We now come to the study of the three-dimensional bolts associated with each rod vector $\xi_{I_a}$ given by  \eqref{eq:xi_A}, with the aim of classifying the possible topologies. 
It is convenient to write down the metric induced at the rod $I_a$ using the adapted coordinates introduced in \eqref{eq:rod_adapt_coords}:
\begin{equation}
\label{eq:induced_metric_bolts}
\diff s^2_{I_a}\,=\,f^{-1}H^{-1}\left(\diff \psi_a-\frac{\beta c_a}{2\pi} \diff \tau_a\right)^2 + f^{-1}H \,\diff z^2 + f^2 \left[\frac{\beta}{2\pi}\diff \tau_a+ \ii \omega_{\psi}\left(\diff \psi_a-\frac{\beta c_a}{2\pi} \diff \tau_a\right)\right]^2, 
\end{equation}
where $c_a$ is the constant given in \eqref{eq:c^A_tau}, and all the functions are evaluated on the rod.

\subsubsection{Local analysis at rod endpoints}

In order classify the possible topologies for the fixed loci of the rod vectors $\xi_{I_a}$, we need to specify the local form of the metric near to the centers $z_a$ and $ z_{a+1}$ where the bolt caps off, as well as the periodic identifications when going around the contracting circles. We first analyze the induced metric \eqref{eq:induced_metric_bolts} near to the rod endpoints.

\paragraph{Bubbling center.} Condition \eqref{eq:smooth_center_condition} guarantees that near the center the metric reads,
\begin{equation}\label{eq:metric_near_B}
\diff s^2_{a}\,=\, \frac{h_a}{f_a}\left[\diff {\tilde r}_a^2+{\tilde r}_a^2\left(\frac{\diff \psi_a}{2h_a}\right)^2\right]+\left(\frac{\beta f_a}{2\pi}\right)^2\,\diff \tau_a^2+\dots \, ,
\end{equation}
where $\tilde r_a = 2\sqrt{r_a}$. This shows that the Killing vector with $2\pi$-periodic orbits contracting at a bubbling center is 
\begin{equation}
    \xi_a \,=\, 2\partial_{\psi_a}\, =\, 2\partial_\psi\,,
\end{equation}
while the  $2\pi$-periodic vector 
\begin{equation}\label{eq:fake_lambda_at_B}
    \tilde\lambda_a \,=\, \partial_{\tau_a} \,=\, \frac{\beta}{2\pi}\partial_\tau + \left( 1 + \frac{\ii\omega_-}{2\pi}\right)\partial_\psi\,
\end{equation}
generates a fixed $S^1$.
The geometry near to the center is then that of $S^1\times {\mathbb R}^2/{\mathbb Z}_{|h_a|}$. The orbifold action is singular (unless $h_a = \pm 1$), reflecting the presence of an $S^1\times \mathbb R^2/\mathbb Z_{|h_a|} \times \mathbb R^2$ orbifold singularity in the full five-dimensional geometry~\cite{Bena:2005va}.

\paragraph{Horizon north pole.} For a horizon rod $I_a$, the following Killing vector contracts at the horizon north pole $z_a$:
\begin{equation}\label{eq:nut_vector_north_pole}
    \begin{aligned}
        \xi_a \,=\, \xi_{I_a}-\xi_{I_{a-1}} \,&=\, \frac{\beta \nn_a}{2\pi}\partial_\tau +\left[ 2\left( p_a - p_{a-1}\right) + \nn_a\left( 1 + \frac{\ii\omega_-}{2\pi}\right)\right]\partial_\psi 
        \\[1mm]
        \,&=\, \nn_a\,\partial_{\tau_a} + 2\left( \pp_a - \pp_{a-1}\right) \partial_{\psi_a}\,. 
    \end{aligned}
\end{equation}
Here, $(\tau_a,\,\psi_a)$ are the adapted rod coordinates introduced in \eqref{eq:rod_adapt_coords}. In order to exhibit the local metric near the pole, we define the new angles
\begin{equation}\label{eq:adapt_north_pole}
   \tilde \psi_a \,=\, \psi_a - 2\left(\pp_a-\pp_{a-1}\right) \frac{\tau_a}{\nn_a}\,,\qquad \tilde\tau_a \,=\, \frac{\tau_a}{\nn_a}\,,
\end{equation}
and expand the metric for $\tilde r_a=2\sqrt{r_a} \rightarrow 0$. This takes the form
\begin{equation}\label{eq:metric_near_poles}
\diff s^2_a\,=\, \frac{h_a}{f_a}\left(\diff {\tilde r}_a^2+{\tilde r}_a^2\diff {\tilde\tau}_a^2\right)+\left(\frac{f_a \,\ii \ww_a}{h_a}\right)^2\,\diff {\tilde\psi}_a^2+\dots \, .
\end{equation}
It is easy to show that it corresponds to an orbifold $\left(S^1\times {\mathbb R}^2\right)/\mathbb Z_{|\nn_a|}$, with the orbifold action specified by the identifications:
\begin{equation}\label{eq:north_pole_orbifold_identifications}
    \left(\tilde\tau_a,\,\tilde\psi_a\right)\ \sim\ \left(\tilde \tau_a +\frac{2\pi}{\nn_a},\,\tilde\psi_a - \frac{4\pi}{\nn_a}\left(\pp_a - \pp_{a-1}\right)\right) \ \sim \  \left( \tilde\tau_a ,\,\tilde\psi_a + 4\pi\right)\,.
\end{equation}
The orbifold is freely acting if 
$(\pp_a- \pp_{a-1})$ and $n_a$ are coprime. Eq.~\eqref{eq:north_pole_orbifold_identifications} identifies a $2\pi$-periodic Killing vector whose orbits describe the fixed $S^1$ at the endpoint:
\begin{equation}
\label{eq:fake_lambda}
\begin{aligned}
    \tilde\lambda_a \,&=\, \frac{1}{\nn_a}\left( \partial_{\tilde\tau_a}- 2\left( \pp_a - \pp_{a-1}\right)\partial_{\tilde\psi_a}\right)\\[1mm]
\,&=\,  \frac{\beta}{2\pi}\partial_\tau+ \left(1 + \frac{\ii\omega_-}{2\pi}\right) \partial_\psi\,.
    \end{aligned}
\end{equation}
Note that this is the same vector as \eqref{eq:fake_lambda_at_B}. Finally, the $2\pi$-periodic contracting vector \eqref{eq:nut_vector_north_pole} in these coordinates reads $\xi_a = \partial_{\tilde\tau_a}$. 

\paragraph{Horizon south pole.} The Killing vector contracting at a horizon south pole reads
\begin{equation}\label{eq:nut_vector_south_pole}
    \begin{aligned}
        \xi_{a}\, =\, \xi_{I_{a-1}}-\xi_{I_{a}} \,&=\,\frac{\beta \nn_{a-1}}{2\pi}\partial_\tau +\left[ 2\left( p_{a-1} - p_{a}\right) + \nn_{a-1}\left( 1 + \frac{\ii\omega_-}{2\pi}\right)\right]\partial_\psi 
        \\[1mm]
        \,&=\, \nn_{a-1}\,\partial_{\tau_a} + 2\left( \pp_{a-1} - \pp_{a}\right) \partial_{\psi_a}\,.
    \end{aligned}
\end{equation}
where $(\tau_a,\,\psi_a)$ are part of a system of adapted rod coordinates for the bubbling rod $I_a$, with $a$ being a south pole. The metric near the pole takes the same form as in \eqref{eq:metric_near_poles}, where $(\tilde\tau_a,\,\tilde\psi_a)$ are now given by
\begin{equation}\label{eq:adapt_south_pole}
   \tilde \psi_a = \psi_a - 2\left(\pp_{a-1}-\pp_{a}\right) \frac{\tau_a}{\nn_{a-1}}\,,\qquad \tilde\tau_a = \frac{\tau_a}{\nn_{a-1}}\,.
\end{equation}
The Killing vector generating the fixed $S^1$ is the same as in the second line of \eqref{eq:fake_lambda}.

\subsubsection{Bolt topologies}

We now turn to the analysis of global aspects.
 Aside from the two non‑compact bolts associated with $I_0$ and $I_s$,\footnote{The rod vectors  $\xi_{I_0} = \partial_\phi- \partial_\psi\,$, $\xi_{I_s} = \partial_\phi + \partial_\psi\,$ associated with the the semi-infinite rods $I_0$, $I_s$ have non-compact bolts with topology ${S}^1 \times \mathbb R^2/\mathbb Z_{|h_a|}$  (with $a\in\{1,s\}$) if they are adjacent to a bubbling rod, or ${S}^1 \times \mathbb R^2$ if they are adjacent to a horizon rod.}
we consider the four types of compact bolts already introduced above, classified by the nature of the adjacent rods.
For each bolt we may write relations between the associated Killing vectors $\xi_a$ and $\xi_{a+1}$ of the form: 
\begin{equation}\label{eq:vectors_lens_topology}
    \mathtt q_1\,\xi_a \,=\, \mathtt q_2\, \xi_{a+1} + \mathtt p\, \tilde\lambda_{a+1}\,,\qquad \mathtt p,\,\mathtt q_{1,2}\in\mathbb Z\,.
\end{equation}
We assume that $\mathtt q_1$ $\mathtt q_2$ are coprime to $\mathtt p$. As  reviewed in  appendix~\ref{sec:lens_space}, when the regions bounding the neighborhoods about the centers $a$ and $a+1$ are glued together, the manifold obtained is topologically the lens space
\begin{equation}
    L\left(\mathtt p,\,\mathtt a\,\mathtt q_2\right)\,,
\end{equation}
where 
\begin{equation}
    \mathtt a\,\mathtt q_1 + \mathtt b\,\mathtt p = 1\,,\qquad \mathtt a,\,\mathtt b \in \,\mathbb Z\,.
\end{equation}
If $|\mathtt p| =1$ and $\mathtt a\,\mathtt q_2\in \mathbb Z$, then the space has $S^3$ topology (up to possible conical singularities at the bubbling centers, as we specify below), while if $\mathtt p=0$ and $|\mathtt a\,\mathtt q_2|=1$, then the space topologically becomes $S^1 \times S^2$ (again, up to possible conical singularities).

\paragraph{bubble$_{a-1}$-HORIZON$_a$-bubble$_{a+1}$.} Let us begin by considering a horizon rod, connecting two horizon poles. The vectors contracting smoothly  at the horizon north and south pole are given by \eqref{eq:nut_vector_north_pole} and \eqref{eq:nut_vector_south_pole}, respectively. Then, we can express 
\begin{equation}
   \left(\pp_a -\pp_{a+1}\right)  \xi_a \,=\, \left(\pp_a - \pp_{a-1}\right)\xi_{a+1} +\nn_a\left(p_{a-1}-p_{a+1}\right)\tilde\lambda_{a+1}\,,
\end{equation}
where
\begin{equation}
    \pp_{a+1}-\pp_{a-1}\,=\, h_a + h_{a+1} \in \mathbb Z\,,
\end{equation}
as it follows from \eqref{eq:ha_horizon_rod}, \eqref{eq:sum_h's}. This shows that event horizons have lens space topology
\begin{equation}\label{eq:hor_topology_1}
    L\Bigl(|\nn_a\left(p_{a+1} - p_{a-1}\right)|\,,\; 1 + \mathtt a\,\left(p_{a+1} - p_{a-1}\right)\Bigr)\,,
\end{equation}
where $\mathtt a $ is an integer such that\footnote{We have used that $\mathtt a (\pp_a-\pp_{a-1})\,=\,\mathtt a (\pp_{a+1}-\pp_{a-1}) +\mathtt a \left(p_a-p_{a+1}\right)\,=\,1 + \mathtt a (\pp_{a+1}-\pp_{a-1}) - \mathtt b \, n_a\left(p_{a-1}-p_{a+1}\right)$, together with the homeomorphism between the lens spaces $L(\mathtt p, \mathtt q)\simeq L(\mathtt p, \mathtt q+ \mathtt p \,\mathbb Z)$ when writing \eqref{eq:hor_topology_1}.} 
\begin{equation}\label{eq:hor_topology_2}
    \mathtt a \left(p_a - p_{a+1}\right) + \mathtt b \,\nn_a\left( \pp_{a-1} - \pp_{a+1}\right) \,=\, 1\,,\qquad \mathtt a,\, \mathtt b \in \mathbb Z\,.
\end{equation}
In the special case where $\pp_{a+1}-\pp_{a-1} =0$, which implies $\xi_a = \xi_{a+1}$, the topology becomes that of ${S}^1 \times {S}^2$. When the orbifold is trivial, namely for $\nn_a = \pm 1$, the horizon reduces to the lens space $L\left(|p_{a+1} - p_{a-1}|,1\right)\simeq S^3/\mathbb Z_{|p_{a+1} - p_{a-1}|}$. If then $\pp_{a+1}-\pp_{a-1}= h_a + h_{a+1} =1$, the horizon has ${S}^3$ topology. 

\paragraph{horizon$_{a-1}$-BUBBLE$_a$-horizon$_{a+1}$.}  Next, we consider a bubbling rod whose endpoints coincide with the south pole of a horizon rod $I_{a-1}$ and the north pole of another horizon rod $I_{a+1}$. In this case the Killing vectors smoothly contracting at the endpoints of the rod are, respectively,
\begin{equation}
    \begin{aligned}
        \xi_a \,&=\, \frac{\beta \nn_{a-1}}{2\pi}\partial_\tau +\left[ 2\left( p_{a-1} - p_{a}\right) + \nn_{a-1}\left( 1 + \frac{\ii\omega_-}{2\pi}\right)\right]\partial_\psi \,,
      \\[1mm]
        \xi_{a+1} \,&=\, \frac{\beta \nn_{a+1}}{2\pi}\partial_\tau +\left[ 2\left( p_{a+1} - p_{a}\right) + \nn_{a+1}\left( 1 + \frac{\ii\omega_-}{2\pi}\right)\right]\partial_\psi \,.
    \end{aligned}
\end{equation}
These are related as
\begin{equation}
   \left(\pp_a - \pp_{a+1}\right) \xi_a = \left(\pp_a - \pp_{a-1}\right) \xi_{a+1} +\left[ \nn_{a-1}\left( \pp_a - \pp_{a+1}\right)-\nn_{a+1}\left( \pp_a - \pp_{a-1}\right) \right]\tilde\lambda_{a+1}\,, 
\end{equation}
where 
\begin{equation}
    \nn_{a-1}\left( \pp_a - \pp_{a+1}\right)-\nn_{a+1}\left( \pp_a - \pp_{a-1}\right)=-\nn_{a+1}\,h_a - \nn_{a-1}\,h_{a+1}\in \mathbb Z\,,
\end{equation}
as it follows from \eqref{eq:ha_horizon_rod}.
According to our previous discussion, this gives the smooth lens space
\begin{equation}
    L\Bigl(|\nn_{a-1}\left( \pp_a - \pp_{a+1}\right)-\nn_{a+1}\left( \pp_a - \pp_{a-1}\right)|,\; 1 + \mathtt a\, \left( \pp_{a+1} - \pp_{a-1}\right)\Bigr)\,,
\end{equation}
with
\begin{equation}
    \mathtt a\left(\pp_a - \pp_{a+1}\right) + \mathtt b \left[\nn_{a-1}\left( \pp_a - \pp_{a+1}\right)-\nn_{a+1}\left( \pp_a - \pp_{a-1}\right)\right] = 1\,,\qquad \mathtt a,\,\mathtt b \in \mathbb Z\,.
\end{equation}

\paragraph{horizon$_{a-1}$-BUBBLE$_a$-bubble$_{a+1}$, or bubble$_{a-1}$-BUBBLE$_a$-horizon$_{a+1}$.}  These bubbling rods connect either a horizon north pole to a bubbling center or a bubbling center to a horizon south pole, respectively. 
We focus on the former case, \textbf{horizon$_{a-1}$-BUBBLE$_a$-bubble$_{a+1}$},  the discussion of the other case being equivalent.
The two $2\pi$-periodic Killing vectors contracting at the endpoints of the rod $I_a$ are 
\begin{equation}
\begin{aligned}
    \xi_a \,&=\, \frac{\beta \nn_{a-1}}{2\pi}\partial_\tau +\left[ 2\left( p_{a-1} - p_{a}\right) + \nn_{a-1}\left( 1 + \frac{\ii\omega_-}{2\pi}\right)\right]\partial_\psi \,,\\[1mm]
    \xi_{a+1} \,&=\, 2\partial_{\psi}\,.
    \end{aligned}
\end{equation}
Using the expression in \eqref{eq:fake_lambda_at_B}, we find the relation
\begin{equation}
   \xi_a \,= \, \left(\pp_{a-1} - \pp_{a}\right) \xi_{a+1} + \nn_{a-1}\tilde\lambda_{a+1}\,,
\end{equation}
giving the lens space
\begin{equation}
    L\Bigl( |\nn_{a-1}|,\,\pp_{a-1}-\pp_{a}\,\Bigr)\,.
\end{equation}
We note, however, that there is a conical singularity at the pole associated to the bubbling center (if $|h_{a+1}| \neq 1$, as it follows from \eqref{eq:metric_near_B}), giving a branched lens space. For $\nn_{a-1} = \pm 1$, the bolt reduces to a branched $S^3$.

\paragraph{bubble$_{a-1}$-BUBBLE$_a$-bubble$_{a+1}$.} Finally, we consider a bubbling rod connecting two bubbling centers. 
The vector contracting at both ends of the rod $I_a$ is given by
\begin{equation}
    \xi_{a} \,=\, \xi_{a+1} \,=\, 2\partial_{\psi}\,,
\end{equation}
corresponding, in the notation of \eqref{eq:vectors_lens_topology}, to the case $\mathtt q_1=\mathtt q_2 = 1$ and $\mathtt p=0$.
From \eqref{eq:metric_near_B} we see that the geometry near the center $a$ exhibits a conical singularity of the form $\mathbb R^2/\mathbb Z_{|h_a|}$.
Similarly, a conical singularity $\mathbb R^2/\mathbb Z_{|h_{a+1}|}$ arises at the other center $a+1$. Then the 
bubbling bolt has the topology of 
\begin{equation}
    S^1\times \Sigma_{[h_a,h_{a+1}]}\,,
\end{equation}
where the circle $S^1$ is generated by $\partial_{\tau_a}$, and $\Sigma_{[h_a,h_{a+1}]}$ denotes a \emph{spindle} geometry. If either $|h_{a}|=1$ or $|h_{a+1}|=1$, then there is no conical singularity and space caps off smoothly at the corresponding center.

\subsection{Global properties of the gauge field}
\label{sec:global_properties_gauge_field}

Since the supergravity gauge field is also turned on, the rod structure as discussed so far is not sufficient to characterize the solution: additional data encoded in certain gauge fluxes through compact submanifolds in the interior of the spacetime should be specified. Indeed, as we showed above, in the Euclidean setup we consider, a five-dimensional space $\mathcal M$ is allowed to have non-trivial compact three-cycles outside the horizons. 
 Therefore, denoting by $\mathcal B_a$ the bolt associated to the rod $I_a$, to any compact bolt ($a = 1,\,2,\,...,\,s-1$) we can associate the electric flux
\begin{equation}\label{eq:fluxes_cycles}
    \mathcal Q^{I_a} \,\sim\, \int_{\mathcal B_a}\left[ \star F - \frac{\ii}{\sqrt{3}}A\wedge F\right]\,.
\end{equation}
We will fix the overall normalization in a convenient way later in this section. The integrand appearing coincides with the Maxwell three-form, which is closed on-shell. In general, when the gauge field cannot be globally defined over ${\cal B}_a$, including the rod endpoints where the bolt caps off, the integral~\eqref{eq:fluxes_cycles} needs to be carefully defined patchwise. 

Any three-dimensional bolt defines a normal two-dimensional non-compact cycle that intersects the bolt once. In our construction this is obtained by foliating the circle generated by $\xi_{I_a}$ over the radial direction $\rho$ ending on the bolt (at $\rho=0$), with metric close to the tip given by~\eqref{eq:normal_space_rod}. In the following, we will denote by $\mathcal N_a \simeq \mathbb R^2$ this normal space. The $\mathcal N_a$ dual to the bolt $\mathcal B_a$ can support a non-trivial gauge flux, defining a potential given by the gauge-invariant expression
\begin{equation}\label{eq:flux_potential}
    \Phi^{I_a} \sim \int_{\mathcal N_a}F\,.
\end{equation}
In general, this integral can be computed by means of the BVAB localization theorem \cite{Atiyah:1984px,BerlineEtAlBook}, and reviewed in sec.~\ref{sec:BVAB}.
To do so, we introduce the equivariantly closed and gauge-invariant polyform 
\begin{equation}
    \Psi_{[F]}^{I_a} = F - \iota_{\xi_{I_a}}A+ \iota_{\xi_{I_a}}A\,\big|_{\rho\rightarrow +\infty}
\end{equation}
such that 
\begin{equation}
    \Phi^{I_a} \sim \int_{\mathcal N_a}\Psi^{I_a}_{[F]}\,.
\end{equation}
By means of the localization theorem we reduce the integral above to the fixed points of $\xi_{I_a}$, namely at the tip of $\mathcal N_a$,
\begin{equation}
    \Phi^{I_a}\, \sim\,  -2\pi\left[\iota_{\xi_{I_a}}A\,\big|_{\rho\rightarrow 0}-\iota_{\xi_{I_a}}A\,\big|_{\rho\rightarrow +\infty}\right] \,\,\sim\,\, 2\sqrt{3}\pi\,\ii\left(\ii \breve A_{I_a}-\ii \breve\omega_{I_a}\right),
\end{equation}
where $\ii \breve A_{I_a}$ denotes the $\phi$-component of the one-form $\ii \breve A$ evaluated along the rod $I_a$, whose explicit expression is
\begin{equation}
    \ii\breve A_{I_a} \,=\, \sum_{b\leq a}\kk_b- \sum_{b>a}\kk_b\, = \,2\sum_{b\leq a}\kk_b\,.
\end{equation}
As we will explain below, if $I_a$ is a horizon rod then the definitions of the electric fluxes \eqref{eq:fluxes_cycles} and the potentials \eqref{eq:flux_potential} reduce to the usual ones of electric charges and electrostatic potentials associated to each horizon. However, also in case $I_a$ is not a horizon rod the above fluxes have been argued to play a role (at least semiclassically) in the first law of thermodynamics \cite{Copsey:2005se, Kunduri:2013vka}. 

For the moment we will specialize the above definitions to the subclass of single-horizon solutions, for which the thermodynamic relations found in~\cite{Kunduri:2013vka} apply. 
The same expressions for fluxes and potentials can also be considered in the case of multi-horizon solutions, though the thermodynamical interpretation in that context is less clear. 

First of all, we should compute the electrostatic potential of the horizon, denoted as $\varphi^{I_a}$, which is related to the boundary condition $\varphi$. 
Analogously to \eqref{eq:def_varphi}, we define the electrostatic potential associated to the horizon rod as
\begin{equation}\label{eq:def_horizon_varphi}
    \varphi^{I_a}\,=\, -\ii\int_{\partial\mathcal N_a}\left(A_a-\sqrt{3}\,\ii\,\diff\tau\right)\,,\qquad I_a \,:\,\text{horizon rod}\,,
\end{equation}
where $A_a$ is a regular gauge field in a patch that contains the horizon and extends up to the boundary. Here, $\partial{\cal N}_a$ is the asymptotic circle generated by the orbits of the rod vector $\xi_{I_a}$, being ${\cal N}_a$ the associated normal space. 
The integral \eqref{eq:def_horizon_varphi} can be mapped, using Stokes' theorem, to a flux integral of the form \eqref{eq:flux_potential}:
\begin{equation}\label{eq:varphi_horizon_rod}
    \varphi^{I_a} + \sqrt{3}\int_{\partial{\cal N}_a}\diff \tau \,=\, -\ii\int_{{\cal N}_a}F\,,
\end{equation}
which can be computed following our equivariant argument discussed above, giving
\begin{equation}
   \varphi^{I_a} + \sqrt{3}\int_{\partial{\cal N}_a}\diff \tau \,=\, 2\sqrt{3}\,\pi\,\left(\ii \breve A_{I_a} - \ii\breve\omega_{I_a}\right) = 4\sqrt{3}\,\pi\Bigg(\sum_{b\leq a} \kk_b + \frac{\nn_a\,\beta}{4\pi}\Bigg)\,.
\end{equation}
By performing the integral equivariantly we then found a relation between $\varphi^{I_a}$, that relates to a boundary integral (on the left-hand side), and a bulk quantity (on the right-hand side) -- the coefficients of the one-forms $\ii\breve A$ and $\ii\breve\omega$ at the horizon rod $I_a$. This relation provides an example of a UV/IR relation (see~\cite{Bobev:2020pjk}) obtained through BVAB localization theorem~\cite{BenettiGenolini:2024xeo}. As a consequence, the electrostatic potential at the horizon is given by
\begin{equation}\label{eq:varphi_single_horizon}
    \varphi^{I_a}\, =\,  4\sqrt{3}\,\pi\sum_{b\leq a} \kk_b\,.
\end{equation}

We now wish to relate $\varphi^{I_a}$ to the boundary condition $\varphi$, defined in \eqref{eq:def_varphi}. The regular gauge field in the patch that contains the horizon rod and extends smoothly up to the asymptotic three-sphere is found by solving the following three conditions
\begin{equation}\label{eq:conditions_for_patch}
    \iota_{\xi_{I_a}}\left( \ii \breve A + \diff\alpha_a \right)\Big|_{I_a} \,=\, 0 \,=\, \iota_{\xi_{I_{0,s}}}\left( \ii\breve A + \diff\alpha_a\right)\Big|_{I_{0,s}}\,,\qquad I_a \,:\,\text{horizon rod}\,,
\end{equation}
for a closed one-form
\begin{equation}
   \diff \alpha_a \,=\, \alpha_a^{(\tau)} \diff \tau + \alpha_a^{(\phi)}\diff \phi + \alpha_a^{(\psi)} \diff \psi\,.
\end{equation}
Eqs. \eqref{eq:conditions_for_patch} are solved by taking 
\begin{equation}\label{eq:alpha_a_horizon}
    \alpha_a^{(\phi)}\,=\,0\,=\,\alpha_a^{(\psi)} \,,\qquad \alpha_a^{(\tau)} \,=\, -\frac{4\pi}{n_a\beta}\sum_{b\leq a}\kk_b\,.
\end{equation}
Then, the regular gauge field asymptotes to
\begin{equation}\label{eq:A_infty}
     A_a \,\rightarrow\,  A_a^{(\infty)} \,=\, \ii \sqrt{3}\left(1 + \frac{4\pi}{n_a\beta}\sum_{b\leq a}\kk_b\right)\diff \tau \,\equiv\, \ii \Phi^{I_a}\diff \tau\,.
\end{equation}
Here, we introduced the horizon potential $\Phi^{I_a}$, measuring the boundary holonomy of the gauge field along the orbits generated by the rod vector $\xi_{I_a}$.\footnote{Note that this quantity could also be computed by an integral of the form \eqref{eq:flux_potential}: 
$$
    \ii\Phi^{I_a} = \frac{1}{n_a\beta}\int_{{\cal N}_a}F\,,\qquad I_a\,:\,\text{horizon rod}\,.
$$
}

Finally, to compare \eqref{eq:varphi_single_horizon} to \eqref{eq:def_varphi}, recall that the rod vector generating the horizon can be expressed as \eqref{eq:combinationU1s}, showing that its orbits wind around the thermal circle generated by the first vector of \eqref{eq:U(1)isometries} $\nn_a$ times. Then, using the expression for the asymptotic gauge field \eqref{eq:A_infty}, we conclude that $\varphi^{I_a}$ is related to the boundary condition $\varphi$ as
\begin{equation}\label{eq:varphi_no_shift}
    \varphi \,=\, \frac{\varphi^{I_a}}{\nn_a} \,=\, \frac{4\sqrt{3}\pi}{\nn_a}\sum_{b\leq a}\kk_b\,.
\end{equation}
In order to compute the flux $\mathcal Q^{I_a}$ \eqref{eq:fluxes_cycles} across the bolt ${\cal B}_a$ when $I_a$ is a horizon rod, we recall that one can always find a spatial hyper-surface $\Sigma_\tau$ (in homological terms, a four-dimensional chain) whose boundary is the disconnected union of the horizon bolt (at $\rho\rightarrow 0$) and an asymptotic three-sphere (at $\rho\rightarrow+\infty$), $\partial\Sigma_\tau = \mathcal B_a\, \cup\, S^3_\infty$ (see e.g.~\cite{Hollands:2010qy}). Therefore, by interpreting the flux integral 
\begin{equation}\label{eq:charges_flux_integral}
    \mathcal Q^{I_a} = -\frac{\ii}{16\pi}\int_{\mathcal B_a}\left[\star F - \frac{\ii}{\sqrt{3}}A_a\wedge F\right]\,, \qquad I_a \,:\,{\rm horizon}
\end{equation}
as a Page-charge, this turns out to be independent of which specific $\rho={\rm const}$ slice of $\Sigma_\tau$ we choose as integration manifold. In particular, the flux can be evaluated across the asymptotic three-sphere, showing the equivalence with the standard definition of the electric charge
\begin{equation}
    Q =  -\frac{\ii}{16\pi}\int_{S^3_\infty}\star F \,,
\end{equation}
using that $A\wedge F$ is suppressed asymptotically. 
\medskip

We next focus on non-horizon rods. In this case we fix the normalization of the integrals \eqref{eq:flux_potential} and \eqref{eq:fluxes_cycles} so as to match the conventions of~\cite{Kunduri:2013vka}:\footnote{Indeed, ${\cal Q}^{I_a}$ is related to the \emph{magnetic flux} introduced in \cite{Kunduri:2013vka}. In the Euclidean finite-$\beta$ setup, the bolt ${\cal B}_a$ is a three-cycle, while in the Lorentzian case it reduces to a two-cycle formed by the spatial sections of ${\cal B}_a$, times the non-compact time direction. As mentioned in section~\ref{sec:rodstrt}, this two-cycle can be a disc, a tube, or a spindle. The magnetic flux is defined in \cite{Kunduri:2013vka} by integrating $\iota_t (\star F - \frac{\ii}{\sqrt{3}}A_a \wedge F)$ over this two-cycle, and corresponds to the Lorentzian counterpart of our integral. In other words, our ${\cal Q}^{I_a}$ provides an uplifted Euclidean version of these fluxes, from which it differs by a factor of $\beta$ due to the integration over $\tau$.
We also note that the integrals~\eqref{eq:potentials_genaral_def} are different from those considered in the Lorentzian analysis of~\cite{Armas:2009dd,Armas:2014gga}.
} 
\begin{equation}\label{eq:potentials_genaral_def}
    \Phi^{I_a} = -\frac{1}{2\pi}\int_{\mathcal N_a}F\,,\qquad \mathcal Q^{I_a} = - \frac{1}{8}\int_{\mathcal B_a}\left[ \star F - \frac{\ii}{\sqrt{3}}A_a \wedge F\right]\,,\qquad I_a \,:\,{\rm bubble}\,.
\end{equation}
The \emph{bubble potential} $\Phi^{I_a}$, whose definition, except for the chosen normalization, is analogous to the one for the electrostatic potential, can be computed using the equivariant localization theorem as above, obtaining
\begin{equation}\label{eq:cycle_potential}
    \Phi^{I_a} = -2\sqrt{3}\,\ii\,\sum_{b\leq a}\kk_b\,.
\end{equation}
Note that the bubble potentials associated with the semi-infinite rods vanish,  $\Phi^{I_0}= \Phi^{I_s}=0$.

Finally, Eq.~\eqref{eq:cycle_potential} may be combined with \eqref{eq:varphi_single_horizon} by recalling that if $I_a$ is a horizon rod, then $I_{a-1}$ is not. Doing so, one finds a relation between an electrostatic potential $\varphi^{I_a}$ and the bubble potential associated to the adjacent bubbling rod, $\Phi^{I_{a-1}}$, 
\begin{equation}
\label{eq:ka}
    \varphi^{I_a} \,=\, 4\sqrt{3}\pi\,\kk_{a}+ 2\pi\ii\,\Phi^{I_{a-1}}\,,\qquad\quad I_a\,:\,\text{horizon},\ \quad I_{a-1}\,:\,\text{bubble}\,.
\end{equation}

\subsection{Thermodynamics of single-horizon solutions}
\label{sec:thermo}

For solutions with a single horizon, a first law of black hole mechanics which takes into account the presence of topologically non-trivial cycles  was presented in \cite{Kunduri:2013vka}. 
In this section, we consider the supersymmetric versions of these thermodynamical relations, and at the same time we extend them to account for the orbifold solutions constructed above (with $n_a\neq \pm1$). In the remaining part of the section we will denote the horizon rod by $I_a= {\cal H}$, and we will also use the label `${\cal H}$' to characterize quantities associated with the horizon.

The Bekenstein-Hawking entropy associated to the horizon rod, defined as $1/4$ the area of the corresponding bolt, reads 
\begin{equation}
    \mathcal S \,=\, \pi\,\beta\,\delta^{\cal H}\,,
\end{equation}
where $\delta^{\cal H}$ is the length of the horizon rod. Translated into our conventions, the first law of~\cite{Kunduri:2013vka} reads
\begin{equation}
\beta\, \diff E\,=\, \diff {\cal S} + \beta\,\Phi^{\cal H}\, \diff Q + \frac{\omega_+^{\cal H}}{n_{\cal H}}\, \diff J_+ + \frac{\omega_{-}^{\cal H}}{n_{\cal H}}\, \diff J_- + \, \sum_{I_a\neq {\cal H}} {\cal Q}^{I_a}\, \diff \Phi^{I_a}\, .
\end{equation}
Here, the charges entering the formula are the asymptotic ones given in \eqref{eq:mass+ang_momenta} and \eqref{eq:electric_charge}, while the horizon angular velocities  can be read from \eqref{eq:ang_vel_hor}. The electrostatic potential $\Phi^{\cal H}$ has been defined in \eqref{eq:A_infty}. The unusual ingredient arising because of the non-trivial topology of the solution is the last sum over the non-horizon rods, $I_a \neq {\cal H}$. Imposing supersymmetry, 
\begin{equation}
E \,=\, \sqrt{3} \,Q\, ,
\end{equation}
gives us a {\it supersymmetric version of the first law},
\begin{equation}
\label{eq:susyfirstlaw}
\diff {\cal S}+\frac{1}{\nn_{\cal H}}\left(\varphi^{\cal H}\, \diff Q+ \omega^{\cal H}_+\, \diff J_{+}+\omega^{\cal H}_-\, \diff J_{-}\right)+\sum_{I_a\neq {\cal H}} {\cal Q}^{I_a}\, \diff \Phi^{I_a}\,=\,0\, .
\end{equation}
The analysis of~\cite{Kunduri:2013vka} also provides a Smarr formula, 
\begin{equation}
    \beta E \,=\, \frac{3}{2}\mathcal S + \frac{3}{2\nn_{\cal H}}\left(\omega_+^{\cal H} \,J_+ + \omega_-^{\cal H} \,J_-\right) + \beta\,\Phi^{\cal H}\,Q + \frac{1}{2}\sum_{I_a\neq {\cal H}} {\cal Q}^{I_a}\,\Phi^{I_a}\,,
\end{equation}
whose supersymmetric version reads
\begin{equation}\label{eq:susy_smarr}
    \mathcal S + \frac{1}{\nn_{\cal H}}\left(\omega_+^{\cal H} \,J_+ + \omega_-^{\cal H}\,J_- + \frac{2}{3}\varphi^{\cal H} \,Q \right)+ \frac{1}{3}\sum_{I_a\neq {\cal H}}{\cal Q}^{I_a} \,\Phi^{I_a} \, =\, 0\,.
\end{equation}
This relation can be combined with the quantum statistical relation satisfied by the orbifold saddle, 
\begin{equation}
\label{eq:QSR_single_horizon}
\begin{aligned}
    I \,&=\, -\mathcal S - \frac{1}{n_{\cal H}}\left(\omega_+^{\cal H} \,J_+ + \omega_-^{\cal H} \,J_- + \varphi^{\cal H}\,Q\right)
    \\[1mm]
 \,&=\, -{\cal S} - \frac{2\pi\ii}{\nn_{\cal H}} \left(J_+ +\left(1-\nn_{\cal H}- 2\pp_{\cal H}\right)J_-\right) - \omega_- J_- - \varphi Q\,,
    \end{aligned}
\end{equation}
to derive a simplified expression for the on-shell action that depends only on gauge field data, namely the conserved electric charge and various fluxes:
\begin{equation}\label{eq:on_shell_action_single_horizon}
    I \,=\, \frac{1}{3}\,\bigg[-\frac{\varphi^{\cal H}}{\nn_{\cal H}} \,Q + \sum_{I_a\neq{\cal H}} {\cal Q}^{I_a}\,\Phi^{I_a} \bigg]\,.
\end{equation}
We will verify the validity of this formula in the explicit examples discussed in the following sections. 

The quantum statistical relation \eqref{eq:QSR_single_horizon} and the first law \eqref{eq:susyfirstlaw} together imply differential relations for the on-shell action. This is obtained by allowing small enough variations of the continuous boundary conditions $\omega_-$ and $\varphi$ while keeping fixed the discrete integers $(\nn_a,\,\pp_a)$ introduced above. According to \eqref{eq:ang_vel_hor} and \eqref{eq:varphi_no_shift}, this corresponds to the variations
\begin{equation}
    \omega_-^{\cal H}\rightarrow \omega_-^{\cal H} + \nn_{\cal H}\,\diff \omega_-\,,\qquad \varphi^{\cal H}\rightarrow \varphi^{\cal H} + \nn_{\cal H}\diff \varphi\,,\qquad \Phi^{I_a} \rightarrow \Phi^{I_a} + \diff \Phi^{I_a}\,.
\end{equation}
Assuming the first law holds, we then find
\begin{equation}\label{eq:semicl_variation_action}
\diff I =  -\,J_-\,\diff\omega_- - Q\,\diff 
\varphi+ \sum_{I_a \neq {\cal H}}{\cal Q}^{I_a}\,\diff\Phi^{I_a}\,.
\end{equation}
Therefore, the charges $Q$, $J_-$ and the electric fluxes ${\cal Q}^{I_a}$ can also be determined from the on-shell action by taking the partial derivatives with respect to the conjugate variables
\begin{equation}
    J_- \,=\, -\frac{\partial I}{\partial\omega_-}\,,\qquad\quad Q \,=\, -\frac{\partial I}{\partial\varphi}\,,\qquad\quad {\cal Q}^{I_a} \,=\, \frac{\partial I}{\partial\Phi^{I_a}}\,,\quad I_a\,\neq\,{\cal H}\,.
\end{equation}

\subsection{Quantization conditions for a compact gauge group}
\label{sec:compact_gauge_conditions}

In this section, we derive quantization conditions for the bubble potentials \eqref{eq:cycle_potential} and allowed shifts for the electrostatic potentials (generalizing \eqref{eq:varphi_no_shift}), arising in case of compact abelian gauge group. This leads us to introduce a new integer, $q_a$,  for each compact rod $I_a$.

In the gravitational path integral we sum over all field configurations ${\cal X}$, either bosonic or fermionic, that satisfy the following periodic identifications when going once around the three independent U(1)'s generated by the Killing vectors in \eqref{eq:U(1)isometries}:
\begin{equation}\label{eq:field_id}
\begin{aligned}
&{\cal X}\left(\tau+\beta,\, \phi+2\pi,\,\psi+\ii \omega_-\right)\ \sim\ (-1)^{\sf F}{\cal X}\left(\tau, \phi, \psi\right)\,, \\[1mm]
&{\cal X}\left(\tau, \phi+ 2\pi, \psi-2\pi\right)\ \sim\ (-1)^{\sf F}{\cal X}\left(\tau, \phi, \psi\right)\,, \\[1mm]
&{\cal X}\left(\tau, \phi, \psi+4\pi\right)\ \sim\ {\cal X}\left(\tau, \phi, \psi\right)\,.
\end{aligned}
\end{equation}
These identifications are intended to be valid when working in a gauge for the vector field $A$ in \eqref{eq:susy_conf} specified by the choice $\alpha\,=\,-\frac{\varphi}{\sqrt{3}\beta}\,\tau$, in which the gauge field $A$ asymptotes to 
\begin{equation}
 A \to \ii \left(\frac{\varphi}{\beta}+\sqrt{3}\right) \diff \tau\, ,\hspace{1cm} \rho\to \infty\,,
\end{equation}
where $\varphi$ is the boundary condition defined in \eqref{eq:def_varphi}. 

The identifications \eqref{eq:field_id} must be compatible with the requirement that the geometry supports a smooth spin structure. In our setup, this means that any field must pick a $(-1)^{\sf F}$ phase when transported around the orbits generated by any rod vector $\xi_{I_a}$, as long as one works in a gauge that is regular in a patch containing the rod $I_a$, where $\xi_{I_a}$ contracts. More concretely, we must impose
\begin{equation}\label{eq:field_id_hor}
{\cal X}_a\left(\tau+n_a \beta , \,\phi+2\pi , \,\psi+2\pi \left[2p_a-1+n_a\left(1+\frac{\ii\omega_-}{2\pi}\right)\right]\right)\ \sim\ (-1)^{\sf F} {\cal X}_a\left(\tau,\, \phi,\,\psi\right)\, ,
\end{equation}
where the subscript ${\cal X}_a$ means that we are working in the aforementioned regular gauge. The gauge transformation relating $A$ and $A_a$ is given by
\begin{equation}
 A_a\,=\,A-\ii\sqrt{3}\left(\diff\alpha_a+\frac{\varphi}{\sqrt{3}\beta}\diff\tau\right)\,, 
\end{equation}
where we recall that $\alpha_a$ is a function of the form $\alpha_a = \alpha_a^{(\tau)} \tau + \alpha_a^{(\phi)}\phi + \alpha_a^{(\psi)} \psi$. This implies that
\begin{equation}\label{eq:field_gauge_map}
 {\cal X}_a \left( \tau,\,\phi,\,\psi\right) \,=\, {\rm exp} \left[\sqrt{3}\,{e_{\cal X}}\left(\alpha_a+\frac{\varphi}{\sqrt{3}\beta}\tau\right)\right]{\cal X}\left( \tau,\phi,\psi\right)\, ,
\end{equation}
where $e_{\cal X}$ is the charge of the field $\cal X$ under the U(1) gauge symmetry. Compatibility between \eqref{eq:field_id_hor} and the 
identifications \eqref{eq:field_id} is equivalent to demanding the transition function in \eqref{eq:field_gauge_map} to be single-valued when going once around the U(1) generated by $\xi_{I_a}$. 
  We note that in ungauged supergravity these conditions are trivially satisfied since all supergravity fields are uncharged under the U(1) gauge symmetry. However, if we consider the UV completion of the theory — e.g., via an embedding in string theory — then a microscopic description must exist in which microstates charged under the gauge symmetry are present to account for the entropy. Typically, in these examples the gauge group turns out to be a compact U(1), then the states have quantized charges of the type $e_{\cal X} \in \mathbb Z\,e$, being $e$ the fundamental unit charge in the theory.  Then, under these assumptions, the requirements above provide a quantization condition for each rod $I_a$, as we will now discuss considering separately bubbling and horizon rods. 

The transition function in \eqref{eq:field_gauge_map} is single-valued when going once around the U(1) generated by $\xi_{I_a}$ if the following condition is obeyed
\begin{equation}\label{eq:trans_funct_single_valued}
n_a\varphi+2\pi\sqrt{3}\,\iota_{\xi_{I_a}}\diff \alpha_a\,=\, 2\pi \ii \,\frac{q_a}{e}\,, \hspace{1cm} q_a\in {\mathbb Z}\, .
\end{equation}
For bubbling rods, $n_a\,=\,0$, we can use \eqref{eq:cycle_potential} to find
\begin{equation}
   \iota_{\xi_{I_a}}\diff\alpha_a\,=\, -2\sum_{b\le a}\kk_b\,=\, \frac{\Phi^{I_a}}{\sqrt{3}\ii} \, .
\end{equation}
Therefore, eq.~\eqref{eq:trans_funct_single_valued} applied to a bubbling rod yields quantization conditions:
\begin{equation}
\label{eq:quantized_bubble_pot}
  e\, \Phi^{I_a}\,=\, q_a \,, \hspace{1cm} I_a:\, \rm{bubble} \, .
\end{equation}
On the other hand, for a horizon rod, $n_a\neq 0$, we use \eqref{eq:varphi_single_horizon} and \eqref{eq:conditions_for_patch} to derive
\begin{equation}
2\pi \sqrt{3}\,\iota_{\xi_{I_a}}\diff \alpha_a\,=\, - \varphi^{I_a}\, .
\end{equation}
Using this in \eqref{eq:trans_funct_single_valued}, we conclude that the electrostatic potentials assigned to each horizon $\varphi^{I_a}$ are related to the boundary condition $\varphi$ via
\begin{equation}
\label{eq:varphi_with_shift}
 \varphi^{I_a}\,=\,n_a\varphi- 2\pi\ii \,\frac{q_a}{e}\, , \hspace{1cm} I_a:\,\rm {horizon}\, .
\end{equation}
For a horizon rod, we can further use \eqref{eq:ka} to find 
\begin{equation}\label{eq:k_from_varphi_shift}
\kk_a\,=\,\frac{-\ii}{2\sqrt{3}e}\left(q_a+q_{a-1}\right) + \frac{n_a\varphi}{4\pi \sqrt{3}}\,, 
\end{equation}
and 
\begin{equation}
\kk_a+\kk_{a+1}\,=\, \frac{q_{a-1}-q_{a+1}}{2\sqrt{3}\ii e} \,,
\end{equation}
analogously to the expressions we found in \eqref{eq:ha_horizon_rod} and \eqref{eq:sum_h's} for the coefficients of the harmonic functions $H$ and $\omega_-$.

We have thus extended the map between the horizon electrostatic potential \eqref{eq:varphi_single_horizon} and the boundary condition $\varphi$, previously given by \eqref{eq:varphi_no_shift}, by including integer shifts $q_a$. We now examine how these quantized parameters $q_a$ modify the quantum statistical relation \eqref{eq:QSR_single_horizon} and its differential form \eqref{eq:semicl_variation_action}. Specializing to the case where there is only a single horizon, the quantum statistical relation takes the form
\begin{equation}
\begin{aligned}
I \,&=\, - \mathcal S - 2\pi \ii\, \tilde J_+ - \omega_-\, J_- - \varphi\, Q\,,\\[1mm]
\tilde J_+ \,&= \,\frac{1}{n_{\cal H}}\Bigl(J_+ + (1 - n_{\cal H} - 2 p_{\cal H}) J_- - \frac{q_{\cal H}}{e}\, Q\Bigr)\,,
\end{aligned}
\end{equation}
where we use the notation of section~\ref{sec:thermo}. Combining this with the first law \eqref{eq:susyfirstlaw} yields its differential version. A key distinction now is that the discrete parameters $n_a$, $p_a$, as well as $q_a/e$ are all held fixed. In particular, the bubble potentials, $\Phi^{I_a}$, are not allowed to vary anymore ($\diff\Phi^{I_a}=0$) due to \eqref{eq:quantized_bubble_pot}, in contrast with the derivation leading to \eqref{eq:semicl_variation_action}. As a result, the on-shell action $I$ satisfies the simple differential relation
\begin{equation}
\diff I = -\,J_-\,\diff\omega_- \,- Q\,\diff\varphi\,,
\end{equation}
which shows that, once all quantization conditions are imposed, the only continuous variables of $I$ are just the chemical potentials $\omega_-$ and $\varphi$.

\section{Revisiting two-center solutions}\label{sec:2centersol}

In this section we revisit the two-center supersymmetric non-extremal black hole solution presented in chap.~\ref{chap:Black_hole}. This comprises just one compact rod, with rod vector the generator of the Euclidean horizon.
Our scope is twofold: on the one hand we want to motivate further the need to consider a complexified version of the solution, and on the other hand we wish to illustrate the discrete shifts and orbifolds introduced in section~\ref{sec:rodstr} in the simplest possible setup.  

\subsection{The complex saddle}\label{sec:complex_saddle}

 We start by considering the basic rod vector \eqref{eq:combinationU1s} with $n=1,p=0$, that is  
\be\label{rod_vector_two_center}
\xi_{\cal H}\,=\, \frac{\beta}{2\pi}\partial_\tau +\partial_\phi + \frac{\ii \omega_-}{2\pi} \partial_\psi \,,
\ee
where we use $\cal{H}$ rather than $I_1$ to label the horizon rod.
Then the horizon temperature and angular velocities coincide with the asymptotic boundary conditions, specified by demanding that the first basis vector in~\eqref{eq:U(1)isometries} has closed orbits. In chap.~\ref{chap:Black_hole},  reality conditions on the parameters were chosen so that the metric is real and positive definite. 
   This makes it straightforward to study regularity of the solution and also gives a real Euclidean on-shell action.
 It was noted in there that this finite-$\beta$ solution can interpolate between the extremal black hole of BMPV (for which $\beta = \infty$) and a horizonless topological soliton (which is reached via a $\beta\to 0$ limit), however in order to see this one needs to implement suitable analytic continuations of the parameters appearing in the real, Euclidean solution (in addition to Wick-rotating the Euclidean time back to Lorentzian signature). Therefore, in order to be able to reach the Lorentzian solutions by a continuous limit, one must give up reality of the metric and the gauge field.
Here we  review this argument and further motivate it. 
 While it is not clear in general what conditions should a complex metric satisfy, the discussion in section~\ref{sec:rodstr} shows that  cancellation of Dirac-Misner strings as well as smoothness along the collapsing orbits of the rod vector  can be achieved even in the complexified setup.

\paragraph{Brief review of the black hole saddle.}The two-center solution is given by choosing the harmonic functions \eqref{eq:harmonicfuncblackholesaddle},
with independent parameters $h_N, \kk, \delta$. We first consider the case where the parameters are all real (with $\delta>0$). Then the functions $H,L$ are real, while $K,M$ are purely imaginary. The metric in~\eqref{eq:susy_conf} is real and positive-definite provided 
\begin{equation}
f^{-1}H \,\equiv\, K^2 + HL>0\,.
\end{equation}
For our two-center solution,
 this condition reduces to
\begin{equation}
\frac{\kk^2}{h_N \,h_S} + h_N \,r_S + h_S \,r_N>0\,,
\end{equation}
which must be satisfied for all $r_{N,S}>0$. 
This requirement imposes that both coefficients of the harmonic function $H$ be positive, which is equivalent to
\begin{equation}
0<h_N <1\,.
\end{equation}


The chemical potentials fixing the boundary conditions are related to the parameters as 
\be\label{eq:omega_2centers}
\omega_+ = 2\pi \ii\,,\qquad\quad \omega_-  = 2\pi \ii(h_S-h_N)\,,\qquad\quad
\varphi  
\,=\,4\sqrt{3}\pi \kk\,,
\ee
\be\label{eq:beta_2centers}
\beta \,=\, 4\pi \ii \ww_N \,=\, -4\pi \ii \ww_S \,=\, -2\pi  \kk \left( 3 + \frac{\kk^2}{h_N^2h_S^2\delta}  \right)\,.
\ee
The conserved charges are:
\be\label{eq:charges_2centers}
J_+ = - 3\pi \ii \kk \delta  \,,\qquad\quad J_- = \pi \ii \kk^3\, \frac{h_N-h_S}{h_N^2h_S^2} \,,\qquad\quad  Q = \frac{\sqrt3\pi\,\kk^2}{h_Nh_S} \,.
\ee
The entropy, defined as 1/4 the area of the bolt of the Killing vector \eqref{rod_vector_two_center}, is given by
\be\label{eq:entropy_2centers}
\mathcal{S} \,=\, \pi \beta\delta  \,=\, 4\sqrt{\pi} \sqrt{\frac{Q^3}{3\sqrt3}-\frac{\pi}{4} J_-^2} - 2\pi \ii J_+\,.
\ee
We see there is a direct correspondence between the free coefficients of the harmonic functions $(h_N,\kk)$, the chemical potentials $(\omega_-,\varphi)$, and the charges $(J_-, Q)$ which play a role in the supersymmetric index. 

However the reality conditions on the parameters 
 assumed above are too limiting. Indeed, they give a real on-shell action $I$,  real $\varphi$ and purely imaginary angular velocities $\omega_\pm$, 
and, as emphasized in~\ref{sec:exampleaction5d}, it is not clear if a purely imaginary value for both angular velocities $\omega_{1,2}$ is acceptable, as the fugacity $\rme^{\omega_2}$ appearing in the index trace~\eqref{eq:microscopicindex} would be just a  phase and the sum over microstates may  not converge. Finally, the angular momenta $J_\pm$ are  purely imaginary in the solution, which then cannot be directly connected to (rotating) Lorentzian solutions. So at this stage we would not  be able to answer the question of whether the extremal solutions can arise as suitable limits of the index saddles. 
 
Therefore, we are led to consider complexifications of the two-center solution. In particular,  the parameters $h_N,\, h_S=1-h_N,\,\kk $ take complex values. 
In this way, $\omega_-$, $\varphi$ and $\beta$ are generically complex, as well as the other thermodynamic quantities, namely, the action, the charges and the entropy. 
 This has the advantage to  straightforwardly allow for the two different limits giving rise to well-behaved Lorentzian solutions reviewed above. The complex solutions have the same asymptotic behavior as \eqref{eq:asymptotics}, however the coordinates $\left( \tau,\,\phi,\,\psi\right)$ should be taken complex now, since the identifications \eqref{eq:coord_identif}  involve complex periodicities. 
As argued in section~\ref{sec:rodstr}, one can nevertheless introduce real, untwisted angular coordinates, cf.~\eqref{eq:rod_adapt_coords}. In the present case these are given by 
\begin{equation}
\tau_{a}\,=\, \frac{2\pi}{\beta}\,\tau - \phi\, , \qquad\quad \psi_a\,=\, \phi+\psi +  \frac{\ii}{\beta}(\omega_+-\omega_-)\,\tau\,, \qquad\quad \phi_a\,=\, \phi\, ,
\end{equation}
and can be used in order to analyze regularity of the metric.

\subsection{Discrete families of saddles from shifts and orbifolds}

We continue revisiting the two-center solutions by elaborating on the infinite family of shifts and orbifolds labelled by the integers $(n,p,q)$ introduced in section~\ref{sec:rodstr}. These $(n,p,q)$ solutions satisfy the very same boundary conditions, hence they potentially contribute to the same gravitational index. 
They are analogous to the shifted and orbifolded solutions first discussed in~\cite{Aharony:2021zkr} in the context of asymptotically AdS$_5\times S^5$ black holes solutions to type IIB supergravity, however they have not been studied in the asymptotically flat  context so far. Also in our case the Killing spinor is not charged under the supergravity gauge field, implying that  the orbifold acts in the five extended dimensions and not in the internal space of any higher-dimensional uplift. This simplifies the study of the topology.   

Recall that the two integers $(n,p)$ specify how the rod vector, corresponding to the horizon generator, is related to the basis of ${\rm U}(1)^3$ isometries. This is illustrated in Eq.~\eqref{eq:combinationU1s}.
 The rod vector~\eqref{eq:velocities_in_rod_vec} reads 
\be
\xi_{\cal H} \,=\, \frac{1}{2\pi}\left(\beta^{\cal H}\,\partial_\tau - \ii\omega_+^{\cal H}\,\partial_\phi + \ii\omega_-^{\cal H}\,\partial_\psi\right)\,,
\ee
where the horizon inverse temperature and angular velocities are given by 
\begin{equation}\label{eq:rel_H_nonH}
 \beta^{\cal H} \,=\, \nn\,\beta\,,\qquad \qquad   \omega_+^{\cal H} \,=\, 2\pi \ii\,,\qquad \qquad\omega_-^{\cal H} \,=\,  \nn\,\omega_- + 2\pi \ii \left(1- \nn -2\pp\right) \,. 
\end{equation}
Also, the horizon electrostatic potential reads
\begin{equation}\label{eq:rel_H_nonH_varphi}
 \varphi^{\cal H}\,=\,n\,\varphi-2\pi\ii q\, ,
\end{equation}
where the shift by the additional integer $q$ should be introduced when the gauge group is ${\rm U}(1)$.\footnote{Compared to the previous section we are setting $e=1$.}

We will assume that both $p$ and $p-1$  are coprime to  $n$.
 According to the analysis in section~\ref{sec:rodstr} (cf.~\eqref{eq:hor_topology_1},  \eqref{eq:hor_topology_2}), the horizon has lens space topology  
\begin{equation}
   L\left(|\nn|\,, \, 1+ \mathtt a \right) \, \simeq \, S^3/{\mathbb Z}_{|\nn|}\,,
\end{equation}
where $\mathtt a$ is an integer defined through the equation
\begin{equation}
    \mathtt a \left(p - 1\right) -   \mathtt b \,\nn \, =\,1\,,\qquad \mathtt a,\,\mathtt b \in \mathbb Z\,.
\end{equation}

The local form of the solution is exactly the same as in subsection~\ref{sec:complex_saddle}, 
however the values taken by the parameters $h_N, h_S$, $\kk$, $\delta$ in terms of the (fixed) boundary conditions now depends on $\nn,\pp, q$. Indeed, the expressions in \eqref{eq:omega_2centers}, \eqref{eq:beta_2centers} now provide the horizon inverse temperature $\beta^{\cal H}$, angular velocities  $\omega_\pm^{\cal H}$, and electrostatic potential $\varphi^{\cal H}$, which are related to the chemical potentials $\beta$, $\omega_\pm$, $\varphi$ as in~\eqref{eq:rel_H_nonH}, \eqref{eq:rel_H_nonH_varphi}. 
Solving for $h_N, h_S$, $\kk$, we obtain the expressions 
\be\label{eq:hNhSk_npq}
 h_N \,=\, \nn\,\frac{\omega_2}{2\pi\ii}+ \pp\,,\qquad\quad   h_{S} \,=\,  \nn\,\frac{\omega_1}{2\pi\ii}+ 1-\nn-\pp\,,\qquad\quad 
 \kk\,=\, \frac{\ii}{2\sqrt 3}\left( \nn\,\frac{\varphi}{2\pi \ii} -  q\right)\,,
\ee
which are the specialization of~\eqref{eq:ha_horizon_rod} and \eqref{eq:k_from_varphi_shift} to the present case.\footnote{Note that in the regime where the metric is real and thus both $\omega_1$, $\omega_2$ are purely imaginary, the condition $0<h_N<1$ ensuring a positive-definite metric can be satisfied by fixing the integer $\pp$ so that $0< \nn\,\frac{\omega_2}{2\pi\ii}+ \pp<1$. This fails if $\frac{\omega_2}{2\pi\ii}\in \mathbb Z$.
}
The remaining parameter, $\delta$, is given by 
\be
\delta \,=\, -\frac{2\pi \kk^3}{h_N^2h_S^2\left(n\beta + 6\pi\kk \right)}\,,
\ee
where one should use~\eqref{eq:hNhSk_npq} in order obtain an expression in terms of the chemical potentials and the integers $(\nn,\pp,q)$.

The saddle-point contribution of the $(\nn,\pp,q)$ solutions to the gravitational index is given by the Euclidean on-shell action. This reads
\be
\begin{aligned}
I_{n,p,q}\left(\omega,\varphi\right)\,&=\,  \frac{\pi}{12\sqrt3}\,\cdot\,\frac{1}{n}\,\cdot\, \frac{(\varphi^{\cal H})^3}{\omega_1^{\cal H} \omega_2^{\cal H}} \\[1mm]
\,&=\, \frac{\pi}{12\sqrt3}\, \frac{\left(n\,\varphi-2\pi\ii q\right)^3}{\nn
\left[\nn\,\omega_1+2\pi\ii \,(1-\nn-\pp)\right]\left[\nn\,\omega_2+2\pi\ii\, \pp\right]}\,,
\end{aligned}
\ee
the overall factor of $\frac{1}{n}$ being due to the action of the orbifold.

The expressions \eqref{eq:charges_2centers} for the angular momenta and the electric charge in terms of the solution parameters remain unchanged, however one should again recall that the map between the parameters and the chemical potentials $\beta,\omega_\pm,\varphi$ depends on $(n,p,q)$.
 The entropy, defined as 1/4 the area of the bolt of the horizon generator, is given by 
\be
\mathcal{S}_{n,p,q} \,=\, \frac{1}{n}\,\pi\,\beta^{\cal H}\,\delta  \,=\, \pi  \beta\delta \,.
\ee
In terms of the charges of the $(n,p,q)$-solution, the same expression reads
\be
\mathcal{S}_{n,p,q} \,=\, \frac{1}{n}\left(4\sqrt{\pi}\, \sqrt{\frac{Q^3}{3\sqrt3}-\frac{\pi}{4} J_-^2} - 2\pi \ii J_+ \right)\,.
\ee
The quantum statistical relation \eqref{eq:QSR_single_horizon} is then satisfied.

It would be interesting to compute  non-perturbative corrections to the action of these $(n,p,q)$ solutions by evaluating the contribution of wrapped branes in a concrete scenario where the solution is uplifted to string or M-theory, as done in~\cite{Aharony:2021zkr} in the asymptotically AdS$_5\times S^5$ Type IIB setup. These corrections may help determine which of the solutions are stable against brane nucleation  and thus be regarded as genuine saddles of the gravitational path integral.


\section{Three-center solutions, black ring and black lens}\label{sec:3centersol}

This section is devoted to the study of solutions with three centers, that is two compact rods. From the general analysis of section~\ref{sec:rodstr}, we know that if one is a horizon rod then the other must be a bubbling rod, since two horizon rods cannot be next to each other.  Hence, the configuration we are going to study is unique up to orientation (the presence of the additional bubbling center breaks the equatorial ${\mathbb Z}_2$ symmetry of the two-center solutions). 

In section~\ref{sec:2centersol} the relevance of regarding the non-extremal two-center solution as a complex solution was emphasized, although there is a choice of the parameters such that the metric is real and positive definite. 
 We find this is even more compelling in the three-center case, since it is not even possible to define an Euclidean section such that the metric is real and positive.

The section is organized as follows. First, we present the solution and provide the relations between its parameters and the grand-canonical variables, specializing the general expressions obtained in section~\ref{sec:rodstr} to the case with three centers. Then, we analyze in detail the thermodynamic properties of the solution; in particular, we compute its on-shell action. 
 Finally, we discuss how the finite-$\beta$ solutions interpolate between extremal black holes with non-spherical horizon topology, such as black rings \cite{Elvang:2004rt} and lenses \cite{Kunduri:2014kja, Tomizawa:2016kjh}, and horizonless topological solitons \cite{Bena:2005va, Berglund:2005vb, Bena:2007kg}, which arise in the $\beta\to0$ limit.


\subsection{Three-center solutions}

Let us begin by specifying the general class of solutions described in section~\ref{sec:gen_ssol} to the case of three centers. We take the first center, placed at $z_1\,= \delta_1$, to be the bubbling center. Instead, the second and third centers, placed at $z_2\,=\,\tfrac{\delta}{2}$ and $z_3\,=\,-\tfrac{\delta}{2}$, will correspond to the north ($N$) and south $(S)$ poles of the Euclidean horizon.

\begin{figure}[h!!]
    \centering
\includegraphics[width=1\linewidth]{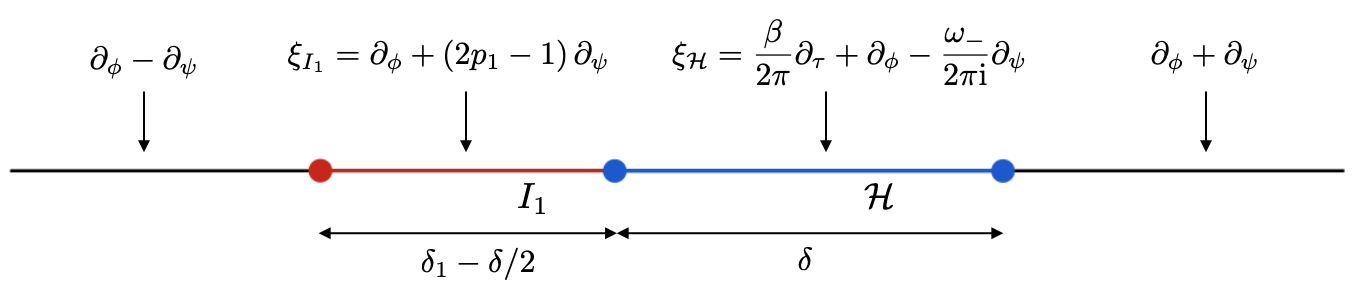}    
    \caption{\it Rod structure of the three-center solution, with the bubbling rod $I_1$ and the horizon rod $\mathcal{H}$. For each rod we indicate the corresponding rod vector.}
\label{fig:rod_str_3centers}
\end{figure}

We next restrict our attention to solutions with $n_{\cal H}\,=\,1$ and $p_{\cal H}=0$, implying that the rod vector \eqref{eq:combinationU1s} degenerating at the horizon, $I_N={\cal H}$, is given by
\begin{equation}
\xi_{\cal H}\,=\, \frac{\beta}{2\pi}\partial_{\tau}+\partial_{\phi}+\frac{\ii\omega_-}{2\pi}\partial_{\psi} \, .
\end{equation}
Instead, the Killing vector contracting at the bubbling rod, $I_1$, is 
\begin{equation}
\xi_{\BR}\,=\,\partial_{\phi}+ \left(2p_{1}-1\right)\partial_{\psi}\, .   
\end{equation}
This rod structure is illustrated in figure~\ref{fig:rod_str_3centers}.
The harmonic functions of the three-center solution are given by
\begin{equation}
\begin{aligned}
H\,=\,&\frac{h_1}{r_1}+\frac{h_N}{r_N}+\frac{h_S}{r_S}\,, \hspace{2.9cm} 
\ii K\,=\,\frac{\kk_1}{r_1}+\frac{\kk_N}{r_N}+\frac{\kk_S}{r_S}\, ,\\[1mm]
L\,=\,&1+\frac{\kk_1^2}{h_1 r_1}+\frac{\kk_N^2}{h_N r_N}+\frac{\kk_S^2}{h_S r_S}\,, \hspace{1cm} \ii M\,=\,-\frac{\kk_1^3}{2h_1^2 r_1}-\frac{\kk_N^3}{2h_N^2 r_N}-\frac{\kk_S^3}{2h_S^2 r_S}\, ,
\end{aligned}
\end{equation}
where we recall that the parameters $h_a$ and $\kk_a$ satisfy
\begin{equation}
\label{eq:constraint_3centersol}
h_1+h_N+h_S\,=\,1 \,,  \hspace{2cm} \kk_1+\kk_N+\kk_S\,=\,0\, .
\end{equation}

\noindent
To express the coefficients of the harmonic functions in terms of the boundary conditions (and the integers characterizing the rod structure) we specify \eqref{eq:ha_bubbling_rod}, \eqref{eq:ha_horizon_rod},  \eqref{eq:varphi_no_shift} and \eqref{eq:cycle_potential} to the case at hands, which yields
\begin{equation}
h_1\,=\, p_{1}\,, \hspace{1cm} h_N\,=\,\frac{1}{2}-p_{1}-\frac{\omega_-}{4\pi \ii} \,, \hspace{1cm} h_S\,=\,\frac{1}{2}+\frac{\omega_-}{4\pi\ii} \, ,
\end{equation}
and 
\begin{equation}
\label{eq:rel_k_varphi}
 \kk_1\,=\,-\frac{\Phi^{\BR}}{2\sqrt{3}\ii}\,, \hspace{1cm} \kk_N\,=\,\frac{\Phi^{\BR}}{2\sqrt{3}\ii}+\frac{\varphi}{4\sqrt{3}\pi}\,, \hspace{1cm} \kk_{S}\,=\,-\frac{\varphi}{4\sqrt{3}\pi}\,.
\end{equation}
 Finally, $\delta$ and $\delta_1$ are determined in terms of $\beta$ and the remaining parameters via  the first in \eqref{eq:DiracMisnergeneral}, which gives the conditions 
\begin{equation}
\ww_1\,=\, 0\, , \hspace{1.5cm} \beta\,=\, 4\pi \ii \,\ww_{N}\,=\, -4\pi \ii \,\ww_{S}\,. 
\end{equation}
These can be explicitly expressed in terms of the parameters of the solution via \eqref{eq:breve_omega_a},
\begin{equation}
\label{eq:non_extremal_bubble}
0\,=\,\frac{3\kk_1}{2h_1}+\frac{h_N}{2\dd_1-\delta}\left(\frac{\kk_1}{h_1}-\frac{\kk_N}{h_N}\right)^3 +\frac{h_S}{2\dd_1+\delta} \left(\frac{\kk_1}{h_1}-\frac{\kk_S}{h_S}\right)^3\,,
\end{equation}
\begin{equation}
\label{eq:beta}
    \beta\,=\, -2\pi \left[3\kk_N + \frac{h_N h_S}{\delta} \left(\frac{\kk_N}{h_N}-\frac{\kk_S}{h_S}\right)^3+ \frac{2\,h_N h_1}{2\delta_1-\delta} \left(\frac{\kk_N}{h_N}-\frac{\kk_1}{h_1}\right)^3\right]\, .
\end{equation}
As already discussed in section~\ref{sec:bolt_topology}, the bolt associated with the bubbling rod is topologically a branched $S^3$. In turn, the allowed topologies for the horizon bolt are $S^1\times S^2$ (corresponding to $p_{1}\,=\,1$), or $L\left(|1-p_{1}|, 1\right)\simeq S^3/{\mathbb Z}_{|1-p_{1}|}$.


\subsection{Thermodynamics}

The mass $E\,=\,\sqrt{3}\, Q$, electric charge $Q$ and angular momenta $J_{\pm}$ of the solutions were already given in \eqref{eq:mass+ang_momenta} and \eqref{eq:electric_charge}. Specializing to the case at hands, one gets
\begin{equation}
\label{eq:3centercharges}
\begin{aligned}
J_{+}\,=\,&-3\pi\ii \left[\kk_1\,\left(\delta_1+\frac{\delta}{2}\right) +\kk_N\, \delta\right]\,, \hspace{5mm} J_{-}\,=\, -\pi\ii \left(\frac{\kk_1^3}{h_1^2}+\frac{\kk_N^3}{h_N^2}+\frac{\kk_S^3}{h_S^2}\right)\, ,\\[1mm]
Q\,=\,&\pi\sqrt{3} \left(\frac{\kk_1^2}{h_1}+\frac{\kk^2_N}{h_N}+\frac{\kk^2_S}{h_S}\right)\,,
\end{aligned}
\end{equation}
where in the expression for $J_{+}$ we have used the second in \eqref{eq:constraint_3centersol} in order to explicitly show that it only depends on the distances between the centers.  The topology of this three-center solution allows us to associate a non-trivial electric flux ${\cal Q}^{\BR}$ to the bubbling rod. This was defined in \eqref{eq:potentials_genaral_def}, reported again here for convenience,
\begin{equation}
\label{def Q_D}
{\cal Q}^{\BR}\,=\,-\frac{1}{8}\int_{{\cal B}_{\BR}} \left[\star F-\frac{\ii}{\sqrt{3}}A_{1}\wedge F\right]\,.
\end{equation}
Let us recall that the notation $A_{1}$ stands for a gauge in which the vector field is regular in a patch containing the integration region, ${\cal B}_{\BR}$. This amounts to fixing the closed one-form $\diff\alpha$ appearing in \eqref{eq:susy_conf} by demanding that the contraction $\iota_{\xi_{I_1}}A_1$ vanishes at the fixed locus of any U(1) isometry degenerating inside the patch. Since this must contain ${\cal B}_{\BR}$, a first condition is
\begin{equation}
\iota_{\xi_{\BR}}A_1\Big|_{\BR}\,=\,\iota_{\xi_{\BR}}\left(\ii {\breve A}+\diff\alpha_{1}\right)\Big|_{\BR}\,=\,0\,.
\end{equation}
On top of this, one has two additional conditions coming from the extra U(1) isometries that degenerate at the endpoints of the bubbling rod, 
\begin{equation}
 \iota_{\xi_{1}} A_1\big|_{\rho\,=\,0, \, z\,=\,\delta_1}\,=\,0\,, \hspace{1cm}    \iota_{\xi_{N}}A_1\big|_{\rho\,=\,0, \, z\,=\,\delta/2}\,=\,0\,,
\end{equation}
where
\begin{equation}
 \xi_{1}\,=\,2\partial_{\psi} \hspace{5mm} \text{and} \hspace{5mm} \xi_{N}\,=\,\frac{\beta}{2\pi}\partial_{\tau}+2h_N\,\partial_{\psi}
\end{equation}
are the $2\pi$-periodic U(1) Killing vectors degenerating at the bubbling center and the north pole, respectively. A closed one-form $\diff\alpha_{1}$ satisfying the above regularity requirements is 
\begin{equation}\label{eq:alpha_1_3c}
\diff\alpha_{1}\,=\, -\frac{\kk_1}{h_1}\left(\diff\psi+\diff\phi-\frac{4\pi h_N}{\beta}\diff \tau\right)-\frac{4\pi\kk_N}{\beta}\diff \tau\, .
\end{equation}
A convenient way of computing ${\cal Q}^{\BR}$ is to note that the integral in \eqref{def Q_D} can be manipulated as follows:
\begin{equation}
\label{eq:int Q_D}
{\cal Q}^{\BR}\,=\, -\frac{1}{8}\int_{{\cal B}_{I_1}} G_1\,=\,  -\frac{\pi}{4} \int_{D}\iota_{\tau_{1}} G_1\,=\,-\frac{\beta}{8} \int_{\partial D} \nu_{V}\, , \hspace{1cm} G_1\equiv \star F-\frac{\ii}{\sqrt{3}}A_{1}\wedge F\,,
\end{equation}
where $D$ denotes the disc parametrized by the coordinate $z \in [\delta/2, \delta_1]$ and $\psi_{_1}$, whose boundary $\partial D$ is the circle parametrized by $\psi_{1}$ at $z\,=\, \delta/2$ (to be precise, this is a disc with a conical deficit). Given a Killing vector $\xi$, the local one-form $\nu_{\xi}$ is defined by \cite{Cassani:2024kjn}
\begin{equation}
\diff \nu_{\xi}\,=\, \iota_{\xi }G\, .
\end{equation}
For the specific choice $\xi\,=\, \partial_{\tau_{1}}$, we have
\begin{equation}
\nu_{\xi}\,=\, \frac{\beta}{2\pi}\nu_{V}+\left(\frac{\ii\omega_-}{2\pi}+1\right)\nu_{U}\,, 
\end{equation}
where $V\,=\, \partial_{\tau}$ and $U\,=\, \partial_{\psi}\,=\, \partial_{\psi_{1}}$. The second term in the above equation cannot contribute to the integral over $\partial D$ in \eqref{eq:int Q_D}, which explains why only $\nu_V$ appears in the last equality. The expression for $\nu_V$ was already computed in \eqref{eq:nuV}, which we just quote:
\begin{equation}
\nu_{V}\,=\, \sqrt{3}\ii f^2 \left(\diff\tau+\ii\omega_{\psi}\left(\diff\psi+\chi\right)+\ii {\breve \omega}\right)-\left(f+\alpha^{(\tau)}_{1}\right)A_{1}\, .
\end{equation}
Thus, we have all the ingredients to evaluate the integral, which yields the following result
\begin{equation}
\label{eq:QD}
{\cal Q}^{\BR}\,=\,2\sqrt{3}\pi^2 \,\ii \,h_{N}\left(\frac{\kk_N}{h_N}-\frac{\kk_1}{h_1}\right)^2\, .
\end{equation}

Finally, the Bekenstein-Hawking entropy ${\cal S}$ is computed as 1/4 the area of the horizon bolt, 
\begin{equation}
\label{eq:BHentropy}
{\cal S}\,=\, \pi \, \beta \, \delta\, .
\end{equation}
Expressed in terms of the charges $Q, J_{\pm}$ and $\Phi^{\BR}$, it reads
\begin{equation}
\label{eq:generalentropy}
\begin{aligned}
&{\mathcal S}  \,=\,  -2\pi\ii \left[ J_+-p_1 J_- + \frac{\Phi^{\BR}}{2} \left(Q - \frac{\pi}{12\sqrt{3}}\frac{\left(\Phi^{\BR}\right)^2}{p_1^2}\right)\right]\\[1mm]
&\!\!+4\sqrt{\pi} \sqrt{\left( 1-p_1\right)\!\left[\frac{Q}{\sqrt{3}}+ \frac{\pi\,(\Phi^{\BR})^2}{12p_1\left(1-p_1\right)} \right]^3-\frac{\pi}{4}\! \left[\left( p_1-1 \right) J_- - \frac{Q\, \Phi^{\BR}}{2}  - \frac{\pi\left( 1+p_1\right)(\Phi^{\BR})^3}{24\sqrt{3}\,p_1^2\left( 1-p_1\right)}\right]^2 }.
\end{aligned}
\end{equation}
This expression will be derived again via the Legendre transform of the on-shell action in appendix~\ref{sec:Legendre}.

Having explicitly computed all thermodynamic quantities, we can explicitly verify that the supersymmetric first law  \eqref{eq:susyfirstlaw},
\begin{equation}\label{eq:susy_first_law_3c}
\diff {\cal S}+\varphi\, \diff Q + \omega_+\, \diff J_{+}+\omega_-\, \diff J_{-}+{\cal Q}^{\BR}\, \diff \Phi^{\BR}\,=\,0\, ,
\end{equation}
is satisfied if $h_1$ (which is an integer) is held fixed in the variation. Assuming now the quantum statistical relation
\begin{equation}
\label{eq:QSR}
I=-{\cal S}-\varphi \,Q -\omega_+ \, J_+ -\omega_-\, J_-\,,
\end{equation}
we can obtain a prediction for the on-shell action prior to embarking ourselves in the direct computation. Plugging the expressions we have obtained for the thermodynamic variables, we find that it is given by
\begin{equation}
\label{eq:I_gen}
I\,=\,\frac{\pi}{12\sqrt{3}}\left[\frac{\varphi^3}{\omega_1\omega_2}-\frac{\left(p_1\varphi- \omega_2\, \Phi^{\BR}\right)^3}{p_1^2\,\omega_2\left(p_1\omega_1 + \left(p_1-1\right)\omega_2\right)}\right]\, ,
\end{equation}
which is the expression advertised in~\eqref{eq:I_gen_intro}.
Moreover, we have verified that 
\begin{equation}
I=\frac{1}{3}\left({\cal Q}^{\BR}\, \Phi^{\BR}-\varphi\, Q\right)\,,
\end{equation} 
consistently with the Smarr formula of \cite{Kunduri:2013vka}, whose supersymmetric version has been given in eq.~\eqref{eq:susy_smarr}. 
In appendix~\ref{sec:action} we show that the above expression for the on-shell action also follows from a direct computation, i.e.\ not relying on the validity of the quantum statistical relation.  This computation requires some care since the gauge Chern-Simons term in the supergravity action needs to be evaluated patchwise.

One can check that regimes of the parameters exist such that the real part of the on-shell action is positive while  the real part of the chemical potentials is negative, as it is required by naive convergence criteria of the partition function.


\subsection{Relation to the two-center black hole saddle}

As a sanity check, we can verify how the relations above reduce to those for the two-center black hole saddle of section \ref{sec:complex_saddle} in the appropriate limit. In order to achieve this, we should demand that the on-shell action (or, equivalently, the entropy) does not depend on $\Phi^{\BR}$. Namely, treating $\Phi^{\BR}$ as a continuous variable, as in \eqref{eq:susy_first_law_3c}, we demand that ${\cal Q}^{\BR}\,=\,\frac{\partial I}{\partial\Phi^{\BR}} = 0$. This condition is solved by
\begin{equation}
\label{eq:blackholemomentum}
\Phi^{\BR} \,=\, p_1\,\frac{\varphi}{\omega_2}\,.
\end{equation}
In this regime, the on-shell action \eqref{eq:I_gen} simplifies to $I = \frac{\pi}{12\sqrt{3}}\frac{\varphi^3}{\omega_1\,\omega_2}$, which reproduces the result for the black hole saddle found in~\eqref{eq:cclpsusyaction}. Notably, in the black hole saddle, the bubble potential $\Phi^{\BR}$ is not set to zero, whereas the flux ${\cal Q}^{\BR}$ is. 

At the level of the parameters of the solutions, imposing \eqref{eq:blackholemomentum} is equivalent to setting 
\begin{equation}
\kk_1 \,=\, -\frac{h_1}{h_1+h_N}\, \kk_S\,.
\end{equation} 
With this choice, the two centers at $z_1\,=\,\delta_1$ and $z_2\,=\,\tfrac{\delta}{2}$  coincide. This follows from eq.~\eqref{eq:non_extremal_bubble}, which is now solved by taking
\begin{equation}
2\dd_1 = \delta\,.
\end{equation}
Thus, the three-center solution collapses into a two-center configuration, which corresponds to the black hole saddle in the non-extremal case. In this limit, the harmonic functions simplify as:
\begin{equation}
\begin{aligned}
H \,&=\, \frac{h_1 + h_N}{r_N}+ \frac{h_S}{r_S}\,,\qquad\qquad\qquad\quad\, \ii K \,=\, -\frac{\kk_S}{r_N}+ \frac{\kk_S}{r_S}\,,\\[1mm]
L\,&=\, 1 + \frac{\kk_S^2}{\left( h_1 + h_N\right) r_N}+ \frac{\kk_S^2}{h_S\,r_S}\,,\qquad \ii M \,=\, \frac{\kk_S^3}{2}\left(\frac{1}{\left(h_1 + h_N\right)^2r_N}- \frac{1}{h_S^2\,r_S}\right)\,.
\end{aligned}
\end{equation}
From these expressions, we identify the black hole parameters as
\begin{equation}
h_N^{\rm BH} = h_1 + h_N\,,\qquad h_S^{\rm BH} = h_S\,,\qquad \kk^{\rm BH} = -\kk_S\,.
\end{equation}


\subsection{Connecting to Lorentzian solutions}
\label{eq:Lorentzian_soln}

In this section we investigate which well-behaved Lorentzian solutions are connected to the Euclidean saddle constructed above. By well-behaved Lorentzian solution we mean a solution admitting a real metric after Wick-rotating back to Lorentzian time, 
and carrying real conserved charges, entropy and fluxes. From the expression of the entropy in terms of the charges \eqref{eq:generalentropy}, it is evident that it is not possible to have real charges and entropy simultaneously unless some constraint on the former is imposed, which is equivalent to fixing $\beta$. As in the two-center case, there are two physically-inequivalent constraints. The first corresponds to taking $\beta\to\infty$, and gives rise to extremal black holes with non-spherical horizon topology \cite{Elvang:2004rt,Elvang:2004ds,Emparan:2006mm, Kunduri:2014kja, Tomizawa:2016kjh}. The second corresponds to taking $\beta\to0$ and leads to horizonless topological solitons \cite{Bena:2005va, Berglund:2005vb} carrying zero entropy. This means that, as we vary $\beta$, the Euclidean saddle continuously interpolates between the two physical configurations (up to a specific analytic continuation of parameters), extending the findings of~\cite{Cassani:2024kjn} to more general classes of solutions. Since the on-shell action \eqref{eq:I_gen} is independent of $\beta$, as expected for the saddle of a supersymmetric index, the same expression continues to hold after both limits. 


\subsubsection{Extremal limit}
\label{sec:extremallimit}

Let us assume for a moment that there exists a complexification of the solution yielding real and finite charges and entropy in the extremal limit. From the expression of the entropy in \eqref{eq:generalentropy}, one concludes that in this limit the charges must satisfy the constraint
\begin{equation}
\label{eq:extconstraint}
\left( 1-p_1\right) J_1 + \left( 1+p_1\right) J_2 + \Phi^{\BR
} \left(Q - \frac{\pi}{12\sqrt{3}}\frac{(\Phi^{\BR})^2}{p_1^2}\right)\,=\,0\,,
\end{equation}
otherwise the entropy would not be real. Using the explicit expression for the charges \eqref{eq:3centercharges}, the potential \eqref{eq:rel_k_varphi} and the inverse temperature \eqref{eq:beta}, one can indeed verify that the above constraint is equivalent to taking the extremal limit, in which the horizon rod contracts to a point,
\begin{equation}
\delta \rightarrow 0 \hspace{1cm} \Rightarrow \hspace{1cm} \beta\to \infty\, .
\end{equation}
As we explicitly show below, the resulting two-center solution describes an extremal and supersymmetric (BPS) black ring/lens. 
 For $p_1 \neq 1$, we can solve \eqref{eq:extconstraint} for $J_1$ and obtain the entropy from \eqref{eq:generalentropy},
\begin{equation}
\mathcal S \,=\, 4\sqrt{\pi} \sqrt{\left(1-\pp_1\right)\left(\frac{Q}{\sqrt{3}}+ \frac{\pi (\Phi^{\BR})^2}{12 p_1\left( 1-\pp_1\right)} \right)^3-\frac{\pi}{4} \left( J_2 - \frac{\pi (\Phi^{\BR})^3}{12\sqrt{3}\,\pp_1^2\left( 1-\pp_1\right)}\right)^2}\,.
\end{equation}
Alternatively, for $\pp_1 \neq -1$ we can solve the constraint for $J_2$, obtaining
\begin{equation}
\begin{aligned}
&\mathcal S =\\
&2\pi \sqrt{\frac{4\left( 1-\pp_1\right)}{\pi} \left[ \frac{Q}{\sqrt{3}}+ \frac{\pi\left(\Phi^{\BR}\right)^2}{12\pp_1\left( 1-\pp_1\right)}\right]^3 - \left[\frac{1-\pp_1}{1+\pp_1}J_1 + \frac{\Phi^{\BR}}{1+\pp_1}\left(Q+ \frac{\pi\,\left(\Phi^{\BR}\right)^2}{6\sqrt{3}\pp_1\left( 1-\pp_1\right)}\right)\right]^2}.
\end{aligned}
\end{equation}

Let us now discuss a complexification of the solution leading to real charges and entropy in the extremal limit. One way to achieve this is by imposing
\begin{equation}
\label{eq:ext_complexification}
h_{S}\,=\, h^*_{N}\,, \hspace{1cm} \kk_S\,=\,-\kk^*_N\, .
\end{equation}
Further using \eqref{eq:constraint_3centersol}, one has that $h_1$ and $\kk_1$ are related to the real and imaginary parts of $h_N$ and $\kk_N$ by
\begin{equation}
\label{eq:ext_complexification2}
{\rm Re}\,h_N\,=\,\frac{1-h_1}{2} \,, \hspace{1cm} \kk_1\,=\,-2\ii\,{\rm Im}\kk_N\, ,
\end{equation}
so that $\RRe\, h_N$ is half-integer quantized. The extremal $\delta\to0$ limit of the harmonic functions is
\begin{equation}
\label{eq:extharmonicfunction}
\begin{aligned}
H \,&=\, \frac{1-h_1}{r}+ \frac{h_1}{r_1}\,,\hspace{3.4cm} K \,=\, 2 \,{\IIm}\kk_N\left(\frac{1}{r}-\frac{1}{r_1}\right)\,,\\[1mm] 
L \,&=\, 1 +  2 \,{\RRe} \frac{\kk_N^2}{h_N} \frac{1}{r}- \frac{4\left(\IIm\kk_N\right)^2}{h_1\,r_1}\,, \hspace{1cm}
M \,=\, -\,\IIm \frac{\kk_N^3}{h_N^2}\,\frac{1}{r}-\frac{4\left(\IIm\kk_N\right)^3}{h_1^2 \,r_1}\,, 
\end{aligned}
\end{equation}
where $r$ is the distance to the origin of ${\mathbb R}^3$, where the horizon poles meet. The extremal solution is then parametrized by three real parameters ($\IIm h_N, \RRe \kk_N, \IIm \kk_N$) and an integer ($h_1$). As we can see, the harmonic functions associated to this extremal solution are real, which implies that the Lorentzian continuation of the solution is also real. The charges of the extremal solution are simply obtained by imposing \eqref{eq:ext_complexification} in the expressions provided in \eqref{eq:3centercharges},
\begin{equation}
\label{eq:extremal_charges}
\begin{aligned}
 J_{+} \,&=\, -6\pi \,{\IIm}\kk_N \,\delta_1 \,, \hspace{1.5cm} J_{-}\,=\, 2\pi \left({\IIm} \frac{\kk_N^3}{h_N^2}+4\frac{\left({\IIm}\kk_N\right)^3}{h_1^2}\right) \, , \\[1mm]
Q\,&=\, 2\sqrt{3}\pi \left({\RRe}\, \frac{\kk_N^2}{h_N}-2\frac{\left({\IIm}\kk_N\right)^2}{h_1}\right)\, ,
\end{aligned}
\end{equation}
where 
\begin{equation}
\label{eq:extrdelta1}
\delta_1\,=\, -\frac{h_1}{6 \IIm \kk_N} \,\IIm \left[h_N\left(\frac{\kk_N}{h_N}+\frac{2\ii \IIm \kk_N}{h_1}\right)^3\right]\, .
\end{equation}
This follows from solving \eqref{eq:non_extremal_bubble} in the $\delta\to0$ limit. In turn, the bubble potential $\Phi^{\BR}$ is given by
\begin{equation}
\label{eq:extrmomentum}
\Phi^{\BR} \,=\, -4\sqrt{3} \,\IIm \kk_N\,.
\end{equation}
This shows that charges $Q, J_{\pm}$, as well as $\Phi^{\BR}$, are real in the extremal limit. Additionally, one can verify that the constraint \eqref{eq:extconstraint} is satisfied, further implying that the entropy is real. Instead, the chemical potentials associated to the extremal solution, which are given by
\begin{equation}
\omega_- \,=\, 4\pi\ii \left(h_N^*-\frac{1}{2}\right)\, ,\hspace{1.5cm} \varphi \,=\,  4\pi \sqrt{3}\,\kk_N^* \,,
\end{equation}
remain complex. Consequently, the corresponding on-shell action is also complex. In contrast, to the extremal black hole with a spherical horizon analyzed in section~\ref{contribindex} one associates real chemical potentials $\omega_-$, $\varphi$, and hence a real on‑shell action, at least in the case where no shifts or orbifolds are considered. This is one of the distinguishing features between these different solutions.

\paragraph{Topology of the Lorentzian solution.} Since in the extremal limit the horizon rod contracts to a point (the origin of $\mathbb R^3$), we are just left with one compact rod, $I_1$. Moreover, since the thermal circle is decompactified we can Wick-rotate back to Lorentzian time, $t=-\ii\tau$. The Lorentzian metric induced on the bolt associated to the surviving rod, ${\cal B}_{\BR}$, is given by
\begin{equation}
\diff s^2_{{\cal B}_{\BR}}\,=\, -f^2\left(\diff t+\omega_{\psi}\,\diff \psi_{1}\right)^2+f^{-1}H^{-1}\,\diff \psi_{1}^2+f^{-1}H\,\diff z^2\, ,
\end{equation}
where $\psi_{1}\,=\,\psi+\left(1-2h_1\right)\phi$. The analysis of the topology of this bolt is analogous, to a large extent, to that of section~\ref{sec:bolt_topology}. The main difference lies on the fact that the time coordinate $t$ is non-compact.
 In particular, the behavior of the metric functions near a bubbling center is the same as in the finite-$\beta$ solution, namely 
\begin{equation}
f \,=\, \mathcal O(r_1^0) \,,\hspace{1cm} H \,=\, \frac{h_1}{r_1}+ \mathcal O(r_1^0)\,,\hspace{1cm} \omega_\psi \,=\, \mathcal O(r_1) \,.
\end{equation}
Thus, we have again that the Killing vector $\partial_{\psi_1} = \partial_{\psi}$ contracts at the bubbling center, where space ends in a conical singularity if $|h_1|\neq 1$. On the contrary, the U(1) generated by this Killing vector has finite size at the horizon center, leading to the conclusion that ${\cal B}_{\BR}$ has ${\mathbb R}\times {\rm Disc}/{\mathbb Z}_{|h_1|}$ topology, with the ${\mathbb R}$ factor being parametrized by the Lorentzian time.

Let us now turn to the analysis of the near-horizon geometry.  To this aim, it is convenient to recast the metric in the following form 
\begin{equation}
\label{eq:adaptedextmetric}
\diff s^2 \,=\, Y \left[ \diff \psi + \chi - Y^{-1}f^2\omega_\psi\left( \diff t + \breve\omega\right)\right]^2 - Y^{-1}H^{-1} f\left( \diff t + \breve\omega\right)^2 + f^{-1}H \,\diff s^2_{\,\mathbb{R}^3}\,,
\end{equation}
where
\begin{equation}
Y \,=\, - f^2 \omega_\psi^2 + \frac{1}{fH}\, ,
\end{equation}
and expand it about the origin of $\mathbb R^3$, for $r\rightarrow 0$. The near-horizon behavior of the harmonic functions depends crucially on $h_1$, which forces us to consider separately the cases $h_1\neq 1$ and $h_1=1$. These describe an extremal black lens and an extremal black ring, respectively. 
\paragraph{Near-horizon of extremal black lens ($h_1\neq 1$).} The behavior of the metric functions near the horizon is
\begin{equation}
f \,=\, \hat f\,r + \mathcal O(r^2)\,, \hspace{1cm} Y\,=\, \frac{1-h_1}{|h_N|^2 {\hat f}}+{\mathcal O}(r)\,, \hspace{1cm} \omega_{\psi}\,=\,\frac{{\hat \omega}_{\psi}}{r}+{\cal O}\left(r^0\right)\,,
\end{equation}
where $\hat f$ and ${\hat\omega}_{\psi}$ are constants. Using this information, one finds that the near-horizon expansion of \eqref{eq:adaptedextmetric} is given by
\begin{equation}
\begin{aligned}
\label{eq:ads2nearcentergeometry}
\diff s^2 \,=\,& \frac{1-h_1}{\hat f}\left[-\frac{|h_N|^2{\hat f}^3}{\left(1-h_1\right)^3}\,r^2\,\diff t^2 + \frac{\diff r^2}{r^2} \right] \\[1mm]
& +\frac{\left(1-h_1\right)^3}{|h_N|^2{\hat f}} \left( \diff\hat \psi + \cos\theta \,\diff \phi\right)^2+ \frac{1-h_1}{\hat f}\left( \diff \theta^2 + \sin^2\theta\,\diff\phi^2\right)+\dots \,,
\end{aligned}
\end{equation}
where 
\begin{equation}
\hat \psi \,=\, \frac{\psi - h_1\,\phi}{1-h_1}\,.
\end{equation}
We recognize the AdS${}_2$ factor characterizing the near-horizon geometry of extremal black holes in the first line of \eqref{eq:ads2nearcentergeometry}. In turn, the second line corresponds to the induced metric at the horizon. Locally, this is the metric of a squashed three-sphere. However, the periodic identifications of the angular coordinates
\begin{equation}
\left( \phi,\,\hat \psi\right) \sim \left( \phi+ 2\pi,\,\hat \psi + 2\pi \right) \sim \left( \phi ,\,\hat \psi + \frac{4\pi}{1-h_1} \right)
\end{equation}
tell us that, as in the non-extremal case, the horizon has lens-space topology 
\begin{equation}
L(|1-h_1|,1) \,\simeq\, S^3/\mathbb Z_{|1-h_1|}\,.
\end{equation}
This corresponds to the extremal black lens of \cite{Kunduri:2014kja, Tomizawa:2016kjh}. A completely smooth geometry in the domain of outer communication is obtained for $h_1=-1$, see below. By writing the explicit expression for ${\hat f}$,
\begin{equation}
    {\hat f}^{-1}\, \equiv\, 2\RRe\frac{\kk_N^2}{h_N}+4\frac{\left(\IIm\kk_N\right)^2}{1-h_1}\,=\,\frac{\left(\RRe\kk_N \left(1-h_1\right)+2\IIm\kk_N \IIm h_N\right)^2}{\left(1-h_1\right)|h_N|^2}\,,
\end{equation}
we can verify that the combination $\frac{1-h_1}{\hat f}$ appearing in the near-horizon geometry \eqref{eq:ads2nearcentergeometry} is positive, ensuring that the metric has the correct signature.

\paragraph{Near-horizon of extremal black ring ($h_1=1$).} The behaviour of the metric functions for $r\rightarrow 0$ is
\begin{equation}
f^{-1}\,=\, \frac{4\left(\IIm \kk_N\right)^2\delta_1}{r^2}+{\cal O}\left(\frac{1}{r}\right)\,, \hspace{5mm} Y\,=\,\frac{\left(\IIm \kk_N\right)^2}{|h_N|^2}+{\cal O}(r)\,, \hspace{5mm} \omega_{\psi}\,=\,\frac{8\left(\IIm\kk_N\right)^3 \delta_1^2}{r^3}+{\cal O}\left(\frac{1}{r^2}\right)\, .
\end{equation}
The near-horizon geometry reads
\begin{equation}
\begin{aligned}
\label{eq:nhgeomBR}
\diff s^2 \,&=\, 4\left(\IIm \kk_N\right)^2\left[-\frac{|h_N|^2}{16\left(\IIm\kk_N\right)^6}\,r^2\,\diff t^2 + \frac{\diff r^2}{r^2} \right] \\[1mm]
&\quad\ +\frac{\left(\IIm \kk_N\right)^2}{|h_N|^2}  \diff\tilde \psi^2+ 4\left(\IIm \kk_N\right)^2\left( \diff \theta^2 + \sin^2\theta\,\diff\phi^2\right)+\dots \,,
\end{aligned}
\end{equation}
where $\tilde \psi\,=\,\psi-\phi$. This corresponds to the near-horizon geometry of the extremal black ring found in \cite{Elvang:2004rt}, whose event horizon has $S^1\times S^2$ topology.

\subsubsection{The black lens}

Among the solutions constructed in this section, two particular cases play a special role as they describe completely smooth supersymmetric non-extremal geometries. These correspond to solutions where $h_1=\pm1$, ensuring that the orbifold singularity at the bubbling center is absent. 
 In the following we specialize the formulae for the on-shell action and chemical potential to these notable cases, focusing on the extremal $\beta\to\infty$ limits. 
Interestingly, the Lorentzian counterpart of these solutions is well known. 

Let us start from the case $h_1=-1$.  From \eqref{eq:ads2nearcentergeometry} we see that the extremal horizon has $L(2,1) \,\simeq\, S^3/\mathbb Z_2$ topology. This corresponds to the extremal black lens constructed in~\cite{Kunduri:2014kja,Kunduri:2016xbo}.
This solution is described by the following harmonic functions~\cite{Kunduri:2014kja}:
\begin{equation}
H \,=\, \frac{2}{r}- \frac{1}{r_1}\,,\qquad K \,=\,\hat k\left( \frac{1}{r} - \frac{1}{r_1}\right)\,,\qquad L \,=\, 1 + \frac{\hat \ell}{r} + \frac{\hat k^2}{r_1}\,,\qquad M \,=\, \frac{\hat m}{r}- \frac{\hat k^3}{2r_1}\,.
\end{equation}
These coefficients are related to the ones we used in section~\ref{sec:extremallimit} by
\begin{equation}\label{eq:extremal_map}
    \hat k \,=\, 2\,{\rm Im}
    \,\kk_N\,,\qquad \hat m \,=\, {\rm Im}\left(\frac{\kk_N^3}{h_N^2}\right)\,,\qquad \hat l \,=\, 2\,{\rm Re}\left(\frac{\kk_N^2}{h_N}\right)\,.
\end{equation}

The conserved charges carried by the solution are obtained by setting $h_1 = -1$ in \eqref{eq:extremal_charges} and \eqref{eq:extrdelta1}. Moreover, the analysis of the extremal limit of section~\ref{sec:extremallimit} implies the following constraint among the charges:
\begin{equation}
2J_1 + \Phi^{\BR}\left( Q - \frac{\pi}{12\sqrt{3}}\left(\Phi^{\BR}\right)^2\right) \,=\,0\,.
\end{equation}
The corresponding entropy is given by
\begin{equation}
\mathcal S \,=\, 4\sqrt{\pi}\sqrt{2\left( \frac{Q}{\sqrt{3}}- \frac{\pi}{24}\left(\Phi^{\BR}\right)^2\right)^3 - \frac{\pi}{4}\left( J_2 - \frac{\pi}{24\sqrt{3}}\left(\Phi^{\BR}\right)^3\right)^2}\,,
\end{equation}
which is consistent with the results of~\cite{Kunduri:2014kja,Kunduri:2016xbo}, up to conventions.  

Similarly, the on-shell action for the supersymmetric extremal black lens is obtained by setting $p_1=-1$ into~\eqref{eq:I_gen}, 
\begin{equation}
I \,=\, \frac{\pi}{12\sqrt{3}}\left[\frac{\varphi^3}{\omega_1\,\omega_2} - \frac{\left( \varphi + \omega_2\,\Phi^{\BR}\right)^3}{\omega_2\left( 2\omega_2 + \omega_1\right)}\right]\,,
\end{equation}
where the chemical potentials turn out to be given by the following complex combinations
\begin{equation}
\begin{aligned}
\omega_1 = 2\pi\left({\rm Im}h_N + \ii\right)\,,\qquad \omega_2 =-2\pi \,{\rm Im}h_N \,,\qquad
\varphi = 4\sqrt{3}\pi \left( {\rm Re}\kk_N + \ii\, {\rm Im}\kk_N\right)\,.
\end{aligned}
\end{equation}
while $\Phi^{\BR}$ reads
\begin{equation}
\Phi^{\BR} \,=\, -4\sqrt{3}\,{\rm Im}\kk_N\,.
\end{equation}

\subsubsection{The black ring}

Let us now set $h_1=1$.  From \eqref{eq:nhgeomBR}, it follows that the extremal horizon has ${S}^1 \times {S}^2$ topology. This corresponds to the supersymmetric black ring studied in~\cite{Elvang:2004rt,Elvang:2004ds,Emparan:2006mm}.
Setting $h_1=1$ in \eqref{eq:extharmonicfunction} indeed gives the harmonic functions that characterize such solution (see for instance~\cite{Gauntlett:2004wh}),
\begin{equation}
H \,=\, \frac{1}{r_1}\,,\qquad K \,=\, \hat k\left( \frac{1}{r}- \frac{1}{r_1}\right)\,,\qquad
L \,=\, 1 + \frac{\hat \ell}{r}- \frac{\hat k^2}{r_1}\,,\qquad M \,=\, \frac{\hat m}{r}- \frac{\hat k^3}{2r_1}\,,
\end{equation}
where the map between $(\hat k,\,\hat l,\,\hat m)$ and the coefficients we used in section~\ref{sec:extremallimit} is the same as in \eqref{eq:extremal_map}.
The charges of the black ring must satisfy the constraint
\begin{equation}
2J_2 + \Phi^{\BR} \left( Q - \frac{\pi}{12\sqrt{3}}\left(\Phi^{\BR}\right)^2 \right) \,=\, 0\,,
\end{equation}
while the entropy takes the simple form
\begin{equation}
\mathcal S \,=\, \frac{\pi}{\sqrt{3}}|\Phi^{\BR}|\sqrt{Q^2 + \frac{\pi^2}{144}\left(\Phi^{\BR}\right)^4 - \frac{\pi}{\sqrt{3}}\Phi^{\BR}\,J_1}\,.
\end{equation}
These expressions are in agreement with the results for the supersymmetric black ring obtained in~\cite{Elvang:2004ds}.

We find that the angular velocity $\omega_1$ is real, while the other potentials are complex, 
\begin{equation}
\omega_1 = 2\pi\, {\rm Im}h_N\,,\qquad \omega_2 = 2\pi\left( \ii- {\rm Im}h_N\right)\,,\qquad \varphi = 4\sqrt{3}\pi\left({\rm Re}\kk_N + \ii\, {\rm Im}\kk_N\right)\,.
\end{equation}
Again, the bubble potential is\footnote{
This corresponds, up to normalization, to the \emph{dipole charge} that characterizes the thermodynamics of the black ring~\cite{Copsey:2005se,Emparan:2006mm}. The relation between these two quantities has been explained in~\cite{Kunduri:2013vka}.}
\begin{equation}\label{eq:bubble_pot_br}
\Phi^{\BR} = -4\sqrt{3}\,{\rm Im}\kk_N\,.
\end{equation}
Finally, the on-shell action for the supersymmetric black ring turns out to be
\begin{equation}
I \,=\, \frac{\pi}{12\sqrt{3}}\frac{\Phi^{\BR}}{\omega_1}\left( 3\varphi^2 - 3\varphi\,\Phi^{\BR}\,\omega_2+ \left(\Phi^{\BR}\right)^2 \omega_2^2\right)\,.
\end{equation}

Therefore, starting from the non-extremal saddles and taking an extremal limit, on the one hand we have correctly reproduced the known extremal entropy of the supersymmetric black lens and black ring, and on the other hand we have assigned for the first time non-trivial chemical potentials and on-shell action to these solutions.

\subsubsection{Topological solitons}
\label{sec:solitoniclimit}

Finally, we discuss a different limit leading to a horizonless topological soliton.
Besides the reality conditions discussed above, another choice that ensures the charges $Q$, $J_\pm$ and the potential $\Phi^{\BR}$ remain real is the obvious one:
\begin{equation}
\kk_N = \ii k_N\,,\qquad\quad \kk_S = \ii k_S\,,\qquad\quad \kk_1 = \ii k_1\,,\qquad\quad k_1 + k_N + k_S =0\,,
\end{equation}
with $k_{N,S,1},\,h_{N,S,1}\in \mathbb R$. However, under this analytic continuation, the entropy given by \eqref{eq:generalentropy} becomes purely imaginary. This follows from the fact that the argument of the square root in \eqref{eq:generalentropy} turns negative:\footnote{An alternative way to see this is to note that the inverse temperature $\beta$  in \eqref{eq:beta} also becomes imaginary under this continuation of the parameters.}
\begin{equation}
\begin{aligned}
&\left( 1-h_1\right)\left(\frac{Q}{\sqrt{3}}+ \frac{\pi\,(\Phi^{\BR})^2}{12h_1\left(1-h_1\right)}\right)^3-\frac{\pi}{4} \left(\left( h_1-1 \right) J_- - \frac{\Phi^{\BR}}{2} Q - \frac{\pi\left( 1+h_1\right)(\Phi^{\BR})^3}{24\sqrt{3}\,h_1^2\left( 1-h_1\right)}\right)^2 =
\\[1mm]
&- \frac{\pi^3}{4h_N^4\,h_S^4}\left( h_N\,k_1+ k_N\left( 1-h_1\right)\right)^6 <0\,.
\end{aligned}
\end{equation}
Thus, the only possibility to get a real quantity is that ${\cal S}$ vanishes. This is equivalent to the following constraint among the charges\footnote{This constraint can be understood as imposing $P_0=0$ in \eqref{eq:p01}.}
\begin{equation}
\label{eq:solitonconstraint}
\begin{aligned}
&\frac{4\left(1- h_1\right)}{3\sqrt{3}\pi}Q^2\left(Q + \frac{\sqrt{3}\pi}{4h_1\left( 1-h_1\right)}(\Phi^{\BR})^2\right) + \left(1- h_1\right)J_1\left(J_2 - \frac{\pi}{12\sqrt{3}\,h_1^2\left(1-h_1\right)}(\Phi^{\BR})^3\right)\\[1mm]
&+h_1\, J_2 \left( J_2 +h_1^{-1}\, \Phi^{\BR}\,Q\right)=0\,,
\end{aligned}
\end{equation}
leading to a horizonless solution. From \eqref{eq:BHentropy}, one further deduces that this condition is equivalent to 
$\beta \rightarrow 0\,,$
which in terms of the parameters reads
\begin{equation}
\label{eq:bubbleequation2}
\frac{3}{2}\frac{k_N}{h_N}- \frac{h_S}{2\delta}\left(\frac{k_N}{h_N} - \frac{k_S}{h_S}\right)^3- \frac{h_1}{2\dd_1 - \delta}\left( \frac{k_N}{h_N}- \frac{k_1}{h_1}\right)^3 =0\,.
\end{equation}
This constraint, together with \eqref{eq:non_extremal_bubble}, which we reproduce here for completeness,
\begin{equation}
\label{eq:bubbleequation1_2}
\frac{3}{2}\frac{k_1}{h_1}- \frac{h_S}{2\dd_1 + \delta}\left(\frac{k_1}{h_1} - \frac{k_S}{h_S}\right)^3-\frac{h_N}{2\dd_1 - \delta} \left(\frac{k_1}{h_1} - \frac{k_N}{h_N}\right)^3=0\,,
\end{equation}
constitutes the system of two \emph{bubble equations} arising in three-center microstate geometries of~\cite{Bena:2005va, Berglund:2005vb}. 

From a global perspective, in the $\beta \to 0$ limit what used to be the thermal isometry in \eqref{eq:U(1)isometries} does not advance the Euclidean time anymore. As a result, the global identifications \eqref{eq:coord_identif} become those relevant for a bubbling geometry,
\begin{equation}
\left( \tau ,\,\phi ,\,\psi \right) \sim \Big( \tau,\,\phi ,\,\psi + 4\pi\left(h_1 +  h_N\right)\Big) \sim \left( \tau ,\,\phi + 2\pi ,\,\psi - 2\pi \right) \sim \left( \tau ,\,\phi ,\,\psi + 4\pi\right)\,,
\end{equation}
where consistency of the angular identifications requires $h_{N,S,1} \in \mathbb Z$~\cite{Bena:2005va, Berglund:2005vb}.  
 Once these conditions are imposed, we can safely Wick-rotate back to Lorentzian time, obtaining a well-behaved horizonless solution. The resulting spacetime is smooth up to certain discrete orbifold singularities, which we discuss below. 

For this solitonic solution, the harmonic functions remain real, ensuring that the metric after Wick-rotating back to Lorentzian time is also real. However, the assigned chemical potentials become imaginary:
\begin{equation}
\omega_1= 2\pi \ii\,h_S\,,\qquad\qquad \omega_2 \,=\, 2\pi\ii\left( h_N + h_1\right)\,,\qquad\qquad \varphi \,=\, -4\sqrt{3}\pi\,\ii\,k_S\,.
\end{equation}
As a result, the on-shell action \eqref{eq:I_gen} is also imaginary. In fact, the angular velocities must be multiples of $2\pi \ii$, i.e. $\omega_{1,2} \in 2\pi\ii\mathbb Z$. This is the same outcome we noticed in the two-center solution discussed in section~\ref{contribindex}, and appears to be a general property of horizonless topological solitons. Moreover, following the general argument developed in section~\ref{sec:compact_gauge_conditions}, we can argue that also $\varphi$ becomes a multiple of $2\pi \ii$. So these solutions appear to be relevant just in a particular limit of the gravitational index where the fugacities trivialize. It would be interesting to investigate the trace over microstates given by~\eqref{eq:microscopicindex} in this limit.

 The horizonless solution obtained  sending $\beta\to0$ has two compact rods, $I_1$ and $I_N$. The associated rod vectors are 
\begin{equation}
\xi_{I_1}\,=\, \partial_{\phi}-\left(1-2h_1\right)\partial_{\psi}\,, \hspace{1.5cm} \xi_{I_N}\,=\, \partial_{\phi}-(h_S-h_N-h_1)\,\partial_{\psi}\,.
\end{equation}
Since both are bubbling rods, we have that the Killing vector $\partial_{\psi}$ contracts at all endpoints, implying that their fixed loci have ${\mathbb R}\times \Sigma_{[h_a, h_{a+1}]}$ topology. The analysis of the topology is analogous to that of section~\ref{sec:bolt_topology}, with the exception that time is no more periodic.
For an odd number of centers we can get rid of orbifold singularities by setting $|h_a| = 1$ for all $a$'s \cite{Gibbons:2013tqa}. In the present three-center case, this leads to three distinct fully regular cases: 
\begin{equation}
\label{eq:smoothsoliton}
h_1 = h_N =- h_S= 1\,, \hspace{1cm} h_1 = -h_N = h_S= 1\,, \hspace{1cm} h_1 = - h_N = -h_S = -1\,.
\end{equation}

\section{Discussion and open problems}\label{sec:conclusions}

In the third part of this Thesis, we addressed two closely related topics:

\paragraph{Equivariant Localization.}

We have studied equivariant localization of the ungauged supergravity on-shell action in five dimensions, with or without supersymmetry. We have  shown that there is  a scheme choice such that the boundary terms cancel out, and one is just left with nut and bolt contributions. While in general this choice depends on the conserved charges of the solution,  if we assume supersymmetry then it only depends on boundary conditions and thus acquires universal value.
 
We have examined in detail the features that are specific to odd dimensions and thus did not appear in previous even-dimensional investigations: the nuts are one-dimensional loci rather than isolated points; their contribution to the action is given by the holonomy of a one-form potential, denoted by $\nu$; global regularity of $\nu$ gives non-trivial conditions, as the fixed locus at the nut may collapse elsewhere on the manifold. 
In particular, we have seen that in black hole solutions with $S^3$ horizon topology and biaxial  ${\rm U}(1)^2$ symmetry, one can combine the Killing vector generating the thermal circle with either one of the axial Killing vectors, obtaining two independent vectors, each with a single nut (generated by the other axial ${\rm U}(1)$). In this context, we have found that the equivariant localization formula for the action using either one of these vectors splits into a sum of two terms, with the second term being naturally associated with the nut of the other vector. The final expression reminds of a sum over gravitational blocks~\cite{Hosseini:2019iad}. This is particularly evident in the supersymmetric case, where the two terms appearing in the final expression can be seen as the contribution from the two centers of the harmonic functions that are used to build the supersymmetric solution.

Besides illustrating the `nuts and bolts' of equivariant localization in five-dimensional supergravity, we have discussed its application to the evaluation of the action of supersymmetric non-extremal black hole solutions (with spherical horizon), which represent saddle-point contributions to the gravitational path integral computing a supersymmetric index. 

\paragraph{Saddles of the supersymmetric index.}

We considered five-dimensional supergravity with boundary conditions such that the gravitational path integral computes a refined supersymmetric index. We investigated candidate saddles of this gravitational index, namely supersymmetric yet non-extremal complexified solutions to the supergravity equations satisfying the boundary conditions and having finite action.

We have constructed multi-charge, doubly-rotating such solutions in supergravity coupled to vector multiplets, and demonstrated  global regularity of the Euclidean section. Then, we have computed their action equivariantly and discussed their supersymmetric thermodynamics. While the supersymmetric action is independent of the inverse temperature $\beta$, the solutions do depend on it. By tuning $\beta$ we have found that the black hole saddles interpolate between two limiting solutions, both having a Lorentzian interpretation: the $\beta\to \infty$ limit gives the extremal supersymmetric black hole, while by taking $\beta\to 0 $ we obtain solitonic solutions carrying a vanishing Bekenstein-Hawking entropy, which are identified with two-center microstate geometries. The on-shell action takes the same functional form in terms of the supersymmetric chemical potentials 
 $\omega_1,\omega_2,\varphi^I$, however $\omega_1-\omega_2$ and $\varphi^I$ take different values in the two solutions as a consequence of different analytic continuations of the parameters being implemented. We have shown how the Legendre transform of the action encompasses both solutions by imposing different conditions on the charges, which in the solitonic case lead to a vanishing entropy. It would be interesting to clarify this intriguing connection in the context of black hole microstate counting and explore it more widely, for instance by comparing the on-shell actions in different regimes of the chemical potentials with a microscopic computation of the index directly in string theory (we will comment more about this in chapter~\ref{chap:conclusions}).\footnote{Further intuition in this direction may come from the results of the author's most recent contribution~\cite{Massai:2025nci} -- which is not further discussed in this Thesis --, that propose an explicit worldsheet description of a five-dimensional (Lorentzian) black hole background, uplifted to the NS$5$-F$1$-P system of Type II string theory and in a suitable decoupling limit, in terms of a null-gauged WZW model. In that work it was shown that this construction is connected to the earlier worldsheet model of the corresponding horizonless geometry through a suitable $\beta \to 0$ limit, together with a change in the gauged current from the non-compact to the compact generator of SL$(2,\mathbb R)$. 
 } 

Furthermore, we classified the solutions using the rod structure formalism, and studied their topology by analyzing the fixed loci of their isometries. We found a rich array of possibilities, which correspond in general to multi-horizon solutions with many bubbles. The fixed loci can have $S^3$, $S^2\times S^1$ and lens space topologies, as well as versions of these geometries with conical singularities.  We illustrated the relation of the angular velocities and electrostatic potential of each horizon with the chemical potentials appearing in the index, and also discussed the role of the specific gauge potentials associated with the bubbles. Since the latter potentials are not associated with an asymptotic symmetry, they are not expected to represent additional variables of the index. Rather, it appears we should sum over their allowed values (when the gauge group is ${\rm U}(1)$, these are indeed integer values).

For the notable case of the three-center solution, we explicitly computed the on-shell action providing the saddle-point contribution to the index. We also illustrated how these solutions are related to known Lorentzian solutions, BPS black rings or lenses and horizonless bubbling solutions, by taking suitable limits.
 We observed that the chemical potentials and on-shell action remain finite in this limit, and verified that the Legendre transform of the on-shell action reproduces the correct Bekenstein-Hawking entropy. While the action of the black ring and black lens can be taken with a positive real part, and the chemical potentials with a negative real part, ensuring naive convergence conditions, both the action and chemical potentials associated with the horizonless solutions are purely imaginary, making it unclear whether the latter should be regarded as true saddles of the grand-canonical index. It would be very interesting to compare these findings with a microscopic evaluation of the index in string theory.
 
 \medskip
 
An important open issue concerns the allowed complexifications of the parameters controlling our solutions. More generally, one would like to establish precise criteria for when a candidate complex saddle of the gravitational index should be regarded as a genuine saddle (different criteria have been proposed and/or assessed in the supersymmetric context in~\cite{Kontsevich:2021dmb,Witten:2021nzp,Aharony:2021zkr,BenettiGenolini:2025jwe}). Another related question is whether the current formulation of equivariant localization can be extended to compute the on-shell action of the five-dimensional solutions we constructed more efficiently. We will return to these important points in chapter~\ref{chap:conclusions}. Once the candidate saddles are filtered according to some allowability criteria, and the actions of the remaining ones have been computed, it should be possible to compare the on-shell actions of the competing saddles with different topologies for given values of the chemical potentials. This would allow us to analyze the different phases of the gravitational index and perhaps develop a systematic recipe for resumming the saddle-point contributions. 

\medskip

While here we provided the general framework for solutions with multiple horizon components, it would be interesting to study in detail explicit examples of these configurations, as well as to consider solutions where the centers are not necessarily aligned along a symmetry axis.
Very recently, saddles of the four-dimensional gravitational index comprising these features have been studied in~\cite{Boruch:2025biv}, and it would be interesting to make the connection between the four-dimensional and the five-dimensional cases more explicit. A further development would be to extend our approach to more general solutions possessing an axial ${\rm U}(1)$ symmetry that does not preserve the Killing spinor, such as those of \cite{Bena:2007ju}. In all these cases, one constructs supersymmetric solutions starting from analytic continuations of the Lorentzian classification of backgrounds with a timelike supersymmetric Killing vector of~\cite{Gauntlett:2002nw}. An even more general approach would be to work directly in Euclidean signature and carry out the supersymmetric analysis from the start treating the spinor and its complex conjugate as independent variables, so as to allow for more general Euclidean complex backgrounds. 

Although while studying the general rod structure we worked in minimal supergravity and thus with a single ${\rm U}(1)$ gauge field, our results straightforwardly generalize to the case where vector multiplets are included. 
Multi-charge saddles of the gravitational index with a single compact rod were indeed presented in chap.~\ref{chap:Black_hole}, see also \cite{Anupam:2023yns} for the three-charge case.
 Recently, the case corresponding to the D1-D5-P brane system (see table~\ref{tab:D1-D5-P}) with one of the three charges turned off was studied  in~\cite{Bandyopadhyay:2025jbc}, where a finite-temperature version of the small black ring was given. We note that the solutions of~\cite{Bandyopadhyay:2025jbc} necessarily have running scalars, so there is no limit in which they reduce to the solutions of minimal supergravity presented here, which in the D1-D5-P system correspond to solutions with the three charges set equal.
It will be interesting to investigate rod structures in the generic three-charge case, thus exploring the index associated with the full D1-D5-P brane system.


%% file: conclusion.tex
\part{Outlook}
\label{part_four}

\chapter{Concluding remarks and future plans}
\label{chap:conclusions}


In this Thesis, we have explored supersymmetric observables mostly from the gravitational side, focusing on setups where the Euclidean gravitational path integral, first introduced by Gibbons and Hawking~\cite{Gibbons:1976ue} only using gravitational variables, takes the form of a refined Witten index~\cite{Witten:1982df} counting microstates, i.e.~a gravitational index. In such cases the gravitational path integral is protected against continuous variations of the string coupling, providing a powerful arena to study quantum gravity: even though a fully well-defined path integral for gravity would in general require new degrees of freedom beyond general relativity, the gravitational index remains robust and captures non-trivial aspects of the quantum theory. As a result, it offers a clean window into the microscopic structure of black holes, bridging the gap between the macroscopic (i.e. geometric) and microscopic (i.e. string-theoretic or CFT) fine-grained description. 

Throughout this Thesis we have mostly worked within the semiclassical approximation, which already captures a number of non-trivial quantum gravity effects (e.g. what is the dominating phase for a given value of the boundary conditions). Nevertheless, direct comparison with microscopic predictions, especially in those cases where microscopic computations provide exact results for the index, offers a controlled setup to understand the gravitational path integral beyond the semiclassical regime, with the final aim of extracting the exact quantum entropy directly from the gravitational description. Hopefully, recent progress on applying localization techniques to (super)gravity, to which this Thesis contributed, strengthen the prospects of this program.

Indeed, incorporating certain corrections to the semiclassical index has already revealed interesting quantum effects. For instance, one-loop effects from the (super)Schwarzian mode around near-BPS geometries (see e.g.~\cite{Boruch:2022tno}) have been shown to be responsible for large quantum corrections to the spectra of absorption and emission of Hawking radiation~\cite{Lin:2025wof}, as well as for a macroscopic energy gap (i.e. a quantum-mechanical decoupling) separating supersymmetric ground states, which account for the BPS black hole degeneracy, from the non-BPS excitations~\cite{Heydeman:2020hhw}.\footnote{In contrast, this gap is absent for extremal black holes that are not protected by supersymmetry, whose quantum entropy is therefore expected to vanish~\cite{Iliesiu:2020qvm}.}

\medskip

In the semiclassical approximation the gravitational index realizes a sum over distinct geometries, as is characteristic of quantum gravity: in this setup, a complete understanding of the structure of the index requires incorporating all geometries compatible with supersymmetry and with finite action. The relevant saddles are spinning gravitational instantons, that always possess an asymptotic Euclidean time circle of period $\beta$, a globally-defined Killing spinor and with chemical potentials constrained by a complex linear relation~\cite{Cabo-Bizet:2018ehj}. This naturally leads to complexified solutions of the field equations, where the complexification can manifest in components of the metric or in the values of certain conserved charges. In this contribution, we encountered many such complex geometries in both gauged (AAdS spaces) and ungauged (AF spaces) supergravity, with and without vector multiplet couplings, and analyzed their role in the gravitational index. 
Physical (Lorentzian) BPS black holes are recovered in the extremal limit, $\beta \to \infty$, after an appropriate Wick-rotation back to real time, thereby clarifying how supersymmetric black holes contribute to the index. This addresses the first of the five general questions about the properties of the gravitational index posed at the end of section~\ref{sec:Grav_path_int}.

The study of the gravitational index advances, for instance, by addressing the other questions raised at the end of the Introduction. In this Thesis, we contributed to this program by taking two concrete steps in this direction:
\begin{description}
\item[Precision holography.] We studied quantum corrections to the thermodynamics of AdS$_5$ black holes arising from four-derivative corrections to the supergravity action, showing perfect consistency with the predictions from Cardy-like limit of the superconformal index of the dual SCFT beyond the leading order in the large-$N$ limit. A detailed summary of results was given in sec.~\ref{sec:Conclusions}.
\item[Euclidean saddles of the supersymmetric index.] We constructed and analyzed five-dimensional Euclidean saddles of the supersymmetric index with asymptotically flat boundary conditions. These saddles have been classified by their rod structure that encodes the relevant topological data. In addition to finite-temperature saddles reproducing the contribution of BPS black holes with non-spherical horizon topology (such as black rings and black lenses), we uncovered saddles with bubbling or spindle topologies outside the horizon, orbifold geometries, and saddles with shifted chemical potentials. These competing contributions point to a potentially intricate phase structure of the index. We studied in detail saddles with one single horizon, and those with one horizon and a bubble outside of it. We found that these Euclidean saddles interpolate between extremal black holes (with $S^3$, $S^1 \times S^2$ or lens space horizons) and horizonless solitons upon varying the inverse temperature and performing suitable analytic continuations. We also explored localization techniques that could provide a systematic and efficient method to compute their on-shell actions. A more detailed summary of our results can be found in section~\ref{sec:conclusions}. 
\end{description}

Further elaborating along these directions, we now conclude by mentioning some interesting open problems in the broader program that call for additional study:
\begin{itemize}

\item It would be valuable to carry out a more precise comparison between the predictions for the gravitational index with asymptotically flat boundary conditions obtained here and computations of the corresponding microscopic index directly in string theory. Such a comparison could shed light on the role (if any) of horizonless solitons in the gravitational path integral, and, more in general, help establish which saddles genuinely contribute to the index and which should be excluded. A possible guide in this direction is the decoupling limit formulated in~\cite{Boruch:2025qdq}. There, the authors constructed a five-dimensional black string by uplifting the four-dimensional saddles of~\cite{Boruch:2023gfn} on a circle of radius $R_M$, and showed that its action reproduces the leading-order behavior of a microscopic index given by the elliptic genus of the MSW CFT$_2$~\cite{Maldacena:1997de}, which is expected to capture at least one sector of the full asymptotically flat index.
Remarkably, this agreement holds even though the gravitational configuration considered does not possess an AdS$_3$ throat (unlike the extremal solution). The reason for this can be understood by taking the following decoupling limit: $\beta \to +\infty$ with $\beta / R_M$ fixed, while zooming onto the near-horizon region. Here, the geometry retains a finite temperature after decoupling the asymptotically flat region, resulting in a complex BTZ black hole (rotating on $S^2$), naturally interpreted as the true gravitational saddle of the dual CFT$_2$ elliptic genus~\cite{Maldacena:1997de,Dijkgraaf:2000fq}. It would be interesting to investigate an analogous decoupling limit for the Euclidean saddles constructed in part~\ref{part_three}, to explore their possible relation to some SCFT$_2$ elliptic genera. This may require uplifting them to six-dimensional black strings following~\cite{Gutowski:2003rg}.

\item

If the gravitational index truly behaves like an ordinary supersymmetric index, its temperature independence should persist even once quantum corrections are included. Testing this requires evaluating corrections from one-loop (and possibly higher-loop) fluctuations of light fields around each saddle and verifying that the result remains independent of $\beta$, a highly non-trivial check of quantum gravity. To do so, two main approaches have been developed in the literature.
The first, pioneered by Sen and collaborators starting with~\cite{Banerjee:2010qc}, employs heat-kernel methods directly in the gravitational description. This technique has been applied to asymptotically flat index saddles in~\cite{H:2023qko,Anupam:2023yns} (see also~\cite{Bobev:2023dwx} for a recent holographic application). The second approach is part of the broader program of applying supersymmetric localization to the supergravity path integral, as mentioned in section~\ref{sec:Grav_path_int}. Here, one exploits index theorems and fixed-point techniques, based on the Atiyah-Singer index theorem, to reduce one-loop determinants to localized contributions from fixed submanifolds (see e.g.\cite{Jeon:2018kec,Hristov:2021zai} for applications in both AF and AdS spaces).
This method, in principle giving access to all-loop corrections, remains technically and conceptually challenging, thus requiring further study. In particular, it relies on an off-shell formulation of supergravity (reviewed in part~\ref{part_two} for unrelated reasons), together with a consistent gauge-fixing of local symmetries (including diffeomorphisms).

\item

As confirmed by our analysis, complex configurations are ubiquitous in index computations as candidate saddles. However, not all of them are expected to actually play a role in the path integral, and it would be important to clarify what are the criteria for a complex geometry to be counted as a genuine saddle, discarding the unphysical ones. A possible criterion to address this issue has been proposed by Witten~\cite{Witten:2021nzp}, building on ideas of Konstevich-Segal~\cite{Kontsevich:2021dmb} (KSW criterion). It states that a metric is allowable if a consistent quantum field theory can be defined on the corresponding spacetime. This requires that all (possibly complex) $p$-forms have positive-definite kinetic terms,
\begin{equation}
{\rm Re} \left[ \sqrt{|g|} \,g^{\mu_1\nu_1} g^{\mu_2\nu_2} ... \,g^{\mu_p\nu_p}F^{(p)}_{\mu_1\mu_2...\mu_p} F^{(p)}_{\nu_1\nu_2...\nu_p} \right] >0\,,
\end{equation} 
where $g_{\mu\nu}$ is a complex-valued metric on a smooth $D$-dimensional manifold ${\cal M}$ and $F^{(p)}$ is a non-trivial $p$-form ($0 \leq p \leq D$). This implies that at each point of ${\cal M}$ one can find a basis of the tangent space that diagonalizes the metric with complex eigenvalues $\lambda_\mu$, which must satisfy
\begin{equation}
\sum_{\mu = 1}^D |{\rm Arg} \lambda_\mu | < \pi\,,
\end{equation}
where ${\rm Arg}\lambda$ is the principal value of the argument of $\lambda$. 

The criterion was analyzed in a supersymmetric setting in~\cite{BenettiGenolini:2025jwe,BenettiGenolini:2026raa,Krishna:2026rma}, where it was tested on a variety of solutions and compared with other physical requirements, such as geometric consistency conditions for Euclidean geometries and the constraints needed to ensure a convergent, well-defined index (which fix the signs of the real or imaginary parts of the associated chemical potentials), finding agreement in many instances.

An alternative criterion was proposed in~\cite{Aharony:2021zkr} in the context of AdS$_5 \times S^5$ compactifications of Type IIB supergravity, dual to ${\cal N}=4$ SYM. There, the authors studied non-perturbative contributions from wrapped Euclidean branes around index saddles, specifically, D$3$-branes wrapping an $S^1$ inside the three-sphere of the external spacetime and an $S^3$ inside the internal five-sphere. Embedding such a probe brane in a supergravity background of action $I$ yields a contribution to the gravitational partition function of the form ${\rm e}^{-I} {\rm e}^{i I_{{\rm D}3}}$, where $I_{\rm D3}$ is the brane action. This new contribution is then suppressed relative to the pure supergravity term ${\rm e}^{-I}$ provided that
\begin{equation}
{\rm Im}\left[ I_{{\rm D}3}\right] >0\,.
\end{equation} 
Applying this condition shows that many candidate saddles are in fact excluded, while the remaining ones agree with the SCFT analysis, where the relevant Bethe roots of the index are known.

The ultimate criterion to decide whether a candidate saddle does or does not contribute to the path integral would be a rigorous study of the path-integral integration contour using Picard-Lefschetz theory, to determine whether the saddle is crossed or not by it.\footnote{For a recent argument involving mini-superspace approximations and contour arguments applied to AdS black holes see~\cite{Mahajan:2025bzo,Singhi:2025rfy,Ailiga:2025osa} and references therein.} However, this is a hard question, and perhaps one may need to address it starting from a simplified setup first. Therefore, no definitive conclusion has yet been reached regarding which allowability criterion most efficiently applies to index saddles, and it is still possible that a more general framework will be required. Analogously to the analysis of~\cite{Aharony:2021zkr}, extending the comparison with microscopic computations to more general constructions, possibly even in the absence of a holographic dual, would help clarify this issue.


\item

Among the Euclidean saddles constructed in part~\ref{part_three}, we explicitly computed the action only for a relatively small subset. A natural question is whether localization techniques can be used to evaluate the on-shell action of our five-dimensional solutions systematically and more efficiently. In principle, this would allow one to determine the action just based on the rod structure and boundary conditions. 
For a general rod structures, however, the method discussed in sec.~\ref{sec:equiv_5d} must be extended. The difficulty arises when different ${\rm U}(1)$ isometries degenerate and the gauge field $A$ is not globally defined, in which case a patch-wise formulation of the localization argument is required~\cite{Colombo:2025yqy}. A further complication arises from the fact that the finite-temperature supersymmetric solutions presented here do not seem to admit a real positive-definite metric, while the standard localization theorem is formulated for Riemannian manifolds.
In the holographic setting, such a framework could help predict, purely from topological data, the action (or entropy) of new geometries, such as the AdS counterparts of the bubbling saddles studied in this Thesis, whose existence in gauged supergravity remains unclear. 

Finally, it would be interesting to apply equivariant localization techniques to off-shell and higher-derivative supergravity. While finding the higher-derivative corrections to two-derivative solutions is in general very hard, use of equivariant localization may allow to bypass this step and evaluate the higher-derivative on-shell action  by knowing just minimal information about the solution, that may possibly be fixed at the two derivative level. 
 Application to supersymmetric non-extremal black hole saddles may allow to extend the recent results of~\cite{Chowdhury:2024ngg,Chen:2024gmc,Cassani:2024tvk} for asymptotically flat black holes (including small ones), and part~\ref{part_two} for asymptotically AdS black holes, beyond first-order in the corrections.
\end{itemize}



\newpage

\subsection*{Acknowledgments}

The writing of a PhD thesis is, for all, a moment to pause for a few weeks and take look backwards at the journey concluded: the things we have learned, the people we have met, the places we have visited. Looking back, I feel grateful for all the experiences and for the personal growth that came with them, and I wish to thank everyone who, in one way or another, played a part in this adventure.

\smallskip

My first thanks go to the University of Padua, where I have spent the last nine years -- almost a third of my life so far. This relationship has not always been simple, but here I found friends and mentors, collected many experiences, felt frustration, worked hard, and had fun. More than anywhere else, here I had the chance to grow and shape myself. I will always carry with me the baggage of lessons learned in these many years, which made me not only the scientist but also the person I am today.

My second thanks go to Davide, who welcomed me more than four years ago as a (probably too chaotic) Master’s student and has guided me ever since, through positive and negative results, papers read and discussed together, seminars and conferences, Master’s and doctoral theses. Day after day, he has supported me with constant care, wisdom, and empathy. Thank you for everything you have taught me, for the soft and hard skills you helped me develop, and for the results we achieved together.

\smallskip

I also wish to thank Alejandro, whose constant collaboration with Davide and me has played a fundamental role in my PhD journey. His insights and contributions have helped me refine my working method and inspired my curiosity toward new directions. Thank you for your support, for your help, and for the mutual trust we have built along the way. 

\smallskip

A sincere thank you goes to all the PhD fellows, postdocs, and professors I had the pleasure of interacting with over these three years, both within the hep-th community and beyond. Among them, a special thanks goes to Stefano for his support and collaboration during the past year. I am also grateful to the entire String Theory group at the University of Padua, and to all its members who have come and gone in these three years: beyond those already mentioned, thanks to Gianguido, Luca, Fabio A., Alessandra, Gianluca, Maxim, Colin, Alfredo, Yixuan, Salvo, Flavio, Matteo, Niccolò, Davide R., Fabio B., Francesco, and Gianmarco, for making the group such a stimulating place to work. I also thank the Theoretical Physics groups at KU Leuven and King's College London for hosting me for some time as a PhD student. 

\smallskip

I would also like to thank Roberto Emparan and Dario Martelli for carefully reading the first version of this Thesis and for their insightful comments and feedback, which helped me refine this work. 

\smallskip

A special mention goes to all the PhD companions who inhabited the infamous {\it Room 382} at the DFA during these years. I cannot name you all -- over three years, with 24 desks, I have seen too many faces for that -- but I am grateful for the friendships that were born among those desks, cemented by our ongoing feud with the Department over the room’s population and by the countless hours we spent together each day. 

Thanks to all the people I have met in Padua over these many years, whom I am proud to call friends. Once again, it would be impossible to mention you all in just a few lines, so let me just name those who have been a constant presence in my life in recent years: thank you Gabriele, Margherita, Tommaso, Giovanni, Alessandro, Francesco I., Francesco F., Giorgio, Elena, Francesca, and also Davide, Pauline, Anna, Chiara, Giacomo, Niccolò, Bogdan, Leonardo, Andrea, Mattia, Gianluca -- hopefully I am not missing anyone. A separate thank you goes to my long-term friend and partner (I know you hate when I use these “grown-up terms”) Michell, whose support has always been unconditional, especially in those many moments when academic life collides less smoothly with personal life.

\smallskip

A final thanks, though surely insufficient, goes to the people who truly made this adventure possible with their support: my family. To my mom Cristiana, my dad Giancarlo, and my brother Tommaso, who, beyond acting as a reliable ``funding trust'' to supplement the scarce PhD income when needed, have always stood by me in difficult moments and while facing important decisions, and who taught me all those aspects of life that the University does not cover, those that really matter. A special thought goes to my late grandmother Marisa, to whom I wish to dedicate this manuscript, for the example of strength and independence in hardship that she has always set, helping me become the man I try to be. Thanks also to my uncle Piero and aunt Raffaella, and to my cousins Gianluca and Marco, for their support and help since my earliest years. And finally, a mention for the youngest addition to the family, Gregorio, who tried so hard to be born just in time to make it into these acknowledgments.

%% file: appendix_A.tex
\chapter{The multi-charge Cardy-like formula from equivariant integration}\label{sec:field_th_section}

In this section we provide a derivation of the flavoured Cardy-like formula \eqref{eq:index_asympt}.
One way to prove~\eqref{eq:index_asympt} would be to extend the three-dimensional effective field theory approach of~\cite{Cassani:2021fyv,ArabiArdehali:2021nsx} to the flavoured case.
Here, however, we choose a different route and present a quicker, though more formal, way to reach the same result, which extends the equivariant integration of the anomaly polynomial presented in \cite{Ohmori:2021dzb} (see also \cite{Nahmgoong:2019hko}) to the flavoured case. 
Equivariant integration of anomaly polynomials is a technique that has already proven effective for different scopes, such as  obtaining the anomaly polynomial of lower-dimensional theories~\cite{Benini:2009mz,Alday:2009qq}, or reproducing the supersymmetric Casimir energy~\cite{Bobev:2015kza}. 
Since the extension we present is straightforward, we will focus on the essential steps of the procedure and make a few comments on its rationale, while referring to the above papers for details.

We place our SCFT in a Euclidean background $\mathcal{M}_4$ comprising Abelian gauge fields $\hat A^I$, $I=1,\ldots,n+1$, coupling to the ${\rm U}(1)^{n+1}$ global currents. The anomaly polynomial $\ext{P}$ is a six-form defined on an extension $\mathcal{Y}_6$ of $\mathcal{M}_4$, that reads: 
\be\label{anomaly_poly}
\ext{P} \,=\, \frac{1}{6}\,{\rm Tr}\,\bigg( \frac{\ext{F}}{2\pi}\bigg)^3 - \frac{1}{24}\,  {\rm Tr}\,\bigg( \frac{\ext{F}}{2\pi}\bigg) \, \ext{p}_1(T\mathcal{Y}_6)\,,
\ee 
where $\ext{F} = \diff \ext{A}$ and $\ext{A} = \ext{A}^I Q_I$ is the extension to $\mathcal{Y}_6$ of the  ${\rm U}(1)^{n+1}$   connection $\hat A = \hat A^IQ_I$ on $\mathcal{M}_4$,
 while
 \be
   \ext{p}_1 = -\frac{1}{8\pi^2} \ext{R}_{ab}\wedge \ext{R}^{ba}
 \ee 
 is the first Pontryagin class defined out of the Riemann curvature two-form $\ext{R}_{ab}$ of $\mathcal{Y}_6$.

We next specify the essential features of the background of interest to us, namely its topology and the group action. We take a space that topologically is $\mathcal{M}_4= S^3\times S^1$. 
 Then we choose a smooth six-dimensional extension with topology $\mathcal{Y}_6 = \mathcal{N}_4 \times D_2$, such that 
   $\partial \mathcal{N}_4 = S^3$ and $\partial D_2 = S^1$. $D_2$ is given the shape of a cigar.
  The gauge fields $\ext{A}^I$ also need to be regular on $\mathcal{Y}_6$. In addition to the ${\rm U}(1)^{n+1}$ symmetry bundle, we consider the ${\rm U}(1)^2$ isometries corresponding to rotation in $S^3$ and the ${\rm U}(1)_{\mathsf T}$ isometry along $S^1$ (where ``${\mathsf T}$'' stands for ``thermal''); the circles defined by the orbits of the respective Killing vectors are non-trivially fibred, and the connection of the fibration contributes to the curvature two-form $\ext{R}_{ab}$. Overall, we thus have a ${\rm U}(1)^2\times {\rm U}(1)_{\mathsf T}\times {\rm U}(1)^{n+1}$ group action on $\mathcal{Y}_6$. We assume that $\mathcal{Y}_6$ can be chosen so that the only fixed point of this action is at the origin of the six-dimensional space; this is possible because $\mathcal{Y}_6$ has {\it two} dimensions more than $\mathcal{M}_4$, so that both $S^3$ and $S^1$ can be made cobordant to the empty set.

The final step is to implement equivariant integration of the anomaly polynomial on $\mathcal{Y}_6$. In order to do so, we assume we can promote the characteristic classes appearing in~\eqref{anomaly_poly} to equivariant classes with respect to the group action. Then we associate complex equivariant parameters $\omega_1,\omega_2$ to the ${\rm U}(1)^2$ rotations, $\mu_{\mathsf T}$ to the ${\rm U}(1)_{\mathsf T}$ shifts and $\varphi^I$ to the ${\rm U}(1)^{n+1}$ action. 
It follows that the Killing vector $K_{\rm eq}$ appearing in the equivariant differential $\diff+ 2\pi \, \iota_{K_{\rm eq}}$ reads
$K_{\rm eq} = \mu_{\mathsf T} \partial_\tau+\omega_1\partial_{\varphi_1} + \omega_2\partial_{\varphi_2}$, where $\partial_{\varphi_1}$, $\partial_{\varphi_2}$ generate the ${\rm U}(1)^2$ rotations in $S^3$ while $\partial_\tau$ advances the coordinate $\tau$ parameterizing $S^1$; here all angular coordinates are taken $2\pi$-periodic. 
We should identify the vector $K_{\rm eq}$ specifying the equivariant action with the Killing vector $K$ obtained by taking suitable bilinears of the Killing spinor ensuring supersymmetry of the background. Up to an irrelevant proportionality constant, the supersymmetric Killing vector in the background of interest reads $K = - 2\pi i\, \partial_\tau+\omega_1\partial_{\varphi_1} + \omega_2\partial_{\varphi_2}$, where $\omega_1,\omega_2$ are precisely the chemical potentials appearing in the definition of the superconformal index~\cite{Cabo-Bizet:2018ehj,Cassani:2021fyv}. Therefore we see that we should fix $\mu_{\mathsf T}=- 2\pi i$, while the remaining equivariant parameters are identified with the chemical potentials appearing in the superconformal index.

We then apply the Atiyah-Bott-Berline-Vergne fixed point theorem, stating that the integral of an equivariantly closed form only receives contributions from the fixed points of the group action, to evaluate the integral
\be\label{integration_anomaly_p}
I_{\rm eq}\equiv -2\pi i \int_{\mathcal{Y}_6}\ext{P} \,=\, -2\pi i\, \frac{\ext{P}|_0}{\ext{e}(T\mathcal{Y}_6)|_0}\,,
\ee
where $\ext{e}(T\mathcal{Y}_6)$ is the equivariant Euler class of $\mathcal{Y}_6$ and $|_0$ denotes the zero-form contribution of the equivariant class at the fixed point.
Close to the fixed point, $\mathcal{Y}_6$ can be modelled as $\mathbb{R}^6$, with the ${\rm U}(1)^2\times {\rm U}(1)_{\mathsf T}$ action rotating the three orthogonal planes. We can then evaluate the equivariant classes in~\eqref{anomaly_poly} using the standard moment map and symplectic form on $\mathbb{R}^6$ (see e.g.~\cite[App.\:A]{Bobev:2015kza}).
This boils down to replacing the Chern roots of the characteristic classes with the equivariant parameters, i.e.\ implementing the rules
\be
{\rm Tr}\,\bigg( \frac{\ext{F}}{2\pi}\bigg)^3\,\bigg|_0\,=\,  k_{IJK}\,\varphi^I\varphi^J\varphi^K\,,  \qquad\quad {\rm Tr}\,\bigg( \frac{\ext{F}}{2\pi}\bigg)\bigg|_0\,=\,  k_I\varphi^I \,,
\ee
\be
\ext{p}_1(T\mathcal{Y}_6)|_0 \,=\, \omega_1^2+\omega_2^2+\mu_{\mathsf T}^2\,,\qquad\quad  \ext{e}(T\mathcal{Y}_6)|_0 \,=\,  \omega_1\omega_2\mu_{\mathsf T}\,,
\ee
where we used the definitions~\eqref{def_anomaly_coeff} of the 't Hooft anomaly coefficients.
Plugging this in \eqref{integration_anomaly_p} and recalling that we are setting $\mu_{\mathsf T} = -2\pi i$, we find that $I_{\rm eq}$ precisely reproduces the expression for $I$ in \eqref{eq:index_asympt}. 
The requirement that the Killing spinor on $\mathcal{M}_4$ extends to a well-defined spinor on $\mathcal{Y}_6$  (that in particular is anti-periodic at the tip of the cigar $D_2$) leads to the constraint~\eqref{eq:linearconstraint}, the argument being analogous to the one given in~\cite{Cabo-Bizet:2018ehj} for the five-dimensional supergravity bulk filling of $\mathcal{M}_4$. This concludes the derivation.


\chapter{Dictionary with superconformal anomalies}
\label{app:B}

\section{R-symmetry anomaly coefficients}
\label{sec:holographicdictionary}

In this appendix, we derive the dictionary between the dimensionless gravitational quantities $g^3 G$ and $\alpha g^2$ and the dual superconformal anomaly coefficients.
We stress that this dictionary is universal, namely it is valid for any holographic $\mathcal{N}=1$ SCFT; the details of the SCFT only affect the explicit expression of the anomaly coefficients in terms of the field theory data. 

$\mathcal{N}=1$ SCFT's have a superconformal anomaly controlled by coefficients $\aa,\cc$. This has been reviewed in chapter~\ref{sec:background}. 
For convenience, we repeat here the relation between the R-symmetry anomalies and the central charges for the four-dimensional superconformal field theories,
\be
\label{relacTrR}
\aa = \frac{3}{32}(3\,{\rm Tr}\mathcal{R}^3 - {\rm Tr}\mathcal{R})\,, \qquad \cc = \frac{1}{32}(9\,{\rm Tr}\mathcal{R}^3 - 5\, {\rm Tr}\mathcal{R})\,,
\ee
with $k_{RRR} = {\rm Tr} {\cal R}^3$ and $k_R = {\rm Tr} {\cal R}$ denoting the cubic and linear 't Hooft anomaly coefficients for the canonically normalized superconformal R-symmetry. In the large-$N$ expansion, these anomaly coefficients are related to the dimensionless gravitational quantities $g^3 G$ and $\alpha g^2$. Let us determine the precise dictionary in our setup. We proceed as in~\cite{Cremonini:2008tw} and use results of \cite{Fukuma:2001uf} (see also~\cite{Baggio:2014hua,Bobev:2021qxx}). 

The Weyl anomaly can be read off from the logarithmically divergent term in the on-shell action~\cite{Henningson:1998gx}. 
It is sufficient to switch off the gauge field and only consider the gravitational part of the bulk action, which in general may take the form
\be
e^{-1} \mathcal{L} \,=\, \frac{1}{16\pi G_{\rm{eff}}} \left(R+12g_{\rm eff}^2 +\alpha_1 R^2 + \alpha_2 R_{\mu\nu}R^{\mu\nu} + \alpha_3 R_{\mu\nu\rho\sigma}R^{\mu\nu\rho\sigma} \right)\,,
\ee
where, importantly, $G_{\rm eff}$ and $g_{\rm eff}$ may also contain terms of order $\alpha$.
It is easy to check that the radius $\ell$ of the AdS$_5$ solution is given by
\be
\ell \,=\, \frac{1}{g_{\rm eff}} \left[1- \frac{g_{\rm eff}^2}{3} (10\alpha_1+2\alpha_2+\alpha_3) \right]\,.
\ee
This equality, as well as the following ones, is meant to hold at linear order in the $\alpha$ corrections.
Identifying the logarithmically divergent piece of the action evaluated on a general solution with the conformal anomaly of the dual SCFT, one obtains the corrected holographic formulae for the anomaly coefficients,
\begin{equation}
\begin{aligned}
\aa \,&=\, \frac{\pi\ell^3}{8G_{\rm{eff}}} \left[ 1-\frac{4}{\ell^2} (10\alpha_1+2\alpha_2+\alpha_3) \right] \,,\\[1mm]
\cc \,&=\, \frac{\pi\ell^3}{8G_{\rm{eff}}} \left[ 1-\frac{4}{\ell^2} (10\alpha_1+2\alpha_2-\alpha_3) \right] \,.
\end{aligned}
\end{equation}

In the main text, we focus on the supersymmetric action~\eqref{eq:4daction2}, where the only purely gravitational four-derivative term is the Gauss-Bonnet one. This fixes 
\be
\alpha_1 =\alpha\lambda_1\,,\quad \alpha_2 =-4\alpha\lambda_1\,,\quad \alpha_3 =\alpha\lambda_1\,,
\ee
while from the Ricci scalar and cosmological constant term we read 
\be
\begin{aligned}\label{eq:Geff}
G_{\rm eff}=\,&G\left(1-4\lambda_2 \alpha g^2\right)\,,\\[1mm]
g_{\rm eff} =\,& g\left(1-5\lambda_1\alpha g^2\right)\,.
\end{aligned}
\ee

Plugging these values in the above expressions, we obtain the AdS radius
\be\label{eq:ads_four_der}
\ell \,=\, \frac{1}{g} \left(1+ 4\lambda_1\alpha g^2 \right)\,,
\ee
and the anomaly coefficients
\begin{equation}\label{a_c_high_der}
\begin{aligned}
\aa \,&=\, \frac{\pi}{8Gg^3} \left( 1 +4\lambda_2 \alpha g^2  \right)\,,\\[1mm]
\cc \,&=\, \frac{\pi}{8Gg^3} \left( 1 +4(2\lambda_1+\lambda_2)\alpha g^2 \right) \,.
\end{aligned}
\end{equation}
Note that at leading order this matches the familiar two-derivative result $\aa = \cc = \frac{\pi}{8G g^3 }$.

Inverting \eqref{relacTrR}, we can also give the holographic expression of the R-symmetry anomaly coefficients:
\be
\begin{aligned}\label{TrRintermsofmu}
{\rm Tr}\,\mathcal{R}^3 \,&=\, \frac{16}{9}(5\aa-3\cc)\,=\, \frac{4\pi}{9G g^3} \left( 1 -4(3\lambda_1-\lambda_2) \alpha g^2  \right)\,, \\[1mm]
{\rm Tr}\,\mathcal{R} \,&=\, 16(\aa-\cc)\,=\, -\frac{16\pi \alpha\lambda_1}{G g}   \,.
\end{aligned}
\ee

We can check the dictionary above by matching the non-invariance of the bulk action~\eqref{eq:4daction2} under a U(1)$_R$ gauge transformation with the R-current anomaly~\cite{Witten:1998qj}. One can see that the boundary value $A_i$, $i=0,\ldots,3$, of the bulk gauge field is to be identified with the background gauge field $\hat A_i$ coupling canonically to the R-current of the dual SCFT as $A_i = \frac{2}{\sqrt{3}\,g}\,\hat A_i$. Making the gauge transformation $\delta A_i = \frac{2}{\sqrt{3}g}\,\partial_i \lambda$, the bulk action~\eqref{eq:4daction2} transforms as
\begin{equation}
\begin{aligned}
\delta_\lambda S
\,=&\,\frac{1}{24 \pi g^3 G}\int_{\partial {\cal M}} \diff^4x \,\hat{e}\, \lambda \left( -\frac{\tilde{c}_3}{9}\,\hat{\epsilon}^{\,ijkl}\hat{F}_{ij}\hat{F}_{kl} \,-\, \frac{\alpha\lambda_1 g^2}{2}\,\hat{\epsilon}^{\,ijkl}\hat{R}_{ijab}\hat{R}_{kl}{}^{ab}\right)\,,
\end{aligned}
\end{equation}
where the hat symbol denotes that the quantities are evaluated at the boundary.

Through the identification $\rme^{i S}=Z_{\rm grav} = Z_{\rm CFT}$, this should match the anomalous variation of the field theory partition function $Z$ under the background gauge transformation $\delta_\lambda\hat A_i =\partial_i\lambda\,$,
\begin{equation}
\begin{aligned}
\delta_\lambda \log Z_{\rm CFT} \,&=\, -i  \int_{\partial {\cal M}} \diff^4 x\, \hat{e} \, \lambda\, \nabla_i J^i \\[1mm]
\,&=\, -i \int_{\partial {\cal M}} \diff^4 x\, \hat{e} \, \lambda \left( \frac{5\aa-3\cc}{27\pi^2} \,\frac{1}{2}\,\epsilon^{ijkl}\hat{F}_{ij} \hat{F}_{kl} \, + \,  \frac{\cc-\aa}{24\pi^2}\, \frac{1}{2}\,\epsilon^{ijkl}\hat{R}_{ijab} \hat{R}_{kl}{}^{ab}\right)\,.
\end{aligned}
\end{equation}
Imposing the holographic matching condition $\delta S= -i\, \delta\log Z_{\rm CFT}$ and recalling the value of $\tilde c_3$ given in \eqref{ctildes}, we obtain the following $\mathcal{O}(\alpha)$ dictionary between the coefficients:
\be
5\aa-3\cc \,=\, \frac{\pi \tilde{c}_3}{4 Gg^3} = \frac{\pi}{4 Gg^3}\left( 1-4(3\lambda_1-\lambda_2)\alpha g^2\right)\,,\qquad \aa-\cc \,=\, -\frac{\pi \alpha \lambda_1}{ Gg}\,,
\ee
which perfectly agrees with the dictionary found by considering the Weyl anomaly. This can be seen as a consistency check for the supersymmetry of our action~\eqref{eq:4daction2}.

\section{$a$-maximization and its four-derivative 5d gravity dual}\label{sec:amaxim_section}

Here, we discuss the gravity dual of $a$-maximization at first order in the higher-derivative corrections. We extend the analysis of~\cite{Tachikawa:2005tq,Hanaki:2006pj} by including the corrections controlled by $\widetilde\lambda_{IJK}$, which allows us to reproduce the most general corrections to the $\aa$ and $\cc$ Weyl anomaly coefficients.

\subsection{$a$-maximization and the large-$N$ expansion}\label{amax_at_largeN}

We start by relating the cubic and linear 't Hooft anomaly coefficients $k_{IJK}$ and $k_I$  to the Weyl anomaly coefficients $\aa$ and $\cc$ of the $\mathcal{N}=1$ SCFT, at first subleading order in the large-$N$ expansion. 

As above, in any $\mathcal{N}=1$ SCFT, the Weyl anomaly coefficients can be expressed as \eqref{relacTrR}
In this section we prefer to denote the superconformal R-symmetry by $Q_R$, to distinguish it more clearly from a generic charge $Q_I$. The exact R-symmetry can be determined via the principle of $a$-maximization \cite{Intriligator:2003jj}. One starts from a trial R-charge $Q_R^{\rm trial}(s) = s^IQ_I$, where the coefficients $s^I$ in the linear combination of the charges need to respect the constraint 
\be\label{eq:constr_rs}
r_Is^I=1\,,
\ee 
so that the supercharge has the canonical R-charge $-1$ (recall our definition of $r_I$ in~\eqref{eq:comm_charges}). Then the values $\bar s^I$ such that $Q_R =\bar{s}^IQ_I $ is the exact superconformal R-charge are identified by maximizing the trial anomaly coefficient
\be\label{atrial}
\aa_{\rm trial}(s) \,=\,  \frac{3}{32}\left( 3k_{RRR}^{\rm trial}(s)  - k_R^{\rm trial} (s)\right) \,=\,    \frac{3}{32}\left(3k_{IJK}s^Is^Js^K   - k_Is^I\right)\,.
\ee

For a holographic $\mathcal{N}=1$ SCFT with a weakly-coupled gravity dual, we can study the problem of $a$-maximization in the large-$N$ expansion by recalling the expansion~\eqref{eq:expansion_k's} of the 't Hooft anomaly coefficients.
Suppose we have solved the maximization problem at leading order, i.e.\ we have found the values $\bar s^I$ such that the function $\aa^{(0)}_{\rm trial} = \frac{9}{32}k^{(0)}_{IJK}  s^Is^Js^K$ is maximized under the constraint $r_I s^I=1$ (here we are using the fact that $k_I$ is subleading in the large-$N$ expansion, which also gives $\aa^{(0)}=\cc^{(0)}$). When including the subleading terms  $k^{(1)}_{IJK}$ and $k^{(1)}_I$, the function $\aa_{\rm trial}$ in \eqref{atrial} will be maximized at some corrected values $\bar s^I + \delta s^I$.  However, the corrections $\delta s^I$ only contribute to the expression for the anomaly coefficient at quadratic order,  hence we immediately conclude that at first order,
\be\label{eq:maximized_a}
\aa  \,=\,   \frac{3}{32}\left(3k_{IJK} \bar s^I \bar s^J\bar s^K  - k_I \bar s^I\right)  + \ldots\,.
\ee
 This argument also allows us to express the anomaly coefficient $\cc$ at the same order,
\be\label{eq:c_for_maximized_a}
\cc  \,=\,   \frac{1}{32}\left(9k_{IJK} \bar s^I \bar s^J\bar s^K  - 5k_I \bar s^I\right) + \ldots\,.
\ee

Nevertheless, we find  it useful to implement $a$-maximization at first order in the corrections, as it provides some relations that will be needed in the main text. Before doing so, let us introduce the projectors on the vector subspaces parallel and orthogonal to any R-symmetry $Q_R^{\rm trial}(s)  = s^I Q_I$,
\be\label{projectors_Rsymm}
P_{\parallel\,I}{}^J = r_I s^J\,,\qquad\ P_{\perp\,I}{}^J = \delta_I{}^J - r_I s^J\,.
\ee  
When implementing these projections, we will denote by ``$R$'' the R-symmetry direction, and will append a tilde on all quantities whose indices are projected along the orthogonal flavour directions. For instance, we can decompose the charges  as $Q_I = r_I Q^{\rm trial}_R + \widetilde{Q}_I$, where the $\widetilde{Q}_I = P_{\perp\,I}{}^J  Q_J$ are flavour charges (as they commute with the supercharge $\mathcal{Q}$). Analogously, we can decompose the 't Hooft anomaly coefficients along the R-symmetry and flavour directions, for instance $P_{\perp\,I}{}^{I'} P_{\parallel\,J}{}^{J'} P_{\parallel\,K}{}^{K'} k_{I'J'K'}  =   \widetilde k_{ I RR}\, r_Jr_K$, and so on.

In order to study the extremization problem, it is convenient introduce a Lagrange multiplier $L$ imposing the constraint \eqref{eq:constr_rs}. Then, the ${\aa}_{\rm trial}$ function is given by
\be
{\aa}_{\rm trial}(s, L)\,=\, \frac{3}{32}\left(3k_{IJK}s^Is^Js^K   - k_Is^I\right)+L \left(r_I s^I-1\right)\,,
\ee
and the extremization equations read
\begin{eqnarray}
\frac{\partial {\aa_{\rm trial}}}{\partial s^I}\Big|_{s={\bar s}+\delta {\bar s}}&\,=\,&\frac{3}{32}\left[9k_{IJK}\left({\bar s}^J+\delta {\bar s}^J\right)\left({\bar s}^K+\delta {\bar s}^K\right)  - k_I\right]+L \,r_I=0\,,\\[1mm]
\label{eq:da/dlambda}
\frac{\partial {\aa_{\rm trial}}}{\partial L}\Big|_{s={\bar s}+\delta {\bar s}}&\,=\,&r_I \left({\bar s}^I+\delta {\bar s}^I\right)-1=0\, .
\end{eqnarray}
Contracting the first with ${\bar s}^I+\delta {\bar s}^I$ and using the second allows us to find the value of $L$. Substituting its value into the first equation, we arrive at
\begin{equation}\label{eq:extremization_a}
\left[\delta_{I}{}^{J}-r_I \left({\bar s}^J+\delta {\bar s}^J\right)\right]\left[9\,k_{JKL}\left({\bar s}^K+\delta {\bar s}^K\right)\left({\bar s}^L+\delta {\bar s}^L\right)-k_{J}\right]\,=\,0\,,
\end{equation}
which just says that
\begin{equation}
9\,{\widetilde k}_{IRR}= {\widetilde k}_{I}\, ,
\end{equation}
in agreement with \cite{Intriligator:2003jj}. Assuming the large-$N$ expansion of the anomaly coefficients reported in \eqref{eq:expansion_k's}, we find that \eqref{eq:extremization_a} at leading order in $N$ reduces to 
\begin{equation}\label{eq:extremization_a_leading_order}
{\widetilde k}^{(0)}_{IRR}\equiv\left[\delta_{I}{}^{J}-r_{I} {\bar s}^J\right]k^{(0)}_{JKL}{\bar s}^K{\bar s}^L\,=\,0\,,
\end{equation}
where the coefficients ${\bar s}^I$ satisfy the constraint $r_I{\bar s}^I=1$. 

Now we consider \eqref{eq:extremization_a} at next-to-leading order. First we note that, because of \eqref{eq:da/dlambda}, the $\delta {\bar s}^I$ satisfy
\begin{equation}
r_I\delta {\bar s}^I=0\,.
\end{equation}
Contracting \eqref{eq:extremization_a_leading_order} with $\delta{\bar s}^I$ and using the above constraint, we deduce
\begin{equation}\label{eq:kdsss}
{k}^{(0)}_{IJK}\delta{\bar s}^I{\bar s}^J{\bar s}^K\,=\,0\, ,
\end{equation}
which justifies why $\delta {\bar s}^I$ does not contribute to the first-order corrections of the anomaly coefficients $\aa$ and $\cc$, as we anticipated. Linearizing \eqref{eq:extremization_a} and using \eqref{eq:kdsss}, one finds that the coefficients $\delta {\bar s}^{I}$ satisfy the following equations
\be\label{eq:ds}
-18k^{(0)}_{IJK}\delta{\bar s}^J{\bar s}^{K}\,=\,\left(\delta_{I}{}^{J}-r_I {\bar s}^J\right)\left(9k^{(1)}_{JKL}{\bar s}^K{\bar s}^L-k^{(1)}_{J}\right)\,,
\ee
where only $n$ are independent, as the contraction with ${\bar s}^{I}$ yields a trivial identity.

In particular, this tells us that $\delta {\bar s}^I=0$ (provided the matrix $k^{(0)}_{IJK}{\bar s}^K$ is invertible) whenever the anomaly coefficients associated to the flavour directions are not corrected. This is precisely the case of the quiver theories considered in sections~\ref{sec:Legendre_transf_gen} and \ref{sec:orbifold_section}, whose anomaly coefficients satisfy \eqref{eq:exp_anomaly_coeff}.

\subsection{Supergravity dual}\label{app:sugradual}

It was demonstrated in~\cite{Tachikawa:2005tq} that the gravity dual of $a$-maximization at the two-derivative level is the extremization of the supergravity prepotential, with the trial $a$-function being proportional to $\frac{1}{({\rm prepotential})^{3}}$; this is equivalent to the conditions for a supersymmetric AdS$_5$ solution. In the following we briefly review the argument, applied to the  U(1) gauging with constant parameters $g_I$ we have been considering in this paper, and provide the $\aa$ and $\cc$ Weyl anomaly coefficients at linear order in the corrections. Our new ingredient is that we include the corrections controlled by $\widetilde\lambda_{IJK}$.

\paragraph{Supersymmetric AdS$_5$ vacuum.}
We start by briefly recalling the conditions for a supersymmetric AdS$_5$ solution in the two-derivative theory. We consider the supergravity theory~\eqref{eq:finalL}, with $\alpha=0$. The susy transformation of the gravitino, found by plugging the on-shell value for the auxiliary fields \eqref{eq:solauxfields} into \eqref{eq:off_shell_grav_vars}, reads 
\be
\begin{aligned}\label{gravitino_var}
\delta\psi_\mu^i \,&=\, \nabla_\mu \epsilon^i + \frac{3}{ 2\sqrt{2}} g_I A_\mu^I \delta^{ij}\epsilon_j + \frac{\ii }{2\sqrt{2}} g_IX^I \delta^{ij}\gamma_\mu\epsilon_j + \ldots  \,,
\end{aligned}
\ee
where the ellipsis denote additional terms that will not be important for the present discussion. 
In order to find an AdS solution we set  $A^I_\mu = 0$ and take  scalars with constant values $X^I = \bar X^I$. 
The gaugino variation requires  that the ``prepotential'' function $g_I X^I$ is extremized with respect to the unconstrained scalars $\phi^x$, $x=1,\ldots,n$, that is 
\be\label{eq_gaugino_var}
g_I  \left.\frac{\partial X^I}{\partial\phi^x} \right|_{X=\bar X} = 0\,.
\ee 
Recall that the constraint satisfied by the scalars can be written as $X_I X^I=1$, with $X_I = C_{IJK}X^JX^K$. 
Then~\eqref{eq_gaugino_var} is solved by requiring that the vectors $g_I$ and $\bar X_I$ are aligned, 
\be\label{scalars_AdS_sol}
g_I =  \sqrt 2\, g \bar X_I  \qquad \Rightarrow \qquad g = \frac{1}{\sqrt2} \,g_I\bar X^I\,.
\ee
The vanishing condition  of the gravitino variation then becomes
\be
\nabla_\mu \epsilon^i + \frac{\ii }{2} g\,  \delta^{ij}\gamma_\mu\epsilon_j \,=\,0\,,
\ee
which is the standard Killing spinor equation in an AdS spacetime with squared radius $\ell^2= 1/g^2$.

\paragraph{Dictionary with field theory quantities.} Studying the gravitino variation \eqref{gravitino_var}, one can see that the supersymmetry spinor parameter at the AdS boundary is only charged (with charge +1 in our conventions) under the symmetry gauged by the linear combination\footnote{See the analysis of~\cite{Cassani:2012ri}, where the bulk susy transformations were expanded near the boundary so as to identify the susy transformations of the non-dynamical conformal supergravity coupling to the field theory at the boundary. In order to compare with that reference, one should use $\delta^{ij}\epsilon_j=-\varepsilon^{ij}\epsilon^j$.
}
\be\label{rel_Acan_AI}
A^{\rm can} \,=\, \frac{3}{2\sqrt2}\, g_I A^I\,.
\ee 
It follows that the spinor parameter has charge  $\frac{3}{2\sqrt2} g_I $ under $A^I$, hence recalling \eqref{eq:comm_charges} we should identify
\be
 r_I =  \frac{3}{2\sqrt2} \frac{g_I}{g}  \,,
\ee
We are going to show below that in the two-derivative supergravity the coefficients $\bar s^I$ determining the superconformal R-symmetry are given by
\be\label{identif_s}
\bar s^I =   \frac{2\sqrt2}{3} \frac{g\bar X^I} {g_J \bar X^J}\,.
\ee
Notice that the identifications  above satisfy the condition $r_I \bar s^I=1$. Using the supersymmetry condition \eqref{scalars_AdS_sol}, the two expressions above can also be written simply as
\be\label{eq:rs_and_X}
 r_I =  \frac{3}{2}  \bar X_I \,,\qquad \quad
\bar s^I =   \frac{2}{3} \bar X^I\,,
\ee
hence the field theory quantities $r_I$ and $\bar s^I$ are related to each other by the field theory dual of the supergravity matrix $\bar a_{IJ}$.

One way to prove \eqref{identif_s} is to look at the supergravity definition of the conserved charges.
Indeed, recalling \eqref{rel_Acan_AI} and the supersymmetric AdS condition~\eqref{scalars_AdS_sol}, we can express
\be
A^I = \frac{2}{3g} \bar X^I A^{\rm can} + (\ldots)^I\,,
\ee
where $(\ldots)^I$ has vanishing contraction with $g_I$ and contains the combinations of gauge fields independent of $A^{\rm can}$. Then the superconformal R-charge is given by integrating the time  component of the R-current obtained by varying the renormalized on-shell action as:
\be
Q_R \,=  \  \chi \int_{\partial \mathcal{M}}\diff^4x\,\frac{\delta S}{\delta A^{\rm can}_t} \,=\, \frac{2}{3g} \bar X^I \chi \int_{\partial \mathcal{M}}\diff^4x\,\frac{\delta S}{\delta A^{I}_t}  \,= \, \frac{2}{3g} \bar X^I  Q_I^{\rm gr}\,,
\ee
where $\chi$ is a coefficient entering in the definition of the charges, that only appears in the intermediate steps. 
Identifying the relation  $Q_R = \frac{2}{3g} \bar X^I  Q_I^{\rm gr}$ thus obtained with the field theory relation $Q_R = \bar s^I Q_I$,  and noting that $Q^{\rm gr}_I = g \,Q_I$ (since $A^I = \hat A^I/g$) we conclude that we must take 
\be
\bar s^I \,=\, \frac{2}{3} \bar X^I  \,= \, \frac{2\sqrt2}{ 3} \frac{ g\bar X^I}{g_J\bar X^J}\,,
\ee
as we wanted to show.

Extending \eqref{identif_s} to hold outside the AdS fixed point, we have
that the leading-order trial anomaly coefficient can be expressed in gravitational variables as
\be
\aa_{\rm trial}^{(0)}(s) \,=\,\frac{9}{32} k^{(0)}_{IJK}s^Is^Js^K   \,=\, \frac{\pi}{2\sqrt2G(g_I X^I)^3}  \,,
\ee
where we used $
k_{IJK}^{(0)} \,=\,  \frac{ 3\pi }{2Gg^3} \, C_{IJK}$ from the dictionary~\eqref{dict_cubic_anom}. This is extremized at the  AdS solution, since the prepotential $g_I X^I$ is; one can also see that it is in fact maximized as a consequence of positive definiteness of the metric on the supergravity scalar manifold~\cite{Tachikawa:2005tq}. The value at the maximum is
\be
\aa^{(0)} = \frac{\pi}{8G g^3}\,,
\ee
which agrees with the leading-order result from the above analysis~\eqref{a_c_high_der}.

\paragraph{Including the first-order corrections.} 
The same argument  used in section~\ref{amax_at_largeN} allows us to extend the result above to include the higher-derivative terms at linear order in $\alpha$.  Namely, as long as we work at first order, we do not need to extremize again, we just have to translate the corrected expressions \eqref{eq:maximized_a}, \eqref{eq:c_for_maximized_a}
in gravitational variables. For $\aa$ we obtain
\be
\aa  \,=\,   \frac{\pi}{8Gg^3} \,C^{(\alpha)}_{IJK} \bar X^I\bar X^J\bar X^K  + \frac{3\pi}{2Gg}\, \alpha\lambda_I \bar X^I     + \mathcal{O}(\alpha^2)\,,
\ee
where we used~\eqref{dict_cubic_anom},  \eqref{dict_linear_anom} for the anomaly coefficients  and \eqref{eq:rs_and_X} for the coefficients $\bar s^I$. Recalling the form of the corrected $C^{(\alpha)}_{IJK}$ in \eqref{correctedCIJK} and \eqref{scalars_AdS_sol}, this becomes
\be\label{a_holo_corrected}
\aa  \,=\,   \frac{\pi}{8Gg^3} \,  \left(1  + \alpha \widetilde\lambda_{IJK}\bar X^I\bar X^J\bar X^K \right)        + \mathcal{O}(\alpha^2)\,.
\ee
Notice that the four-derivative invariant controlled by $\lambda_I$ does not contribute to $\aa$. By setting $\widetilde\lambda_{IJK}=0$ we recover the findings of~\cite{Hanaki:2006pj}, where only the Weyl$^2$ invariant (i.e.\ the one controlled by our coefficients $\lambda_I$) was considered. Here we see that the corrections controlled by $\widetilde\lambda_{IJK}$, which were not included in the analysis of~\cite{Hanaki:2006pj}, also contribute to the result.
Finally, recalling that 
\be
\cc-\aa \,=\, -\frac{1}{16}{\rm Tr}{\cal R} \,=\,  \frac{\pi}{Gg}\, \alpha\lambda_I  \bar X^I   + \mathcal{O}(\alpha^2)\,,
\ee 
we obtain the holographic expression for the corrected $\cc$,
\be\label{c_holo_corrected}
\cc \,=\, \frac{\pi}{8Gg^3} \,  \left(1  +  \alpha \widetilde\lambda_{IJK}\bar X^I\bar X^J\bar X^K   + 8\alpha g^2 \lambda_I  \bar X^I     \right)    + \mathcal{O}(\alpha^2)\,,
\ee
which depends on both types of corrections.

\paragraph{Examples.} When discussing specific examples of AdS/CFT dual pairs, we need to choose the coefficients controlling the higher-derivative corrections in our five-dimensional effective theory in such a way that they match their field theory counterpart, according to the dictionary we have derived. For instance, in order to describe the gravity dual of $\mathcal{N}=4$ SYM we must fix 
\be
\text{$\mathcal{N}=4$ SYM}\quad\longleftrightarrow\qquad \widetilde\lambda_{IJK} =  - g^2 C_{IJK}\,, \qquad \lambda_I=0\,,
\ee
where the overall numerical coefficient is immaterial as it can be set to any other non-zero value by redefining $\alpha$.
We see that the $\widetilde\lambda_{IJK}$ coefficient plays a crucial role here.
 In this way our formulae give the Weyl anomaly coefficients $\aa=\cc=  \frac{\pi}{8Gg^3} \,  (1  -   \alpha g^2 )$; these match the exact expressions $\aa=\cc=\frac{N^2-1}{4}$ upon identifying  $\frac{\pi}{2Gg^3}  = N^2$ and $  \alpha g^2 = 1/N^2$.
 
If instead we wish to describe the quiver theories with 't Hooft anomaly coefficients of the form \eqref{eq:exp_anomaly_coeff}, including the $\mathbb{C}^3/\Gamma$ orbifold theories of section~\ref{sec:orbifold_section}, then we should  take 
\be
   \text{quivers satisfying  \eqref{eq:exp_anomaly_coeff}}\quad\longleftrightarrow\qquad \widetilde\lambda_{IJK} \,=\,   - 12   \lambda_{(I} g_J g_{K)} \,,\qquad \lambda_I = \frac{ g_I}{8\sqrt2 \,g} = \frac18\bar X_I\,,
\ee 
where again we have arbitrarily chosen a convenient overall normalization.
In this case the holographic Weyl anomaly coefficients read
\be
\begin{aligned}
\aa  \,&=\,   \frac{\pi}{8Gg^3} \,  \left(1  -3\alpha g^2   \right)        + \mathcal{O}(\alpha^2)\,,\\[1mm]
\cc \,&=\, \frac{\pi}{8Gg^3} \,  \left(1  - 2 \alpha g^2 \right)        + \mathcal{O}(\alpha^2)\,
\end{aligned}
\ee
and the dictionary with the field theory quantities is $\aa^{(0)}=\frac{\pi}{8Gg^3}$ and $\nu = \frac{2\pi \alpha}{Gg}$. In particular, for the $\mathbb{C}^3/\mathbb{Z}_\nu$ orbifold theories, $\aa^{(0)}= \frac{\nu N^2}{4}$, hence  $\frac{\pi}{2Gg^3}= \nu N^2$ and $\alpha g^2 = \frac{1}{4N^2}$.

\paragraph{Consistency check from Weyl anomaly.} As a consistency check, we can compare the formulae for $\aa$ and $\cc$ given above with those obtained from the analysis of the holographic Weyl anomaly in the presence of four-derivative terms, see e.g.~\cite{Fukuma:2001uf}.

In order to determine $\aa$ and $\cc$, we can ignore all gauge fields and fix the scalars to the value they take in the two-derivative AdS$_5$ solution, $X = \bar X$. This value receives corrections at linear order in $\alpha$, however these only affect the Lagrangian at $\mathcal{O}(\alpha^2)$ since the scalar potential of the two-derivative theory, $\mathcal{V}(X(\phi))$, is extremized with respect to the physical scalars $\phi^x$.  Then the Lagrangian at first order in $\alpha$ takes the form
\be
e^{-1}\mathcal{L} = \frac{1}{16\pi G}\left( R - 2 ( \mathcal{V}(\bar X) + \alpha\mathcal{V}^{(1)}(\bar X)) +  \alpha_1 R^2+\alpha_2 R_{\mu\nu}R^{\mu\nu} + \alpha_3 R_{\mu\nu \rho\sigma}R^{\mu\nu \rho\sigma} \right)\,,
\ee
where $\mathcal{V}^{(1)}$ is the correction to the scalar  potential dictated by four-derivative supergravity, and $\alpha_1,\alpha_2,\alpha_3$ are some coefficients. In our Lagrangian given in section~\ref{sec:final_Lagr}, the only term quadratic in the Riemann curvature after setting to zero the gauge fields is the Gauss-Bonnet term, and the coefficients read 
\be
\alpha_1= \alpha \lambda_I \bar X^I\,, \quad \alpha_2= - 4 \alpha \lambda_I \bar X^I \,, \quad\alpha_3= \alpha \lambda_I \bar X^I\,. 
\ee
 Recalling that $\mathcal{V}(\bar  X)=-6g^2$ and defining $g_{\rm eff}^2$ such that   
 \be
 -6 g_{\rm eff}^2 = (\mathcal{V}(\bar X) + \alpha\mathcal{V}^{(1)}(\bar X)) = -6 g^2 \left (1 - 10 \alpha g^2  \lambda_I \bar X^I  -\frac23 \alpha  \widetilde \lambda_{IJK}\bar X^I\bar X^J\bar X^K\right) \,,
 \ee
 we end up with
\be
e^{-1}\mathcal{L} = \frac{1}{16\pi G}\left( R + 12 g_{\rm eff}^2 +  \alpha_1 R^2+\alpha_2 R_{\mu\nu}R^{\mu\nu} + \alpha_3 R_{\mu\nu\rho\sigma}R^{\mu\nu\rho\sigma} \right)\,.
\ee
One also finds the corrected AdS radius, 
\begin{equation}\label{eq:corrected_AdS_rad}
\ell = \frac{1}{g}\left(1+ 4\alpha\,g^2\,\lambda_I \bar X^I + \frac{1}{3}\alpha\, \widetilde\lambda_{IJK}\bar X^I \bar X^J \bar X^K \right)\,.
\end{equation}
The analysis now continues similarly to the minimal gauged supergravity case discussed in  appendix~A of~\cite{Cassani:2022lrk}. One eventually arrives at the expressions for $\aa$, $\cc$ obtained in \eqref{a_holo_corrected}, \eqref{c_holo_corrected}. These agree with those given in Eq.~\eqref{a_c_high_der} upon identifying $\lambda^{\rm there}_1 = \lambda_I\bar X^I$ and $\lambda_2^{\rm there} = \frac14 \widetilde{\lambda}_{IJK}\bar X^I\bar X^J\bar X^K$.


\chapter{Review: ${\cal N}=2$, $D=5$ conformal supergravity}
\label{chap:conformal_sugra}

\section{Superconformal multiplets in five dimensions}

 We start by reviewing the five-dimensional ${\cal N}=2$ superconformal multiplets to be used in the construction of off-shell Poincar\'e supergravity, with or without curvature squared actions. We follow the presentation of~\cite{Ozkan:2013nwa}. 

 \paragraph{Weyl multiplets.}

Weyl multiplets contain all the gauge fields associated with the superconformal algebra, as well as the proper matter fields required to obtain off-shell closure of the algebra. There exists two different choices for matter fields, leading to two different Weyl multiplets: the \emph{Standard Weyl multiplet} and the \emph{Dilaton Weyl multiplet}. Although we will mostly employ the formulation of the standard Weyl multiplet in the following discussion, we will briefly review both of them. 

To construct Weyl multiplets, we start assigning to any generator of the superconformal algebra a gauge field:

\begin{table}[h!]
    \centering
    \begin{tabular}{|c|ccccccc|}
    \hline
         Generators\;& \;  $P_a$\;& \; $M_{ab}$\;& \; ${\cal D}$\;& \; $K_a$ \;& \;$U_{ij}$ \;& \; ${\cal Q}_\alpha{}_i$ \; & \; ${\cal S}_{\alpha}{}_i$\\
         \hline 
         Gauge fields \;&\; $e^a{}_\mu$ \;&\; $\omega^{ab}{}_\mu$ \; & \; $b_\mu$ \; & \; $f^a_\mu$ \; & \; $V_\mu^{ij}$ \; & \; $\psi_\mu^i$ \; & \; $\phi_\mu^i$
         \\
         \hline
\end{tabular}
    \caption{\it Gauge fields of the superconformal algebra.}
    \label{tab:superconformal_table}
\end{table}
\noindent
Here, $a,b,...$ are Lorentz indices, $\alpha$ is a spinor index and $i =1,2$ is an ${\rm SU}(2)$ index. Here we follow the conventions of \cite{Bergshoeff:2004kh} for the SU(2) indices. Any SU(2) triplet $A_{i}{}^{j}$ can be expanded in terms of the Pauli matrices ${\vec \sigma}_{i}{}^{j}$ as follows
\begin{equation}
{\mathnormal A}_{i}{}^{j} \,=\, \ii\,{\vec A}\cdot {\vec \sigma}_{i}{}^{j}\, ,
\end{equation}
and the indices $i, j$ are raised (lowered) with $\varepsilon^{ij}$ ($\varepsilon_{ij}$), following the NW-SE convention:
\begin{equation}
A^{ij}=\varepsilon^{ik}A_{k}{}^{j}\, , \hspace{1cm} A_{ij}=A_{i}{}^{k}\varepsilon_{kj}\, ,\qquad \varepsilon_{12} = \varepsilon^{12} = - \varepsilon_{21} = 1\,.
\end{equation}
Since $A^{[ij]}=0$, we can always split $A^{ij}$ into its traceless ${\tilde A}^{ij}$ and trace $A$ contributions, in a way such that
\begin{equation}
A^{ij}={\tilde A}^{ij}+\frac{1}{2} \,\delta^{ij} A\, .
\end{equation}
In table~\ref{tab:superconformal_table}, $P_a$ and $M_{ab}$ are generators of the Poincar\'e algebra, $K_a$ is the generator of special conformal transformations, ${\cal D}$ the dilatation and $U_{ij}$ are the ${\rm SU}(2)_R$ generators. The fermionic symmetries, ${\cal Q}_\alpha{}_i$ and ${\cal S}_\alpha{}_i$, denote the supersymmetry and special supersymmetry, respectively, and the corresponding variables are symplectic Majorana spinors. A Standard Weyl multiplet is obtained by introducing a scalar $D$, an antisymmetric tensor $T_{ab}$ and a symplectic Majorana spinor $\chi^i$ as additional auxiliary fields. 

The set of transformation rules are given by\footnote{Here, the spinor $\epsilon^i$ is a parameter for ${\cal Q}$-transformations, while $\eta^i$ generates ${\cal S}$-transformations. Finally, $\Lambda_\mu^{(K)}$ is a parameter for special conformal transformations.}
\begin{equation}
\label{eq:susy_weyl}
\begin{aligned}
\delta e^a_\mu &= \frac{1}{2}\bar\epsilon^i \gamma^a\psi_{\mu}{}_i\,,
\\[1mm]
\delta \psi^i_\mu &= \left[ \partial_\mu + \frac{1}{2}b_\mu + \frac{1}{4}\omega^{ab}_\mu \gamma_{ab}\right] \epsilon^i - V_\mu^{ij}\epsilon_j + \ii \gamma_{ab}T^{ab}\gamma_\mu \epsilon^i - \ii \gamma_\mu \eta^i\,,
\\[1mm]
\delta V_\mu^{ij} &= -\frac{3}{2}\ii \bar\epsilon^{(i}\phi_\mu^{j)} + 4\bar\epsilon^{(i} \gamma_\mu \chi^{j)} + \ii \bar\epsilon^{(i}\gamma_{ab}T^{ab}\psi_\mu^{j)} + \frac{3}{2}\ii \bar \eta^{(i} \psi_\mu^{j)}\,,
\\[1mm]
\delta T_{ab} &= \frac{1}{2}\ii\bar \epsilon^i \gamma_{ab} \chi_i - \frac{3}{32}\ii\bar\epsilon^i \hat R_{ab}{}_i\left({\cal Q}\right)\,,
\\[1mm]
\delta\chi^i &= \frac{1}{4}\epsilon^i D - \frac{1}{64}\gamma_{ab} \hat R^{abij}\left(V\right) \epsilon_j + \frac{1}{8}\ii\gamma^{ab} \gamma^\mu {\cal D}_\mu T_{ab} \epsilon^i - \frac{1}{8}\ii \gamma^a {\cal D}^b T_{ab} \epsilon^i
\\
&- \frac{1}{4}\gamma^{abcd} T_{ab} T_{cd} \epsilon^i + \frac{1}{6}T_{ab}^2 \epsilon^i + \frac{1}{4}\gamma_{ab}T^{ab} \eta^i\,,
\\[1mm]
\delta D &= \bar\epsilon^i \gamma^\mu {\cal D}_\mu \chi_i - \frac{5}{3}\ii\bar\epsilon^i \gamma_{ab}T^{ab}\chi_i -\ii\bar\eta^i\chi_i\,,
\\[1mm]
\delta b_\mu &= \frac{1}{2}\ii \bar\epsilon^i \phi_\mu{}_i -2\bar\epsilon^i \gamma_\mu \chi_i + \frac{1}{2}\ii\bar\eta^i\psi_\mu{}_i + 2 \Lambda^{(K)}_\mu\,,
\end{aligned}
\end{equation}
where the relevant supercovariant derivatives read
\begin{equation}
\begin{aligned}
{\cal D}_\mu \chi^i &= \left[\partial_\mu - \frac{7}{2}b_\mu + \frac{1}{4}\omega^{ab}_\mu \gamma_{ab}\right] \chi^i - V_\mu^{ij} \chi_j - \frac{1}{4}\psi_\mu^i D + \frac{1}{64}\gamma_{ab}\hat R^{abij} \left(V\right) \psi_{\mu}{}_j 
\\
&\quad - \frac{1}{8}\ii \gamma^{ab}\gamma^\mu{\cal D}_\mu T_{ab} \psi_\mu^i + \frac{1}{8}\ii \gamma^a{\cal D}^b T_{ab} \psi_\mu^i + \frac{1}{4}\gamma^{abcd}T_{ab}T_{cd} \psi_\mu^i - \frac{1}{6}T_{ab}^2 \psi_\mu^i - \frac{1}{4}\psi_{ab}T^{ab} \phi_\mu^i\,,
\\[1mm]
{\cal D}_\mu T_{ab} &= \partial_\mu T_{ab} - b_\mu T_{ab} - 2\omega^c_\mu{}_{[a}T_{b]c} - \frac{1}{2}\ii \bar\psi^i_\mu \gamma_{ab} \chi_i + \frac{3}{32}\ii \bar\psi_\mu^i \hat R_{ab}{}_i\left( {\cal Q}\right)\,.
\end{aligned}
\end{equation}
Here, the supercovariant curvatures appearing in the transformation rules are given by
\begin{equation}
\begin{aligned}
\hat R_{\mu\nu}{}^{ab} \left( M\right) &= 2\partial_{[\mu} \omega^{ab}_{\nu]} + 2\omega^{ac}_{[\mu} \omega_{\nu]}{}_{c}{}^b + 8 f_{[\mu}^{[a}e_{\nu]}^{b]} + \text{fermionic terms} \,,
\\[1mm]
\hat R_{\mu\nu}{}^{ij}\left(V\right) &= 2\partial_{[\mu}V_{\nu]}^{ij} - 2 V_{[\mu}^{k(i}V_{\nu]}{}_k^{j)} + \text{fermionic terms}\,,
\\[1mm]
\hat R_{\mu\nu}^i\left( {\cal Q}\right) &= \text{fermionic terms}\,,
\end{aligned}
\end{equation}
where we are not showing terms that depend on fermionic fields. Finally, notice that not all the gauge fields appearing in table \ref{tab:superconformal_table} are independent, rather the spin connection $\omega^{ab}{}_\mu$, the conformal boosts gauge field $f^a_\mu$ and the special supersymmetry gauge field $\phi_\mu^i$ are composite fields, expressed in terms of the other ones, by
\begin{equation}
    \begin{aligned}
        \omega_\mu{}^{ab} &= 2e^{\nu[a}\partial_{[\mu}e_{\nu]}{}^{b]}- e^{\nu[a}e^{b]\sigma}e_{\mu c}\partial_\nu e_\sigma{}^c+ 2e_\mu{}^{[a}b^{b]}- \frac{1}{2}\bar\psi^{[b}_i\gamma^{a]}\psi_{\mu}^i - \frac{1}{4}\bar\psi^b_i \gamma_\mu \psi^a{}^i\,,
        \\[1mm]
\phi_\mu^i &= \frac{\ii}{3}\gamma^a\tilde R_{\mu a}^i\left({\cal Q}\right) - \frac{\ii}{24}\gamma_\mu\gamma^{ab}\tilde R_{ab}^i\left( {\cal Q}\right) \,,
\\[1mm]
f^a_\mu &= -\frac{1}{6}\tilde R_\mu{}^a + \frac{1}{48}e_\mu{}^a \tilde R\,,\qquad \tilde R = \tilde R_\mu{}^\mu\,,\qquad \tilde R_{\mu\nu} = \tilde R_{\mu\rho}{}^{ab}\left(M\right) e_b{}^\rho\,e_{\nu a}\,,
    \end{aligned}
\end{equation}
where $\tilde R_{ab}^i\left( {\cal Q}\right)$ and $\tilde R_{\mu\nu}{}^{ab}\left(M\right)$ denote the quantities obtained by setting $\phi_\mu^i=f_\mu^a=0$ in $\hat R_{ab}^i\left( {\cal Q}\right)$ and $\hat R_{\mu\nu}{}^{ab}\left( M\right)$, respectively. These relations follow from imposing a set of so-called \emph{conventional constraints}, corresponding to the requirement that some of supercovariant curvatures vanish, namely
\begin{equation}
\label{eq:conventional_constraints}
    \hat R_{\mu\nu}{}^a\left(P\right) = 0\,,\qquad \gamma^\mu\hat R_{\mu\nu}^i\left( {\cal Q}\right)=0\,,\qquad e^\nu{}_b\,\hat R_{\mu\nu}{}^{ab}\left(M\right)=0\,.
\end{equation}
Here, the supercovariant curvatures are reported explicitly in~\cite{Bergshoeff:2001hc}, and we will not repeat them here to keep the presentation light. The constraints \eqref{eq:conventional_constraints} can be understood as a generalization of the torsion‐free condition of general relativity, which uniquely expresses the spin connection $\omega^{ab}_\mu$ in terms of the vielbein $e^a_\mu$. Imposing them guarantees manifest covariance under diffeomorphisms. For further details see the classic reference~\cite{Freedman:2012zz}.

\bigskip

The gauge sector of the \emph{dilaton Weyl multiplet} is the same as the one summarized in the table \ref{tab:superconformal_table}. However, here the matter sector contains a different set of fields, given by physical vector $C_\mu$, an antisymmetric two-form $B_{\mu\nu}$, a dilaton $\sigma$, and a dilatino $\psi^i$. Additional details regarding the transformation rules and regarding how you can map between the formulations for Weyl multiplets can be found in~\cite{Bergshoeff:2001hc,Ozkan:2013nwa}. In the following constructions, we will mainly consider the standard Weyl formalism.

\paragraph{Abelian vector multiplet.}

The transformation rules for an Abelian vector multiplet are given by
\begin{equation}
\label{eq:susy_vector}
\begin{aligned}
\delta A_\mu &= -\frac{1}{2}\ii X\bar\epsilon^i\psi_\mu{}_i + \frac{1}{2}\bar\epsilon^i \gamma_\mu \lambda_i \,,
\\[1mm]
\delta Y^{ij} &= -\frac{1}{2} \bar\epsilon^{(i} \gamma^\mu{\cal D}_\mu \lambda^{j)} + \frac{1}{2}\ii \bar\epsilon^{(i}\gamma_{ab} T^{ab} \lambda^{j)} -4\ii\sigma\bar\epsilon^{(i}\chi^{j)} + \frac{1}{2}\ii\bar\eta^{(i}\lambda^{j)} \,,
\\[1mm]
\delta\lambda^i &= -\frac{1}{4}\gamma_{ab}\hat F^{ab} \epsilon^i - \frac{1}{2}\ii \gamma^\mu {\cal D}_\mu X \epsilon^i + X \gamma_{ab} T^{ab} \epsilon^i - Y^{ij} \epsilon_j + X \eta^i\,,
\\[1mm]
\delta X &= \frac{1}{2}\ii\bar\epsilon^i \lambda_i\,.
\end{aligned}
\end{equation}
Here, the superconformally covariant derivatives are defined as
\begin{equation}
\begin{aligned}
{\cal D}_\mu X &= \left[ \partial_\mu - b_\mu\right] X - \frac{1}{2}\ii \bar\psi_\mu^i \lambda_i \,,
\\[1mm]
{\cal D}_\mu\lambda^i &= \left[ \partial_\mu - \frac{3}{2}b_\mu + \frac{1}{4}\omega_\mu^{ab} \gamma_{ab}\right] \lambda^i - V_\mu^{ij} \lambda_j
\\
&\quad + \frac{1}{4}\gamma_{ab} \hat F^{ab} \psi_\mu^i + \frac{1}{2}\ii \gamma^\mu {\cal D}_\mu X \psi_\mu^i + Y^{ij} \psi_\mu{}_j - X \gamma_{ab} T^{ab} \psi_\mu^i - X \phi_\mu^i\,,
\end{aligned}
\end{equation}
with supercovariant field strength given by
\begin{equation}
\hat F_{\mu\nu} = F_{\mu\nu} + \text{fermionic terms} \,,\qquad F_{\mu\nu} = 2\partial_{[\mu} A_{\nu]}\,.
\end{equation}

\paragraph{Linear multiplet.}

The transformation rules for the off-shell linear multiplet are given by
\begin{equation}
\label{eq:susy_linear}
\begin{aligned}
\delta L^{ij} &= \ii \bar\epsilon^{(i} \varphi^{j)}\,,
\\[1mm]
\delta\varphi^i &= -\frac{1}{2}\ii \gamma^\mu{\cal D}_\mu L^{ij} \epsilon_j - \frac{1}{2}\ii \gamma^a E_a \epsilon^i + \frac{1}{2}N\epsilon^i - \gamma_{ab} T^{ab} L^{ij} \epsilon_j + 3 L^{ij} \eta_j\,,
\\[1mm]
\delta E_a &= -\frac{1}{2}\ii \bar\epsilon^i\gamma_{ab} {\cal D}^b \varphi_i - 2\bar\epsilon^i \gamma^b \varphi_i T_{ba} - 2\bar \eta^i \gamma_a \varphi_i\,,
\\[1mm]
\delta N &= \frac{1}{2}\bar\epsilon^i \gamma^\mu {\cal D}_\mu \varphi_i + \frac{3}{2}\ii \bar\epsilon^i \gamma_{ab} T^{ab} \varphi_i + 4\ii \bar\epsilon^i \chi^j L_{ij} + \frac{3}{2}\ii \bar\eta^i \varphi_i \,,
\end{aligned}
\end{equation}
with the following superconformally covariant derivatives 
\begin{equation}
\begin{aligned}
{\cal D}_\mu L^{ij} &= \left[ \partial_\mu -3 b_\mu\right] L^{ij} + 2 V_\mu^{(i} L^{j)k} - \ii \bar\psi_\mu^{(i}\varphi^{j)} \,,
\\[1mm]
{\cal D}_\mu \varphi^i &= \left[ \partial_\mu - \frac{7}{2}b_\mu + \frac{1}{4}\omega_\mu^{ab}\gamma_{ab} \right] \varphi^i - V_\mu^{ij} \varphi_j + \frac{1}{2}\ii \gamma^\mu {\cal D}_\mu L^{ij} \psi_\mu{}_j + \frac{1}{2}\ii \gamma^a E_a \psi_\mu^i\,,
\\[1mm]
{\cal D}_\mu E_a &= \left[ \partial_\mu - 4b_\mu\right] E_a + \omega_{\mu ab}E^b + \frac{1}{2}\ii \bar\psi_\mu^i \gamma_{ab}{\cal D}^b \varphi_i + 2\bar\psi_\mu^i \gamma^b \varphi_i T_{ba} + 2\bar\phi_\mu^i \gamma_a \varphi_i\,.
\end{aligned}
\end{equation}
Here, $E_a$ is a constrained vector satisfying ${\cal D}^aE_a =0$.

\subsection{Superconformal actions}

The starting point for the construction of superconformally invariant actions is the vector-linear Lagrangian 
\begin{equation}
    \label{eq:L_VL}
        {\cal L}_{VL} = Y^{ij}L_{ij} + A_a E^a + X N + .\,.\,.\,,
    \end{equation}
The ellipsis denotes that we are neglecting all fermionic terms, since we focus on the bosonic sector of the theory. 

To construct an action for the linear multiplet using the superconformal tensor calculus, we employ the embedding formulae that allow to express the fields of a vector multiplet in terms of those of a linear multiplet coupled to a Weyl multiplet~\cite{Coomans:2012cf}:
\begin{equation}
\label{eq:vector_to_linear}
\begin{aligned}
X &= 2L^{-1} N + \,.\,.\,.\,,
\\[1mm]
Y_{ij} &= L^{-1}\Box_c L_{ij} -L^{-3} {\cal D}_a L_{k(i}{\cal D}^a L_{j)m} L^{km} - N^2 L^{-3} L_{ij} - L^{-3} E_a^2 L_{ij} 
\\
&\quad + \frac{8}{3}L^{-1} T_{ab}^2 L_{ij} + 4 L^{-1} D L_{ij} + 2 L^{-3}E_a L_{k(i} {\cal D}^a L_{j)}{}^k + \,.\,.\,.\,,
\\[1mm]
\hat F_{\mu\nu} &= 4{\cal D}_{[\mu}L^{-1} E_{\nu]} + 2 L^{-1} \hat R_{\mu\nu}{}^{ij}\left(V\right) L_{ij} - 2 L^{-3} L_k^l{\cal D}_{[\mu} L^{kp}{\cal D}_{\nu]} L_{lp} + \,.\,.\,.\,,
\end{aligned}
\end{equation}
where $L = \sqrt{L_{ij}L^{ij}}$ and the superconformal d'Alambertian acts on $L_{ij}$ as 
\begin{equation}
\label{eq:dAlambertL}
    \Box_c L^{ij} =\left( \partial^a - 4 b^a + \omega_b{}^{ba}\right) {\cal D}_a L_{ij} + 2 V_a^{(i}{}_k {\cal D}_a L^{j)k} + 6 L^{ij} f_a^a + .\,.\,.\,.
\end{equation}
Inserting the composite expression into the vector-linear action \eqref{eq:L_VL} one obtains an action for the linear multiplet coupled to the Weyl multiplet:
\begin{equation}
\label{eq:L_linear}
\begin{aligned}
    {\cal L}_{L}^{\rm S}&= L^{-1} \,L_{ij} \Box_c L^{ij} - L^{-3}\,L^{ij}{\cal D}_{\mu}L_{k(i}{\cal D}^\mu L_{j)m}L^{km} + L^{-1}\,N^2 - L^{-1}\,E_a^2 + \frac{8}{3}LT_{ab}^2 + 4 LD 
    \\[1mm]
    &\quad - \frac{1}{2}L^{-3}E^{\mu\nu} L_k{}^l\partial_\mu L^{kp}\partial_\nu L_{pl} + 2 E^{\mu\nu}\partial_\mu\left( L^{-1}E_\nu + L^{-1}V_\nu^{ij} L_{ij}\right)\,,
    \end{aligned}
\end{equation}
being $E_{\mu\nu} = 2{\cal D}_{[\mu} E_{\nu]}$. 

Similarly, the elements of a linear multiplet can be constructed in terms of the fields in a vector multiplet and a Weyl multiplet~\cite{Fujita:2001kv}:
\begin{equation}
\begin{aligned}
L_{ij} &= 2X Y_{ij} + \,.\,.\,.\,,
\\[1mm]
E^a &= {\cal D}_b\left[ -X F^{ab} + 8X^2 T^{ab}\right] - \frac{1}{8}\epsilon^{abcde} F_{bc} F_{de} + \,.\,.\,.\,,
\\[1mm]
N &= X \Box_c X + \frac{1}{2}\left({\cal D}_a X\right)^2 - \frac{1}{4}F_{ab}^2 + Y_{ij}^2 + 8 F_{ab} T^{ab} - 4X^2 \left( D + \frac{26}{3}T_{ab}^2\right) + \,.\,.\,.\,,
\end{aligned}
\end{equation}
where the superconformal d'Alambertian is
\begin{equation}\label{eq:dAlambertianR}
    \Box_c X = \left( \partial^a - 2b^a + \omega_b{}^{ba}\right) {\cal D}_a X + 2 f_a^a\,X + .\,.\,.\,,
\end{equation}
where all fermionic contributions are suppressed, again. For $n+1$ coupled vector multiplets, the relevant superconformal action is then given by: 
\begin{equation}
\label{eq:L_vector}
\begin{aligned}
    {\cal L}_V^{\rm S} &=C_{IJK}\Bigl[ -\frac{1}{4}X^I F_{\mu\nu}^J F^K{}^{\mu\nu} + \frac{1}{3}X^I X^J \Box_c X^K +\frac{1}{6}X^I {\cal D}_\mu X^J \,{\cal D}^\mu X^K  + X^I Y_{ij}^JY^K{}^{ij} \\[1mm]
    &\quad - \frac{4}{3}X^I X^J X^K \left( D + \frac{26}{3}T^2\right) + 4 X^I X^J F^K_{\mu\nu}T^{\mu\nu} - \frac{1}{24}\epsilon^{\mu\nu\rho\sigma\lambda}F^I_{\mu\nu}F^J_{\rho\sigma} A^K_\lambda\Bigr]\,,
    \end{aligned}
\end{equation}
where $I,J,K = 1,2,\,...\,,n+1$, and the coefficient $C_{IJK}$ is symmetric in $I,J,K$ and determines the coupling of the $n+1$ vector multiplets. 

\paragraph{Recovering the universal case} To obtain a superconformal action for a single vector multiplet ($n=0$) coupled to a standard Weyl multiplet, we use the map $X^I = \bar X^I X$, $A_\mu^I = \bar X^I A_\mu$ and $Y_{ij}^I = \bar X^I Y_{ij}$, where $C_{IJK} \bar X^I \bar X^J \bar X^K = \bar C$. Then, by introducing the cubic polynomial ${\cal C} = \bar C X^3$, we can rewrite \eqref{eq:L_vector} as 
\begin{equation}
\label{eq:L_vector2}
\begin{aligned}
    {\cal L}_V^{\rm S} &=\frac{1}{3}\Bigl[ -\frac{{\cal C}''}{8}F_{\mu\nu}^2  + \frac{{\cal C}'}{3} \Box_c X +\frac{{\cal C}''}{12} \left({\cal D}_\mu X\right)^2  + \frac{{\cal C}''}{2} Y_{ij}^2 - 4{\cal C}\left( D + \frac{26}{3}T^2\right) 
    \\[1mm]
    &\quad + 4{\cal C}' F_{\mu\nu}T^{\mu\nu} - \frac{{\cal C}'''}{48}\epsilon^{\mu\nu\rho\sigma\lambda}F^I_{\mu\nu}F^J_{\rho\sigma} A^K_\lambda\Bigr] \,.
    \end{aligned}
\end{equation}

\subsubsection{Gauge-fixing and off-shell Poincar\'e supergravity}

To require the gauge-fixing conditions \eqref{eq:gauge_fixing_cond} to be maintained by superconformal transformations (i.e. $\delta b_\mu = \delta L_{ij} = \delta \varphi^i = 0$, with transformations given in \eqref{eq:susy_vector} and \eqref{eq:susy_linear}), the following compensating transformations are needed 
\begin{equation}
\label{eq:compensation_gauge_fixing}
\begin{aligned}
\eta_i& = \frac{1}{3}\left[\gamma_{ab}T^{ab} \epsilon_i - \frac{1}{\sqrt{2}}N\delta_{ik} \epsilon^k + \frac{\ii}{\sqrt{2}}\gamma^aE_a \delta_{ik} \epsilon^k + \ii \gamma^a \tilde V_a^{(k}{}_l \delta^{j)l} \delta_{ki} \epsilon_j \right]\,,
\\[1mm]
\Lambda^{(K)}_\mu &= \,.\,.\,.\,,
\end{aligned}
\end{equation}
where $\tilde V_\mu{}^{ij}$ corresponds to the traceless part of the ${\rm SU}(2)_R$ vector $V_\mu^{ij}$, namely
\begin{equation}
    V_\mu^{ij} = \tilde V_\mu^{ij} + \frac{1}{2}\delta^{ij} V_\mu\,,\qquad V_\mu = V_\mu^{ij}\delta_{ij}\,,\qquad \tilde V_\mu^{ij} \delta_{ij} =0\,.
\end{equation}

After gauge-fixing, the bosonic part of the \emph{off-shell} Poincar\'e gauged supergravity is given by
\begin{equation}
\label{eq:off_shell_poinc_general}
\begin{aligned}
{\cal L}_{gR}^{\rm S} &= -  {\cal L}_L^{\rm S} - 3{\cal L}_V^{\rm S} - 3g_I{\cal L}_{VL}^I 
\\[1mm]
&=\frac{1}{8}\left({\cal C}+3\right)R+\frac{1}{8}{\cal C}_{IJ}  F^I_{\mu\nu}F^J{}^{\mu\nu}+\frac{8}{3}\left(13{\cal C}-1\right)T_{\mu\nu}T^{\mu\nu}-12 X_I F^I_{\mu\nu}T^{\mu\nu}\\
&\quad +4\left({\cal C}-1\right)D+{\tilde V}_{\mu}{}^{ij}{\tilde V}^{\mu}{}_{ij}-\sqrt{2}V_{\mu} E^{\mu}- E_{\mu}^2-N^2+\frac{1}{4}{\cal C}_{IJ} \partial_\mu X^J\partial^\mu X^K-\frac{1}{2}{\cal C}_{IJ} Y^I{}^{ij}Y^J_{ij}\\
&\quad +\frac{1}{8}C_{IJK}\epsilon^{\mu\nu\rho\sigma\lambda}F^I_{\mu\nu}F^J_{\rho\sigma} A^K_{\lambda}-\frac{3}{\sqrt{2}}\,g_I Y^I_{ij}\delta^{ij}-3\,g_I A^I_{\mu}E^{\mu}-3\,g_I X^I N \, ,
\end{aligned}
\end{equation}
where
\begin{equation}
{\cal C} = C_{IJK} X^I X^J X^K\,,\qquad X_{I} = C_{IJK} X^J X^K\,,\qquad {\cal C}_{IJ} = 6 C_{IJK} X^K\,.
\end{equation}
Following the steps described around \eqref{eq:L_vector2}, we can reduce \eqref{eq:off_shell_poinc_general} to \eqref{eq:offshell2daction}, by taking $g_I \bar X^I = \sqrt{\frac{2}{3}} \,g$.  

The Killing spinor equations for this off-shell theory can be obtained by setting to zero the gravitino supersymmetry variations obtained after plugging \eqref{eq:gauge_fixing_cond} and \eqref{eq:compensation_gauge_fixing} into \eqref{eq:susy_weyl}, obtaining
\begin{equation}\label{eq:off_shell_grav_vars}
    \delta\psi_\mu^i = \left(\partial_\mu +\frac{1}{4}\omega_\mu{}^{ab}\gamma_{ab}\right) \epsilon^i - \frac{1}{2}V_{\mu}\,\delta^{ij}\epsilon_j  + \frac{2\ii}{3}\left(\gamma_{\mu\nu\rho} - 4 g_{\mu\nu}\gamma_\rho\right) T^{\nu\rho} - \frac{\ii}{3\sqrt{2}}N\gamma_\mu \delta^{ij}\epsilon_j +\,.\,.\,.\,.
\end{equation}
where we are neglecting some terms that depend on $E_\mu$ and $\tilde V_\mu^{ij}$ that are always vanishing when the auxiliary fields are evaluated on-shell. 

In the minimal ($n=0$) theory, the Killing spinor equations \eqref{eq:Killing_spinor_equations} can be derived by setting to zero the gravitino variations obtained after inserting into \eqref{eq:off_shell_grav_vars} the on-shell values of the auxiliary fields \eqref{eq:auxfields}, and setting $X=-\sqrt{3}$. After these substitutions one finds
\begin{equation}
\left(\partial_\mu +\frac{1}{4}\omega_\mu{}^{ab}\gamma_{ab}\right) \epsilon^i +\frac{\sqrt{3}}{2}g A_{\mu}\,\delta^{ij}\epsilon_j  -\frac{\ii}{8\sqrt{3}}\left(\gamma_{\mu\nu\rho} - 4 g_{\mu\nu}\gamma_\rho\right) F^{\nu\rho} - \frac{\ii\,g}{2}\gamma_\mu \delta^{ij}\epsilon_j  =0 \,,
\end{equation}
which reproduces \eqref{eq:Killing_spinor_equations} upon a change of conventions, namely $g \to -g$ and $A \to \tfrac{2}{\sqrt{3}g}A$.

\section{Curvature-square invariants}
\label{app:offshell_invariants}

Using superconformal tensor calculus, an off-shell Weyl-squared action in the standard Weyl multiplet was constructed in~\cite{Hanaki:2006pj} as a supersymmetric completion of the mixed gauge-gravitational Chern-Simons term, $\epsilon^{\mu\nu\rho\sigma\lambda}R_{\mu\nu}{}^{\gamma\delta} R_{\rho\sigma\gamma\delta} A_\lambda$. The procedure to obtain it is similar to the one we followed above: first, one should find the embedding formulae for the Weyl multiplet into a linear multiplet, then by plugging this embedding into the linear-vector action \eqref{eq:L_VL} one would obtain a superconformal action coupling Weyl and vector multiplets, including curvature-squared terms. The bosonic part of the supersymmetric action obtained after gauge-fixing \eqref{eq:gauge_fixing_cond} reads 
\be
\label{eq:offshell_Weyl_squared}
\begin{aligned}
{\cal L}_{C^2}^{\rm{off-shell}}=\,&8\,\lambda_I\left[\frac{1}{8}X^I C^{\mu\nu\rho\sigma}C_{\mu\nu\rho\sigma}+\frac{64}{3}X^I D^2+\frac{1024}{9}X^I D \,T_{\mu\nu}T^{\mu\nu}-\frac{32}{3}D\,T_{\mu\nu}F^{I}{}^{\mu\nu}\right.\\[1mm]
&-\frac{16}{3}\, X^I\, C_{\mu\nu\rho\sigma}T^{\mu\nu}T^{\rho\sigma}+2C_{\mu\nu\rho\sigma}T^{\mu\nu}F^{I}{}^{\rho\sigma}+\frac{1}{16}\,\epsilon^{\mu\nu\rho\sigma\lambda}A^I_{\mu}C_{\nu\rho\alpha\beta}C_{\sigma\lambda}{}^{\alpha\beta}\\[1mm]
&-\frac{1}{12}\epsilon^{\mu\nu\rho\sigma\lambda}A^I_{\mu}V^{ij}_{\nu\rho}V_{ij}{}_{\sigma\lambda}+\frac{16}{3}Y^I_{ij}V^{ij}_{\mu\nu}T^{\mu\nu}-\frac{1}{3}X^I V^{ij}_{\mu\nu}V_{ij}^{\mu\nu}+\frac{64}{3}X^I\nabla_{\nu}T_{\mu\rho}\nabla^{\mu}T^{\nu\rho}\\[1mm]
&-\frac{128}{3}X^IT_{\mu\nu}\nabla^{\nu}\nabla_{\rho}T^{\mu\rho}-\frac{256}{9}X^IR^{\nu\rho}T_{\mu\nu}T^{\mu}{}_{\rho}+\frac{32}{9}X^I R \,T_{\mu\nu}T^{\mu\nu}\\[1mm]
&-\frac{64}{3}X^I\nabla_{\mu}T_{\nu\rho}\nabla^{\mu}T^{\nu\rho}+1024 X^I T^{\mu\nu}T_{\nu\rho}T^{\rho\sigma}T_{\sigma\mu}-\frac{2816}{27}X^I \left(T_{\mu\nu}T^{\mu\nu}\right)^2\\[1mm]
&-\frac{64}{9}T^{\mu\nu}F^{I}_{\mu\nu}T_{\rho\sigma}T^{\rho\sigma}-\frac{256}{3}T_{\mu\rho}T^{\rho\lambda}T_{\nu\lambda}F^{I}{}^{\mu\nu}-\frac{32}{3}\epsilon^{\mu\nu\rho\sigma\lambda}T_{\rho\alpha}\nabla^{\alpha}T_{\sigma\lambda}F^I_{\mu\nu}\\[1mm]
&\left.-16\epsilon^{\mu\nu\rho\sigma\lambda}T_{\rho}{}^{\alpha}\nabla_{\sigma}T_{\lambda\alpha}F^I_{\mu\nu}-\frac{128}{3}X^I\epsilon^{\mu\nu\rho\sigma\lambda}T_{\mu\nu}T_{\rho\sigma}\nabla^{\alpha}T_{\lambda\alpha}\right]\,,
\end{aligned}
\ee
where $V_{\mu\nu}^{ij}=2\partial_{[\mu}V^{ij}_{\nu]}-2V^{k(i}_{[\mu}V_{\nu]}{}_{k}{}^{j)}$ is the field strength associated to the SU(2) connection $V_{\mu}^{ij}$.

To obtain the $R^2$ invariant we start from the superconformal action \eqref{eq:L_vector}, choose $C_{I11} = \sigma_I$ and zero otherwise, where $\sigma_I$ are some dimensionless constants, and use the composite expressions \eqref{eq:vector_to_linear} allowing us to express the elements of a vector multiplet in terms of those of a linear multiplet coupled to the Weyl multiplet. Therefore, the invariant is given by~\cite{Ozkan:2013nwa}
\be
\label{eq:offshell_R^2}
\begin{aligned}
{\cal L}_{R^2}^{\rm{off-shell}}=\,&\sigma_I \left(X^I {\underline Y}_{ij}{\underline Y}^{ij}+2 {\underline X} \,{\underline Y}^{ij}Y_{ij}^I-\frac{1}{8}X^I {\underline X}^2 R-\frac{1}{4}X^I {\underline F}_{\mu\nu}{\underline F}^{\mu\nu}-\frac{1}{2}{\underline X}\,{\underline F}^{\mu\nu}F_{\mu\nu}^I\right.\\[1mm]
&+\frac{1}{2}X^I \partial_{\mu}{\underline X}\partial^{\mu}{\underline X}+X^I {\underline X}\nabla^2 {\underline X}-4 X^I {\underline X}^2\left(D+\frac{26}{3}T_{\mu\nu}T^{\mu\nu}\right)+4 {\underline X}^2 F^I_{\mu\nu}T^{\mu\nu}\\[1mm]
&\left.+8X^I {\underline X}\, {\underline F}_{\mu\nu}T^{\mu\nu}-\frac{1}{8}\epsilon^{\mu\nu\rho\sigma\lambda}A^I_{\mu}{\underline F}{}_{\nu\rho} {\underline F}{}_{\sigma\lambda}\right)\,,
\end{aligned}
\ee
where the underlined fields are given by the gauge-fixed version of \eqref{eq:vector_to_linear}, that is
\be
\begin{aligned}
{\underline X}=\,&2N\,, \\[1mm]
{\underline Y}^{ij}=\,&\frac{1}{\sqrt{2}}\,\delta^{ij}\left(-\frac{3}{8}R-N^2-P_{\mu}P^{\mu}+\frac{8}{3}T_{\mu\nu}T^{\mu\nu}+4D-{\tilde V}^{kl}_{\mu}{\tilde V}_{kl}^{\mu}\right)+2P^{\mu}{\tilde V}_{\mu}^{ij}\\[1mm]
&-\sqrt{2}\nabla^{\mu}{\tilde V}_{\mu}^{k(i}\delta^{j)}{}_{k}\,,\\[1mm]
{\underline F}{}_{\mu\nu}=\,& 2\sqrt{2}\partial_{[\mu}\left(V_{\nu]}+\sqrt{2}P_{\nu]}\right)\, .
\end{aligned}
\ee

Let us emphasize that, for the applications of chapter \ref{chap:flavour} with $n\neq 0$, the scalars $X^I$ in this appendix do not coincide with the ones in the main text, as they satisfy a modified cubic constraint when going on-shell. We refer to the discussion at the beginning of section~\ref{sec:4der_action} for a detailed explanation. 


\chapter{Useful formulae in four-derivative supergravity}

\section{Equations of motion from four-derivative action}\label{app:eoms}

In this appendix we derive the equations of motion from the higher-derivative action. We start with a general discussion, that may also be useful in other contexts, for a higher-derivative action constructed out of the metric and a gauge field, also allowing for Chern-Simons terms. Then we specialize to the action \eqref{eq:4daction2} studied in the main text.

Let us consider the following general action (where we set $16\pi G=1$ for simplicity), 
\begin{equation}
S=\int \diff^5x \,e\, {\mathcal L}'\left(R_{\mu\nu\rho\sigma}, F_{\mu\nu}\right)+S_{\rm{CS}}\, ,
\end{equation}
where ${\mathcal L}'\left(R_{\mu\nu\rho\sigma}, F_{\mu\nu}\right)$ denotes a Lagrangian constructed out of arbitrary contractions (via the inverse metric $g^{\mu\nu}$) of the Riemann tensor $R_{\mu\nu\rho\sigma}$ and the field strength $F_{\mu\nu}$, while
\begin{equation}
S_{\rm{CS}}=\int \diff^5x \, e\, \left[-\frac{\tilde c_3}{12\sqrt{3}}\epsilon^{\mu\nu\rho\sigma\lambda}F_{\mu\nu}F_{\rho\sigma}A_{\lambda}-\frac{\lambda_1\,\alpha}{2\sqrt{3}}\epsilon^{\mu\nu\rho\sigma\lambda}R_{\mu\nu\alpha\beta}R_{\rho\sigma}{}^{\alpha\beta}A_{\lambda}\right]\, ,
\end{equation}
are the Chern-Simons terms, which will be treated separately for convenience. Here the choice of the coefficients $\tilde{c}_3,\lambda_1$ reflects the one in the main text. The variation of the Lagrangian with respect to the inverse metric $g^{\mu\nu}$ and the gauge field $A_{\mu}$~is
\begin{equation}
\begin{aligned}
\delta S=\,&\int \diff^5x\, e\, \left [\left(\frac{\partial {\cal L}'}{\partial g^{\mu\nu}}-\frac{1}{2}g_{\mu\nu}{\cal L}'\right)\delta g^{\mu\nu}+P^{\mu\nu\rho\sigma}\delta R_{\mu\nu\rho\sigma}-2\nabla_{\mu}\left(\frac{\partial {\cal L}'}{\partial F_{\mu\nu}}\right)\delta A_{\nu}\right. \\[1mm]
&+\nabla_\mu \Xi^\mu\bigg]+\delta S_{\rm{CS}}\, ,
\end{aligned}
\end{equation}
where $\Xi^\mu$ is a boundary term
\begin{equation}\label{Theta0}
\Xi^\mu=2\frac{\partial {\cal L}'}{\partial F_{\mu\nu}}{\delta A}_{\nu}\, ,
\end{equation}
and the tensor $P^{\mu\nu\rho\sigma}$ is defined as
\begin{equation}\label{eq:Pmunurhosigma}
P^{\mu\nu\rho\sigma}=\frac{\partial {\cal L}'}{\partial R_{\mu\nu\rho\sigma}}\, ,
\end{equation}
assuming it inherits the following symmetries of the Riemann tensor
\begin{equation}
P_{\mu\nu\rho\sigma}=-P_{\nu\mu\rho\sigma}\, ,\hspace{0.75cm} P_{\mu\nu\rho\sigma}=-P_{\mu\nu\sigma\rho}\, ,\hspace{0.75cm}P_{\mu\nu\rho\sigma}=P_{\rho\sigma\mu\nu}\, .
\end{equation}
Recalling the well-known variation of the Riemann tensor
\be
\delta R^\mu{}_{\nu \rho\sigma} = 2 \nabla_{[\rho} \,\delta \Gamma^\mu_{\sigma]\nu}\,,
\ee
\be
\delta \Gamma^\mu_{\sigma\nu} = \frac{1}{2}g^{\mu\kappa}\left(\nabla_\sigma \delta g_{\kappa\nu} + \nabla_\nu \delta g_{\kappa\sigma} - \nabla_{\kappa}\delta g_{\sigma\nu} \right)\,,
\ee
and performing integrations by parts, we obtain that 
\begin{align}
\int \!\diff^5 x\,e\, P^{\mu\nu\rho\sigma}\, \delta R_{\mu\nu\rho\sigma}\,& =\, \int \!\diff^5 x\,e \left[\left(-P^{\alpha\beta\gamma}{}_{\mu}R_{\alpha\beta\gamma \nu}-2\nabla^{\alpha}\nabla^{\beta} P_{\beta \mu\nu\alpha}\right) \delta g^{\mu\nu} + \nabla_{\nu} \upsilon^\nu \right]\,,
\end{align}
where the boundary term reads
\begin{align}\label{bdry_term_metric_var}
\upsilon^\nu \,&=\, 2 P_\mu{}^{\lambda\nu\sigma}\,  \delta \Gamma^\mu_{\lambda\sigma} -  2\nabla_{\lambda} P^{\rho\nu\lambda\sigma} \delta g_{\rho\sigma}\nn\\[1mm]
&=\,  2P^{\rho\lambda\nu\sigma}\,   \nabla_\lambda \delta g_{\rho\sigma}    -  2\nabla_{\lambda} P^{\rho\nu\lambda\sigma} \delta g_{\rho\sigma}\,.
\end{align}
Then, 
\begin{equation}\label{eq:deltaS}
\begin{aligned}
\delta S=\,&\int \diff^5x\, e\, \left [\left(\frac{\partial {\cal L}'}{\partial g^{\mu\nu}}-\frac{1}{2}g_{\mu\nu}{\cal L}'-P^{\alpha\beta\gamma}{}_{\mu}R_{\alpha\beta\gamma \nu}-2\nabla^{\alpha}\nabla^{\beta} P_{\beta \mu\nu\alpha}\right)\delta g^{\mu\nu}-2\nabla_{\mu}\left(\frac{\partial {\cal L}'}{\partial F_{\mu\nu}}\right)\delta A_{\nu}\right. \\[1mm]
&\left.+\nabla_\mu (\upsilon^\mu+\Xi^\mu)\right]+\delta S_{\rm{CS}}\, ,
\end{aligned}
\end{equation}
This first line can be expressed exclusively in terms of $\frac{\partial {\cal L}'}{\partial F^{\mu\nu}}$ and $P_{\mu\nu\rho\sigma}$, once $\frac{\partial {\cal L}'}{\partial g^{\mu\nu}}$ is expressed in terms of the latter \cite{Padmanabhan:2011ex}. To this aim, let us write the Lie derivative of the Lagrangian in two different ways. First, we can write it as 
\begin{equation}\label{eq:LieL1}
\begin{aligned}
{\mathsterling}_{\xi}{\mathcal L}'=\,\xi^{\alpha}\partial_{\alpha}{\mathcal L}'=\,&\xi^{\alpha}\left(\frac{\partial {\cal L}'}{\partial R_{\mu\nu\rho\sigma}}\nabla_{\alpha}R_{\mu\nu\rho\sigma}+\frac{\partial {\cal L}'}{\partial g^{\mu\nu}}\nabla_{\alpha}g^{\mu\nu}+\frac{\partial {\cal L}'}{\partial F_{\mu\nu}}\nabla_{\alpha}F_{\mu\nu}\right)\\[1mm]
=\,&\xi^{\alpha}\left(P^{\mu\nu\rho\sigma}\nabla_{\alpha}R_{\mu\nu\rho\sigma}+\frac{\partial {\cal L}'}{\partial F_{\mu\nu}}\nabla_{\alpha}F_{\mu\nu}\right)\,.
\end{aligned}
\end{equation}
However, another possibility is 
\begin{equation}\label{eq:LieL2}
{\mathsterling}_{\xi}{\mathcal L}'=\frac{\partial {\cal L}'}{\partial R_{\mu\nu\rho\sigma}}\mathsterling_{\xi} R_{\mu\nu\rho\sigma}+\frac{\partial {\cal L}'}{\partial g^{\mu\nu}}\mathsterling_{\xi} g^{\mu\nu}+\frac{\partial {\cal L}'}{\partial F_{\mu\nu}}\mathsterling_{\xi} F_{\mu\nu}\, .
\end{equation}
After a bit of algebra, we can rewrite each of the terms appearing in this equation as follows,
\begin{eqnarray}
\frac{\partial {\cal L}'}{\partial R_{\mu\nu\rho\sigma}}\mathsterling_{\xi} R_{\mu\nu\rho\sigma}&=&\xi^{\alpha}P^{\mu\nu\rho\sigma}\nabla_{\alpha}R_{\mu\nu\rho\sigma}+4 P_{\mu}{}^{\alpha\beta\gamma}R_{\nu\alpha\beta\gamma}\nabla^{\mu}\xi^{\nu}\, ,\\[1mm]
\frac{\partial {\cal L}'}{\partial g^{\mu\nu}}\mathsterling_{\xi} g^{\mu\nu}&=&-2\frac{\partial {\cal L}'}{\partial g^{\mu\nu}}\nabla^{(\mu}\xi^{\nu)}\, ,\\[1mm]
\frac{\partial {\cal L}'}{\partial F_{\mu\nu}}\mathsterling_{\xi} F_{\mu\nu}&=&\xi^{\alpha}\frac{\partial {\cal L}'}{\partial F_{\mu\nu}}\nabla_{\alpha}F_{\mu\nu}+2\frac{\partial {\cal L}'}{\partial F^{\mu\rho}} F_{\nu}{}^{\rho}\nabla^{\mu}\xi^{\nu}\, .
\end{eqnarray}
Substituting in \eqref{eq:LieL2} and making use of \eqref{eq:LieL1}, we get the following identity
\begin{equation}
\nabla^{(\mu}\xi^{\nu)}\left(2P_{\mu}{}^{\alpha\beta\gamma}R_{\nu\alpha\beta\gamma}-\frac{\partial {\cal L}'}{\partial g^{\mu\nu}}+\frac{\partial {\cal L}'}{\partial F^{\mu\rho}} F_{\nu}{}^{\rho}\right)+\nabla^{[\mu}\xi^{\nu]}\left(2P_{\mu}{}^{\alpha\beta\gamma}R_{\nu\alpha\beta\gamma}+\frac{\partial {\cal L}'}{\partial F^{\mu\rho}} F_{\nu}{}^{\rho}\right)=0\,.
\end{equation}
Since this equality must be true for an arbitrary vector $\xi^{\mu}$, we conclude that the terms in brackets must vanish, which leads to 
\begin{eqnarray}
\label{eq:dL/dg}
\frac{\partial {\cal L}'}{\partial g^{\mu\nu}}&=&2P_{(\mu}{}^{\alpha\beta\gamma}R_{\nu)\alpha\beta\gamma}+\frac{\partial {\cal L}'}{\partial F^{(\mu|\rho}} F_{|\nu)}{}^{\rho}\, ,\\[1mm]
\frac{\partial {\cal L}'}{\partial F^{[\mu|\rho}} F_{|\nu]}{}^{\rho}&=&-2P_{[\mu}{}^{\alpha\beta\gamma}R_{\nu]\alpha\beta\gamma}\,.
\end{eqnarray}
Then, we can eliminate $\frac{\partial {\cal L}'}{\partial g^{\mu\nu}}$ in \eqref{eq:deltaS} by using \eqref{eq:dL/dg}, which yields 
\begin{equation}
\begin{aligned}
\delta S=\,&\int \diff^5x\, e\, \left [\left(-\frac{1}{2}g_{\mu\nu}{\cal L}'+P^{\alpha\beta\gamma}{}_{\mu}R_{\alpha\beta\gamma \nu}-2\nabla^{\alpha}\nabla^{\beta} P_{\beta \mu\nu\alpha}+\frac{\partial {\cal L}'}{\partial F^{\mu\rho}}F_{\nu}{}^{\rho}\right)\delta g^{\mu\nu}\right. \\[1mm]
&\left.-2\nabla_{\mu}\left(\frac{\partial {\cal L}'}{\partial F_{\mu\nu}}\right)\delta A_{\nu}+\nabla_\mu (\upsilon^\mu+\Xi^\mu)\right]+\delta S_{\rm{CS}}\, .
\end{aligned}
\end{equation}
Now let us perform the variation of the Chern-Simons terms explicitly. They yield,
\begin{equation}
\begin{aligned}
\delta S_{\rm{CS}}=\,&\int \diff^5x\, e \left[\left(-\frac{\tilde c_3}{4\sqrt{3}}\epsilon^{\nu\alpha\beta\gamma\delta}F_{\alpha\beta}F_{\gamma\delta}-\frac{\lambda_1\,\alpha}{2\sqrt{3}}\epsilon^{\nu\alpha\beta\gamma\delta}R_{\alpha\beta\rho\sigma}R_{\gamma\delta}{}^{\rho\sigma}\right){\delta A}_{\nu}-2\nabla^{\alpha}\nabla^{\beta}\Pi_{\beta \mu\nu \alpha}{\delta g}^{\mu\nu}\right.\\[1mm]
&\left.+\nabla_{\mu}\left(\Theta^{\mu}_{\rm{CS}}+\Xi^{\mu}_{\rm{CS}}\right)\right]\,,
\end{aligned}
\end{equation}
where the boundary terms are given by 
\begin{eqnarray}
\label{Theta1}
\Theta^{\mu}_{\rm{CS}}&=&2\Pi^{\mu\sigma\rho\lambda}\,   \nabla_\lambda \delta g_{\rho\sigma}-2\nabla_{\lambda} \Pi^{\lambda\sigma\rho\mu} \delta g_{\rho\sigma}\, ,\\
\Xi^{\mu}_{\rm{CS}}&=&\frac{\tilde c_3}{3\sqrt 3}\epsilon^{\nu\mu\rho\sigma\lambda}F_{\rho\sigma}A_\lambda{\delta A}_{\nu} \, .
\label{Theta2}
\end{eqnarray}
and where we have defined
\begin{equation}
\Pi^{\mu\nu\rho\sigma}=-\frac{\lambda_1\,\alpha}{\sqrt{3}}\epsilon^{\mu\nu\alpha\beta\gamma}R_{\alpha\beta}{}^{\rho\sigma}A_{\gamma}\, .
\end{equation}
Arranging all the terms together, we arrive to the final form for the variation of the action,
\begin{equation}
\begin{aligned}
\delta S=\,&\int \diff^5x\, e\, \left \{\left[-\frac{1}{2}g_{\mu\nu}{\cal L}'+P^{\alpha\beta\gamma}{}_{\mu}R_{\alpha\beta\gamma \nu}-2\nabla^{\alpha}\nabla^{\beta} (P_{\beta \mu\nu\alpha}+\Pi_{\beta \mu\nu\alpha})+\frac{\partial {\cal L}'}{\partial F^{\mu\rho}}F_{\nu}{}^{\rho}\right]\delta g^{\mu\nu}\right. \\[1mm]
&\left.\left[-2\nabla_{\mu}\left(\frac{\partial {\cal L}'}{\partial F_{\mu\nu}}\right)-\frac{\tilde c_3}{4\sqrt{3}}\epsilon^{\nu\alpha\beta\gamma\delta}F_{\alpha\beta}F_{\gamma\delta}-\frac{\lambda_1\,\alpha}{2\sqrt{3}}\epsilon^{\nu\alpha\beta\gamma\delta}R_{\alpha\beta\rho\sigma}R_{\gamma\delta}{}^{\rho\sigma}\right]\delta A_{\nu}+\nabla_\mu \Theta^{\mu}\right\}\,,
\end{aligned}
\end{equation} 
where 
\begin{equation}\label{Theta_final}
\Theta^\mu=\upsilon^{\mu}+\Theta^{\mu}_{\rm{CS}}+\Xi^{\mu}+\Xi^{\mu}_{\rm{CS}}\, .
\end{equation}
Therefore, the equations of motion of  the metric and gauge field are given by
\begin{eqnarray}
-\frac{1}{2}g_{\mu\nu}{\cal L}'+P^{\alpha\beta\gamma}{}_{(\mu|}R_{\alpha\beta\gamma |\nu)}-2\nabla^{\alpha}\nabla^{\beta} (P_{\beta (\mu\nu)\alpha}+\Pi_{\beta (\mu\nu)\alpha})+\frac{\partial {\cal L}'}{\partial F^{(\mu|\rho}}F_{|\nu)}{}^{\rho}&=&0\, \label{EinsteinEqCorrected},\\[2mm]
2\nabla_{\mu}\left(\frac{\partial {\cal L}}{\partial F_{\mu\nu}}\right)+\frac{\tilde c_3}{4\sqrt{3}}\epsilon^{\nu\alpha\beta\gamma\delta}F_{\alpha\beta}F_{\gamma\delta}+\frac{\lambda_1\,\alpha}{2\sqrt{3}}\epsilon^{\nu\alpha\beta\gamma\delta}R_{\alpha\beta\rho\sigma}R_{\gamma\delta}{}^{\rho\sigma}&=&0\, .
\label{MaxwellEqCorrected}
\end{eqnarray}

We now specialize to the action \eqref{eq:4daction2} that we study in the main text.
The concrete expressions for $P_{\mu\nu\rho\sigma}$ and $\frac{\partial {\cal L}}{\partial F_{\mu\nu}}$ are the following,
\begin{equation}\label{Ptensor_explicit}
\begin{aligned}
P_{\mu\nu\rho\sigma}=&{\tilde c}_{0}g_{\mu[\rho}g_{\sigma]\nu}+\lambda_1\,\alpha\left[2R_{\mu\nu\rho\sigma}-4\left(R_{\mu[\rho}g_{\sigma]\nu}-R_{\nu[\rho}g_{\sigma]\mu}\right)+2 g_{\mu[\rho}g_{\sigma]\nu} R \right.\\[1mm]
&\left.-\frac{1}{2}F_{\mu\nu}F_{\rho\sigma}-\frac{1}{12}g_{\mu[\rho}g_{\sigma]\nu} F^2+\frac{1}{3}\left(F_{\mu\alpha}F_{[\rho}{}^{\alpha}g_{\sigma]\nu}-F_{\nu\alpha}F_{[\rho}{}^{\alpha}g_{\sigma]\mu}\right)\right]\, ,\\[1mm]
\frac{\partial {\cal L}'}{\partial F_{\mu\nu}}=&-\frac{\tilde c_2}{2}F^{\mu\nu}+\lambda_1\,\alpha \left(-C^{\mu\nu\rho\sigma}F_{\rho\sigma}+\frac{1}{2}F^{\mu\rho}F_{\rho\sigma}F^{\nu\sigma}\right)\, .
\end{aligned}
\end{equation}


\section{Field redefinitions}

\subsection{Field redefinitions in the minimal theory}\label{app:field_red_minimal}

In order to implement field redefinitions in a systematic fashion, it is convenient to derive first a set of replacement rules for all the terms that can appear in (\ref{eq:4daction}) modulo total derivatives and the use of Bianchi and Ricci identities. Using (\ref{eq:masterrule1}) and (\ref{eq:masterrule2}), one can derive the following rules for particular choices of the tensors $K_{\mu\nu}$ and $L_{\mu}$: 
\begin{equation}
\begin{aligned}
R^{\mu\nu}R_{\mu\nu}&\ \rightarrow\  \frac{1}{4}F^4-\frac{7}{144}\left(F^2\right)^2-4g^2 R-\frac{1}{3}g^2 F^2\, ,\\
R^2&\ \rightarrow\  \frac{1}{144}\left(F^2\right)^2-20g^2 R-\frac{5}{3}g^2 F^2\, ,\\
RF^2&\ \rightarrow\  \frac{1}{2}\left(F^2\right)^2-20g^2 F^2\, ,\\
R^{\mu\nu}F_{\mu\rho}F_{\nu}{}^{\rho}&\ \rightarrow\  \frac{1}{2}F^4-\frac{1}{12}\left(F^2\right)^2-4g^2 F^2\, ,\\
\nabla_{\rho}F^{\rho\mu}\nabla_{\sigma}F^{\sigma}{}_{\mu} &\ \rightarrow\  \frac{1}{3}F^4-\frac{1}{6}\left(F^2\right)^2\, ,\\
\epsilon^{\mu\nu\rho\sigma\lambda}F_{\nu\rho}F_{\sigma\lambda}\nabla_{\delta}F^{\delta}{}_{\mu} &\ \rightarrow\  \frac{4}{\sqrt{3}}F^4-\frac{2}{\sqrt{3}}\left(F^2\right)^2\, .
\end{aligned}
\end{equation}
Using Bianchi identities, integrations by parts and Ricci identities\footnote{Our conventions for the Riemann tensor are such that $[\nabla_{\mu}, \nabla_{\nu}]\xi^{\sigma}=-R_{\mu\nu\alpha}{}^{\sigma}\xi^{\alpha}$.}, one can further derive the following rules 
\begin{equation}
\begin{aligned}
F^{\nu \rho}[\nabla_{\mu}, \nabla_{\nu}]F^{\mu}{}_{\rho}&\ \rightarrow\ -\frac{1}{2}R_{\mu\nu\rho\sigma}F^{\mu\nu}F^{\rho\sigma}+\frac{1}{2}F^4-\frac{1}{12}\left(F^2\right)^2-4g^2 F^2\,,\\
\nabla_{\mu}F_{\nu\rho}\nabla^{\mu}F^{\nu\rho}&\ \rightarrow\  R_{\mu\nu\rho\sigma}F^{\mu\nu}F^{\rho\sigma}-\frac{1}{3}F^4-\frac{1}{6}\left(F^2\right)^2+8g^2 F^2\,,\\
\epsilon^{\mu\nu\rho\sigma\lambda}F_{\mu\nu}F_{\rho}{}^{\alpha}\nabla_{\sigma}F_{\lambda\alpha} &\ \rightarrow\  -\frac{1}{\sqrt{3}}F^4+\frac{1}{2\sqrt{3}}\left(F^2\right)^2\,.
\end{aligned}
\end{equation}
This set of rules is enough since the four-derivative part of the action \eqref{eq:4daction} can always be written (up to total derivatives) as a linear combination of the following terms
\begin{equation}\label{eq:genexpL4d}
\begin{aligned}
{\mathcal L}_{4\partial}=\,&{a}_{1}\,R_{\mu\nu\rho\sigma}R^{\mu\nu\rho\sigma}+{a}_{2}\,R_{\mu\nu\rho\sigma}F^{\mu\nu}F^{\rho\sigma}+a_3 \,\left(F^2\right)^2+a_4\,F^4+a_5\, \epsilon^{\mu\nu\rho\sigma\lambda}R_{\mu\nu\alpha\beta}R_{\rho\sigma}{}^{\alpha\beta} A_{\lambda}\\
&+b_1\,R_{\mu\nu}R^{\mu\nu}+b_2\,R^2+b_3 \,RF^2+b_4\, R^{\mu\nu}F_{\mu\rho}F_{\nu}{}^{\rho}+b_5\,\nabla_{\mu}F_{\nu\rho}\nabla^{\mu}F^{\nu\rho}+b_6\,\nabla_{\rho}F^{\rho\mu}\nabla_{\sigma}F^{\sigma}{}_{\mu}\\
&+b_7\, F^{\nu \rho}[\nabla_{\mu}, \nabla_{\nu}]F^{\mu}{}_{\rho}+b_8 \,\epsilon^{\mu\nu\rho\sigma\lambda}F_{\nu\rho}F_{\sigma\lambda}\nabla_{\delta}F^{\delta}{}_{\mu}+b_9\,\epsilon^{\mu\nu\rho\sigma\lambda}F_{\mu\nu}F_{\rho}{}^{\alpha}\nabla_{\sigma}F_{\lambda\alpha}\, .
\end{aligned}
\end{equation}
Making use of the replacement rules that we have just derived, one can show that there is a field redefinition such that the action (\ref{eq:4daction}), with ${\mathcal L}_{4\partial}$ given by (\ref{eq:genexpL4d}), in terms of the new fields reads
\begin{equation}\label{eq:new4daction}
S=\frac{1}{16\pi G}\int \diff^5x \, e \left[c_{0}R+12c_{1} g^2-\frac{c_{2}}{4}F^2-\frac{c_{3}}{12\sqrt{3}}\epsilon^{\mu\nu\rho\sigma\lambda}F_{\mu\nu}F_{\rho\sigma} A_{\lambda}+\alpha \,{\mathcal L}'_{4\partial}\right]\, ,
\end{equation}
where 
\begin{equation}\label{eq:L4dprime}
{\mathcal L}'_{4\partial}=\,{a}'_{1}\,R_{\mu\nu\rho\sigma}R^{\mu\nu\rho\sigma}+{a}'_{2}\,R_{\mu\nu\rho\sigma}F^{\mu\nu}F^{\rho\sigma}+a'_3 \,\left(F^2\right)^2+a'_4\,F^4+a'_5\, \epsilon^{\mu\nu\rho\sigma\lambda}R_{\mu\nu\alpha\beta}R_{\rho\sigma}{}^{\alpha\beta} A_{\lambda}\, ,
\end{equation}
and 
\begin{equation}\label{eq:rulesa'}
\begin{aligned}
a'_1=\,&a_1\,,\\
a'_2=\,&a_2+b_5-\frac{b_7}{2}\,,\\
a'_3=\,&a_3-\frac{7b_1}{144}+\frac{b_2}{144}+\frac{b_3}{12}-\frac{b_4}{12}-\frac{b_5}{6}-\frac{b_6}{6}-\frac{b_7}{12}-\frac{2b_8}{\sqrt{3}}+\frac{b_9}{2\sqrt{3}}\,,\\
a'_4=\,&a_4+\frac{b_1}{4}+\frac{b_4}{2}-\frac{b_5}{3}+\frac{b_6}{3}+\frac{b_7}{2}+\frac{4b_8}{\sqrt{3}}-\frac{b_9}{\sqrt{3}}\,,\\
a'_5=\,&a_5\, .
\end{aligned}
\end{equation}
Finally, the $c_{i}=1+\alpha g^2 \delta c_{i}$ coefficients are given by
\begin{equation}\label{eq:rulesc}
\begin{aligned}
{\delta c}_{1}=\,&{\delta d}_{1}+\frac{5}{3} \left(4b_{1}+20 b_2+{\delta c}_{0}-\delta d_0\right)\,,\\
{\delta c}_{2}=\,&{\delta d}_{2}+\frac{8b_1}{3}+\frac{40b_2}{3}+80b_3+16b_4-32b_5+16b_7+\frac{\delta c_0}{3}-\frac{\delta d_0}{3}\,,\\
{\delta c}_{3}=\,&{\delta d}_{3}\,.
\end{aligned}
\end{equation}
When using the basis of four-derivative invariants in \eqref{eq:4daction2}, the resulting action is 
\begin{equation}\label{eq:new4daction2}
S=\frac{1}{16\pi G}\int \diff^5x \, e \left[{\tilde c}_{0}R+12{\tilde c}_{1} g^2-\frac{{\tilde c}_{2}}{4}F^2-\frac{{\tilde c}_{3}}{12\sqrt{3}}\epsilon^{\mu\nu\rho\sigma\lambda}F_{\mu\nu}F_{\rho\sigma} A_{\lambda}+\alpha \,{\mathcal L}''_{4\partial}\right]\, ,
\end{equation}
where 
\begin{equation}\label{eq:L4dprime}
{\mathcal L}''_{4\partial}=\,{a}''_{1}\,{\cal X}_{\rm {GB}}+{a}''_{2}\,C_{\mu\nu\rho\sigma}F^{\mu\nu}F^{\rho\sigma}+a''_3 \,\left(F^2\right)^2+a''_4\,F^4+a''_5\, \epsilon^{\mu\nu\rho\sigma\lambda}R_{\mu\nu\alpha\beta}R_{\rho\sigma}{}^{\alpha\beta} A_{\lambda}\, ,
\end{equation}
with 
\begin{equation}\label{eq:rulesa''}
\begin{aligned}
a''_1=\,&a_1\,,\\
a''_2=\,&a_2+b_5-\frac{b_7}{2}\,,\\
a''_3=&\frac{1}{144} \!\Bigl(-29 a_1-18 a_2+144 a_3-7 b_1+b_2+12 b_3-12 b_4-42 b_5-24 b_6-3 b_7
\\&-96 \sqrt{3} b_8+24 \sqrt{3} b_9\Bigr),\\
a''_4=\,&\frac{1}{12} \left(12 a_1+8 a_2+12 a_4+3 b_1+6 b_4+4 b_5+4 b_6+2 b_7+16 \sqrt{3} b_8-4 \sqrt{3} b_9\right)\,,\\
a''_5=\,&a_5\, .
\end{aligned}
\end{equation}
and 
\begin{equation}\label{eq:rulesctilde}
\begin{aligned}
{\delta {\tilde c}}_{1}=\,&\delta d_{1}-\frac{5}{3} \left(4 a_1-4 b_1-20 b_2-{\delta \tilde c}_0+\delta d_{0}\right)\,,\\
{\delta {\tilde c}}_{2}=\,& \delta d_2+\frac{1}{3} \left(-8 a_1+24 a_2+8 b_1+40 b_2+240 b_3+48 b_4-72 b_5+36 b_7+{\delta \tilde c}_0-\delta d_0\right)\,,\\
{\delta {\tilde c}}_{3}=\,&{\delta d}_{3}\,.
\end{aligned}
\end{equation}
Let us note that the coefficients in front of the Ricci scalar, controlled by ${\delta c}_{0}$ and ${\delta \tilde c}_{0}$, are arbitrary parameters that we can fix at will. This corresponds to adding a term $\sim \alpha g^2{\cal E}$ to the action through a perturbative field redefinition consisting of a constant rescaling of the metric.


\subsection{Field redefinitions with vector multiplets}\label{app:field_redefinitions}

Perturbative field redefinitions of the form
\begin{equation}
g_{\mu\nu}\to g_{\mu\nu}+\alpha \,\Delta_{\mu\nu}\,, \hspace{1cm} A^I_{\mu}\to A^I_{\mu}+\alpha \,\Delta^I_{\mu}\,,
\end{equation}
are very helpful to simplify the final form of the Lagrangian, permitting us to reduce the number of terms. This is due to the fact that they allow us to eliminate terms proportional to $\alpha$ and to the two-derivative equations of motion. Therfore, this is equivalent to using the two-derivative equations of motion in the part of the Lagrangian multiplied by $\alpha$. Doing so, one finds the following set of replacement rules:
\begin{equation}\label{eq:redriccisquare}
\begin{aligned}
R_{\mu\nu}R^{\mu\nu}\rightarrow&\, a_{IJ}a_{KL}\left(-\frac{7}{16}\, F^I_{\mu\nu}F^J{}^{\mu\nu}  F^K_{\rho\sigma}F^L{}^{\rho\sigma}+ \frac{9}{4}\,F^I{}_{\mu\nu}F^{J\,\nu\rho}F^K{}_{\rho\sigma}F^{L\,\sigma\mu}- \frac{3}{4}\, \partial_{\mu} X^I\partial^{\mu} X^J \, F^K_{\rho\sigma} F^L{}^{\rho\sigma}\right.\\[1mm]
&\left.+\frac{9}{2}\,\partial_\mu X^I \,\partial^\nu X^J \,F^{K\,\mu\rho}F^L_{\nu\rho}\right) + \frac{9}{4}a_{K(I}a_{J)L} \,\partial_{\mu} X^I \partial^{\mu} X^J \partial_{\nu} X^K \partial^{\nu} X^L \\[1mm]
&+ \frac{20}{9}{\cal V}^2 + \frac{1}{3}{\cal V}a_{IJ}\,F^I_{\mu\nu} F^J{}^{\mu\nu}+ 2{\cal V}a_{IJ}\, \partial_{\mu} X^I \partial^{\mu} X^J\,\,,
\end{aligned}
\end{equation}
\begin{equation}\label{eq:redscalarcurvsquare}
\begin{aligned}
R^2\rightarrow& \,a_{IJ}a_{KL}\left[\frac{1}{16}F^I_{\mu\nu} F^J{}^{\mu\nu}F^K_{\mu\nu} F^L{}^{\mu\nu} + \frac{9}{4}\,\partial_{\mu} X^I \partial^{\mu} X^J \left(\partial_{\nu} X^K \partial^{\nu} X^L+ \frac{1}{3}F^K_{\rho\sigma} F^L{}^{\rho\sigma}\right)\right]\\[1mm]
&+\frac{100}{9}\,{\cal V}^2 +10\,{\cal V} \,a_{IJ}\, \partial_{\mu} X^I  \partial^{\mu} X^J+ \frac{5}{3}\,{\cal V}\,a_{IJ}\,F^I_{\mu\nu} F^J{}^{\mu\nu}\,,
\end{aligned}
\end{equation}

\begin{equation}\label{eq:redricciFF}
\begin{aligned}
R^\mu{}_{\nu}\,\mathcal F_{\mu\rho}\mathcal F^{\nu\rho} \rightarrow &
\,-\frac{1}{4}\,a_{IJ}\, F^I_{\mu\nu} F^J{}^{\mu\nu}\mathcal F^2 + \frac{3}{2}\,a_{IJ}\,\mathcal F_{\mu\nu}\mathcal F^{\nu\rho}F^I_{\rho\sigma}F^{J\,\sigma\mu}\\[1mm]
&+\frac{3}{2}\,a_{KL}\, \partial_\mu X^K\, \partial^\nu X^L\, \mathcal F^{\mu\rho}\mathcal F_{\nu\rho}+\frac{2}{3}\,{\cal V}\,\mathcal F^2\,\,,
\end{aligned}
\end{equation}

\begin{equation}\label{eq:redscalarcurvFF}
\begin{aligned}
R\, \mathcal F^2 \rightarrow& 
\,\left(\frac{1}{4}\,a_{IJ} F^I_{\mu\nu} F^J{}^{\mu\nu} + \frac{3}{2}\,a_{IJ}\, \partial_{\mu} X^I \partial^{\mu}X^J+\frac{10}{3}{\cal V}\right)\mathcal F^2\,.
\end{aligned}
\end{equation}

\begin{equation}\label{eq:rednablaFsquare}
\begin{aligned}
\nabla_\mu\mathcal F^{\mu\nu}\, \nabla^\rho \mathcal F_{\rho\nu} \rightarrow& -\frac{1}{8}\,{\mathcal C}_{IK} {\mathcal C}_{JL} \left(F^I_{\mu\nu} F^J{}^{\mu\nu}F^K_{\rho\sigma} F^L{}^{\rho\sigma}-2 F^I_{\mu\nu}F^{J\,\nu\rho}F^K_{\rho\sigma} F^{L\,\sigma\mu}\right) \\[1mm]
&+ 9a_{IK}a_{JL}\,\partial_\mu X^I\, \partial^\nu X^J\, F^{K\,\mu\rho}F^L_{\nu\rho} - 3{\mathcal C}_{IJ} a_{KL}\, \epsilon^{\mu\nu\rho\sigma\lambda} F^I_{\mu\nu} F^J_\rho{}^{\alpha}F^K_{\sigma\alpha}\,\partial_\lambda X^L\,,
\end{aligned}
\end{equation}

\begin{equation}\label{eq:redepsilonFFnablaF}
\begin{aligned}
\epsilon^{\mu\nu\rho\sigma\lambda} F^I_{\mu\nu} F^J_{\rho\sigma}\,\nabla^\alpha\mathcal F_{\alpha\lambda}
\rightarrow& \, {\mathcal C}_{KL} \left(F^I_{\mu\nu} F^K{}^{\mu\nu}F^J_{\rho\sigma}F^L{}^{\rho\sigma}- 2\, F^I_{\mu\nu}F^{K\,\nu\rho}F^J_{\rho\sigma} F^{L\,\sigma \mu}\right) \\[1mm]
&+12a_{KL}\, \epsilon^{\mu\nu\rho\sigma\lambda} F^{(I}{}_{\mu\nu}F^{J)}{}_{\rho\alpha}F_\sigma^K{}^{\alpha}\,\partial_\lambda X^L \,.
\end{aligned}
\end{equation}
Using them we can remove the last four terms in \eqref{eq:Weyl2}, as well as the Ricci-squared terms. However, instead of removing the latter, we are going to fix the coefficient in front of them in a way such that the Weyl-squared invariant is completed to the Gauss-Bonnet term ${\cal X}_{\rm{GB}}=R_{\mu\nu\rho\sigma}R^{\mu\nu\rho\sigma}-4R_{\mu\nu}R^{\mu\nu}+R^2$, just for convenience. The only price to pay is that we have to update the couplings of the four-derivative terms in \eqref{eq:Weyl2} and of the two-derivative corrections in \eqref{eq:DeltaL2dC2}. Finally, a perturbative constant rescaling of the metric allows us to fix to 1 the coefficient in front of the Ricci scalar, so as to be in the Einstein frame. The resulting Lagrangian that is obtained upon implementing these field redefinitions is the one reported in section~\ref{sec:final_Lagr}.


\chapter{Black hole thermodynamics with four derivatives and Chern-Simons terms}\label{sec:aladsthermodynamics}

In this section we consider stationary spacetimes with an event horizon and derive the first law of black hole mechanics \cite{Bardeen:1973gs} and the quantum statistical relation \cite{Gibbons:1976ue} in the class of theories for which the results of the previous section apply.  For these purposes, we will follow Wald's approach~\cite{Wald:1993nt}, suitably taking into account the contribution of the gauge field. The strategy will be to define a conserved charge which takes the same value when integrated at the horizon and at infinity, and then identify the various contributions in terms of the conserved charges and entropy of the solution.

The first law in the Einstein-Maxwell theory has been extensively studied in the past making use of Wald's formalism, see e.g.~\cite{Gao:2001ut, Gao:2003ys, Compere:2006my, Compere:2007vx, Prabhu:2015vua,  Elgood:2020svt, Hajian:2022lgy} and references therein. Depending on the gauge choice one makes, the contribution of the gauge field can arise either from the integral at the horizon or from the integral at infinity (or from both if one works in a general gauge). In all cases one obtains exactly the same term, $-\Phi\,\delta Q$, with the gauge choice just manifesting in the identification of the electrostatic chemical potential $\Phi$.
The reason behind is that the Noether charge associated to U$(1)$ transformations in the Einstein-Maxwell theory, $Q\sim\int \star F$,  is gauge invariant and satisfies Gauss law, which implies that the charges computed at the horizon and at infinity coincide.\footnote{This naturally extends to theories whose Lagrangians only depend on the gauge field $A_{\mu}$ through its field strength $F_{\mu\nu}$.} As we have seen in section~\ref{sec:conservedchargesgeneral}, in presence of Chern-Simons terms the Noether charge is neither gauge invariant nor it satisfies the Gauss law. Indeed, defining a charge obeying the Gauss law was our main motivation for introducing the Page charge \eqref{eq:pagecharge}, which in any case fails to be gauge invariant as well. This raises the question of which charge appears in the first law (and quantum statistical relation). 
In this section, we try to address this question. Our strategy to deal with this issue will be the following. First, we will set ourselves in the regular gauge, which means that we will impose 
\begin{equation}\label{eq:regulargauge}
\iota_{\xi}A|_{\cal H}=0\,,
\end{equation}
where $\xi=t+\Omega_k\, \phi_k$ is the generator of the horizon, with $t$ the Killing vector generating time translations, $\Omega_k$ the angular velocity, and ${\cal H}$ is an arbitrary slice of the horizon. In this gauge, the contributions of the gauge field appear at infinity, where we further assume that
\begin{equation}\label{eq:assumptions}
\iota_{\phi_k}A|_{\partial{\cal M}\, \cap \, {\cal C}}=0\, , \hspace{1cm}{\boldsymbol \Delta}|_{\partial{\cal M}\, \cap \, {\cal C}}=0\, .
\end{equation}
Crucially, this implies that the Page charge \eqref{eq:pagecharge} matches the Noether charge computed asymptotically, namely
\begin{equation}\label{eq:equiv_charges}
Q=\int_{{\partial\cal M}\, \cap \, {\cal C}}{\bf Q}_{0, 1}\, ,
\end{equation}
and also that the angular momenta \eqref{eq:defJi} can be computed at infinity by the following integral,
\begin{equation}\label{eq:Ji_inf}
J_{k}=-\int_{\partial {\cal M}\, \cap \, {\cal C}}{\bf Q}^{\rm g}_{\phi_k}\, .
\end{equation}
Let us notice that the condition \eqref{eq:assumptions} holds for any solution that is asymptotically AdS, as opposed to asymptotically {\it locally} AdS. 

Eventually, we will show that the asymptotic integrals \eqref{eq:equiv_charges} and \eqref{eq:Ji_inf}, where the notions of Page, Komar and Noether charges coincide, are the ones that enter the first law and the quantum statistical relation.

In what follows, we specify $\xi$ to be the generator of the horizon and take $\chi$ to be constant, which trivializes the gauge transformations. In particular, one can take $\chi=0$ without loss of generality.

\section{First law of black hole mechanics}\label{sec:firstlaw}

The variation of the Noether current \eqref{eq:Noethercurrent} under a generic perturbation of the fields (keeping $\xi$ and $\chi$ fixed) is given by,
\begin{equation}
\delta {\bf J}_{\xi, \chi}=\delta {\bf S}_{\xi, \chi}+\delta {\boldsymbol \Theta_{\xi, \chi}}-\iota_{\xi}\delta {\bf L}-\chi \delta {\boldsymbol \Lambda}\, .
\end{equation}
Restricting to perturbations obeying the linearized equations of motion and going on-shell, using \eqref{eq:generaldL} one finds
\begin{equation}
\begin{aligned}
\delta {\bf J}_{\xi, \chi}=\,&\delta {\boldsymbol \Theta_{\xi, \chi}}-\iota_{\xi}\diff {\boldsymbol \Theta}-\chi \delta {\boldsymbol \Lambda}\\[1mm]
=\,&\delta {\boldsymbol \Theta_{\xi, \chi}}-\delta_{\xi, \chi} {\boldsymbol \Theta}+\diff\left(\iota_{\xi}{\boldsymbol \Theta}-\chi \delta {\boldsymbol \Delta}\right)\\[1mm]
=\,& \diff \left(\delta {\bf Q}_{\xi, \chi}\right)\, ,
\end{aligned}
\end{equation}
so that
\begin{equation}
\diff\left(\delta {\bf Q}_{\xi, \chi}-\iota_{\xi}{\boldsymbol \Theta}+\chi \delta {\boldsymbol \Delta}\right)=\delta {\boldsymbol \Theta_{\xi, \chi}}-\delta_{\xi, \chi} {\boldsymbol \Theta}\, .
\end{equation}
When further restricting to field symmetries, the right-hand side vanishes, which means that the surface charge,
\begin{equation}\label{eq:surfacecharge1stlaw}
\delta {\bf Q}_{\xi, \chi}-\iota_{\xi}{\boldsymbol \Theta}+\chi \delta {\boldsymbol \Delta}\,, 
\end{equation}
is conserved on-shell. 

Let $\cal C$ be a Cauchy slice and denote as $\cal H$ its intersection with the event horizon. In particular, one can think of $\cal H$ as the bifurcation surface, as originally assumed by Wald \cite{Wald:1993nt}. However, according to \cite{Jacobson:1993vj}, the results should extend to any slice of the horizon, as long as regularity at the bifurcation surface is guaranteed. Assuming regularity of the fields on and outside the horizon so that it is allowed to apply Stokes theorem, we have that
\begin{equation}\label{eq:horizonintegral}
\int_{\partial {\cal M}\, \cap \, {\cal C}}\left(\delta {\bf Q}_{\xi, \chi}-\iota_{\xi}{\boldsymbol \Theta}+\chi \delta {\boldsymbol \Delta}\right)=\int_{\cal H}\left(\delta {\bf Q}_{\xi, \chi}-\iota_{\xi}{\boldsymbol \Theta}+\chi \delta {\boldsymbol \Delta}\right)\,,
\end{equation}
from which stems the first law, as we shall see. Setting $\chi=0$ and splitting the asymptotic contributions into the gravitational and electromagnetic pieces using \eqref{eq:surfacecharge2} and \eqref{eq:split_theta}, we have 
\begin{equation}
\int_{\partial {\cal M}\, \cap \, {\cal C}}\left(\delta {\bf Q}_{\xi, 0}-\iota_{\xi}{\boldsymbol \Theta}\right)=\, \int_{\partial {\cal M}\, \cap \, {\cal C}}\left(\delta{\bf Q}^{\rm g}_{\xi}-\iota_{\xi}{\boldsymbol \Theta}^{\rm g}\right)+\int_{\partial {\cal M}\, \cap \, {\cal C}}\left(\iota_{\xi}\delta A \,{\bf Q}_{0, 1}+\iota_{\xi}A \,\delta{\bf Q}_{0, 1}-\iota_{\xi}{\boldsymbol \Theta}^{\rm{em}}\right)\,.
\end{equation}
The first term yields \cite{Wald:1993nt, Iyer:1994ys},
\begin{equation}
\int_{\partial {\cal M}\, \cap \, {\cal C}}\left(\delta{\bf Q}^{\rm g}_{\xi}-\iota_{\xi}{\boldsymbol \Theta}^{\rm g}\right)=\delta E-\Omega_k\, \delta J_{k}\, .
\end{equation}
Instead the second term gives us the contribution of the electric charge to the first law, which by our gauge choice appears from the integral at infinity, as already anticipated. In order to see this explicitly, let us massage it as follows 
\begin{equation}
\begin{aligned}
\int_{\partial {\cal M}\, \cap \, {\cal C}}\left(\iota_{\xi}\delta A \,{\bf Q}_{0, 1}+\iota_{\xi}A \,\delta{\bf Q}_{0, 1}-\iota_{\xi}{\boldsymbol \Theta}^{\rm {em}}\right)=\,& \int_{\partial {\cal M}\, \cap \, {\cal C}}\left(-\Phi \,\delta {\bf Q}_{0, 1}+\iota_{\xi}{\bf Q}_{0, 1}\wedge \delta A\right)\\[1mm]
=&\int_{\partial {\cal M}\, \cap \, {\cal C}}\left(-\Phi \,\delta {\bf Q}_{0, 1}+\iota_{\xi}{\bf k}_{0, 1}\wedge \delta A-\iota_{\xi}{\boldsymbol\Delta}\wedge \delta A\right)\,.
\end{aligned}
\end{equation}
Now we notice that the second and third term in the second line do not contribute, provided the assumptions \eqref{eq:assumptions} hold.\footnote{The second term has been recently understood in \cite{Ortin:2022uxa, Ballesteros:2023iqb,Ortin:2024mmg} to give rise (after integration by parts) to a term $\sim-{\Phi}_{\rm m}\,{\delta Q}_{\rm{m}}$, where $Q_{\rm m}$ are the magnetic charges and $\Phi_{\rm m}$ the associated chemical potentials. However, our initial assumptions \eqref{eq:assumptions} forbid this kind of term.} In such case, we obtain
\begin{equation}
\int_{\partial {\cal M}\, \cap \, {\cal C}}\left(\iota_{\xi}\delta A \,{\bf Q}_{0, 1}+\iota_{\xi}A \,\delta{\bf Q}_{0, 1}-\iota_{\xi}{\boldsymbol \Theta}^{\rm {em}}\right)=-\Phi\, \delta\int_{\partial {\cal M}\, \cap \, {\cal C}}{\bf Q}_{0, 1}=-\Phi \,\delta Q\,,
\end{equation}
where we have made use of \eqref{eq:equiv_charges}. Putting together all the asymptotic contributions,

\begin{equation}\label{eq:1stlaw_inf}
\int_{\partial {\cal M}\, \cap \, {\cal C}}\left(\delta {\bf Q}_{\xi, 0}-\iota_{\xi}{\boldsymbol \Theta}\right)=\delta E-\Omega_k \,\delta J_k -\Phi \,\delta Q\, .
\end{equation}
If the first law holds in its standard form, we should be able to identify the horizon contribution with $\beta^{-1}\delta {\cal S}$, where $\beta$ is the inverse Hawking temperature and ${\cal S}$ is the black hole entropy \cite{Wald:1993nt}. Indeed, when regularity of the fields at the bifurcation surface is assumed, the horizon integral reduces to \cite{Wald:1993nt, Iyer:1994ys, Jacobson:1993vj},
\begin{equation}
\int_{\cal H}\left(\delta {\bf Q}_{\xi, 0}-\iota_{\xi}{\boldsymbol \Theta}\right)\,=\,\delta\int_{\cal H}{\bf Q}^{\rm g}_{\xi}\,=\,\beta^{-1}\delta \left(-2\pi\int_{\cal H} \diff^3x\, \sqrt{\gamma}\,{\cal P}^{\mu\nu\rho\sigma}\, n_{\mu\nu}\, n_{\rho\sigma}\right)\, ,
\end{equation}
where $\gamma$ is the metric induced at ${\cal H}$ and $n_{\mu\nu}$ the binormal normalized so that $n_{\mu\nu}n^{\mu\nu}=-2$. This allows us to infer the expression for the black hole entropy,
\begin{equation}\label{eq:Waldentropy}
{\cal S}=-2\pi\int_{\cal H} \diff^3x\, \sqrt{\gamma}\,{\cal P}^{\mu\nu\rho\sigma}\, n_{\mu\nu}\, n_{\rho\sigma}\,,
\end{equation}
which coincides with Wald's prescription \cite{Wald:1993nt}. We will apply this formula in the next Sections, to compute the corrected entropy of supersymmetric AdS$_5$ black holes, obtaining a perfect agreement with the Euclidean methods employed in~\ref{sec:chargesandBPSlimit}.


\section{Quantum statistical relation} \label{sec:qsr}

In order to derive the quantum statistical relation \cite{Gibbons:1976ue} in our four-derivative setup, we start from the equality between the integral of the surface charge ${\bf k}_{\xi, \chi}$ at infinity and at the horizon,
\begin{equation}\label{eq:raw-qsr}
\int_{\Sigma_{\epsilon} \,\cap \,{\cal C}}{\bf k}_{\xi, \chi}=\int_{\cal H}{\bf k}_{\xi, \chi}\,.
\end{equation}
We recall that $\xi=t+\Omega_k\, \phi_k$ is the generator at the horizon and $\chi$ is assumed to be constant. Making use of \eqref{eq:improvedcurrent}, we find
\begin{equation}
\begin{aligned}
\int_{\Sigma_{\epsilon} \,\cap \,{\cal C}}{\bf Q}_{\xi, \chi}-\int_{\cal H}{\bf Q}_{\xi, \chi}=\,&-\int_{\Sigma_{\epsilon} \,\cap \,{\cal C}}{\boldsymbol \Xi}_{\xi, \chi}+\int_{\cal H}{\boldsymbol \Xi}_{\xi, \chi}\\[1mm]
=\,&-\int_{\cal C}\diff{\boldsymbol \Xi}_{\xi, \chi}=-\int_{\cal C} \left[\iota_{\xi}{\bf L}+\diff\left(\chi{\boldsymbol \Delta}\right)\right]\,.
\end{aligned}
\end{equation}
Next, we add suitable boundary terms,
\begin{equation}\label{eq:QSR}
\int_{\Sigma_{\epsilon} \,\cap \,{\cal C}}\left({\bf Q}_{\xi, \chi}+\iota_{\xi}{\bf B}+\chi {\boldsymbol \Delta}\right)-\int_{\cal H}\left({\bf Q}_{\xi, \chi}+\chi {\boldsymbol \Delta}\right)=-\int_{\cal C}\iota_{\xi}{\bf L}+\int_{\Sigma_{\epsilon} \,\cap \,{\cal C}} \iota_{\xi}{\bf B}\,,
\end{equation}
so that we can identify the right-hand side with $\beta^{-1}I$ (where $I$ is the renormalized Euclidean on-shell action) when the cutoff is removed~\cite{Papadimitriou:2005ii}. Then, 
\begin{equation}\label{eq:qsr2}
\beta^{-1}I=\int_{\partial {\cal M} \,\cap \,{\cal C}}\left({\bf Q}_{\xi, \chi}+\iota_{\xi}{\bf B}+\chi {\boldsymbol \Delta}\right)-\int_{\cal H}\left({\bf Q}_{\xi, \chi}+\chi {\boldsymbol \Delta}\right)\, .
\end{equation}

Let us first consider the integral at infinity. Setting $\chi=0$ and using  \eqref{eq:surfacecharge2} again, we have
\begin{equation}
\int_{\partial {\cal M} \,\cap \,{\cal C}}\left({\bf Q}_{\xi, 0}+\iota_{\xi}{\bf B}\right)=\,\int_{\partial {\cal M} \,\cap \,{\cal C}}\left({\bf Q}^{\rm g}_{t}+\iota_{t}{\bf B}\right)+\Omega_k\int_{\partial {\cal M} \,\cap \,{\cal C}}{\bf Q}^{\rm g}_{\phi_k}-\Phi\int_{\partial {\cal M} \,\cap \,{\cal C}}{\bf Q}_{0, 1}\,.
\end{equation}
As discussed in \cite{Wald:1993nt, Papadimitriou:2005ii}, the first term can be interpreted as the mass of the solution,
\begin{equation}\label{eq:mass}
E=\int_{\partial {\cal M} \,\cap \,{\cal C}}\left({\bf Q}^{\rm g}_{t}+\iota_{t}{\bf B}\right)\,,
\end{equation}
while the second and third yield (minus) the angular momenta and Page charge, as we showed in and \eqref{eq:Ji_inf} and \eqref{eq:equiv_charges}. Hence,
\begin{equation}\label{eq:qsr_inf}
\int_{\partial {\cal M} \,\cap \,{\cal C}}\left({\bf Q}_{\xi, 0}+\iota_{\xi}{\bf B}\right)=\,E-\Omega_{k} J_{k}-\Phi \,Q\, .
\end{equation}
On the other hand, the horizon contribution in \eqref{eq:qsr2} boils down to
\begin{equation}\label{eq:qsr_hor}
\int_{\cal H}{\bf Q}_{\xi, 0}=\int_{\cal H}{\bf Q}^{\rm g}_{\xi}=\beta^{-1}{\cal S}\, ,
\end{equation}
where ${\cal S}$ is given in \eqref{eq:Waldentropy}. Finally, plugging \eqref{eq:qsr_inf} and \eqref{eq:qsr_hor} into \eqref{eq:qsr2}, we get the quantum statistical relation
\begin{equation}\label{eq:qsr}
\beta^{-1}I=E-\beta^{-1}{\cal S}-\Omega_k\, J_k-\Phi\, Q\,,
\end{equation}
as we wanted to show.



\chapter{Other useful formulae}


\section{Expression for $\breve \omega$}
\label{app:omega_eq}

The purpose of this appendix is to derive eq.~\eqref{eq:omega_final}. Namely, we want to solve
\begin{equation}
\star_3 \diff\breve\omega\,=\, H\diff M - M \diff H +\frac{3}{2}\left(K\diff L - L\diff K \right)\,, 
\end{equation}
for the general choice of harmonic functions made in \eqref{eq:multicenter_ansatz}, under the assumption that all centers are placed on the $z$-axis. We start noticing that
\begin{equation}
\label{eq:def_omega_*}
\ii{\breve \omega}=\sum_{a=1}^{s}\left[h_0\mm_a-\mm_0 h_a+\frac{3}{2}\left(\kk_0 \ell_a-\ell_0\kk_a\right)\right]\cos\theta_a \diff \phi+\ii{\breve \omega}_{*}\,,
\end{equation}
where ${\breve \omega}_{*}$ satisfies the same equation as $\breve\omega$ but ignoring the constant term in all the harmonic functions. This implies that
\begin{equation}
\label{eq:omega_*}
\star_3 \diff\, \ii {\breve\omega}_{*}=\sum_{a}\sum_{b}C_{ab} \frac{\diff r_b^{-1}}{r_a}=\sum_{a}\sum_{b>a}C_{ab} \left(\frac{\diff r_b^{-1}}{r_a}-\frac{\diff r_a^{-1}}{r_b}\right)\,,
\end{equation}
where 
\begin{equation}
C_{ab}=h_a \mm_b- h_b \mm_a+ \frac{3}{2}\left(\kk_a \ell_b-\kk_b \ell_a\right)\, .
\end{equation}
We can solve \eqref{eq:omega_*} for each pair of centers labelled by $a,b$ separately, and then sum up all the contributions. The result is
\begin{equation}
\ii\breve\omega_{*}=\left[c_{\omega}- \sum_a\sum_{b>a}\frac{C_{ab}}{\delta_{ab}}\frac{r_a+\delta_{ab}\cos\theta_a}{r_b} \right]\diff \phi \,,
\end{equation}
where $c_{\omega}$ is an arbitrary integration constant and $\delta_{ab}\equiv z_a-z_b$ (emphasis is put on the fact that there is no absolute value). Now we note that
\begin{equation}
\begin{aligned}
\frac{r_a+\delta_{ab}\cos\theta_a}{r_b} \,=\,&\left(1+\cos\theta_a\right)\frac{r_a+\delta_{ab}}{r_b}-\frac{\delta_{ab}+r_a\cos\theta_a}{r_b}\\[1mm]
\,=\,&\left(1+\cos\theta_a\right)\left(-1+\frac{r_a+\delta_{ab}}{r_b}\right)+\cos\theta_a-\cos\theta_b +1\, .
\end{aligned}
\end{equation}
Hence, fixing $c_{\omega}=\sum_a\sum_{b>a}\frac{C_{ab}}{\delta_{ab}}$, we have
\begin{equation}
\label{eq:sol_omega_*}
\ii \breve\omega_{*}=\sum_{a}\sum_{b>a}\frac{C_{ab}}{\delta_{ab}}\left(\cos\theta_b-\cos\theta_a\right)\diff \phi+ \ii\breve \omega_{\rm reg}\,,
\end{equation}
where 
\begin{equation}
\label{eq:omega_reg}
\ii {\breve \omega}_{\rm reg}\,=\,\sum_{a}\sum_{b>a}\frac{C_{ab}}{\delta_{ab}}\left(1+\cos\theta_a\right)\left(1-\frac{r_a+\delta_{ab}}{r_b}\right)\diff \phi\, .
\end{equation}
One can check that 
$\breve \omega_{\rm reg}$ is regular in the entire $z$-axis, being this statement independent of the sign of $\delta_{ab}$. Finally, we want to re-write the first term in \eqref{eq:sol_omega_*} in the form $\sum_{a}{\rm coeff}_a\cos\theta_a \diff \phi$, so that it combines with the first term in \eqref{eq:def_omega_*}. To this aim, we assume that the centers are ordered in a way such that $\delta_{ab}>0$ if $b>a$. Then, 
\begin{equation}
\begin{aligned}
&\sum_{a}\sum_{b>a}\frac{C_{ab}}{\delta_{ab}}\left(\cos\theta_b-\cos\theta_a\right)\,=\,   \sum_a\sum_{b<a} \frac{C_{ba}}{\delta_{ba}} \cos\theta_a - \sum_a\sum_{b>a} \frac{C_{ab}}{\delta_{ab}} \cos\theta_a  \\[1mm]
&\,=\, -\sum_{a} \sum_{b\neq a}\frac{C_{ab}}{|\delta_{ab}|} \cos\theta_a \,,
\end{aligned}
\end{equation}
where in the first equality we have exchanged the labels $a,b$ in the first term, while in the second equality we have used that $C_{ba}=-C_{ab}$ and that $\delta_{ab} = |\delta_{ab}|$ if $b>a$ and $\delta_{ba} = - \delta_{ab} =  |\delta_{ab}|$ if $b<a$.
 All in all, the general expression for $\breve \omega$ is
\begin{equation}
\ii\breve \omega=\sum_a \ii\ww_a \cos\theta_a \diff \phi +\sum_{a}\sum_{b>a}\frac{C_{ab}}{\delta_{ab}}\left(1+\cos\theta_a\right)\left(1-\frac{r_a+\delta_{ab}}{r_b}\right)\diff \phi\,, 
\end{equation}
where 
\begin{equation}
\ii\ww_{a}=h_0\mm_a-\mm_0 h_a+\frac{3}{2}\left(\kk_0 \ell_a-\ell_0\kk_a\right)-\sum_{b\neq a}\frac{C_{ab}}{|\delta_{ab}|}\, .
\end{equation}
Note that because of the assumption $\delta_{ab}>0$ if $a<b$, only the absolute value of $\delta_{ab}$ appears in \eqref{eq:omega_final}.

$\,$
\vskip0.1cm


\section{Absence of Dirac-Misner strings in two-center solutions}
\label{sec:DiracMisner}

The aim of this appendix is to show that the regularity conditions \eqref{eq:DiracMisner} automatically imply the absence of Dirac-Misner singularities. The latter arise as a consequence of the fact that the 1-forms $\chi$ and $\ii \breve \omega$, given respectively in \eqref{eq:chi} and \eqref{eq:breve_omega}, are not well defined along the $z$-axis, where the coordinate $\phi$ is not defined. It is convenient to distinguish the following three regions: \textbf{I} ($z>\delta/2$), \textbf{II} ($-\delta/2<z<\delta/2$) and \textbf{III} ($z<-\delta/2$). Our analysis focuses on the saddles studied in \ref{sec:gen_saddles}, which satisfy $w_N=-w_S$. Hence, 1-form $\ii \breve \omega$ is well defined in \textbf{I} and \textbf{III}, but not in \textbf{II}. In turn, $\chi$ is ill-defined in the entire $z$-axis. This can be fixed by a coordinate change of $\tau$ and $\psi$. However, since the coordinate change is different in each region, one has to check that they are compatible in the overlaps. We shall see that this is automatically guaranteed once \eqref{eq:DiracMisner} is assumed. The coordinate transformations in each of the regions are the following.
\begin{itemize}
\item \textbf{Region I ($z>\delta/2$).}  A coordinate transformation that renders the 1-form $\diff \psi +\chi$ regular is
\begin{equation}
 \left(\tau^I, \psi^I\right)=\left(\tau, \psi+\phi +\ii \left(\Omega_+-\Omega_-\right)\tau\right)\, .
\end{equation}
Explicitly, one has
\begin{equation}
\diff{\psi}+\chi = \diff \psi^I-\ii \left(\Omega_+-\Omega_-\right)\diff\tau^I+ \chi-\diff \phi\, ,
\end{equation}
which is now regular in \textbf{I}. The condition $\beta \Omega_{+}=2\pi \ii$ implies that the coordinates $(\tau^I, \psi^I, \phi)$ have untwisted identifications, inherited from \eqref{eq:twisted_bdry_conditions}:
\begin{equation}
\label{eq:untwisted_id_I}
\left(\tau^I, \psi^I, \phi\right)\sim \left(\tau^I+\beta, \psi^I, \phi\right)\sim \left(\tau^I, \psi^I+4\pi, \phi\right)\sim \left(\tau^I, \psi^I, \phi+2\pi\right).
\end{equation}
\item \textbf {Region II ($-\delta/2<z<\delta/2$).} A coordinate transformation rendering regular both $\diff \tau + \ii {\breve \omega}$ and $\diff \psi +\chi$ in this region is 
\begin{equation}
\begin{aligned}
\left(\tau^{II}, \psi^{II}\right)=&\left(\tau-2\ii w_{N}\phi, \psi+\phi+\frac{h_S-h_N-1}{2\ii w_N}\tau\right)\\
=&\left(\tau-\frac{\beta}{2\pi} \phi, \psi+\phi+\ii \left(\Omega_+-\Omega_-\right)\tau\right)\, ,
\end{aligned}
\end{equation}
where in the second line we have made use of \eqref{eq:DiracMisner}. Written in this way, it is straightforward to check that $(\tau^{II}, \psi^{II}, \phi)$ satisfy the same untwisted identifications of region \textbf{I}, \eqref{eq:untwisted_id_I}. In the overlap \textbf{I-II}, one has
\begin{equation}
\tau^{II}=\tau^{I}-\frac{\beta}{2\pi}\phi\,, \hspace{1cm} \psi^{II}=\psi^{I} \,,
\end{equation}
which is consistent with the global identifications of the coordinates.

\item \textbf {Region III ($z<-\delta/2$).} Finally, the coordinate transformation needed in this region is 
\begin{equation}
 \left(\tau^{III}, \psi^{III}\right)=\left(\tau, \psi-\phi -\ii \left(\Omega_++\Omega_-\right)\tau\right).
\end{equation}
As in region \textbf{I}, the condition $\beta\Omega_{+}=2\pi \ii$ implies that $(\tau^{III}, \psi^{III}, \phi)$ satisfy the untwisted identifications of \eqref{eq:untwisted_id_I}. In the overlap \textbf{II-III}, we have 
\begin{equation}
\tau^{II}=\tau^{III}-\frac{\beta}{2\pi}\phi\,, \hspace{1cm} \psi^{II}=\psi^{III}+2\phi -\frac{4\pi}{\beta}\tau^{III} \,,
\end{equation}
which is again consistent with the global identifications of the coordinates.
\end{itemize}


\section{Lens spaces}
\label{sec:lens_space}

We recall here the definition of a Lens space and highlight some properties that are needed in the main text. 

Given a unit $S^3$ parameterized by $(Z_1,Z_2)\in\mathbb{C}^2$, with 
$|Z_1|^2+|Z_2|^2=1$, and given coprime integers $(\mathtt p,\mathtt q)$, the Lens space $L(\mathtt p,\mathtt q)\simeq S^3/\mathbb{Z}_{\mathtt p}$ is defined by the orbifold identification
\be
(Z_1,Z_2)\,\sim\, \left( \rme^{\frac{2\pi \ii}{\mathtt p}} Z_1\,,\,\rme^{\frac{-2\pi \ii \mathtt q}{\mathtt p}} Z_2 \right)\,.
\ee
Since $\mathtt p$ and $\mathtt q$ are coprime, the orbifold is freely acting and the resulting space is a smooth manifold. 

We can take $Z_1=\sin\vartheta \,\rm e^{\ii \phi_1}$, $Z_2=\cos\vartheta \,\rm e^{\ii \phi_2}$, with $\vartheta\in [0,\frac{\pi}{2}]$ and $\phi_1,\phi_2$ satisfying the identifications
\be
(\phi_1,\phi_2) \,\sim \, (\phi_1+2\pi,\phi_2) \,\sim \, (\phi_1,\phi_2+2\pi)\,.
\ee
These are the angular coordinates of footnote~\ref{foot:phi1phi2basis}.
 The orbits of $\partial_{\phi_1}$ and $\partial_{\phi_2}$ are $2\pi$-periodic and smoothly collapse to zero size at either endpoint of the $\vartheta$ interval.
The orbifold introduces the new identification
\be\label{identif_pq}
(\phi_1,\phi_2) \,\sim \, \left(\phi_1+\frac{2\pi}{\mathtt p}\,,\,\phi_2-\frac{2\pi \mathtt q}{\mathtt p}\right)\,,
\ee
hence the nowhere vanishing vector 
\be
\lambda \,=\, \frac{1}{\mathtt p}\, \partial_{\phi_1} - \frac{\mathtt q}{\mathtt p}\,\partial_{\phi_2}
\ee
also has $2\pi$-periodic orbits. It describes an $S^1$ fibration over a two-dimensional base which in general is a spindle. The same expression can be written as
\be
 \partial_{\phi_1}  \,=\, \mathtt q\,\partial_{\phi_2} + \mathtt p\,\lambda \,,
\ee
showing that going once around the circle that smoothly collapses at $\vartheta=0$ is the same as going $\mathtt q$ times around the circle that smoothly collapses at $\vartheta=\frac{\pi}{2}$ and $\mathtt p$ times around the non-vanishing fibre.

In the main text, we deal with orbifold identifications of the type
\be\label{identif_pq1q2}
(\phi_1,\phi_2) \,\sim \, \left(\phi_1+\frac{2\pi \mathtt q_1}{\mathtt p}\,,\,\phi_2-\frac{2\pi \mathtt q_2}{\mathtt p}\right)\,,
\ee
where both $\mathtt q_1$, $\mathtt q_2$ are coprime to $\mathtt p$. We thus have a $2\pi$-periodic nowhere-vanishing vector $\tilde \lambda$ such that 
\be\label{vectors_pq1q2}
\mathtt q_1\, \partial_{\phi_1}  \,=\, \mathtt q_2\,\partial_{\phi_2} +\mathtt  p\,\tilde\lambda \,.
\ee
It is not hard to see that this orbifold in fact defines the lens space $L(\mathtt p,\mathtt a \,\mathtt q_2)$, where $\mathtt a$ is an integer determined by the equation 
\be
\mathtt a\, \mathtt q_1 + \mathtt b\, \mathtt p \,=\, 1\,, \qquad \mathtt a,\mathtt b\in\,\mathbb{Z}\,,
\ee
which can always be solved for coprime  $\mathtt q_1,\mathtt p$ by the B\'ezout lemma. The equivalence is seen as follows:
 applying $\mathtt a$ times the identification \eqref{identif_pq1q2} and using $\frac{\mathtt a \,\mathtt q_1}{\mathtt p} = \frac{1}{\mathtt p}- \mathtt b$, we land on \eqref{identif_pq} with $\mathtt q=\mathtt a\, \mathtt q_2$; conversely, applying $\mathtt q_1$ times the identification \eqref{identif_pq}  with $\mathtt q=\mathtt a\,\mathtt q_2$, and using $\mathtt q_1 \frac{\mathtt a\, \mathtt q_2}{\mathtt p} =  \frac{\mathtt q_2}{\mathtt p}- \mathtt b\,\mathtt q_2$, we arrive at \eqref{identif_pq1q2}.

Starting from \eqref{vectors_pq1q2}, it is straightforward to see that the vector $\lambda$ specifying the lens space circle fibration is given by
\be
\lambda \,=\, \frac{1}{\mathtt q_1}\, \tilde \lambda + \frac{\mathtt b\, \mathtt q_2}{\mathtt q_1}\, \partial_{\phi_2}\,.
\ee


\chapter{Localization of the $4D$ supergravity action}
\label{app:4Dexample}


Equivariant localization of the on-shell action of four-dimensional (gauged) supergravity has been studied in a series of papers~\cite{BenettiGenolini:2023kxp,BenettiGenolini:2024kyy,BenettiGenolini:2024xeo,BenettiGenolini:2024hyd,BenettiGenolini:2024lbj,Crisafio:2024fyc}. In this appendix, we provide a non-supersymmetric discussion in the ungauged case, apply it to the Euclidean Kerr-Newman solution and take its supersymmetric limit at the end. The results are used in section~\ref{sec:exampleaction4d}, where the relation between the five-dimensional localization discussed in chapter~\ref{chap:Black_hole} and the four-dimensional version discussed in previous literature is investigated.

\section{Einstein-Maxwell on-shell action as an equivariant integral}
\label{sec:4Dexample}

We consider Einstein-Maxwell theory in four dimensions, which is also the bosonic sector of pure $\mathcal{N}=2$ supergravity, and show that the Euclidean on-shell action of solutions admitting a U(1) symmetry can be computed equivariantly. 

Using the trace of the Einstein equation, the bulk contribution to the on-shell action can be written as\footnote{We take a non-canonical normalization for the four-dimensional gauge field  in order to align with the conventions used in section~\ref{sec:exampleaction4d}.}
\begin{equation}
\label{eq:EM}
I_{\rm bulk}\,=\,-\frac{1}{16\pi G_4}\int_{{\cal M}} \left(R\star_4 1-\frac{2}{3}F\wedge\star_4 F\right)\,=\,\frac{1}{24\pi G_4}\int_{{\cal M}} F\wedge \star F\, ,
\end{equation}
which equals the integral on ${\cal M}$ of the following $\xi$-equivariantly-closed polyform
\begin{equation}
\label{eq:polyform_EM}
{\Psi}\,=\,F\wedge \star_4 F\,+\,\Bigl(-\left(\iota_{\xi}A+c\right)\star_4 F+\nu\, F\Bigr)\,+\,\Bigl(-\left(\iota_{\xi}A +c\right) \nu\Bigr)\,,\end{equation}
where $\xi$ is the Killing vector field generating the ${\rm U}(1)$ symmetry and  $\nu$ is the zero-form satisfying the differential equation
\begin{equation}
\diff{\nu}=\iota_{\xi}\star_4 F\,.
\end{equation}
This implies ${\cal L}_\xi \nu = 0$, and we have assumed a gauge such that ${\cal L}_{\xi}A=0$, so that ${\cal L}_\xi \Psi =0$.

We can thus evaluate $I_{\rm bulk}$ using the equivariant localization formula~\eqref{eq:BVAB_bdry}. 
The full on-shell action is
\begin{equation}
I = I_{\rm bulk} + I_{\rm GHY}\,,
\end{equation}
where, as in the main text, the GHY term is
\begin{equation}
I_{\rm GHY} = -\frac{1}{8\pi G_4}\int_{\partial {\cal M}} \diff^3 x\left( \sqrt{h}\, \mathcal K - \sqrt{h_{\rm bkg}}\,\mathcal K_{\rm bkg}\right)\,.
\end{equation}

We can then use the localization formula~\eqref{eq:BVAB_bdry} to compute the on-shell action of solutions with a U(1) symmetry. We illustrate this with the Kerr-Newman solution.

\subsection{On-shell action of Kerr-Newman} \label{sec:KNaction} 

The Kerr-Newman solution (with no magnetic charge) in Euclidean signature reads
\begin{eqnarray}
\label{eq:KN4dsol}
\diff s^2&=&\frac{\Delta_r}{r^2-\aa^2\cos^2 \theta}\left(\diff \tau+\aa \sin^2\theta\diff \phi\right)^2+ \left(r^2-\aa^2\cos^2 \theta\right)\left(\frac{\diff r^2}{\Delta_r}+\diff \theta^2\right)\quad\\[1mm]
&&+\,\frac{\sin^2\theta}{r^2-\aa^2\cos^2 \theta}\left(\left(r^2-\aa^2\right)\diff\phi-\aa\diff \tau\right)^2\,, \nonumber\\[1mm]
\ii A&=&\frac{\sqrt{3}\,q\,r}{r^2-\aa^2\cos^2 \theta}\left(\diff \tau+\aa\sin^2\theta \diff \phi\right)-\Phi\, \diff \tau\,,
\end{eqnarray}
where $\Phi$ is the electrostatic potential, $\Delta_r=\left(r-r_+\right)\left(r-r_{-}\right)$, and $r_{\pm}=m \pm \sqrt{m^2+\aa^2-q^2}$.
We demand $r_+^2>\aa^2$, so that the metric is positive-definite for $r>r_+$.
 The parameter $m$ is directly identified with the ADM mass, namely $E = m/G_4$, while the electric charge reads $Q=q/(\sqrt{3}G_4)$ and the angular momentum is $J=\ii \aa E$. The expressions for the inverse temperature $\beta$, the electrostatic potential $\Phi$ and the angular velocity $\Omega$ are
\begin{equation}
\beta=\frac{4\pi\left(r_+^2-\aa^2\right)}{r_+-r_-}\,, \hspace{1.5cm} \Phi= \frac{\sqrt{3}\,q\,r_+}{r_+^2-\aa^2}\,, \hspace{1.5cm} \Omega=\frac{\ii \,\aa}{r_+^2-\aa^2}\, . 
\end{equation}

We consider the Killing vector,
\begin{equation}\label{eq:localizing_vector_4d}
\xi=\partial_{\tau}\, ,
\end{equation}
which has a nut at the north pole of the horizon ($r=r_+, \theta=0$) and an anti-nut at the south pole ($r=r_+, \theta=\pi$)~\cite{Gibbons:1979xm}. The equivariant parameters are given by
\begin{equation}
\epsilon^{\pm}_{1}=\frac{2\pi}{\beta}\,, \hspace{1cm} \epsilon^{\pm}_{2}=\pm \ii \Omega\, ,
\end{equation}
where the $\pm$ refers to the nut ($+$) and anti-nut ($-$).  All the information required to compute \eqref{eq:polyform_EM} is encoded in $\iota_\xi A$ and $\nu$. An explicit computation yields
\begin{equation}\label{eq:iotaAandnu4d}
\iota_{\xi}A \,=\,-\frac{\ii \sqrt{3} \,q\, r}{r^2-\aa^2\cos^2\theta}+\ii \Phi\,, \hspace{1.5cm}
\nu\,=\,-\frac{\ii\sqrt{3}\, \aa \,q\cos\theta}{r^2-\aa^2 \cos^2\theta}\, .
\end{equation}

We can now evaluate the action. From the boundary terms in~\eqref{eq:BVAB_bdry} we get:
\begin{equation}
-\frac{1}{24\pi G_4}\int \eta \wedge \Psi_{(2)}\,=\, \frac{\beta Q}{2}\left( -\Phi + \ii c\right) \, ,
\end{equation}
whereas the boundary term involving $\Psi_{(0)}$ vanishes. From the fixed points we obtain the contribution
\begin{equation}
\label{eq:nut_contr_KN}
\frac{(2\pi)^2}{24\pi G_4} \left(\frac{\Psi_{(0)}|_{+}}{\epsilon_1^+\, \epsilon_2^+}+\frac{\Psi_{(0)}|_{-}}{\epsilon_1^- \,\epsilon_2^-}\right)\,=\,  -\ii c\,\frac{\beta Q}{2}\, .
\end{equation}
Putting together the contributions to the equivariant integral, we obtain
\begin{equation}
\frac{1}{24\pi G_4}\int_{\cal M}\Psi\,=\, -\frac{\beta \Phi Q}{2}\, ,
\end{equation}
that is independent of $c$.
Finally, the contribution from the GHY term is
\begin{equation}
I_{\rm GHY}\,=\,\frac{\beta E}{2}\, .
\end{equation}
Thus, we arrive at the final expression for the on-shell action
\begin{equation}
I\,=\,\frac\beta2\left(E-\Phi Q\right)\, .
\end{equation}
We note that when fixing the constant $c$ as
\begin{equation}
\label{eq:c4d}
\ii c\,=\, \Phi - \frac{E}{Q}\, ,
\end{equation}
the contribution to the on-shell action coming from the boundary terms vanishes, and we are just left with the contribution \eqref{eq:nut_contr_KN} from the fixed points of the isometry:
\begin{equation}\label{eq:susyaction4dsum}
I\,=\,\frac{(2\pi)^2}{24\pi G_4}\left(\frac{\Psi_{(0)}|_{+}}{\epsilon^{+}_{1}\, \epsilon^{+}_ {2}}+\frac{\Psi_{(0)}|_{-}}{\epsilon^{-}_{1}\, \epsilon^{-}_ {2}}\right)\,=\,-\frac{\beta}{2}  \left(\Phi-E/Q\right)Q\, =\,\frac{\beta}{2}\left(E-\Phi Q\right)\, .
\end{equation}

We now consider the supersymmetric (non-extremal) limit. This amounts to taking
\begin{equation}
E=\sqrt{3}Q\, ,
\end{equation}
or equivalently $m=q$, as dictated by the supersymmetry algebra. In terms of the potentials, this corresponds to fixing $\beta\Omega = \pm 2\pi \ii$. 
The special value of $\crho$ given by \eqref{eq:c4d} is fixed to $\crho = -\ii\left( \Phi - \sqrt{3}\right)$, which depends on the electrostatic potential and thus on boundary conditions, only. We also note that the Killing vector \eqref{eq:localizing_vector_4d} used to localize coincides with the one arising as a bilinear of the Killing spinor.
The supersymmetric limit of the on-shell action \eqref{eq:susyaction4dsum}  reads in terms of the parameters,
\begin{equation}\label{eq:susyaction4d}
I \,=\, \frac{\pi q^2}{G_4} \,.
\end{equation}
It can be conveniently written in terms of the supersymmetric chemical potential $\varphi$, defined as
\begin{equation}
\varphi = \beta\big( \Phi - \sqrt{3}\,\big) = -6\pi G_4Q \,,
\end{equation}
which gives
\begin{equation}
I \,=\, \frac{\varphi^2}{12\pi G_4}\,.
\end{equation}
Using this expression, one can study the supersymmetric thermodynamics of the solution~\cite{Iliesiu:2021are,Hristov:2022pmo}.

\section{Correspondence between localization in 4 and 5D}\label{sec:exampleaction4d}

We now consider the uplift to five dimensions of  four-dimensional supersymmetric Euclidean black hole saddles~\cite{Whitt:1984wk,Yuille:1987vw,Boruch:2023gfn}, and compute their action equivariantly using the general results derived in chapter \ref{chap:Black_hole}. On the one hand, this will provide us with a class of five-dimensional ALF solutions, which displays a different asymptotic behaviour from the example considered in section~\ref{sec:exampleaction5d}. On the other hand, it will give us the opportunity to illustrate how equivariant localization in five dimensions matches localization in four dimensions upon dimensional reduction.
 In fact, an alternative approach to localization in odd dimensions for solutions possessing a  ${\rm U}(1)$ symmetry with no fixed points would consist of dimensionally reducing to an even-dimensional spacetime, and then applying localization there.\footnote{For a similar application of the localization theorem by dimensional reduction in the context of gauged supergravity see the recent work~\cite{BenettiGenolini:2025icr}.} 
 
\paragraph{5D $\to$ 4D reduction.}

We take the following reduction ansatz for the metric and gauge field of five-dimensional minimal supergravity \hbox{along a Killing direction~$\partial_\psi$:}
\begin{equation}\label{eq:ansatzdimred}
\begin{aligned}
\diff s^2\, &=\, \diff s^2_{(4D)} + \left( \diff\psi + \mathcal A\right)^2\,,\\
A\, &=\, A_{(4D)}\,.
\end{aligned}
\end{equation}
Note that this is not the most general ansatz we could consider: while the generic ${\rm U}(1)$ reduction of five-dimensional minimal supergravity yields four-dimensional $\mathcal N=2$ supergravity coupled to a vector multiplet, here we want to focus on a reduction to pure $\mathcal{N}=2$ four-dimensional supergravity, truncating away the vector multiplet. As far as the bosonic sector is concerned, the latter truncation consists of setting to zero two four-dimensional  scalars, the dilaton from the metric and an axion  from the gauge field, as well as imposing the following duality relation between the gauge fields appearing in~\eqref{eq:ansatzdimred}:\footnote{In general in this section we denote four-dimensional quantities with the subscript $(4D)$. However, since the four-dimensional gauge field turns out to be equal to the five-dimensional one, 
 we will drop the subscript in $A_{(4D)}$ from now on.}
\begin{equation}\label{eq:dualization}
 \mathcal F = \frac{\ii}{\sqrt{3}}\star_4 F\,\,,
\end{equation}
where
${\cal F}=\diff {\cal A}$, $F = \diff A$, and  $\star_4$ is the Hodge-star built using the four-dimensional Euclidean metric. Substituting \eqref{eq:ansatzdimred}, \eqref{eq:dualization} into the five-dimensional field equations, we are left with four-dimensional field equations which are derived from the (Euclidean) pure $\mathcal{N}=2$ supergravity action,
\begin{equation}
I_{(4D)} \,=\, -\frac{\Delta_\psi}{16\pi}\int\left( R_{(4D)} \,\star_41 - \frac{2}{3}\,F\wedge \star_4 F\right) + I_{\rm GHY}\,,
\end{equation}
where $\Delta_\psi$ is the length of the reduction circle.\footnote{This factor is usually reabsorbed in the definition of Newton's gravitational constant of the reduced theory. However, since we chose to work in conventions such that $G_5=1$, we have $G_4= 1/\Delta_\psi$.}
 Note that, due to the need of dualizing away the KK gauge field $\mathcal A$ through~\eqref{eq:dualization}, substituting the ansatz \eqref{eq:ansatzdimred} into the five-dimensional action \eqref{eq:actionminimal5d} does not directly yield the correct four-dimensional action $I_{(4D)}$ of the consistent truncation. In order to address this, one should add to the reduced action  obtained by substituting \eqref{eq:ansatzdimred} a term proportional to $\int F \wedge  \mathcal{F}$, implementing the dualization upon varying with respect to $\mathcal{F}$.  More precisely, the actions are related as
\begin{equation}
I - \frac{\ii \Delta_\psi}{16\sqrt{3}\pi}\int F\wedge  \mathcal F\,=\, I_{(4D)} \,.
\end{equation}
When comparing the Euclidean on-shell actions in five and four dimensions, as we are going to do, it is important to include this dualization term.
Using Einstein's equations and \eqref{eq:dualization}, we can express the relation between the on-shell actions as
\begin{equation}\label{eq:action5dto4d}
I \,=\, \frac{I_{(4D)}}{2}\,.
\end{equation}

\paragraph{4D supersymmetric solutions.} Following~\cite{Gauntlett:2002nw}, a five-dimensional supersymmetric solution of the form \eqref{eq:ansatzdimred} is obtained by considering a solution with a Gibbons-Hawking base-space such that
\begin{equation}\label{eq:LandMIWP}
L = H\,,\quad\quad M = -\frac{1}{2}K\,.
\end{equation} 
Then, the four-dimensional metric and gauge field can be expressed as
\begin{equation}\label{eq:Eucl4d1}
\begin{aligned}
\diff s^2_{(4D)} \,=&\,\, H^{-1}f\left( \diff\tau + \ii \breve\omega\right)^2 + Hf^{-1} \diff s^2_{\mathbb R^3} \,,\\[1mm]
\ii A \,=\,&\, \sqrt{3}\left(-f\left(\diff\tau + \ii \breve\omega\right) + \ii\breve A + \alpha^{(\tau)} \diff\tau \right)\,,
\end{aligned}
\end{equation}
where, according to \eqref{eq:defchi}, \eqref{eq:susysolminimalghbase2} and \eqref{eq:susysolminimalghbase3}, one finds that
\be\label{eq:Eucl4d2}
Hf^{-1}\,=\, K^2 + H^2\,\,,\quad\quad
\star_3\, \diff \breve\omega \,=\, 2\left( K \diff H - H \diff K\right)\,\,,\quad\quad \star_3\, \diff\breve A \,=\, - \diff K\,.
\ee

In Lorentzian signature, the supersymmetric solutions just described precisely correspond to the class of timelike supersymmetric solutions of four-dimensional pure $\mathcal{N}=2$ supergravity~\cite{Gauntlett:2002nw}.
 As first shown in~\cite{TOD1983241}, such four-dimensional solutions can be expressed in the form of the Israel-Wilson-Perjes (IWP) solution~\cite{Perjes:1971gv,Israel:1972vx} and are controlled by one complex function. 
 In the same way, the Euclidean solutions \eqref{eq:Eucl4d1}, \eqref{eq:Eucl4d2} match the Euclidean version of the timelike supersymmetric four-dimensional solutions, first discussed in \cite{Whitt:1984wk,Yuille:1987vw}. These solutions are determined by two independent harmonic functions, $\mathcal V$ and $\widetilde{\mathcal V}$. While $\mathcal V$, $\widetilde{\mathcal V}$ can a priori be complex \cite{Whitt:1984wk}, the metric is real only if both are taken real.  The agreement with \eqref{eq:Eucl4d1}, \eqref{eq:Eucl4d2} follows from identifying
\be\label{eq:mapV_H_K}
\mathcal V \,=\, H - \ii K\,,\quad\quad \widetilde{\mathcal V}\,=\, H + \ii K\,.
\ee
Recall that our reality conditions are such that $H$ is real while $K$ is purely imaginary, hence $\mathcal{V}$ and $\widetilde{\mathcal{V}}$ are real and independent.
Then the four-dimensional metric and gauge field can be written as
\begin{equation}
\begin{aligned}
\label{eq:IWPmetric}
\diff s_{(4D)}^2\,&=\, \frac{1}{\mathcal V\widetilde{\mathcal V}}\left( \diff \tau + \ii\breve\omega\right)^2 + \mathcal V\widetilde{\mathcal V}\,\delta_{ij}\, \diff x^i \diff x^j\,,\\[1mm]
\ii A \,&=\, \sqrt{3}\,\bigg(-\frac{\mathcal V + \widetilde{\mathcal V}}{2\mathcal V\widetilde{\mathcal V}}\left( \diff \tau + \ii \breve\omega\right) +\ii \breve A   +\zeta\, \diff\tau\bigg)\,, 
\end{aligned}
\end{equation}
where the local one-forms $\breve\omega$, $\breve A$ are expressed in terms of $\mathcal V$ and $\widetilde{\mathcal V}$ as
\begin{equation}
\begin{aligned}
\star_3\, \diff \breve\omega= -\ii\,\big(\,\widetilde{\mathcal V}\,\diff \mathcal V-\mathcal V\, \diff \widetilde{\mathcal V}\,\big) \,\,,\qquad\qquad 
\star_3\,\diff\breve A = -\frac{\ii}{2}\,\diff\big(\,\mathcal V - \widetilde{\mathcal V}\,\big)  \,.
\end{aligned}
\end{equation}

\paragraph{The 4D Euclidean saddle and its action.} A smooth and non-extremal Euclidean black hole saddle of the four-dimensional supergravity theory is obtained by choosing the harmonic functions as~\cite{Whitt:1984wk,Yuille:1987vw,Boruch:2023gfn}
\begin{equation}\label{eq:choiceVtildeV}
\mathcal V = 1 + \frac{q}{r_N}\,\,,\qquad\qquad \widetilde{\mathcal V} = 1 + \frac{q}{r_S}\,.
\end{equation}
Then, smoothness of the metric and the gauge field requires that the period of the compactified Euclidean time $\beta$ and the electrostatic potential $\Phi$ read
\begin{equation}\label{eq:beta_Phi_4D}
\beta = \frac{4\pi q\left( q+\delta\right)}{\delta}\,,\qquad\qquad\Phi=\frac{\sqrt{3}\left(2q+\delta\right)}{2\left(q+ \delta\right)}\,.
\end{equation}

As  verified in~\cite{Whitt:1984wk,Boruch:2023gfn,Hegde:2023jmp}, there is a change of coordinates that maps the Euclidean IWP solution \eqref{eq:IWPmetric}--\eqref{eq:choiceVtildeV} to the supersymmetric non-extremal limit of the Euclidean Kerr-Newman solution.

The four-dimensional on-shell action is given by a sum of a nut and an anti-nut contribution~\cite{Whitt:1984wk}. In appendix~\ref{sec:4Dexample}, we computed the action of the generic electrically-charged Euclidean Kerr-Newman solution via equivariant localization, and recover the action of the solution above by taking the supersymmetric limit. 
There, we use the Killing vector $V = \partial_\tau$, which has two isolated fixed points at the poles of the horizon's two-sphere. The four-dimensional on-shell action  can then be written as the sum
\begin{equation}\label{eq:action4dsum}
I_{(4D)} = \frac{\pi\Delta_\psi}{6}\left(\frac{\Psi_{(0)}\big|_+}{\epsilon_1^+\,\epsilon_2^+}\,+\,\frac{\Psi_{(0)}\big|_-}{\epsilon_1^-\,\epsilon_2^-} \right)\,,
\end{equation}
where $\Psi_{(0)}$ is the zero-form component of the equivariantly-closed polyform whose top-form is the four-dimensional on-shell action, and the $\pm$ symbols denote the north and south pole of the horizon's two-sphere (see appendix~\ref{sec:4Dexample} for details). 

\paragraph{Comparison with 5D computation.} In order to compare equivariant localization in five and four dimensions, we now compute the same four-dimensional action $I_{(4D)}$ using the five-dimensional uplifted   solution. In section~\ref{sec:actionnut} we derived the expression for the five-dimensional on-shell action for two-center solutions as a sum of two terms denoted as $\mathcal I_N$ and $\mathcal I_S$, with their expression being given in \eqref{eq:gravblocks}. We are going to show that these map, through the relation \eqref{eq:action5dto4d}, into the four-dimensional fixed-point contributions in \eqref{eq:action4dsum} as 
\begin{equation}\label{eq:map5d4dblocks}
\mathcal I_N \,=\, \frac{\pi\Delta_\psi}{12}\,\frac{\Psi_{(0)}\big|_+}{\epsilon_1^+\,\epsilon_2^+}\,\,,\qquad\qquad \mathcal I_S \,=\, \frac{\pi\Delta_\psi}{12}\,\frac{\Psi_{(0)}\big|_-}{\epsilon_1^-\,\epsilon_2^-}\,.
\end{equation} 

The five-dimensional solution is of the form \eqref{eq:ansatzdimred}, \eqref{eq:LandMIWP}--\eqref{eq:Eucl4d2}. From~\eqref{eq:mapV_H_K}, \eqref{eq:choiceVtildeV}, we obtain the harmonic functions
\begin{equation}\label{eq:HandKIWP}
H =\, 1 + \frac{q}{2}\left(\frac{1}{r_N}+ \frac{1}{r_S}\right)\,,\qquad\qquad \ii K = -\frac{q}{2}\left( \frac{1}{r_N}-\frac{1}{r_S}\right)\,.
\end{equation}
While the four-dimensional metric is asymptotically $S^1_\beta\times \mathbb{R}^3$, the five-dimensional metric asymptotically for $r\to\infty$ reads
\begin{equation}
\diff s^2 \ \to\ \diff\tau^2 + \diff r^2 + r^2 \left( \diff\theta^2 + \sin^2\theta \diff\phi^2\right)+ \left( \diff\psi + h_+ \cos\theta \diff\phi\right)^2\,,
\end{equation}
where one can take the period of $\psi$ as $\Delta_\psi = 4\pi h_+$ so that the hypersurfaces at fixed $\tau$ and $r$ topologically are three-spheres. Thus, this is an example of solution with ALF Gibbons-Hawking base-space, which is different from the one considered in section~\ref{sec:exampleaction5d}.

The expressions \eqref{eq:beta_Phi_4D} for the temperature and the electrostatic potential are consistent with \eqref{eq:inversehawkingtemperaturesusy} and \eqref{eq:gaugechoice}, meaning that a smooth five-dimensional Euclidean saddle corresponds to a smooth four-dimensional one. 

Unlike what happens for the four-dimensional geometry, in the five-dimensional solution $V=\partial_\tau$ has no fixed points. However, we can combine it with the generator $\partial_\psi$ of the compactified direction to form a pair of Killing vectors with a one-dimensional nut, in accordance with the discussion of section~\ref{sec:onshellactionpreliminaries}. These two vectors were defined in \eqref{eq:kvsusy} and read
\begin{equation}
\xi_N= \partial_\tau + \frac{2\pi q}{\beta}\,\partial_\psi \,\,,\qquad\qquad \xi_S = \partial_\tau - \frac{2\pi q}{\beta}\,\partial_\psi \,.
\end{equation}

It is now straightforward to verify that \eqref{eq:map5d4dblocks} holds. Indeed, the equivariant parameters can be mapped as $\epsilon_{1,2}^N= \epsilon_{1,2}^+$ and $\epsilon_{1,2}^S= \epsilon_{1,2}^-$, while the parameter $c$ specifying the localization scheme takes the same expression both in five and four dimensions, $\ii c= \Phi - \sqrt{3}$. A less straightforward correspondence holds between the integral of the pullbacks of $\nu_V$ and $\nu_U$ on $\mathcal M_N$, computed in \eqref{eq:pullbacknususy}, and the value of its four-dimensional analog $\nu_{(4D)}$  (cf.\ \eqref{eq:iotaAandnu4d}) at the nut and anti-nut. One obtains
\begin{equation}
\int_{\mathcal M_N} \!\!\!\iota^*\nu_V = \Delta_\psi\,\nu_{(4D)}\Big|_+ = - 2\sqrt{3}\pi\ii \Delta_\psi\frac{q}{\beta}\,,\quad\quad\Omega_N\! \int_{\mathcal M_N} \!\!\!\iota^*\nu_{U} = -\Delta_\psi \,\nu_{(4D)}\Big|_- = - 2\sqrt{3}\pi\ii \Delta_\psi\frac{q}{\beta}\,.
\end{equation}
As a consequence, we have found that \eqref{eq:map5d4dblocks} holds, with
\begin{equation}
\mathcal I_N \, =\, \frac{\pi}{4}\,\Delta_\psi\, q^2\,=\,\mathcal I_S \,.
\end{equation}
The on-shell action can also be expressed in terms of the coefficients of the harmonic function as in eq.~\eqref{eq:onshellactionfinal},
\begin{equation}
I \,=\,\mathcal I_N + \mathcal I_S\,=\,  \frac{\pi}{2}\,\Delta_\psi\,q^2\,.
\end{equation}

We thus confirm that the two blocks we have identified in five dimensions match the contributions from the nut and anti-nut of the supersymmetric Killing vector in four dimensions.


\chapter{On-shell action for asymptotically flat three-center saddles}

\label{sec:action}

\section{Evaluation of the on-shell action}

Here we compute the on-shell action $I$ of the three-center solution studied in section~\ref{sec:3centersol}. This is given by the sum
\begin{equation}
\label{eq:actionintegral}
\begin{aligned}
I &\,=\, I_{\rm bulk} + I_{\rm GHY} \\
&\,=\, -\frac{1}{16\pi}\int_{\mathcal M}  \left(R\star_51 - \frac{1}{2}F\wedge \star_5 F + \frac{\ii}{3\sqrt{3}}\,F\wedge F \wedge A \right) \\[1mm]
&\ \,\quad  -\frac{1}{8\pi}\int_{\partial \mathcal M}\diff^4 x\left( \sqrt{h}\,\mathcal K - \sqrt{h_{\rm bkg}}\,\mathcal K_{\rm bkg}\right)\,,
\end{aligned}
\end{equation}
where $\partial\mathcal M$ is the cutoff boundary of the regulated spacetime at large but finite distance, $h$ denotes the determinant of the induced metric at such boundary and $\mathcal K$ its extrinsic curvature. To remove divergences, we perform a background subtraction using a reference flat background with the same asymptotics, characterized by $h_{\rm bkg}$ and $\mathcal K_{\rm bkg}$.

A subtle issue arises in the computation of the on-shell action due to the presence of the Chern-Simons term. Since $A$ is not a globally defined one-form, the integral $\int F\wedge F \wedge A$ is ill-defined and must be treated with care. A similar issue was encountered in early studies of black rings \cite{Hanaki:2007mb}, where it was noted that the Page charge, used to compute the electric charge and angular momentum, is not well-defined for the same reason. The way out of this problem involves introducing an appropriate number of patches and summing their contributions while including a compensating term along the patch overlaps.

To address this issue, we introduce the relevant patches. We define the open set 
\begin{equation}
\label{eq:patchB}
{\mathcal U}_{\BR} \equiv \left\{\left(\tau,\,\psi,\, \rho,\,\phi,\,z\right)\,:\,\, \frac{\rho^2}{\varepsilon^2} + \frac{(z- \frac{2\dd_1+\delta}{4})^2}{\left( \varepsilon + \frac{2\dd_1-\delta}{4}\right)^2} < 1\,,\quad \varepsilon >0\right\}\,,
\end{equation}
while ${\mathcal U}_{\cal H}$ is defined as the complement of ${\mathcal U}_{\BR}$. The patch ${\cal U}_{\cal H}$ coincides with the one defined in \eqref{eq:conditions_for_patch}; this covers the horizon rod and extends to the asymptotic region. Instead, ${\mathcal U}_{\BR}$ does not extend to the asymptotic region. The parameter $\varepsilon$, which controls the size of the region ${\mathcal U}_{\BR}$, is a positive small quantity, whose exact value should not affect the final result. Indeed, many expressions simplify in the limit $\varepsilon \rightarrow 0$, as we discuss below. 

The boundary of the closure of ${\mathcal U}_{\BR}$, denoted by $\partial {\mathcal U}_{\BR}$, is the surface parametrized by
\begin{equation}
\rho = \varepsilon \,\sin \theta\,,\qquad z =\frac{2\dd_1+\delta}{4}+ \left(\varepsilon + \frac{2\dd_1-\delta}{4}\right) \cos\theta\,.
\end{equation}
Taking the limit $\varepsilon \rightarrow 0$, this region collapses to $\rho = 0$, with $ z = \frac{2\dd_1\left( 1 + \cos\theta\right)+\delta\left( 1-\cos\theta\right)}{4} $. This shows that in this limit, the set ${\mathcal U}_{\BR}$ shrinks onto the bolt ${\mathcal B}_{\BR}$, leading to the simplifications we will exploit below.

The gauge field can be constructed patch-by-patch. The gauge fields in the two patches are related by the transformation
\begin{equation}
\label{eq:gaugemap}
A_{1} = A_{\cal H} - \ii \sqrt{3}\,\diff \left(\alpha_{1}-\alpha_{\cal H}\right)\,,
\end{equation}
where 
\begin{equation}
\label{eq:lambda}
\alpha_{1} = - \frac{\kk_1}{h_1}\left(\phi + \psi - \frac{4\pi h_N}{\beta } \tau \right) -\frac{4\pi\kk_N}{\beta}\tau\,, \hspace{1cm} \alpha_{\cal H}\,=\,\frac{4\pi\kk_S}{\beta}\tau\, ,
\end{equation}
as we showed in \eqref{eq:alpha_a_horizon} and \eqref{eq:alpha_1_3c}.
Having introduced the two patches, we now extend the approach of~\cite{Hanaki:2007mb} to properly define the integral of the Chern-Simons form in our setup. In order to define the integral properly, we start from a gauge invariant quantity. The natural quantity to consider is the six-form $F\wedge F \wedge F = \diff \left(F \wedge F\wedge A\right)$. To define an appropriate integration manifold, we assume the existence of a smooth six-dimensional extension of the spacetime, denoted by $\mathcal M_6$, such that its boundary coincides with the five-dimensional spacetime under consideration. This ensures that $\mathcal M$ is cobordant to a point. Additionally, we extend the two patches ${\mathcal U}_{\cal H}$ and ${\mathcal U}_{\BR}$ smoothly into $\mathcal M_6$, denoting these extensions by $\widehat {\mathcal U}_{\cal H}$ and $\widehat {\mathcal U}_{\BR}$. We can then write a well-defined integral as
\begin{equation}
\int_{\mathcal M_6} F\wedge F \wedge F = \int_{\widehat{\mathcal U}_{\cal H}\cap {\mathcal M_6}}  F\wedge F \wedge F + \int_{\widehat{\mathcal U}_{\BR}\cap {\mathcal M_6}} F\wedge F \wedge F\,.
\end{equation}
Integrating by parts gives
\begin{equation}
\int_{\mathcal M_6} F\wedge F \wedge F= \int_{{\mathcal U}_{\cal H}}F\wedge F \wedge A_{\cal H} + \int_{\mathcal U_{\BR}} F \wedge F \wedge A_{1} + \int_{\partial{\widehat{\mathcal  U}}_{\BR} \cap \mathcal M_6} \left(A_{\cal H} - A_{1} \right)\wedge F \wedge F\, .
\end{equation}
Using the gauge map \eqref{eq:gaugemap}, this expression simplifies to
\begin{equation}
\int_{\mathcal M_6} F\wedge F \wedge F= \int_{{\mathcal U}_{\cal H}}F\wedge F \wedge A_{\cal H} + \int_{{\mathcal U}_{\BR}} F \wedge F \wedge A_{1} - \ii\sqrt{3}\int_{\partial{\mathcal U}_{\BR}} \diff\left(\alpha_1 - \alpha_{\cal H}\right) \wedge A_1 \wedge F\,.
\end{equation}
Following~\cite{Hanaki:2007mb}, we take this sum of three terms as the formal definition of the Chern-Simons integral in five dimensions:
\begin{equation}
\int_{\mathcal M} F\wedge F\wedge A \,\equiv
\,\int_{\mathcal M_6} F\wedge F \wedge F\,.
\end{equation}

We now turn to the explicit evaluation of the on-shell action \eqref{eq:actionintegral}. Using the trace of Einstein equations, the bulk term simplifies to
\begin{equation}
\label{eq:onshellaction0}
I_{\rm bulk} = \frac{1}{48\pi}\int_{\mathcal U_{\cal H}}F\wedge G_{\cal H} + \frac{1}{48\pi}\int_{\mathcal U_{\BR}} F\wedge G_1 - \frac{1}{48\pi}\int_{\partial{\mathcal U}_{\BR}}\diff \left(\alpha_1-\alpha_{\cal H}\right) \wedge A_1 \wedge F\,,
\end{equation}
where $G \equiv \star_5 F - \frac{\ii}{\sqrt{3}}A \wedge F$.
The size of the patch ${\mathcal U}_{\BR}$ is controlled by the parameter $\varepsilon$ introduced in \eqref{eq:patchB}. In the limit $\varepsilon \rightarrow 0$, the volume of ${\mathcal U}_{\BR}$ collapses to zero. Consequently, any smooth five-form integrated over ${\mathcal U}_{\BR}$ must vanish. In particular,
\begin{equation}
\lim_{\varepsilon\rightarrow 0}\,\int_{{\mathcal U}_{\BR}} F\wedge G_1\, =\, 0.
\end{equation}
This follows from the fact that any smooth form in $\mathcal U_{\BR}$, such as $F\wedge G_1$, must have a component along the direction dual to $\xi_{\BR}$ that vanishes on $\mathcal B_{\BR}$, where $\xi_{\BR}$ contracts. 
We may integrate by parts the first term in \eqref{eq:onshellaction0} and reduce it to a boundary term by using the Maxwell equations,
\begin{equation}
\begin{aligned}
\int_{\mathcal U_{\cal H}}F\wedge G_{\cal H} =\int_{\partial\mathcal M} A_{\cal H} \wedge \star_5 F -\int_{\partial{\mathcal U}_{\BR}}A_1 \wedge \star_5 F -\ii\sqrt{3}\int_{\partial{\mathcal U}_{\BR}}\diff\left(\alpha_1-\alpha_{\cal H}\right)\wedge \star_5 F\,.
\end{aligned}
\end{equation}
For the same reason as above, we conclude that
\begin{equation}
\lim_{\varepsilon\rightarrow 0}\,\int_{\partial{\mathcal U}_{\BR}} A_{1}\wedge \star_5 F \,=\, 0\,.
\end{equation}
However, $\diff\left(\alpha_1-\alpha_{\cal H}\right)$ is not a smooth one-form in ${\mathcal U}_{\BR}$, in particular its norm diverges at the bolt ${\mathcal B}_{\BR}$. As a result, terms involving $\diff \alpha$ contribute additional non-trivial corrections to the on-shell action, which would be absent in the case of a globally defined gauge field. 
Taking this contribution into account, the bulk action simplifies to
\begin{equation}
I_{\rm bulk} = \frac{1}{48\pi}\int_{\partial \mathcal M} A_{\cal H} \wedge \star_5 F - \frac{\ii\sqrt{3}}{48\pi}\int_{\partial{\mathcal U}_{\BR}} \diff\left(\alpha_1-\alpha_{\cal H}\right)\wedge G_{\cal H} + \mathcal O(\varepsilon) \,.
\end{equation}
It is straightforward to show that the contribution from the asymptotic region of the spacetime yields
\begin{equation}
\frac{1}{48\pi}\int_{\partial \mathcal M} A_{\cal H} \wedge \star_5 F\,=\, -\frac{\beta\Phi}{3}Q\,.
\end{equation}
The second term can be computed explicitly in the $\varepsilon \rightarrow 0$ limit, giving
\begin{equation}
\begin{aligned}
-\frac{\ii\sqrt{3}}{48\pi}\lim_{\varepsilon \rightarrow 0} \int_{\partial{\mathcal U}_{\BR}}\diff \left(\alpha_1-\alpha_{\cal H}\right)\wedge G_1 &= -\frac{\ii\sqrt{3}}{24}\left[\iota_{\xi_{\BR}} \diff\left(\alpha_1-\alpha_{\cal H}\right)\right] \int_{\mathcal B_{\BR}} G_1 \,,\\
&= \frac{\Phi^{\BR}}{3}{\cal Q}^{\BR}\,,
\end{aligned}
\end{equation}
where to obtain the second line we used \eqref{eq:potentials_genaral_def}, \eqref{eq:lambda}, \eqref{eq:rel_k_varphi}, \eqref{eq:QD}.

We are now left to compute the boundary terms of \eqref{eq:actionintegral}, following the computation in section~\ref{sec:actionboundaryterms}:
\begin{equation}
I_{\rm GHY} \,=\, - \frac{1}{8\pi}\lim_{r\rightarrow + \infty}\int_{\partial\mathcal M} \diff^4 x\,\frac{r^2}{2}\,\partial_r f^{-1}\,\sin\theta \,=\, \frac{\beta}{\sqrt{3}}Q\,.
\end{equation}
Putting everything together, we finally obtain that the on-shell action is given by
\begin{equation}
\label{eq:onshellaction}
\begin{aligned}
I &= \frac{1}{3}\left(\Phi^{\BR}{\cal Q}^{\BR} - \varphi Q\right)\\
&=\frac{\pi}{12\sqrt{3}}\left[\frac{\varphi^3}{\omega_1\omega_2}-\frac{\left(p_1\varphi- \omega_2 \Phi^{\BR}\right)^3}{p_1^2\omega_2\left(p_1\omega_1 + \left(p_1-1\right)\omega_2\right)}\right]\,,
\end{aligned}
\end{equation}
which is consistent with \eqref{eq:I_gen}.


\section{Legendre transform}
\label{sec:Legendre}

The entropy \eqref{eq:BHentropy}, expressed as a microcanonical function of the conserved charges, is related to the on-shell action \eqref{eq:I_gen} by a Legendre transform. This transformation is performed with respect to the thermodynamic variables $\varphi$ and $\omega_{1,2}$, subject to the constraint $\omega_+ = 2\pi \ii$~\cite{Cabo-Bizet:2018ehj}. According to the quantum statistical relation \eqref{eq:QSR}, the Legendre transform is implemented by extremizing the following functional:
\begin{equation}
\mathcal S = {\rm ext}_{\{\varphi,\,\omega_1,\,\omega_2,\,\Lambda\}}\left[ -I - \varphi Q- \omega_1 J_1- \omega_2 J_2 -\Lambda\left( \omega_1 + \omega_2 - 2\pi \ii\right)\right]\,,
\end{equation}
where $\Lambda$ is a Lagrange multiplier enforcing the constraint on the angular velocities. Since $I$ is an homogeneous function of degree one with respect to $\varphi$, $\omega_{1,2}$, it follows that
\begin{equation}
\mathcal S = 2\pi \ii\Lambda\,.
\end{equation}
The extremization equations
\begin{equation}
Q = - \frac{\partial \,I}{\partial \varphi}\,,\qquad\quad J_{1,2} = -\frac{\partial \,I}{\partial\omega_{1,2}} - \Lambda \,,
\end{equation}
yield a quadratic formula for $\Lambda$,
\begin{equation}
P_0 + P_1 \Lambda + \Lambda^2=0\,,
\end{equation}
where the coefficients are given by
\begin{equation}
\label{eq:p01}
\begin{aligned}
P_1 &= \left( 1-p_1\right) J_1 + \left( 1+p_1\right) J_2 + \Phi^{\BR} \left( Q - \frac{\pi}{12\sqrt{3}}\frac{\left(\Phi^{\BR}\right)^2}{p_1^2}\right)
\,,\\[1mm]
P_0 &= \frac{4\left(1-p_1\right)}{\pi}\left(\frac{Q}{\sqrt{3}}+ \frac{\pi}{12p_1\left( 1-p_1\right)}\left(\Phi^{\BR}\right)^2 \right)^3- \left( J_2 - \frac{\pi}{12\sqrt{3}\,p_1^2\left( 1-p_1\right)}\left(\Phi^{\BR}\right)^3\right)^2\\[1mm]
& + \left( J_2 - \frac{\pi}{12\sqrt{3}\,p_1^2\left( 1-p_1\right)}\left(\Phi^{\BR}\right)^3\right) P_1\,.
\end{aligned}
\end{equation}
Solving for $\Lambda$,
\begin{equation}
\label{eq:lagrangemultiplier}
\begin{aligned}
\Lambda &= -\frac{\ii}{2}\sqrt{4P_0 - P_1^2 } - \frac{P_1}{2}\,,
\end{aligned}
\end{equation}
we find the entropy
\begin{equation}
\label{eq:generalentropy0}
\begin{aligned}
&\mathcal S =\\
&4\sqrt{\pi} \sqrt{\left( 1-p_1\right)\left(\frac{Q}{\sqrt{3}}+ \frac{\pi\,\left(\Phi^{\BR}\right)^2}{12p_1\left( 1-p_1\right)} \right)^3-\frac{\pi}{4} \left(\left( p_1-1 \right) J_- - \frac{\Phi^{\BR}}{2} Q - \frac{\pi\left( 1+p_1\right)\left(\Phi^{\BR}\right)^3}{24\sqrt{3}\,p_1^2\left( 1-p_1\right)}\right)^2 }\\[1mm]
&-2\pi\ii \left[ J_+-p_1 J_- + \frac{\Phi^{\BR}}{2} \left( Q - \frac{\pi}{12\sqrt{3}}\frac{\left(\Phi^{\BR}\right)^2}{p_1^2}\right)\right]\,.
\end{aligned}
\end{equation}
Using the explicit expression for the charges \eqref{eq:3centercharges}, one can verify that this expression agrees with the Bekenstein-Hawking entropy \eqref{eq:BHentropy}.

Finally, the electric flux \eqref{eq:QD} can be determined by differentiating the action with respect to $\Phi^{\BR}$:
\begin{equation}
\label{eq:blackholelimit}
{\cal Q}^{\BR} \,=\, \frac{\partial I}{\partial \Phi^{\BR}}\,=\, -\frac{\partial\mathcal S}{\partial\Phi^{\BR}}\,=\, \frac{\pi}{4\sqrt{3}}\frac{\left(p_1\,\varphi - \Phi^{\BR}\omega_2 \right)^2}{p_1^2(p_1\omega_1 + \left(p_1-1\right)\omega_2)}\,.
\end{equation}
